# STATISTICAL CONSEQUENCES OF FAT TAILS

Real World Preasymptotics, Epistemology, and Applications

Papers and Commentary

Third Edition

NASSIM NICHOLAS TALEB

This format is based on André Miede's *ClassicThesis*, with adaptation from Lorenzo Pantieri's *Ars Classica*. With immense gratitude to André and Lorenzo.

**This black and white economy version** is designed for print-on-demand aiming to satisfy the needs of the student population. Some graphs may not be as clear as the colored rendering but this simplification should not carry any practical consequence on the understanding of the material.

# COAUTHORS [1]

Yaneer Bar-Yam (Chapter 12)
Pasquale Cirillo (Chapters 15, 17, and 18 )
Raphael Douady (Chapter 16)
Andrea Fontanari (Chapter 15)
Hélyette Geman ( Chapter 30)
Donald Geman (Chapter 30)
Espen Haug (Chapter 26 )
The Universa Investments team (Chapter 27 and 29)
The Uncertainty Reading Group (Chapter 13)

[1] Papers relied upon here are [52, 54, 55, 56, 57, 113, 124, 147, 177, 194, 270, 273, 274, 275, 277, 278, 279, 280, 289, 293, 294, 295, 296]

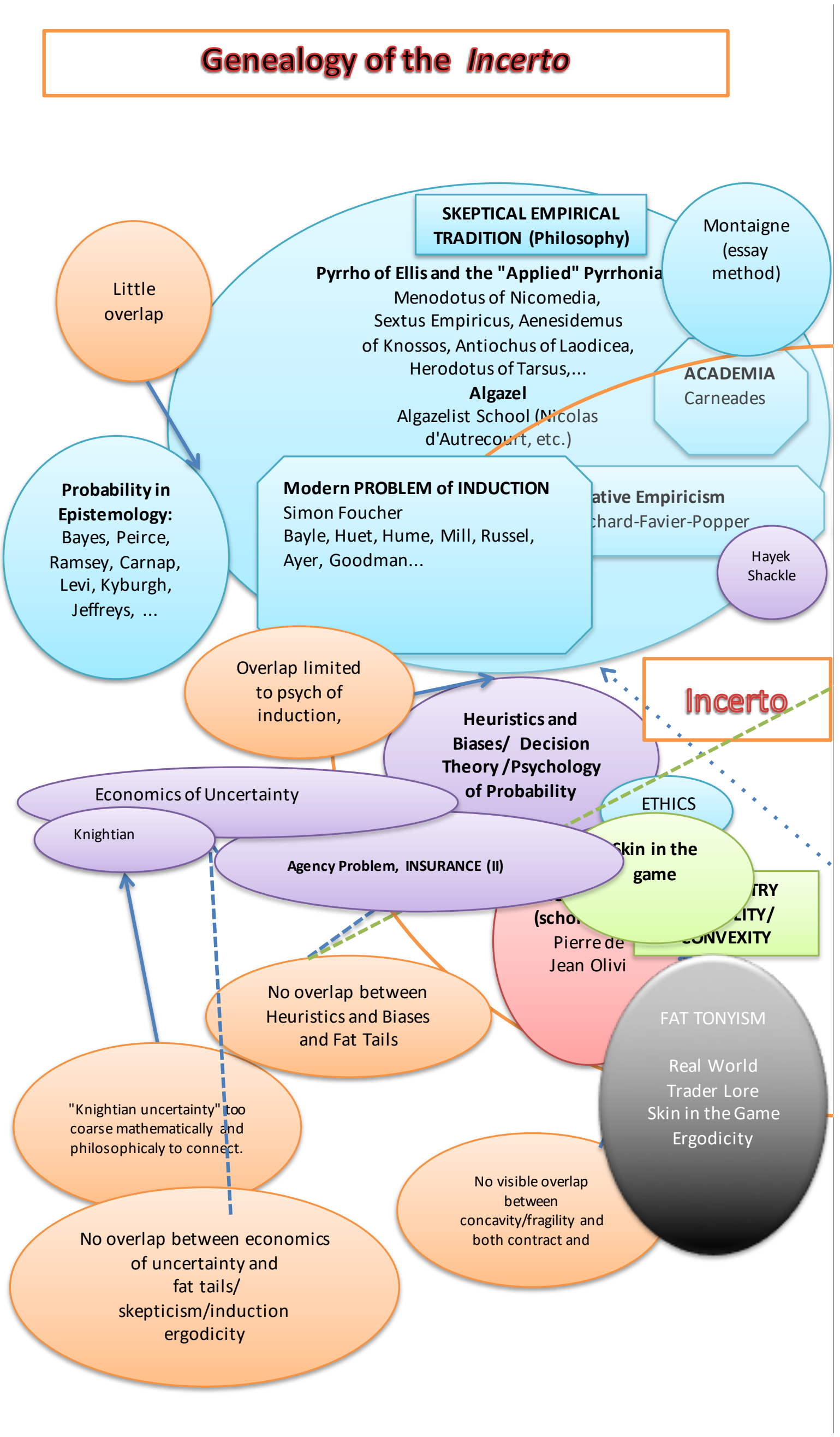

*Genealogy of the Incerto project with links to the various research traditions.*

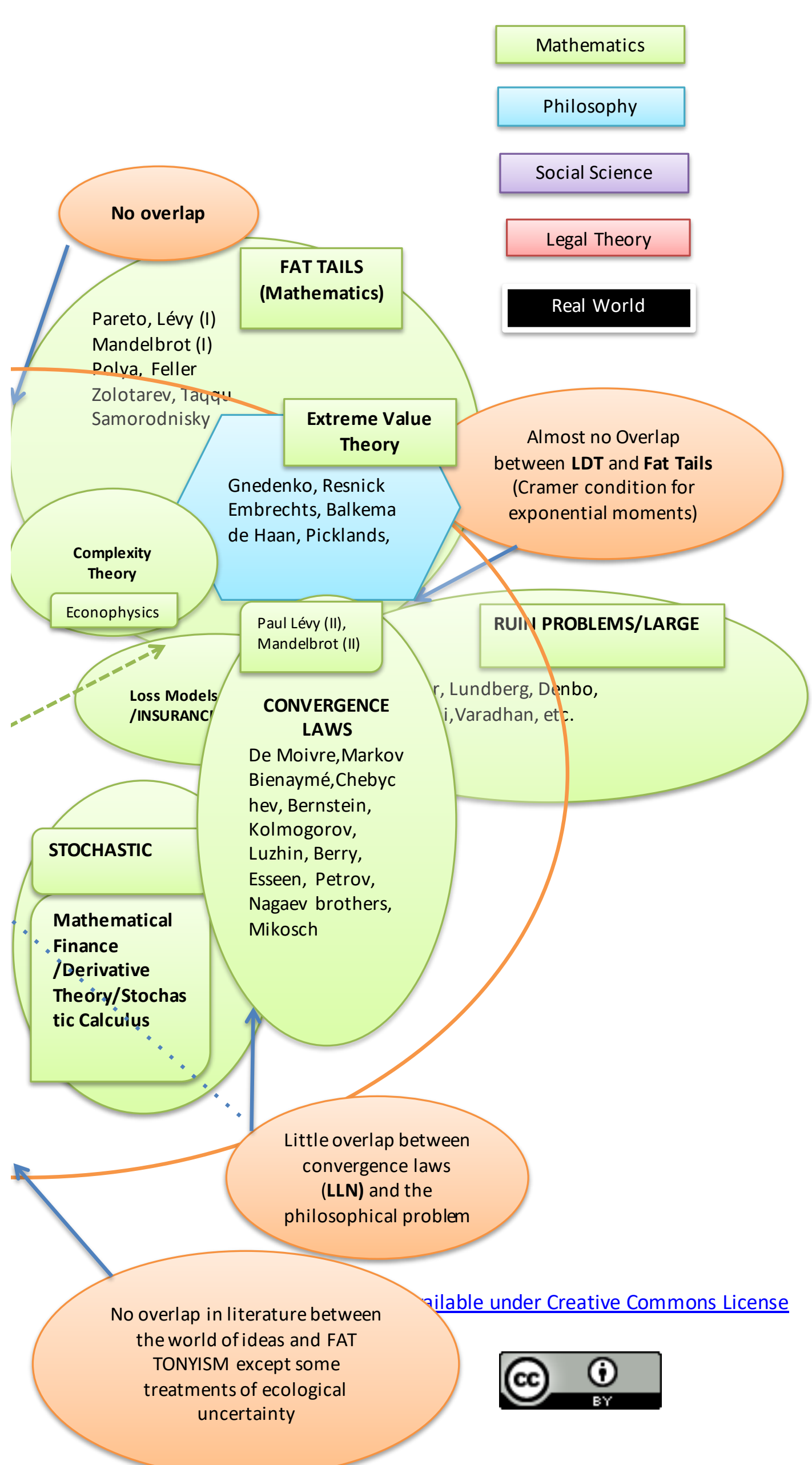

*(cont from left page).*

# CONTENTS

*Nontechnical chapters are indicated with a star *; Discussion chapters are indicated with a †; adaptation from published ("peer-reviewed") papers with a ‡.*
*While chapters are indexed by Arabic numerals, expository and very brief mini-chapters (half way between appendices and full chapters) use letters such as A, B, etc.*

# NOTES FOR THE THIRD EDITION

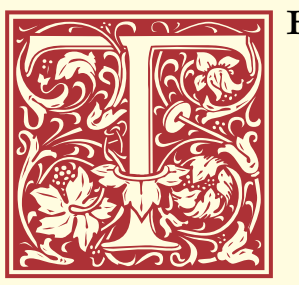

THIS EDITION contains three main additions. Chapter 12 explains how single point forecasting fails decision making under fat tailed threats such as pandemics. Chapter 13 discusses mistakes in the "forecasting" industry, written with the collaboration of Ronald Richman, Marcos Carreira, and James Sharpe, thanks to the Online Discussion Group on Uncertainty. Chapter 23 explains how Lindy can be derived from distance away from absorption. Appendix B shows how maximum entropy distributions generate fat tails, which includes the nonextensive Tsallis entropy and an exposition of the associated q-distributions.The author thanks Constantino Tsallis for the gentle reminder.

# 1 PROLOGUE*,†

The less you understand the world, the easier it is to make a decision.

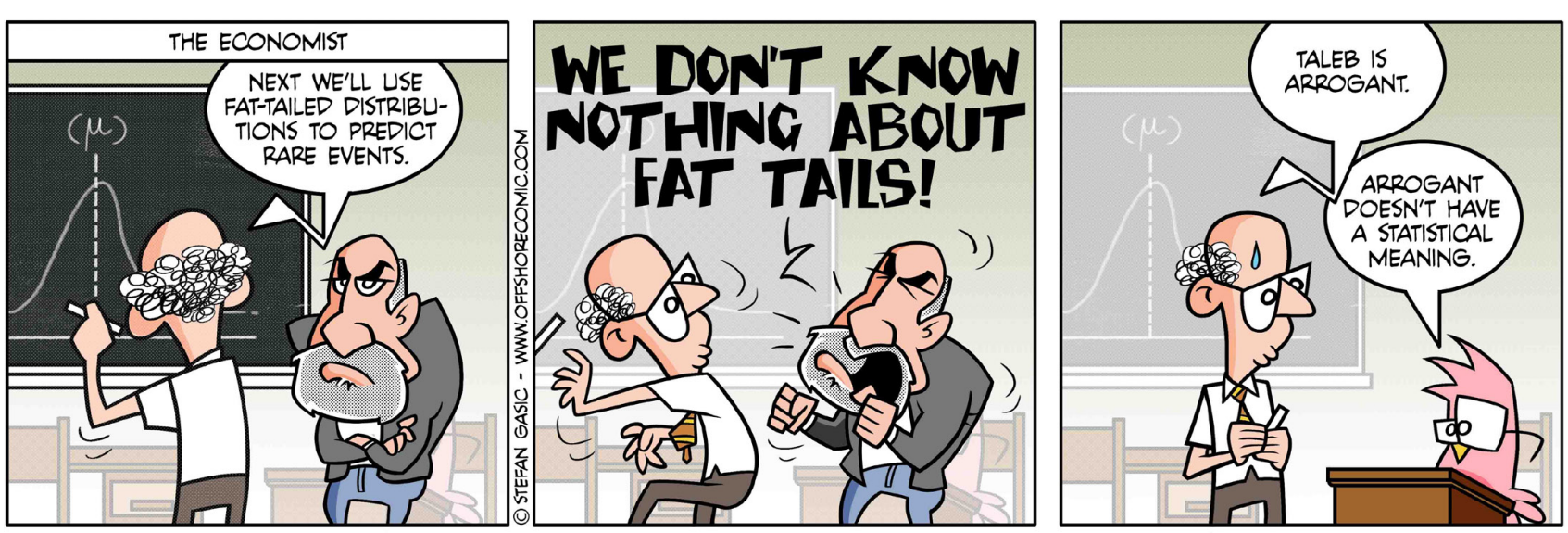

Figure 1.1: *The problem is not awareness of "fat tails", but the lack of understanding of their consequences. Saying "it is fat tailed" implies much more than changing the name of the distribution, but a general overhaul of the statistical tools and types of decisions made. Credit Stefan Gasic.*

The main idea behind the *Incerto* project is that while there is a lot of uncertainty and opacity about the world, and an incompleteness of information and understanding, there is little, if any, uncertainty about what actions should be taken based on such an incompleteness, in any given situation.

THIS BOOK consists in 1) published papers and 2) (uncensored) commentary, about classes of statistical distributions that deliver extreme events, and how we should deal with them for both statistical inference and decision making. Most "standard" statistics come from theorems designed for thin tails: they need to be adapted preasymptotically to fat tails, which is not trivial –or abandoned altogether.

Discussion chapter.

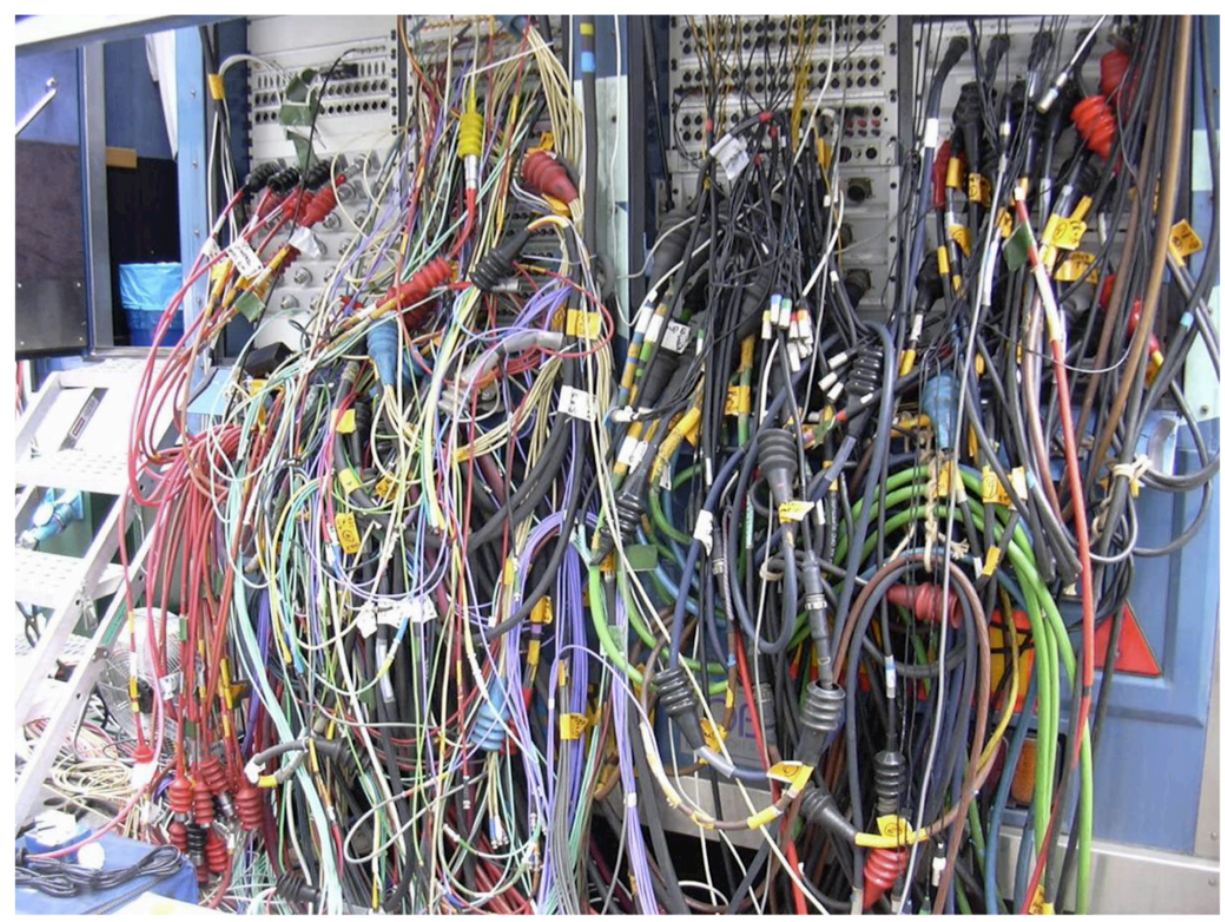

Figure 1.2: *Complication without insight: the clarity of mind of many professionals using statistics and data science without an understanding of the core concepts, what it is fundamentally about.*
Credit: Wikimedia.

So many times this author has been told *of course we know this* or the beastly portmanteau *nothing new* about fat tails by a professor or practitioner who just produced an analysis using "variance", "GARCH", "kurtosis" , "Sharpe ratio", or "value at risk", or produced some "statistical significance" that is clearly not significant.

More generally, this book draws on the author's multi-volume series, *Incerto* [272] and associated technical research program, which is about how to live in the real world, a world with a structure of uncertainty that is too complicated for us.

The *Incerto* tries to connect five different fields related to tail probabilities and extremes: mathematics, philosophy, social science, contract theory, decision theory, and the real world. If you wonder why contract theory, the answer is: option theory is based on the notion of contingent and probabilistic contracts designed to modify and share classes of exposures in the tails of the distribution; in a way option theory is mathematical contract theory. Decision theory is not about understanding the world, but getting out of trouble and ensuring survival. This point is the subject of the next volume of the *Technical Incerto*, with the temporary working title *Convexity, Risk, and Fragility*.

## A WORD ON TERMINOLOGY

"Thick tails" is often used in academic contexts. For us, here, it maps to much "higher kurtosis than the Gaussian" –to conform to the finance practitioner's lingo. As to "Fat Tails", we prefer to reserve it both extreme thick tails or membership in the power law class (which we show in Chapter 8 cannot be disentangled). For many it is meant to be a narrower definition, limited to "power laws" or "regular variations" – but we prefer to call "power laws" "power laws" (when we are quite certain about the process), so what we call "fat tails" may sometimes be more technically "extremely thick tails" for many.

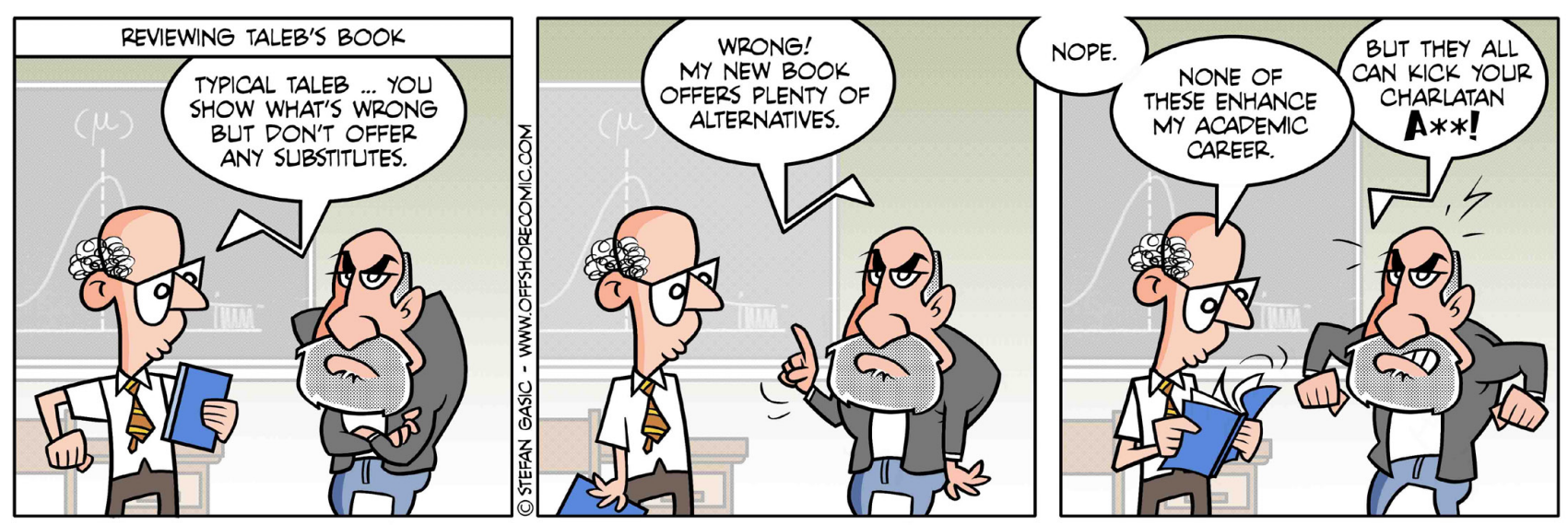

Figure 1.3: *The classic response: a "substitute" is something that does not harm rent-seeking. Credit: Stefan Gasic.*

To avoid ambiguity, we stay away from designations such as "heavy tails" or "long tails".

The next two chapters will clarify.

## ACKNOWLEDGMENTS

In addition to coauthors mentioned earlier, the author is indebted to Zhuo Xi, Jean-Philippe Bouchaud, Robert Frey, Spyros Makridakis, Mark Spitznagel, Brandon Yarkin, Raphael Douady, Peter Carr, Marco Avellaneda, Didier Sornette, Paul Embrechts, Bruno Dupire, Jamil Baz, Damir Delic, Yaneer Bar-Yam, Diego Zviovich, Ole Peters, Chitpuneet Mann, Harry Crane –and of course endless, really endless discussions with the great Benoit Mandelbrot.

Social media volunteer editors such as Antonio Catalano, Maxime Biette, Caio Vinchi, Jason Thorell, Marco Alves, and Petri Helo cleared many typos. I am most thankful to Edmond Shami who found many such errors.

Some of the papers that turned into chapters have been presented at conferences; the author thanks Lauren de Haan, Bert Zwart, and others for comments on extreme value related problems. More specific acknowledgements will be made within individual chapters. As usual, the author would like to express his gratitude to the staff at Naya restaurant in NY.

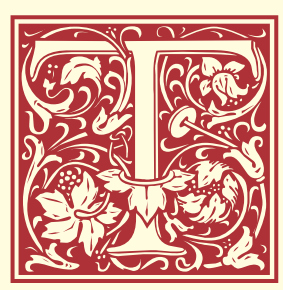

HIS AUTHOR presented the present book and the main points at the monthly Bloomberg Quant Conference in New York in Sep-tember 2018. After the lecture, a prominent mathematical finance professor came to see me. "This is very typical Taleb", he said. "You show what's wrong but don't offer too many substitutes".

Clearly, in business or in anything subjected to the rigors of the real world, he would have been terminated. People who never had any skin in the game [282] cannot figure out the necessity of circumstantial suspension of belief and the informational value of unreliability for decision making: *don't give a pilot a faulty metric; learn to provide only reliable information; letting the pilot know that the plane is defective saves lives*. Nor can they get the outperformance of *via negativa* –Popperian science works by removal. The late David Freedman had tried unsuccessfully to tame vapid and misleading statistical modeling vastly *outperformed* by "nothing".

But it is the case that the various chapters and papers here *do* offer solutions and alternatives, except that these aren't the most comfortable for some as they require some mathematical work for re-derivations for fat tailed conditions.

# 2 GLOSSARY, DEFINITIONS, AND NOTATIONS

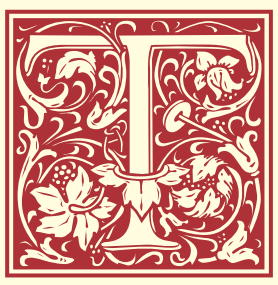

HIS IS A catalogue raisonné of the main topics and notations. Notations are redefined in the text every time; this is an aid for the random peruser. Some chapters extracted from papers will have specific notations, as specified. Note that while our terminology may be at variance with that of some research groups, it aims at remaining consistent.

## 2.1 GENERAL NOTATIONS AND FREQUENTLY USED SYMBOLS

$\mathbb{P}$ is the probability symbol; typically in $\mathbb{P}(X > x)$, $X$ is the random variable, $x$ is the realization. More formal measure-theoretic definitions of events and other French matters are in Chapter 11 and other places where such formalism makes sense.

$\mathbb{E}$ is the expectation operator.

$\mathbb{V}$ is the Variance operator.

$\mathbb{M}$ is the mean absolute deviation which is, when centered, centered around the mean (rather than the median).

$\varphi(.)$ and $f(.)$ are usually reserved for the PDF (probability density function) of a pre-specified distribution. In some chapters, a distinction is made for $f_x(x)$ and $f_y(y)$, particularly when $X$ and $Y$ follow two separate distributions.

$n$ is usually reserved for the number of summands.

$p$ is usually reserved for the moment order.

r.v. is short for a random variable.

$F(.)$ is reserved for the CDF (cumulative distribution function $\mathbb{P}(X < x)$, $\overline{F}(.)$, or $S$ is the survival function $\mathbb{P}(X > x)$.

$\sim$ indicates that a random variable is distributed according to a certain specified law.

$\chi(t) = \mathbb{E}(e^{itX_s})$ is the characteristic function of a distribution. In some discussions, the argument $t \in \mathbb{R}$ is represented as $\omega$. Sometimes $\Psi$ is used.

$\xrightarrow{D}$ denotes convergence in distribution, as follows. Let $X_1, X_2, \dots, X_n$ be a sequence of random variables; $X_n \xrightarrow{D} X$ means the CDF $F_n$ for $X_n$ has the following limit:

$$\lim_{n\to\infty} F_n(x) = F(x)$$

for every real $x$ for which $F$ is continuous.

$\xrightarrow{P}$ denotes convergence in probability, that is for $\varepsilon > 0$, we have, using the same sequence as before

$$\lim_{n\to\infty} \mathbb{P}((|X_n - X| > \varepsilon) = 0.$$

$\xrightarrow{a.s.}$ denotes almost sure convergence, the stronger form:

$$\mathbb{P}\left(\lim_{n\to\infty} X_n = X\right) = 1.$$

$S_n$ is typically a sum for $n$ summands.

$\alpha$ and $\alpha_S$: we shall typically try to use $\alpha_s \in (0,2]$ to denote the tail exponent of the limiting and Platonic stable distribution and $\alpha_p \in (0,\infty)$ the corresponding Paretian (preasymptotic) equivalent but only in situations where there could be some ambiguity. Plain $\alpha$ should be understood in context.

$\mathcal{N}(\mu_1, \sigma_1)$ the Gaussian distribution with mean $\mu_1$ and variance $\sigma_1^2$.

$\mathcal{L}(.,.)$ or $\mathcal{LN}(.,.)$ is the Lognormal distribution, with PDF $f^{(L)}(.)$ typically parameterized here as $\mathcal{L}(X_0 - \frac{1}{\sigma^2}, \sigma)$ to get a mean $X_0$, and variance $\left(e^{\sigma^2} - 1\right) X_0^2$.

$\mathcal{S}(\alpha_S, \beta, \mu, \sigma)$ is the stable distribution with tail index $\alpha_s$ in $(0,2]$, symmetry index $\beta$ in $-1,1)$, centrality parameter $\mu$ in $\mathbb{R}$ and scale $\sigma > 0$.

$\mathfrak{P}$ is the power law class (see below).

$\mathfrak{S}$ is the subexponential class (see below).

$\delta(.)$ is the Dirac delta function.

$\theta(.)$ is the Heaviside theta function.

erf(.), the error function, is the integral of the Gaussian distribution $\text{erf}(z) = \frac{2}{\sqrt{\pi}} \int_0^z e^{-t^2} dt$. erfc(.), is the complementary error function $1 - erf(.)$.

$\|.\|_p$ is a norm defined for (here a real vector) $\mathbf{X} = (X_1, \dots, X_n)^T$,

$\|X\|_p \triangleq \left(\frac{1}{n} \sum_{i=1}^n |x_i|^p\right)^{1/p}$. Note the absolute value in this text.

${}_1F_1(.;.;.)$ is the Kummer confluent hypergeometric function:

$${}_1F_1(a;b;z) = \sum_{k=0}^{\infty} \frac{a_k \frac{z^k}{k}!}{b_k}.$$

${}_2\tilde{F}_2$ is the generalized hypergeometric function regularized:

$$ {}_2\tilde{F}_2(.,.;.,.;.) = \frac{{}_2F_2(a;b;z)}{(\Gamma(b_1)\ldots\Gamma(b_q))} $$

and ${}_pF_q(a;b;z)$ has series expansion $\sum_{k=0}^{\infty} \frac{(a_1)_k \ldots (a_p)_k}{(b_1)_k \ldots (b_p)_k} z^k/k!$, were $(a_q)_{(.)}$ is the Pockhammer symbol.

$(a_q)_{(.)}$ is the Q-Pochhammer symbol $(a_q)_n = \prod_{i=1}^{n-1}\left(1 - aq^i\right)$.

## 2.2 CATALOGUE RAISONNÉ OF GENERAL & IDIOSYNCRATIC CONCEPTS

Next is the duplication of the definition of some central topics.

### 2.2.1 Power Law Class $\mathfrak{P}$

The power law class is conventionally defined by the property of the survival function, as follows. Let $X$ be a random variable belonging to the class of distributions with a "power law" right tail, that is:

$$ \mathbb{P}(X > x) = L(x)\, x^{-\alpha} \tag{2.1} $$

where $L : [x_{\min}, +\infty) \to (0, +\infty)$ is a slowly varying function, defined as

$$ \lim_{x \to +\infty} \frac{L(kx)}{L(x)} = 1 $$

for any $k > 0$ [26].

The survival function of $X$ is called to belong to the "regular variation" class $RV_\alpha$. More specifically, a function $f : \mathbb{R}^+ \to \mathbb{R}^+$ is index varying at infinity with index $\rho$ ($f \in RV_\rho$) when

$$ \lim_{t \to \infty} \frac{f(tx)}{f(t)} = x^\rho. $$

More practically, there is a point where $L(x)$ approaches its limit, $l$, becoming a constant –which we call the "Karamata constant" and the point is dubbed the "Karamata point". Beyond such value the tails for power laws are calibrated using such standard techniques as the Hill estimator. The distribution in that zone is dubbed the strong Pareto law by B. Mandelbrot[191],[88].

The same applies, when specified, to the left tail.

### 2.2.2 Law of Large Numbers (Weak)

The standard presentation is as follows. Let $X_1, X_2, \dots X_n$ be an infinite sequence of independent and identically distributed (Lebesgue integrable) random variables with expected value $\mathbb{E}(X_n) = \mu$ (though one can somewhat relax the i.i.d. assumptions). The sample average $\overline{X}_n = \frac{1}{n}(X_1 + \cdots + X_n)$ converges to the expected value, $\overline{X}_n \to \mu$, for $n \to \infty$.

Finiteness of variance is not necessary (though of course the finite higher moments accelerate the convergence).

The strong law is discussed where needed.

### 2.2.3 The Central Limit Theorem (CLT)

*The Standard (Lindeberg-Lévy) version of CLT* is as follows. Suppose a sequence of i.i.d. random variables with $\mathbb{E}(X_i) = \mu$ and $\mathbb{V}(X_i) = \sigma^2 < +\infty$, and $\overline{X}_n$ the sample average for $n$. Then as $n$ approaches infinity, the sum of the random variables $\sqrt{n}(\overline{X}_n \mu)$ converges in distribution to the Gaussian [24] [25]:

$$\sqrt{n}\left(\overline{X}_n - \mu\right) \xrightarrow{d} N\left(0, \sigma^2\right).$$

Convergence in distribution here means that the CDF (cumulative distribution function) of $\sqrt{n}$ converges pointwise to the CDF of $\mathcal{N}(0, \sigma)$ for every real $z$,

$$\lim_{n \to \infty} \mathbb{P}\left(\sqrt{n}(\overline{X}_n - \mu) \leq z\right) = \lim_{n \to \infty} \mathbb{P}\left[\frac{\sqrt{n}(\overline{X}_n - \mu)}{\sigma} \leq \frac{z}{\sigma}\right] = \Phi\left(\frac{z}{\sigma}\right), \ \sigma > 0$$

where $\Phi(z)$ is the standard normal CDF evaluated at $z$.

There are many other versions of the CLT, presented as needed.

### 2.2.4 Expected Excess Deviation Over a Threshold

Let $X$ be a random variable that lives in either $(0, \infty)$ or $(-\infty, \infty)$ ,$K$ a positive constant and $\mathbb{E}$ the expectation operator under "real world" (physical) distribution. By classical results [97]:

$$\lim_{K \to \infty} \mathbb{E}(X|_{X>K}) = K\,\lambda \tag{2.2}$$

- If $\lambda = 1$ , $X$ is said to be in the thin tailed class and has a characteristic scale.(Case 1)
- If $\lambda > 1$ , $X$ is said to be in the fat tailed regular variation class and has no characteristic scale.
- If
$$\lim_{K \to \infty} \mathbb{E}(X|_{X>K}) - K = \lambda$$
where $\lambda > 0$, then $X$ is in the borderline exponential class.(Case 3)

This property can be used in diagnostics with the "Mean Excess Plot".

Note that when the variable is time, Case 2 is the famous "Lindy Effect" of reverse aging, case 3 corresponds to memorylessness, typically from the interarrival time between events under a Poisson process.

### 2.2.5 Law of Medium Numbers or Preasymptotics

This is pretty much the central topic of this book. We are interested in the behavior of the random variable for $n$ large but not too large or asymptotic. While it is not a big deal for the Gaussian owing to extremely rapid convergence (by both LLN and CLT), this is not the case for other random variables.

See **Kappa** next.

### 2.2.6 Kappa Metric

Metric here should not be interpreted in the mathematical sense of a distance function, but rather in its engineering sense, as a quantitative measurement.

Kappa, in $[0,1]$, developed by this author here, in Chapter 8, and in a paper [281], gauges the preasymptotic behavior or a random variable; it is 0 for the Gaussian considered as benchmark, and 1 for a Cauchy or a r.v. that has no mean.

Let $X_1, \ldots, X_n$ be i.i.d. random variables with finite mean, that is $\mathbb{E}(X) < +\infty$. Let $S_n = X_1 + X_2 + \ldots + X_n$ be a partial sum. Let $\mathbb{M}(n) = \mathbb{E}(|S_n - \mathbb{E}(S_n)|)$ be the expected mean absolute deviation from the mean for $n$ summands (recall we do not use the median but center around the mean). Define the "rate" of convergence for $n$ additional summands starting with $n_0$:

$$\kappa_{n_0,n} : \frac{\mathbb{M}(n)}{\mathbb{M}(n_0)} = \left(\frac{n}{n_0}\right)^{\frac{1}{2-\kappa_{n_0,n}}}, n_0, n = 1, 2, ..., \tag{2.3}$$

$n > n_0 \geq 1$, hence

$$\kappa(n_0, n) = 2 - \frac{\log(n) - \log(n_0)}{\log\left(\frac{\mathbb{M}(n)}{\mathbb{M}(n_0)}\right)}. \tag{2.4}$$

Further, for the baseline values $n = n_0 + 1$, we use the shorthand $\kappa_{n_0}$.

### 2.2.7 Elliptical Distribution

$\mathbf{X}$, a $p \times 1$ random vector is said to have an elliptical (or elliptical contoured) distribution with location parameters $\mu$, a non-negative matrix $\Sigma$ and some scalar function $\Psi$ if its characteristic function is of the form $\exp(it'\mu)\Psi(t\Sigma t')$.

In practical words, one must have a single covariance matrix for the joint distribution to be elliptical. Regime switching, stochastic covariances (correlations), all these prevent the distributions from being elliptical. So we will show in Chapter 6 that a linear combination of variables following thin-tailed distributions can produce explosive thick-tailed properties when ellipticality is violated. This (in addition to fat tailedness) invalidates much of modern finance.

### 2.2.8 Statistical independence

Independence between two variables $X$ and $Y$ with marginal PDFs $f(x)$ and $f(y)$ and joint PDF $f(x,y)$ is defined by the identity:

$$\frac{f(x,y)}{f(x)f(y)} = 1,$$

regardless of the correlation coefficient (or covariation in case of distributions in the infinite variance class). In the class of elliptical distributions, the bivariate Gaussian with correlation coefficient 0 is both independent and uncorrelated. This does not apply to the Student T or the Cauchy in their multivariate forms.

### 2.2.9 Stable (Lévy stable) Distribution

This is a generalization of the CLT.

Let $X_1, \ldots, X_n$ be independent and identically distributed random variables. Consider their sum $S_n$. We have

$$\frac{Sn - a_n}{b_n} \xrightarrow{D} X_s, \tag{2.5}$$

where $X_s$ follows a stable distribution $\mathcal{S}$, $a_n$ and $b_n$ are norming constants, and, to repeat, $\xrightarrow{D}$ denotes convergence in distribution (the distribution of $X$ as $n \to \infty$). The properties of $\mathcal{S}$ will be more properly defined and explored in the next chapter. Take it for now that a random variable $X_s$ follows a stable (or $\alpha$-stable) distribution, symbolically $X_s \sim S(\alpha_s, \beta, \mu, \sigma)$, if its characteristic function$\chi(t) = \mathbb{E}(e^{itX_s})$ is of the form:

$$\chi(t) = e^{\left(i\mu t - |t\sigma|_s^{\alpha}\left(1 - i\beta \tan\left(\frac{\pi\alpha - s}{2}\right)\text{sgn}(t)\right)\right)} \text{ when } \alpha_s \neq 1. \tag{2.6}$$

The constraints are $-1 \leq \beta \leq 1$ and $0 < \alpha_s \leq 2$.

### 2.2.10 Multivariate Stable Distribution

A random vector $\mathbf{X} = (X_1, \ldots, X_k)^T$ is said to have the multivariate stable distribution if every linear combination of its components $\mathbf{Y} = a_1 X_1 + \cdots + a_k X_k$ has a stable distribution. That is, for any constant vector $\mathbf{a} \in \mathbb{R}^k$, the random variable $\mathbf{Y} = a^T \mathbf{X}$ should have a univariate stable distribution.

### 2.2.11 Karamata Point

See **Power Law Class**

### 2.2.12 Subexponentiality

The natural boundary between Mediocristan and Extremistan occurs at the subexponential class which has the following property:

Let $\mathbf{X} = X_1, \ldots, X_n$ be a sequence of independent and identically distributed random variables with support in ($\mathbb{R}^+$), with cumulative distribution function $F$. The subexponential class of distributions is defined by (see [300], [229]):

$$\lim_{x \to +\infty} \frac{1 - F^{*2}(x)}{1 - F(x)} = 2 \tag{2.7}$$

where $F^{*2} = F' * F$ is the cumulative distribution of $X_1 + X_2$, the sum of two independent copies of $X$. This implies that the probability that the sum $X_1 + X_2$ exceeds a value $x$ is twice the probability that either one separately exceeds $x$. Thus, every time the sum exceeds $x$, for large enough values of $x$, the value of the sum is due to either one or the other exceeding $x$—the maximum over the two variables—and the other of them contributes negligibly.

More generally, it can be shown that the sum of $n$ variables is dominated by the maximum of the values over those variables in the same way. Formally, the following two properties are equivalent to the subexponential condition [47],[99]. For a given $n \geq 2$, let $S_n = \Sigma_{i=1}^n x_i$ and $M_n = \max_{1 \leq i \leq n} x_i$

a) $\lim_{x \to \infty} \frac{\mathbb{P}(S_n > x)}{\mathbb{P}(X > x)} = n$,

b) $\lim_{x\to\infty} \frac{P(S_n>x)}{P(M_n>x)} = 1.$

Thus the sum $S_n$ has the same magnitude as the largest sample $M_n$, which is another way of saying that tails play the most important role.

Intuitively, tail events in subexponential distributions should decline more slowly than an exponential distribution for which large tail events should be irrelevant. Indeed, one can show that subexponential distributions have no exponential moments:

$$\int_0^\infty \mathbf{e}^{\epsilon x}\, dF(x) = +\infty \tag{2.8}$$

for all values of $\varepsilon$ greater than zero. However, the converse isn't true, since distributions can have no exponential moments, yet not satisfy the subexponential condition.

### 2.2.13 Student T as Proxy

We use the student T with $\alpha$ degrees of freedom as a convenient two-tailed power law distribution. For $\alpha = 1$ it becomes a Cauchy, and of course Gaussian for $\alpha \to \infty$.

The student T is the main bell-shaped power law, that is, the PDF is continuous and smooth, asymptotically approaching zero for large negative/positive $x$, and with a single, unimodal maximum (further, the PDF is quasiconcave but not concave).

### 2.2.14 Citation Ring

A highly circular mechanism by which academic prominence is reached thanks to discussions where papers are considered prominent because other people are citing them, with no external filtering, thus causing research to concentrate and get stuck around "corners", focal areas of no real significance. This is linked to the operation of the academic system in the absence of adult supervision or the filtering of skin in the game.

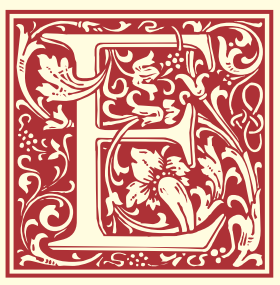

XAMPLE of fields that are, practically, frauds in the sense that their results are not portable to reality and only serve to feed additional papers that, in turn, will produce more papers: Modern Financial Theory, econometrics (particularly for macro variables), GARCH processes, psychometrics, stochastic control models in finance, behavioral economics and finance, decision making under uncertainty, macroeconomics, and a bit more.

### 2.2.15 Rent seeking in academia

There is a conflict of interest between a given researcher and the subject under consideration. The objective function of an academic department (and person) becomes collecting citations, honors, etc. at the expense of the purity of the subject: for instance many people get stuck in research corners because it is more beneficial to their careers and to their department.

### 2.2.16 Pseudo-empiricism or Pinker Problem

Discussion of "evidence" that is not statistically significant, or use of metrics that are uninformative because they do not apply to the random variables under consideration –like for instance making inferences from the means and correlations for fat tailed variables. This is the result of:

i) the focus in statistical education on Gaussian or thin-tailed variables,

ii) the absence of probabilistic knowledge combined with memorization of statistical terms,

iii) complete cluelessness about dimensionality,

all of which are prevalent among social scientists.

Example of pseudo-empiricism: comparing death from terrorist actions or epidemics such as ebola (fat tailed) to falls from ladders (thin tailed). This confirmatory "positivism" is a disease of modern science; it breaks down under both dimensionality and fat-tailedness.

Actually one does not need to distinguish between fat tailed and Gaussian variables to get the lack of rigor in these activities: simple criteria of statistical significance are not met –nor do these operators grasp the notion of such a concept as significance.

### 2.2.17 Preasymptotics

Mathematical statistics is largely concerned with what happens with $n = 1$ (where $n$ is the number of summands) and $n = \infty$. What happens in between is what we call the real world –and the major focus of this book. Some distributions (say those with finite variance) are Gaussian in behavior asymptotically, for $n = \infty$, but not for extremely large but not infinite $n$.

### 2.2.18 Stochasticizing

Making a deterministic parameter stochastic, (i) in a simple way, or (ii) via a more complex continuous or discrete distribution.

(i) Let $s$ be the deterministic parameter; we stochasticize (entry-level style) by creating a two-state Bernouilli with $p$ probability of taking a value $s_1$, $1 - p$ of taking value $s_2$. A transformation is mean-preserving when $ps_1 + (1 - p)s_2 = s$, that is, preserves the mean of the $s$ parameter. More generally, it can be in a similar manner variance preserving, etc.

(ii) We can use a full probability distribution, typically a Gaussian if the variable is two-tailed, and the Lognormal or the exponential if the variable is one-tailed (rarely a power law). When $s$ is standard deviation, one can stochasticize $s^2$, where it becomes "stochastic volatility", with a variance or standard deviation typically dubbed "Vvol".

### 2.2.19 Value at Risk, Conditional VaR

The mathematical expression of the Value at Risk, VaR, for a random variable $X$ with distribution function $F$ and threshold $\lambda \in [0, 1]$

$$\text{VaR}_\lambda(X) = -\inf\{x \in \mathbb{R} : F(x) > \lambda\},$$

and the corresponding CVar or Expected Shortfall ES at threshold $\lambda$:

$$\text{ES}_\lambda(X) = \mathbb{E}\left(-X \,|_{X \leq -\text{VaR}_\lambda(X)}\right)$$

or, in the positive domain, by considering the tail for $X$ instead of that of $-X$.

More generally the expected shortfall for threshold $K$ is $\mathbb{E}(X|_{X>K})$.

### 2.2.20 Skin in the Game

A filtering mechanism that forces cooks to eat their own cooking and be exposed to harm in the event of failure, thus throws dangerous people out of the system. Fields that have skin in the game: plumbing, dentistry, surgery, engineering, activities where operators are evaluated by tangible results or subjected to ruin and bankruptcy. Fields where people have no skin in the game: circular academic fields where people rely on peer assessment rather than survival pressures from reality.

### 2.2.21 MS Plot

The MS plot, "maximum to sum", allows us to see the behavior of the LLN for a given moment consider the contribution of the maximum observation to the total, and see how it behaves as $n$ grows larger. For a r.v. $X$, an approach to detect if $\mathbb{E}(X^p)$ exists consists in examining convergence according to the law of large numbers (or, rather, absence of), by looking the behavior of higher moments in a given sample. One convenient approach is the Maximum-to-Sum plot, or MS plot as shown in Figure 19.1.

The MS Plot relies on a consequence of the law of large numbers [95] when it comes to the maximum of a variable. For a sequence $X_1, X_2, ..., X_n$ of non-negative i.i.d. random variables, if for $p = 1, 2, 3, \ldots$, $\mathbb{E}[X^p] < \infty$, then

$$R_n^p = M_n^p / S_n^p \rightarrow^{a.s.} 0$$

as $n \to \infty$, where $S_n^p = \sum_{i=1}^{n} X_i^p$ is the partial sum, and $M_n^p = \max(X_1^p, ..., X_n^p)$ the partial maximum. (Note that we can have $X$ the absolute value of the random variable in case the r.v. can be negative to allow the approach to apply to odd moments.)

### 2.2.22 Maximum Domain of Attraction, MDA

The extreme value distribution concerns that of the maximum r.v., when $x \to x^*$, where $x^* = \sup\{x : F(x) < 1\}$ (the right "endpoint" of the distribution) is in the maximum domain of attraction, MDA [136]. In other words,

$$\max(X_1, X_2, \ldots X_n) \xrightarrow{P} x^*.$$

### 2.2.23 Substitution of Integral in the psychology literature

The verbalistic literature makes the following conflation. Let $K \in \mathbb{R}^+$ be a threshold, $f(.)$ a density function and $p_K \in [0, 1]$ the probability of exceeding it, and $g(x)$ an impact function. Let $I_1$ be the expected payoff above $K$:

$$I_1 = \int_K^\infty g(x) f(x) \, \mathrm{d}x,$$

and Let $I_2$ be the impact at $K$ multiplied by the probability of exceeding $K$:

$$I_2 = g(K) \int_K^\infty f(x) \, dx = g(K) p_K.$$

The substitution comes from conflating $I_1$ and $I_2$, which becomes an identity if and only if $g(.)$ is constant above $K$ (say $g(x) = \theta_K(x)$, the Heaviside theta function). For $g(.)$ a variable function with positive first derivative, $I_1$ can be close to $I_2$ only under thin-tailed distributions, not under the fat tailed ones.

### 2.2.24 Inseparability of Probability (another common error)

Let $F : \mathcal{A} \to [0, 1]$ be a probability distribution (with derivative $f$) and $g : \mathbb{R} \to \mathbb{R}$ a measurable function, the "payoff"". Clearly, for $\mathcal{A}'$ a subset of $\mathcal{A}$:

$$\int_{\mathcal{A}'} g(x)\mathrm{d}F(x) = \int_{\mathcal{A}'} f(x)g(x)\mathrm{d}x$$
$$\neq \int_{\mathcal{A}'} f(x)\mathrm{d}x \; g\left(\int_{\mathcal{A}'} \mathrm{d}x\right)$$

In discrete terms, with $\pi(.)$ a probability mass function:

$$\sum_{x \in \mathcal{A}'} \pi(x)g(x) \neq \sum_{x\in\mathcal{A}'} \pi(x)\; g\left(\frac{1}{n}\sum_{x\in\mathcal{A}'} x\right) \tag{2.9}$$
$$= \text{probability of event } \times \text{ payoff of average event}$$

The general idea is that probability is a kernel into an equation not an end product by itself outside of explicit bets.

Note that for $X$ a non-negative random variable:

$$\int_K^\infty x\, f(x)\, dx = \int_K^\infty \mathbb{P}(X > x)\, dx + K\, \mathbb{P}(X > K) \tag{2.10}$$

which is useful for the pricing of contingent claims.[1]

### 2.2.25 Wittgenstein's Ruler

"Wittgenstein's ruler" is the following riddle: are you using the ruler to measure the table or using the table to measure the ruler? Well, it depends on the results. Assume there are only two alternatives: a Gaussian distribution and a Power Law one. We show that a large deviation, say a "six sigma" indicates the distribution is power law.

### 2.2.26 Black Swans

*Black Swans result from the incompleteness of knowledge with effects that can be very consequential in fat tailed domains.*

Basically, they are things that fall outside what you can expect and model, and carry large consequences. The idea is to not predict them, but be convex (or at least not concave) to their impact: fragility to a certain class of events is detectable, even measurable (by gauging second order effects and asymmetry of responses), while the statistical attributes of these events may remain elusive.

It is hard to explain to modelers that we need to learn to work with things we have never seen (or imagined) before, but it is what it is[2].

---

1 The proof is is 13.6.

2 As Paul Portesi likes to repeat (attributing or perhaps misattributing to this author): "You haven't seen the other side of the distribution".

**Note the epistemic dimension:** Black swans are observer-dependent: a Black Swan for the turkey is a White Swan for the butcher. September 11 was a Black Swan for the victims, but not to the terrorists. This observer dependence is a central property. An "objective" probabilistic model of Black Swan isn't just impossible, but defeats the purpose, owing to the incomplete character of information and its dissemination.

> ***Grey Swans:*** *Large deviations that are are both consequential and have a very low frequency but remain consistent with statistical properties are called "Grey Swans". But of course the "greyness" depends on the observer: a Grey Swan for someone using a power law distribution will be a Black Swan to naive statisticians irremediably stuck within, and wading into, thin-tailed frameworks and representations.*

Let us repeat: no, it is not about fat tails; it is just that fat tails make them worse. The connection between fat-tails and Black Swans lies in the exaggerated impact from large deviations in fat tailed domains.

### 2.2.27 The Empirical Distribution is Not Empirical

The empirical distribution $\widehat{F}(t)$ is as follows: Let $X_1, \ldots X_n$ be independent, identically distributed real random variables with the common cumulative distribution function $F(t)$.

$$\widehat{F}_n(t) = \frac{1}{n}\sum_{i=1}^{n} \mathbb{1}_{x_i \geq t},$$

where $\mathbb{1}_A$ is the indicator function.

By the Glivenko-Cantelli theorem, we have uniform convergence of the max norm to a specific distribution –the Kolmogorov-Smirnoff –regardless of the initial distribution. We have:

$$\sup_{t\in\mathbb{R}} \left|\widehat{F}_n(t) - F(t)\right| \xrightarrow{\text{as.}} 0; \tag{2.11}$$

this distribution-independent convergence concerns probabilities of course, not moments –a result this author has worked on and generalized for the "hidden moment" above the maximum.

We note the main result (further generalized by Donsker into a Brownian Bridge since we know the extremes are 0 and 1)

$$\sqrt{n}\left(\widehat{F}_n(t) - F(t)\right) \xrightarrow{D} \mathcal{N}\left(0, F(t)(1-F(t))\right) \tag{2.12}$$

"The empirical distribution is not empirical" means that since the empirical distributions are necessarily censured on the interval $[x_{min}, x_{max}]$, for fat tails this can carry huge consequences because we must not analyze fat tails in probability space but rather in payoff space.

Further see the entry on **the hidden tail** (next).

### 2.2.28 The Hidden Tail

Consider $K_n$ the maximum of a sample of $n$ independent identically distributed variables; $K_n = \max\left(X_1, X_2, \ldots, X_n\right)$. Let $\phi(.)$ be the density of the underlying distribution. We can decompose the moments in two parts, with the "hidden" moment above $K_n$.

$$\mathbb{E}(X^p) = \underbrace{\int_L^{K_n} x^p \phi(x)\, dx}_{\mu_{L,p}} + \underbrace{\int_{K_n}^{\infty} x^p \phi(x)\, dx}_{\mu_{K,p}}$$

where $\mu_L$ is the observed part of the distribution and $\mu_K$ the hidden one (above $K$). By Glivenko-Cantelli the distribution of $\mu_{K,0}$ should be independent of the initial distribution of $X$, but higher moments do not, hence there is a bit of a problem with Kolmogorov-Smirnoff-style tests.

### 2.2.29 Shadow Moment

This is called in this book "plug-in" estimation. It is not done by measuring the directly observable sample mean which is biased under fat-tailed distributions, but by using maximum likelihood parameters, say the tail exponent $\alpha$, and deriving the shadow mean or higher moments.

### 2.2.30 Tail Dependence

Let $X_1$ and $X_2$ be two random variables not necessarily in the same distribution class. Let $F^{\leftarrow}(q)$ be the inverse CDF for probability $q$, that is $F^{\leftarrow}(q) = \inf\{x \in \mathbb{R} : F(x) \geq q\}$, $\lambda_u$ the upper tail dependence is defined as

$$\lambda_u = \lim_{q \to 1} \mathrm{P}\left(X_2 > F_2^{\leftarrow}(q) | X_1 > F_1^{\leftarrow}(q)\right) \tag{2.13}$$

Likewise for the lower tail dependence index.

### 2.2.31 Metaprobability

Comparing two probability distributions via some tricks which includes stochasticizing parameters. Or stochasticize a parameter to get the distribution of a call price, a risk metric such as VaR (see entry), CVaR, etc., and check the robustness or convexity of the resulting distribution.

### 2.2.32 Dynamic Hedging

The payoff of a European call option $C$ on an underlying $S$ with expiration time indexed at $T$ should be replicated with the following stream of dynamic hedges, the limit of which can be seen here, between present time $t$ and $T$:

$$\lim_{\Delta t \to 0} \left( \sum_{i=1}^{n=T/\Delta t} \frac{\partial C}{\partial S} \Big|_{S=S_{t+(i-1)\Delta t}, t=t+(i-1)\Delta t,} \left( S_{t+i\Delta t} - S_{t+(i-1)\Delta t} \right) \right) \tag{2.14}$$

We break up the period into n increments $\Delta$t. Here the hedge ratio $\frac{\partial C}{\partial S}$ is computed as of time $t$ +(i-1) $\Delta t$, but we get the nonanticipating difference between the price at the time the hedge was initiatied and the resulting price at $t+ i\, \Delta t$.

This is supposed to make the payoff deterministic *at the limit of* $\Delta t \to 0$. In the Gaussian world, this would be an Ito-McKean integral.

We show where this replication is never possible in a fat-tailed environment, owing to the special presamptotic properties.

# Part I

# FAT TAILS AND THEIR EFFECTS, AN INTRODUCTION

# 3 A NON-TECHNICAL OVERVIEW - THE DARWIN COLLEGE LECTURE [*,‡]

Abyssus abyssum invocat
תהום אל תהום קורא
*Psalms*

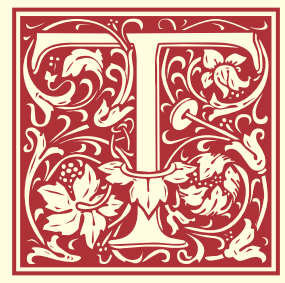HIS chapter presents a nontechnical yet comprehensive presentation of the entire *statistical consequences of thick tails* project. It compresses the main ideas in one place. Mostly, it provides a list of more than a dozen consequences of thick tails on statistical inference.

## 3.1 ON THE DIFFERENCE BETWEEN THIN AND THICK TAILS

We begin with the notion of thick tails and how it relates to extremes using the two imaginary domains of Mediocristan (thin tails) and Extremistan (thick tails).

- In Mediocristan, when a sample under consideration gets large, no single observation can really modify the statistical properties.
- In Extremistan, the tails (the rare events) play a disproportionately large role in determining the properties.

Research and discussion chapter.

A shorter version of this chapter was presented at Darwin College, Cambridge (UK) on January 27 2017, as part of the Darwin College Lecture Series on Extremes. The author extends the warmest thanks to D.J. Needham and Julius Weitzdörfer, as well as their invisible assistants who patiently and accurately transcribed the lecture into a coherent text. The author is also grateful towards Susan Pfannenschmidt and Ole Peters who corrected some mistakes. Jamil Baz prevailed upon me to add more commentary to the chapter to accommodate economists and econometricians who, one never knows, may eventually identify with some of it.

Another way to view it:

Assume a large deviation $X$.

- In Mediocristan, the probability of sampling higher than $X$ twice in a row is greater than sampling higher than $2X$ once.
- In Extremistan, the probability of sampling higher than $2X$ once is greater than the probability of sampling higher than $X$ twice in a row.

Let us randomly select two people in Mediocristan; assume we obtain a (very unlikely) combined height of 4.1 meters – a tail event. According to the Gaussian distribution (or, rather its one-tailed siblings), the most likely combination of the two heights is 2.05 meters and 2.05 meters. Not 10 centimeters and 4 meters.

Simply, the probability of exceeding 3 sigmas is 0.00135. The probability of exceeding 6 sigmas, twice as much, is $9.86 \times 10^{-10}$. The probability of two 3-sigma events occurring is $1.8 \times 10^{-6}$. Therefore the probability of two 3-sigma events occurring is considerably higher than the probability of one single 6-sigma event. This is using a class of distribution that is not fat-tailed.

Figure 3.1 shows that as we extend the ratio from the probability of two 3-sigma events divided by the probability of a 6-sigma event, to the probability of two 4-sigma events divided by the probability of an 8-sigma event, i.e., the further we go into the tail, we see that a large deviation can only occur via a combination (a sum) of a large number of intermediate deviations: the right side of Figure 3.1. In other words, for something bad to happen, it needs to come from a series of very unlikely events, not a single one. This is the logic of Mediocristan.

Let us now move to Extremistan and randomly select two people with combined wealth of \$ 36 million. The most likely combination is not \$18 million and \$ 18 million. It should be approximately \$ 35,999,000 and \$ 1,000.

This highlights the crisp distinction between the two domains; for the class of subexponential distributions, ruin is more likely to come from a single extreme event than from a series of bad episodes. This logic underpins classical risk theory as outlined by the actuary Filip Lundberg early in the $20^{th}$ Century [184] and formalized in the 1930s by Harald Cramer [62], but forgotten by economists in recent times. For insurability, losses need to be more likely to come from many events than a single one, thus allowing for diversification.

This indicates that insurance can only work in Mediocristan; you should never write an uncapped insurance contract if there is a risk of catastrophe. The point is called the catastrophe principle.

As we saw earlier, with thick tailed distributions, extreme events away from the centre of the distribution play a very large role. Black Swans are not "more frequent" (as it is commonly misinterpreted), they are more consequential. The fattest tail distribution has just one very large extreme deviation, rather than many departures form the norm. Figure 4.4 shows that if we take a distribution such as the Gaussian and start fattening its tails, then the number of departures away from one standard deviation drops. The probability of an event staying within one standard deviation of the mean is 68 percent. As the tails fatten, to mimic what happens in financial markets for example, the probability of an event staying within one standard deviation of the mean rises to between 75 and 95 percent. So note that as we fatten the tails we get higher peaks, smaller shoulders, and a higher incidence of a very large deviation. Because probabilities need to add up to 1 (even in France) increasing mass in one area leads to decreasing it in another.

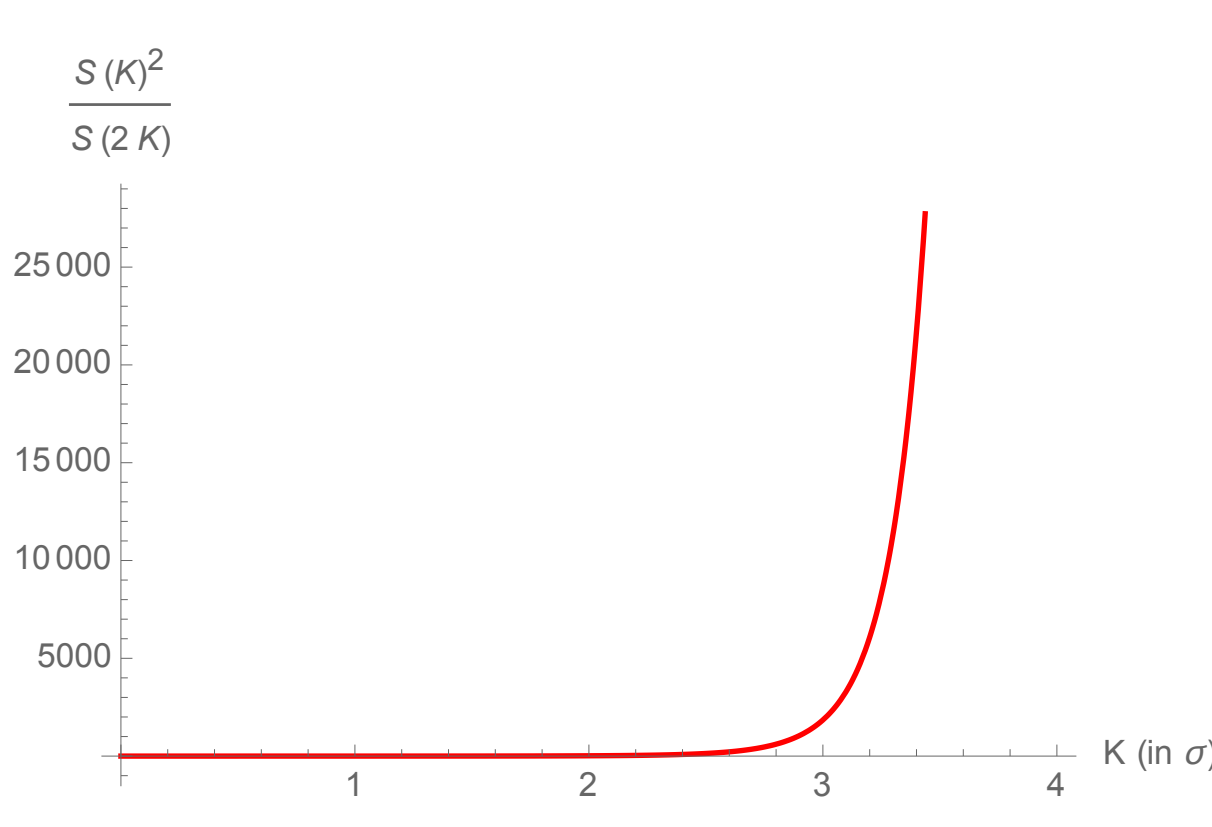

Figure 3.1: *Ratio of S(.) survival functions for two occurrences of size K by one of 2K for a Gaussian distribution**. *The larger the K, that is, the more we are in the tails, the more likely the event is to come from two independent realizations of K (hence* $P(K)^2$*, and the less likely from a single event of magnitude 2K.*

*This is fudging for pedagogical simplicity. The more rigorous approach would be to compare 2 occurrences of size $K$ to 1 occurrence of size $2K$ plus 1 regular deviation – but the end graph would not change at all.

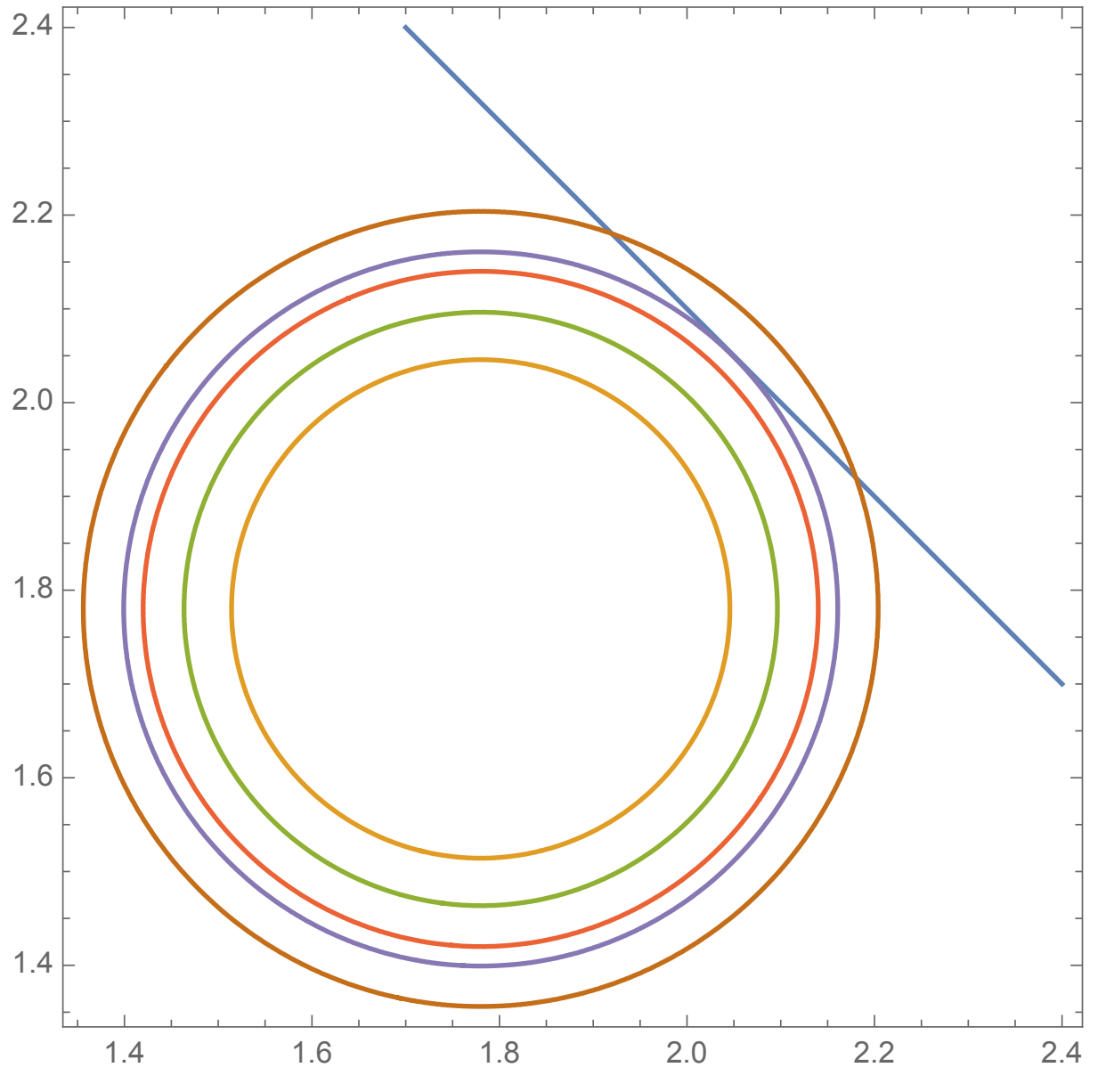

Figure 3.2: *Iso-densities for two independent Gaussian distributions. The line shows* $x + y = 4.1$. *Visibly the maximal probability is for* $x = y = 2.05$.

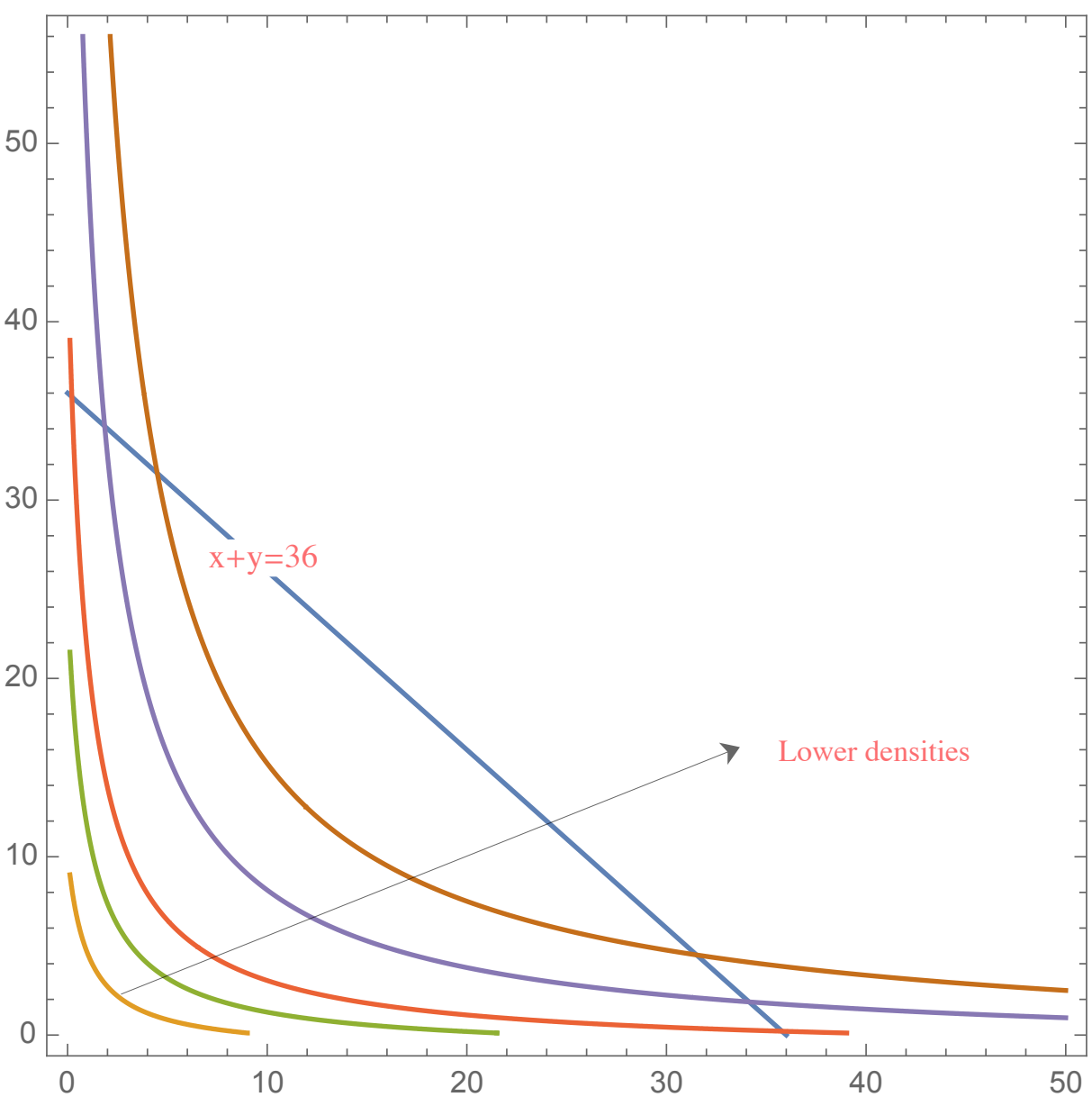

Figure 3.3: *Iso-densities for two independent thick tailed distributions (in the power law class). The line shows $x + y = 36$. Visibly the maximal probability is for either $x = 36 - \epsilon$ or $y = 36 - \epsilon$, with $\epsilon$ going to 0 as the sum $x + y$ becomes larger.*

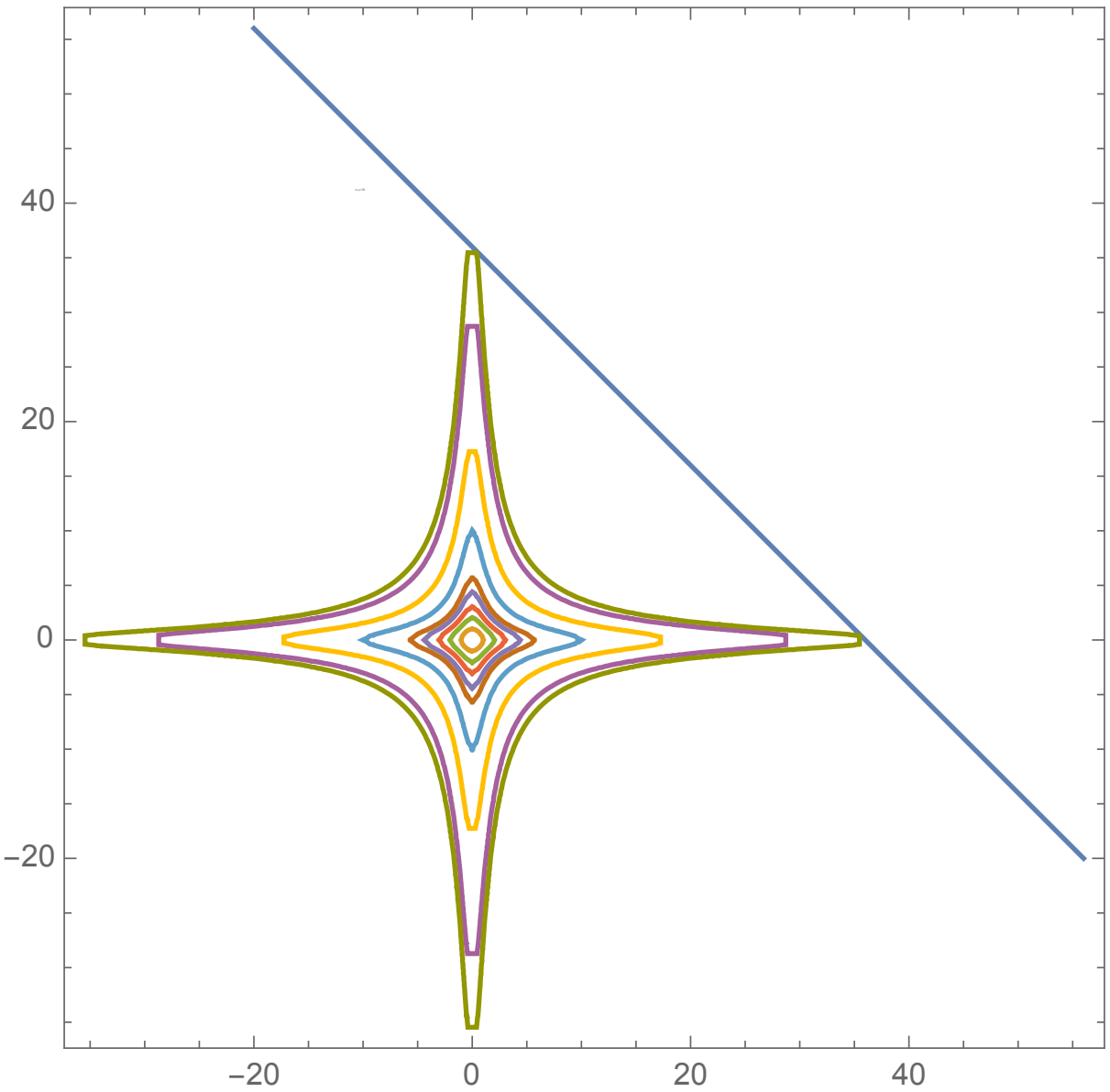

Figure 3.4: *Same representation as in Figure 3.1, but concerning power law distributions with support on the real line; we can see the iso-densities looking more and more like a cross for lower and lower probabilities. More technically, there is a loss of ellipticality.*

## 3.2 TAIL WAGGING DOGS: AN INTUITION

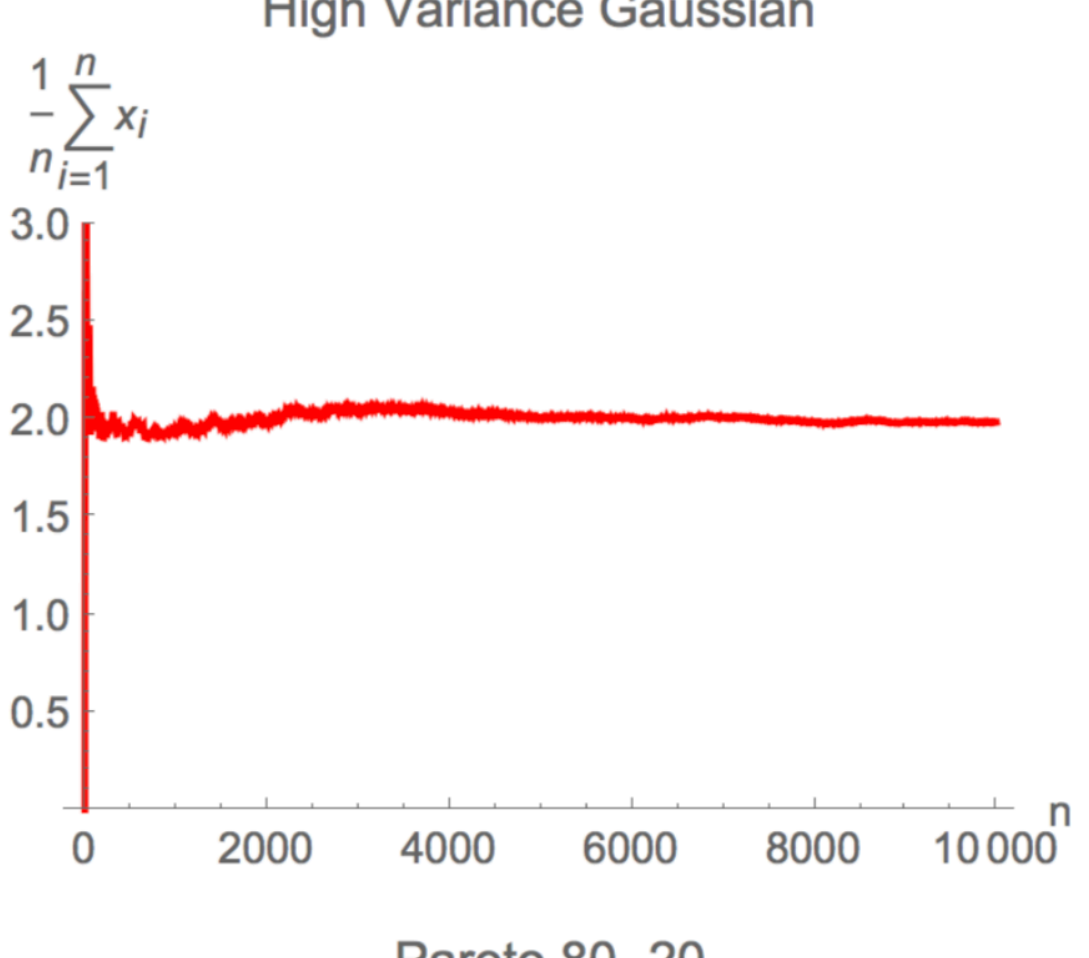

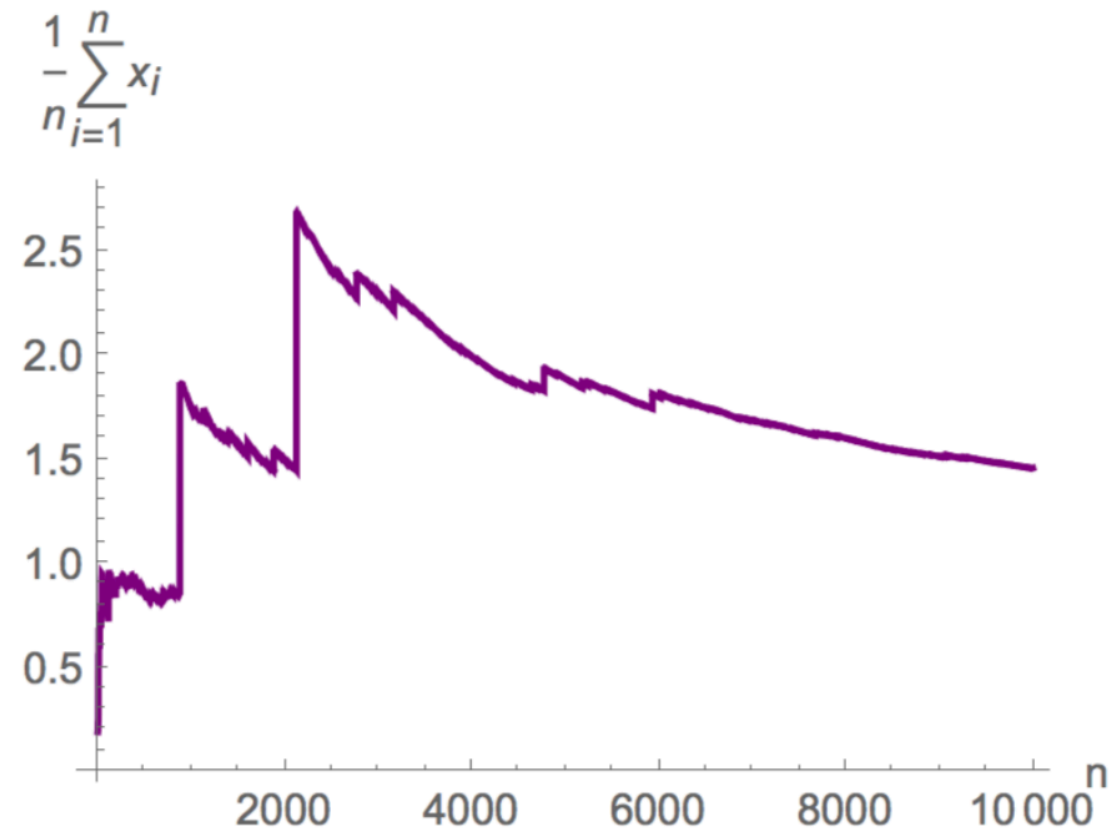

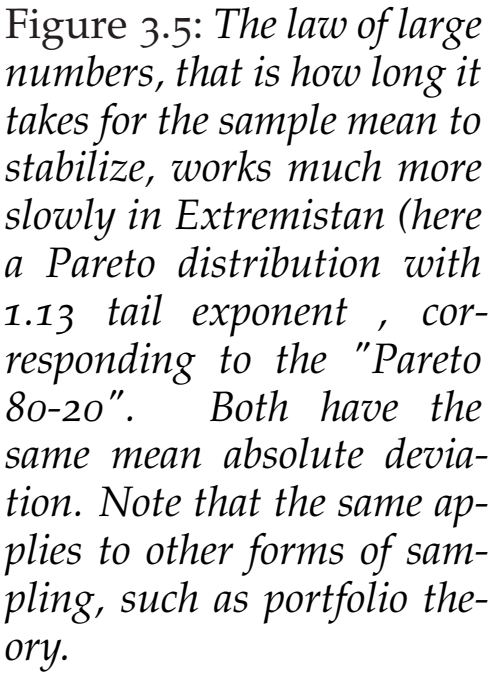
Figure 3.5: *The law of large numbers, that is how long it takes for the sample mean to stabilize, works much more slowly in Extremistan (here a Pareto distribution with 1.13 tail exponent , corresponding to the "Pareto 80-20". Both have the same mean absolute deviation. Note that the same applies to other forms of sampling, such as portfolio theory.*

**The tail wags the dog effect**

Centrally, the thicker the tails of the distribution, the more *the tail wags the dog*, that is, the information resides in the tails and less so in the "body" (the central part) of the distribution. Effectively, for very fat tailed phenomena, all deviations become informationally sterile except for the large ones.

The center becomes just noise. Although the "evidence based" science might not quite get it yet, under such conditions, there is no evidence in the body.

This property also explains the slow functioning of the law of large numbers in certain domains as tail deviations, where the information resides, are –by definition– rare.

The property explains why, for instance, a million observations of white swans do not confirm the non-existence of black swans, or why a million confirmatory observations count less than a single disconfirmatory one. We will link it to the Popper-style asymmetries later in the chapter.

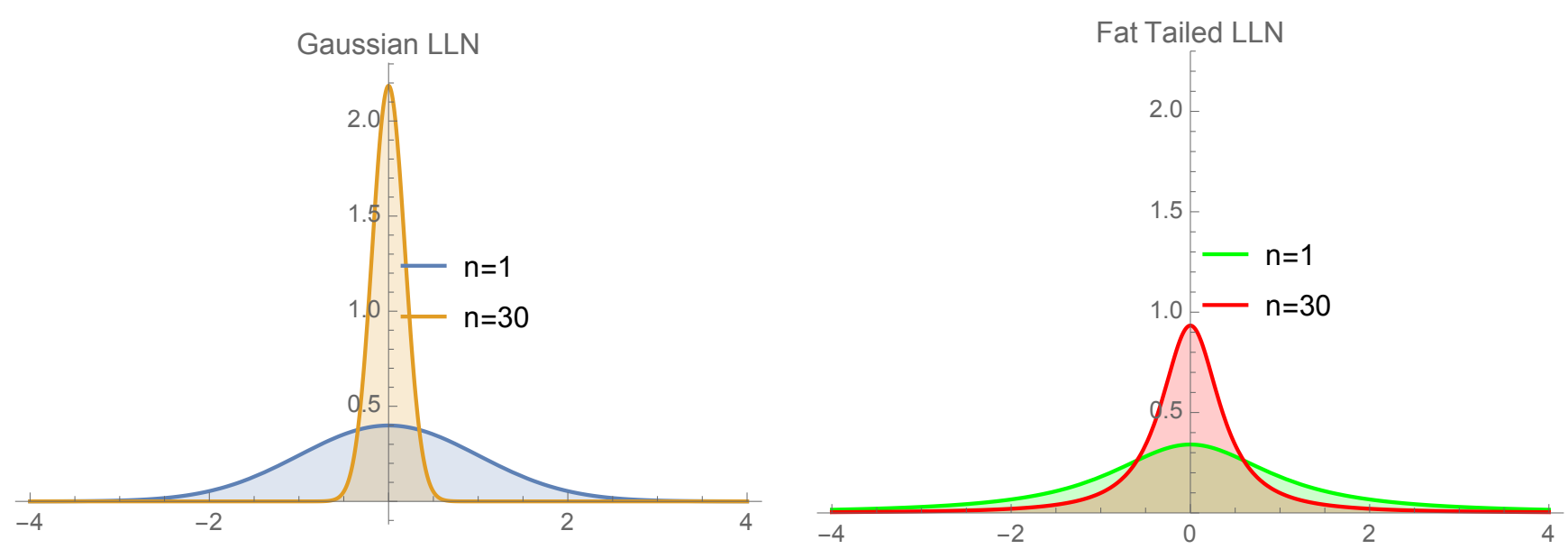

Figure 3.6: *What happens to the distribution of an average as the number of observations n increases? This is the same representation as in Figure 3.5 seen in distribution/probability space. The fat tailed distribution does not compress as easily as the Gaussian. You need a much, much larger sample. It is what it is.*

It also explains why one should never compare random variables driven by the tails (say, pandemics) to ones driven by the body (say, number of people who drown in their swimming pool). See Cirillo and Taleb (2020) [57] for the policy implications of systemic risks.

## 3.3 A (MORE ADVANCED) CATEGORIZATION AND ITS CONSEQUENCES

Let us now consider the degrees of thick tailedness in a casual way for now (we will get deeper and deeper later in the book). The ranking is by severity.

Distributions:

**Thick Tailed ⊃ Subexponential ⊃ Power Law (Paretian)**

First there are entry level thick tails. This is any distribution with fatter tails than the Gaussian i.e. with *more* observations within $\pm 1$ standard deviation than $\text{erf}\left(\frac{1}{\sqrt{2}}\right) \approx 68.2\%$[3] and with kurtosis (a function of the fourth central moment) higher than 3 [4].

Second, there are subexponential distributions satisfying our thought experiment earlier (the one illustrating the catastrophe principle). Unless they enter the class of power laws, distributions are not really thick tailed because they do not have monstrous impacts from rare events. In other words, they can have all the moments .

Level three, what is called by a variety of names, power law, or member of the regular varying class, or the "Pareto tails" class; these correspond to real thick tails but the fattailedness depends on the parametrization of their tail index. Without getting into a tail index for now, consider that there will be *some* moment that is infinite, and moments higher than that one will also be infinite.

[3] The error function erf is the integral of the Gaussian distribution $\text{erf}(z) = \frac{2}{\sqrt{\pi}} \int_0^z dt e^{-t^2}$.

[4] The moment of order $p$ for a random variable $X$ is the expectation of a $p$ power of $X$, $\mathbb{E}(X^p)$.

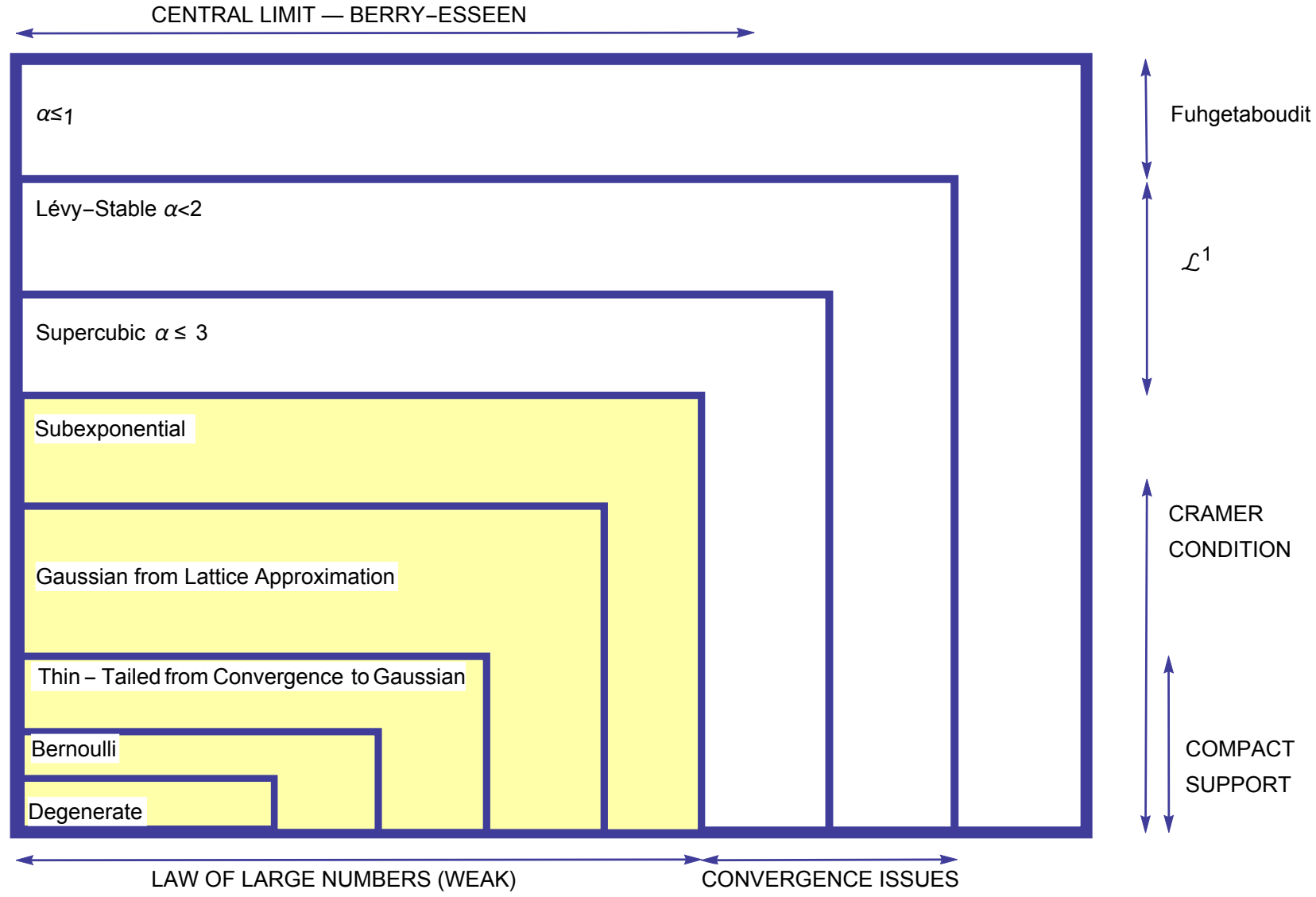

Figure 3.7: *The tableau of thick tails, along the various classifications for convergence purposes (i.e., convergence to the law of large numbers, etc.) and gravity of inferential problems. Power Laws are in white, the rest in yellow. See Embrechts et al [97].*

Let us now work our way from the bottom to the top of the central tableau in Figure 3.7. At the bottom left we have the degenerate distribution where there is only one possible outcome i.e. no randomness and no variation. Then, above it, there is the Bernoulli distribution which has two possible outcomes, not more. Then above it there are the two Gaussians. There is the natural Gaussian (with support on minus and plus infinity), and Gaussians that are reached by adding random walks (with compact support, sort of, unless we have infinite summands)[5]. These are completely different animals since one can deliver infinity and the other cannot (except asymptotically). Then above the Gaussians sit the distributions in the subexponential class that are not members of the power law class. These members have all moments. The subexponential class includes the lognormal, which is one of the strangest things in statistics because sometimes it cheats and fools us. At low variance, it is thin-tailed; at high variance, it behaves like the very thick tailed. Some people take it as good news that the data is not Paretian but lognormal; it is not necessarily so. Chapter 8 gets into the weird properties of the lognormal.

5 Compact support means the real-valued random variable $X$ takes realizations in a bounded interval, say $[a,b]$,$(a,b]$, $[a,b)$, etc. The Gaussian has an exponential decline $e^{-x^2}$ that accelerates with deviations, so some people such as Adrien Douady consider it effectively of compact support.

Membership in the subexponential class does not satisfy the so-called Cramer condition, allowing insurability, as we illustrated in Figure 3.1, recall out thought experiment in the beginning of the chapter. More technically, the Cramer condition means that the expectation of the exponential of the random variable exists.[6]

Once we leave the yellow zone, where the law of large numbers (LLN) largely works[7], and the central limit theorem (CLT) eventually ends up working[8], then we encounter convergence problems. So here we have what are called power laws. We rank them by their tail index $\alpha$, on which later; take for now that the lower the tail index, the fatter the tails. When the tail index is $\alpha \leq 3$ we call it supercubic ($\alpha = 3$ is cubic). That's an informal borderline: the distribution has no moment other than the first and second, meaning both the laws of large number and the central limit theorem apply in theory.

Then there is a class with $\alpha \leq 2$ we call the Levy-Stable to simplify (though it includes similar power law distributions with $\alpha$ less than 2 not explicitly in that class; but in theory, as we add add up variables, the sum ends up in that class rather than in the Gaussian one thanks to something called the generalized central limit theorem, GCLT ). From here up we are increasingly in trouble because there is no variance. For $1 \leq \alpha \leq 2$ there is no variance, but mean absolute deviation (that is, the average variations taken in absolute value) exists.

Further up, in the top segment, there is no mean. We call it the Fuhgetaboudit. If you see something in that category, you go home and you don't talk about it.

The traditional statisticians approach to thick tails has been to claim to assume a different distribution but keep doing business as usual, using same metrics, tests, and statements of significance. Once we leave the yellow zone, for which statistical techniques were designed (even then), things no longer work as planned. The next section presents a dozen issues, almost all terminal. We will get a bit more technical and use some jargon.

## 3.4 THE MAIN CONSEQUENCES AND HOW THEY LINK TO THE BOOK

Here are some consequences of moving out of the yellow zone, the statistical comfort zone:

---

[6] Technical point: Let $X$ be a random variable. The Cramer condition: for all $r > 0$,

$$\mathbb{E}(e^{rX}) < +\infty,$$

where $\mathbb{E}$ is the expectation operator.

[7] Take for now the following definition for the law of large numbers: it roughly states that if a distribution has a finite mean, and you add independent random variables drawn from it —that is, your sample gets larger— you eventually converge to the mean. How quickly? that is the question and the topic of this book.

[8] We will address ad nauseam the central limit theorem but here is the initial intuition. It states that $n$-summed independent random variables with finite second moment end up looking like a Gaussian distribution. Nice story, but how fast? Power laws on paper need an infinity of such summands, meaning they never really reach the Gaussian. Chapter 7 deals with the limiting distributions and answers the central question: "how fast?" both for CLT and LLN. How fast is a big deal because in the real world we have something different from $n$ equals infinity.

**Summary of the problem with overstandardized statistics**

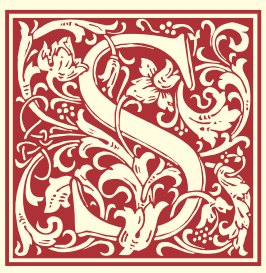

TATISTICAL ESTIMATION is based on two elements: the central limit theorem (which is assumed to work for "large" sums, thus making about everything conveniently normal) and that of the law of large numbers, which reduces the variance of the estimation as one increases the sample size. However, things are not so simple; there are caveats. In Chapter 8, we show how sampling is distribution dependent, and varies greatly within the same class. As shown by Bouchaud and Potters in [31] and Sornette in [254], the tails for some finite variance but infinite higher moments can, under summation, converge to the Gaussian within $\pm\sqrt{n \log n}$, meaning the center of the distribution inside such band becomes Gaussian, but remote parts, those tails, don't –and the remote parts determine so much of the properties.

Life happens in the preasymptotics.

Sadly, in the entry on estimators in the monumental *Encyclopedia of Statistical Science* [175], W. Hoeffding writes:

> "The exact distribution of a statistic is usually highly complicated and difficult to work with. Hence the need to approximate the exact distribution by a distribution of a simpler form whose properties are more transparent. The limit theorems of probability theory provide an important tool for such approximations. In particular, the classical central limit theorems state that the sum of a large number of independent random variables is approximately normally distributed under general conditions. In fact, the normal distribution plays a dominating role among the possible limit distributions. To quote from Gnedenko and Kolmogorov's text [[129], Chap. 5]:
>
> > *"Whereas for the convergence of distribution functions of sums of independent variables to the normal law only restrictions of a very general kind, apart from that of being infinitesimal (or asymptotically constant), have to be imposed on the summands, for the convergence to another limit law some very special properties are required of the summands"*.
>
> Moreover, many statistics behave asymptotically like sums of independent random variables. All of this helps to explain the importance of the normal distribution as an asymptotic distribution."

Now what if we do not reach the normal distribution, as life happens before the asymptote? This is what this book is about.[a]

[a] The reader is invited to consult a "statistical estimation" entry in any textbook or online encyclopedia. Odds are that the notion of "what happens if we do not reach the asymptote" will never be discussed –as in the 9500 pages of the monumental *Encyclopedia of Statistics*. Further, ask a regular user of statistics about how much data one needs for such and such distributions, and don't be surprised at the answer. The problem is that people have too many prepackaged statistical tools in their heads, ones they never had to rederive themselves. The motto here is: "statistics is never standard".

Figure 3.8: *In the presence of thick tails, we can fit markedly different regression lines to the same story (the Gauss-Markov theorem —necessary to allow for linear regression methods —doesn't apply anymore). Left: a regular (naïve) regression. Right: a regression line that tries to accommodate the large deviation —a "hedge ratio" so to speak, one that protects the agent from a large deviation, but mistracks small ones. Missing the largest deviation can be fatal. Note that the sample doesn't include the critical observation, but it has been guessed using "shadow mean" methods.*

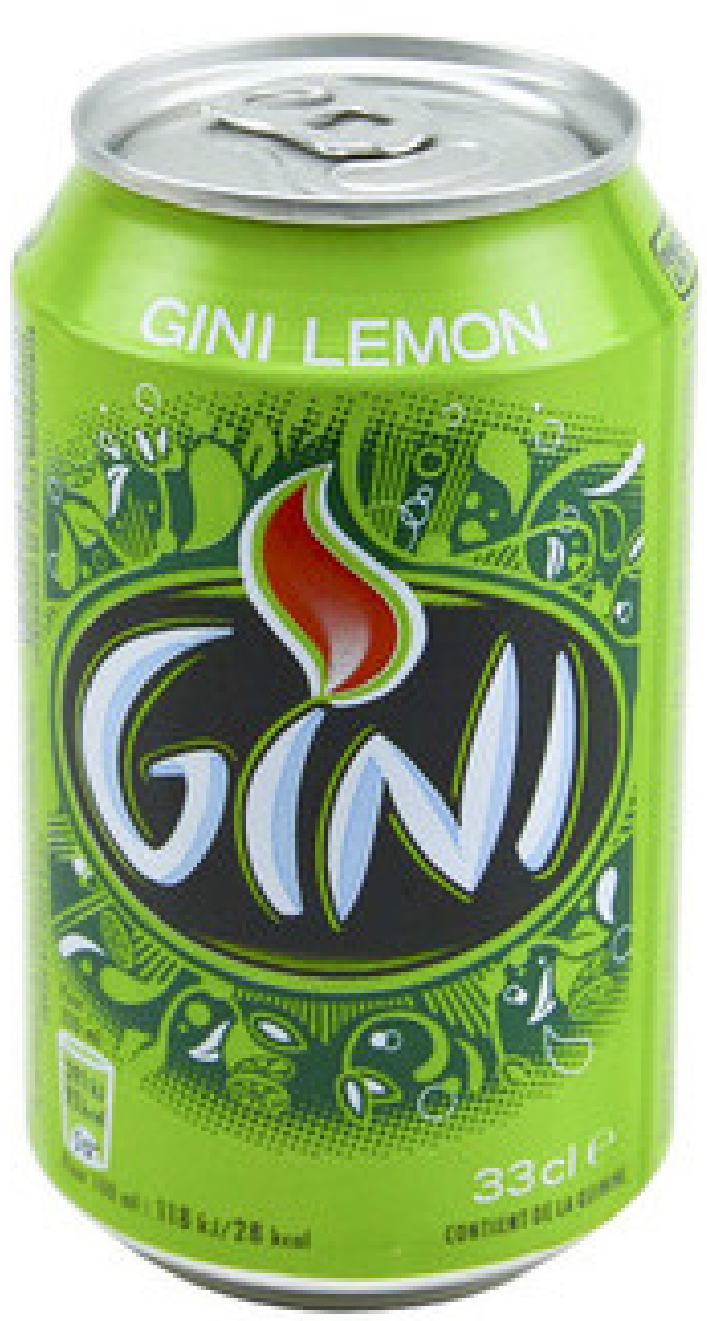

Figure 3.9: *Inequality measures such as the Gini coefficient require completely different methods of estimation under thick tails, as we will see in Part III. Science is hard.*

**Consequence 1**

*The law of large numbers, when it works, works too slowly in the real world.*

This is more shocking than you think as it cancels most statistical estimators. See Figure 3.5 in this chapter for an illustration. The subject is treated in Chapter 8 and distributions are classified accordingly.[9]

**Consequence 2**

*The mean of the distribution will rarely correspond to the sample mean; it will have a persistent small sample effect (downward or upward) particularly when the distribution is skewed (or one-tailed).*

This is another problem of sample insufficiency. In fact, there is no very thick tailed- one tailed distribution in which the population mean can be properly estimated directly from the sample mean –rare events determine the mean, and these, *being rare*, take a lot of data to show up[10]. Consider that some power laws (like the one described as the "80/20" in common parlance have 92 percent of the observations falling below the true mean). For the sample average to be informative, we need orders of magnitude more data than we do (people in economics still do not understand this, though traders have an intuitive grasp of the point). The problem is discussed briefly further down in 3.8, and more formally in the "shadow mean" chapters, Chapters 17 and 18. Further, we will introduce the notion of hidden properties are in 3.8. Clearly by the same token, variance will be likely to be underestimatwd.

**Consequence 3**

*Metrics such as standard deviation and variance are not useable.*

They fail out of sample –even when they exist; even when all moments exist. Discussed in ample details in Chapter 4. It is a scientific error that the notion of standard deviation (often mistaken for average deviation by its users) found its way as a measure of variation as it is very narrowly accurate in what it purports to do, in the best of circumstances.

**Consequence 4**

*Beta, Sharpe Ratio and other common hackneyed financial metrics are uninformative.*

This is a simple consequence of the previous point.[11] Either they require much more data, many more orders of magnitude, or some different model than the

---

[9] What we call preasymptotics is the behavior of a sum or sequence when $n$ is large but not infinite. This is (sort of) the focus of this book.

[10] The population mean is the average if we sampled the entire population. The sample mean is, obviously, what we have in front of us. Sometimes, as with wealth or war casualties, we can have the entire population, yet the population mean isn't that of the sample. In these situations we use the concept of "shadow mean", which is the expectation as determined by the data generating process or mechanism.

[11] Roughly, Beta is a metric showing how much an asset $A$ is expected to move in response to a move in the general market (or a given benchmark or index), expressed as the ratio of the covariance between $A$ and the market over the variance of the market.

The Sharpe ratio expresses the average return (or excess return) of an asset or a strategy divided by its standard deviation.

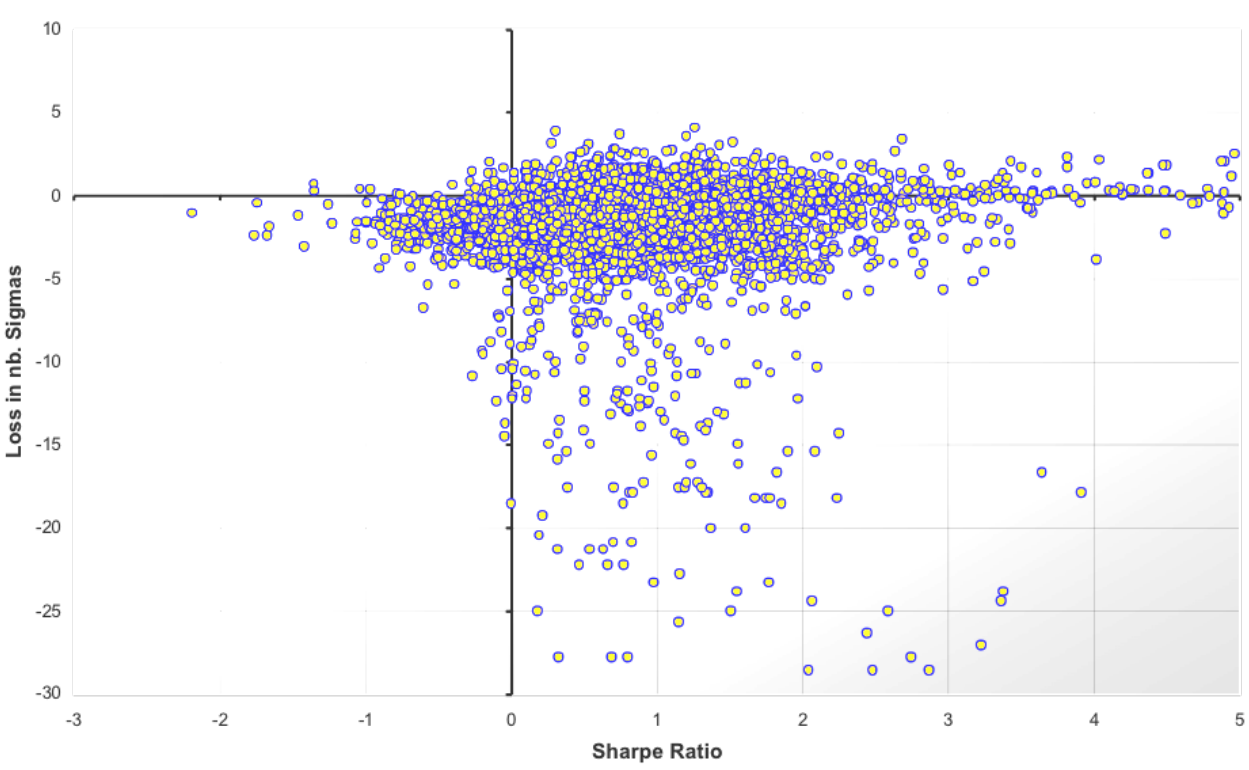

Figure 3.10: *We plot the Sharpe ratio of hedge funds on the horizontal axis as computed up to crisis of 2008 and their subsequent losses expressed in standard deviation during the crisis. Not only does the Sharpe ratio completely fail to predict out of sample performance, but, if anything, it can be seen as a weak predictor of failure. Courtesy Raphael Douady.*

one being used, of which we are not yet aware. Figure 3.4 show the Sharpe ratio, supposed to predict performance, fails out of sample — it acts in exact reverse of the intention. Yet it is still used because people can be suckers for numbers.

Practically every single economic variable and financial security is thick tailed. Of the 40,000 securities examined, not one appeared to be thin-tailed. This is the main source of failure in finance and economics.

Financial theorists claim something highly unrigorous like "if the first two moments exist, then mean-variance portfolio theory works, even if the distribution has thick tails" (they add some conditions of ellipticality we will discuss later). The main problem is that *even if variance exists, we don't know what it can be with acceptable precision; it obeys a slow law of large numbers because the second moment of a random variable is necessarily more thick tailed than the variable itself*. Further, stochastic correlations or covariances also represent a form of thick tails (or loss of ellipticality), which invalidates these metrics.

Practically any paper in economics using covariance matrices is suspicious.

Details are in Chapter 4 for the univariate case and Chapter 6 for multivariate situations.

**Consequence 5**

*Robust statistics is not robust and the empirical distribution is not empirical.*

The story of my life. Much like the Soviet official journal was named *Pravda* which means "truth" in Russian, almost as a joke, robust statistics are like a type of prank, except that most professionals aren't aware of it.

First, robust statistics shoots for measures that can handle tail events —large observations —without changing much. This the wrong idea of robustness: a metric that doesn't change in response of a tail event may be doing so precisely because it is uninformative. Further, these measures do not help with expected payoffs. Second, robust statistics are usually associated with a branch called "nonparametric" statistics, under the impression that the absence of parameters will make the analysis less distribution dependent. This book shows all across that it makes things worse.

The winsorization of the data, by removing outliers, distorts the expectation operation and actually reduces information –though it would be a good idea to check if the outlier is real or a fake outlier of the type we call in finance a "bad print" (some clerical error or computer glitch).

The so-called (nonparametric) "empirical distribution" is not empirical at all (as it misrepresents the expected payoffs in the tails), as we will show in Chapter 10 —this is at least the case for the way it is used in finance and risk management. Take for now the following explanation: future maxima are poorly tracked by past data without some intelligent extrapolation.

Consider someone looking at building a flood protection system with levees. The naively obtained "empirical" distribution will show the worst past flood level, the past maxima. Any worse level will have zero probability (or so). But by definition, if it was a past maxima, it had to have exceeded what was a past maxima before it to become one, and the empirical distribution would have missed it. For thick tails, the difference between past maxima and future expected maxima is much larger than thin tails.

**Consequence 6**

*Linear least-square regression doesn't work (failure of the Gauss-Markov theorem).*

See Figure 3.8 and the commentary. The logic behind the least-square minimization method is the Gauss-Markov theorem which explicitly requires a thin-tailed distribution to allow the line going through the data points to be unique. So either we need a lot, a lot of data to minimize the squared deviations (in other words, the Gauss-Markov theorem applies, but not for our preasymptotic situations as the real world has finite, not infinite data), or we can't because the second moment does not exist. In the latter case, if we minimize mean absolute deviations (MAD), as we see in 4.1, not only we may still be facing an insufficiency of data for proper convergence, but the deviation slope may not be unique.

We discuss the point in some details in 6.7 and show how thick tails produce an in-sample higher coefficient of determination ($R^2$) than the real one because of the small sample effect of thick tails. When variance is infinite, $R^2$ should be 0. But because samples are necessarily finite, it will show, deceivingly, higher numbers than 0. Effectively, to conclude, under thick tails, $R^2$ is useless, uninformative, and often (as with IQ studies) downright fraudulent.

**Consequence 7**

*Maximum likelihood methods can work well for some parameters of the distribution (good news).*

Take a power law. We may estimate a parameter for its shape, the tail exponent (for which we use the symbol $\alpha$ in this book[12]), which, adding some other parameter (the scale) connects us back to its mean considerably better than going about it directly by sampling the mean.

**Example:** The mean of a simple Pareto distribution with minimum value $L$ and tail exponent $\alpha$ and PDF $\alpha L^{\alpha} x^{-\alpha-1}$ is $L\frac{\alpha}{\alpha-1}$, a function of $\alpha$. So we can get it from these two parameters, one of which may already be known. This is what we call "plug-in" estimator. One can estimate $\alpha$ with a low error with visual aid (or using maximum likelihood methods with low variance — it is inverse-gamma distributed), then get the mean. It beats the direct observation of the mean.

The logic is worth emphasizing:

> The tail exponent $\alpha$ captures, by extrapolation, the low-probability deviation not seen in the data, but that plays a disproportionately large share in determining the mean.

This generalized approach to estimators is also applied to Gini and other inequality estimators.

So we can produce more reliable (or at least less unreliable) estimators for, say, a function of the tail exponent in *some* situations. But, of course, not all.

Now a real-world question is warranted: what do we do when we do not have a reliable estimator? Better stay home. We must not expose ourselves to harm in the presence of fragility, but can still take risky decisions if we are bounded for maximum losses (Figure 3.4).

> **Consequence 8**
> *The gap between disconfirmatory and confirmatory empiricism is wider than in situations covered by common statistics i.e., the difference between absence of evidence and evidence of absence becomes larger. (What is called "evidence based" science, unless rigorously disconfirmatory, is usually interpolative, evidence-free, and unscientific.)*

From a controversy the author had with the cognitive linguist and science writer Steven Pinker: making pronouncements (and generating theories) from recent variations in data is not acceptable, unless one meets some standards of significance, which requires more data under thick tails (the same logic as that of the slow LLN). Stating "violence has dropped" because the number of people killed in wars has declined from the previous year or decade is not a scientific statement: a scientific claim distinguishes itself from an anecdote as it aims at affecting what happens out of sample, hence the concept of statistical significance.

Let us repeat that nonstatistically significant statements are not the realm of science. However, saying violence has risen upon a single observation may be a rigorously scientific claim. The practice of reading into descriptive statistics may be acceptable under thin tails (as sample sizes do not have to be large), but never so under thick tails, except, to repeat, in the presence of a large deviation.

[12] To clear up the terminology: in this book, the tail exponent, commonly written $\alpha$ is the limit of quotient of the log of the survival function in excess of K over log K, which would be 1 for Cauchy. Some researchers use $\alpha - 1$ from the corresponding density function.

**Consequence 9**

*Principal component analysis (PCA) and factor analysis are likely to produce spurious factors and loads.*

This point is a bit technical; it adapts the notion of sample insufficiency to large random vectors seen via the dimension reduction technique called principal component analysis (PCA) . The issue a higher dimensional version of our law of large number complications. The story is best explained in Figure 3.26, which shows the accentuation of what is called the "Wigner Effect", from insufficiency of data for the PCA. Also, to be technical, note that the Marchenko-Pastur distribution is not applicable in the absence of a finite fourth moment (or, has been shown in [27], for tail exponent in excess of 4).[13]

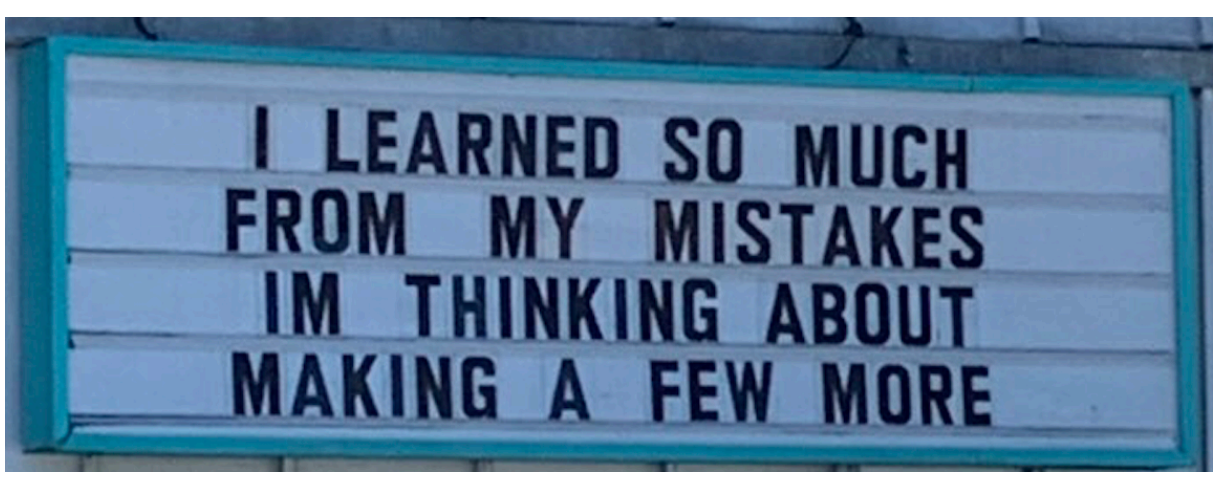

Figure 3.11: *Under thick tails (to the left), mistakes are terminal. Under thin tails (to the left) they can be great learning experiences. Source: You had one Job.*

**Consequence 10**

*The method of moments (MoM) fails to work. Higher moments are uninformative or do not exist.*

The same applies to the GMM, the generalized method of moment, crowned with a Bank of Sweden Prize known as a Nobel. This is a long story, but take for now that the estimation of a given distribution by moment matching fails if higher moments are not finite, so every sample delivers a different moment –as we will soon see with the $4^{th}$ moment of the SP500.

Simply, higher moments for thick tailed distributions are explosive. Particularly in economics.

**Consequence 11**

*There is no such thing as a typical large deviation.*

Conditional on having a "large" move, the magnitude of such a move is not convergent, especially under serious thick tails (the Power Law tails class). This is associated with the catastrophe principle we saw earlier. In the Gaussian world, the expectation of a movement, conditional that the movement exceeds 4 standard deviations, is about 4 standard deviations. For a Power Law it will be a multiple of that. We call this the Lindy property and it is discussed in 5 and particularly in Chapter 11.

[13] To be even more technical, principal components are independent when correlations are 0. However, for fat tailed distributions, as we will see more technically in 6.3.1, absence of correlation does not imply independence.

**Consequence 12**

*The Gini coefficient ceases to be additive..*

Methods of measuring sample data for Gini are interpolative –they in effect have the same problem we saw earlier with the sample mean underestimating or overestimating the true mean. Here, an additional complexity arises as the Gini becomes super-additive under thick tails. As the sampling space grows, the conventional Gini measurements give an illusion of large concentrations of wealth. (In other words, inequality in a continent, say Europe, can be higher than the weighted average inequality of its members). The same applies to other measures of concentration such as the top 1% has $x$ percent of the total wealth, etc.

It is not just Gini, but other measures of concentration such as the top 1% owns $x$% of the total wealth, etc. The derivations are in Chapters 15 and 16.

**Consequence 13**

*Large deviation theory fails to apply to thick tails. I mean, it really doesn't apply.*

I really mean it doesn't apply[14]. The methods behind the large deviation principle (Varadan [314] , Dembo and Zeituni [71], etc.) will be very useful in the thin-tailed world. And there only. See discussion and derivations in Appendix D as well as the limit theorem chapters, particularly Chapter 7.

**Consequence 14**

*Risks of financial options are never mitigated by dynamic hedging.*

This might be technical and uninteresting for nonfinance people but the entire basis of financial hedging behind Black-Scholes rests on the possibility and necessity of dynamic hedging, both of which will be shown to be erroneous in Chapters 24 and25 ,and 26. The required exponential decline of deviates away from the center requires the probability distribution to be outside the sub-exponential class. Again, we are talking about something related the Cramer condition –it all boils down to that exponential moment.

Recall the author has been an option trader and to option traders dynamic hedging is not the way prices are derived —and it has been so, as shown by Haug and the author, for centuries.

**Consequence 15**

*Forecasting in frequency space diverges from expected payoff.*

[14] Do not confuse large deviation theory LDT, with extreme value theory, EVT, which covers all major classes of distributions

And also:

**Consequence 16**

*Much of the claims in the psychology and decision making literature concerning the "overestimation of tail probability" and irrational behavior with respect of rare events comes form misunderstanding by researchers of tail risk, conflation of probability and expected payoffs, misuse of probability distributions, and ignorance of extreme value theory (EVT).*

These point is explored in the next section here and in an entire chapter (Chapter **??**): the foolish notion of focus on frequency rather than expectation can carry a mild effect under thin tails; not under thick tails. Figures 3.12 and 3.13 show the effect.

**Consequence 17**

*Ruin problems are more acute and ergodicity is required under thick tails.*

This is a bit technical but explained in the end of this chapter.

Let us discuss some of the points.

### 3.4.1 Forecasting

In *Fooled by Randomness* (2001/2005), the character is asked which was *more probable* that a given market would go higher or lower by the end of the month. Higher, he said, much more probable. But then it was revealed that he was making trades that benefit if that particular market *goes down*. This of course, appears to be paradoxical for the nonprobabilist but very ordinary for traders, particularly under nonstandard distributions (yes, the market is more likely to go up, but should it go down it will fall much much more). This illustrates the common confusion between a *forecast* and an exposure (a forecast is a binary outcome, an exposure has more nuanced results and depends on full distribution). This example shows one of the extremely elementary mistakes of talking about *probability presented* as single numbers not distributions of outcomes, but when we go deeper into the subject, many less obvious, or less known paradox-style problems occur. Simply, it is of the opinion of the author, that it is not rigorous to talk about "probability" as a final product, or even as a "foundation"of decisions.

In the real world one is not paid in probability, but in dollars (or in survival, etc.). The fatter the tails, the more one needs to worry about payoff space – the saying goes: "payoff swamps probability" (see box). One can be wrong very frequently if the cost is low, so long as one is convex to payoff (i.e. make large gains when one is right). Further, one can be forecasting with 99.99% accuracy and still go bust (in fact, more likely to go bust: funds with impeccable track records were those that went bust during the 2008-2009 rout [15]). A point that may be technical for those outside quantitative finance: it is the difference between a vanilla option

[15] R. Douady, data from Risk Data about funds that collapsed in the 2008 crisis, personal communication

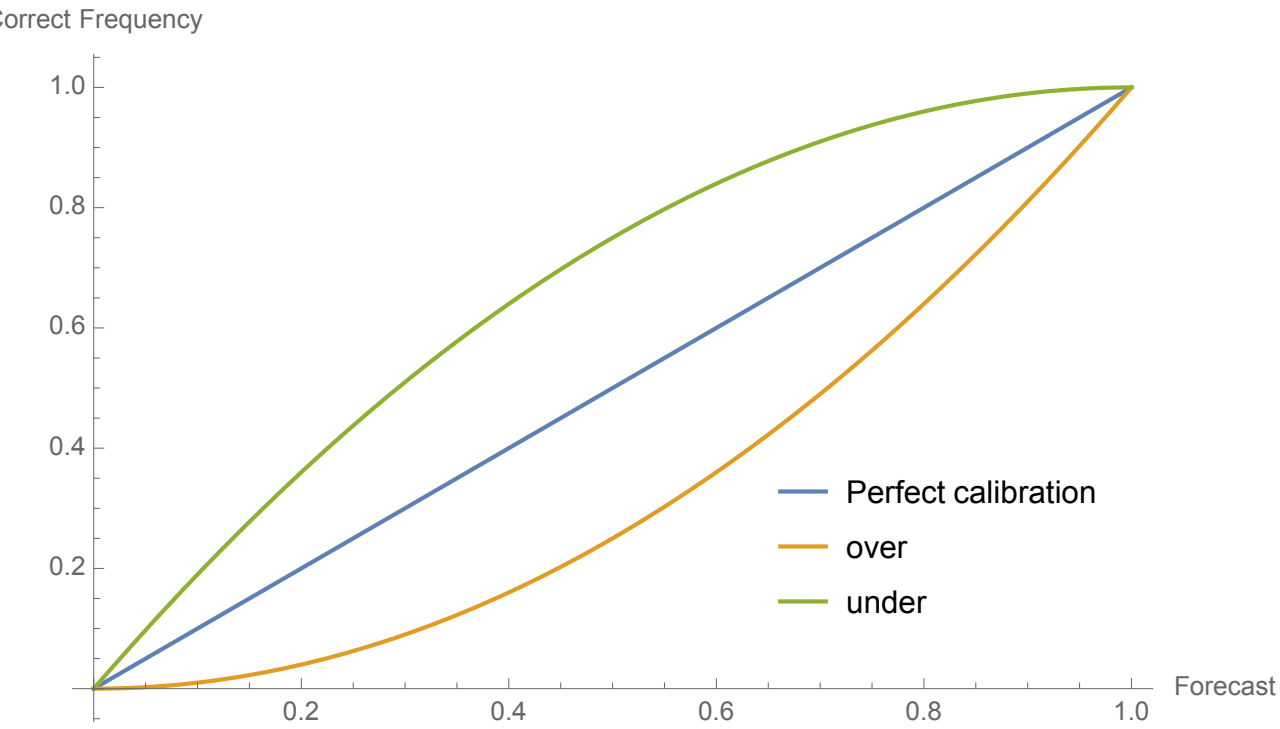

Figure 3.12: *Probabilistic calibration as seen in the psychology literature. The x axis shows the estimated probability produced by the forecaster, the y axis the actual realizations, so if a weather forecaster predicts* 30% *chance of rain, and rain occurs* 30% *of the time, they are deemed "calibrated". We hold that calibration in frequency (probability) space is an academic exercise (in the bad sense of the word) that mistracks real life outcomes outside narrow binary bets. It is particularly fallacious under thick tails. The point is discussed at length in Chapter 11.*

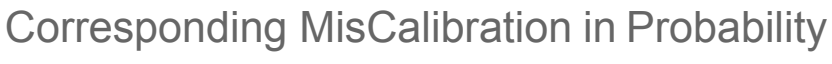

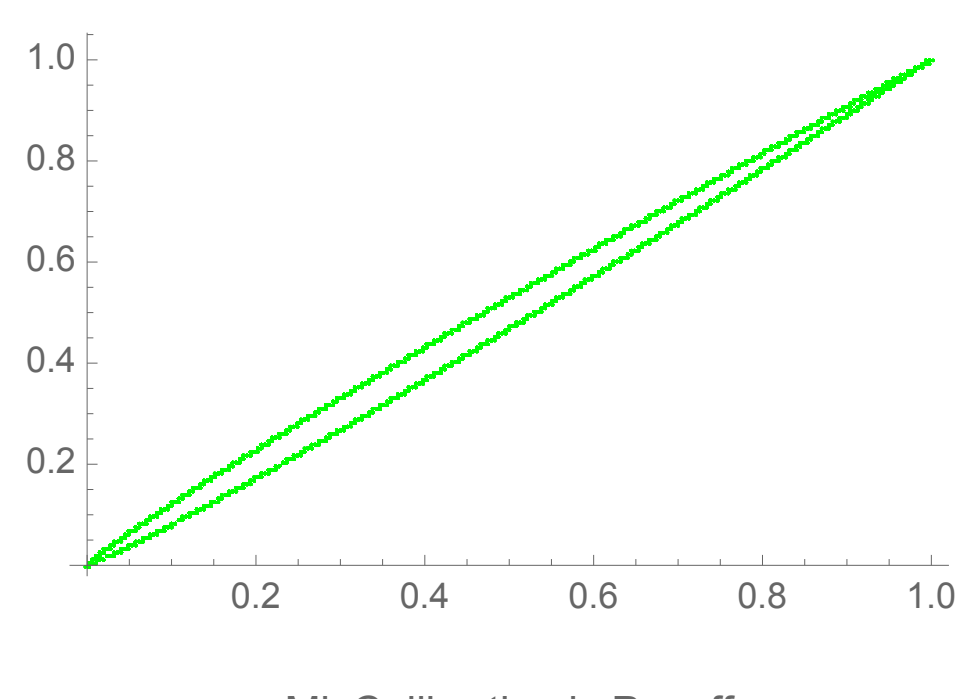

MisCalibration in Payoff

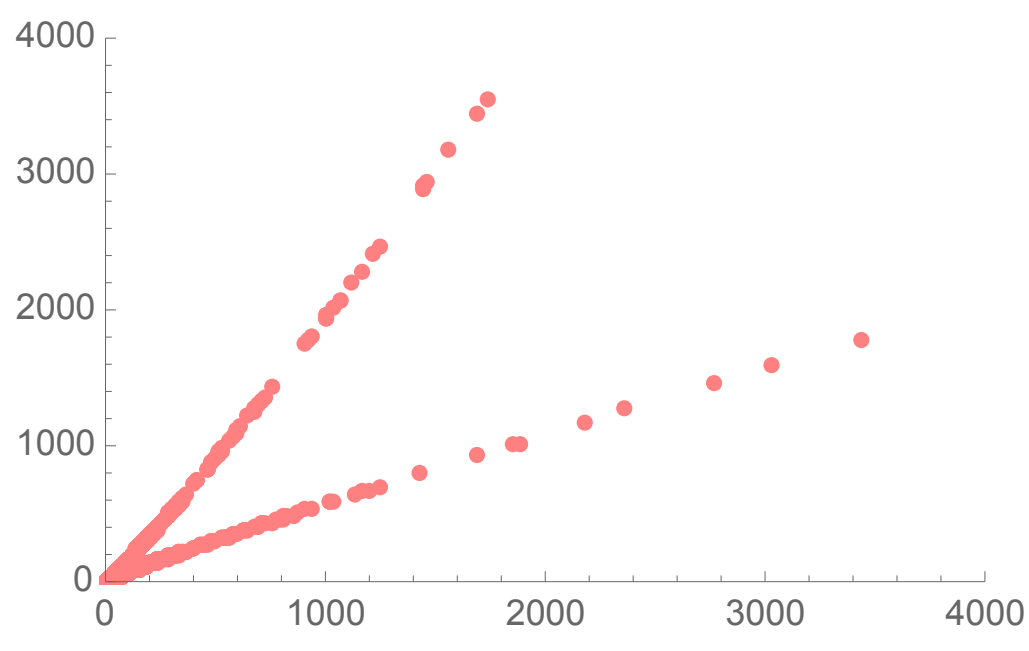

Figure 3.13: *How miscalibration in probability corresponds to miscalibration in payoff under power laws. The generated distribution is Pareto with tail index* $\alpha = 1.15$.

and a corresponding binary of the same strike, as discussed in *Dynamic Hedging* [271]: counterintuitively, thick tailedness *lowers* the value of the binary and raise that of the vanilla. This is expressed by the author's adage: "I've never seen a rich forecaster." We will examine in depth in 4.3.1 where we show that fattening the tails cause the probability of events higher than 1 standard deviations to drop –but the consequences to rise (in term of contribution to moments, say effect on the mean or other metrics).

Figure 3.12 shows the extent of the problem.

**Remark 3.1**

*Probabilistic forecast errors ("calibration") are in a different probability class from that true real-world P/L variations (or true payoffs).*

*"Calibration", which is a measure of how accurate one's predictions, lies in probability space –between 0 and 1. Any standard measure of such calibration will necessarily be thin-tailed (and, if anything, extra-thin tailed since it is bounded) –whether the random variable under such prediction is thick tailed or not. On the other hand, payoffs in the real world can be thick tailed, hence the distribution of such "calibration" will follow the property of the random variable.*

We show full derivations and proofs in Chapter 11.

### 3.4.2 The Law of Large Numbers

Let us now discuss the law of large numbers which is the basis of much of statistics. The law of large numbers tells us that as we add observations the mean becomes more stable, the rate being around $\sqrt{n}$. Figure 3.5 shows that it takes many more observations under a fat-tailed distribution (on the right hand side) for the mean to stabilize.

The "equivalence" is not straightforward.

One of the best known statistical phenomena is Pareto's 80/20 e.g. twenty percent of Italians own 80 percent of the land. Table 3.1 shows that while it takes 30 observations in the Gaussian to stabilize the mean up to a given level, it takes $10^{11}$ observations in the Pareto to bring the sample error down by the same amount (assuming the mean exists).

Despite this being trivial to compute, few people compute it. You cannot make claims about the stability of the sample mean with a thick tailed distribution. There are other ways to do this, but not from observations on the sample mean.

## 3.5 EPISTEMOLOGY AND INFERENTIAL ASYMMETRY

**Definition 3.1** (Asymmetry in distributions)
*It is much easier for a criminal to fake being an honest person than for an honest person*

**Payoff swamps probability in Extremistan:** To see the main difference between Mediocristan and Extremistan, consider the *event* of a plane crash. A lot of people will lose their lives, something very sad, say between 100 and 400 people, so the event is counted as a bad episode, a single one. For forecasting and risk management, we work on minimizing such a probability to make it negligible.

Now, consider a type of plane crashes that will kill *all* the people who ever rode the plane, even all passengers who ever rode planes in the past. All. Is it the same type of event? The latter event is in Extremistan and, for these, we don't talk about probability but focus instead on the magnitude of the event.

- For the first type, management consists in reducing the probability – the frequency – of such events. Remember that we count events and aim at reducing their counts.
- For the second type, it consists in reducing the effect should such an event take place. We do not count events, we measure impact.

If you think the thought experiment is a bit weird, consider that the money center banks lost in 1982 more money than they ever made in their history, the Savings and Loans industry (now gone) did so in 1991, and the entire banking system lost every penny ever made in 2008-9. One can routinely witness people lose everything they earned cumulatively in a single market event. The same applies to many, many industries (e.g. automakers and airlines).

But banks are only about money; consider that for wars we cannot afford the naive focus on event frequency without taking into account the magnitude, as done by the science writer Steven Pinker in [227], discussed in Chapter 18. This is without even examining the ruin problems (and nonergodicity) presented in the end of this section. More technically, one needs to meet the Cramer condition of non-subexponentiality for a tally of events (taken at face value) for raw probability to have any meaning at all. The plane analogy was proposed by the insightful Russ Robert during one of his *Econtalk* podcasts with the author.

*to fake being a criminal. Likewise it is easier for a fat-tailed distribution to fake being thin tailed than for a thin tailed distribution to fake being thick tailed.*

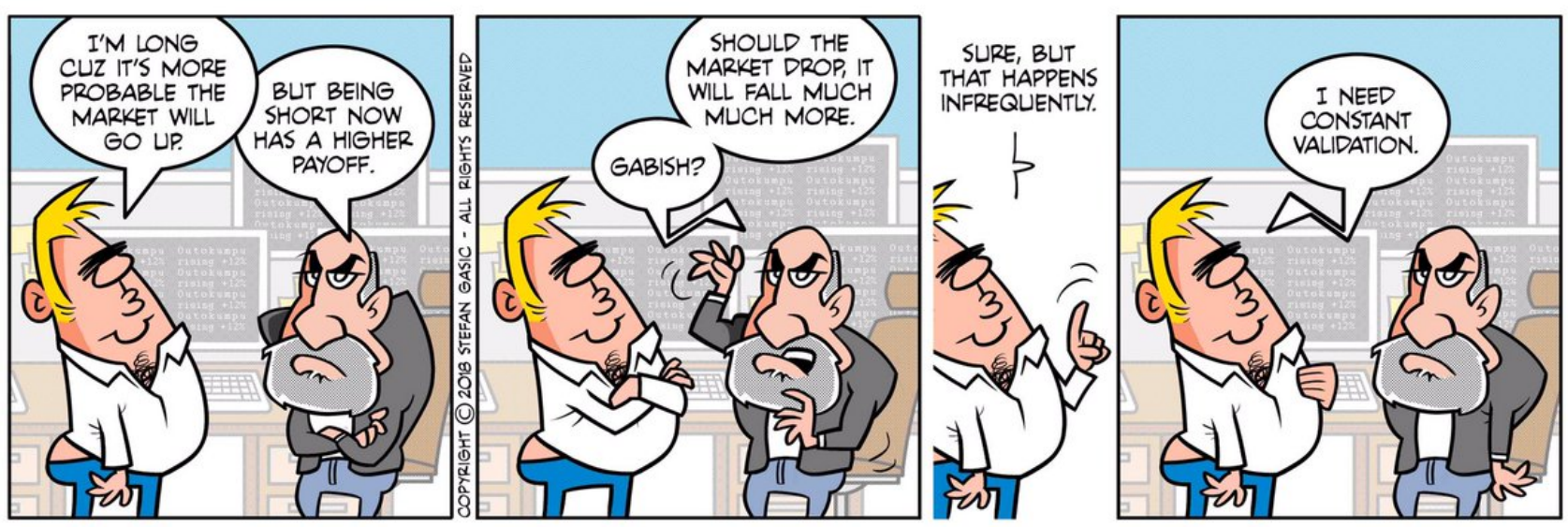

Figure 3.14: *Life is about payoffs, not forecasting, and the difference increases in Extremistan. (Why "Gabish" rather than "capisce"? Gabish is the recreated pronunciation of Siculo-Galabrez (Calabrese); the "p" used to sound like a "b" and the "g" like a Semitic kof, a hard K, from Punic. Much like capicoli is "gabagool".)*

Table 3.1: *Corresponding $n_\alpha$, or how many observations to get a drop in the error around the mean for an equivalent $\alpha$-stable distribution (the measure is discussed in more details in Chapter 8). The Gaussian case is the $\alpha = 2$. For the case with equivalent tails to the 80/20 one needs at least $10^{11}$ more data than the Gaussian.*

| $\alpha$ | $n_\alpha$ Symmetric | $n_\alpha^{\beta=\pm\frac{1}{2}}$ Skewed | $n_\alpha^{\beta=\pm 1}$ One-tailed |
|---|---|---|---|
| 1 | *Fughedaboudit* | - | - |
| $\frac{9}{8}$ | $6.09 \times 10^{12}$ | $2.8 \times 10^{13}$ | $1.86 \times 10^{14}$ |
| $\frac{5}{4}$ | 574,634 | 895,952 | $1.88 \times 10^{6}$ |
| $\frac{11}{8}$ | 5,027 | 6,002 | 8,632 |
| $\frac{3}{2}$ | 567 | 613 | 737 |
| $\frac{13}{8}$ | 165 | 171 | 186 |
| $\frac{7}{4}$ | 75 | 77 | 79 |
| $\frac{15}{8}$ | 44 | 44 | 44 |
| 2 | 30. | 30 | 30 |

**Principle 3.1: Epistemology: the invisibility of the generator.**

- *We do not observe probability distributions, just realizations.*
- *A probability distribution cannot tell you if the realization belongs to it.*
- *You need a meta-probability distribution to discuss tail events (i.e., the conditional probability for the variable to belong to a certain distributions vs. others).*

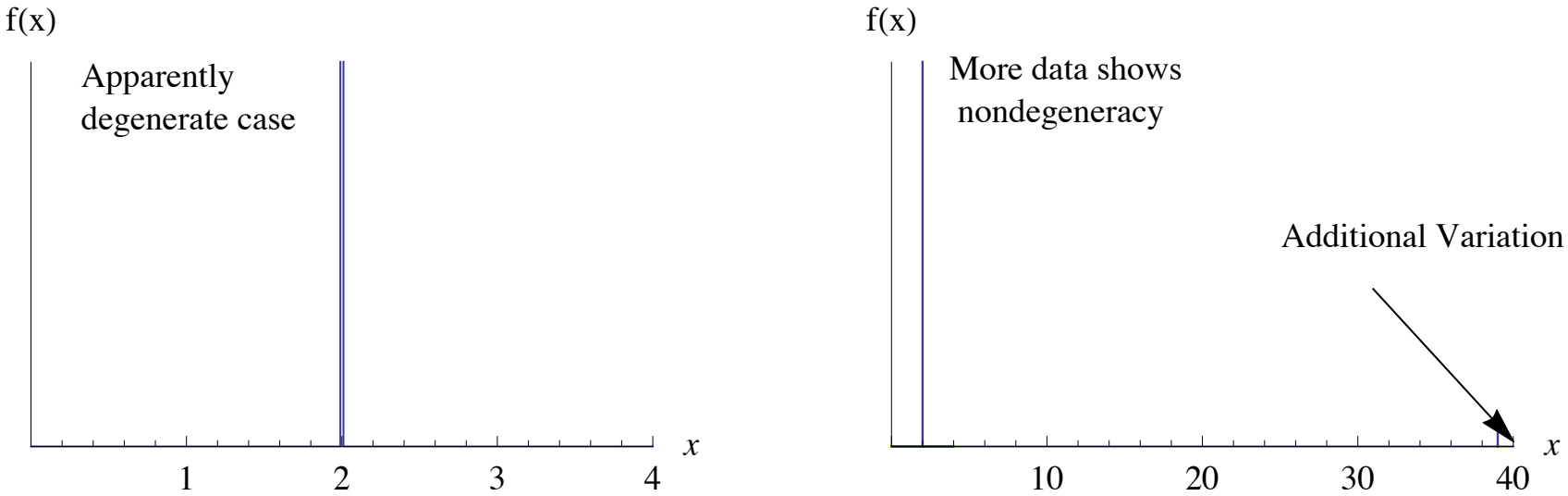

Figure 3.15: ***The Masquerade Problem (or Central Asymmetry in Inference)****. To the left, a degenerate random variable taking seemingly constant values, with a histogram producing a Dirac stick. One cannot rule out nondegeneracy. But the right plot exhibits more than one realization. Here one can rule out degeneracy. This central asymmetry can be generalized and put some rigor into statements like "failure to reject" as the notion of what is rejected needs to be refined. We can use the asymmetry to produce rigorous rules.*

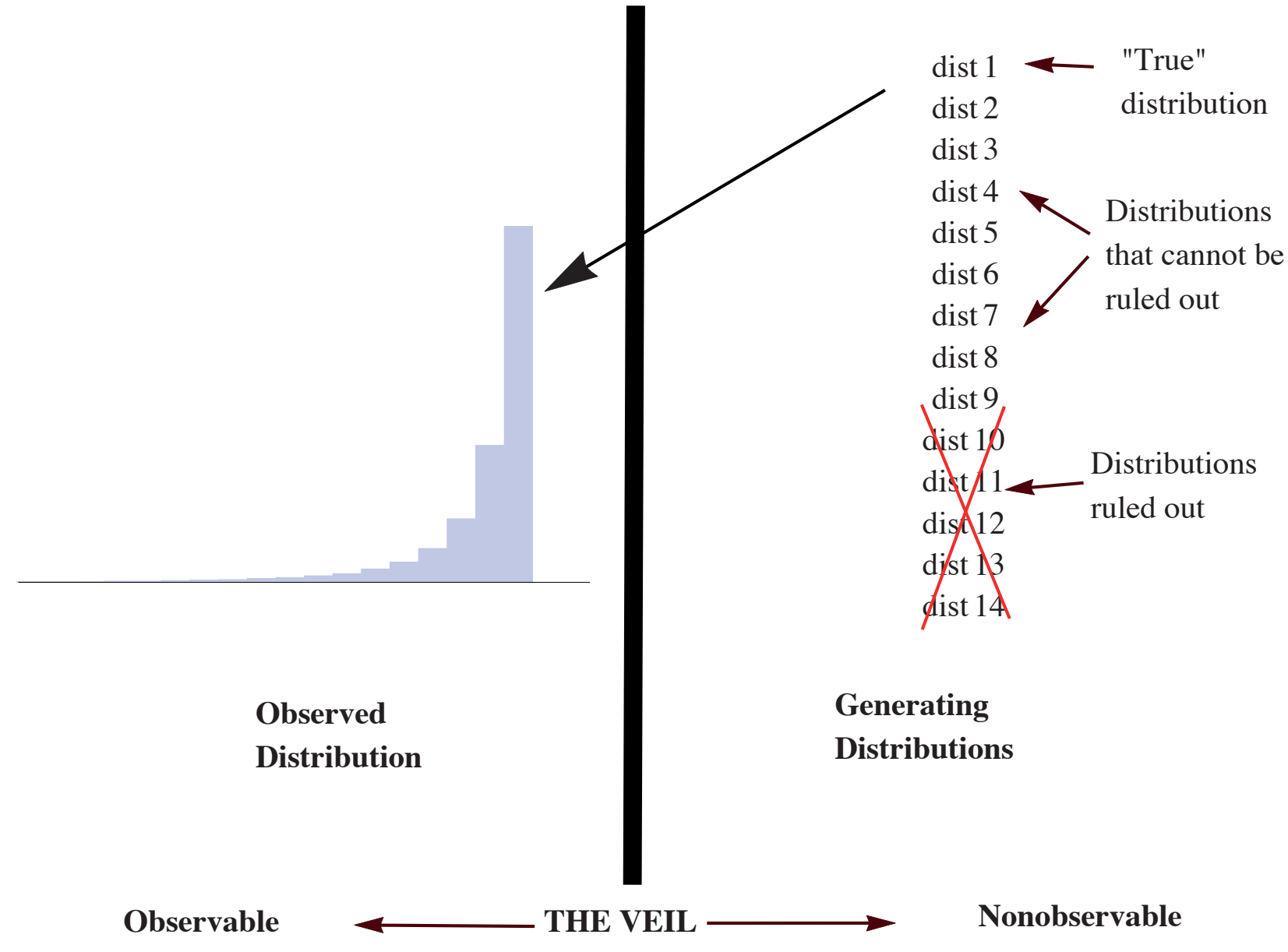

Figure 3.16: ***"The probabilistic veil"****. Taleb and Pilpel [295] cover the point from an epistemological standpoint with the "veil" thought experiment by which an observer is supplied with data (generated by someone with "perfect statistical information", that is, producing it from a generator of time series). The observer, not knowing the generating process, and basing his information on data and data only, would have to come up with an estimate of the statistical properties (probabilities, mean, variance, value-at-risk, etc.). Clearly, the observer having incomplete information about the generator, and no reliable theory about what the data corresponds to, will always make mistakes, but these mistakes have a certain pattern. This is the central problem of risk management.*

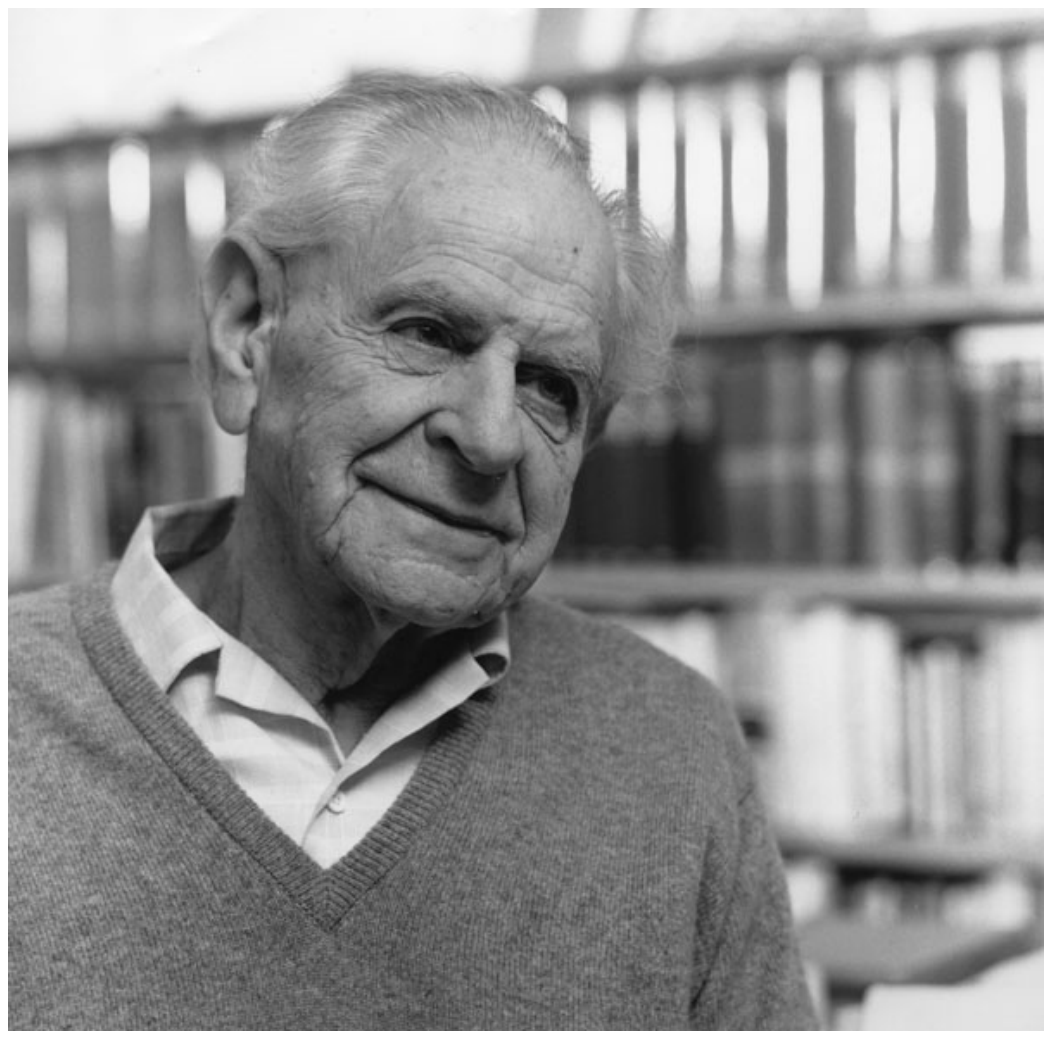

Figure 3.17: *Popper's solution of the problem of induction is in asymmetry: relying on confirmatory empiricism, that is focus on "ruling out" what fails to work, via negativa style. We extend this approach to statistical inference with the probabilistic veil by progressively ruling out entire classes of distributions.*

**Scientific Rigor and Asymmetries by The Russian School of Probability**

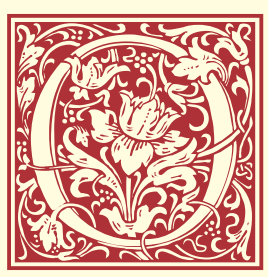NE CAN BELIEVE in the rigor of mathematical statements about probability without falling into the trap of providing naive computations subjected to model error. There is a wonderful awareness of the asymmetry throughout the works of the Russian school of probability –and asymmetry here is the analog of Popper's idea in mathematical space.

Members across three generations: P.L. Chebyshev, A.A. Markov, A.M. Lyapunov, S.N. Bernshtein (ie. Bernstein), E.E. Slutskii, N.V. Smirnov, L.N. Bol'shev, V.I. Romanovskii, A.N. Kolmogorov, Yu.V. Linnik, and the new generation: V. Petrov, A.N. Nagaev, A. Shyrayev, and a few more.

They had something rather potent in the history of scientific thought: they thought in inequalities, not equalities (most famous: Markov, Chebyshev, Bernstein, Lyapunov). They used bounds, not estimates. Even their central limit version was a matter of bounds, which we exploit later by seeing what takes place *outside the bounds*. They were world apart from the new generation of users who think in terms of precise probability –or worse, mechanistic social scientists. Their method accommodates skepticism, one-sided thinking: "$A$ is $> x$, $AO(x)$ [Big-O: "of order" $x$], rather than $A = x$.

For those working on integrating the mathematical rigor in risk bearing they provide a great source. We always know one-side, not the other. We know the lowest value we are willing to pay for insurance, not necessarily the upper bound (or vice versa).[a]

[a] The way this connects asymmetry to robustness is as follows. Is robust what does not produce variability across perturbation of parameters of the probability distribution. If there is change, but with an asymmetry, i.e. a concave or convex response to such perturbations, the classification is fragility and antifragility, respectively, see [269].

Let us now examine the epistemological consequences. Figure 3.15 illustrates the Masquerade Problem (or Central Asymmetry in Inference). On the left is a

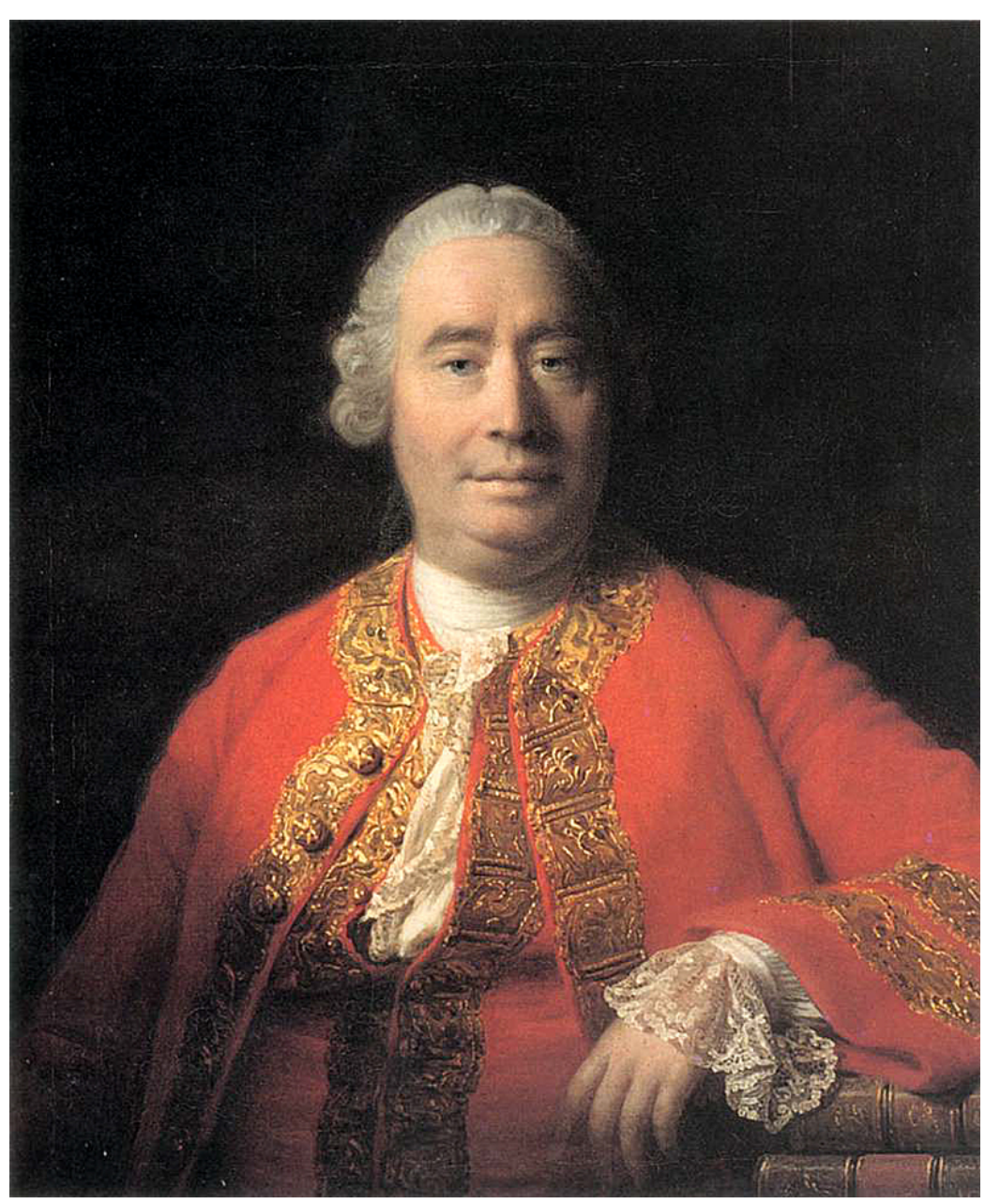

Figure 3.18: *The Problem of Induction. The philosophical problem of enumerative induction, expressed in the question:*
***"How many white swans do you need to count before ruling out the future occurrence of a black one?"***
*maps surprisingly perfectly to our problem of the working of the law of large numbers:*
***"How much data do you need before making a certain claim with an acceptable error rate?"***
*It turns out that the very nature of statistical inference reposes on a clear definition and quantitative measure of the inductive mechanism. It happens that, under thick tails, we need considerably more data; as we will see in Chapters 7 and 8 there is a way to gauge the relative speed of the inductive mechanism, even if ultimately the problem of induction cannot be perfectly solved.*
*The problem said of induction is generally misattributed to Hume, [273] .*

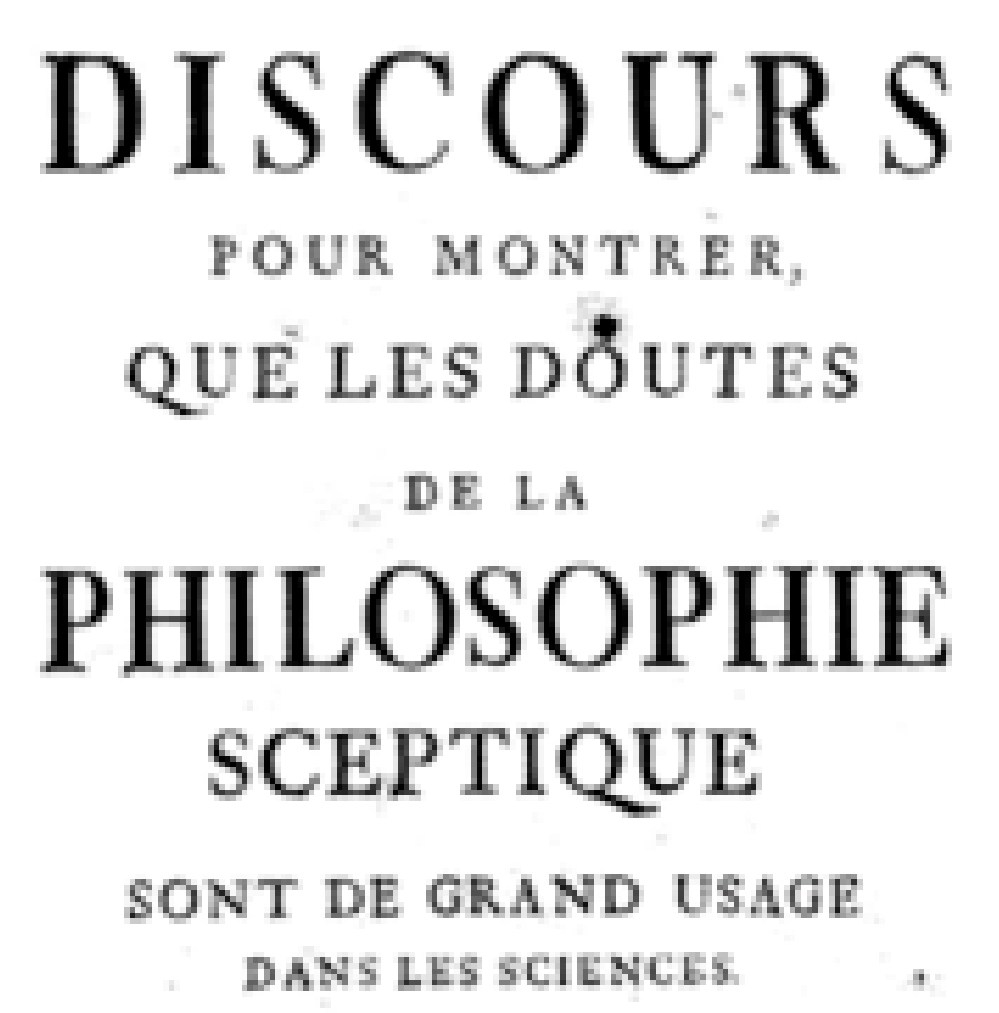

DISCOURS
POUR MONTRER,
QUE LES DOUTES
DE LA
PHILOSOPHIE
SCEPTIQUE
SONT DE GRAND USAGE
DANS LES SCIENCES.

Figure 3.19: *A Discourse to Show that Skeptical Philosophy is of Great Use in Science by François de La Mothe Le Vayer (1588-1672), apparently Bishop Huet's source. Every time I find a IJoriginal thinkerİ who figured out the skeptical solution to the Black Swan problem, it turns out that he may just be cribbing a predecessor –not maliciously, but we forget to dig to the roots. As we insist, "Hume's problem" has little to do with Hume, who carried the heavy multi-volume Dictionary of Pierre Bayle (his predecessors) across Europe. I thought it was Huet who was as one digs, new predecessors crop up*

degenerate random variable taking seemingly constant values with a histogram producing a Dirac stick.

We have known at least since Sextus Empiricus that we cannot rule out non-degeneracy but there are situations in which we can rule out degeneracy. If I see a distribution that has no randomness, I cannot say it is not random. That is, we cannot say there are no Black Swans. Let us now add one observation.

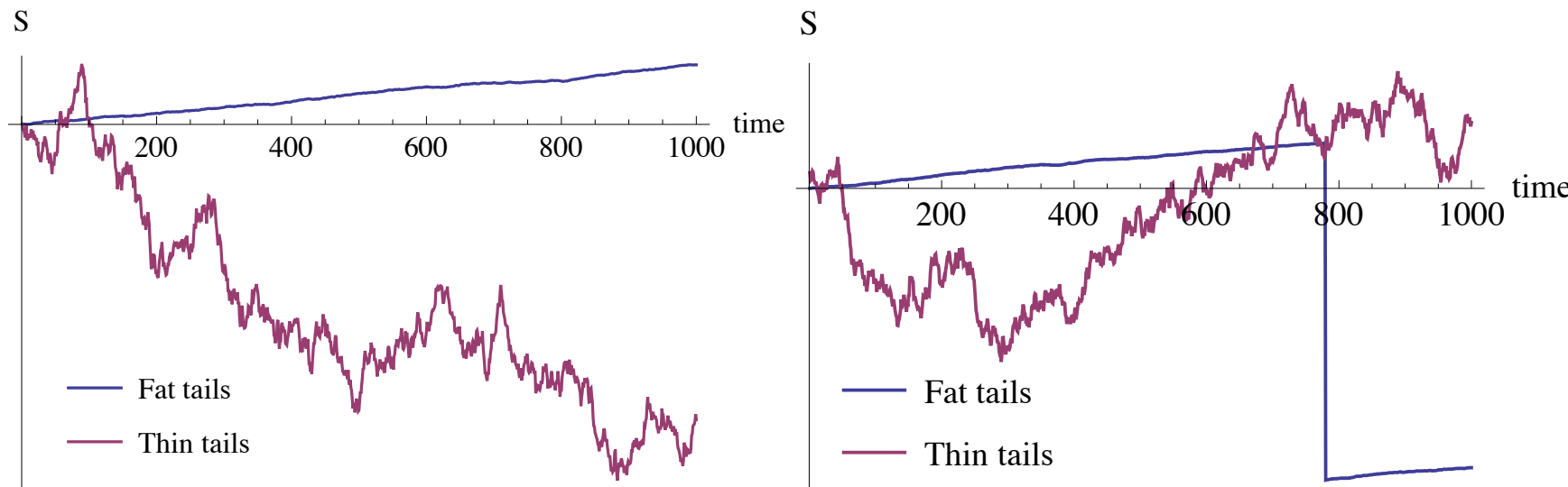

Figure 3.20: *It is not possible to "accept" thin tails, very easy to reject thintailedness. One distribution can produce jumps and quiet days will not help rule out their occurrence.*

I can now see it is random, and I can rule out degeneracy. I can say it is not "not random". On the right hand side we have seen a Black Swan , therefore the statement that there are no Black Swans is wrong. This is the negative empiricism that underpins Western science. As we gather information, we can rule things out. The distribution on the right can hide as the distribution on the left, but the distribution on the right cannot hide as the distribution on the left (check). This gives us a very easy way to deal with randomness. Figure 3.16 generalizes the problem to how we can eliminate distributions.

If we see a 20 sigma event, we can rule out that the distribution is thin-tailed. If we see no large deviation, we can not rule out that it is not thick tailed unless we understand the process very well. This is how we can rank distributions. If we reconsider Figure 3.7 we can start seeing deviations and ruling out progressively from the bottom. These ranks are based on how distributions can deliver tail events. Ranking distributions (by order or priority for the sake of inference) becomes very simple. Consider the logic: if someone tells you there is a ten-sigma event, it is much more likely that they have the wrong distribution than it is that you really have ten-sigma event (we will refine the argument later in this chapter). Likewise, as we saw, thick tailed distributions do not deliver a lot of deviation from the mean. But once in a while you get a big deviation. So we can now rule out what is not mediocristan. We can rule out where we are not Ş we can rule out mediocristan. I can say this distribution is thick tailed by elimination. But I can not certify that it is thin tailed. This is the Black Swan problem.

**Application of the Maquerade Problem: Argentina's stock market before and after Aug 12, 2019** For an illustration of the asymmetry of inference applied to parameters of a distribution, or how a distribution can masquerade as having thinner tails than it actually has, consider what we knew about the Argentinian market before and after the large drop of Aug 12, 2019 (shown in Figure 3.21). Using this reasoning, any future parameter uncertainty should make tails fatter, not thinner. Rafal Weron, in [319], showed how we are more likely to overestimate the tail index when fitting a stable distribution (lower means fatter tails).

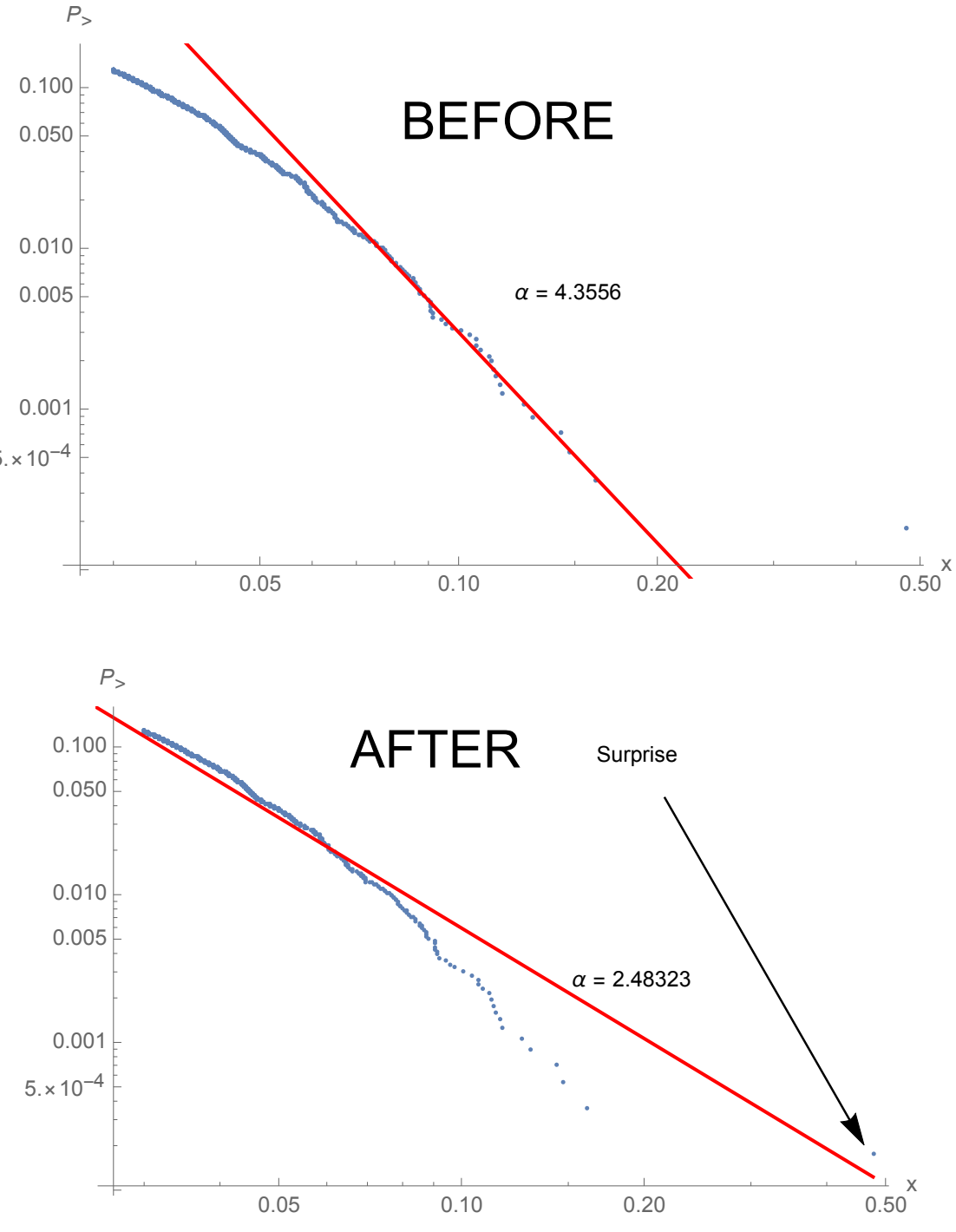

Figure 3.21: *A single day reveals the true tails of a distribution. Argentina's stock market before and after Aug 12, 2019. You may suddenly revise the tails as thicker (lower parameter α), never the reverse –it would take a long, long time for that to happen. Data obtained thanks to Diego Zviovich.*

## 3.6 NAIVE EMPIRICISM: EBOLA DIFFERS FROM FALLS FROM LADDERS

Let us illustrate one of the problem of thin-tailed thinking in the fat-tailed domain with a real world example. People quote so-called "empirical" data to tell us we are foolish to worry about ebola when only two Americans died of ebola in 2016. We are told that we should worry more about deaths from diabetes or people tangled in their bedsheets. Let us think about it in terms of tails. If we were to read in the newspaper that 2 billion people have died suddenly, it is far more likely that they died of ebola than smoking or diabetes or tangled in their bedsheets?

**Principle 3.2**

*Thou shalt not compare a multiplicative fat-tailed process in Extremistan in the subexponential class to a thin-tailed process from Mediocristan, particularly one that has Chernoff bounds..*

This is a simple consequence of the catastrophe principle we saw earlier, as illustrated in Figure 3.1.

Alas few "evidence based" people get (at the time of writing) the tail wag the dog effect.

It is naïve empiricism to compare these processes, to suggest that we worry too much about ebola (epidemics or pandemics) and too little about diabetes. In fact it is the other way round. We worry too much about diabetes and too little about ebola and other ailments with multiplicative effects. This is an error of reasoning that comes from not understanding thick tails –sadly it is more and more common. What is worse, such errors of reasoning are promoted by **empirical psychology** which does not appear to be empirical. It is also used by shills for industry passing for "risk communicators" selling us pesticides and telling us not to worry because harm appears to be minimal in past data (see Figure ).

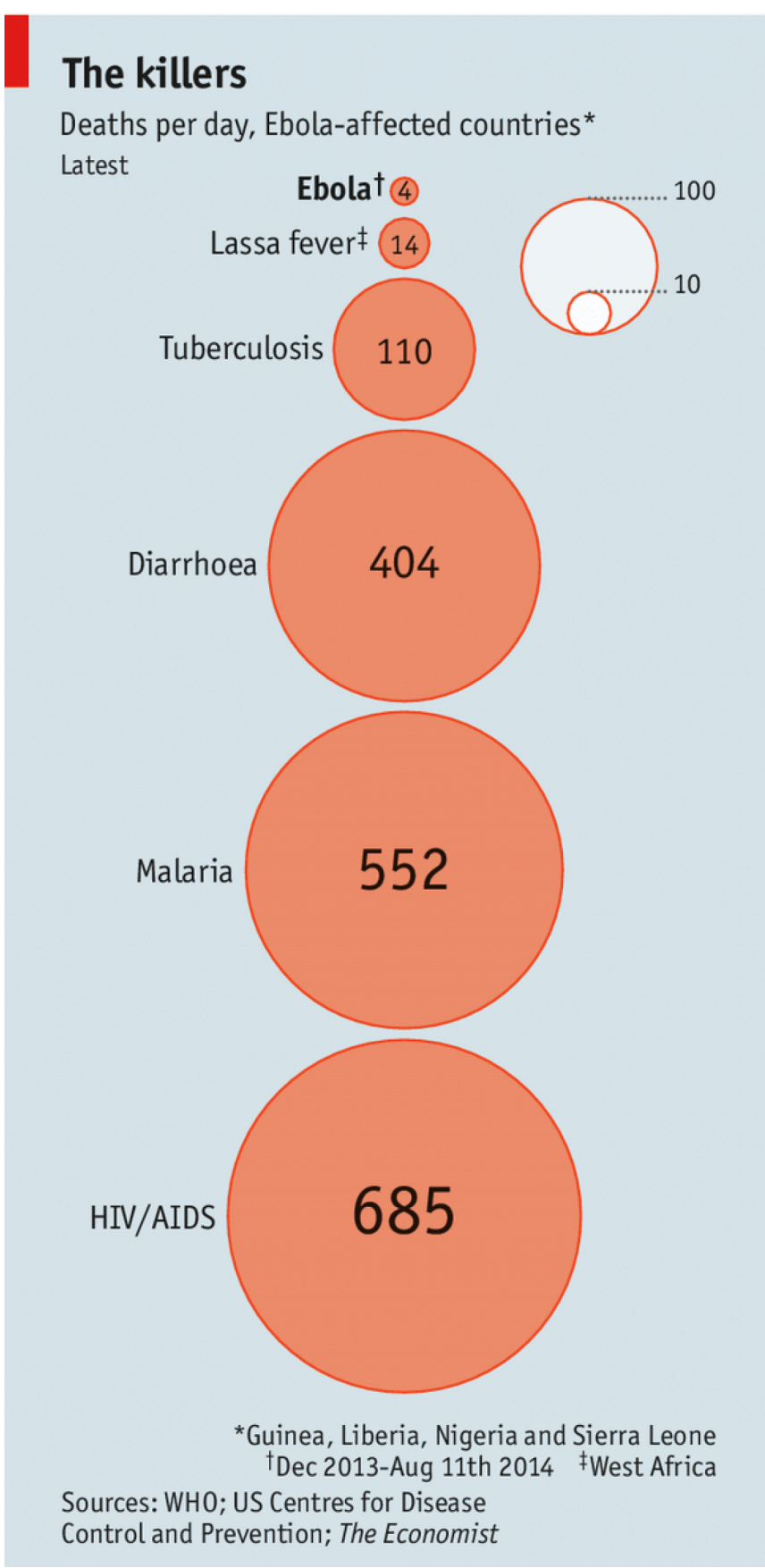

Figure 3.22: *Naive empiricism: never compare thick tailed variables to thin tailed ones, since the means do not belong to the same class of distributions. This is a generalized mistake made by The Economist, but very common in the so-called learned discourse. Even the Royal Statistical Society fell for it once they hired a "risk communication" person with a sociology or journalism background to run it.*

The correct reasoning is generally absent in decision theory and risk circles outside of the branches of extreme value theory and the works of the ABC group in Berlin's Max Planck directed by Gerd Gigerenzer [126] which tells you that your grandmother's instincts and teachings are not to be ignored and, when her recommendations clash with psychologists and decision theorists, it is usually the psychologists and decision theorists who are unrigorous. A simple look at the summary by "most cited author" Baruch Fishhoff's in *Risk: a Very Short Introduction* [111] shows no effort to disentangle the two classes of distribution. The problem linked to the "risk calibration" and "probabilistic calibration" misunder-

# Causes of death in the US

What Americans die from, what they search on Google, and what the media reports on

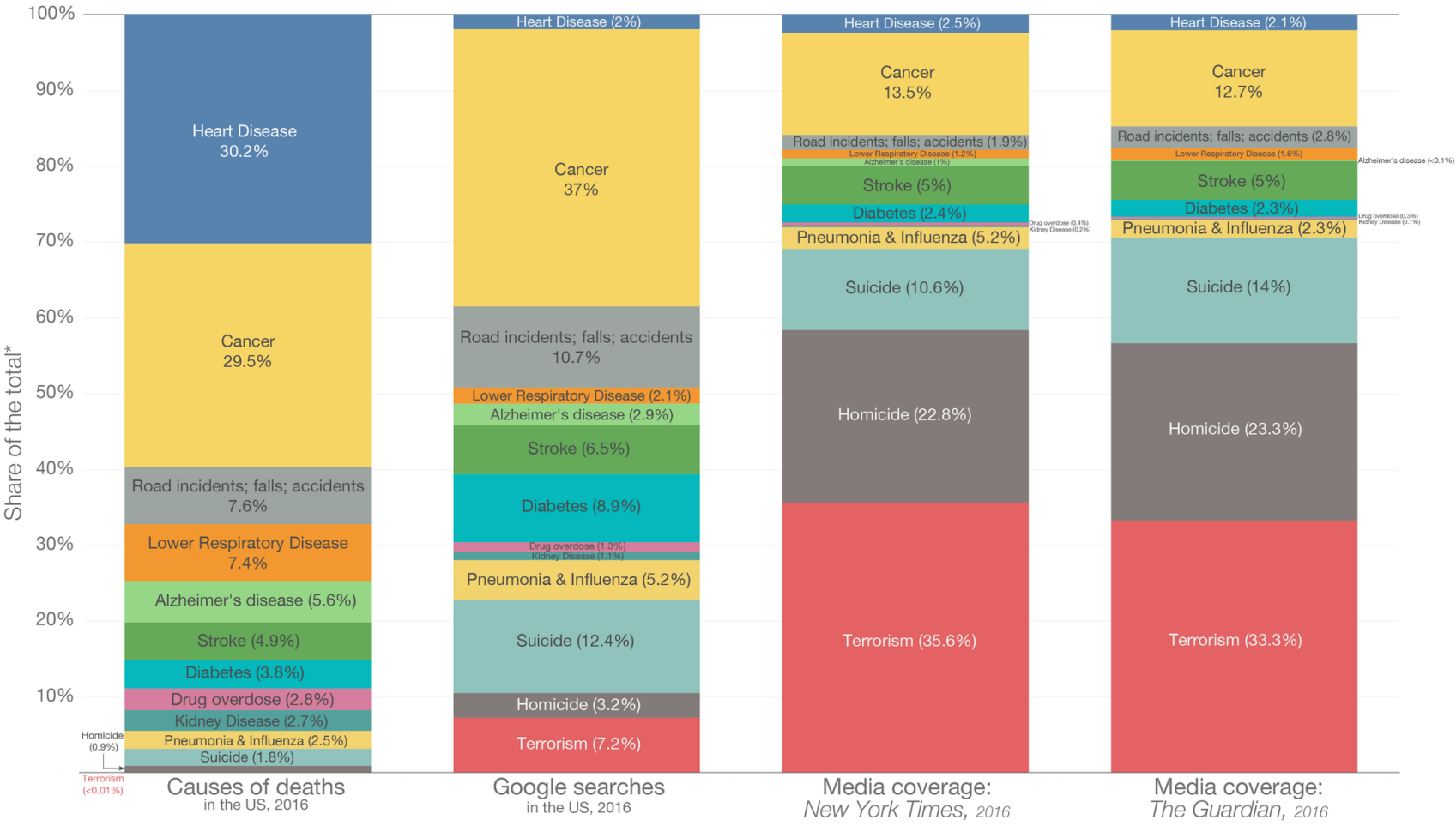

Figure 3.23: *Bill Gates's Naive (Non-Statistical) Empiricism: the founder of Microsoft*[1] *is promoting and financing the development of the above graph, yet at the same time claiming that the climate is causing an existential risk, not realizing that his arguments conflict since existential risks are necessarily absent in past data. Furthermore, a closer reading of the graphs shows that cancer, heart disease, and Alzheimer, being ailments of age, do not require the attention on the part of young adults and middle-aged people something terrorism and epidemics warrant.*

*Another logical flaw is that terrorism is precisely low because of the attention it commands. Relax your vigilance and it may go out of control. The same applies to homicide: fears lead to safety.*

*If this map shows something, it is the rationality of common people with a good tail risk detector, compared to the ignorance of "experts". People are more calibrated to consequences and properties of distributions than psychologists claim.*

[1] *Microsoft is a technology company still in existence at the time of writing.*

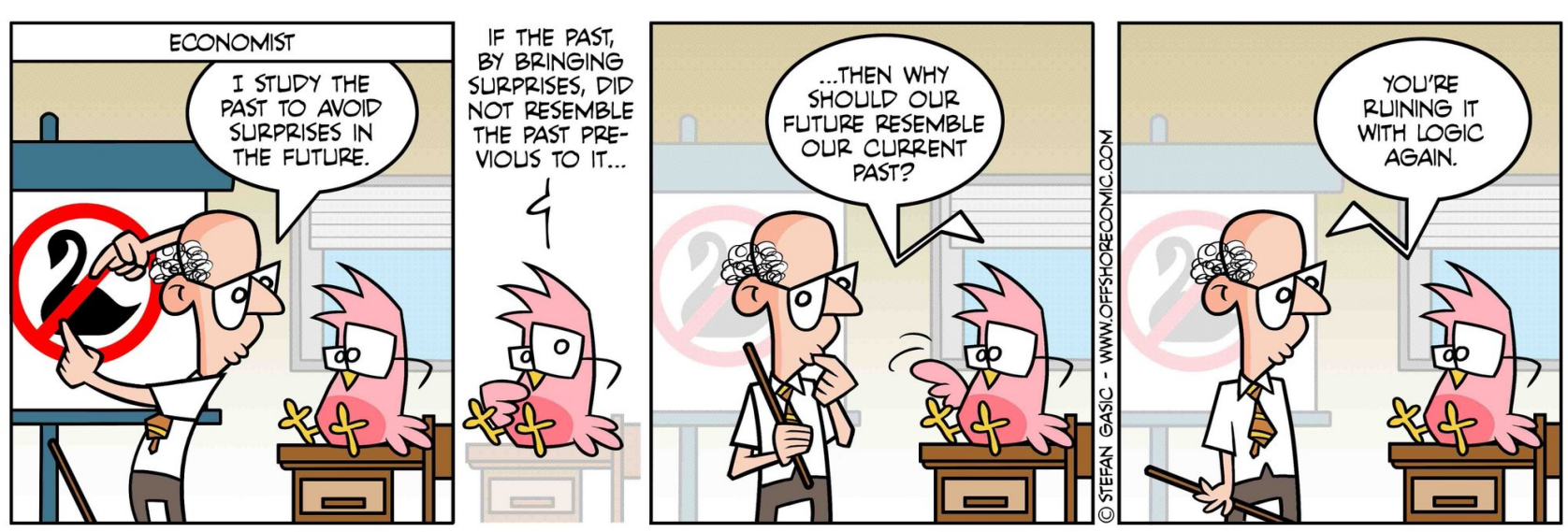

Figure 3.24: *Because of the slowness of the law of large numbers, under thick tails, the past's past doesn't resemble the past's future; accordingly, today's past will not resemble today's future. Things are easier under thin tails. Credit Stefan Gasic.*

stood by psychologists and discussed more technically in Chapter 11 discussing expert calibration under thick tails.[16]

[16] The Gigerenzer school is not immune to mistakes, as evidenced by their misunderstanding of the risks of COVID-19 in early 2020 –the difference between Mediocristan and Extremistan has not reached them yet. But this author is optimistic that it will.

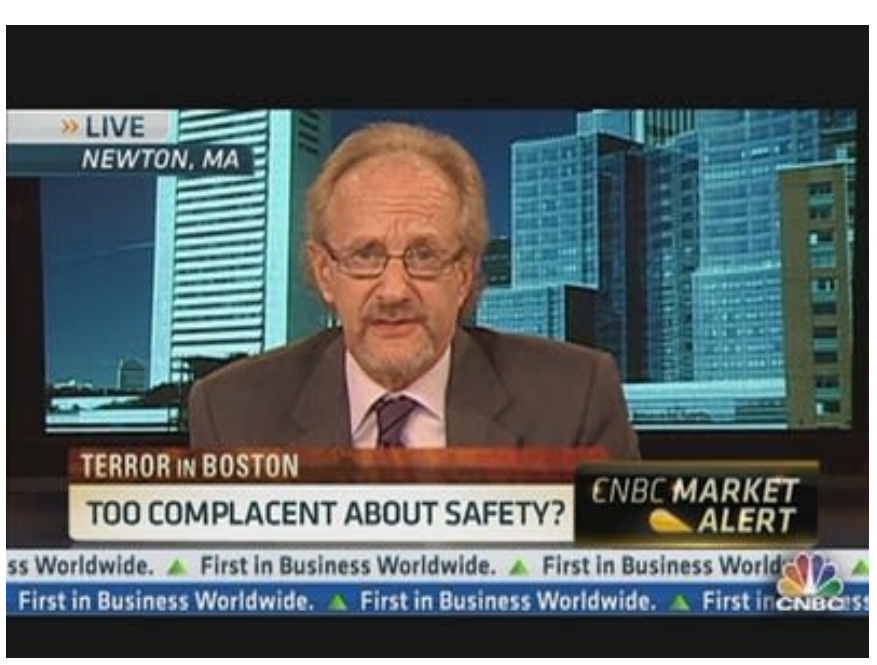

Figure 3.25: *Beware the lobbyist using pseudo-empirical arguments. "Risk communications" shills such as the fellow here, with a journalism background, are hired by firms such as Monsanto (and cars and Tobacco companies) to engage in smear campaigns on their behalf using "science", "empirical arguments" and "evidence", and downplay "public fears" they deem irrational. Lobbying organizations penetrate such centers as "Harvard Center for Risk Analysis" with a fancy scholarly name that helps convince the layperson. The shills' line of argument, commonly, revolves around "no evidence of harm" and "rationality". Other journalists, in turn, espouse such arguments owing to their ability to sway the statistically naive. Probabilistic and risk literacy, statistical knowledge and journalism have suffered greatly from the spreading of misconceptions by nonscientists, or, worse, nonstatisticians.*

### 3.6.1 How some multiplicative risks scale

The "evidence based" approach is still too primitive to handle second order effects (and risk management) and has certainly caused way too much harm with the COVID-19 pandemic to remain useable outside of single patient issues. One of the problems is the translation between individual and collective risk (another is the mischaracterization of evidence and conflation with absence of evidence).

At the beginning of the COVID-19 pandemic, many epidemiologists innocent of probability compared the risk of death from it to that of drowning in a swimming pool. For a single individual, this might have been true (although COVID-19 turned out rapidly to be the main source of fatality in many parts, and later even caused 80% of the fatalities New York City). But conditional on having 1000 deaths, the odds of the cause being drowning in swimming pools is slim.

This is because your neighbor having COVID *increases* the chances that you get it, whereas your neighbor drowning in her or his swimming pool does not increase your probability of drowning (if anything, like plane crashes, it *decreases* other people's chance of drowning ).

This aggregation problem is discussed in more technical terms with ellipticality, see Section 6.8 –joint distributions are no longer elliptical, causing the sum to be fat-tailed even when individual variables are thin-tailed.

It is also discussed as a problem in ethics [298]: by contracting the disease you cause more deaths than your own. Although the risk of death from a contagious disease can be smaller than, say, that from a car accident, it becomes psychopathic to follow "rationality" (that is, first order rationality models) as you will eventually cause systemic harm and even, eventually, certain self-harm.

## 3.7 PRIMER ON POWER LAWS (ALMOST WITHOUT MATHEMATICS)

Let us now discuss the intuition behind the Pareto Law. It is simply defined as: say $X$ is a random variable. For a realization $x$ of $X$ sufficiently large, the probability of exceeding $2x$ divided by the probability of exceeding $x$ is "not too different" from the probability of exceeding $4x$ divided by the probability of exceeding $2x$, and so forth. This property is called "scalability".[17]

Table 3.2: *An example of a power law*

| | |
|---|---|
| Richer than 1 million | 1 in 62.5 |
| Richer than 2 million | 1 in 250 |
| Richer than 4 million | 1 in 1,000 |
| Richer than 8 million | 1 in 4,000 |
| Richer than 16 million | 1 in 16,000 |
| Richer than 32 million | 1 in ? |

So if we have a Pareto (or Pareto-style) distribution, the ratio of people with $ 16 million compared to $ 8 million is the same as the ratio of people with $ 2 million and $ 1 million. There is a constant inequality. This distribution has no characteristic scale which makes it very easy to understand. Although this distribution often has no mean and no standard deviation we can still understand it –in fact we can understand it much better than we do with more standard statistical distributions. But because it has no mean we have to ditch the statistical textbooks and do something more solid, more rigorous, even if it seems less mathematical.

A Pareto distribution has no higher moments: moments either do not exist or become statistically more and more unstable. So next we move on to a problem with economics and econometrics. In 2009 I took 55 years of data and looked at how much of the kurtosis (a function of the fourth moment) came from the largest observation –see Table 3.3. For a Gaussian the maximum contribution over the same time span should be around .008 $\pm$ .0028. For the S&P 500 it was about 80 percent. This tells us that we don't know anything about the kurtosis of these securities. Its sample error is huge; or it may not exist so the measurement is heavily sample dependent. If we don't know anything about the fourth moment, we know nothing about the stability of the second moment. It means we are not in a class of distribution that allows us to work with the variance, even if it exists. Science is hard; quantitative finance is hard too.

For silver, in 46 years 94 percent of the kurtosis came from one single observation. We cannot use standard statistical methods with financial data. GARCH (a method popular in academia) does not work because we are dealing with squares.

[17] To put some minimum mathematics: let $X$ be a random variable belonging to the class of distributions with a "power law" right tail:

$$\mathbb{P}(X > x) = L(x)\, x^{-\alpha} \tag{3.1}$$

where $L : [x_{\min}, +\infty) \to (0, +\infty)$ is a slowly varying function, defined as $\lim_{x \to +\infty} \frac{L(kx)}{L(x)} = 1$ for any $k > 0$. We can transform and apply to the negative domain.

Table 3.3: *Kurtosis from a single observation for financial data* $\frac{Max\left(X^4_{t-\Delta ti}\right)^n_{i=0}}{\sum^n_{i=0} X^4_{t-\Delta ti}}$

| Security | Max Q | Years. |
| --- | --- | --- |
| Silver | 0.94 | 46. |
| SP500 | 0.79 | 56. |
| CrudeOil | 0.79 | 26. |
| Short Sterling | 0.75 | 17. |
| Heating Oil | 0.74 | 31. |
| Nikkei | 0.72 | 23. |
| FTSE | 0.54 | 25. |
| JGB | 0.48 | 24. |
| Eurodollar Depo 1M | 0.31 | 19. |
| Sugar | 0.3 | 48. |
| Yen | 0.27 | 38. |
| Bovespa | 0.27 | 16. |
| Eurodollar Depo 3M | 0.25 | 28. |
| CT | 0.25 | 48. |
| DAX | 0.2 | 18. |

The variance of the squares is analogous to the fourth moment. We do not know the variance. But we can work very easily with Pareto distributions. They give us less information, but nevertheless, it is more rigorous if the data are uncapped or if there are any open variables.

Table 3.3, for financial data, debunks all the college textbooks we are currently using. A lot of econometrics that deals with squares goes out of the window. This explains why economists cannot forecast what is going on –they are using the wrong methods and building the wrong confidence intervals. It will work within the sample, but it will not work outside the sample –and samples are by definition finite and will always have finite moments. If we say that variance (or kurtosis) is infinite, we are not going to observe anything that is infinite within a sample.

Principal component analysis, PCA (see Figure 3.26) is a dimension reduction method for big data and it works beautifully with thin tails (at least sometimes). But if there is not enough data there is an illusion of what the structure is. As we increase the data (the $n$ variables), the structure becomes flat (something called in some circles the "Wigner effect" for random matrices, after Eugene Wigner — do not confuse with Wigner's discoveries about the dislocation of atoms under radiation). In the simulation, the data that has absolutely no structure: principal components (PCs) should be all equal (asymptotically, as data becomes large); but the small sample effect causes the ordered PCs to show a declining slope. We have zero correlation on the matrix. For a thick tailed distribution (the lower section), we need a lot more data for the spurious correlation to wash out i.e., dimension reduction does not work with thick tails.

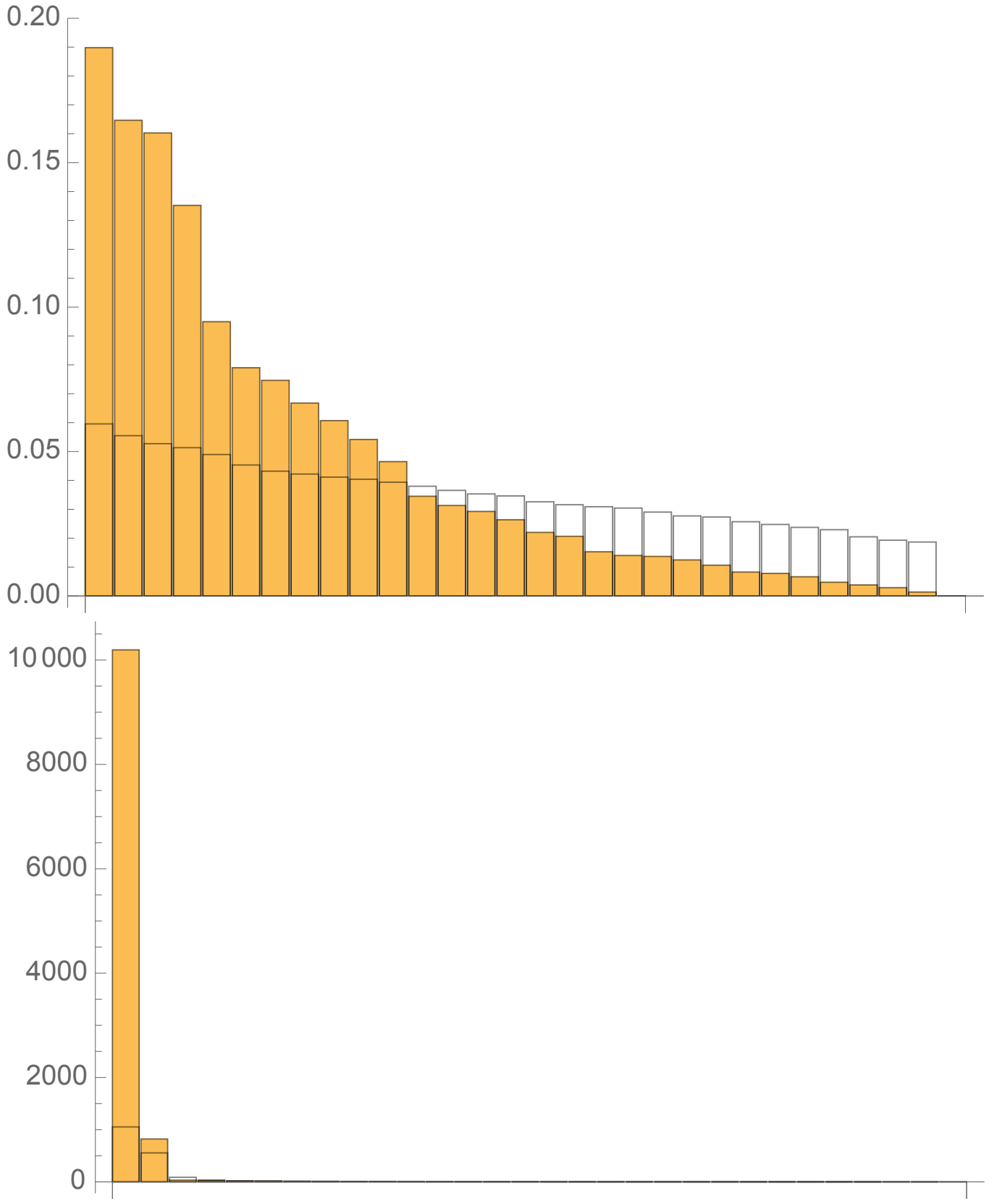

Figure 3.26: ***Spurious PCAs Under Thick Tails:*** *A Monte Carlo experiment that shows how spurious correlations and covariances are more acute under thick tails. Principal Components ranked by variance for 30 Gaussian uncorrelated variables (above),* $n = 100$ *(shaded) and* $1000$ *data points (transparent), and principal components ranked by variance for 30 stable distributed ( below, with tail* $\alpha = \frac{3}{2}$ *, symmetry* $\beta = 1$*, centrality* $\mu = 0$*, scale* $\sigma = 1$*), with same* $n = 100$ *(shaded) and* $n = 1000$ *(transparent). Both are "uncorrelated" identically distributed variables. We can see the "flatter" PCA structure with the Gaussian as n increases (the difference between PCAs shrinks). Such flattening does not occur in reasonable time under fatter tails.*

## 3.8 WHERE ARE THE HIDDEN PROPERTIES?

The following summarizes everything that I wrote in *The Black Swan* (a message that somehow took more than a decade to go through without distortion). Distributions can be one-tailed (left or right) or two-tailed. If the distribution has a thick tail, it can be thick tailed one tail or it can be thick tailed two tails. And if is thick tailed one tail, it can be thick tailed left tail or thick tailed right tail.

See Figure 3.28 for the intuition: if it is thick tailed and we look at the sample mean, we observe fewer tail events. The common mistake is to think that we can naively derive the mean in the presence of one-tailed distributions. But there are unseen rare events and with time these will fill in. But by definition, they are low probability events.

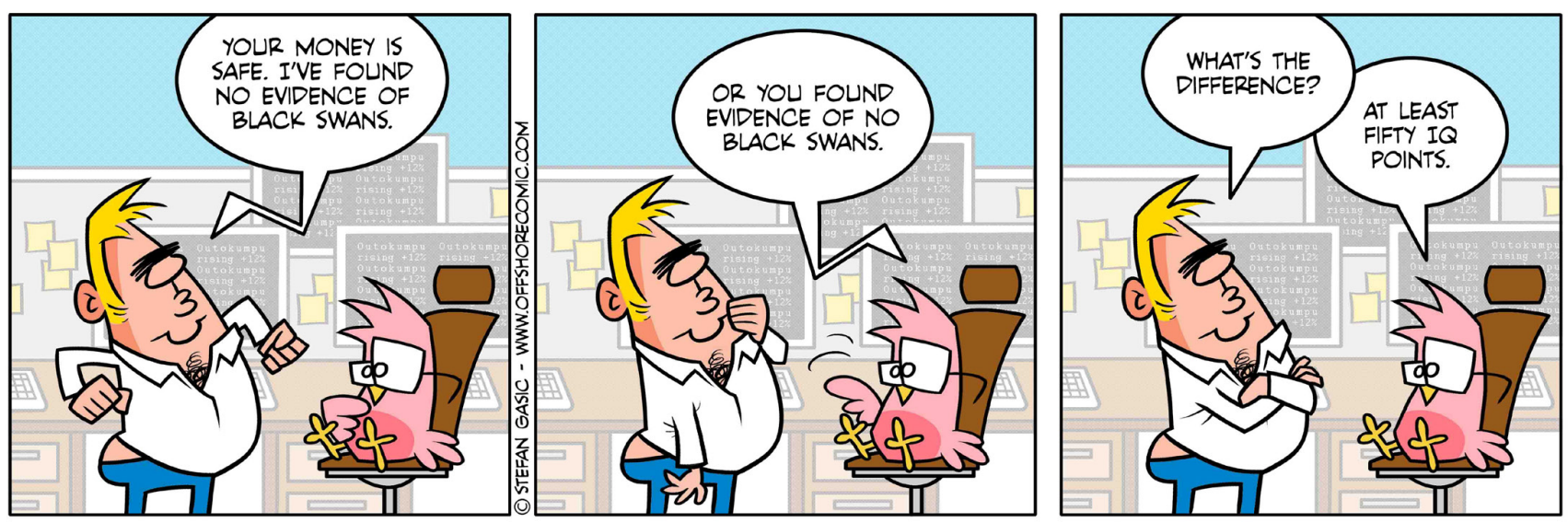

Figure 3.27: ***A central asymmetry***: *the difference between absence of evidence and evidence of absence is compounded by thick tails. It requires a more elaborate understanding of random events —or a more naturalistic one. (Please do not attribute IQ points here as equivalent to the ones used in common psychometrics: the suspicion is that high scoring people on IQ tests fail to get the asymmetry. IQ here should be interpreted as "real" intelligence, not the one from that test. ) Courtesy Stefan Gasic.*

> It is easier to be *fooled by randomness* about the quality of the performance with a short volatility time series (left skewed, exposed to sharp losses) than with a long tail volatility one (right skewed, exposed to sharp gains). Simply short volatility overestimate the performance (while the other underestimates it (see Fig 3.28). This is another version of the asymmetry attributed to Popper that we saw earlier in the chapter.

The trick is to estimate the distribution and then derive the mean (which implies extrapolation). This is called in this book "plug-in" estimation, see Table 3.4. It is not done by measuring the directly observable sample mean which is biased under fat-tailed distributions. This is why, outside a crisis, banks seem to make large profits. Then once in a while they lose everything and more and have to be bailed out by the taxpayer. The way we handle this is by differentiating the true mean (which I call "shadow") from the realized mean, as in the Tableau in Table 3.4.

We can also do that for the Gini coefficient to estimate the "shadow" one rather than the naively observed one.

This is what we mean when we say that the "empirical" distribution is not "empirical". In other words: 1) there is a wedge between population and sample attributes and, 2) even exhaustive historical data must be seen as mere sampling from a broader phenomenon (the past is in sample; inference is what works out of sample).

Once we have figured out the distribution, we can estimate the statistical mean. This works much better than directly measuring the sample mean. For a Pareto distribution, for instance, 98% of observations are below the mean. There is a bias in the observed mean. But once we know that we have a Pareto distribution, we

**WITTGENSTEIN'S RULER: WAS IT REALLY A "10 SIGMA EVENT"?**

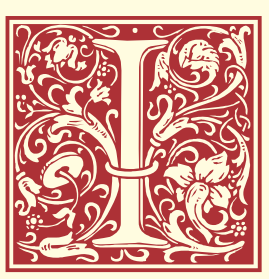N THE SUMMER OF 1998, the hedge fund called "Long Term Capital Management" (LTCM) proved to have a very short life; it went bust from some deviations in the markets –those "of an unexpected nature". The loss was a *yuuuge deal* because two of the partners received the Swedish Riksbank Prize, marketed as the "Nobel" in economics. More significantly, the fund harbored a large number of finance professors; LTCM had imitators among professors (at least sixty finance PhDs blew up during that period from trades similar to LTCM's, and owing to risk management methods that were identical). At least two of the partners made the statement that it was a "10 sigma" event (10 standard deviations), hence they should be absolved of all accusations of incompetence (I was first hand witness of two such statements).

Let us apply what the author calls "Wittgenstein's ruler": are you using the ruler to measure the table or using the table to measure the ruler?

Assume to simplify there are only two alternatives: a Gaussian distribution and a Power Law one. For the Gaussian, the "event" we define as the survival function of 10 standard deviations is 1 in $1.31 \times 10^{23}$. For the Power law of the same scale, a student T distribution with tail exponent 2, the survival function is 1 in 203.

What is the probability of the data being Gaussian conditional on a 10 sigma event, compared to that alternative?

We start with Bayes' rule. $\mathbb{P}(A|B) = \frac{\mathbb{P}(A)\mathbb{P}(B|A)}{\mathbb{P}(B)}$. Replace $\mathbb{P}(B) = \mathbb{P}(A)\mathbb{P}(B|A) + \mathbb{P}(\overline{A})\mathbb{P}(B|\overline{A})$ and apply to our case.

$$P(\text{Gaussian}|\text{Event}) = \frac{\mathbb{P}(\text{Gaussian})P(\text{Event}|\text{Gaussian})}{(1 - \mathbb{P}(\text{Gaussian}))P(\text{Event}|\text{NonGaussian}) + \mathbb{P}(\text{Gaussian})P(\text{Event}|\text{Gaussian})}$$

| $\mathbb{P}$(Gaussian) | $P$(Gaussian\|Event) |
|---|---|
| 0.5 | $2 \times 10^{-21}$ |
| 0.999 | $2 \times 10^{-18}$ |
| 0.9999 | $2 \times 10^{-17}$ |
| 0.99999 | $2 \times 10^{-16}$ |
| 0.999999 | $2 \times 10^{-15}$ |
| 1 | 1 |

**Moral:** If there is a tiny probability, $< 10^{-10}$ that the data might not be Gaussian, one can firmly reject Gaussianity in favor of the thick tailed distribution. The heuristic is to reject Gaussianity in the presence of any event $> 4$ or $> 5$ STDs –we will see throughout the book why patches such as conditional variance are inadequate and can be downright fraudulent.[a]

[a] The great Benoit Mandelbrot used to be extremely critical of methods that relied on a Gaussian and added jumps or other ad hoc tricks to explain what happened in the data (say Merton's jump diffusion process [202]) –one can always fit back jumps ex post. He used to cite the saying attributed to John von Neumann: "With four parameters I can fit an elephant, and with five I can make him wiggle his trunk."

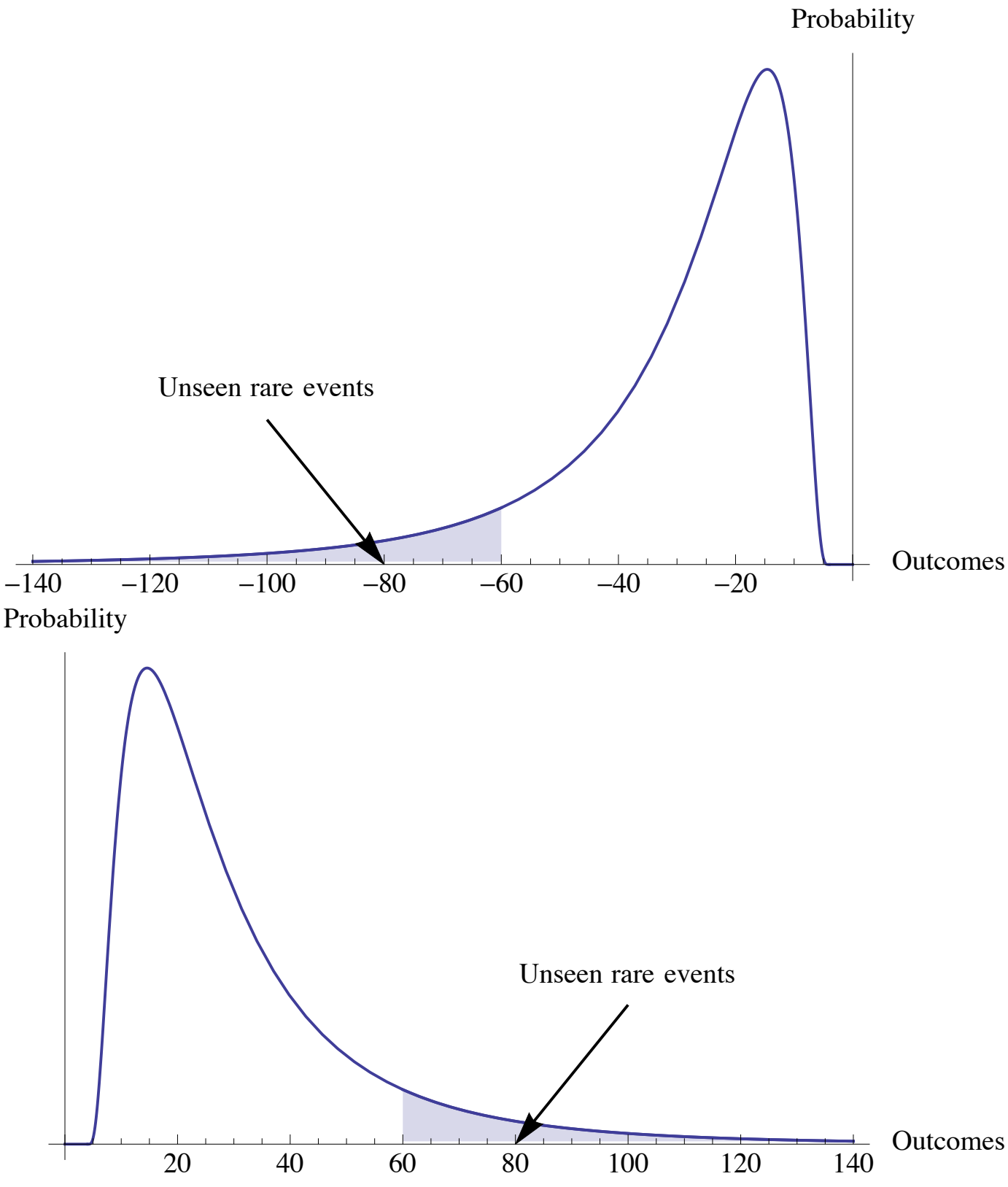

Figure 3.28: *Shadow Mean at work: Below: Inverse Turkey Problem – The unseen rare event is positive. When you look at a positively skewed (antifragile) time series and make (nonparametric) inferences about the unseen, you miss the good stuff and underestimate the benefits. Above: The opposite problem. The filled area corresponds to what we do not tend to see in small samples, from insufficiency of data points. Interestingly, the shaded area increases with model error (owing to the convexity of tail probabilities to uncertainty).*

Table 3.4: *Shadow mean vs. Sample mean and their ratio for different minimum thresholds. The shadow mean is obtained via maximum likelihood, ML (from plug-in estimators) . In bold the values for the 145k threshold. Rescaled data. From Cirillo and Taleb [55]. Details are explained in Chapters 18 and 15.*

| **L** | **Sample Mean** | **ML Mean** | **Ratio** |
|---|---|---|---|
| 10K | $9.079 \times 10^6$ | $3.11 \times 10^7$ | 3.43 |
| 25K | $9.82 \times 10^6$ | $3.62 \times 10^7$ | 3.69 |
| 50K | $1.12 \times 10^7$ | $4.11 \times 10^7$ | 3.67 |
| 100K | $1.34 \times 10^7$ | $4.74 \times 10^7$ | 3.53 |
| 200K | $1.66 \times 10^7$ | $6.31 \times 10^7$ | 3.79 |
| 500K | $2.48 \times 10^7$ | $8.26 \times 10^7$ | 3.31 |

should ignore the sample mean and look elsewhere. Chapters 15 and 17 discuss the techniques.

Note that the field of Extreme Value Theory [135] [97] [136] focuses on tail properties, not the mean or statistical inference.

## 3.9 BAYESIAN SCHMAYESIAN

In the absence of reliable information, Bayesian methods can be of little help. This author has faced since the publication of *The Black Swan* numerous questions concerning the use of something vaguely Bayesian to solve problems about the unknown under thick tails. Since one cannot manufacture information beyond what's available, no technique, Bayesian nor Schmayesian can help. The key is that one needs a reliable prior, something not readily observable (see Diaconis and Friedman [78] for the difficulty for an agent in formulating a prior).

A problem is the speed of updating, as we will cover in Chapter 7, which is highly distribution dependent. The mistake in the rational expectation literature is to believe that two observers supplied with the same information would necessarily converge to the same view. Unfortunately, the conditions for that to happen in real time or to happen at all are quite specific.

One of course can use Bayesian methods (under adequate priors) for the estimation of parameters if 1) one has a clear idea about the range of values (say from universality classes or other stable basins) and 2) these parameters follow a tractable distribution with low variance such as, say, the tail exponent of a Pareto distribution (which is inverse-gamma distributed), [14].

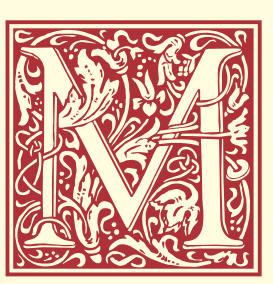

ORAL HAZARD AND RENT SEEKING in financial education: One of the most depressing experience this author had was when teaching a course on Fat Tails at the University of Massachusetts Amherst, at the business school, during a very brief stint there. One PhD student in finance bluntly said that he liked the ideas but that a financial education career commanded "the highest salary in the land" (that is, among all other specialties in education). He preferred to use Markowitz methods (even if they failed in fat-tailed domains) as these were used by other professors, hence allowed him to get his papers published, and get a high paying job.
I was disgusted, but predicted he would subsequently have a very successful career writing non-papers. He did.

## 3.10 $x$ VS $f(x)$: EXPOSURES TO $x$ CONFUSED WITH KNOWLEDGE ABOUT $x$

Take $X$ a random or nonrandom variable, and $F(X)$ the exposure, payoff, the effect of $X$ on you, the end bottom line. ($X$ is often is higher dimensions but let's assume to simplify that it is a simple one-dimensional variable).

Practitioners and risk takers often observe the following disconnect: people (nonpractitioners) talking $X$ (with the implication that practitioners should care about $X$ in running their affairs) while practitioners think about $F(X)$, nothing but $F(X)$. And the straight confusion since Aristotle between $X$ and $F(X)$ has been chronic as discussed in *Antifragile* [276] which is written around that theme. Sometimes people mention $F(X)$ as utility but miss the full payoff. And the confusion is at two level: one, simple confusion; second, in the decision-science literature, seeing the difference and not realizing that action on $F(X)$ is easier than action on $X$.

- The variable $X$ can be unemployment in Senegal, $F_1(X)$ is the effect on the bottom line of the IMF, and $F_2(X)$ is the effect on your grandmother (which I assume is minimal).
- $X$ can be a stock price, but you own an option on it, so $F(X)$ is your exposure an option value for $X$, or, even more complicated the utility of the exposure to the option value.
- $X$ can be changes in wealth, $F(X)$ the convex-concave way it affects your well-being. One can see that $F(X)$ is vastly more stable or robust than $X$ (it has thinner tails).

**Convex vs. linear functions of a variable $X$** Consider Fig. 3.30; confusing $F(X)$ (on the vertical) and $X$ (the horizontal) is more and more significant when $F(X)$ is nonlinear. The more convex $F(X)$, the more the statistical and other properties of $F(X)$ will be divorced from those of X. For instance, the mean of $F(X)$ will be different from $F$(Mean of$X$), by Jensen's inequality. But beyond Jensen's inequality, the difference in risks between the two will be more and more considerable. When it comes to probability, the more nonlinear $F$, the less the probabilities of $X$ matters compared to that of $F$. Moral of the story: focus on $F$, which we can alter, rather than on the measurement of the elusive properties of $X$.

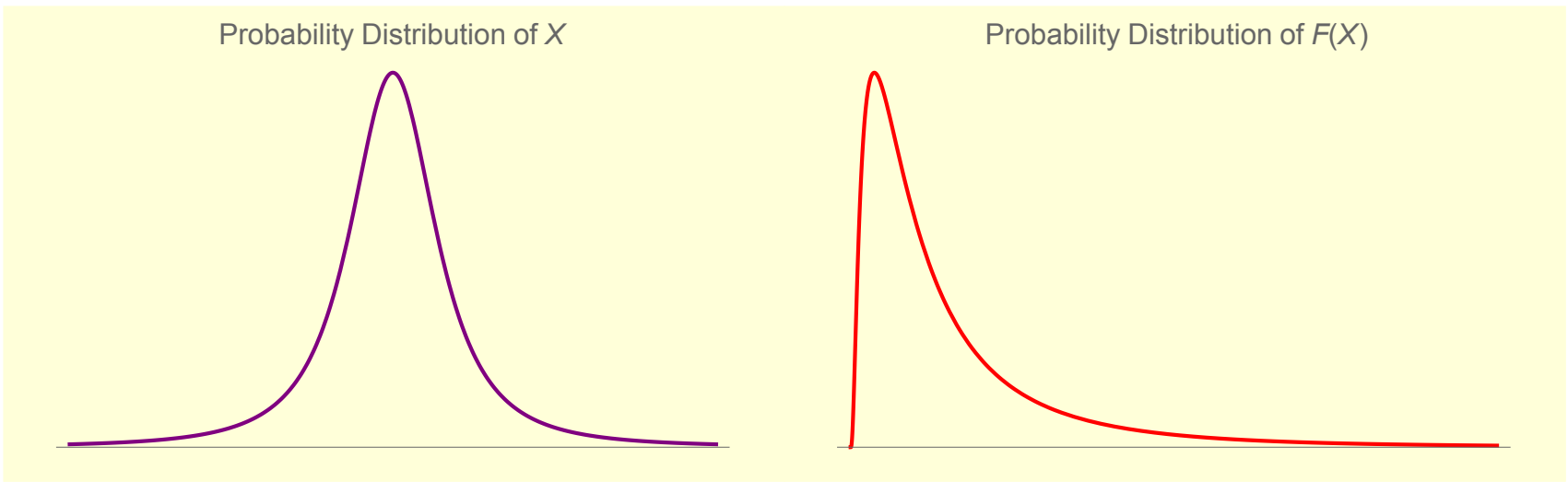

Figure 3.29: *The Conflation Problem X (random variable) and F(X) a function of it (or payoff). If F(X) is convex we don't need to know much about it –it becomes an academic problem. And it is safer to focus on transforming F(X) than X.*

**Limitations of knowledge** What is crucial, our limitations of knowledge apply to $X$ not necessarily to $F(X)$. We have no control over $X$, some control over $F(X)$. In some cases a very, very large control over $F(X)$.

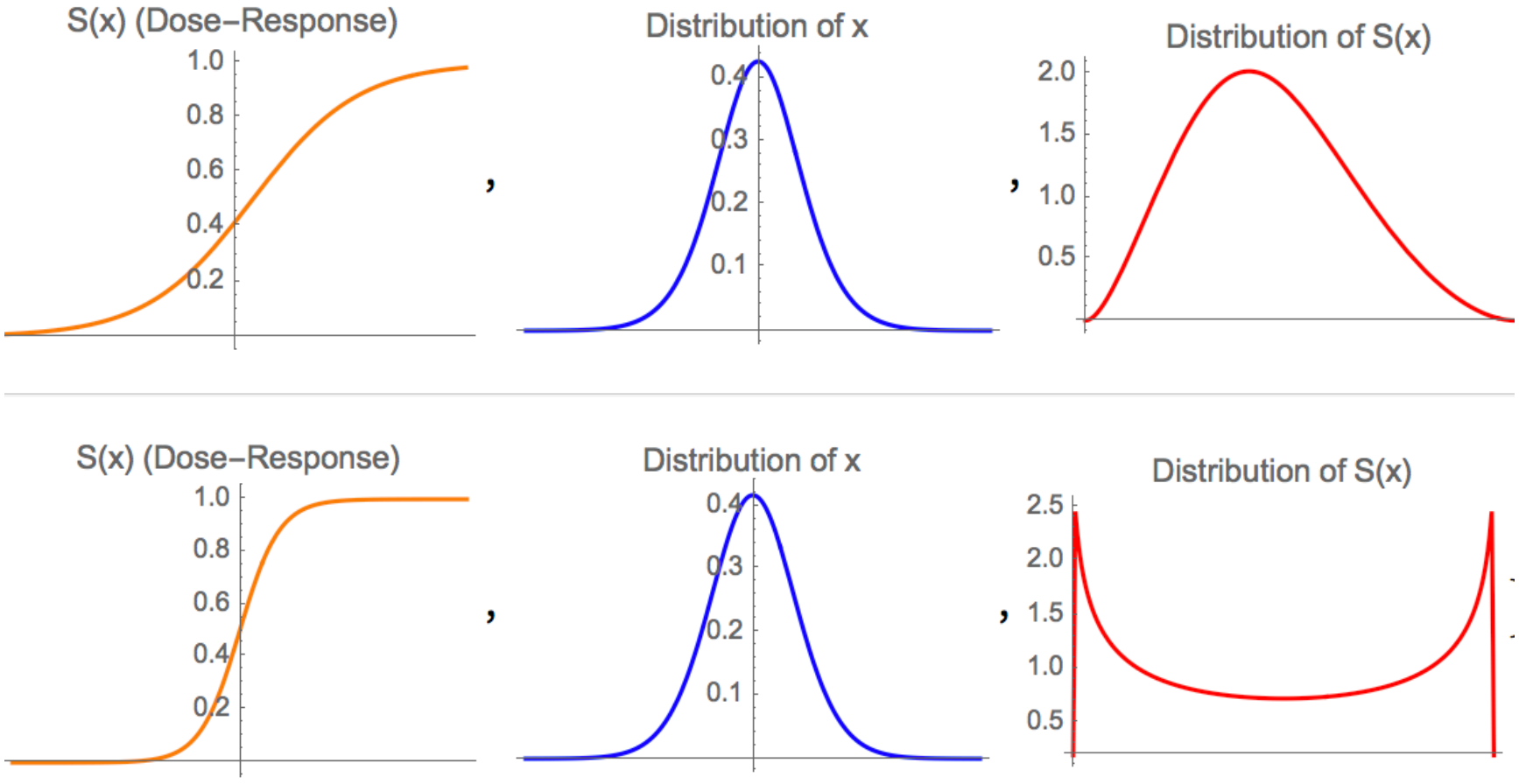

Figure 3.30: *The Conflation Problem: a convex-concave transformation of a thick tailed X produces a thin tailed distribution (above). A sigmoidal transformation (below) that is bounded on a distribution in* $(-\infty, \infty)$ *produces an ArcSine distribution, with compact support.*

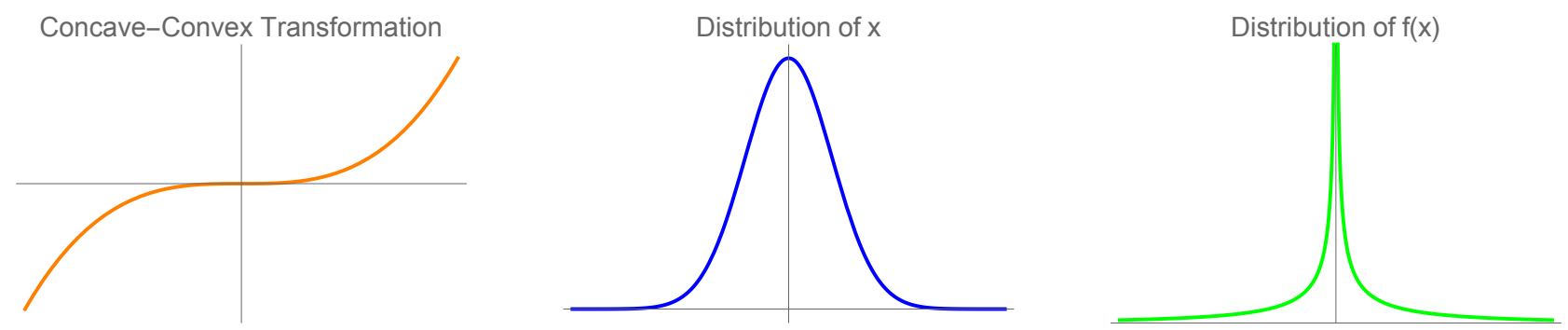

Figure 3.31: *A concave-convex transformation (of the style of a probit –an inverse CDF for the Gaussian– or of a logit) makes the tails of the distribution of* $f(x)$ *thicker*

The danger with the treatment of the Black Swan problem is as follows: people focus on $X$ ("predicting $X$"). My point is that, although we do not understand $X$, we can deal with it by working on F which we can understand, while others work on predicting $X$ which we can't because small probabilities are incomputable, particularly in thick tailed domains. $F(x)$ is how the end result affects you.

The probability distribution of $F(X)$ is markedly different from that of $X$, particularly when $F(X)$ is nonlinear. We need a nonlinear transformation of the distribution of $X$ to get $F(X)$. We had to wait until 1964 to start a discussion on "convex transformations of random variables", Van Zwet (1964)[313] –as the topic didn't seem important before.

**Ubiquity of S curves** $F$ is almost always nonlinear (actually I know of no exception to nonlinearity), often "S curved", that is convex-concave (for an increasing function). See the longer discussion in G.

**Fragility and Antifragility** When $F(X)$ is concave (fragile), errors about $X$ can translate into extreme negative values for $F(X)$. When $F(X)$ is convex, one is largely immune from severe negative variations. In situations of trial and error, or with an option, we do not need to understand $X$ as much as our exposure to the risks. Simply the statistical properties of $X$ are swamped by those of $H$. The point of *Antifragile* is that exposure is more important than the naive notion of "knowledge", that is, understanding $X$.
The more nonlinear $F$ the less the probabilities of $X$ matters in the probability distribution of the final package $F$.
Many people confuse the probabilites of $X$ with those of $F$. I am serious: the *entire* literature reposes largely on this mistake. For Baal's sake, focus on $F$, not $X$.

## 3.11 RUIN AND PATH DEPENDENCE

Let us finish with path dependence and time probability. Our greatgrandmothers did understand thick tails. These are not so scary; we figured out how to survive by making rational decisions based on deep statistical properties.

Path dependence is as follows. If I iron my shirts and then wash them, I get vastly different results compared to when I wash my shirts and then iron them. My first work, *Dynamic Hedging* [271], was about how traders avoid the "absorbing barrier" since once you are bust, you can no longer continue: anything that will eventually go bust will lose *all* past profits.

The physicists Ole Peters and Murray Gell-Mann [218] shed new light on this point, and revolutionized decision theory showing that a key belief since the development of applied probability theory in economics was wrong. They pointed out that all economics textbooks make this mistake; the only exception are by information theorists such as Kelly and Thorp.

Let us explain ensemble probabilities.

Assume that 100 of us, randomly selected, go to a casino and gamble. If the $28^{th}$ person is ruined, this has no impact on the $29^{th}$ gambler. So we can compute the casino's return using the law of large numbers by taking the returns of the 100 people who gambled. If we do this two or three times, then we get a good estimate of what the casino's "edge" is. The problem comes when ensemble probability is applied to us as individuals. It does not work because if one of us goes to the casino and on day 28 is ruined, there is no day 29. This is why Cramer showed insurance could not work outside what he called "the Cramer condition", which excludes possible ruin from single shocks. Likewise, no individual investor will achieve the alpha return on the market because no single investor has infinite pockets (or, as Ole Peters has observed, is running his life across branching parallel universes). We can only get the return on the market under strict conditions.

Time probability and ensemble probability are not the same. This only works if the risk takers has an allocation policy compatible with the Kelly criterion [170],[303] using logs. Peters wrote three papers on time probability (one with Murray Gell-Mann) and showed that a lot of paradoxes disappeared.

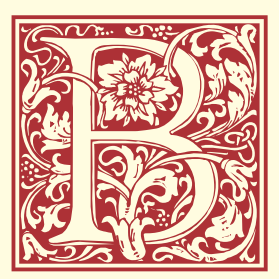**etter be convex than right:** In the fall of 2017, a firm went bust betting against volatility –they were predicting lower real market volatility (rather, variance) than "expected" by the market. *They were correct in the prediction, but went bust nevertheless.* They were just very concave in the payoff function. Recall that $x$ is not $f(x)$ and that in the real world there are almost no linear $f(x)$.

The following example can show us how. Consider the following payoff in the figure below. The payoff function is $f(x) = 1 - x^2$ daily, meaning if $x$ moves by up to 1 unit (say, standard deviation), there is a profit, losses beyond. This is a typical contract called "variance swap".

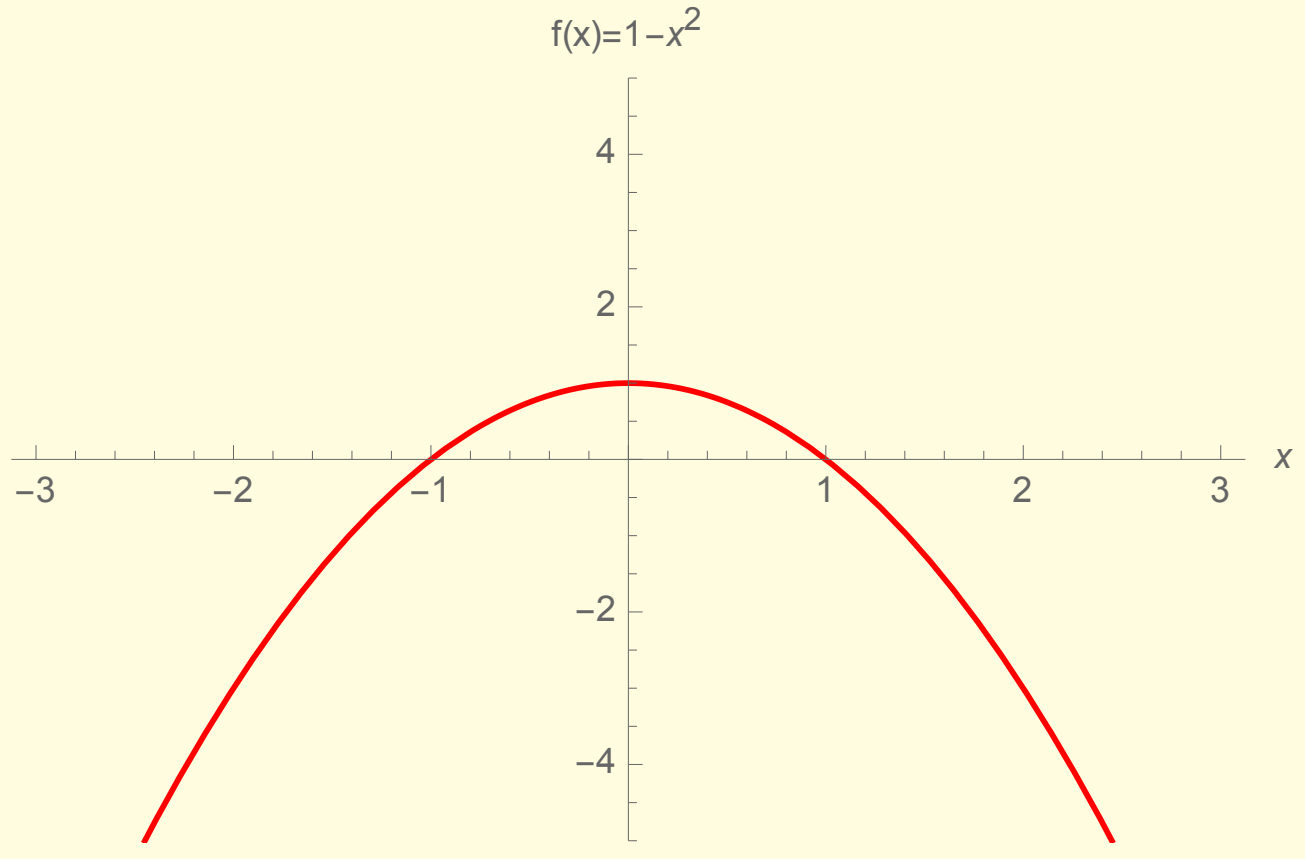

Now consider the following two types successions of deviations of $x$ for 7 days (expressed in standard deviations).

Succession 1 (thin tails): $\{1, 1, 1, 1, 1, 0, 0\}$. Mean variation= 0.71. P/L= 2.

Succession 2 (thick tails): $\{0, 0, 0, 0, 0, 0, 5\}$. Mean variation= 0.71 (same). P/L=−18 (bust, really bust).

In both cases they forecast right, but the lumping of the volatility –the fatness of tails– made a huge difference.

This in a nutshell explains why, in the real world, "bad" forecasters can make great traders and decision makers and vice versa –something every operator knows but that the mathematically and practically unsophisticated "forecasting" literature, centuries behind practice, misses.

Let us see how we can work with these and what is wrong with the literature. If we visibly incur a tiny risk of ruin, but have a frequent exposure, it will go to probability one over time. If we ride a motorcycle we have a small risk of ruin, but if we ride that motorcycle a lot then we will reduce our life expectancy. The way to measure this is:

**Principle 3.3: Repetition of exposures**

*Focus only on the reduction of life expectancy of the unit assuming repeated exposure at a certain density or frequency.*

Behavioral finance so far makes conclusions from statics not dynamics, hence misses the picture. It applies trade-offs out of context and develops the consensus that people irrationally overestimate tail risk (hence need to be "nudged" into taking more of these exposures). But the catastrophic event is an absorbing barrier. No risky exposure can be analyzed in isolation: risks accumulate. If we ride a motorcycle, smoke, fly our own propeller plane, and join the mafia, these risks add up to a near-certain premature death. Tail risks are not a renewable resource.

Every risk taker who managed to survive understands this. Warren Buffett understands this. Goldman Sachs understands this. They do not want small risks, they want zero risk because that is the difference between the firm surviving and not surviving over twenty, thirty, one hundred years. This attitude to tail risk can explain that Goldman Sachs is 149 years old –it ran as partnership with unlimited liability for approximately the first 130 years, but was bailed out once in 2009, after it became a bank. This is not in the decision theory literature but we (people with skin in the game) practice it every day. We take a unit, look at how long a life we wish it to have and see by how much the life expectancy is reduced by *repeated* exposure.

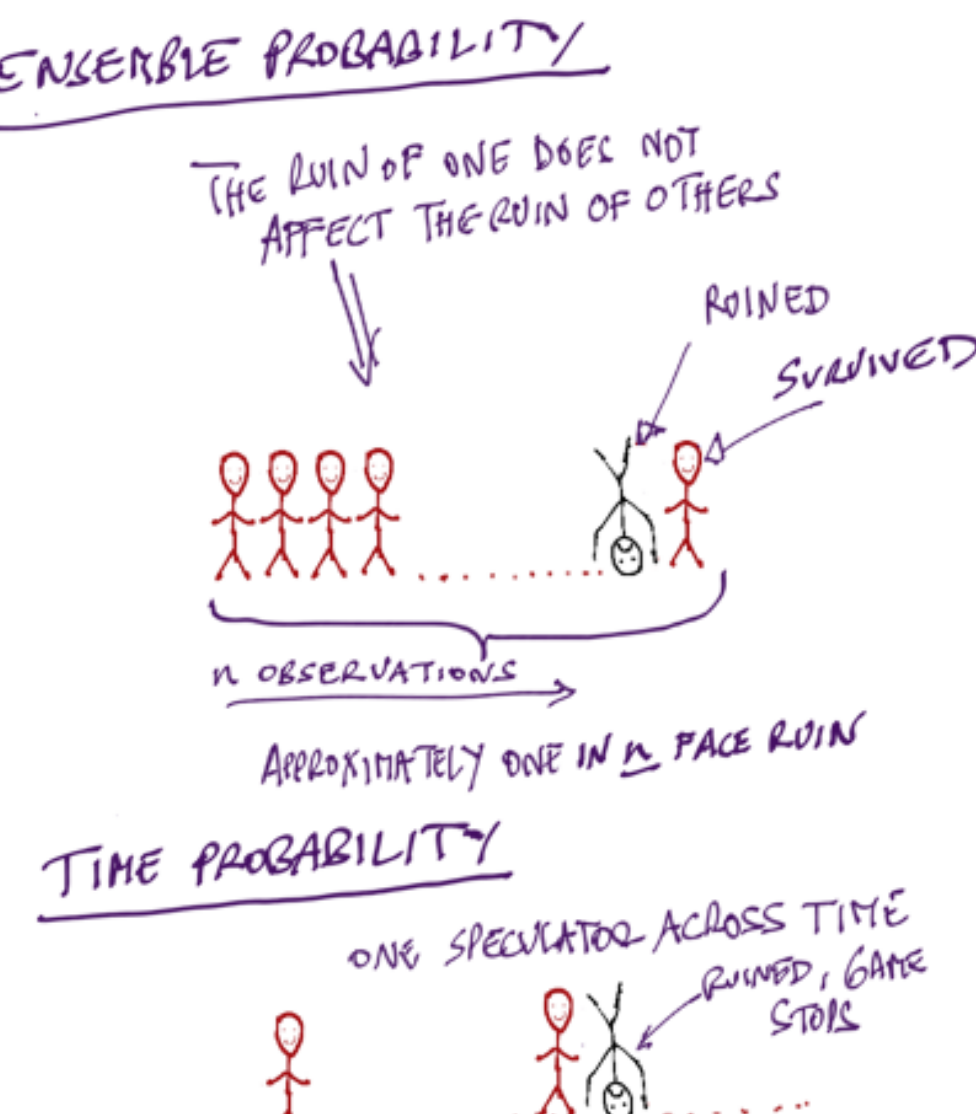

Figure 3.32: *Ensemble probability vs. time probability. The treatment by option traders is done via the absorbing barrier. I have traditionally treated this in Dynamic Hedging [271] and Antifragile[269] as the conflation between X (a random variable) and $f(X)$ a function of said r.v., which may include an absorbing state.*

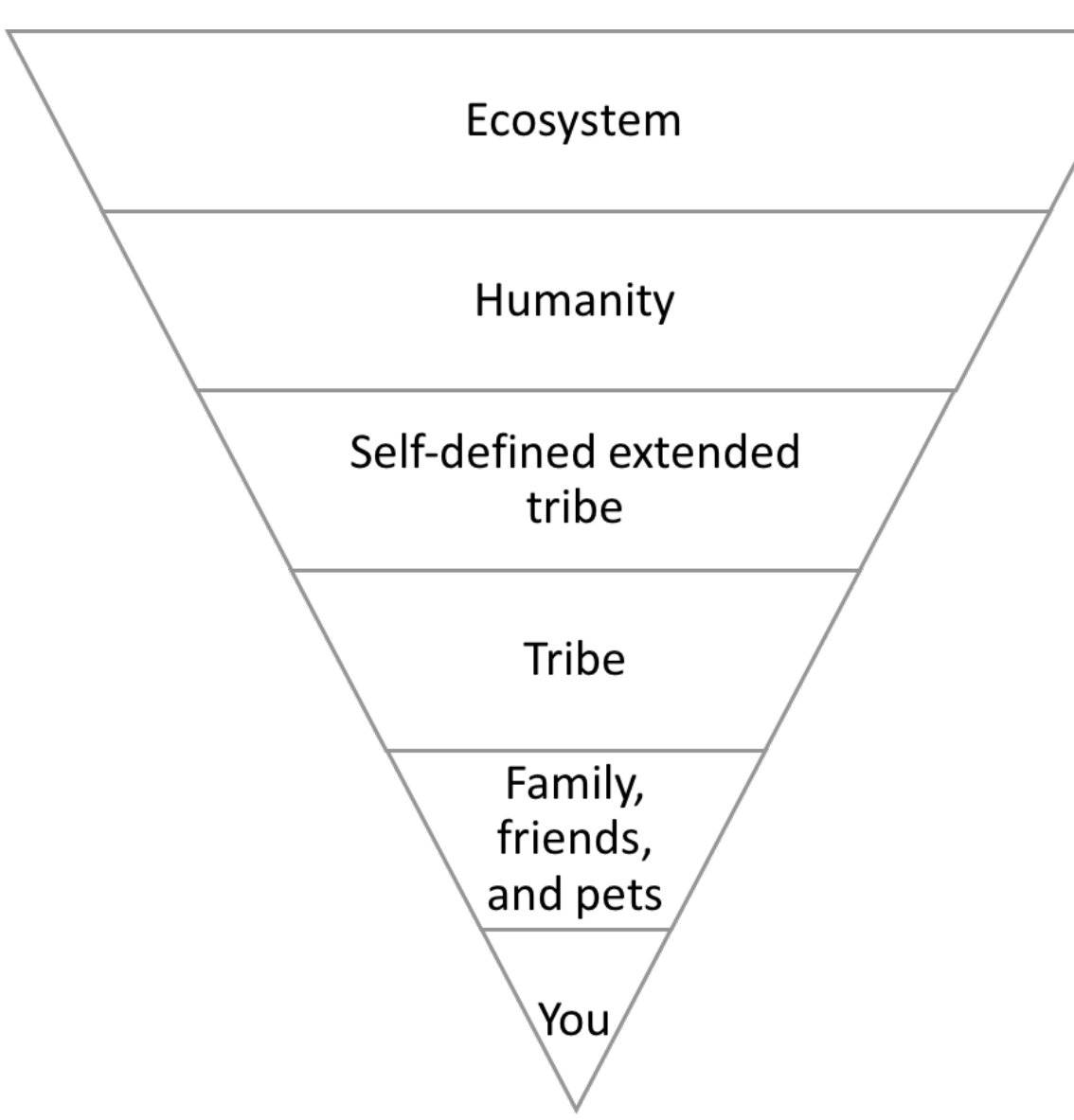

Figure 3.33: *A hierarchy for survival. Higher entities have a longer life expectancy, hence tail risk matters more for these. Lower entities such as you and I are renewable.*

**Remark 3.2: Psychology of decision making**

*The psychological literature focuses on one-single episode exposures and narrowly defined cost-benefit analyses. Some analyses label people as paranoid for overestimating small risks, but don't get that if we had the smallest tolerance for collective tail risks, we would not have made it for the past several million years.*

Next let us consider layering, why systemic risks are in a different category from individual, idiosyncratic ones. Look at the (inverted) pyramid in Figure 3.33: the worst-case scenario is not that an individual dies. It is worse if your family, friends and pets die. It is worse if you die and your arch enemy survives. They collectively have more life expectancy lost from a terminal tail event.

So there are layers. The biggest risk is that the entire ecosystem dies. The precautionary principle puts structure around the idea of risk for units expected to survive.

Ergodicity in this context means that your analysis for ensemble probability translates into time probability. If it doesn't, ignore ensemble probability altogether.

## 3.12 WHAT TO DO?

To summarize, we first need to make a distinction between mediocristan and Extremistan, two separate domains that about never overlap with one another. If we fail to make that distinction, we don't have any valid analysis. Second, if we don't make the distinction between time probability (path dependent) and ensemble probability (path independent), we don't have a valid analysis.

The next phase of the *Incerto* project is to gain understanding of fragility, robustness, and, eventually, anti-fragility. Once we know something is fat-tailed, we can use heuristics to see how an exposure there reacts to random events: how much is a given unit harmed by them. It is vastly more effective to focus on being insulated from the harm of random events than try to figure them out in the required details (as we saw the inferential errors under thick tails are huge). So it is more solid, much wiser, more ethical, and more effective to focus on detection heuristics and policies rather than fabricate statistical properties.

The beautiful thing we discovered is that everything that is fragile has to present a concave exposure [269] similar –if not identical –to the payoff of a short option, that is, a negative exposure to volatility. It is nonlinear, necessarily. It has to have harm that accelerates with intensity, up to the point of breaking. If I jump 10m I am harmed more than 10 times than if I jump one meter. That is a necessary property of fragility. We just need to look at acceleration in the tails. We have built effective stress testing heuristics based on such an option-like property [290].

In the real world we want simple things that work [127]; we want to impress our accountant and not our peers. (My argument in the latest instalment of the *Incerto*, *Skin in the Game* is that systems judged by peers and not evolution rot from overcomplication). To survive we need to have clear techniques that map to our procedural intuitions.

The new focus is on how to detect and measure convexity and concavity. This is much, much simpler than probability.

## 3.13 KAHNEMAN'S PARTING GIFT

JOSEPH WALKER, in a podcast, asked the great Daniel Kahneman, shortly before the latter's passing: "So, as you know, Nassim Taleb argues that we underestimate tail risks. Does that contradict prospect theory?" Kahneman's answer was:

> "Prospect theory is not a realistic description of how one would think in Nassim Taleb's world. Certainly not a description of how one should think in Nassim Taleb's world."[288]

Clearly, there is no such thing as "Taleb's world"; Kahneman here means the *real* world as described to him by this author and presented in this book, in addition to real-world dynamics that entail ruin in the event of model error.

## NEXT

The next three chapters will examine the technical intuitions behind thick tails in discussion form, in not too formal a language. Derivations and formal proofs come later with the adaptations of the journal articles.

# 4 UNIVARIATE FAT TAILS, LEVEL 1, FINITE MOMENTS†

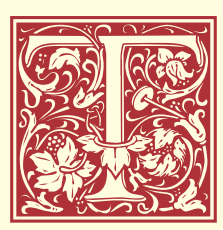

THE NEXT TWO chapters organized as follows. We look at three levels of fat-tails with more emphasis on the intuitions and heuristics than formal mathematical differences, which will be pointed out later in the discussions of limit theorems. The three levels are:

- Fat tails, entry level (sort of), i.e., finite moments
- Subexponential class
- Power Law class

Level one will be the longest as we will use it to build intuitions. While this approach is the least used in mathematics papers (fat tails are usually associated with power laws and limit behavior), it is relied upon the most analytically and practically. We can get the immediate consequences of fat-tailedness with little effort, the equivalent of a functional derivative that provides a good grasp of local sensitivities. For instance, as a trader, the author was able to get most of the effect of fattailedness with a simple heuristic of averaging option prices across two volatilities, which proved sufficient in spite of its simplicity.

## 4.1 A SIMPLE HEURISTIC TO CREATE MILDLY FAT TAILS

A couple of reminders about convexity and Jensen's inequality:

Let $\mathscr{A}$ be a convex set in a vector space in $\mathbb{R}$, and let $\varphi : \mathscr{A} \to \mathbb{R}$ be a function; $\varphi$ is called convex if $\forall x_1, x_2 \in \mathscr{A}, \forall t \in [0,1]$ :

$$\varphi\left(t x_1 + (1-t) x_2\right) \leq t \varphi\left(x_1\right) + (1-t)\varphi\left(x_2\right)$$

† Discussion chapter.

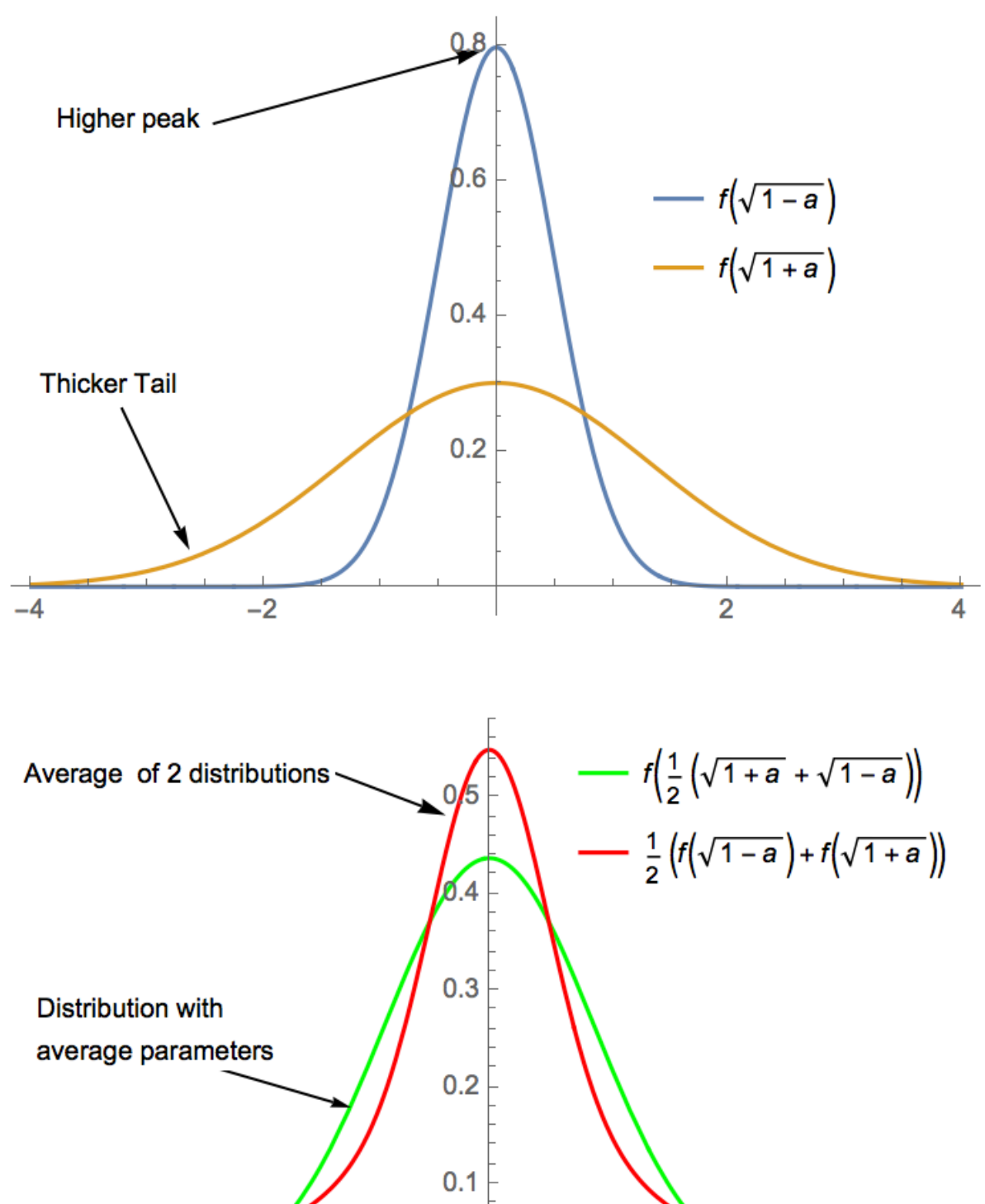

Figure 4.1: *How random volatility creates fatter tails owing to the convexity of some parts of the density to the scale of the distribution.*

For a random variable $X$ and $\varphi(.)$ a convex function, by Jensen's inequality[161]:

$$\varphi(\mathbb{E}[X]) \leq \mathbb{E}[\varphi(X)].$$

**Remark 4.1: Fat Tails and Jensen's inequality**

*For a Gaussian distribution (and, members of the location-scale family of distributions), tail probabilities are convex to the scale of the distribution, here the standard deviation $\sigma$ (and to the variance $\sigma^2$). This allows us to fatten the tails by "stochasticizing" either the standard deviation or the variance, hence checking the effect of Jensen's inequality on the probability distribution.*

Heteroskedasticity is the general technical term often used in time series analysis to characterize a process with fluctuating scale. Our method "stochasticizes", that is, perturbates the variance or the standard deviation[2] of the distribution under the constraint of conservation of the mean.

[2] "Volatility" in the quant language means standard deviation, but "stochastic volatility" is usually stochastic variance.

But note that *any* heavy tailed process, even a power law, can be described *in sample* (that is finite number of observations necessarily discretized) by a simple Gaussian process with changing variance, a regime switching process, or a combination of Gaussian plus a series of variable jumps (though not one where jumps are of equal size, see the summary in [203]).[3]

This method will also allow us to answer the great question: "where do the tails start?" in 4.3.

Let $f(\sqrt{a}, x)$ be the density of the normal distribution (with mean 0) as a function of the variance for a given point $x$ of the distribution.

Compare $f\left(\frac{1}{2}\left(\sqrt{1-a}+\sqrt{a+1}\right), x\right)$ to $\frac{1}{2}\left(f\left(\sqrt{1-a}, x\right)+f\left(\sqrt{a+1}, x\right)\right)$; the difference between the two will be owed to Jensen's inequality. We assume the average $\sigma^2$ constant, but the discussion works just as well if we just assumed $\sigma$ constant —it is a long debate whether one should put a constraint on the average variance or on that of the standard deviation, but 1) doesn't matter much so long as one remains consistent, 2) for our illustrative purposes here there is no real fundamental difference.

Since higher moments increase under fat tails, though not necessarily lower ones, it should be possible to simply increase fat tailedness (via the fourth moment) while keeping lower moments (the first two or three) invariant. [4]

### 4.1.1 A Variance-preserving heuristic

Keep $\mathbb{E}(X^2)$ constant and increase $\mathbb{E}(X^4)$, by "stochasticizing" the variance of the distribution, since $\mathbb{E}(X^4)$is itself analog to the variance of $\mathbb{E}(X^2)$ measured across samples – $\mathbb{E}(X^4)$ is the noncentral equivalent of $\mathbb{E}\left(\left(X^2-\mathbb{E}\left(X^2\right)\right)^2\right)$ so we will focus on the simpler version outside of situations where it matters. Further, we will do the "stochasticizing" in a more involved way in later sections of the chapter.

An effective heuristic to get some intuition about the effect of the fattening of tails consists in simulating a random variable set to be at mean 0, but with the following variance-preserving tail fattening trick: the random variable follows a distribution $N(0, \sigma\sqrt{1-a})$ with probability $p=\frac{1}{2}$ and $N\left(0, \sigma\sqrt{1+a}\right)$ with the remaining probability $\frac{1}{2}$, with $0 \leqslant a < 1$.

---

3 The jumps for such a process can be simply modeled as a regime that is characterized by a Gaussian with low variance and extremely large mean (and a low-probability of occurrence), so, technically, Poisson jumps are mixed Gaussians.

4 To repeat what we stated in the previous chapter, the literature sometimes separates "Fat tails" from "Heavy tails", the first term being reserved for power laws, the second to subexponential distribution (on which, later). Fughedaboutdit. We simply call "Fat Tails" something with a higher kurtosis than the Gaussian, even when kurtosis is not defined. The definition is functional as used by practioners of fat tails, that is, option traders and lends itself to the operation of "fattening the tails", as we will see in this section.

The characteristic function[5] is

$$\phi(t,a) = \frac{1}{2} e^{-\frac{1}{2}(1+a)t^2\sigma^2} \left(1 + e^{a t^2 \sigma^2}\right) \tag{4.1}$$

Odd moments are nil. The second moment is preserved since

$$M(2) = (-i)^2 \partial^{t,2} \phi(t)|_0 \ = \sigma^2 \tag{4.2}$$

and the fourth moment

$$M(4) = (-i)^4 \partial^{t,4} \phi|_0 = 3\left(a^2+1\right)\sigma^4 \tag{4.3}$$

which puts the traditional kurtosis at $3\left(a^2+1\right)$ (assuming we do not remove 3 to compare to the Gaussian). This means we can get an "implied $a$ from kurtosis. The value of $a$ is roughly the mean deviation of the stochastic volatility parameter "volatility of volatility" or Vvol in a more fully parametrized form.

**Limitations of the simple heuristic** This heuristic, while useful for intuition building, is of limited powers as it can only raise kurtosis to twice that of a Gaussian, so it should be used only pedagogically, to get some intuition about the effects of the convexity. Section 4.1.2 will present a more involved technique.

**Remark 4.2: Peaks**

*As Figure 4.4 shows: fat tails manifests themselves with higher peaks, a concentration of observations around the center of the distribution.*

This is usually misunderstood.

### 4.1.2 Fattening of Tails With Skewed Variance

We can improve on the fat-tail heuristic in 4.1, (which limited the kurtosis to twice the Gaussian) as follows. We Switch between Gaussians with variance:

$$\begin{cases} \sigma^2(1+a), & \text{with probability } p \\ \sigma^2(1+b), & \text{with probability } 1-p \end{cases} \tag{4.4}$$

with $p \in [0,1)$ and $b = -a\frac{p}{1-p}$, giving a characteristic function:

$$\phi(t,a) = p\, e^{-\frac{1}{2}(a+1)\sigma^2 t^2} - (p-1)\, e^{-\frac{\sigma^2 t^2 (ap+p-1)}{2(p-1)}}$$

with Kurtosis $\frac{3\left(\left(1-a^2\right)p-1\right)}{p-1}$ thus allowing polarized states and high kurtosis, all variance preserving.

Thus with, say, $p = 1/1000$, and the corresponding maximum possible $a = 999$, kurtosis can reach as high a level as 3000.

[5] Note there is no difference between characteristic and moment generating functions when the mean is 0, a property that will be useful in later, more technical chapters.

This heuristic approximates quite well the effect on probabilities of a lognormal weighting for the characteristic function

$$\phi(t,V)=\int_0^\infty \frac{e^{-\frac{t^2 v}{2}-\frac{\left(\log(v)-v0+\frac{Vv^2}{2}\right)^2}{2Vv^2}}}{\sqrt{2\pi v Vv}}\,dv \tag{4.5}$$

where $v$ is the variance and $Vv$ is the second order variance, often called volatility of volatility. Thanks to integration by parts we can use the Fourier transform to obtain all varieties of payoffs (see Gatheral [120]). But the absence of a closed-form distribution can be remedied as follows, with the use of distributions for the variance that are analytically more tractable.

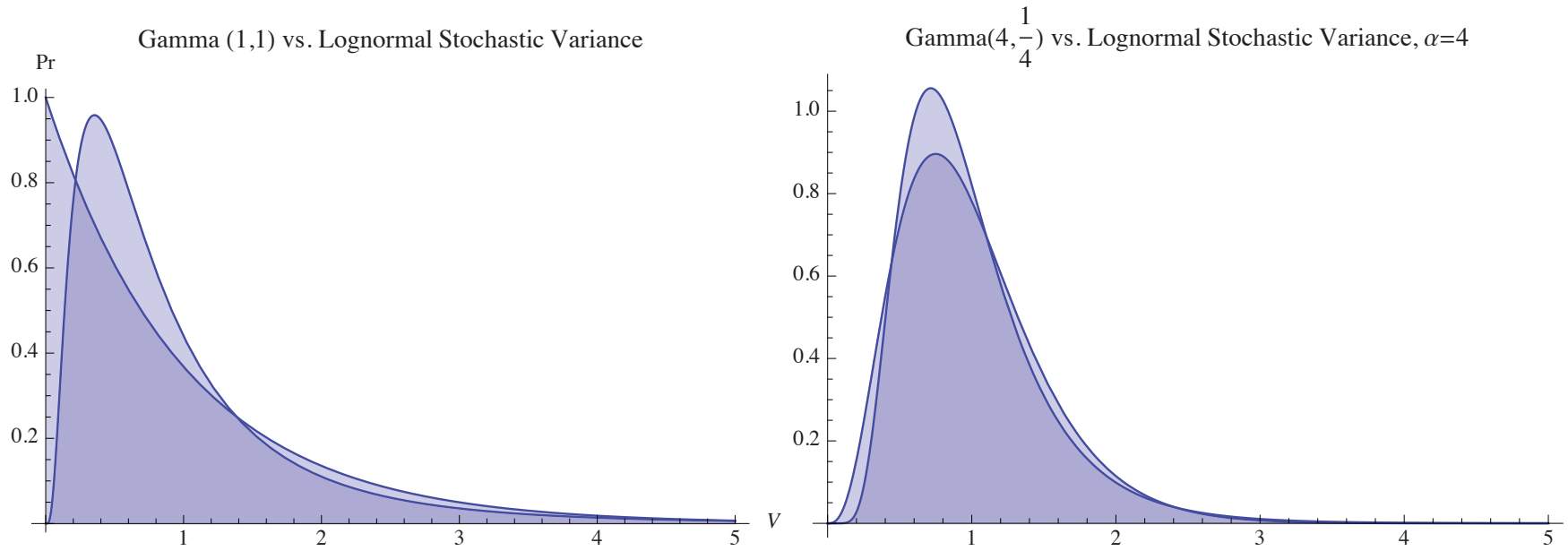

Figure 4.2: *Stochastic Variance: Gamma distribution and Lognormal of same mean and variance.*

**Gamma Variance** The gamma distribution applied to the variance of a Gaussian is is a useful shortcut for a full distribution of the variance, which allows us to go beyond the narrow scope of heuristics [40]. It is easier to manipulate analytically than the Lognormal.

Assume that the variance of the Gaussian follows a gamma distribution.

$$\Gamma_a(v)=\frac{v^{a-1}\left(\frac{V}{a}\right)^{-a} e^{-\frac{av}{V}}}{\Gamma(a)}$$

with mean $V$ and variance $\frac{V}{\sqrt{a}}$. Figure 4.2 shows the matching to a lognormal with same first two moments where we calibrate the lognormal to mean $\frac{1}{2}\log\left(\frac{aV^3}{aV+1}\right)$ and standard deviation $\sqrt{-\log\left(\frac{aV}{aV+1}\right)}$. The final distribution becomes (once again, assuming the same mean as a fixed volatility situation:

$$f_{a,V}(x)=\int_0^\infty \frac{e^{-\frac{(x-\mu)^2}{2v}}}{\sqrt{2\pi}\sqrt{v}}\Gamma_a(v)\mathrm{d}v, \tag{4.6}$$

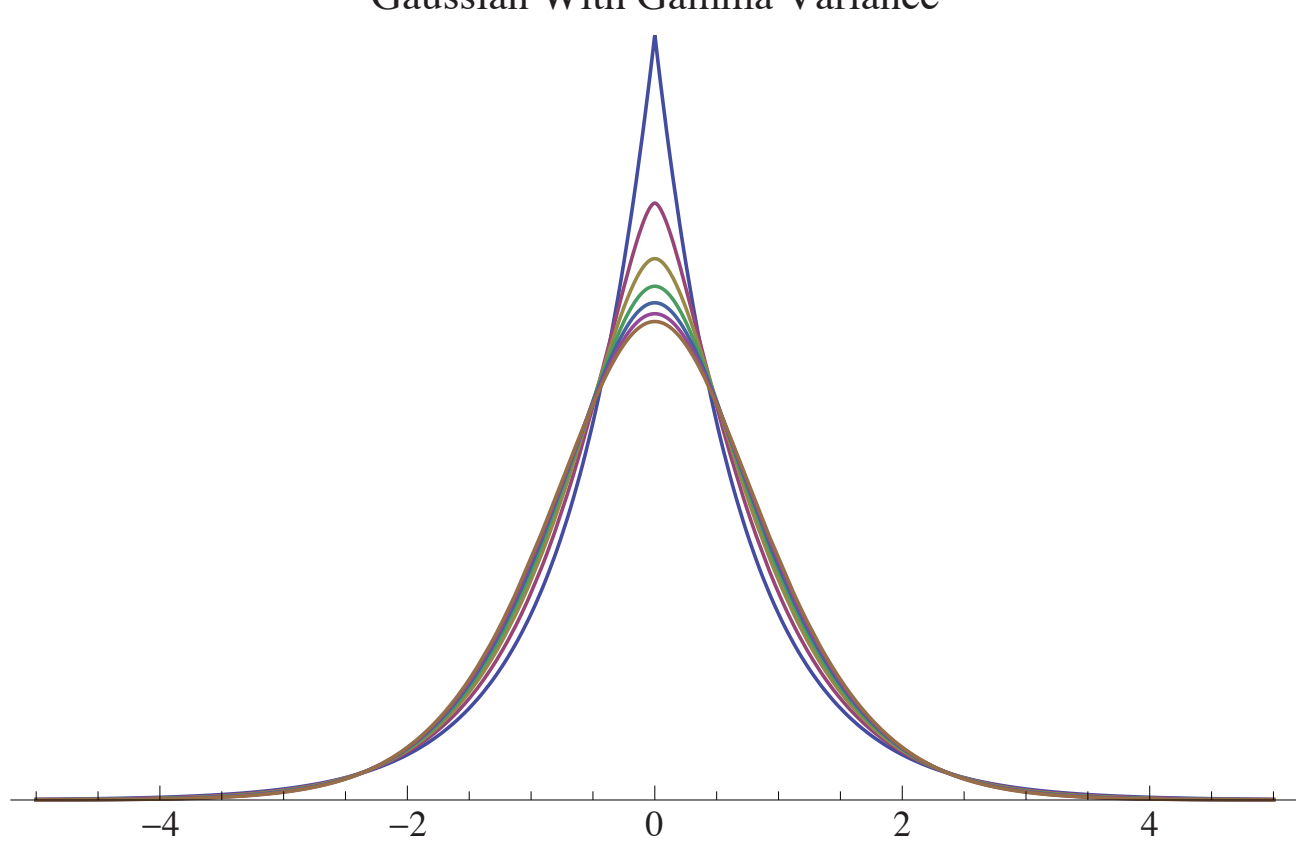

Figure 4.3: *Stochastic Variance using Gamma distribution by perturbating α in equation 4.7.*

allora:

$$f_{\alpha,V}(x) = \frac{2^{\frac{3}{4}-\frac{a}{2}} a^{\frac{a}{2}+\frac{1}{4}} V^{-\frac{a}{2}-\frac{1}{4}} \left|x-\mu\right|^{a-\frac{1}{2}} K_{a-\frac{1}{2}}\left(\frac{\sqrt{2}\sqrt{a}|x-\mu|}{\sqrt{V}}\right)}{\sqrt{\pi}\Gamma(a)}. \tag{4.7}$$

where $K_n(z)$ is the Bessel $K$ function, which satisfies the differential equation

$$-y\left(n^2+z^2\right)+z^2y''+zy'=0.$$

Let us now get deeper into the different forms of stochastic volatility.

## 4.2 DOES STOCHASTIC VOLATILITY GENERATE POWER LAWS?

We have not yet defined power laws; take for now the condition that least one of the moments is infinite.

And the answer: depend on whether we are stochasticizing $\sigma$ or $\sigma^2$ on one hand, or $\frac{1}{\sigma}$ or $\frac{1}{\sigma^2}$ on the other.

Assume the base distribution is the Gaussian, the random variable $X \sim \mathcal{N}(\mu, \sigma)$. Now there are different ways to make $\sigma$, the scale, stochastic. Note that since $\sigma$ is non-negative, we need it to follow some one-tailed distribution.

- We can make $\sigma^2$ (or, possibly $\sigma$) follow a Lognormal distribution. It does not yield closed form solutions, but we can get the moments and verify it is not a power law.
- We can make $\sigma^2$ (or $\sigma$) follow a gamma distribution. It does yield closed form solutions, as we saw in the example above, in Eq. 4.7.
- We can make $\frac{1}{\sigma^2}$ —the precision parameter—follow a gamma distribution.
- We can make $\frac{1}{\sigma^2}$ follow a lognormal distribution.

The results shown in Table 4.1 come from the following simple properties of density functions and expectation operators. Let $X$ be any random variable with PDF $f(.)$ in the location-scale family, and $\lambda$ any random variable with PDF $g(.)$; $X$ and $\lambda$ are assumed to be independent. Since by standard results, the moments of order $p$ for the product and the ratio $\frac{X}{\lambda}$ are:

$$\mathbb{E}\left((X\lambda)^p\right) = \mathbb{E}\left(X^p\right)\mathbb{E}\left(\lambda^p\right)$$

and

$$\mathbb{E}\left(\left(\frac{X}{\lambda}\right)^p\right) = \mathbb{E}\left(\left(\frac{1}{\lambda}\right)^p\right)\mathbb{E}\left(X^p\right).$$

Table 4.1: *Transformations for stochastic volatility. We can see from the density of the transformations $\frac{1}{x}$ or $\frac{1}{\sqrt{x}}$ if we have a power law on hand. $\mathcal{LN}$, $\mathcal{N}$,$\mathcal{G}$ and $\mathcal{P}$ are the Lognormal, Normal, Gamma, and Pareto distributions, respectively.*

| distr | $p(x)$ | $p\left(\frac{1}{x}\right)$ | $p\left(\frac{1}{\sqrt{x}}\right)$ |
|---|---|---|---|
| $\mathcal{LN}(m,s)$ | $\frac{e^{-\frac{(m-\log(x))^2}{2s^2}}}{\sqrt{2\pi}sx}$ | $\frac{e^{-\frac{(m+\log(x))^2}{2s^2}}}{\sqrt{2\pi}sx}$ | $\frac{\sqrt{\frac{2}{\pi}}e^{-\frac{(m+2\log(x))^2}{2s^2}}}{sx}$ |
| $\mathcal{N}(m,s)$ | $\frac{e^{-\frac{(m-x)^2}{2s^2}}}{\sqrt{2\pi}s}$ | $\frac{e^{-\frac{\left(m-\frac{1}{x}\right)^2}{2s^2}}}{\sqrt{2\pi}sx^2}$ | $\frac{\sqrt{\frac{2}{\pi}}e^{-\frac{\left(m-\frac{1}{x^2}\right)^2}{2s^2}}}{sx^3}$ |
| $\mathcal{G}(a,b)$ | $\frac{b^{-a}x^{a-1}e^{-\frac{x}{b}}}{\Gamma(a)}$ | $\frac{b^{-a}x^{-a-1}e^{-\frac{1}{bx}}}{\Gamma(a)}$ | $\frac{2b^{-a}x^{-2a-1}e^{-\frac{1}{bx^2}}}{\Gamma(a)}$ |
| $\mathcal{P}(1,\alpha)$ | $\alpha x^{-\alpha-1}$ | $\alpha x^{\alpha-1}$ | $2\alpha x^{2\alpha-1}$ |

Table 4.2: *The p-moments of possible distributions for variance*

| distr | $\mathbb{E}\left(X^p\right)$ | $\mathbb{E}\left((\frac{1}{X})^p\right)$ | $\mathbb{E}\left((\frac{1}{\sqrt{X}})^p\right)$ |
|---|---|---|---|
| $\mathcal{LN}(m,s)$ | $e^{mp+\frac{p^2s^2}{2}}$ | $e^{\frac{1}{2}p\left(ps^2-2m\right)}$ | $e^{\frac{1}{8}p\left(ps^2-4m\right)}$ |
| $\mathcal{G}(a,b)$ | $b^p(a)_p$ | $\frac{(-1)^p b^{-p}}{(1-a)_p},\ p<a$ | fughedaboudit |
| $\mathcal{P}(1,\alpha)$ | $\frac{\alpha}{\alpha-p},\ p<\alpha$ | $\frac{\alpha}{\alpha+p}$ | $\frac{2\alpha}{2\alpha+p}$ |

(via the Mellin transform).

Note that as proprety of location-scale family, $\frac{1}{\lambda}f_{\frac{x}{\lambda}}(\frac{x}{\lambda}) = f_x(\frac{x}{\lambda})$ so, for instance, if $x \sim \mathcal{N}(0,1)$ (that is, normally distributed), then $\frac{x}{\sigma} \sim \mathcal{N}(0,\sigma)$.

## 4.3 THE BODY, THE SHOULDERS, AND THE TAILS

Where do the tails start?

We assume the tails start at the level of convexity of the segment of the probability distribution to the scale of the distribution –in other words, affected by the stochastic volatility effect.

### 4.3.1 The Crossovers and Tunnel Effect.

Notice in Figure 4.4 a series of crossover zones, invariant to $a$. Distributions called "bell shape" have a convex-concave-convex shape (or quasi-concave shape).

Let $X$ be a random variable with distribution with PDF $p(x)$ from a general class of all unimodal one-parameter continuous pdfs $p_\sigma$ with support $\mathcal{D} \subseteq \mathbb{R}$ and scale parameter $\sigma$. Let $p(.)$ be quasi-concave on the domain, but neither convex nor concave. The density function $p(x)$ satisfies: $p(x) \geq p(x+\epsilon)$ for all $\epsilon > 0$, and $x > x^*$ and $p(x) \geq p(x-\epsilon)$ for all $x < x^*$ with $x^* = argmax_x p(x)$

$$p\left(\omega\, x + (1-\omega)\; y\right) \geq \min\left(p(x), p(y)\right).$$

A- If the variable is "two-tailed", that is, its domain of support $\mathcal{D}$= (-∞,∞), and where $p^\delta(x) \triangleq \frac{p(x,\sigma+\delta)+p(x,\sigma-\delta)}{2}$,

1. There exist a "high peak" inner tunnel, $A_T$= $(a_2, a_3)$ for which the $\delta$-perturbed $\sigma$ of the probability distribution $p^\delta(x){\geq}p(x)$ if $x \in (a_2, a_3)$
2. There exists outer tunnels, the "tails", for which $p^\delta(x){\geq}p(x)$ if $x \in (-\infty, a_1)$ or $x \in (a_4, \infty)$
3. There exist intermediate tunnels, the "shoulders", where $p^\delta(x){\leq}\, p(x)$ if $x \in (a_1, a_2)$ or $x \in (a_3, a_4)$

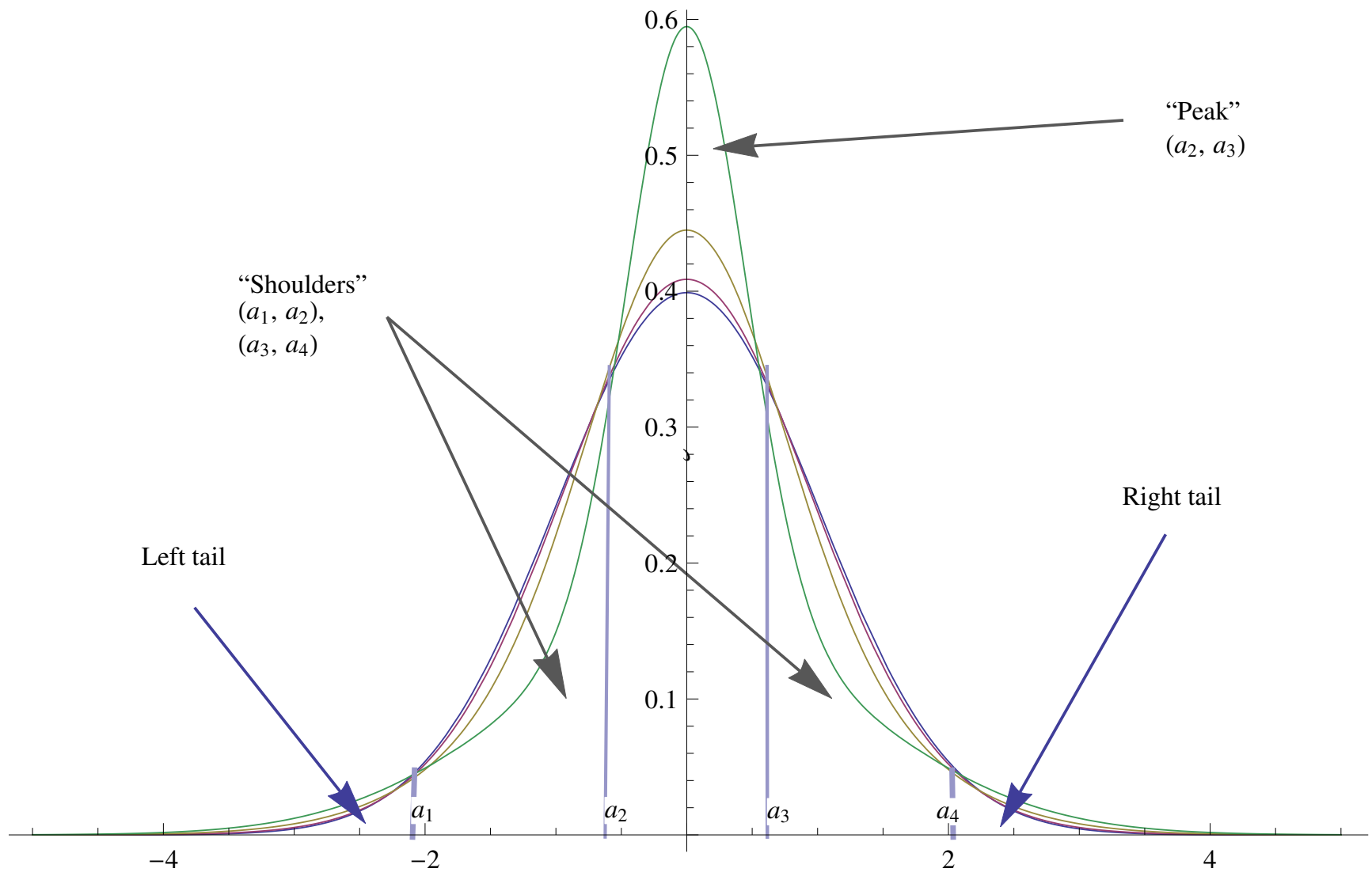

Figure 4.4: *Where do the tails start? Fatter and fatter fails through perturbation of the scale parameter $\sigma$ for a Gaussian, made more stochastic (instead of being fixed). Some parts of the probability distribution gain in density, others lose. Intermediate events are less likely, tails events and moderate deviations are more likely. We can spot the crossovers $a_1$ through $a_4$. The "tails" proper start at $a_4$ on the right and $a_1$on the left.*

**The Black Swan Problem:** As we saw, it is not merely that events in the tails of the distributions matter, happen, play a large role, etc. The point is that these events play the major role *and* their probabilities are not (easily) computable, not reliable for any effective use. The implication is that Black Swans do not necessarily come from fat tails; le problem can result from an incomplete assessment of tail events.

Let $A = \{a_i\}$ the set of solutions $\left\{x : \frac{\partial^2 p(x)}{\partial \sigma^2}|_a = 0\right\}$.

For the Gaussian $(\mu, \sigma)$, the solutions obtained by setting the second derivative with respect to $\sigma$ to 0 are:

$$\frac{e^{-\frac{(x-\mu)^2}{2\sigma^2}}\left(2\sigma^4 - 5\sigma^2(x-\mu)^2 + (x-\mu)^4\right)}{\sqrt{2\pi}\sigma^7} = 0,$$

which produces the following crossovers:

$$\{a_1, a_2, a_3, a_4\} = \left\{\mu - \sqrt{\frac{1}{2}\left(5+\sqrt{17}\right)}\sigma, \mu - \sqrt{\frac{1}{2}\left(5-\sqrt{17}\right)}\sigma,\right.$$
$$\left.\mu + \sqrt{\frac{1}{2}\left(5-\sqrt{17}\right)}\sigma, \mu + \sqrt{\frac{1}{2}\left(5+\sqrt{17}\right)}\sigma\right\} \tag{4.8}$$

In figure 4.4, the crossovers for the intervals are numerically: $-2.13\sigma$, $-.66\sigma$ , $.66\sigma$ , $2.13\sigma$.

As to a symmetric power law(as we will see further down), the Student T Distribution with scale $s$ and tail exponent $\alpha$:

$$p(x) \triangleq \frac{\left(\frac{\alpha}{\alpha+\frac{x^2}{s^2}}\right)^{\frac{\alpha+1}{2}}}{\sqrt{\alpha}sB\left(\frac{\alpha}{2}, \frac{1}{2}\right)}$$

$$\{a_1, a_2, a_3, a_4\} = \left\{-\frac{\sqrt{\frac{5\alpha+\sqrt{(\alpha+1)(17\alpha+1)}+1}{\alpha-1}}s}{\sqrt{2}}, -\frac{\sqrt{\frac{5\alpha-\sqrt{(\alpha+1)(17\alpha+1)}+1}{\alpha-1}}s}{\sqrt{2}},\right.$$
$$\left.\frac{\sqrt{\frac{5\alpha-\sqrt{(\alpha+1)(17\alpha+1)}+1}{\alpha-1}}s}{\sqrt{2}}, \frac{\sqrt{\frac{5\alpha+\sqrt{(\alpha+1)(17\alpha+1)}+1}{\alpha-1}}s}{\sqrt{2}}\right\}$$

where $B(.)$ is the Beta function $B(a,b) = \frac{\Gamma(a)\Gamma(b)}{\Gamma(a+b)} = \int_0^1 dt t^{a-1}(1-t)^{b-1}$.

When the Student is "cubic", that is, $\alpha = 3$:

$$\{a_1, a_2, a_3, a_4\} = \left\{-\sqrt{4+\sqrt{13}}s, -\sqrt{4-\sqrt{13}}s, \sqrt{4-\sqrt{13}}s, \sqrt{4+\sqrt{13}}s\right\}$$

**In Summary, Where Does the Tail Start?**

For a general class of symmetric distributions with power laws, the tail starts at: $\pm \frac{\sqrt{\frac{5\alpha+\sqrt{(\alpha+1)(17\alpha+1)}+1}{\alpha-1}}s}{\sqrt{2}}$, with $\alpha$ infinite in the stochastic volatility Gaussian case where $s$ is the standard deviation. The "tail" is located between around 2 and 3 standard deviations. This flows from our definition: which part of the distribution is convex to errors in the estimation of the scale.
But in practice, because historical measurements of STD will be biased lower because of small sample effects (as we repeat fat tails accentuate small sample effects), the deviations will be > 2-3 STDs.

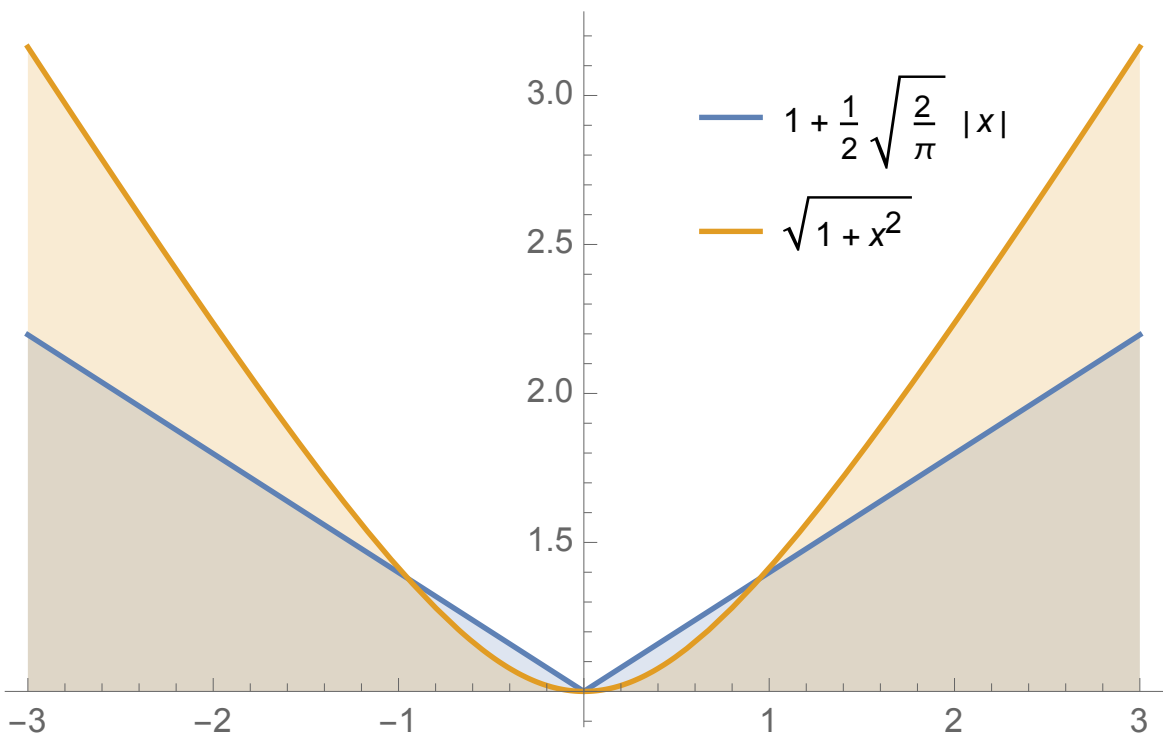

Figure 4.5: *We compare the behavior of $\sqrt{K+x^2}$ and $K+|x|$. The difference between the two weighting functions increases for large values of the random variable $x$, which explains the divergence of the two (and, more generally, higher moments) under fat tails.*

We can verify that when $\alpha \to \infty$, the crossovers become those of a Gaussian. For instance, for $a_1$:

$$\lim_{\alpha\to\infty} -\frac{\sqrt{\frac{5\alpha-\sqrt{(\alpha+1)(17\alpha+1)}+1}{\alpha-1}}s}{\sqrt{2}} = -\sqrt{\frac{1}{2}\left(5-\sqrt{17}\right)}s$$

B- For some one-tailed distribution that have a "bell shape" of convex-concave-convex shape, under some conditions, the same 4 crossover points hold. The Lognormal is a special case.

$$\{a_1,a_2,a_3,a_4\} = \left\{e^{\frac{1}{2}\left(2\mu-\sqrt{2}\sqrt{5\sigma^2+\sqrt{17}\sigma^2}\right)},\right.$$
$$\left. e^{\frac{1}{2}\left(2\mu-\sqrt{2}\sqrt{-\sqrt{17}\sigma^2+5\sigma^2}\right)}, e^{\frac{1}{2}\left(2\mu+\sqrt{2}\sqrt{5\sigma^2-\sqrt{17}\sigma^2}\right)}, e^{\frac{1}{2}\left(2\mu+\sqrt{2}\sqrt{\sqrt{17}\sigma^2+5\sigma^2}\right)}\right\}$$

**Stochastic Parameters** The problem of elliptical distributions is that they do not map the return of securities, owing to the absence of a single variance at any point in time, see Bouchaud and Chicheportiche (2010) [46]. When the scales of the distributions of the individuals move but not in tandem, the distribution ceases to be elliptical. Figure 6.2 shows the effect of applying the equivalent of stochastic volatility methods: the more annoying stochastic correlation. Instead of perturbating the correlation matrix Σ as a unit as in section 6, we perturbate the correlations with surprising effect.

## 4.4 FAT TAILS, MEAN DEVIATION AND THE RISING NORMS

Next we discuss the beastly use of standard deviation and its interpretation.

### 4.4.1 The Common Errors

We start by looking at standard deviation and variance as the properties of higher moments. Now, What is standard deviation? It appears that the same confusion about fat tails has polluted our understanding of standard deviation.

> The difference between standard deviation (assuming mean and median of 0 to simplify) $\sigma = \sqrt{\frac{1}{n}\sum x_i^2}$ and mean absolute deviation $MAD = \frac{1}{n}\sum|x_i|$ increases under fat tails, as one can see in Figure 4.5 . This can provide a conceptual approach to the notion.

Dan Goldstein and the author [133] put the following question to investment professionals and graduate students in financial engineering –people who work with risk and deviations all day long.

> A stock (or a fund) has an average return of 0%. It moves on average 1% a day in absolute value; the average up move is 1% and the average down move is 1%. It does not mean that all up moves are 1% –some are .6%, others 1.45%, and so forth.
>
> Assume that we live in the Gaussian world in which the returns (or daily percentage moves) can be safely modeled using a Normal Distribution. Assume that a year has 256 business days. What is its standard deviation of returns (that is, of the percentage moves), the ĂIJsigma that is used for volatility in financial applications?
>
> What is the daily standard deviation?

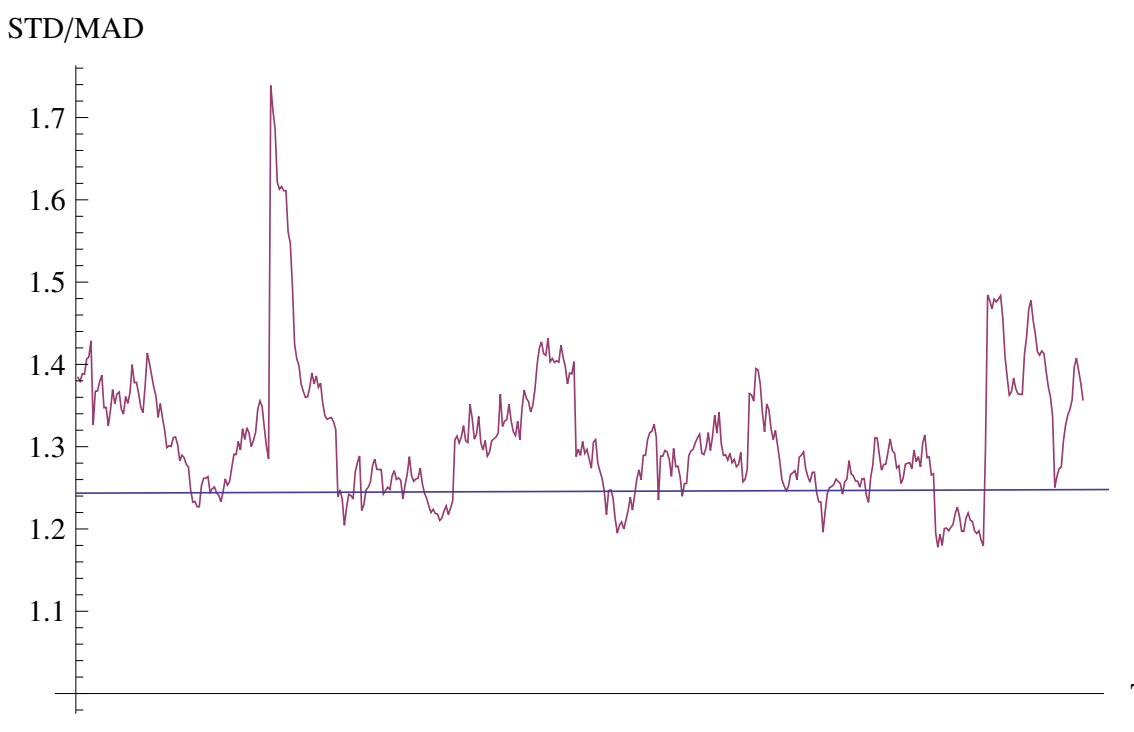

Figure 4.6: *The Ratio STD/MAD for the daily returns of the SP500 over the past 47 years, seen with a monthly rolling window. We can consider the level* $\sqrt{\frac{\pi}{2}} \approx 1.253$ *(as approximately the value for Gaussian deviations), as the cut point for fat tailedness.*

What is the yearly standard deviation?

As the reader can see, the question described mean deviation. And the answers were overwhelmingly wrong. For the daily question, almost all answered 1%. Yet a Gaussian random variable that has a daily percentage move in absolute terms of 1% has a standard deviation that is higher than that, about 1.25%. It should be up to 1.7% in empirical distributions. The most common answer for the yearly question was about 16%, which is about 80% of what would be the true answer. The professionals were scaling daily volatility to yearly volatility by multiplying by $\sqrt{256}$ which is correct provided one had the correct daily volatility.

So subjects tended to provide MAD as their intuition for STD. When professionals involved in financial markets and continuously exposed to notions of volatility talk about standard deviation, they use the wrong measure, mean absolute deviation (MAD) instead of standard deviation (STD), causing an average underestimation of between 20 and 40%. In some markets it can be up to 90%. Further, responders rarely seemed to immediately understand the error when it was pointed out to them. However when asked to present the equation for standard deviation they effectively expressed it as the mean root mean square deviation. Some were puzzled as they were not aware of the existence of MAD.

Why this is relevant: Here you have decision-makers walking around talking about "volatility" and not quite knowing what it means. We note some clips in the financial press to that effect in which the journalist, while attempting to explain the "VIX", i.e., volatility index, makes the same mistake. Even the website of the department of commerce misdefined volatility.

Further, there is an underestimation as MAD is by Jensen's inequality lower (or equal) than STD.

**How the ratio rises** For a Gaussian the ratio $\sim 1.25$, and it rises from there with fat tails.

**Example**: Take an extremely fat tailed distribution, with $n$=$10^6$, observations are all -1 except for a single one of $10^6$,

$$X = \left\{-1, -1, ..., -1, 10^6\right\}.$$

The mean absolute deviation, MAD $(X) = 2$. The standard deviation STD $(X)$=1000. The ratio standard deviation over mean deviation is 500.

### 4.4.2 Some Analytics

**The ratio for thin tails** As a useful heuristic, consider the ratio $h$:

$$h = \frac{\sqrt{\mathbb{E}\left(X^2\right)}}{\mathbb{E}(|X|)},$$

where $\mathbb{E}$ is the expectation operator (under the probability measure of concern and $X$ is a centered variable such $\mathbb{E}(x) = 0$); the ratio increases with the fat tailedness of the distribution; (The general case corresponds to $\frac{(\mathbb{E}(x^p))^{\frac{1}{p}}}{\mathbb{E}(|x|)}$ , $p > 1$, under the condition that the distribution has finite moments up to $n$, and the special case here $n = 2$).[6]

Simply, $x^p$ is a weighting operator that assigns a weight, $x^{p-1}$, which is large for large values of $X$, and small for smaller values.

The effect is due to the convexity differential between both functions, $|X|$ is piecewise linear and loses the convexity effect except for a zone around the origin.

**Mean Deviation vs Standard Deviation, more technical** Why the [REDACTED] did statistical science pick STD over Mean Deviation? Here is the story, with analytical derivations not seemingly available in the literature. In Huber [153]:

> There had been a dispute between Eddington and Fisher, around 1920, about the relative merits of $dn$ (mean deviation) and $Sn$ (standard deviation). Fisher then pointed out that for **exactly normal** observations, $Sn$ is 12% more efficient than $dn$, and this seemed to settle the matter. (My emphasis)

Let us rederive and see what Fisher meant.

Let $n$ be the number of summands:

$$\text{Asymptotic Relative Efficiency (ARE)} = \lim_{n\to\infty} \left( \frac{\mathbb{V}(Std)}{\mathbb{E}(Std)^2} \Big/ \frac{\mathbb{V}(Mad)}{\mathbb{E}(Mad)^2} \right)$$

Assume we are certain that $X_i$, the components of the sample follow a Gaussian distribution, normalized to mean=0 and a standard deviation of 1.

**Relative Standard Deviation Error** The characteristic function $\Psi_1(t)$ of the distribution of $x^2$: $\Psi_1(t) = \int_{-\infty}^{\infty} \frac{e^{-\frac{x^2}{2}+itx^2}}{\sqrt{2\pi}}\, dx = \frac{1}{\sqrt{1-2it}}$. With the squared deviation $z = x^2$, $f$, the pdf for $n$ summands becomes:

$$f_Z(z) = \frac{1}{2\pi} \int_{-\infty}^{\infty} \exp(-itz) \left( \frac{1}{\sqrt{1-2it}} \right)^n dt = \frac{2^{-\frac{n}{2}} e^{-\frac{z}{2}} z^{\frac{n}{2}-1}}{\Gamma\left(\frac{n}{2}\right)}, \; z > 0.$$

Now take $y = \sqrt{z}$, $f_Y(y) = \frac{2^{1-\frac{n}{2}} e^{-\frac{z^2}{2}} z^{n-1}}{\Gamma\left(\frac{n}{2}\right)}$, $z > 0$, which corresponds to the Chi Distribution with $n$ degrees of freedom. Integrating to get the variance: $\mathbb{V}_{std}(n) = n - \frac{2\Gamma\left(\frac{n+1}{2}\right)^2}{\Gamma\left(\frac{n}{2}\right)^2}$. And, with the mean equalling $\frac{\sqrt{2}\Gamma\left(\frac{n+1}{2}\right)}{\Gamma\left(\frac{n}{2}\right)}$, we get $\frac{\mathbb{V}(Std)}{\mathbb{E}(Std)^2} = \frac{n\Gamma\left(\frac{n}{2}\right)^2}{2\Gamma\left(\frac{n+1}{2}\right)^2} - 1$.

**Relative Mean Deviation Error** Characteristic function again for $|x|$ is that of a folded Normal distribution, but let us redo it:

[6] The word "infinite" moment is a big ambiguous, it is better to present the problem as "undefined" moment in the sense that it depends on the sample, and does not replicate outside. Say, for a two-tailed distribution (i.e. with support on the real line), the designation"infinite" variance might apply for the fourth moment, but not to the third.

$\Psi_2(t) = \int_0^\infty \sqrt{\frac{2}{\pi}} e^{-\frac{x^2}{2}+itx} = e^{-\frac{t^2}{2}} \left(1 + i \operatorname{erfi}\left(\frac{t}{\sqrt{2}}\right)\right)$, where erfi is the imaginary error function $erf(iz)/i$.

The first moment: $M_1 = -i \frac{\partial}{\partial t^1} \left( e^{-\frac{t^2}{2n^2}} \left(1 + i \operatorname{erfi}\left(\frac{t}{\sqrt{2}n}\right)\right)\right)^n \Big|_{t=0} = \sqrt{\frac{2}{\pi}}$.

The second moment, $M_2 = (-i)^2 \frac{\partial^2}{\partial t^2} \left( e^{-\frac{t^2}{2n^2}} \left(1 + i \operatorname{erfi}\left(\frac{t}{\sqrt{2}n}\right)\right)\right)^n \Big|_{t=0} = \frac{2n+\pi-2}{\pi n}$.

Hence, $\frac{\mathbb{V}(Mad)}{\mathbb{E}(Mad)^2} = \frac{M_2 - M_1^2}{M_1^2} = \frac{\pi-2}{2n}$.

**Finalmente, the Asymptotic Relative Efficiency For a Gaussian**

$$\text{ARE} = \lim_{n\to\infty} \frac{n\left(\frac{n\Gamma\left(\frac{n}{2}\right)^2}{\Gamma\left(\frac{n+1}{2}\right)^2} - 2\right)}{\pi - 2} = \frac{1}{\pi - 2} \approx .875$$

which means that the standard deviation is 12.5% more "efficient" than the mean deviation *conditional on the data being Gaussian* and these blokes bought the argument. Except that the slightest contamination blows up the ratio. We will show later why Norm $\ell^2$ is not appropriate for about anything; but for now let us get a glimpse on how fragile the STD is.

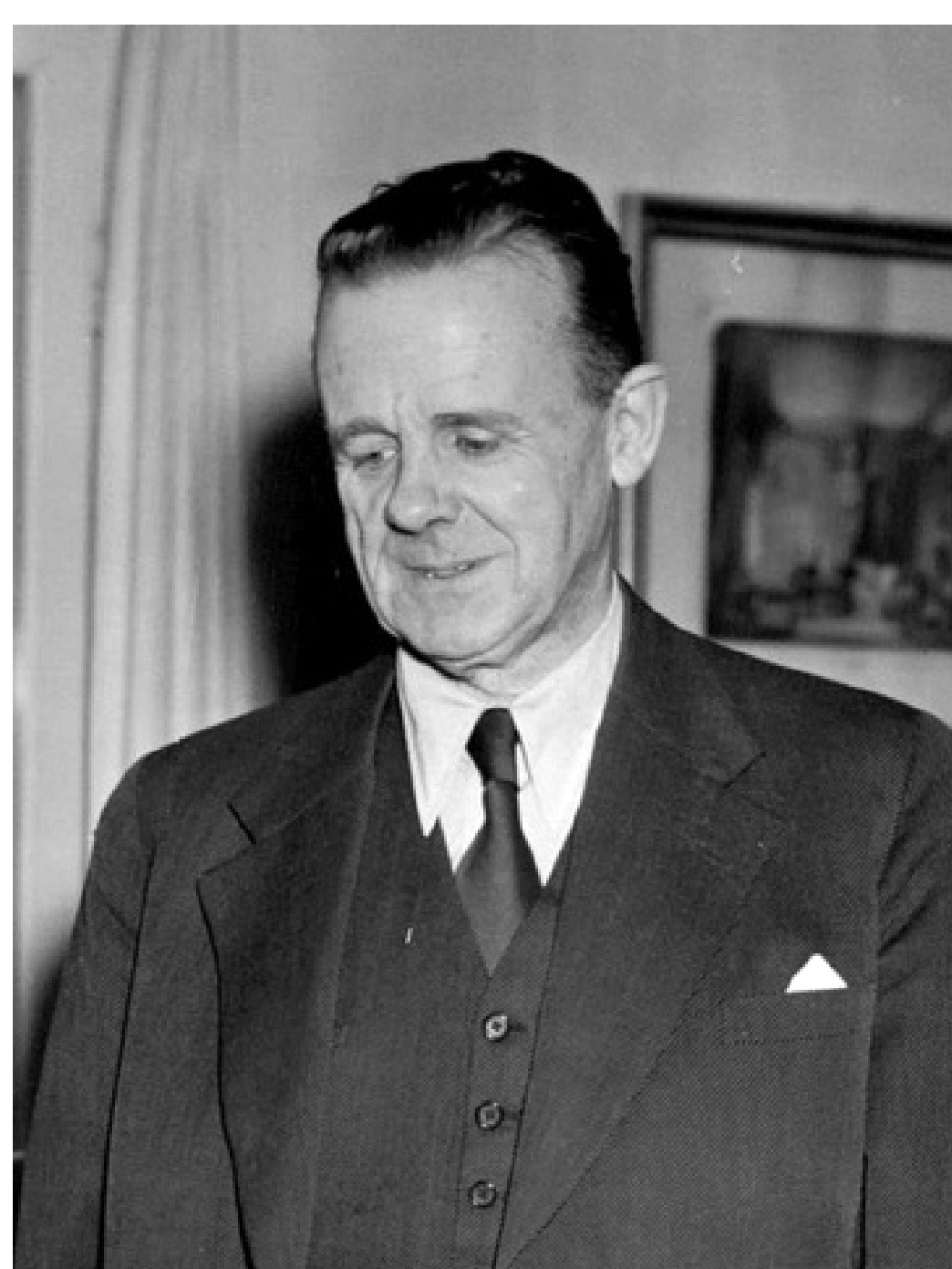

Figure 4.7: *Harald Cramér, of the Cramer condition, and the ruin problem.*

### 4.4.3 Effect of Fatter Tails on the "efficiency" of STD vs MD

Consider a standard mixing model for volatility with an occasional jump with a probability $p$. We switch between Gaussians (keeping the mean constant and central at 0) with:

$$\mathbb{V}(x) = \begin{cases} \sigma^2(1+a) & \text{with probability } p \\ \sigma^2 & \text{with probability } (1-p) \end{cases}$$

For ease, a simple Monte Carlo simulation would do. Using $p = .01$ and $n = 1000$... Figure 4.8 shows how a=2 causes degradation. A minute presence of outliers makes MAD more "efficient" than STD. Small "outliers" of 5 standard deviations cause MAD to be five times more efficient.[7]

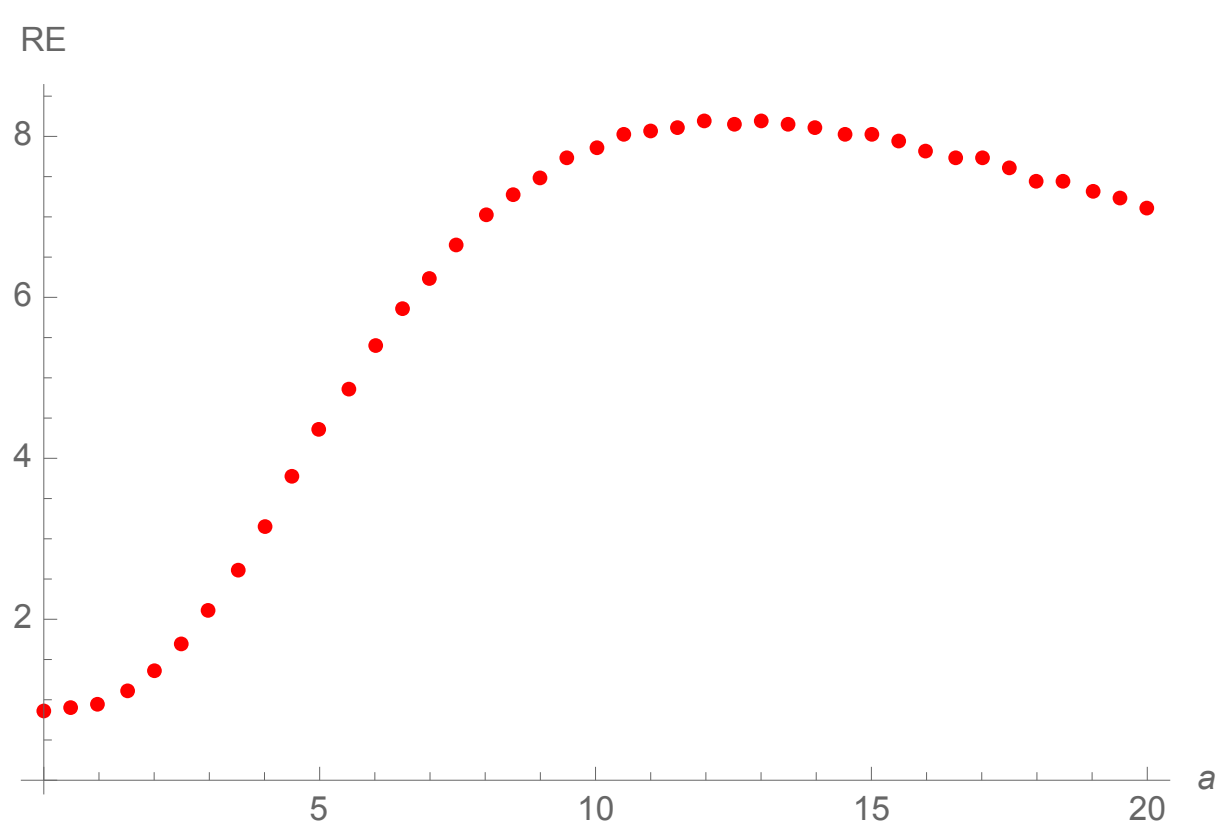

Figure 4.8: *A simulation of the Relative Efficiency ratio of Standard deviation over Mean deviation when injecting a jump size $\sqrt{(1+a)} \times \sigma$, as a multiple of $\sigma$ the standard deviation.*

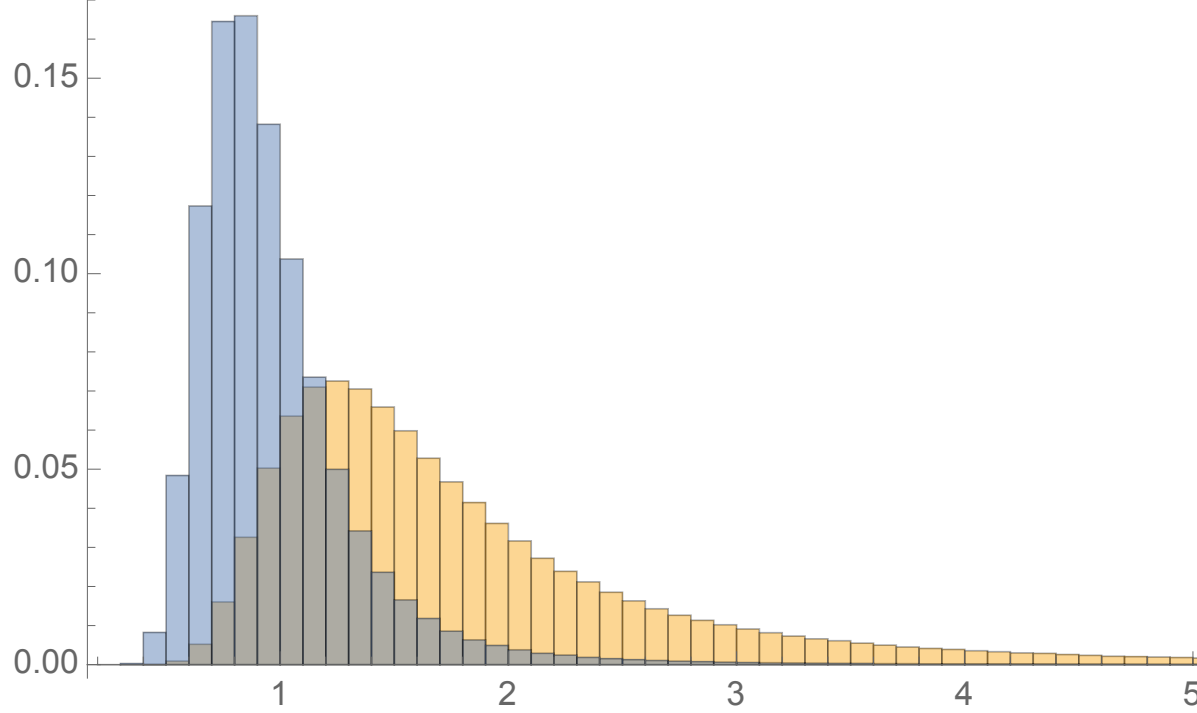

Figure 4.9: *Mean deviation (blue) vs standard deviation (yellow) for a finite variance power law. The result is expected (MD is the thinner distribution), complicated by the fact that standard deviation has an infinite variance since the square of a Paretian random variable with exponent $\alpha$ is Paretian with an exponent of $\frac{1}{2}\alpha$. In this example the mean deviation of standard deviation is 5 times higher.*

[7] The natural way is to center MAD around the median; we find it more informative for many of our purposes here (and decision theory) to center it around the mean. We will make note when the centering is around the mean.

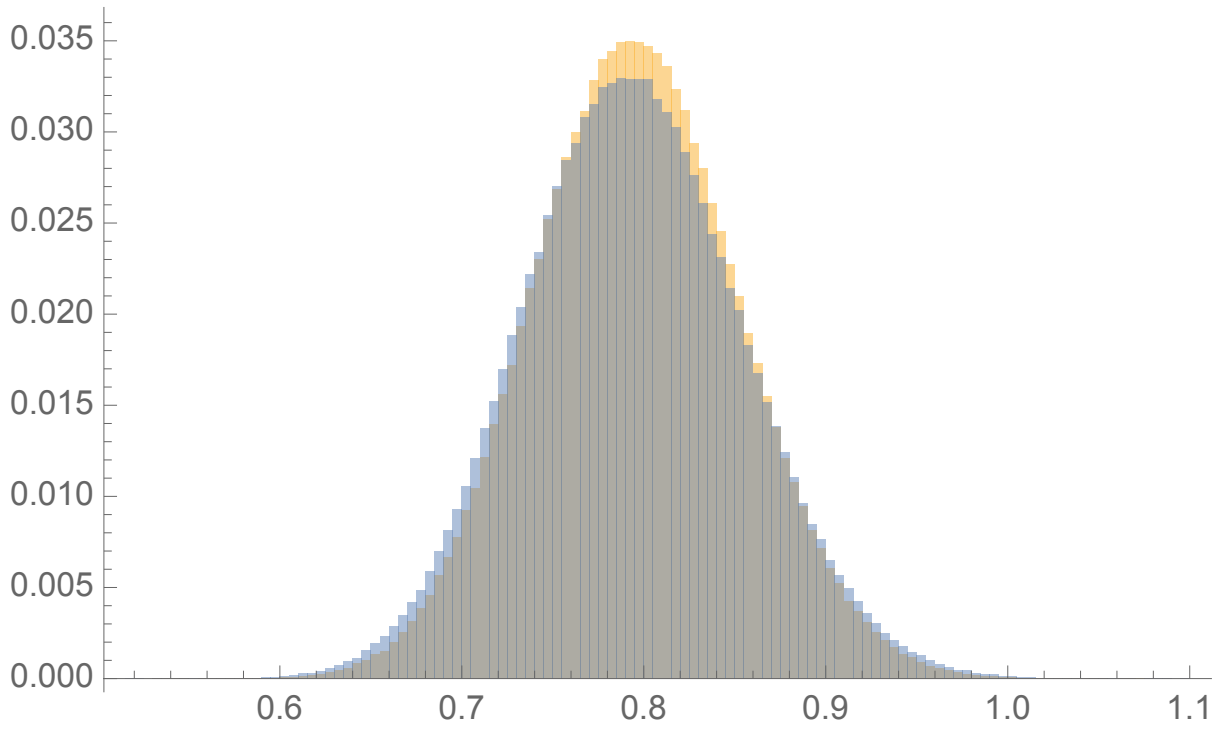

Figure 4.10: *For a Gaussian, there is small difference in distribution between MD and STD (adjusting for the mean for the purpose of visualization).*

### 4.4.4 Moments and The Power Mean Inequality

Let $X \triangleq (x_i)_{i=1}^n$,

$$\|X\|_p \triangleq \left( \frac{\sum_{i=1}^n |x_i|^p}{n} \right)^{1/p}$$

For any $1 \leq p < q$ the following inequality holds:

$$\sqrt[p]{\sum_{i=1}^n w_i |x_i|^p} \leq \sqrt[q]{\sum_{i=1}^n w_i |x_i|^q} \tag{4.9}$$

where the positive weights $w_i$ sum to unity. (Note that we avoid $p < 1$ because it does not satisfy the triangle inequality).

*Proof.* The proof for positive $p$ and $q$ is as follows: Define the following function: $f : R^+ \to R^+; f(x) = x^{\frac{q}{p}}$. $f$ is a power function, so it does have a second derivative:

$$f''(x) = \left(\frac{q}{p}\right)\left(\frac{q}{p} - 1\right) x^{\frac{q}{p}-2}$$

which is strictly positive within the domain of $f$, since $q > p$, $f$ is convex. Hence, by Jensen's inequality : $f\left(\sum_{i=1}^n w_i x_i^p\right) \leq \sum_{i=1}^n w_i f(x_i^p)$, so $\sqrt[\frac{p}{q}]{\sum_{i=1}^n w_i x_i^p} \leq \sum_{i=1}^n w_i x_i^q$ after raising both side to the power of $1/q$ (an increasing function, since $1/q$ is positive) we get the inequality. □

What is critical for our exercise and the study of the effects of fat tails is that, for a given norm, dispersion of results increases values. For example, take a flat distribution, X= $\{1, 1\}$. $\|X\|_1 = \|X\|_2 = ... = \|X\|_n = 1$. Perturbating while preserving $\|X\|_1$, $X = \left\{\frac{1}{2}, \frac{3}{2}\right\}$ produces rising higher norms:

$$\{\|X\|_n\}_{n=1}^5 = \left\{1, \frac{\sqrt{5}}{2}, \frac{\sqrt[3]{7}}{2^{2/3}}, \frac{\sqrt[4]{41}}{2}, \frac{\sqrt[5]{61}}{2^{4/5}}\right\}. \tag{4.10}$$

Trying again, with a wider spread, we get even higher values of the norms, $X = \left\{\frac{1}{4}, \frac{7}{4}\right\}$,

$$\{||X||_n\}_{n=1}^5 = \left\{1, \frac{5}{4}, \frac{\sqrt[3]{\frac{43}{2}}}{2}, \frac{\sqrt[4]{1201}}{4}, \frac{\sqrt[5]{2101}}{2 \times 2^{3/5}}\right\}. \tag{4.11}$$

So we can see (removing constraints and/or allowing for negative values) how higher moments become rapidly explosive.

One property quite useful with power laws with infinite moment:

$$\|X\|_\infty = \sup\left(|x_i|\right)_{i=1}^n \tag{4.12}$$

**Gaussian Case** For a Gaussian, where x $\sim N(0, \sigma)$, as we assume the mean is 0 without loss of generality,

Let $\mathbb{E}(X)$ be the expectation operator for $X$,

$$\frac{\mathbb{E}\left(X^{1/p}\right)}{\mathbb{E}(|X|)} = 2^{\frac{p-3}{2}}\left((-1)^p + 1\right)\sigma^{p-1}\Gamma\left(\frac{p+1}{2}\right)$$

or, alternatively

$$\frac{\mathbb{E}\left(X^p\right)}{\mathbb{E}(|X|)} = 2^{\frac{1}{2}(p-3)}\left(1 + (-1)^p\right)\left(\frac{1}{\sigma^2}\right)^{\frac{1}{2}-\frac{p}{2}}\Gamma\left(\frac{p+1}{2}\right) \tag{4.13}$$

where $\Gamma$(z) is the Euler gamma function; $\Gamma(z) = \int_0^\infty t^{z-1}e^{-t}dt$. For odd moments, the ratio is 0. For even moments:

$$\frac{\mathbb{E}\left(X^2\right)}{\mathbb{E}\left(|X|\right)} = \sqrt{\frac{\pi}{2}}\,\sigma$$

hence

$$\frac{\sqrt{\mathbb{E}\left(X^2\right)}}{\mathbb{E}\left(|X|\right)} = \frac{STD}{MD} = \sqrt{\frac{\pi}{2}}$$

As to the fourth moment, it equals $3\sqrt{\frac{\pi}{2}}\sigma^3$ .

For a Power Law distribution with tail exponent $\alpha$=3, say a Student T

$$\frac{\sqrt{\mathbb{E}\left(X^2\right)}}{\mathbb{E}\left(|X|\right)} = \frac{STD}{MD} = \frac{\pi}{2}$$

We will return to other metrics and definitions of fat tails with Power Law distributions when the moments are said to be "infinite", that is, do not exist. Our heuristic of using the ratio of moments to mean deviation works only in sample, not outside.

**Pareto Case** For a standard Pareto distribution with minimum value (and scale) $L$, PDF $f(x) = \alpha L^{\alpha} x^{-\alpha-1}$ and standard deviation $\frac{\sqrt{\frac{\alpha}{\alpha-2}}L}{\alpha-1}$, we have

$$\frac{STD}{MD} = \frac{1}{2\sqrt{\alpha-2}(\alpha-1)^{\alpha-1}\alpha^{\frac{1}{2}-\alpha}}, \tag{4.14}$$

by centering around the mean.

**"Infinite" moments** Infinite moments, say infinite variance, always manifest themselves as computable numbers in observed sample, yielding finite moments of all orders, simply because the sample is finite. A distribution, say, Cauchy, with undefined means will always deliver a measurable mean in finite samples; but different samples will deliver completely different means. Figures 4.11 and 4.12 illustrate the "drifting" effect of the moments with increasing information.

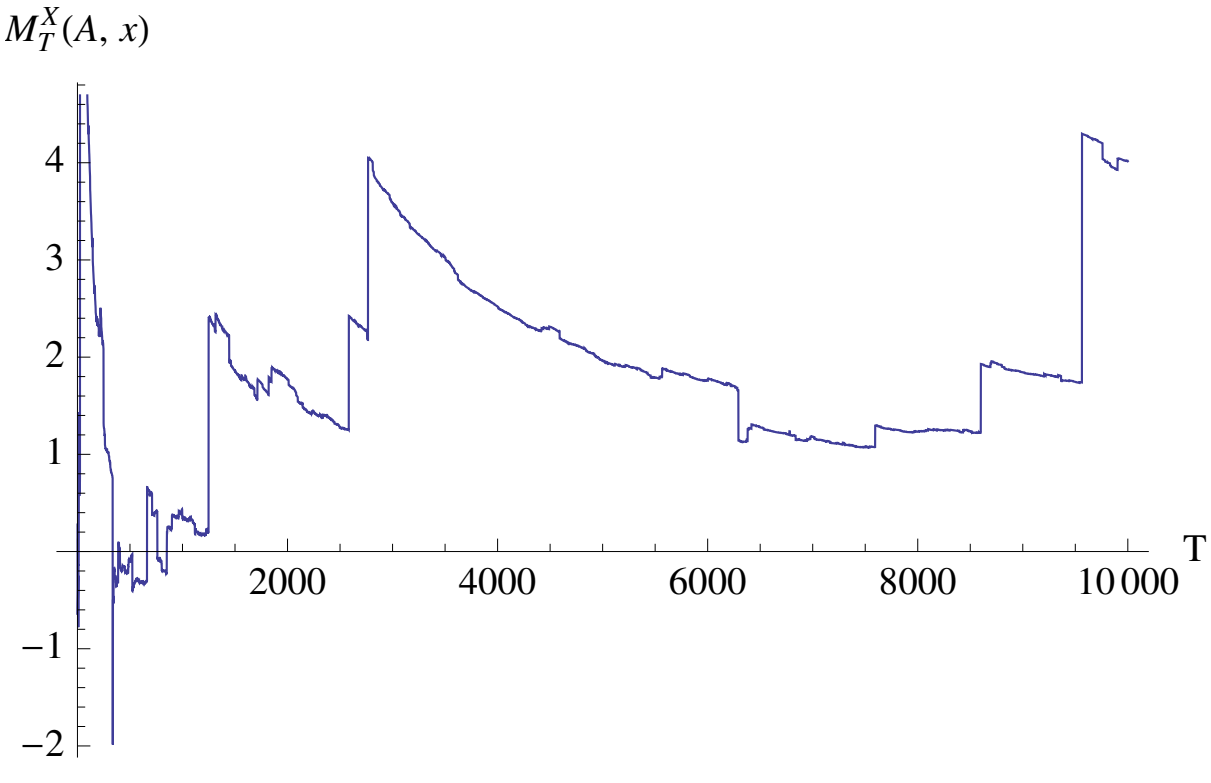

Figure 4.11: *The mean of a series with undefined mean (Cauchy).*

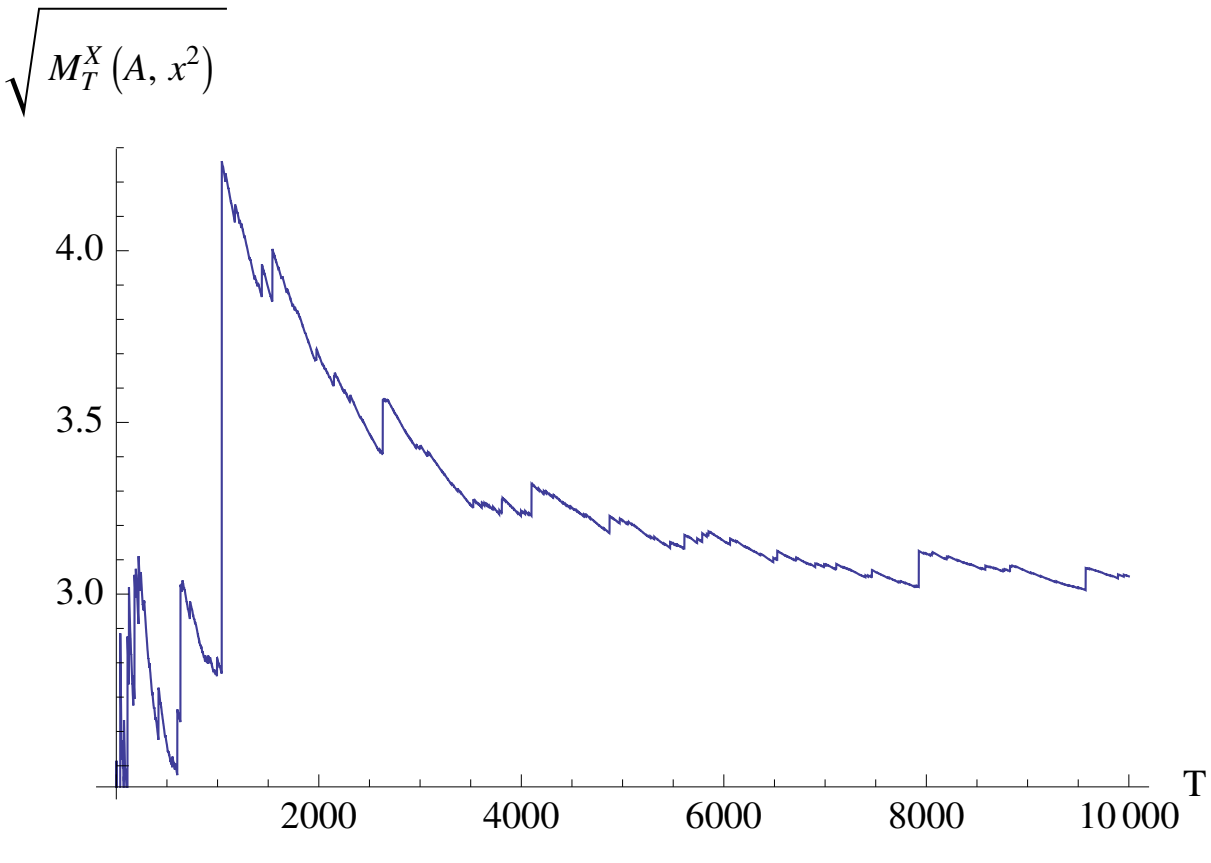

Figure 4.12: *The square root of the second moment of a series with infinite variance. We observe pseudoconvergence before a jump.*

### 4.4.5 Comment: Why we should retire standard deviation, now!

The notion of standard deviation has confused hordes of scientists; it is time to retire it from common use and replace it with the more effective one of mean deviation. Standard deviation, STD, should be left to mathematicians, physicists and mathematical statisticians deriving limit theorems. There is no scientific reason to use it in statistical investigations in the age of the computer, as it does more harm than good-particularly with the growing class of people in social science mechanistically applying statistical tools to scientific problems.

Say someone just asked you to measure the "average daily variations" for the temperature of your town (or for the stock price of a company, or the blood pressure of your uncle) over the past five days. The five changes are: (-23, 7, -3, 20, -1). How do you do it?

Do you take every observation: square it, average the total, then take the square root? Or do you remove the sign and calculate the average? For there are serious differences between the two methods. The first produces an average of 15.7, the second 10.8. The first is technically called the root mean square deviation. The second is the mean absolute deviation, MAD. It corresponds to "real life" much better than the first-and to reality. In fact, whenever people make decisions after being supplied with the standard deviation number, they act as if it were the expected mean deviation.

It is all due to a historical accident: in 1893, the great Karl Pearson introduced the term "standard deviation" for what had been known as "root mean square error". The confusion started then: people thought it meant mean deviation. The idea stuck: every time a newspaper has attempted to clarify the concept of market "volatility", it defined it verbally as mean deviation yet produced the numerical measure of the (higher) standard deviation.

But it is not just journalists who fall for the mistake: I recall seeing official documents from the department of commerce and the Federal Reserve partaking of the conflation, even regulators in statements on market volatility. What is worse, Goldstein and I found that a high number of data scientists (many with PhDs) also get confused in real life.

It all comes from bad terminology for something non-intuitive. By a psychological phenomenon called attribute substitution, some people mistake MAD for STD because the former is easier to come to mind – this is "Lindy"[8] as it is well known by cheaters and illusionists.

1) MAD is more accurate in sample measurements, and less volatile than STD since it is a natural weight whereas standard deviation uses the observation itself as its own weight, imparting large weights to large observations, thus overweighing tail events.

2) We often use STD in equations but really end up reconverting it within the process into MAD (say in finance, for option pricing). In the Gaussian world, STD is about 1.25 time MAD, that is, $\sqrt{\frac{\pi}{2}}$. But we adjust with stochastic volatility where STD is often as high as 1.6 times MAD.

---

[8] See a definition of "Lindy" in 23

3) Many statistical phenomena and processes have "infinite variance" (such as the popular Pareto 80/20 rule) but have finite, and sometimes very well behaved, mean deviations. Whenever the mean exists, MAD exists. The reverse (infinite MAD and finite STD) is never true.

4) Many economists have dismissed "infinite variance" models thinking these meant "infinite mean deviation". Sad, but true. When the great Benoit Mandelbrot proposed his infinite variance models fifty years ago, economists freaked out because of the conflation.

It is sad that such a minor point can lead to so much confusion: our scientific tools are way too far ahead of our casual intuitions, which starts to be a problem with science. So I close with a statement by Sir Ronald A. Fisher: 'The statistician cannot evade the responsibility for understanding the process he applies or recommends.'

**Note** The usual theory is that if random variables $X_1, \dots, X_n$ are independent, then

$$\mathbb{V}(X_1 + \cdots + X_n) = \mathbb{V}(X_1) + \cdots + \mathbb{V}(X_n).$$

by the linearity of the variance. But then it assumes that one cannot use another metric then by simple transformation make it additive[9]. As we will see, for the Gaussian $\text{md}(X) = \sqrt{\frac{2}{\pi}}\sigma$ —for the Student T with 3 degrees of freedom, the factor is $\frac{2}{\pi}$, etc.

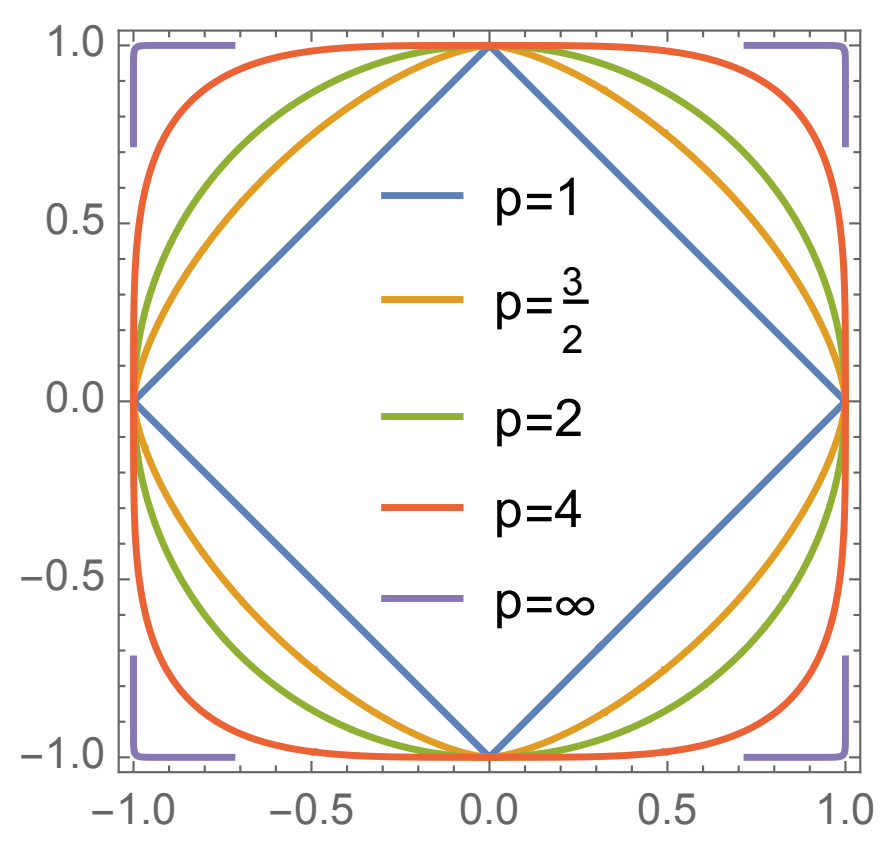

Figure 4.13: *Rising norms and the unit circle/square: values of the iso-norm* $\left(|x_1|^p + |x_2|^p\right)^{1/p} = 1$. *We notice the area inside the norm (i.e. satisfying norm* $\leq 1$*),* $v(p) = \frac{4\Gamma\left(\frac{p+1}{p}\right)^2}{\Gamma\left(\frac{p+2}{p}\right)}$, *with* $v(1) = 2$ *and* $v(\infty) = 4$.

## 4.5 VISUALIZING THE EFFECT OF RISING $p$ ON ISO-NORMS

Consider the region $\mathscr{R}_{(p)}^{(n)}$ defined as $\mathbf{X} = (\mathbf{x_1}, \dots, \mathbf{x_n}) :\in \left(\sum_{\mathbf{i=1}}^{\mathbf{n}} \mathbf{x_i^p}\right)^{\mathbf{1/p}} \leq \mathbf{1}$, with the border defined by the identity.

[9] For instance option pricing in the Black-Scholes formula is done using variance, but the price maps directly to MAD; an at-the-money straddle is just a conditional mean deviation. So we translate MAD into standard deviation, then back to MAD

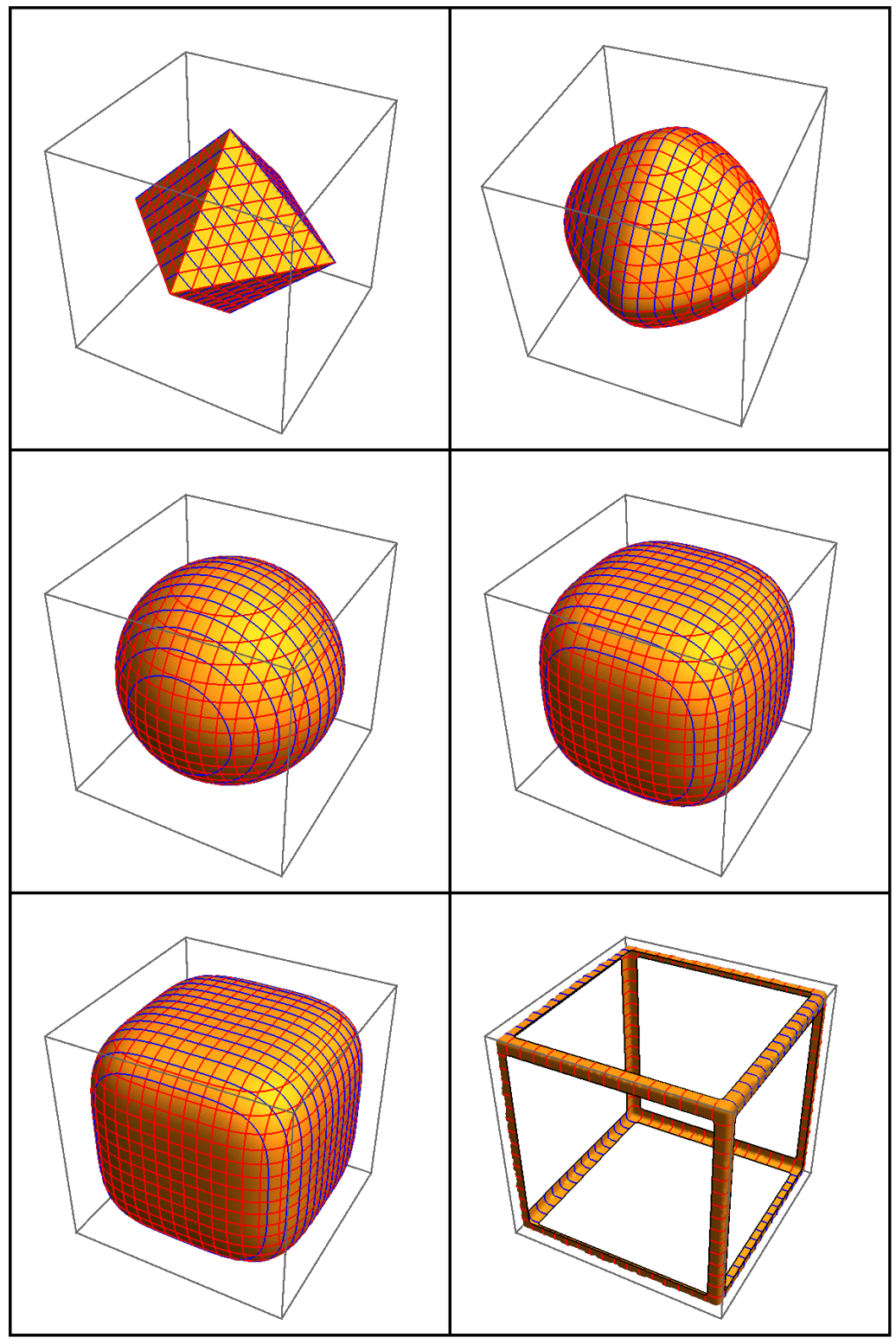

Figure 4.14: *Rising norms and the unit cube: values of the iso-norm* $\left(|x_1|^p + |x_2|^p + |x_3|^p\right)^{1/p} = 1$ *for* $p = 1, \frac{3}{2}, 2, 3, 4,$ *and* $\infty$. *The volume satisfying the inequality norm* $\leq 1$ *increases for* $\frac{4}{3}$ *for* $p = 1$, $\frac{4\pi}{3}$ *for* $p = 2$ *(the unit sphere), to* $2^3$ *for* $p = \infty$ *(the unit cube), a much higher increase than in Figure 4.13 . We can see the operation of the curse of dimensionality in the smaller and smaller volume for* $p = 1$, *relative to the maximum when* $p = \infty$.

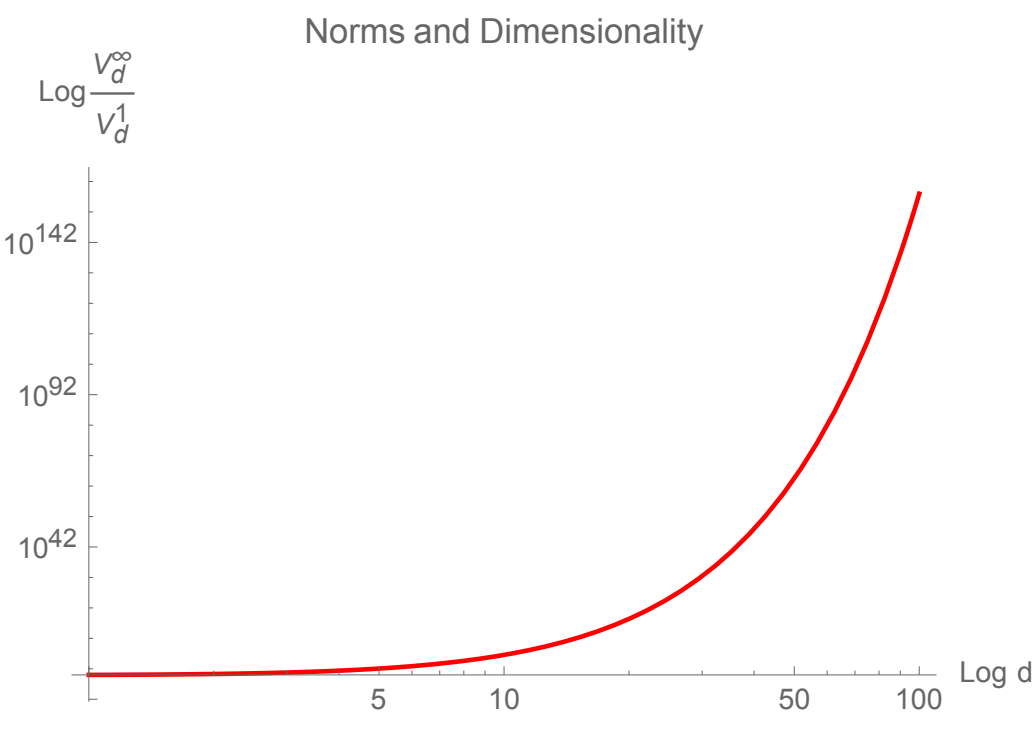

Figure 4.15: *The curse of dimensionality, with yuuuge applications across statistical areas, particularly model error in higher dimentions. As d increases, the ratio of* $V^\infty$ *over* $V^1$ *blows up. If for* $d = 2$, *it is 2 it is already six figures for* $d = 9$.

As the norm rises, we calculate the following measure of the ball:

$$V_n^p = \int \cdots \int_{\mathbf{X} \in \mathscr{R}^{(\mathbf{n})}_{(\mathbf{n})}} 1 d\mathbf{X} = \frac{\left(4\Gamma\left(1+\frac{1}{\mathbf{p}}\right)\right)^{\mathbf{n}}}{\Gamma\left(\frac{\mathbf{n}}{\mathbf{p}}+1\right)} \tag{4.15}$$

Figures 4.13 and 4.14 show two effects.

The first is how rising norms occupy a larger share of the space.

The second gives us a hint of the curse of dimensionality, useful in many circumstances (and, centrally, for model error). Compare figures 4.13 and 4.14: you will notice that in the first case, for $d = 2$, $p = 1$, $m$ occupies half the area of the square, with $p = \infty$ all of it. The ratio of norms is $\frac{1}{2}$. But for $d = 3$, $p = 1$ occupies $\frac{4/3}{2^3} = \frac{1}{6}$ of the space (again, $p = \infty$ occupies all of it). The ratio of higher moments to lower moments increases with dimensionality, as seen in Figure 4.15.

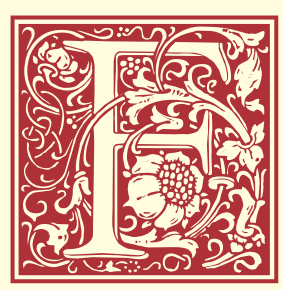
URTHER READING: We stop here and present probability books in general. For more general intuition about probability, the indispensable Borel's [100]. Kolmogorov [173], Loeve [183], Feller [108],[107]. For measure theory, Billingsley [24].

**For subexponentiality** Pitman [229], Embrechts and Goldie (1982) [98], Embrechts (1979, which seems to be close to his doctoral thesis) [99], Chistyakov (1964) [47], Goldie (1978) [132], and Teugels [300].

**For extreme value distributions** Embrechts et al [97], De Haan and Ferreira [136].

**For stable distributions** Uchaikin and Zolotarev [310], Zolotarev [326], Samorindsky and Taqqu [246].

**Stochastic processes** Karatsas and Shreve [168], Oksendal [214], Varadhan [315].

# 5 LEVEL 2: SUBEXPONENTIALS AND POWER LAWS

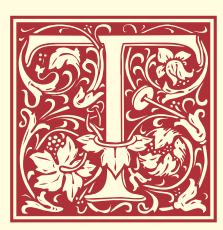
HIS CHAPTER briefly presents the subexponential vs. the power law classes as "true fat tails" (already defined in Chapter 3) and presents some wrinkles associated with them. Subexponentiality (without scalability), that is membership in the subexponential but not power law class is a small category (of the common distributions, only the borderline exponential –and gamma associated distributions such as the Laplace – and the the lognormal fall in that class).

### 5.0.1 Revisiting the Rankings

Table 5.1 reviews the rankings of Chapter 3. Recall that probability distributions range between extreme thin-tailed (Bernoulli) and extreme fat tailed. Among the categories of distributions that are often distinguished due to the convergence properties of moments are:

1. Having a support that is compact (but not degenerate)
2. Subgaussian
3. Subexponential
4. Power Law with exponent greater than 2
5. Power Law with exponent less than or equal to 2. In particular, Power Law distributions have a finite mean only if the exponent is greater than 1, and have a finite variance only if the exponent exceeds 2
6. Power Law with exponent less than 1

Our interest is in distinguishing between cases where tail events dominate impacts, as a formal definition of the boundary between the categories of distributions to be considered as mediocristan and Extremistan.

Centrally, a subexponential distribution is the cutoff between "thin" and "fat" tails. It is defined as follows.

The mathematics is crisp: the excedance probability or survival function needs to be exponential in one not the other. Where is the border?

Table 5.1: *Ranking distributions*

| Class | Description |
|---|---|
| True Thin Tails | Compact support (e.g. : Bernouilli, Binomial) |
| Thin tails | Gaussian reached organically through summation of true thin tails, by Central Limit; compact support except at the limit $n \to \infty$ |
| Conventional Thin tails | Gaussian approximation of a natural phenomenon |
| Starter Fat Tails | Higher kurtosis than the Gaussian but rapid convergence to Gaussian under summation |
| Subexponential | (e.g. lognormal) |
| Supercubic $\alpha$ | Cramer conditions do not hold for $t > 3, \int e^{-tx}\, d(Fx) = +\infty$ |
| Infinite Variance | Levy Stable $\alpha < 2$ , $\int e^{-tx} dF(x) = +\infty$ |
| Undefined First Moment | Fuhgetaboutdit |

The natural boundary between Mediocristan and Extremistan occurs at the subexponential class which has the following property:

Let $\mathbf{X} = X_1, \ldots, X_n$ be a sequence of independent and identically distributed random variables with support in ($\mathbb{R}^+$), with cumulative distribution function $F$. The subexponential class of distributions is defined by (see [300], [229]):

$$\lim_{x \to +\infty} \frac{1 - F^{*2}(x)}{1 - F(x)} = 2 \tag{5.1}$$

where $F^{*2} = F' * F$ is the cumulative distribution of $X_1 + X_2$, the sum of two independent copies of $X$. This implies that the probability that the sum $X_1 + X_2$ exceeds a value $x$ is twice the probability that either one separately exceeds $x$. Thus, every time the sum exceeds $x$, for large enough values of $x$, the value of the sum is due to either one or the other exceeding $x$—the maximum over the two variables—and the other of them contributes negligibly.

More generally, it can be shown that the sum of $n$ variables is dominated by the maximum of the values over those variables in the same way. Formally, the following two properties are equivalent to the subexponential condition [47],[99]. For a given $n \geq 2$, let $S_n = \Sigma_{i=1}^n x_i$ and $M_n = \max_{1 \leq i \leq n} x_i$

a) $\lim_{x \to \infty} \frac{P(S_n > x)}{P(X > x)} = n$,

b) $\lim_{x \to \infty} \frac{P(S_n > x)}{P(M_n > x)} = 1$.

Thus the sum $S_n$ has the same magnitude as the largest sample $M_n$, which is another way of saying that tails play the most important role.

Intuitively, tail events in subexponential distributions should decline more slowly than an exponential distribution for which large tail events should be irrelevant. Indeed, one can show that subexponential distributions have no exponential moments:

$$\int_0^\infty \mathbf{e}^{\epsilon x}\, dF(x) = +\infty \tag{5.2}$$

for all values of $\varepsilon$ greater than zero. However,the converse isn't true, since distributions can have no exponential moments, yet not satisfy the subexponential condition.

We note that if we choose to indicate deviations as negative values of the variable $x$, the same result holds by symmetry for extreme negative values, replacing $x \to +\infty$ with $x \to -\infty$. For two-tailed variables, we can separately consider positive and negative domains.

### 5.0.2 What is a Borderline Probability Distribution?

The best way to figure out a probability distribution is to... invent one. In fact in the next section, 5.0.3, we will build one that is the exact borderline between thin and fat tails *by construction*. Consider for now that the properties are as follows:

Let $\overline{F}$ be the survival function. We have $\overline{F} : \mathbb{R} \to [0, 1]$ that satisfies

$$\lim_{x \to +\infty} \frac{\overline{F}(x)^n}{\overline{F}(nx)} = 1, \tag{5.3}$$

and

$$\lim_{x \to +\infty} \overline{F}(x) = 0$$

$$\lim_{x \to -\infty} \overline{F}(x) = 1$$

Note : another property of the demarcation is the absence of Lucretius fallacy from *The Black Swan*, mentioned earlier (i.e. future extremes will not be similar to past extremes under fat tails, and such dissimilarity increases with fat tailedness):

Let us look at the demarcation properties for now. Let $X$ be a random variable that lives in either $(0, \infty)$ or $(-\infty, \infty)$ and $\mathbb{E}$ the expectation operator under "real world" (physical) distribution. By classical results [97]:

$$\lim_{K \to \infty} \frac{1}{K} \mathbb{E}(X|_{X>K}) = \lambda \tag{5.4}$$

- If $\lambda = 1$ , $X$ is said to be in the thin tailed class $\mathcal{D}_1$ and has a characteristic scale
- If $\lambda > 1$ , $X$ is said to be in the fat tailed regular variation class $\mathcal{D}_2$ and has no characteristic scale
- If
$$\lim_{K \to \infty} \mathbb{E}(X|_{X>K}) - K = \mu$$
where $\mu > 0$, then $X$ is in the borderline exponential class

The second case is called the "Lindy effect" when the random variable $X$ is time survived. The subject is examined outside of this fat-tails project. See Iddo Eliazar's exposition [90].

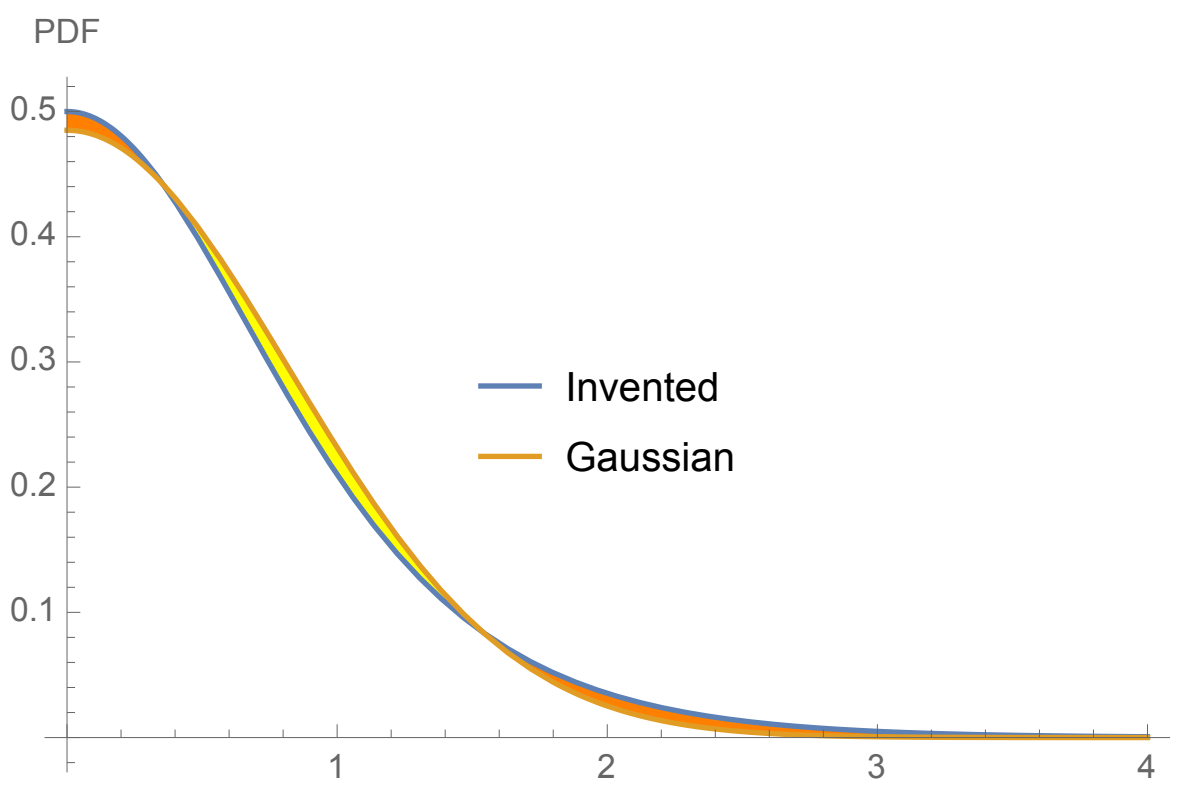

Figure 5.1: *Comparing the invented distribution (at the cusp of subexponentiality) to the Gaussian of the same variance ($k = 1$). It does not take much to switch from Gaussian to subexponential properties.*

### 5.0.3 Let Us Invent a Distribution

While the exponential distribution is at the cusp of the subexponential class but with support in $[0, \infty)$, we can construct a borderline distribution with support in $(-\infty, \infty)$, as follows [1]. Find survival functions $\overline{F} : \mathbb{R} \to [0, 1]$ that satisfy:

$$\forall x \geq 0, \lim_{x \to +\infty} \frac{\overline{F}(x)^2}{\overline{F}(2x)} = 1, \; \overline{F}'(x) \leq 0$$

and

$$\lim_{x \to +\infty} \overline{F} = 0.$$

$$\lim_{x \to -\infty} \overline{F} = 1.$$

[1] The Laplace distribution, which doubles the exponential on both sides, does not fit the property as the ratio of the square to the double is $\frac{1}{2}$.

Let us assume a candidate function a sigmoid, using the hyperbolic tangent

$$\overline{F}^{\kappa}(x) = \frac{1}{2}\left(1 - \tanh(kx)\right)\ , \kappa > 0.$$

We can use this as a kernel distribution (we mix later to modify the kurtosis).

Let $f(.)$ be the density function:

$$f(x) = -\frac{\partial \overline{F}(x)}{\partial x} = \frac{1}{2} k \operatorname{sech}^2(kx). \tag{5.5}$$

The characteristic function:

$$\phi(t) = \frac{\pi t \operatorname{csch}\left(\frac{\pi t}{2k}\right)}{2k}. \tag{5.6}$$

Given that it is all real, we can guess that the mean is 0 –so are all odd moments.

The second moment will be $\lim_{t \to 0} (-i)^2 \frac{\partial^2}{\partial t^2} \frac{\pi t \operatorname{csch}\left(\frac{\pi t}{2k}\right)}{2k} = \frac{\pi^2}{12k^2}$ And the fourth moment: $\lim_{t \to 0} (-i)^4 \frac{\partial^4}{\partial t^4} \frac{\pi t \operatorname{csch}\left(\frac{\pi t}{2k}\right)}{2k} = \frac{7\pi^4}{240k^4}$, hence the Kurtosis will be $\frac{21}{5}$. The distribution we invented has slightly fatter tails than the Gaussian.

## 5.1 LEVEL 3: SCALABILITY AND POWER LAWS

Now we get into the serious business.

**Why power laws?** There are a lot of theories on why things should be power laws, as sort of exceptions to the way things work probabilistically. But it seems that the opposite idea is never presented: power laws should be the norm, and the Gaussian a special case ([269]), effectively the topic of *Antifragile* and the next volume of the *Technical Incerto*), owing to concave-convex responses (sort of dampening of fragility and antifragility, bringing robustness, hence thinning the tails).

### 5.1.1 Scalable and Nonscalable, A Deeper View of Fat Tails

So far for the discussion on fat tails we stayed in the finite moments case. For a certain class of distributions, those with finite moments, $\frac{P_{X>nK}}{P_{X>K}}$ depends on n and K. For a scale-free distribution, with K "in the tails", that is, large enough, $\frac{P_{X>nK}}{P_{X>K}}$ depends on n not K. These latter distributions lack in characteristic scale and will end up having a Paretian tail, i.e., for $x$ large enough, $P_{X>x} = Cx^{-\alpha}$ where $\alpha$ is the tail and $C$ is a scaling constant.

Note: We can see from the scaling difference between the Student and the Pareto the conventional definition of a Power Law tailed distribution is expressed more

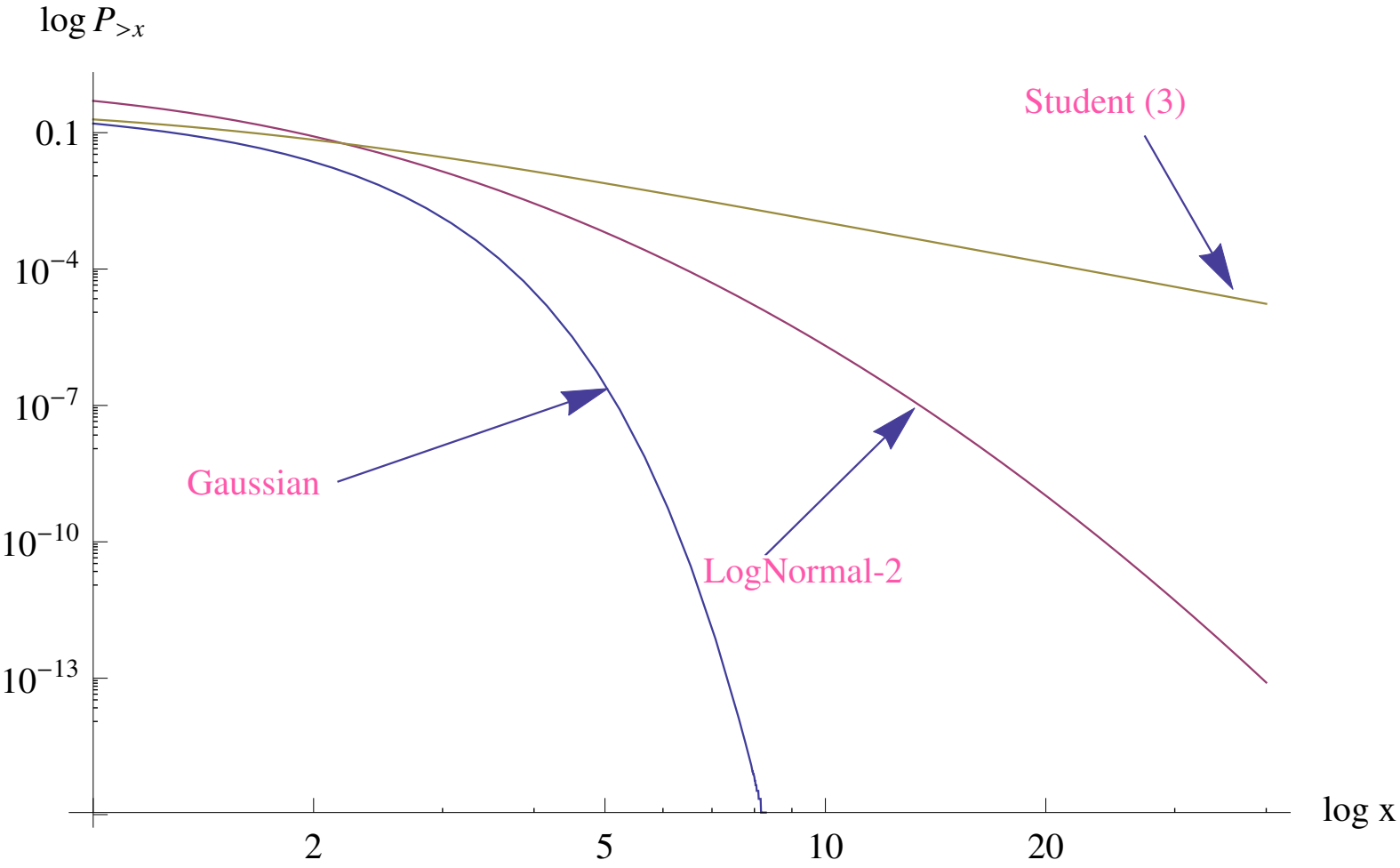

Figure 5.2: *Three Types of Distributions. As we hit the tails, the Student remains scalable while the Standard Lognormal shows an intermediate position before eventually ending up getting an infinite slope on a log-log plot. But beware the lognormal as it may have some surprises (Chapter 8)*
.

Table 5.2: *Scalability, comparing regularly varying functions/powerlaws to other distributions*

| k | $\mathbb{P}(X > k)^{-1}$ (Gaussian) | $\frac{\mathbb{P}(X>k)}{\mathbb{P}(X>2\ k)}$ (Gaussian) | $\mathbb{P}(X > k)^{-1}$ Student(3) | $\frac{\mathbb{P}(X>k)}{\mathbb{P}(X>2\ k)}$ Student (3) | $\mathbb{P}(X > k)^{-1}$ Pareto(2) | $\frac{\mathbb{P}(X>k)}{\mathbb{P}(X>2\ k)}$ Pareto (2) |
|---|---|---|---|---|---|---|
| 2 | 44 | 720 | 14.4 | 4.9 | 8 | 4 |
| 4 | 31600. | $5.1 \times 10^{10}$ | 71.4 | 6.8 | 64 | 4 |
| 6 | $1.01 \times 10^{9}$ | $5.5 \times 10^{23}$ | 216 | 7.4 | 216 | 4 |
| 8 | $1.61 \times 10^{15}$ | $9 \times 10^{41}$ | 491 | 7.6 | 512 | 4 |
| 10 | $1.31 \times 10^{23}$ | $9 \times 10^{65}$ | 940 | 7.7 | 1000 | 4 |
| 12 | $5.63 \times 10^{32}$ | fughedaboudit | 1610 | 7.8 | 1730 | 4 |
| 14 | $1.28 \times 10^{44}$ | fughedaboudit | 2530 | 7.8 | 2740 | 4 |
| 16 | $1.57 \times 10^{57}$ | fughedaboudit | 3770 | 7.9 | 4100 | 4 |
| 18 | $1.03 \times 10^{72}$ | fughedaboudit | 5350 | 7.9 | 5830 | 4 |
| 20 | $3.63 \times 10^{88}$ | fughedaboudit | 7320 | 7.9 | 8000 | 4 |

formally as $\mathbb{P}(X > x) = L(x)x^{-\alpha}$ where $L(x)$ is a "slow varying function", which satisfies the following:

$$\lim_{x\to\infty} \frac{L(t\ x)}{L(x)} = 1$$

for all constants $t > 0$.

For x large enough, $\frac{\log P_{>x}}{\log x}$ converges to a constant, namely the tail exponent -$\alpha$. A scalable should produce the slope $\alpha$ in the tails on a log-log plot, as $x \to \infty$. Compare to the Gaussian (with STD $\sigma$ and mean $\mu$) , by taking the PDF this time instead of the exceedance probability $\log\left(f(x)\right) = \frac{(x-\mu)^2}{2\sigma^2} - \log(\sigma\sqrt{2\pi}) \approx -\frac{1}{2\sigma^2}x^2$ which goes to $-\infty$ faster than $-\log(x)$ for $\pm x \to \infty$.

So far this gives us the intuition of the difference between classes of distributions. Only scalable have "true" fat tails, as others turn into a Gaussian under summation. And the tail exponent is asymptotic; we may never get there and what we may see is an intermediate version of it. The figure above drew from Platonic off-the-shelf distributions; in reality processes are vastly more messy, with switches between exponents as deviations get larger.

**Definition 5.1** (the class $\mathfrak{P}$)
*The $\mathfrak{P}$ class of power laws (regular variation) is defined for r.v. X as follows:*

$$\mathfrak{P} = \{X : \mathbb{P}(X > x) \sim L(x)\, x^{-\alpha}\} \tag{5.7}$$

### 5.1.2 Grey Swans

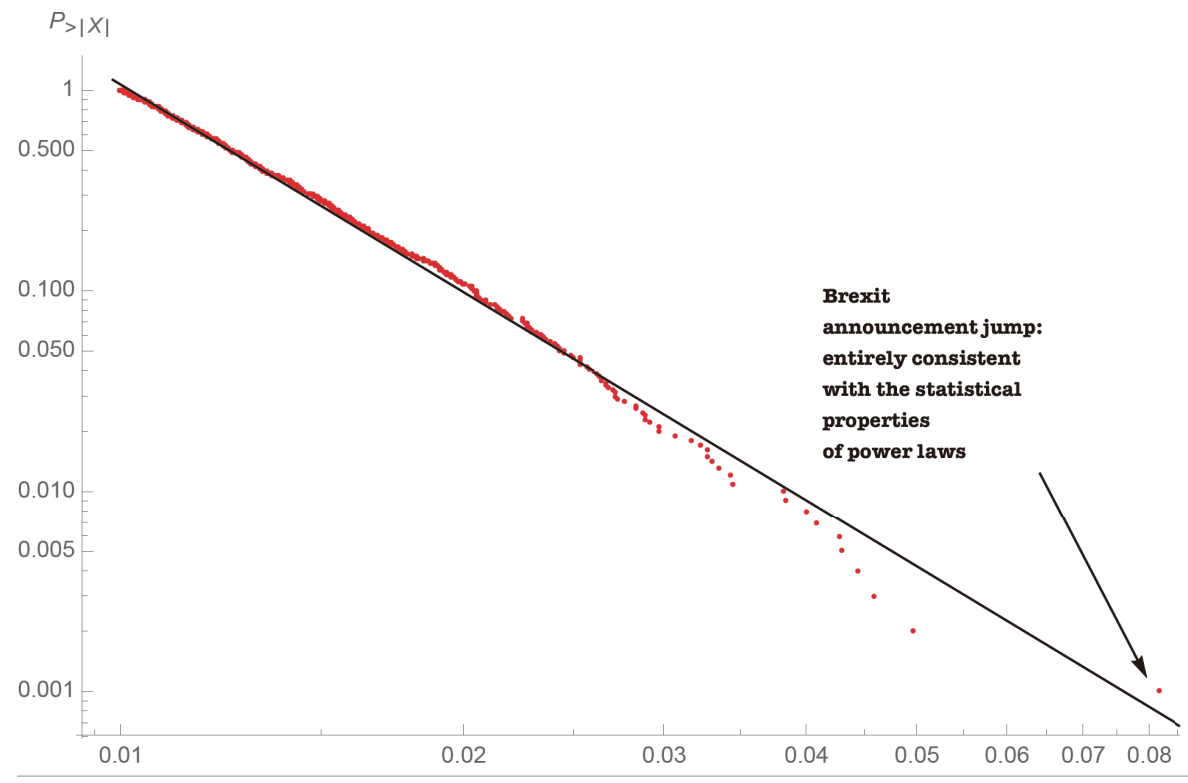

Figure 5.3: *The graph represents the log log plot of GBP, the British currency. We can see the "Grey Swan" of Brexit (that is, the jump in the currency when the unexpected referendum results came out); when seen using a power law the large deviation is rather consistent with the statistical properties.*

**Why do we use Student T to simulate symmetric power laws?** For convenience, only for convenience. It is not that we *believe* that the generating process is Student T. Simply, the center of the distribution does not matter much for the properties involved in certain classes of decision making.

The lower the exponent, the less the center plays a role. The higher the exponent, the more the student T resembles the Gaussian, and the more justified its use will be accordingly.

More advanced methods involving the use of Levy laws may help in the event of asymmetry, but the use of two different Pareto distributions with two different exponents, one for the left tail and the other for the right one would do the job (without unnecessary complications).

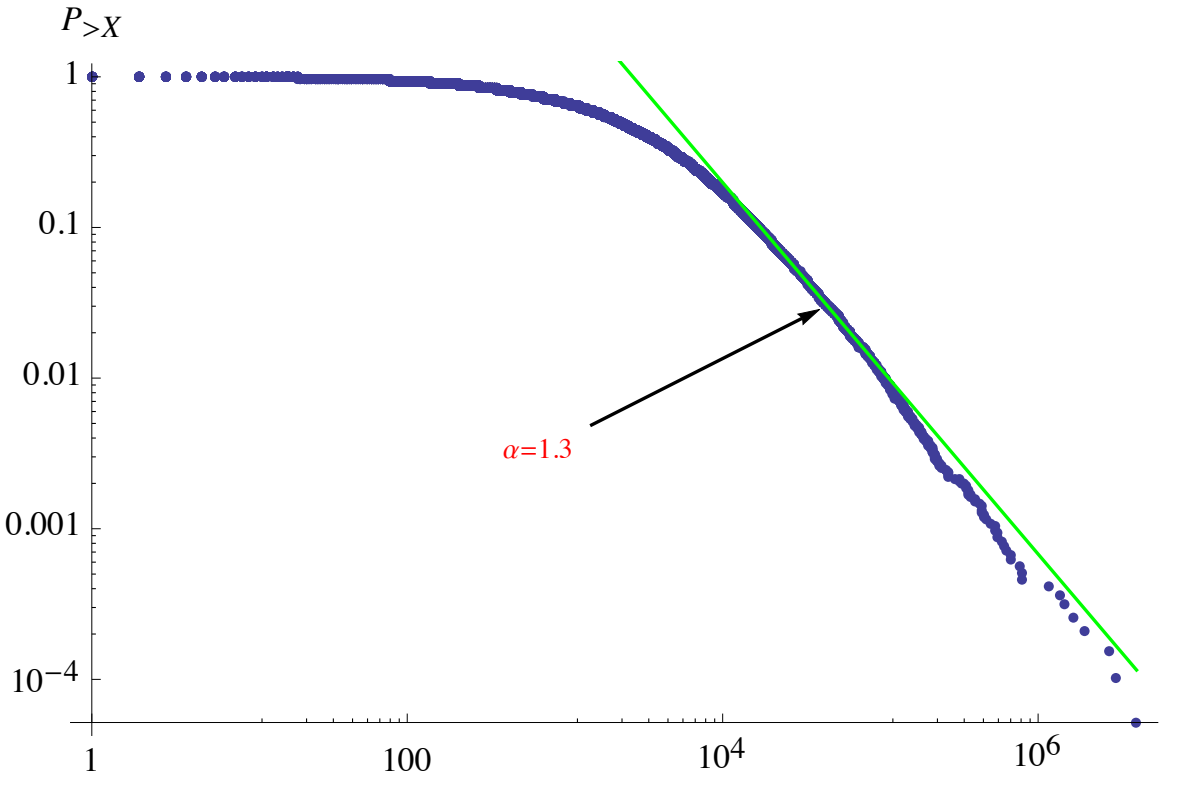

Figure 5.4: *Book Sales: the near tail can be robust for estimation of sales from rank and vice versa –it works well and shows robustness so long as one doesn't compute general expectations or higher non-truncated moments.*

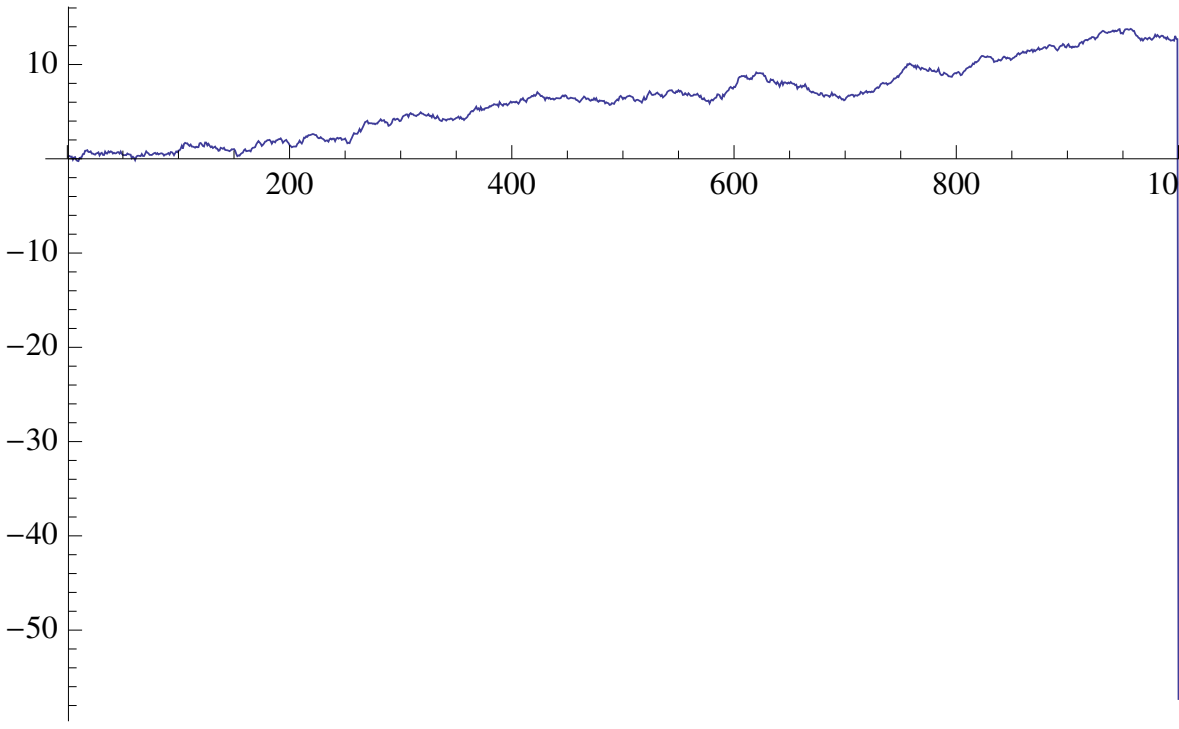

Figure 5.5: *The Turkey Problem, where nothing in the past properties seems to indicate the possibility of the jump.*

**Estimation issues** Note that there are many methods to estimate the tail exponent $\alpha$ from data, what is called a "calibration. However, we will see, the tail exponent is rather hard to guess, and its calibration marred with errors, owing to the insufficiency of data in the tails. In general, the data will show thinner tail than it should.

We will return to the issue in more depth in later chapters.

## 5.2 SOME PROPERTIES OF POWER LAWS

Two central properties.

### 5.2.1 Sums of variables

**Property 5.1: Tail exponent of a sum**

*Let $X_1, X_2, \ldots X_n$ be random variables neither independent nor identically distributed, each $X_i$ following a distribution with a different asymptotic tail exponent $\alpha_i$ (we assume that random variables outside the power law class will have an asymptotic alpha = $+\infty$). Assume further we are concerned with the right tail of the distribution (the argument remains identical when we apply it to the left tail). See [117] for further details.*

*Consider the weighted sum $S_n = \sum_{i=1}^{n} w_i X_i$, with all weights $w_i$ strictly positive. Consider $\alpha_s$ the asymptotic tail exponent for the sum.*

*For all $w_i > 0$,*

$$\alpha_s = \min(\alpha_i).$$

Clearly, if $\alpha_2 \leq \alpha_1$ and $w_2 > 0$,

$$\lim_{z \to \infty} \frac{\log\left(w_1 z^{-\alpha_1} + w_2 z^{-\alpha_2}\right)}{\log(z)} = \alpha_2.$$

The implication is that adding a single summand with undefined (or infinite) mean, variance, or higher moments leads to the total sum to have undefined (or infinite) mean, variance, or higher moments.

**Principle 5.1: Fat Tails + Thin Tails = Fat Tails**

*Mixing power law distributed and thin tailed variables results in power laws, no matter the composition.*

### 5.2.2 Product of variables

The same effect is observed with the product of variables with Paretian tail.

Consider the weighted product $P_n = \prod_{i=1}^{n} w_i X_i$, with all weights $w_i$ strictly positive (to simplify). Consider $\alpha_p$ the asymptotic tail exponent for the product.

For all $w_i > 0$,

$$\alpha_p = \min(\alpha_i).$$

We can illustrate with the product of two Paretian variables $X_1$ and $X_2$ with tail index $\alpha_1$ and $\alpha_2$ respectively, the pdf of the product $f(z)$ becomes:

$$f(z) \propto \frac{\alpha_1 \alpha_2 z^{-\alpha_1 - \alpha_2 - 1}\left(z^{\alpha_1} - z^{\alpha_2}\right)}{\alpha_1 - \alpha_2}$$

with survival function: $\propto \frac{\alpha_1 z^{-\alpha_2} - \alpha_2 z^{-\alpha_1}}{\alpha_1 - \alpha_2}$.

We note that for both variables having the same exponent, $\lim_{\alpha_2 \to \alpha_1} \frac{\alpha_1 \alpha_2 x^{-\alpha_1 - \alpha_2 - 1}\left(x^{\alpha_1} - x^{\alpha_2}\right)}{\alpha_1 - \alpha_2} = \alpha_1^2 x^{-\alpha_1 - 1} \log(x)$.

If, again, $\alpha_2 \leq \alpha_1$:

$$\lim_{x \to \infty} \frac{\log\left(\alpha_1 z^{-\alpha_2} - \alpha_2 z^{-\alpha_1}\right) - \log\left(\alpha_1 - \alpha_2\right)}{\log(z)} = \alpha_2,$$

the same asymptotic dominance of the fatter tail.

Note: for the exact distribution of the product of stable variables, see **??**–though the "exact" part entails unwieldy Fox $H$ functions.

**Principle 5.2: Fat Tails × Thin Tails = Fat Tails**

*The multiplication of power law distributed and thin tailed variables results in power laws, no matter the composition.*

### 5.2.3 Transformations

The second property while appearing benign, can be vastly more annoying:

**Property 5.2**

*Let $X$ be a random variable with tail exponent $\alpha$. The tail exponent of $X^p$ is $\frac{\alpha}{p}$.*

This tells us that the variance of a finite variance random variable with tail exponent $< 4$ will be infinite. In fact we will see it does cause problems for stochastic volatility models, when the real process can actually be of infinite variance.

This gives us a hint, without too much technical effort, on how a convex transformation of a random variable thickens the tail.

*Proof.* The general approach is as follows. Let $p(.)$ be a probability density function and $\phi(.)$ a transformation (with some restrictions). We have the distribution of the transformed variable (assuming the support is conserved –stays the same):

$$p\left(\phi(x)\right) = \frac{p\left(\phi^{(-1)}(x)\right)}{\phi'\left(\phi^{(-1)}(x)\right)}. \tag{5.8}$$

Assume that $x > l$ and $l$ is large (i.e. a point where the slowly varying function "ceases to vary" within some order of $x$). The PDF for these values of $x$ can be written as $p(x) \propto K x^{-\alpha-1}$. Consider $y = \phi(x) = x^p$: the inverse function of $y = x^p$ is $x = y^{\frac{1}{p}}$. Applying to the denominator in Eq. 5.8, we get $\frac{1}{p} x^{\frac{1-p}{p}}$. □

Integrating above $l$, the survival function will be: $\mathbb{P}(Y > y) \propto y^{-\frac{\alpha}{p}}$.

## 5.3 BELL SHAPED VS NON BELL SHAPED POWER LAWS

**The slowly varying function effect, a case study** The fatter the tails, the less the "body" matters for the moments (which become infinite, eventually). But for power laws with thinner tails, the zone that is not power law (the slowly moving part) plays a role –"slowly varying" is more or less formally defined in 5.1.1,21.2.2 and 5.1.1. This section will show how apparently equal distributions can have different shapes.

Let us compare a double Pareto distribution with the following PDF:

$$f_P(x) = \begin{cases} \alpha(1+x)^{-\alpha-1} & x \geq 0 \\ \\ \alpha(1-x)^{-\alpha-1} & x < 0 \end{cases}$$

to a Student T with same centrality parameter 0, scale parameter $s$ and PDF $f_S(x) = \frac{\alpha^{\alpha/2}\left(\alpha+\frac{x^2}{s^2}\right)^{\frac{1}{2}(-\alpha-1)}}{sB\left(\frac{\alpha}{2},\frac{1}{2}\right)}$ where $B(.)$ is the Euler beta function, $B(a,b) = \frac{(\Gamma(a))(\Gamma(b))}{\Gamma(a+b)} = \int_0^1 t^{a-1}(1-t)^{b-1}\, dt$.

We have two ways to compare distributions.

- Equalizing by tail ratio: setting $\lim_{x\to\infty} \frac{f_p(x)}{f_s(x)} = 1$ to get the same tail ratio, we get the equivalent "tail" distribution with $s = \left(\alpha^{1-\frac{\alpha}{2}} B\left(\frac{\alpha}{2},\frac{1}{2}\right)\right)^{1/\alpha}$.
- Equalizing by standard deviations (when finite): we have, with $\alpha > 2$, $\mathbb{E}(X_P^2) = \frac{2}{\alpha^2-3\alpha+2}$ and $\mathbb{E}(X_S^2) = \frac{\alpha\left(\alpha^{1-\frac{\alpha}{2}} B\left(\frac{\alpha}{2},\frac{1}{2}\right)\right)^{2/\alpha}}{\alpha-2}$.

  So we could set $\sqrt{\mathbb{E}(X_P^2)} = \sqrt{k}\sqrt{\mathbb{E}(X_S^2)}$ $k \to \frac{2\alpha^{-2/\alpha}B\left(\frac{\alpha}{2},\frac{1}{2}\right)^{-2/\alpha}}{\alpha-1}$ }.

Finally, we have the comparison "bell shape" semi-concave vs the angular double-convex one as seen in Figure 5.6.

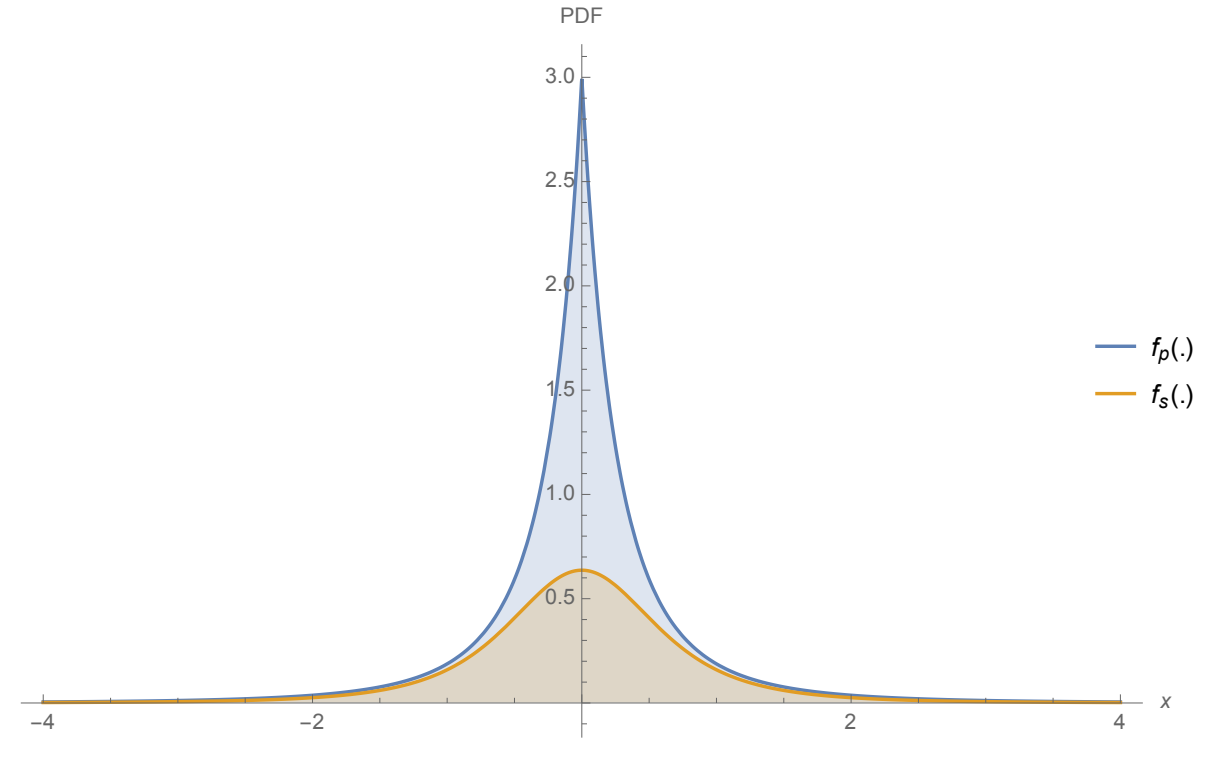

Figure 5.6: *Comparing two symmetric power laws of same exponent, one with a brief slowly varying function, the other with an extended one. All moments eventually become the same in spite of the central differences in their shape for small deviations.*

## 5.4 INTERPOLATIVE POWERS OF POWER LAWS: AN EXAMPLE

Consider Jobless Claims during the COVID-19 pandemic: unemployment jumped many so-called standard deviations in March of 2020. But was the jump an outlier? Maybe if you look at 5.7 and think like someone trained in thin tails. But not really. As Figure 5.8 shows, the tail exponent is hardly changed. The scale of the distribution could perhaps vary, but the exponent is patently robust to out-of-sample observations.

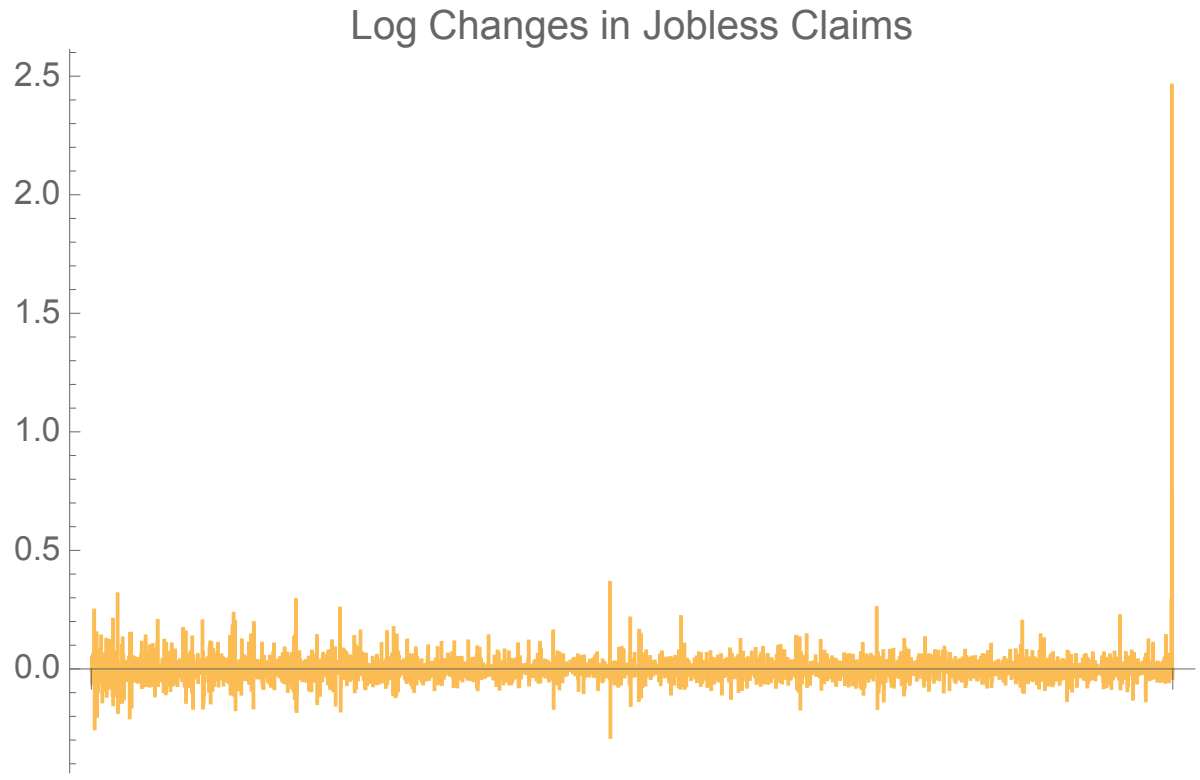

Figure 5.7: *Jobless claims: looks like the jump is a surprise... but only to untrained economists. As Fig. 5.8 shows, it shouldn't be. And to the trained eyes (a la Benoit Mandelbrot), variations were mild but certainly never Gaussian.*

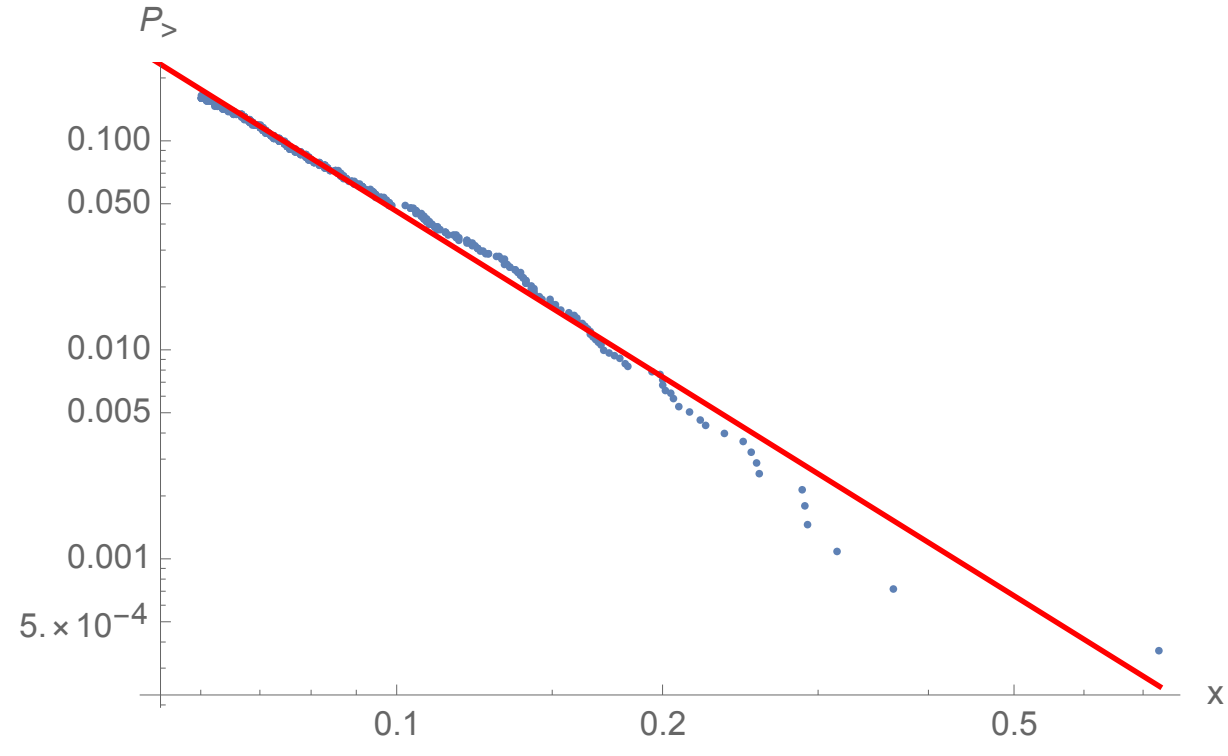

Figure 5.8: *Zipf plot for jobless claims: we did not need the abrupt jump during the COVID-19 pandemic (last point on the right) to realize it was a power law.*

## 5.5 SUPER-FAT TAILS: THE LOG-PARETO DISTRIBUTION

The mother of all fat tails, the log-Pareto distribution, is not present in common lists of distributions but we can rederive it here. The log-Pareto is the Paretian analog of the lognormal distribution.

**Remark 5.1: Rediscovering the log-Pareto distribution**

*If $X \sim \mathcal{P}(L, \alpha)$ the Pareto distribution with PDF $f^{(P)}(x) = \alpha L^{\alpha} x^{-\alpha-1}$ , $x \geq L$ and survival function $S^{(P)}(x) = L^{\alpha} x^{-\alpha}$, then:*

*$e^X \sim \mathcal{LP}(L, \alpha)$ the log-Pareto distribution with PDF*

$$f^{(LP)}(x) = \frac{\alpha L^{\alpha} \log^{-\alpha-1}(x)}{x} \, , x \geq e^L$$

*and survival function*

$$S^{(LP)}(x) = L^{\alpha} \log^{-\alpha}(x)$$

While for a regular power law, we have an asymptotic linear slope on the log-log plot, i.e.,

$$\lim_{x \to \infty} \frac{\log\left(L^{\alpha} x^{-\alpha}\right)}{\log(x)} = -\alpha,$$

the slope for a log-Pareto goes to 0:

$$\lim_{x \to \infty} \frac{\log\left(L^{\alpha} \log(x)^{-\alpha}\right)}{\log(x)} = 0,$$

and clearly no moment can exist regardless of the value of the tail parameter $\alpha$. The difference between asymptotic behaviors is visible is Fig 5.9.

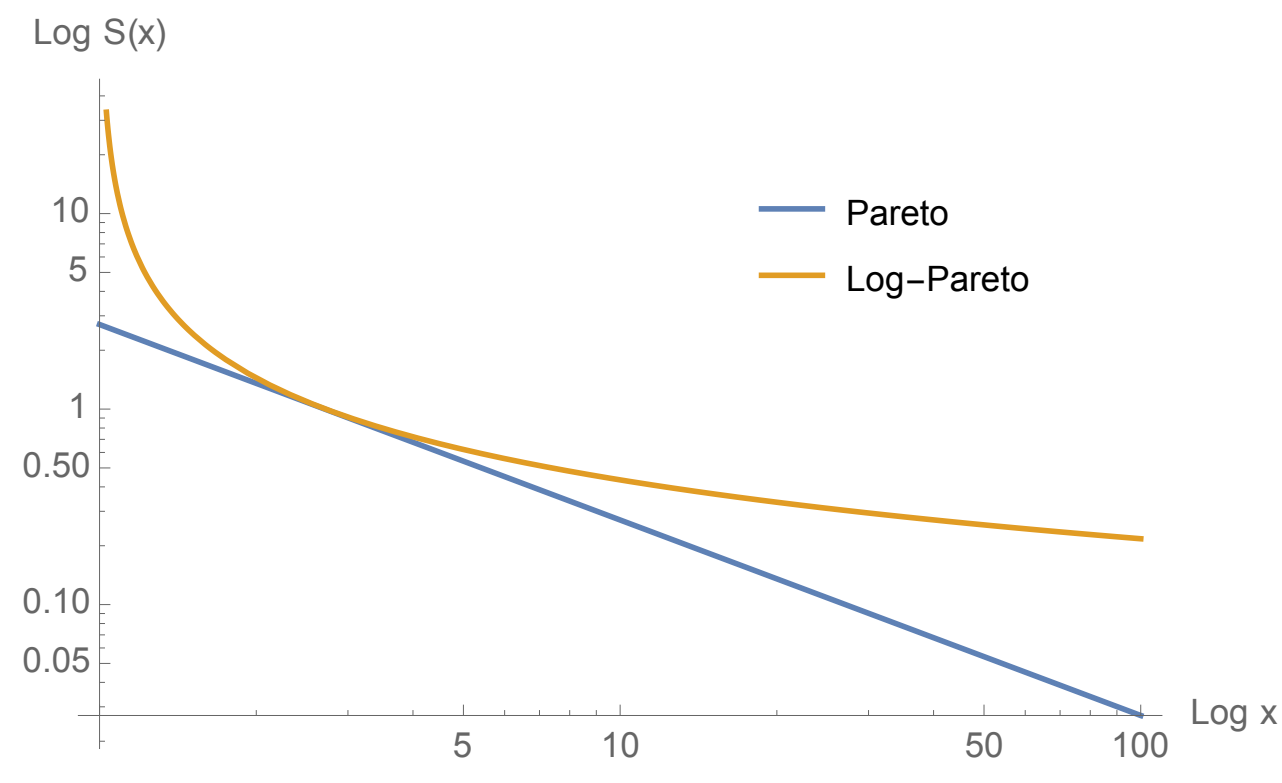

Figure 5.9: *Comparing log-log plots for the survival functions of the Pareto and log-Pareto*

## 5.6 PSEUDO-STOCHASTIC VOLATILITY: AN INVESTIGATION

We mentioned earlier in Chapter 3 that a "10 sigma" statement means we are not in the Gaussian world. We also discussed the problem of nonobservability of probability distributions: we observe data, not generating processes.

It is therefore easy to be fooled by a power law by mistaking it for a heteroskedastic process. In hindsight, we can always say: "conditional volatility was high, at such standard deviation it is no longer a 10 sigma, but a mere 3 sigma deviation".

The way to debunk these claims is to reason with the aid of an inverse problem: how a power law with a constant scale can masquerade as a heteroskedastic process. We will see in Appendix F how econometrics' reliance on heteroskedasticity (i.e. moving variance) has severe defects since the variance of that variance doesn't have a structure. (We are not even considering the even more realistic case of a power law with varying scale!)

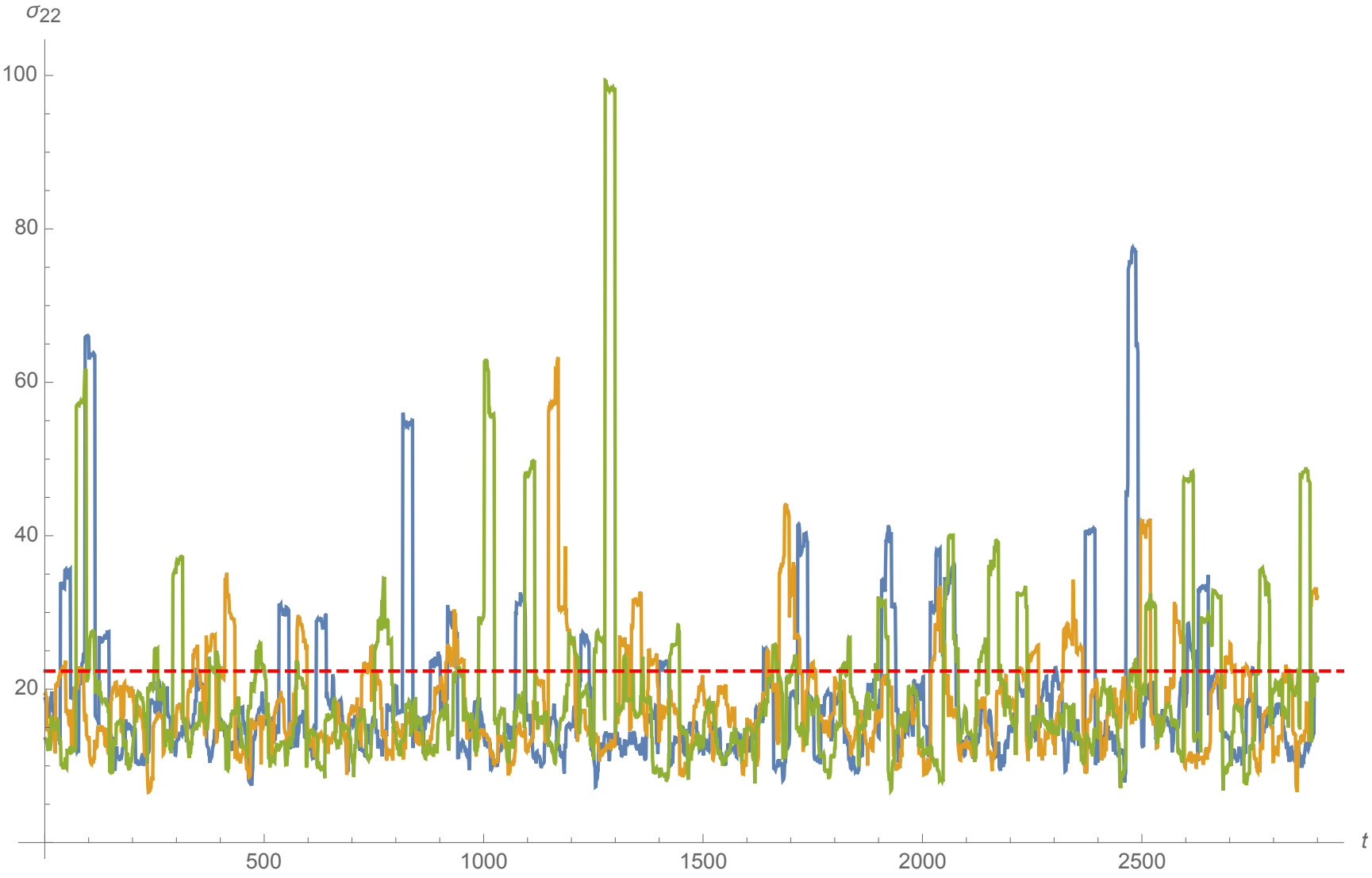

Figure 5.10: *Running 22-day (i.e., corresponding to monthly) realized volatility (standard deviation) for a Student T distributed returns sampled daily. It gives the impression of stochastic volatility when in fact the scale of the distribution is constant.*

Fig. 5.10 shows the volatility of returns of a market that greatly resemble ones should one use a standard simple stochastic volatility process. By stochastic volatility we assume the variance is distributed randomly [2].

Let $X$ be the returns with mean 0 and scale $\sigma$, with PDF $\varphi(.)$:

$$\varphi(x) = \frac{\left(\frac{\alpha}{\alpha+\frac{x^2}{\sigma^2}}\right)^{\frac{\alpha+1}{2}}}{\sqrt{\alpha}\,\sigma B\left(\frac{\alpha}{2},\frac{1}{2}\right)}, x \in (-\infty,\infty).$$

Transforming to get $Y = X^2$ (to get the distribution of the second moment), $\psi$, the PDF for $Y$ becomes,

$$\psi(y) = \frac{\left(\frac{\alpha\sigma^2}{\alpha\sigma^2+y}\right)^{\frac{\alpha+1}{2}}}{\sigma B\left(\frac{\alpha}{2},\frac{1}{2}\right)\sqrt{\alpha y}}, y \in (0,\infty),$$

[2] One can have models with either stochastic variance or stochastic standard deviation. The two have different expectations.

which we can see transforms into a power law with asymptotic tail exponent $\frac{\alpha}{2}$. The characteristic function $\chi_y(\omega) = \mathbb{E}(\exp(i\omega Y))$ can be written as

$$\begin{aligned}\chi_y(\omega) = \frac{1}{2B\left(\frac{\alpha}{2},\frac{1}{2}\right)} & \left(\pi\sqrt{\alpha}\sigma\sqrt{\frac{1}{\alpha\sigma^2}}((\pi\alpha)\csc)\left(\frac{\sqrt{\pi}\,{}_1\tilde{F}_1\left(\frac{1}{2};1-\frac{\alpha}{2};-i\alpha\sigma^2\omega\right)}{\Gamma\left(\frac{\alpha+1}{2}\right)}\right.\right. \\ & \left.\left. -\left(\frac{1}{\alpha\sigma^2}\right)^{-\frac{\alpha}{2}}(-i\omega)^{\alpha/2}\,{}_1\tilde{F}_1\left(\frac{\alpha+1}{2};\frac{\alpha+2}{2};-i\alpha\sigma^2\omega\right)\right)\right)\end{aligned} \tag{5.9}$$

From which we get the mean deviation of the second moment as follows[3]:

| $\alpha$ | MD of the second moment |
|---|---|
| $\frac{5}{2}$ | $\frac{\sqrt[4]{\frac{5}{3}}\,2^{3/4}\left(2\,{}_2F_1\left(\frac{1}{4},\frac{7}{4};\frac{5}{4};-\frac{5}{6}\right)+3\left(\frac{6}{11}\right)^{3/4}\right)\sigma^2\Gamma\left(\frac{7}{4}\right)}{\sqrt{\pi}\Gamma\left(\frac{5}{4}\right)}$ |
| $3$ | $\frac{6\sigma^2}{\pi}$ |
| $\frac{7}{2}$ | $\frac{5\ 7^{3/4}\left(7\,{}_2F_1\left(\frac{3}{4},\frac{9}{4};\frac{7}{4};-\frac{7}{6}\right)-3\,{}_2F_1\left(\frac{7}{4},\frac{9}{4};\frac{11}{4};-\frac{7}{6}\right)\right)\sigma^2\Gamma\left(\frac{5}{4}\right)}{6\ 6^{3/4}\sqrt{\pi}\Gamma\left(\frac{7}{4}\right)}$ |
| $4$ | $\frac{1}{7}\left(3\sqrt{21}-7\right)\sigma^2$ |
| $\frac{9}{2}$ | $\frac{3\sqrt[4]{\frac{3}{2}}\left(6\left(\frac{2}{5}\right)^{3/4}-6\,{}_2F_1\left(\frac{5}{4},\frac{11}{4};\frac{9}{4};-\frac{3}{2}\right)\right)\sigma^2\Gamma\left(\frac{11}{4}\right)}{5\sqrt{\pi}\Gamma\left(\frac{9}{4}\right)}$ |
| $5$ | $\frac{\sigma^2\left(7\sqrt{15}-16\tan^{-1}\left(\sqrt{\frac{5}{3}}\right)\right)}{6\pi}$ |

NEXT

The next chapter will venture in higher dimensions. Some consequences are obvious, others less so –say correlations exist even when covariances do not.

[3] As customary, we do not use standard deviation as a metric owing to its instability and its lack of information, but prefer mean deviation.

# 6 THICK TAILS IN HIGHER DIMENSIONS†

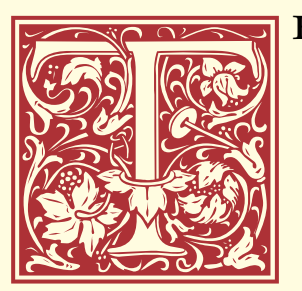
his discussion is about as simplified as possible handling of higher dimensions. We will look at 1) the simple effect of fat-tailedness for multiple random variables, 2) Ellipticality and distributions, 3) random matrices and the associated distribution of eigenvalues , 4) How we can look at covariance and correlations when moments don't exist (say, as in the Cauchy case).

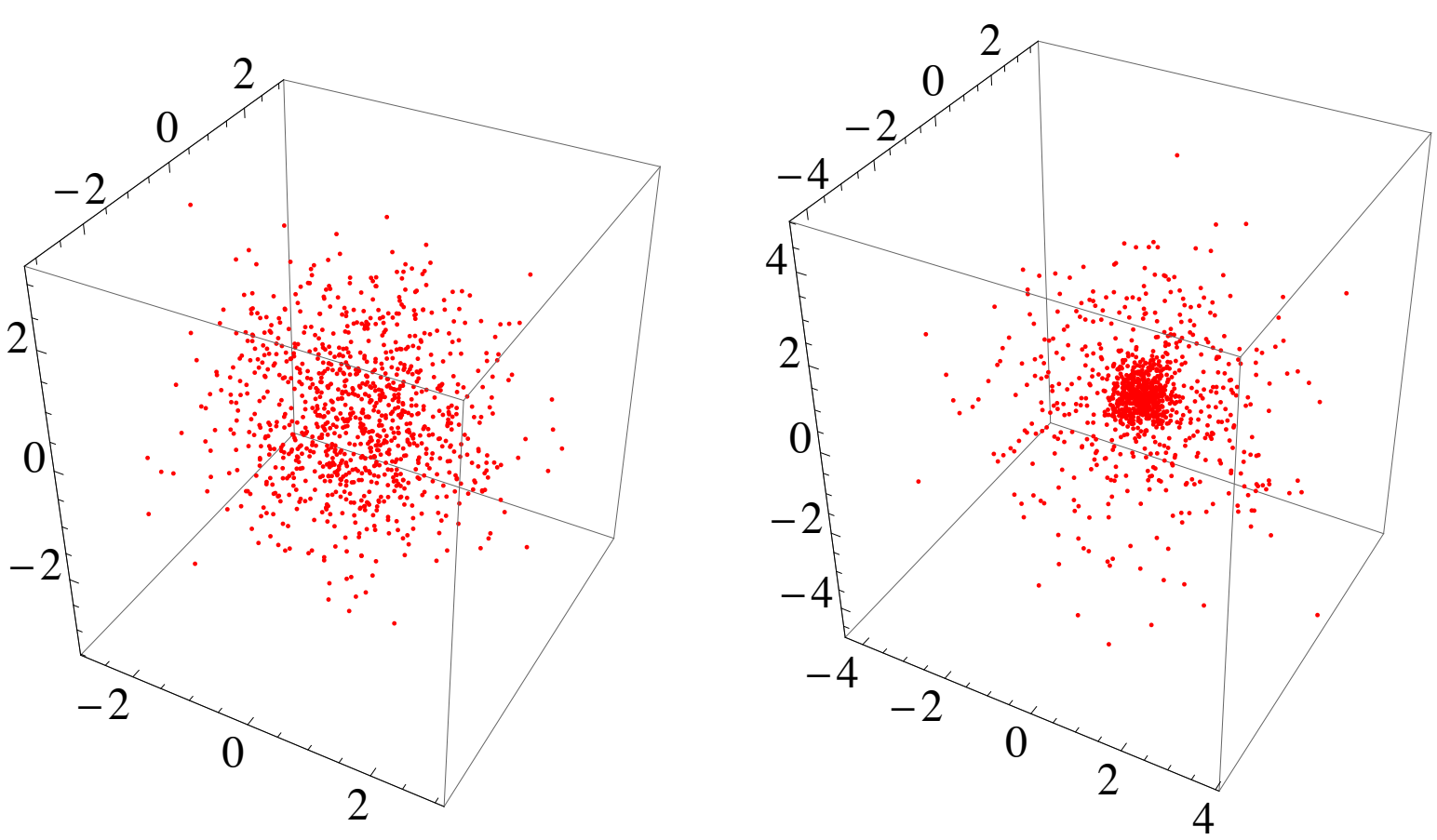

Figure 6.1: *Thick tails in higher dimensions: For a 3 dimentional vector, thin tails (left) and thick tails (right) of the same variance. In place of a bell curve with higher peak (the "tunnel") of the univariate case, we see an increased density of points towards the center.*

† Discussion chapter.

## 6.1 THICK TAILS IN HIGHER DIMENSION, FINITE MOMENTS

We will build the intuitions of thick tails from convexity to scale as we did in the previous chapter, but using higher dimensions.

Let $\vec{X} = (X_1, X_2, \ldots, X_m)$ be a $p \times 1$ random vector with the variables assumed to be drawn from a multivariate Gaussian. Consider the joint probability distribution $f(x_1, \ldots, x_m)$. We denote the $m$-variate multivariate Normal distribution by $\mathcal{N}(\vec{\mu}, \Sigma)$, with mean vector $\vec{\mu}$, variance-covariance matrix $\Sigma$, and joint pdf,

$$f\left(\vec{x}\right) = (2\pi)^{-m/2}|\Sigma|^{-1/2}\exp\left(-\frac{1}{2}\left(\vec{x}-\vec{\mu}\right)^T \Sigma^{-1}\left(\vec{x}-\vec{\mu}\right)\right) \tag{6.1}$$

where $\vec{x} = (x_1, \ldots, x_m) \in \mathbb{R}^m$, and $\Sigma$ is a symmetric, positive definite $(m \times m)$ matrix.

We can apply the same simplied variance preserving heuristic as in 4.1 to fatten the tails:

$$\begin{aligned} f_a\left(\vec{x}\right) =& \frac{1}{2}(2\pi)^{-m/2}|\Sigma_1|^{-1/2}\exp\left(-\frac{1}{2}\left(\vec{x}-\vec{\mu}\right)^T \Sigma_1{}^{-1}\left(\vec{x}-\vec{\mu}\right)\right) \\ &+ \frac{1}{2}(2\pi)^{-m/2}|\Sigma_2|^{-1/2}\exp\left(-\frac{1}{2}\left(\vec{x}-\vec{\mu}\right)^T \Sigma_2{}^{-1}\left(\vec{x}-\vec{\mu}\right)\right) \end{aligned} \tag{6.2}$$

where $a$ is a scalar that determines the intensity of stochastic volatility, $\Sigma_1 = \Sigma(1+a)$ and $\Sigma_2 = \Sigma(1-a)$.[2]

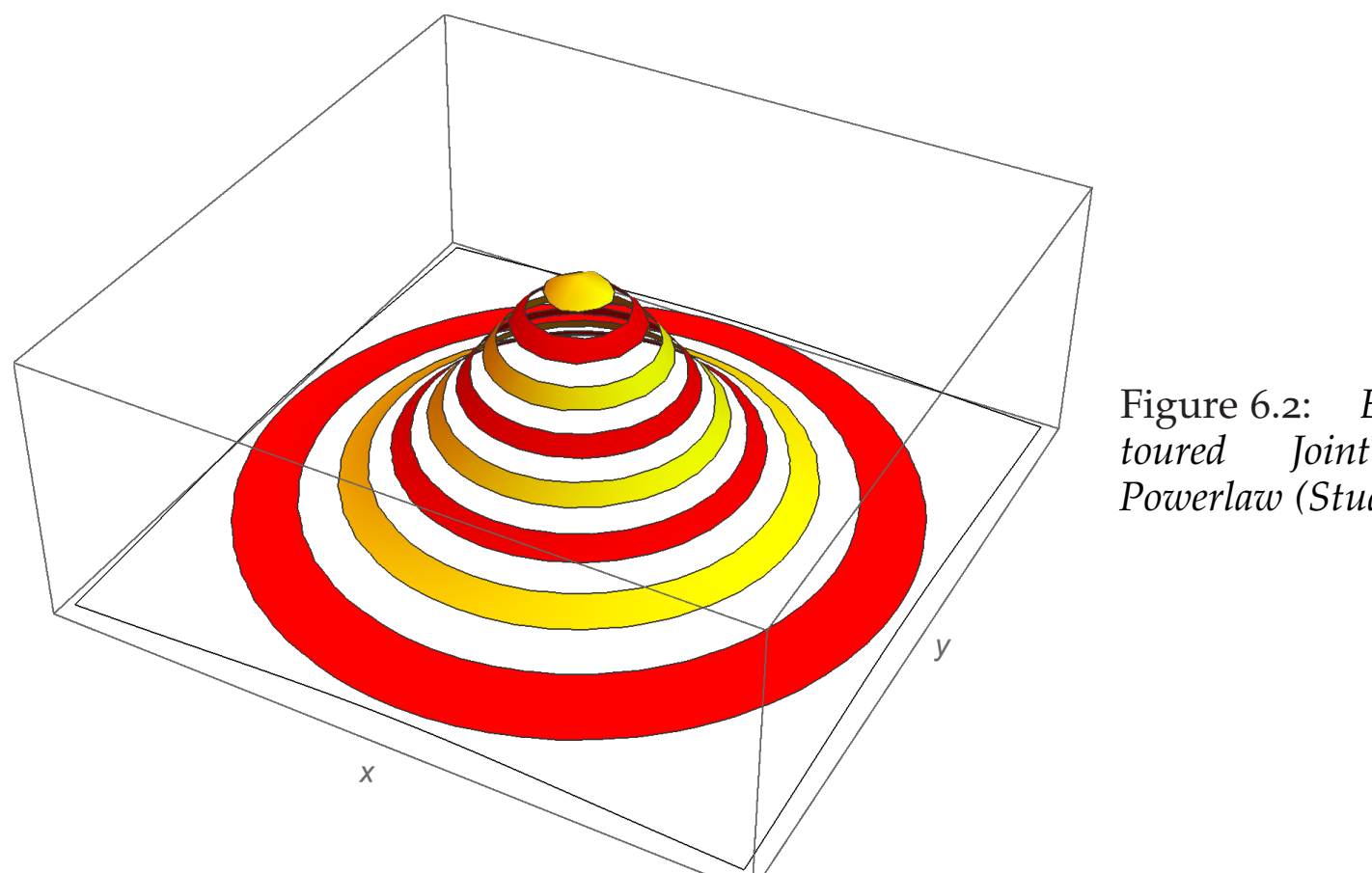

Figure 6.2: *Elliptically Contoured Joint Returns of Powerlaw (Student T).*

Notice in Figure 6.1, as with the one-dimensional case, a concentration in the middle part of the distribution.[3]

[2] We can simplify by assuming as we did in the single dimension case, without any loss of generality, that $\vec{\mu} = (0, \ldots, 0)$.

[3] We created thick tails making the variances stochastic while keeping the correlations constant; this is to preserve the positive definite character of the matrix.

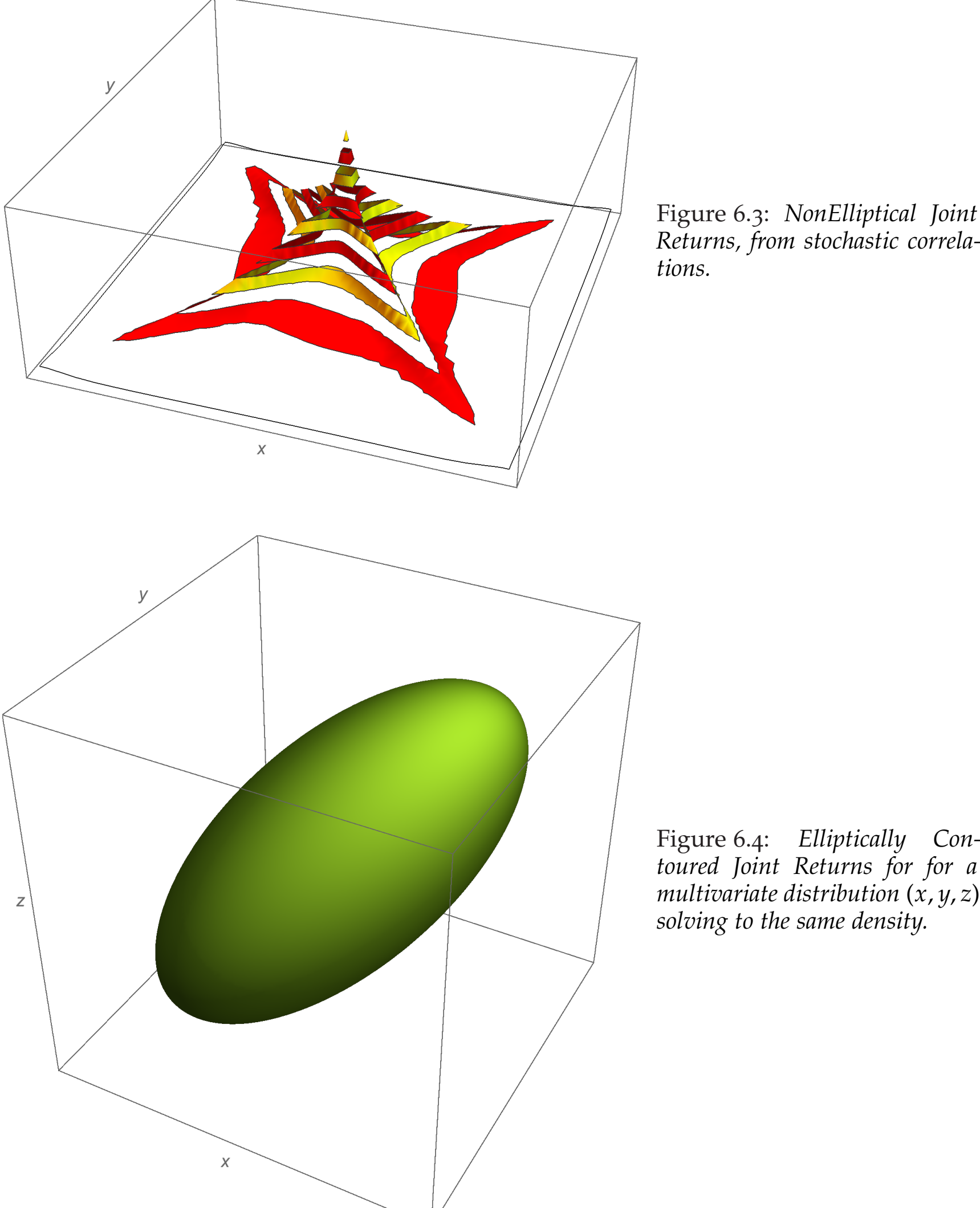

Figure 6.3: *NonElliptical Joint Returns, from stochastic correlations.*

Figure 6.4: *Elliptically Contoured Joint Returns for for a multivariate distribution* $(x, y, z)$ *solving to the same density.*

## 6.2 joint fat-tailedness and ellipticality of distributions

There is another aspect, beyond our earlier definition(s) of fat-tailedness, once we increase the dimensionality into random vectors:

**What is an Elliptically Contoured Distribution?** From the standard definition, [104], $\mathbf{X}$, a $p \times 1$ random vector is said to have an elliptical (or elliptical contoured) distribution with location parameters $\mu$, a non-negative matrix $\Sigma$, and some scalar function $\Psi$ if its characteristic function $\varphi$ is of the form

$$\varphi(t) = \exp(it'\mu)\Psi(t\Sigma t'). \tag{6.3}$$

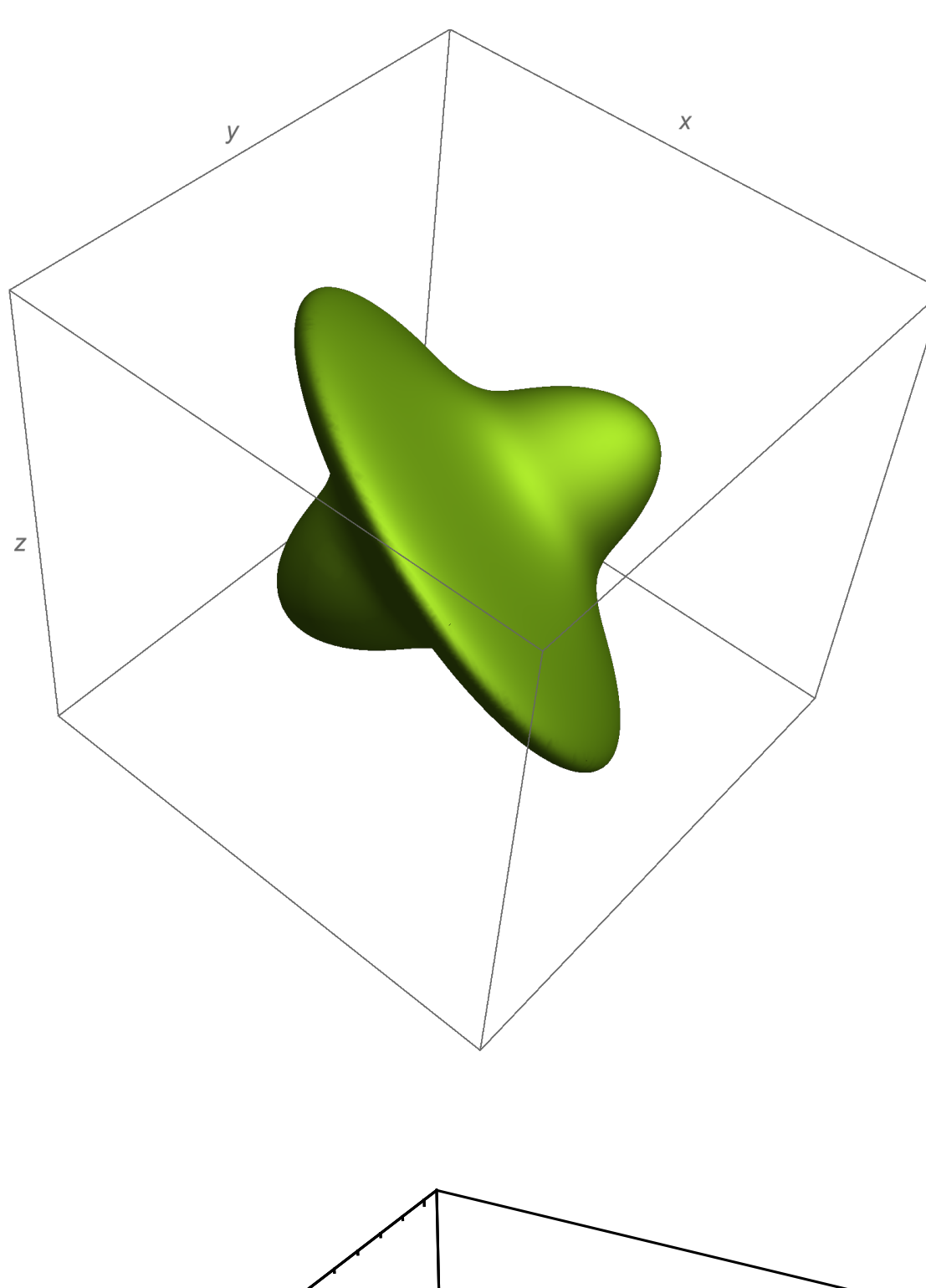

Figure 6.5: *NonElliptical Joint r.v., from stochastic correlations, for a multivariate distribution* $(x, y, z)$*, solving to the same density.*

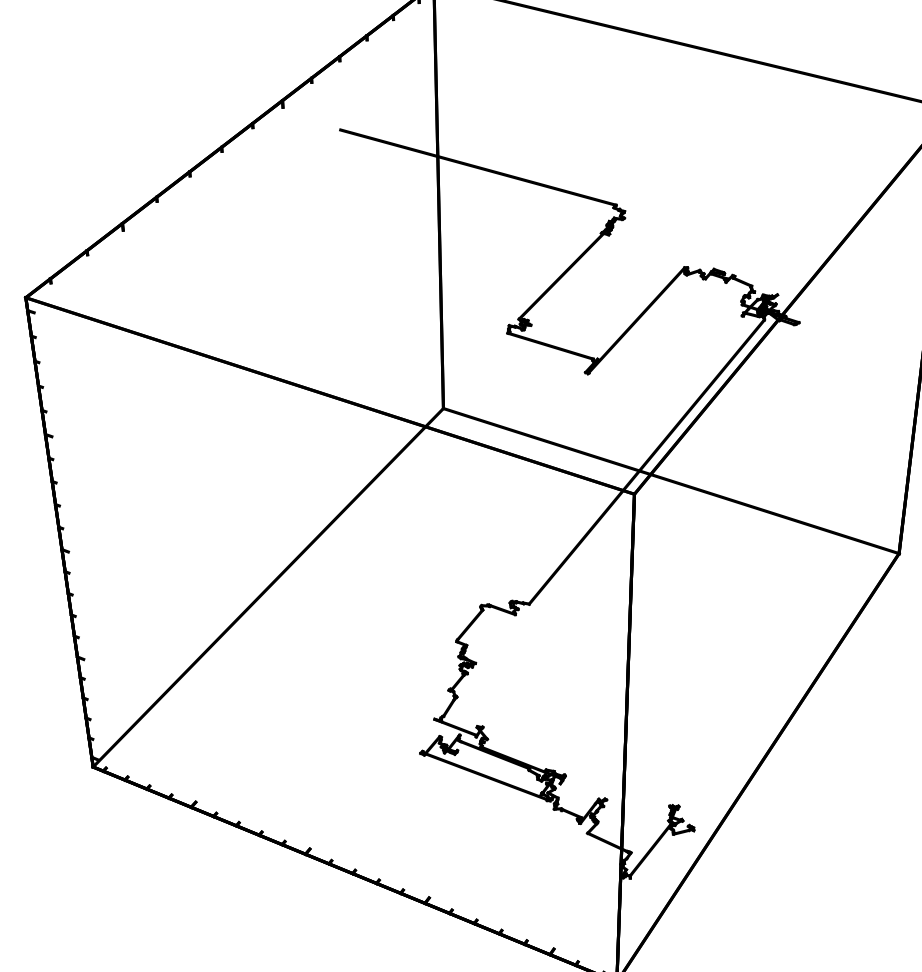

Figure 6.6: ***History moves by jumps**: A thick tailed historical process, in which events are distributed according to a power law that corresponds to the "80/20", with* $\alpha \simeq 1.13$*, represented as a 3-D Levy process.*

There are equivalent definitions focusing on the density; take for now that the main attribute is that $\Psi$ is a function of *a single* covariance matrix $\Sigma$.

Intuitively, an elliptical distribution should show an ellipse for iso-density plots; see how we represented in 2-D (for a bivariate) and 3-D (for a trivariate) in Figures 6.2 and 6.4. A noneliptical distribution would violate the shape as shown in Figures 6.3 and 6.5.

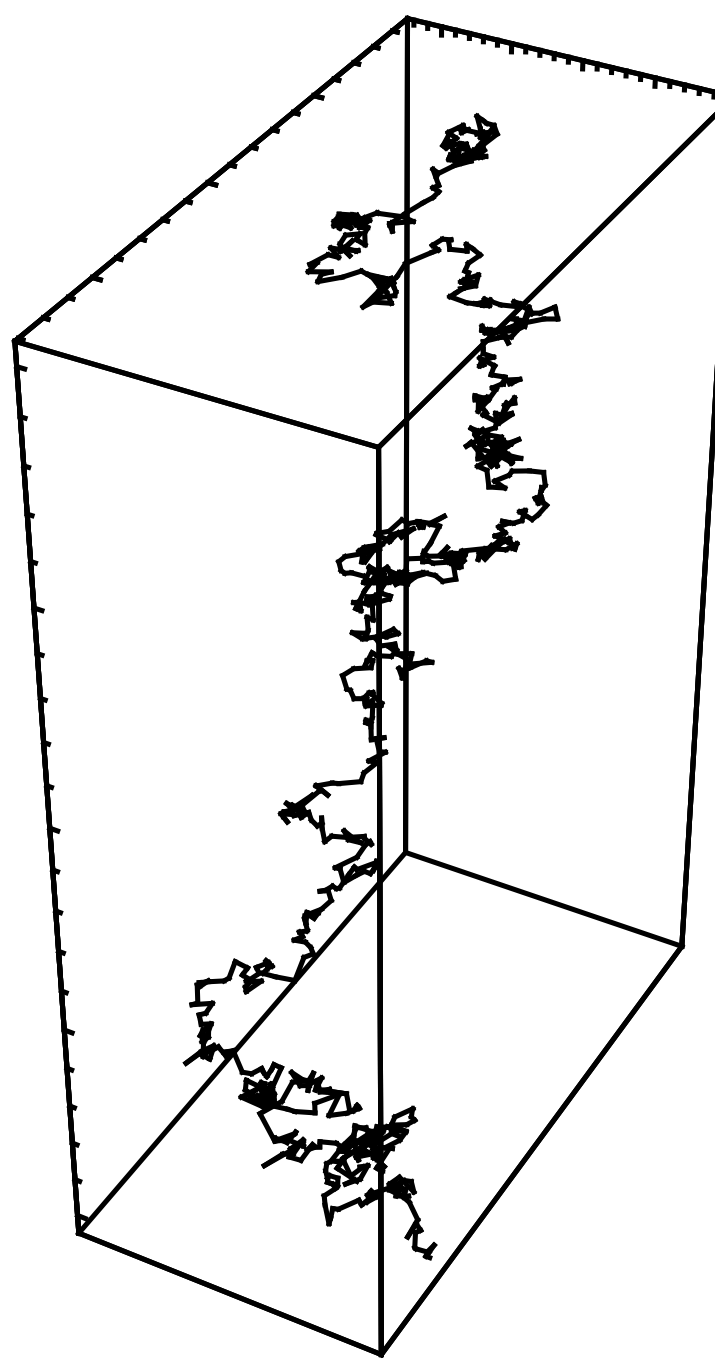

Figure 6.7: *What the proponents of "great moderation" or "long peace" have in mind: history as a thin-tailed process.*

The main property of the class of elliptical distribution is that it is closed under linear transformation. Intuitively, as we saw in Chapter 3 with the example of height vs wealth, it means (in a bivariate situation) that tails are less likely to come from one than two marginal deviations.

**Ellipticality and Central Flaws in Financial Theory** This closure under linear transformation leads to attractive properties in the building of portfolios, and in the results of portfolio theory (in fact one cannot have portfolio theory without ellipticality of distributions).

Under ellipticality, all portfolios can be characterized completely by their location and scale and any two portfolios with identical location and scale (in return space) have identical distributions returns.

Note that (ironically) Lévy-Stable distributions are elliptical –but only in the way they are defined.

So ellipticality (under the condition of finite variance) allows the extension of the results of modern portfolio theory (MPT) under the so-called "nonnormality", initally discovered by[216], also see[141]. However it appears (from those of us who work with stochastic covariances) that returns are not elliptical by any conceivable measure, see Chicheportiche and Bouchaud [46] and simple visual graphs of stability of correlation as in F.8.

A simple pedagogical example using the $1 \pm a$ heuristic we presented in 4.1. Consider the bivariate normal with characteristic function $\Psi(t_1, t_2) = e^{-\rho t_2 t_1 - \frac{t_1^2}{2} - \frac{t_2^2}{2}}$. Now let us stochasticize the $\rho$ parameter, with $p$ probability of $\rho_1$ and $(1-p)$ probability of $rho_2$:

$$\Psi(t_1, t_2) = p e^{-\rho_1 t_2 t_1 - \frac{t_1^2}{2} - \frac{t_2^2}{2}} + (1-p) p e^{-\rho_2 t_2 t_1 - \frac{t_1^2}{2} - \frac{t_2^2}{2}} \tag{6.4}$$

Figure 6.8 shows the result with $p = \frac{1}{2}$ and $\rho_1 = \rho_2$.

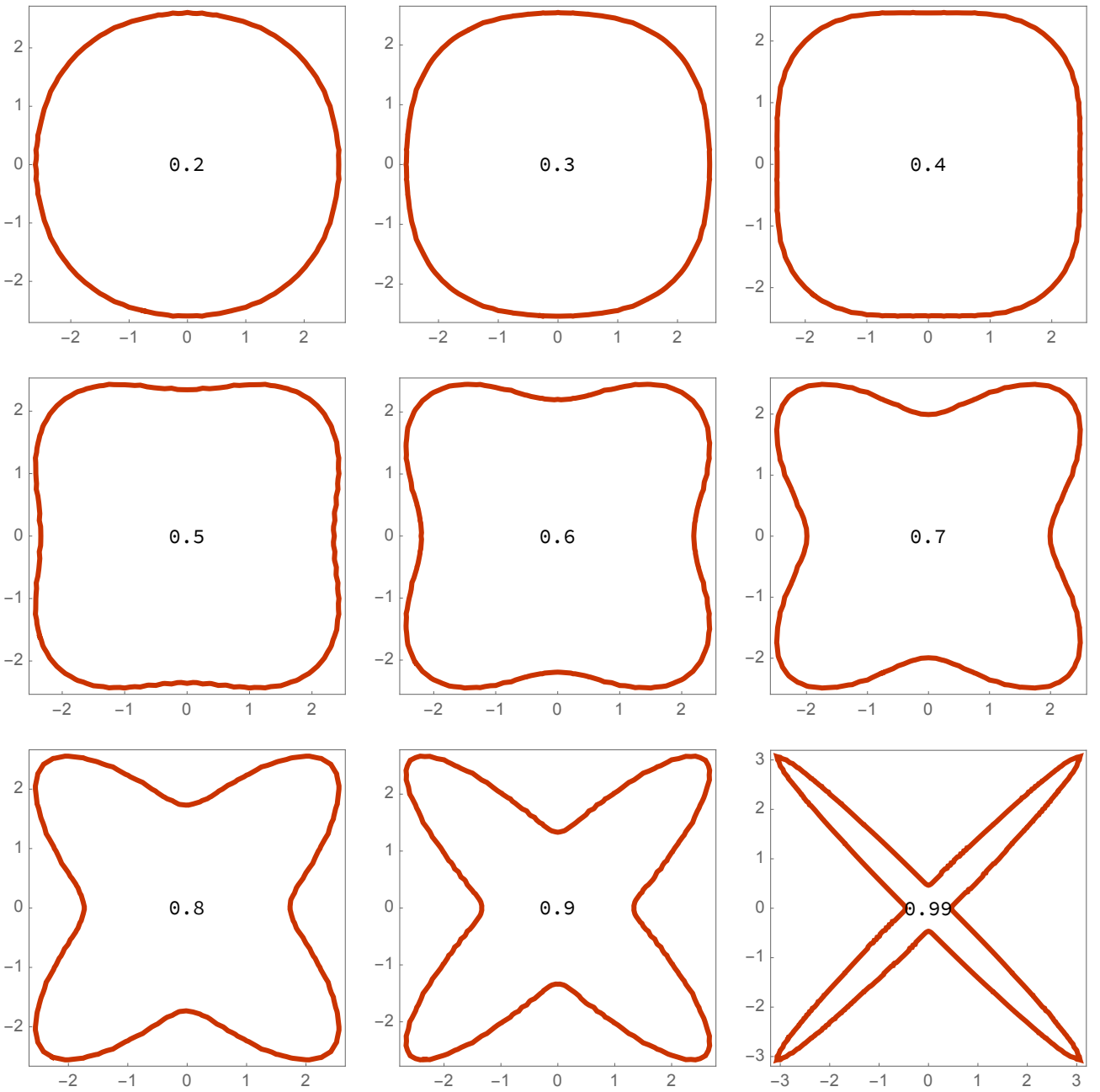

Figure 6.8: *Stochastic correlation for a standard binormal distribution: isodensities for different combinations. We use a very simple technique of Eq. 6.4, with switch between $\rho_1 = \rho$ and $\rho_2 = -\rho$ over the span with probability $p = \frac{1}{2}$.*

We can be more formal and show the difference, when $\Sigma$ is stochastic, between $\Psi\left(t\mathbb{E}(\Sigma)t'\right)$ and $\mathbb{E}\left(\Psi(t\Sigma t')\right)$ in Eq. 29.2.

**Diversification**

Recall that financial theory fails under thick tails (and no patches have fixed the issue outside of the "overfitting" we discussed in earlier chapters). Absence of ellipticality closes the matter. The implication is that *all* methods based on Markowitz-style portfolio construction, that is, grounded in the idea of diversification, fail to reduce the risk, while managing to deceivingly smooth out daily volatility. Adding leverage makes blowups certain in the long run [a].

[a] This includes an abhorrent approach called "risk parity" largely used to raise money via pseudotheoretical and pseudoacademic smoke, a method called "asset gathering".

## 6.3 MULTIVARIATE STUDENT T

The multivariate Student T is a convenient way to model, as it collapses to the Cauchy for $\alpha = 1$. The alternative would be the multivariate stable, which, we will see, is devoid of density.

Let $\mathbf{X}$ be a $(p \times 1)$ vector following a multivariate Student T distribution, $\mathbf{X} \sim \mathcal{S}_t(\mathbf{M}, \mathbf{\Sigma}, \alpha)$, where $\mathbf{\Sigma}$ is a $(p \times p)$ matrix, $\mathbf{M}$ a $p$ length vector and $\alpha$ a Paretian tail exponent with PDF

$$f(\mathbf{X}) = \left( \frac{(\mathbf{X}-\mathbf{M}).\mathbf{\Sigma}^{-1}.(\mathbf{X}-\mathbf{M})}{\nu} + 1 \right)^{-\frac{1}{2}(\nu+p)}. \tag{6.5}$$

In the most simplified case, with $p = 2$, $M = (0,0)$, and $\mathbf{\Sigma} = = \begin{pmatrix} 1 & \rho \\ \rho & 1 \end{pmatrix}$,

$$f(x_1, x_2) = \frac{\nu\sqrt{1-\rho^2}\left(\frac{-\nu\rho^2+\nu-2\rho x_1 x_2+x_1^2+x_2^2}{\nu-\nu\rho^2}\right)^{-\frac{\nu}{2}-1}}{2\pi\left(\nu-\nu\rho^2\right)}. \tag{6.6}$$

### 6.3.1 Ellipticality and Independence under Thick Tails

Take the product of two Cauchy densities for $x$ and $y$ (what we used in Figure 3.1):

$$f(x)f(y) = \frac{1}{\pi^2\left(x^2+1\right)\left(y^2+1\right)} \tag{6.7}$$

which, patently, as we saw in Chapter 3 (with the example of the two randomly selected persons with a total net worth of \$36 million), is not elliptical. Compare to the joint distribution $f_\rho(x, y)$:

$$f_\rho(x,y) = \frac{1}{2\pi\sqrt{1-\rho^2}\left(y\left(\frac{y}{1-\rho^2}-\frac{\rho x}{1-\rho^2}\right)+x\left(\frac{x}{1-\rho^2}-\frac{\rho y}{1-\rho^2}\right)+1\right)^{3/2}}, \tag{6.8}$$

and setting $\rho = 0$ to get no correlation,

$$f_0(x,y) = \frac{1}{2\pi\left(x^2+y^2+1\right)^{3/2}} \tag{6.9}$$

which is elliptical. This illustrates how absence of correlation is not independence as:

> Independence between two variables $X$ and $Y$ is defined by the identity:
>
> $$\frac{f(x,y)}{f(x)f(y)} = 1,$$
>
> regardless of the correlation coefficient. In the class of elliptical distributions, the bivariate Gaussian with coefficient 0 is both independent and uncorrelated. This does not apply to the Student T or the Cauchy.

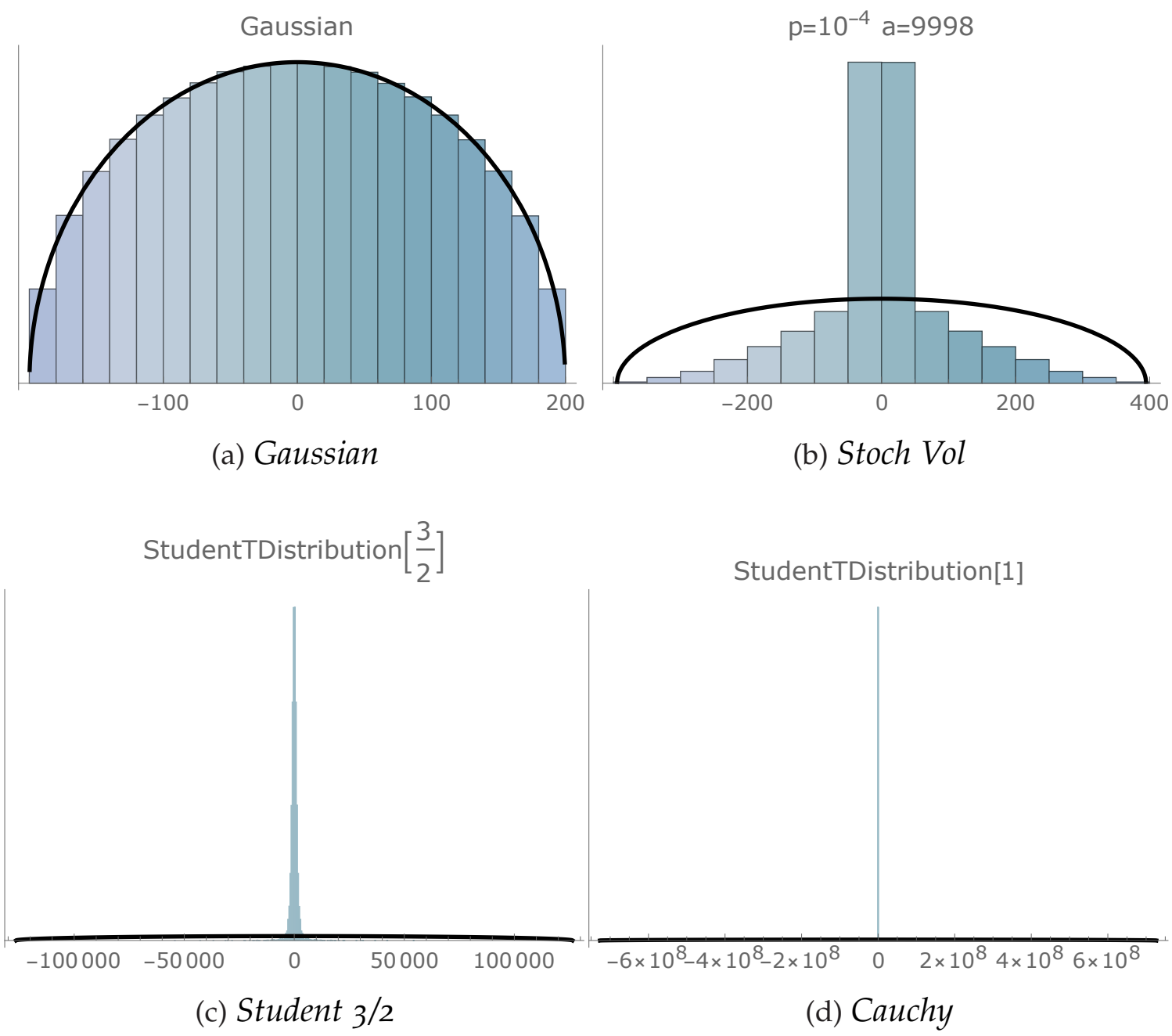

(a) *Gaussian*

(b) *Stoch Vol*

(c) *Student 3/2*

(d) *Cauchy*

Figure 6.9: *The various shapes of the distribution of the eigenvalues for random matrices, which in the Gaussian case follow the Wigner semicircle distribution. The Cauchy case corresponds to the Student parametrized to have 1 degrees of freedom.*

The reason the multivariate stable distribution with correlation coefficient set to 0 is not independent is the following.

A random vector $\mathbf{X} = (X_1, \ldots, X_k)'$ is said to have the multivariate stable distribution if every linear combination of its components $\mathbf{Y} = a_1 X_1 + \cdots + a_k X_k$ has a stable distribution. That is, for any constant vector $\mathbf{a} \in \mathbb{R}^k$, the random variable $\mathbf{Y} = a^T \mathbf{X}$ should have a univariate stable distribution. And to have a linear combination remain within the same class requires ellipticality. Hence *by construction*, $f_0(x, y)$ is not necessarily equal to $f(x) f(y)$. Consider the Cauchy case that has an explicit density function. The denominator of the product of densities includes an additional term, $x^2 y^2$, which pushes the iso-densities in one direction or another, as we saw in the introductory examples of Chapter 3.

## 6.4 FAT TAILS AND MUTUAL INFORMATION

We notice that because of the artificiality in constructing multivariate distributions, for the distributions under considerations, mutual information is not 0 in the absence of "correlation", since the ratio of joint densities/product of densities $\neq 1$ under 0 "correlation" $\rho$.

What is the mutual information of a Student T (which includes the Cauchy)?

$$\mathbb{I}(X,Y) = \mathbb{E} \log \left( \frac{f(x,y)}{f(x)f(y)} \right)$$

where the expectation is taken under the joint distribution for $X$ and $Y$. The mutual information thanks to the log is additive (Note that one can use any logarithmic base and translate by dividing by log(2)).

So $\mathbb{I}(X,Y) = \mathbb{E}\left(log f(x,y)\right) - \mathbb{E} log\left(f(x)\right) - \mathbb{E} log\left(f(y)\right)$ or $\mathbb{H}(X)$+$\mathbb{H}(Y)$ -$\mathbb{H}(X,Y)$ where $\mathbb{H}$ is the entropy and $\mathbb{H}(X,Y)$ the joint entropy.

We note that $-\frac{1}{2}\log(1-\rho^2)$ is the mutual information of a Gaussian regardless of parametrization. So for $X,Y \sim$ Multivariate Student T $(\alpha,\rho)$, the mutual information $\mathbb{I}_\alpha(X,Y)$:

$$\mathbb{I}_\alpha(X,Y) = -\frac{1}{2}\log\left(1-\rho^2\right) + \lambda_\alpha \tag{6.10}$$

where

$$\lambda_\alpha = -\frac{2}{\alpha} + \log(\alpha) + 2\pi(\alpha+1)\csc(\pi\alpha) + 2\log\left(B\left(\frac{\alpha}{2},\right.\right.$$
$$\left.\left.\frac{1}{2}\right)\right) - (\alpha+1)H_{-\frac{\alpha}{2}} + (\alpha+1)H_{-\frac{\alpha}{2}-\frac{1}{2}} - 1 - \log(2\pi) \tag{6.11}$$

where csc(.) is the cosecant of the argument, $B(.,.)$ is the beta function and $H(.)^{(r)}$ is the harmonic number $H_n^r = \sum_{i=1}^n \frac{1}{i^r}$ with $H_n = H_n^{(1)}$. We note that $\lambda_\alpha \underset{\alpha\to\infty}{\to} 0$.

To conclude this brief section metrics linked to entropy such as mutual information are vastly more potent than correlation; mutual information can detect nonlinearities.

## 6.5 FAT TAILS AND RANDOM MATRICES, A RAPID INTERLUDE

The eigenvalues of matrices themselves have an analog to Gaussian convergence: the semi-circle distribution, as shown in Figure 6.9.

Let $\mathbf{M}$ be a $(n,n)$ symmetric matrix. We have the eigenvalues $\lambda_i$, $1 \leq i, \leq n$ such that $\mathbf{M}.\mathbf{V}_i = \lambda_i \mathbf{V}_i$ where $\mathbf{V}_i$ is the $i^{th}$ eigenvector.

The Wigner semicircle distribution with support $[-R,R]$ has for PDF $f$ presenting a semicircle of radius $R$ centered at (0, 0) and then suitably normalized :

$$f(\lambda) = \frac{2}{\pi R^2}\sqrt{R^2-\lambda^2} \text{ for } -R \leq \lambda \leq R. \tag{6.12}$$

This distribution arises as the limiting distribution of eigenvalues of $(n,n)$ symmetric matrices with finite moments as the size $n$ of the matrix approaches infinity.

We will tour the "fat-tailedness" of the random matrix in what follows as well as the convergence.

This is the equivalent of thick tails for matrices. Consider for now that the $4^{th}$ moment reaching Gaussian levels (i.e. 3) for an univariate situation is equivalent to the eigenvalues reaching Wigner's semicircle.

## 6.6 CORRELATION AND UNDEFINED VARIANCE

Next we examine a paradox: while covariances can be infinite, correlation is finite. However, it will have a huge sampling error to be informative –same problem we discussed with PCA in Chapter 3.

Question: Why it is that a fat tailed distribution in the power law class $\mathfrak{P}$ with infinite or undefined mean (and higher moments) would have, in higher dimensions, undefined (or infinite) covariance but finite correlation?

Consider a distribution with support in $(-\infty, \infty)$. It has no moments: $\mathbb{E}(X)$ is indeterminate, $\mathbb{E}(X^2) = \infty$, no covariance, $\mathbb{E}(XY)$ is indeterminate. But the (noncentral) correlation for n variables is bounded by $-1$ and 1.

$$r \triangleq \frac{\sum_{i=1}^{n} x_i y_i}{\sqrt{\sum_{i=1}^{n} x_i^2}\sqrt{\sum_{i=1}^{n} y_i^2}}, \; n = 2, 3, ...$$

By the subexponentiality property, we have

$$\mathbb{P}\left(X_1 + \ldots + X_n \rangle\, x\right) \sim \mathbb{P}\left(\max\left(X_1, \ldots X_n\right) > x\right)$$

as $x \to \infty$. We note that the power law class is included in the subexponential class $\mathfrak{S}$.

Order the variables in absolute values such that $|x_1| \leq |x_2| \leq \ldots \leq |x_n|$

Let $\kappa_1 = \sum_{i=1}^{n-1} x_i y_i$, $\kappa_2 = \sum_{i=1}^{n-1} x_i^2$, and $\kappa_3 = \sum_{i=1}^{n-1} y_i^2$.

$$\lim_{x_n \to \infty} \frac{x_n y_n + \kappa_1}{\sqrt{x_n^2 + \kappa_2}\sqrt{y_n^2 + \kappa_3}} = \frac{y_n}{\sqrt{\kappa_3 + y_n^2}},$$

$$\lim_{y_n \to \infty} \frac{x_n y_n + \kappa_1}{\sqrt{x_n^2 + \kappa_2}\sqrt{y_n^2 + \kappa_3}} = \frac{x_n}{\sqrt{\kappa_2 + x_n^2}}$$

$$\lim_{\substack{x_n \to +\infty \\ y_n \to +\infty}} \frac{x_n y_n + \kappa_1}{\sqrt{x_n^2 + \kappa_2}\sqrt{y_n^2 + \kappa_3}} = 1$$

$$\lim_{\substack{x_n \to +\infty \\ y_n \to -\infty}} \frac{x_n y_n + \kappa_1}{\sqrt{x_n^2 + \kappa_2}\sqrt{y_n^2 + \kappa_3}} = -1$$

and

$$\lim_{\substack{x_n \to -\infty \\ y_n \to +\infty}} \frac{x_n y_n + \kappa_1}{\sqrt{x_n^2 + \kappa_2}\sqrt{y_n^2 + \kappa_3}} = -1$$

for all values of $n \geq 2$.

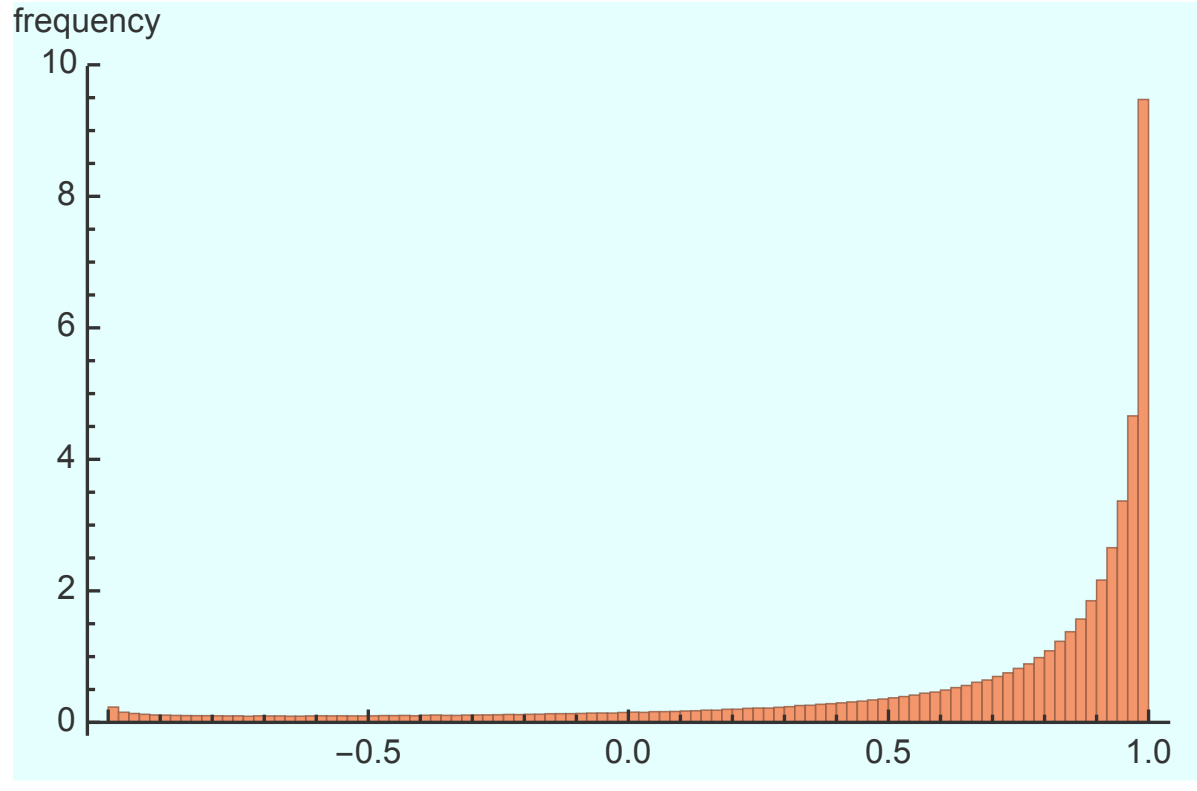

Figure 6.10: *Sample distribution of correlation for a sample of* $10^3$. *The correlation exists for a bivariate T distribution (exponent* $\frac{2}{3}$, *correlation* $\frac{3}{4}$ *) but... not useable.*

An example of the distribution of correlation is shown in Fig. 6.10. Finite correlation doesn't mean low variance: it exists, but may not be useful for statistical purpose owing to the noise and slow convergence.

## 6.7 FAT TAILED RESIDUALS IN LINEAR REGRESSION MODELS

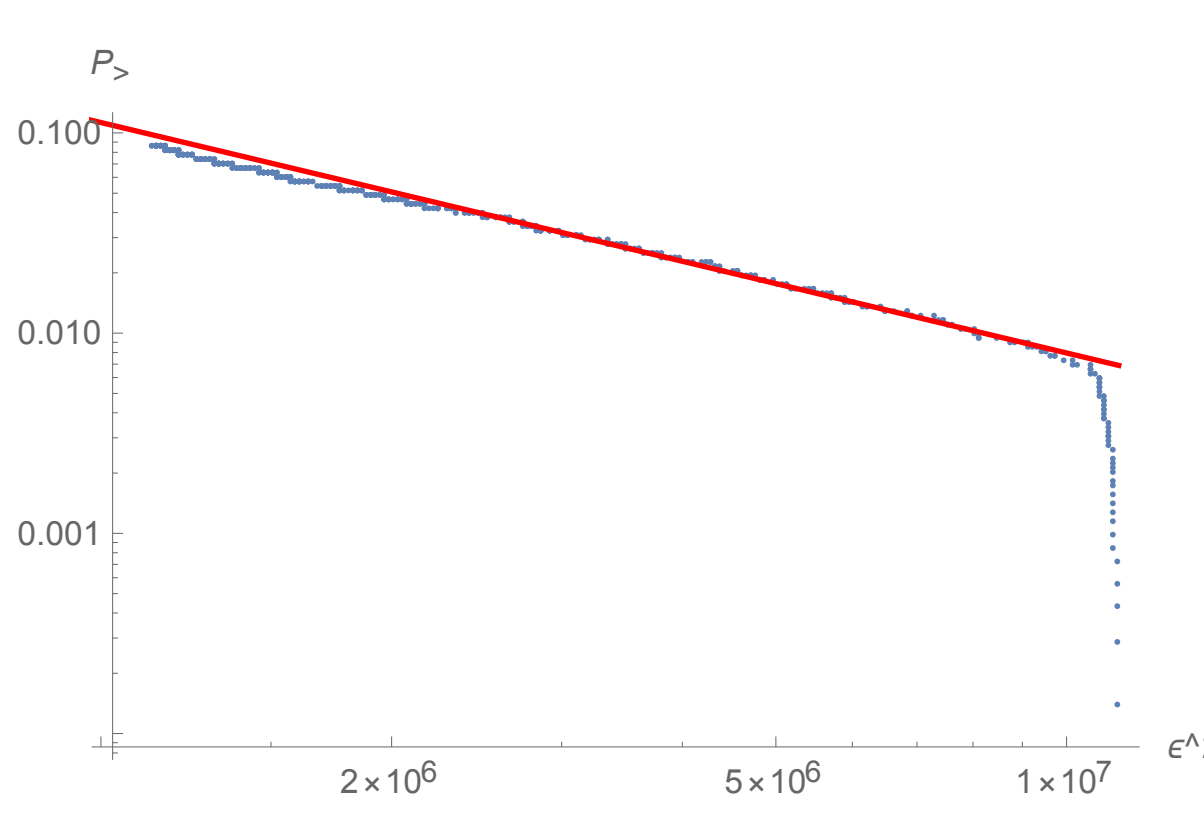

Figure 6.11: *The log-log-plot of the survival function of the squared residuals* $\epsilon^2$ *for the IQ-income linear regression using the standard Winsconsin Longitudinal Studies (WLS) data. We notice that the income variables are winsorized. Clipping the tails creates the illusion of a high* $R^2$. *Actually, even without clipping the tail, the coefficient of determination will show much higher values owing to the small sample properties for the variance of a power law.*

We mentioned in Chapter 3 that linear regression fails to inform under fat tails. Yet it is practiced. For instance, it is patent that income and wealth variables are power law distributed (with a spate of problems, see our Gini discussions in 15). However IQ scores are Gaussian (seemingly by design). Yet people regress one on the other failing to see that it is improper.

Consider the following linear regression in which the independent and independent are of different classes:

$$Y = aX + b + \epsilon,$$

where $X$ is standard Gaussian ($\mathcal{N}(0,1)$) and $\epsilon$ is power law distributed, with $\mathbb{E}(\epsilon) = 0$ and $\mathbb{E}(\epsilon^2) < +\infty$. There are no restrictions on the parameters.

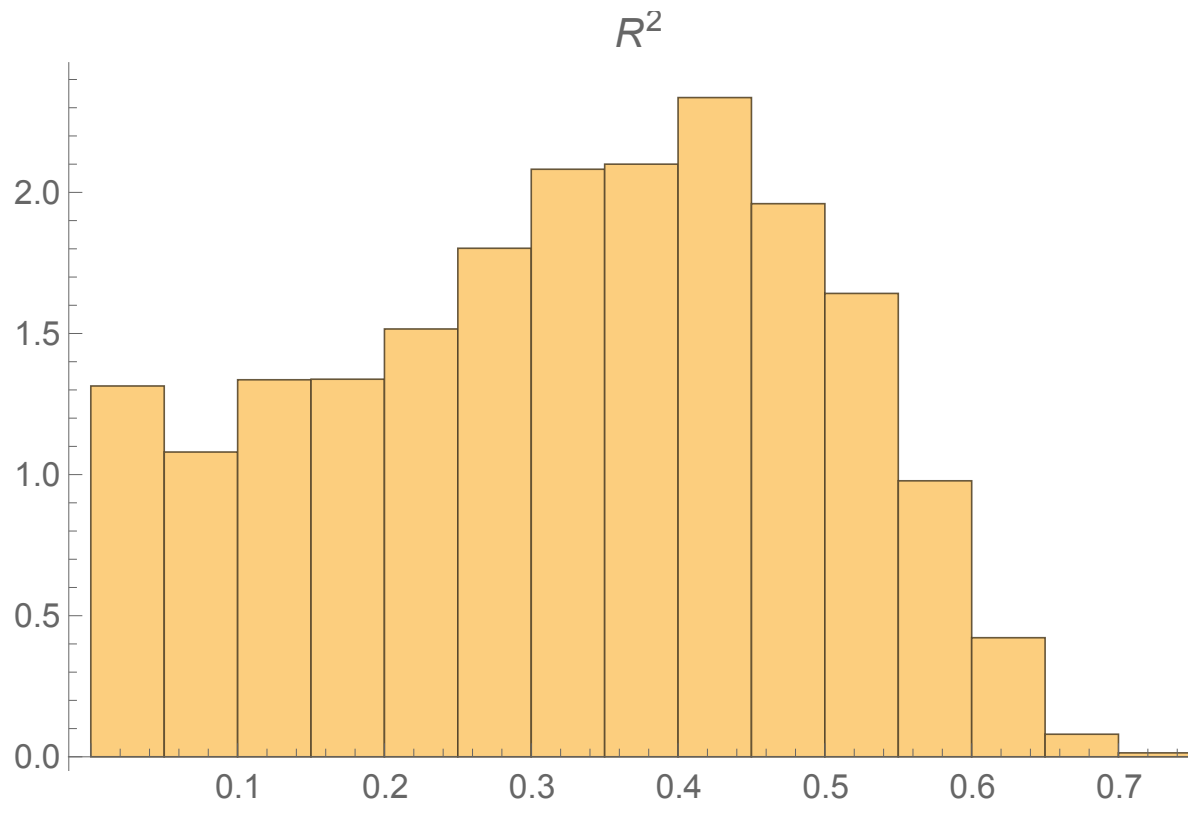

Figure 6.12: *An infinite variance case that shows a high $R^2$ in sample; but it ultimately has a value of 0. Remember that $R^2$ is stochastic. The problem greatly resembles that of P values in Chapter 22 owing to the complication of a metadistribution in* $[0,1]$.

Clearly we can compute the coefficient of determination $R^2$ as 1 minus the ratio of the expectation of the sum of residuals over the total squared variations, so we get the more general answer to our idiosyncratic model. Since $X \sim \mathcal{N}(0,1)$, $aX+b \sim \mathcal{N}(b,|a|)$, we have

$$R^2 = 1 - \frac{SS_{res}}{SS_{tot}} = 1 - \frac{\sum_{i=1}^{n}\left(y_i - (ax_i+b)\right)^2}{\sum_{i=1}^{n}\left(y_i - \overline{y}\right)^2}.$$

We can show that, for large $n$

$$R^2 = \frac{a^2}{a^2 + \mathbb{E}(\epsilon_i^2)} + O\left(\frac{1}{n^2}\right). \tag{6.13}$$

And of course, for infinite variance:

$$\underset{E(\epsilon^2)\to+\infty}{lim} \mathbb{E}(R^2) = 0.$$

When $\epsilon$ is T-distributed with $\alpha$ degrees of freedom, clearly $\epsilon^2$ will follow an FRatio distribution $(1,\alpha)$ –a power law with exponent $\frac{\alpha}{2}$.

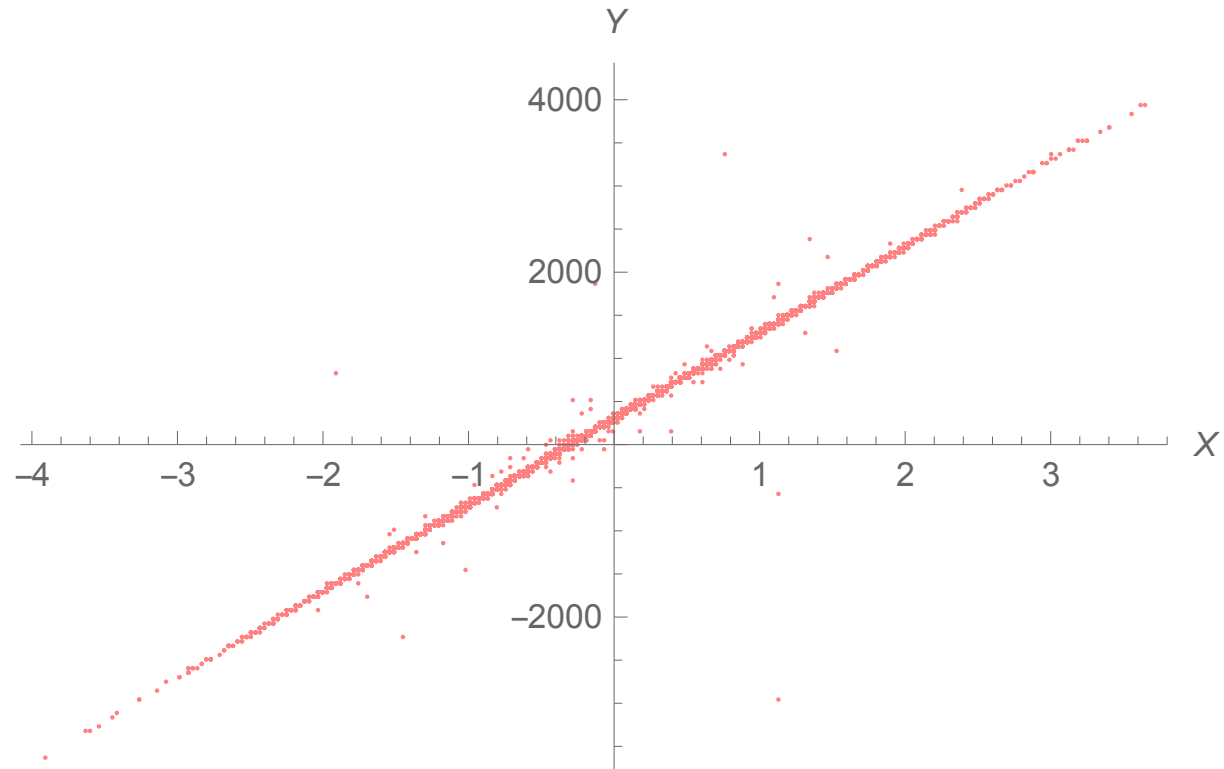

Figure 6.13: *A Cauchy regression with an expected $R^2 = 0$, faking it but showing higher values in small samples (here .985).*

Note that we can also compute the same "expectation" by taking, simply, the square of the correlation between $X$ and $Y$. For instance, assume the distribution for $\epsilon$ is the Student T distribution with zero mean, scale $\sigma$ and tail exponent $\alpha > 2$ (as we saw earlier, we get identical results with other ones so long as we constrain the mean to be 0). Let's start by computing the correlation: the numerator is the covariance $Cov(X, Y) = \mathbb{E}\left((aX + b + \epsilon)X\right) = a$. The denominator (standard deviation for $Y$) becomes $\sqrt{\mathbb{E}\left(((aX + \epsilon) - a)^2\right)} = \sqrt{\frac{2\alpha a^2 - 4a^2 + \alpha\sigma^2}{\alpha - 2}}$. So

$$\mathbb{E}(R^2) = \frac{a^2(\alpha - 2)}{2(\alpha - 2)a^2 + \alpha\sigma^2} \tag{6.14}$$

And the limit from above:

$$\lim_{\alpha \to 2^+} \mathbb{E}(R^2) = 0.$$

We are careful here to use $\mathbb{E}(R^2)$ rather than the seemingly deterministic $R^2$ because it is a stochastic variable that will be extremely sample dependent, and only stabilize for large $n$, perhaps even astronomically large $n$. Indeed, recall that *in sample* the expectation will always be finite, even if the $\epsilon$ are Cauchy! The point is illustrated in Figures 6.12 and 6.13. Actually, when one uses the maximum likelihood estimation of $R^2$ via $\mathbb{E}\left(\epsilon^2\right)$ using $\alpha$, (the "shadow mean" method in Chapters 15 and 16, among others) we notice that in the IQ example used in the graph, the mean of the sample residuals are about half of the maximum likelihood one, making $R^2$ even lower (that is, virtually 0)[4].

The point invalidates much studies of the relations IQ-wealth and IQ-income of the kind [323]; we can see the striking effect in Figure 6.11. Given that $R$ is bounded in $[0, 1]$, it will reach its true value very slowly – see the P-Value problem in Chapter 22.

**Property 6.1**

*When a fat tailed random variable is regressed against a thin tailed one, the coefficient of determination $R^2$ will be biased higher, and requires a much larger sample size to converge (if it ever does).*

Note that sometimes people try to solve the problem by some nonlinear transformation of a random variable (say, the logarithm) to try to establish a linear relationship. If the required transformation is exact, things will be fine –but only if exact. Errors can arise from the discrepancy. For correlation is extremely delicate and unlike mutual information, non-additive and often uninformative. The point has been explored by this author in [284].

4 $2.2\ 10^9$ vs $1.24\ 10^9$.

## NEXT

We will examine in chapter 8 the slow convergence of power laws distributed variables under the law of large numbers (LLN): it can be as much as $10^{13}$ times slower than the Gaussian.

# A | SPECIAL CASES OF THICK TAILS

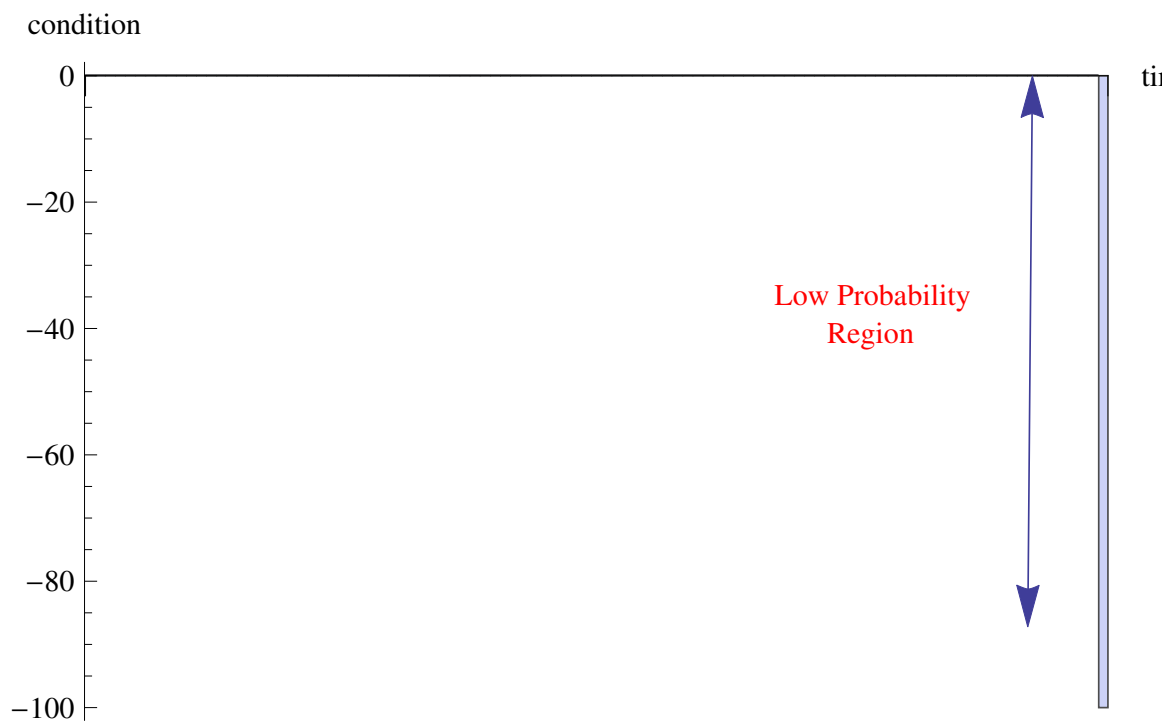

Figure A.1: *A coffee cup is less likely to incur "small" than large harm. It shatters, hence is exposed to (almost) everything or nothing. The same type of payoff is prevalent in markets with, say, (reval)devaluations, where small movements beyond a barrier are less likely than larger ones.*

For unimodal distributions, thick tails are the norm: one can look at tens of thousands of time series of the socio-economic variables without encountering a single episode of "platykurtic" distributions. But for multimodal distributions, some surprises can occur.

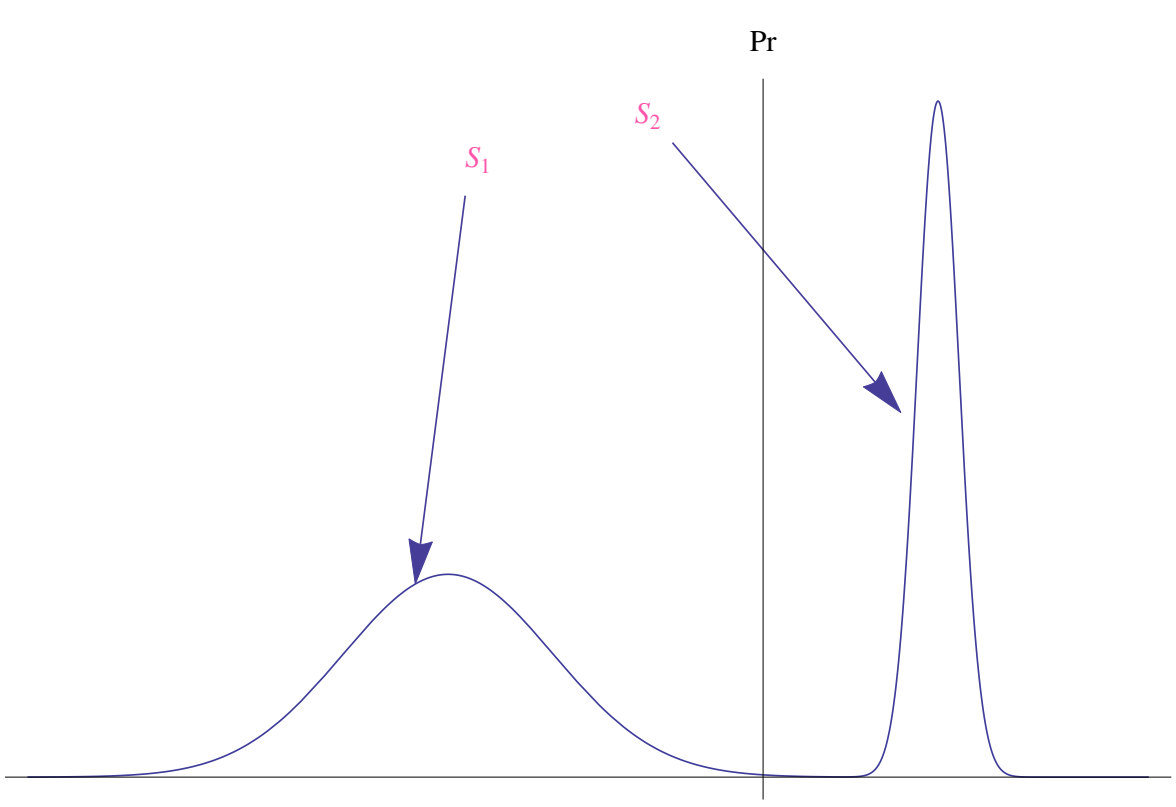

Figure A.2: *The War and peace model. Kurtosis =1.7, much lower than the Gaussian.*

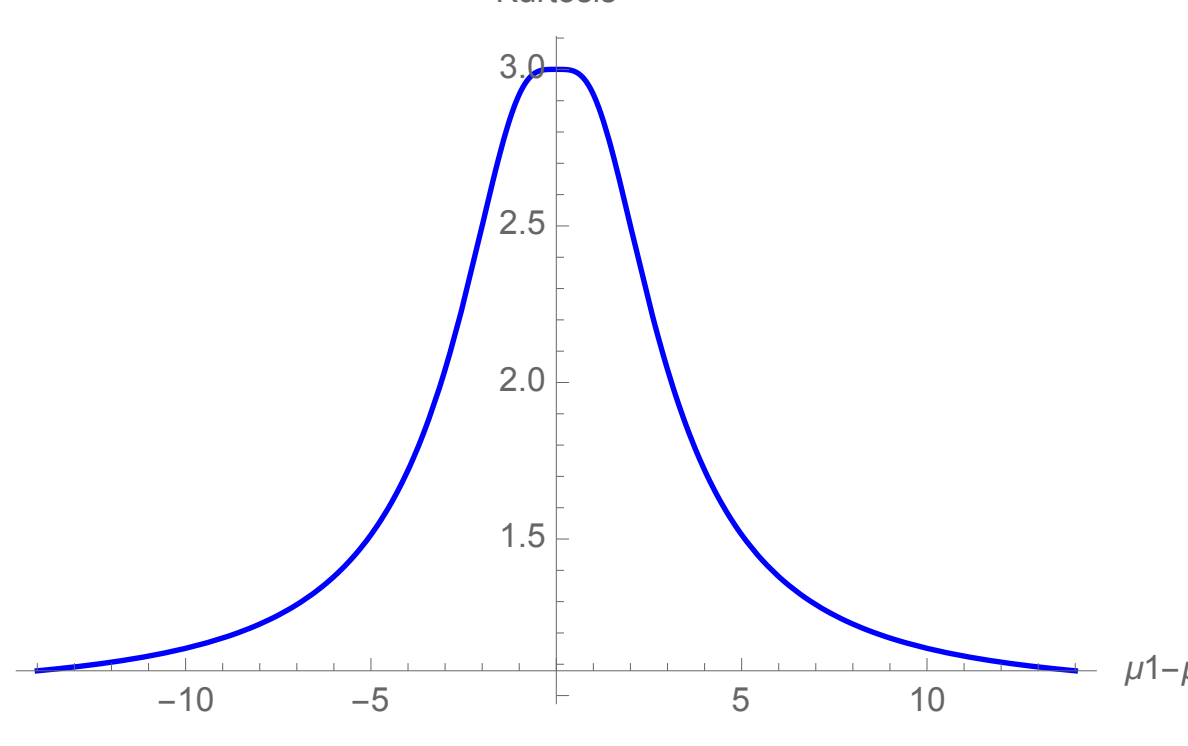

Figure A.3: *Negative (relative) kurtosis and bimodality (3 is the Gaussian).*

## A.1 MULTIMODALITY AND THICK TAILS, OR THE WAR AND PEACE MODEL

We noted earlier in 4.1 that stochasticizing (that is, making a deterministic variable stochastic), ever so mildly, variances, the distribution gains in thick tailedness (as expressed by kurtosis). But we maintained the same mean.

But should we stochasticize the mean as well (while preserving the initial average), and separate the potential outcomes wide enough, so that we get many modes, the "kurtosis" (as measured by the fourth moment) would drop. And if we associate different variances with different means, we get a variety of "regimes", each with its set of probabilities.

Either the very meaning of "thick tails" loses its significance under multimodality, or takes on a new one where the "middle", around the expectation ceases to matter.[10, 185].

Now, there are plenty of situations in real life in which we are confronted to many possible regimes, or states. Assuming finite moments for all states, consider the following structure: $s_1$ a calm regime, with expected mean $m_1$ and standard deviation $\sigma_1$, $s_2$ a violent regime, with expected mean $m_2$ and standard deviation $\sigma_2$, or more such states. Each state has its probability $p_i$.

Now take the simple case of a Gaussian with switching means and variance: with probability $\frac{1}{2}$, $X \sim \mathcal{N}(\mu_1, \sigma_1)$ and with probability $\frac{1}{2}$, $X \sim \mathcal{N}(\mu_2, \sigma_2)$. The kurtosis will be

$$Kurtosis = 3 - \frac{2\left((\mu_1 - \mu_2)^4 - 6\left(\sigma_1^2 - \sigma_2^2\right)^2\right)}{\left((\mu_1 - \mu_2)^2 + 2\left(\sigma_1^2 + \sigma_2^2\right)\right)^2} \tag{A.1}$$

As we see the kurtosis is a function of $d = \mu_1 - \mu_2$. For situations where $\sigma_1 = \sigma_2$, $\mu_1 \neq \mu_2$ , the kurtosis will be below that of the regular Gaussian, and our measure will naturally be negative. In fact for the kurtosis to remain at 3,

$$|d| = \sqrt[4]{6}\sqrt{\max(\sigma_1, \sigma_2)^2 - \min(\sigma_1, \sigma_2)^2},$$

the stochasticity of the mean offsets the stochasticity of volatility.

Assume, to simplify a one-period model, as if one was standing in front of a discrete slice of history, looking forward at outcomes. (Adding complications (transition matrices between different regimes) doesn't change the main result.)

The characteristic function $\phi$(t) for the mixed distribution becomes:

$$\phi(t) = \sum_{i=1}^{N} p_i e^{-\frac{1}{2}t^2\sigma_i^2 + itm_i}$$

For $N = 2$, the moments simplify to the following:

$$M_1 = p_1 m_1 + (1 - p_1)\, m_2$$
$$M_2 = p_1 \left(m_1^2 + \sigma_1^2\right) + (1 - p_1)\left(m_2^2 + \sigma_2^2\right)$$
$$M_3 = p_1 m_1^3 + (1 - p_1)\, m_2 \left(m_2^2 + 3\sigma_2^2\right) + 3 m_1 p_1 \sigma_1^2$$
$$M_4 = p_1 \left(6 m_1^2 \sigma_1^2 + m_1^4 + 3\sigma_1^4\right) + (1 - p_1)\left(6 m_2^2 \sigma_2^2 + m_2^4 + 3\sigma_2^4\right)$$

Let us consider the different varieties, all characterized by the condition $p_1 < (1 - p_1)$, $m_1 < m_2$, preferably $m_1 < 0$ and $m_2 > 0$, and, at the core, the central property: $\sigma_1 > \sigma_2$.

**Variety 1: War and Peace.** Calm period with positive mean and very low volatility, turmoil with negative mean and extremely high volatility.

**Variety 2: Conditional deterministic state** Take a bond $B$, paying interest $r$ at the end of a single period. At termination, there is a high probability of getting $B(1 + r)$, and a small a possibility of default. Getting exactly $B$ is very unlikely. Think that there are no intermediary steps between war and peace: these are separable and discrete states. Bonds don't just default "a little bit". Note the divergence, the probability of the realization being at or close to the mean is about nil. Typically, $p(\mathbb{E}(x))$ the PDF of the expectation are smaller than at the different means of regimes, so $\mathbb{P}(x = \mathbb{E}(x)) < \mathbb{P}\left(x = m_1\right)$ and $< \mathbb{P}\left(x = m_2\right)$, but in the extreme case (bonds), $\mathbb{P}(x = \mathbb{E}(x))$ becomes increasingly small. The tail event is the realization around the mean.

The same idea applies to currency pegs, as devaluations cannot be "mild", with all-or-nothing type of volatility and low density in the "valley" between the two distinct regimes.

With option payoffs, this bimodality has the effect of raising the value of at-the-money options and lowering that of the out-of-the-money ones, causing the exact opposite of the so-called "volatility smile".

Note the coffee cup has no state between broken and healthy. And the state of being broken can be considered to be an absorbing state (using Markov chains for transition probabilities), since broken cups do not end up fixing themselves.

Nor are coffee cups likely to be "slightly broken", as we see in figure A.1.

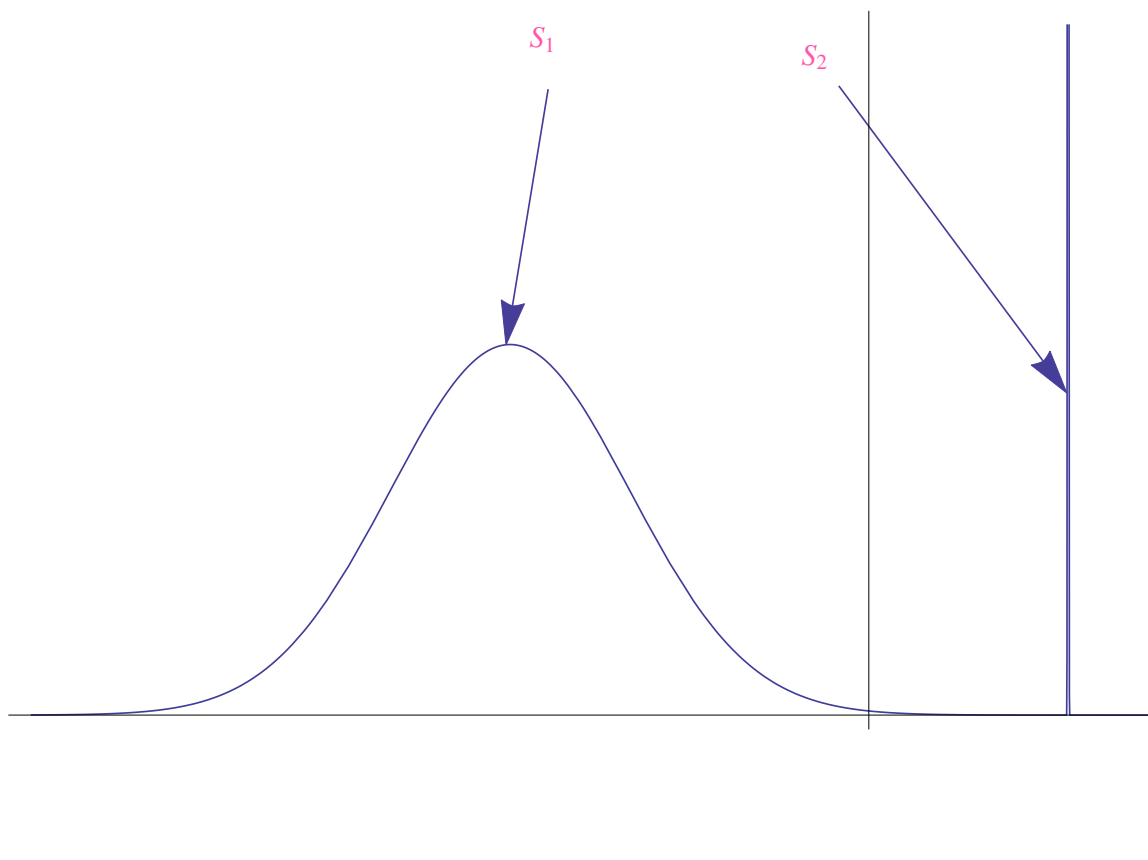

Figure A.4: *The Bond payoff/Currency peg model. Absence of volatility stuck at the peg, deterministic payoff in regime 2, mayhem in regime 1. Here the kurtosis K=2.5. Note that the coffee cup is a special case of both regimes 1 and 2 being degenerate.*

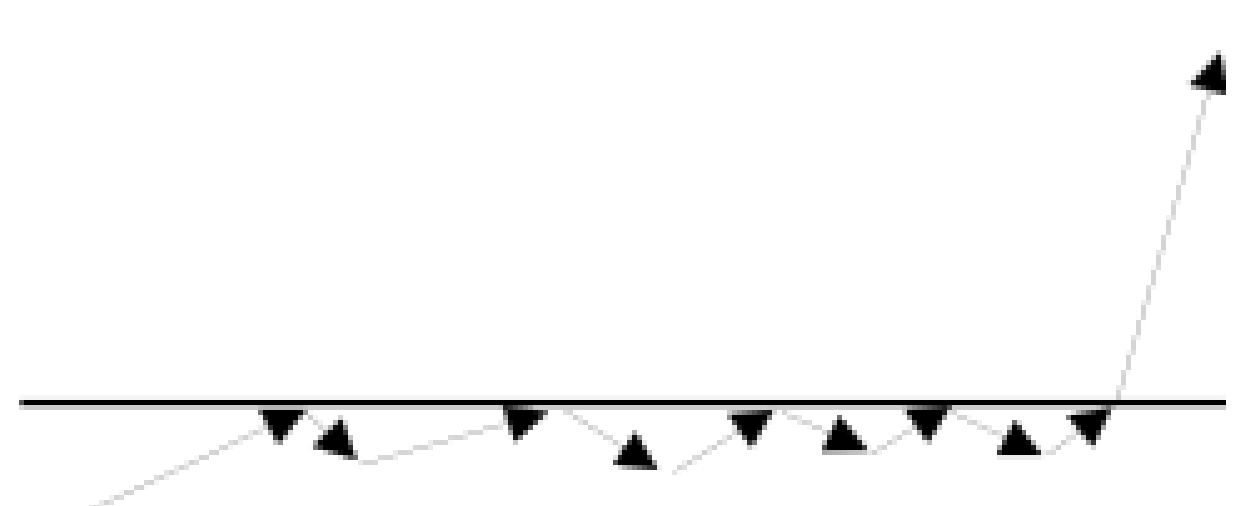

Figure A.5: *Pressure on the peg which may give a Dirac PDF in the "no devaluation" regime (or, equivalently,low volatility). It is typical for finance imbeciles to mistake regime $S_2$ for low volatility.*

**A brief list of other situations where bimodality is encountered:**

1. Currency pegs
2. Mergers
3. Professional choices and outcomes
4. Conflicts: interpersonal, general, martial, any situation in which there is no intermediary between harmonious relations and hostility.
5. Conditional cascades

## A.2 TRANSITION PROBABILITIES: WHAT CAN BREAK WILL BREAK

So far we looked at a single period model, which is the realistic way since new information may change the bimodality going into the future: we have clarity over one-step but not more. But let us go through an exercise that will give us an idea about fragility. Assuming the structure of the model stays the same, we can look at the longer term behavior under transition of states. Let $P$ be the matrix of transition probabilitites, where $p_{i,j}$is the transition from state $i$ to state $j$ over $\Delta$t, (that is, where S(t) is the regime prevailing over period t, $P\left(S(t+\Delta t) = s_j \,\middle|\, S(t) = s_i\right)$)

$$P = \begin{pmatrix} p_{1,1} & p_{1,2} \\ p_{2,1} & p_{2,2} \end{pmatrix}$$

After $n$ periods, that is, $n$ steps,

$$P^n = \begin{pmatrix} a_n & b_n \\ c_n & d_n \end{pmatrix}$$

Where

$$a_n = \frac{(p_{1,1}-1)\,(p_{1,1}+p_{2,2}-1)^{\,n}+p_{2,2}-1}{p_{1,1}+p_{2,2}-2}$$

$$b_n = \frac{(1-p_{1,1})\,((p_{1,1}+p_{2,2}-1)^{\,n}-1)}{p_{1,1}+p_{2,2}-2}$$

$$c_n = \frac{(1-p_{2,2})\,((p_{1,1}+p_{2,2}-1)^{\,n}-1)}{p_{1,1}+p_{2,2}-2}$$

$$d_n = \frac{(p_{2,2}-1)\,(p_{1,1}+p_{2,2}-1)^{\,n}+p_{1,1}-1}{p_{1,1}+p_{2,2}-2}$$

The extreme case to consider is the one with the absorbing state, where $p_{1,1} = 1$, hence (replacing $p_{i,\neq i|i=1,2} = 1 - p_{i,i}$).

$$P^n = \left( \begin{array}{cc} 1 & 0 \\ 1 - p_{2,2}^N & p_{2,2}^N \end{array} \right)$$

and the "ergodic" probabilities:

$$\lim_{n\to\infty} P^n = \left( \begin{array}{cc} 1 & 0 \\ 1 & 0 \end{array} \right)$$

The implication is that the absorbing state regime 1, $S(1)$ will end up dominating with probability 1: what can break and is irreversible will eventually break.

With the "ergodic" matrix,

$$\lim_{n\to\infty} P^n = \pi.\mathbf{1}^{\top}$$

where $\mathbb{1}^{\top}$ is the transpose of unitary vector $\{1,1\}$, $\pi$ the matrix of eigenvectors.

The eigenvalues become $\lambda = \left( \begin{array}{c} 1 \\ p_{1,1} + p_{2,2} - 1 \end{array} \right)$ and associated eigenvectors $\pi = \left( \begin{array}{cc} 1 & 1 \\ \frac{1-p_{1,1}}{1-p_{2,2}} & 1 \end{array} \right)$.

# B | FAT TAILS FROM MAXIMUM ENTROPY

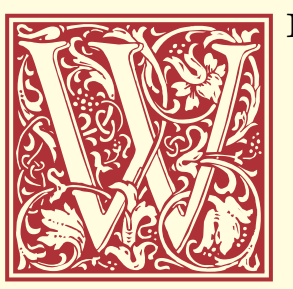

WHERE do fat tails come from? There are many ways to generate power law distributions, with theories (discussed intuitively in *The Black Swan* mostly from a relaxation of the independence in successive random variables), such as that of the preferential attachment (more leads to more) or the Matthew Effect (the rich gets richer). Also, for a standard stochastic process, it emerges naturally from the time between re-visits to the origin, as it relates to the Lindy Effect (Chapter 23).

An Entropy approach is more adapted to the reasoning in this book. We show the derivation of how a certain probability distribution emerges naturally when one maximizes entropy, physical or epistemic, that is, ignorance, under constraints.

The maximum Shannon entropy distribution under specified (or known) mean and central variance, that is first and second moment, $\mathbb{E}(X)$ and $\mathbb{E}(X^2)$ is the Gaussian. We will show detailed derivations.

Shannon Entropy is additive. For nonextensive systems (e.g., systems with long-range interactions, memory effects, and/or fractal structures), additivity fails –the same for the Kahneman-Tversky value function (pseudo-utility) used in Prospect Theory. So, a generalized entropy, that is Tsallis entropy, is presented, which shows us how fat tailed distributions emerge naturally from it.

A more research-oriented discussion is in Chapter 30 where we discuss the maximum entropy distribution under tail risk constraints.

In this discussion we assume continuous probability distributions.

## B.1 SHANNON/BOLTZMAN-GIBBS ENTROPY

We start with differential entropy for a continuous distribution. Let $p(.)$ be a probability density function from a distribution with support $\mathcal{D}$ in $\mathbb{R}$.

$$\mathcal{H}(X) = -\int_{\mathcal{D}} p(x) \log p(x)\, dx\ . \tag{B.1}$$

We need to maximize $\mathcal{H}(X)$ under constraints, such as the required normalization:

$$\int_{\mathcal{D}} p(x)\, dx = 1,$$

and for $j \geq 1$, $f_j(x)$ a function of the random variable, and $c_j$ a constant,

$$\int_{\mathcal{D}} f_j(x) p(x)\, dx = c_j$$

Using the calculus of variation approach, we construct the functional with $\lambda_i$ as Lagrange multipliers, by the standard Euler-Lagrange equation:

$$J\left[p(x)\right] = \int_{\mathcal{D}} p(x) \log p(x)\, dx - \lambda_0 \left( \int_{\mathcal{D}} p(x)\, dx - 1 \right) - \sum_{i=1}^{n} \lambda_j \left( \int_{\mathcal{D}} f_i(x) p(x)\, dx - c_j \right) \tag{B.2}$$

Now consider the variational derivative $\frac{\delta J[p(x)]}{\delta p(x)}$:

$$\frac{\delta J\left[p(x)\right]}{\delta p(x)} = \log p(x) + 1 - \lambda_0 - \sum_{j=1}^{n} \lambda_j f_j(x); \tag{B.3}$$

the maximum entropy under constraints is obtained for the derivative at 0 because of the concavity of the function, see [61].

### B.1.1 Constraining the Mean and Variance leads to the Gaussian

If the support is on the real line, and constraints are such that:
$f_1 = x$, $c_1 = m$, $f_2 = x^2$, $c2 = \sigma^2 + m^2$, we have

$$\frac{\delta J[p(x)]}{\delta p(x)} = \log(p(x)) - \lambda_0 - \lambda_1 x + 1 - \lambda_2 x^2, \tag{B.4}$$

with solution $p(x) = e^{\lambda_0 + \lambda_2 x^2 + \lambda_1 x - 1}$. Plowing back the constraints $\int p(x) d(x) = 1$, $\int p(x) d(x) = m$, etc., we end up with

$$\frac{\sqrt{\pi} e^{-\frac{\lambda_1^2}{4\lambda_2} + \lambda_0 - 1}}{\sqrt{-\lambda_2}} = 1, m = \frac{\sqrt{\pi} e^{-\frac{\lambda_1^2}{4\lambda_2} + \lambda_0 - 1} \lambda_1}{2\left(-\lambda_2\right)^{3/2}}, m^2 + \sigma^2 = \frac{\sqrt{\pi} e^{-\frac{\lambda_1^2}{4\lambda_2} + \lambda_0 - 1} \left(\lambda_1^2 - 2\lambda_2\right)}{4\left(-\lambda_2\right)^{5/2}}$$

Solving for the system of equations we arrive to:

$$p(x) = \frac{e^{-\frac{(m-x)^2}{2\sigma^2}}}{\sqrt{2\pi\sigma}}.$$

Et voila: This, according to the great Benoit Mandelbrot, explains the emergence of the Gaussian in systems constrained by energy ($x^2$).

### B.1.2 Power Law Constraints

A power law in $[L, \infty]$ emerges from the maximization with, for only constraint, a finite multiplicative outcome, that is, $\mathbb{E}\left(\log(x)\right) + \log(L)$ equals a constant. So we solve for $-\lambda_0 + \log(p(x)) - \lambda_1 \left(\log(x) + \log(L)\right) + 1 = 0$ which yields $e^{\lambda_0 - 1} L^{\lambda_1} x^{\lambda_1}$, and normalizing for $\int_L^\infty p(x) dx = 1$ obtains $\lambda_0 = \log\left(-\lambda_1 - 1\right) - \left(2\lambda_1 + 1\right) \log(L) + 1\}$. With $\lambda_1 = -1 - \alpha$ we obtain the familiar

$$p(x) = \alpha L^\alpha x^{-\alpha - 1}.$$

Finito.

### B.1.3 A Good Epistemic Intuition

Let us ignore all non central constraints, just the normalization (summing to 1). Assume the domain is $[0, 1]$. The maximum distribution is $p(x) = e^{\lambda_0 - 1}$. Solving $\int_0^1 e^{\lambda_0 - 1} dx = 1$ gives $\lambda_0 = 1$, with $p(x) = 1$. This explains why, in the absence of knowledge about a probability distribution, that is, maximum ignorance, it is assumed to be uniform in $[0, 1]$. That applies of course to the distribution of probability itself!

Remarkably, we get the same exact result using the **Probability Integral Transform**. If any random variable $X$ has continuous CDF $F_X$, then $U = F_X(X) \sim \mathcal{U}(0, 1)$ because for any $u \in [0, 1]$, $\mathbb{P}(U \leq u) = \mathbb{P}(F_X(X) \leq u) = \mathbb{P}(X \leq F_X^{-1}(u)) = F_X(F_X^{-1}(u)) = u$. See discussions on probability calibration in Chapters 11 and 13, and section 14.4 where this is used.

## B.2 TSALLIS ENTROPY

For a continuous probability distribution $p(x)$ with support $\mathcal{D}$, the Tsallis Entropy (a generalization of Boltzman-Gibbs entropy, is defined as [309]:

$$S_q[p] = \frac{1}{q-1} \left( 1 - \int_{\mathcal{D}} p(x)^q \, dx \right), \quad q \in \mathbb{R} \tag{B.5}$$

where $p(.)$ is the probability density, $q$ is the entropic index, for which the Boltzman-Gibbs and Shannon versions of entropy become a special case at the limit of $q \to 1$ (L'Hôpital), where entropy is additive. For $q < 1$, it is subadditive, for $q > 1$ it is superadditive.

Non-additivity is explained as follows,[311]: Given two independent systems $A$ and $B$, that is, with joint pdf $p(A, B) = p(A)p(B)$, the Tsallis entropy of the sum is $S_q(A + B) = S_q(A) + S_q(B) + (1 - q)\, S_q(A)S_q(B)$.

We note that $p(x)^q > p(x)$ when $q < 1$ and $< p(x)$ otherwise, meaning it increases small probabilities. [308]

**Definition B.1** (Escort Distribution)
*Given a probability density function $p(x)$ with support $\mathcal{D}$ and a real parameter $q > 0$, the escort distribution of order $q$, $P_q(x)$, is defined as:*

$$P_q(x) = \frac{p(x)^q}{\int_{\mathcal{D}} p(y)^q\, dy}.$$

When $q = 1$, we recover the original distribution: $P_1(x) = p(x)$.

**Definition B.2** (Escort Mean (q-Expectation))
*Given an observable (function) $f(x)$, the* ***escort mean*** *or* ***q-expectation value*** *of $f(x)$ under the distribution $p(x)$ is defined as:*

$$\langle f(x) \rangle_q = \int_{\mathcal{D}} f(x)\, P_q(x)\, dx = \frac{\int_{\mathcal{D}} f(x)\, p(x)^q\, dx}{\int_{\mathcal{D}} p(x)^q\, dx}$$

Now, for the Euler-Lagrange equation, we impose the following constraint on top of normalization: $\langle x \rangle_q = m$.

Allora, the Lagragian:

$$J\left[p(x)\right] = \underbrace{\frac{1 - \int_0^\infty p(x)^q\, dx}{q - 1}}_{A} - \underbrace{\lambda_0 \left( \int_0^\infty p(x)\, dx - 1 \right)}_{B} - \underbrace{\lambda_1 \frac{\int_0^\infty x p(x)^q\, dx}{\int_0^\infty p(x)^q\, dx}}_{C}. \tag{B.6}$$

**Variational Derivative of $C$:** Let: the numerator: $N = \int_0^\infty x p(x)^q\, dx$, the denominator: $D = \int_0^\infty p(x)^q\, dx$

The term is $-\lambda_1 \frac{N}{D}$. We compute the variation of $\frac{N}{D}$, the escort mean.

Perturb $p(x) \to p(x) + \epsilon \delta p(x)$.

- **Numerator variation:**

$$p(x)^q \to (p(x) + \epsilon \delta p(x))^q \approx p(x)^q + \epsilon q p(x)^{q-1} \delta p(x) + O(\epsilon^2)$$

$$\Rightarrow \int_0^\infty x(p(x)^q + \epsilon q p(x)^{q-1} \delta p(x))\, dx = N + \epsilon q \int_0^\infty x p(x)^{q-1} \delta p(x)\, dx$$

- **Denominator variation:**

$$D \to \int_0^\infty (p(x)^q + \epsilon q p(x)^{q-1} \delta p(x))\, dx = D + \epsilon q \int_0^\infty p(x)^{q-1} \delta p(x)\, dx$$

The variation of the fraction $\frac{N}{D}$:

$$\frac{N + \delta N}{D + \delta D} \approx \frac{N}{D} + \frac{\epsilon q \int_0^\infty x p(x)^{q-1} \delta p(x) \, dx}{D} - \frac{N \epsilon q \int_0^\infty p(x)^{q-1} \delta p(x) \, dx}{D^2}$$

Using $\frac{N+\delta N}{D+\delta D} \approx \frac{N}{D} + \frac{\delta N D - N \delta D}{D^2}$:

$$\delta N = \epsilon q \int_0^\infty x p(x)^{q-1} \delta p(x) \, dx$$

$$\delta D = \epsilon q \int_0^\infty p(x)^{q-1} \delta p(x) \, dx$$

$$\delta \left( \frac{N}{D} \right) = \epsilon q \left( \int_0^\infty x p(x)^{q-1} \delta p(x) \, dx \cdot \frac{D - \frac{N}{D} \int_0^\infty p(x)^{q-1} \delta p(x) \, dx}{D^2} \right)$$

$$= \epsilon q \int_0^\infty \left( x - \frac{N}{D} \right) p(x)^{q-1} \delta p(x) \, dx$$

Since $D = \int_0^\infty p(x)^q \, dx$ and $N = \int_0^\infty x p(x)^q \, dx$, we have:

$$\frac{xD - N}{D^2} = \frac{x \int_0^\infty p(y)^q \, dy - \int_0^\infty y p(y)^q \, dy}{D^2} = \frac{1}{D} \left( x - \frac{\int_0^\infty y p(y)^q \, dy}{\int_0^\infty p(y)^q \, dy} \right) = \frac{x - \langle x \rangle_q}{D}$$

$$\frac{\delta}{\delta p(x)} \left( \frac{N}{D} \right) = q \frac{x - \langle x \rangle_q}{D} p(x)^{q-1}$$

For the full term:

$$\frac{\delta}{\delta p(x)} \left( -\lambda_1 \frac{N}{D} \right) = -\lambda_1 q \frac{x - \langle x \rangle_q}{\int_0^\infty p(y)^q \, dy} p(x)^{q-1}$$

$$\frac{\delta \mathcal{J}[p(x)]}{\delta p(x)} = -q p(x)^{q-1} \left( \frac{1}{q-1} + \lambda_1 \frac{x - \langle x \rangle_q}{\int_0^\infty p(y)^q \, dy} \right) - \lambda_0$$

**Solving For Maximum Entropy** To find the extremum of the functional $\mathcal{J}[p(x)]$, we set the variational derivative to zero:

$$\frac{\delta \mathcal{J}[p(x)]}{\delta p(x)} = 0 \quad \Rightarrow \quad -q \, p(x)^{q-1} \left( \frac{1}{q-1} + \lambda_1 \frac{x - \langle x \rangle_q}{\int_0^\infty p(y)^q \, dy} \right) - \lambda_0 = 0. \tag{B.7}$$

Solving for $p(x)^{q-1}$, we obtain:

$$p(x)^{q-1} = -\frac{\lambda_0}{q} \left( \frac{1}{q-1} + \lambda_1 \frac{x - \langle x \rangle_q}{\int_0^\infty p(y)^q \, dy} \right)^{-1}. \tag{B.8}$$

Raising both sides to the power $\frac{1}{q-1}$, and denoting $D = \int_0^\infty p(y)^q \, dy$, we have:

$$p(x) = \left[ -\frac{\lambda_0}{q} \left( \frac{1}{q-1} + \lambda_1 \frac{x - \langle x \rangle_q}{D} \right)^{-1} \right]^{\frac{1}{q-1}}. \tag{B.9}$$

Let us reparametrize using the standard notation:

$$\begin{aligned} \mu_q &:= \langle x \rangle_q, \\ \beta &:= \lambda_1 / D, \\ C_q &:= \text{normalization constant.} \end{aligned}$$

Then we define the $q$-exponential function as:

**Definition B.3** (q-Exponential Function)

$$\exp_q(-\beta(x - \mu_q)) := \begin{cases} \left[1 - (1-q)\beta(x - \mu_q)\right]^{1/(1-q)} & \textit{if } 1 - (1-q)\beta(x - \mu_q) > 0, \\ 0 & \textit{otherwise.} \end{cases}$$

Thus, the maximum entropy distribution under the escort mean constraint is:

$$p(x) = C_q \exp_q\left(-\beta(x - \mu_q)\right). \tag{B.10}$$

This is the $q$-exponential distribution, which generalizes the exponential law:

- For $q \to 1$, it recovers the standard exponential distribution.
- For $q > 1$, it has power-law (heavy) tails.
- For $q < 1$, it has compact support.

**Normalization Constant on $[0, \infty)$** To determine $C_q$, we normalize $p(x)$ over $x \in [0, \infty)$. Let us fix $\mu_q = 0$ without loss of generality (by shifting coordinates). Then:

$$p(x) = C_q \left[1 - (1-q)\beta x\right]^{1/(1-q)}, \qquad \text{for } x \in [0, x_{\max}]$$

where the support cutoff $x_{\max}$ is:

$$x_{\max} = \begin{cases} \infty & \text{if } q > 1, \\ \frac{1}{(1-q)\beta} & \text{if } q < 1. \end{cases}$$

Then:

$$\int_0^{x_{\max}} p(x) \, dx = 1 \quad \Rightarrow \quad C_q \int_0^{x_{\max}} \left[1 - (1-q)\beta x\right]^{1/(1-q)} dx = 1.$$

$$\frac{C_q\left(\left(\beta(q-1)x_{\max}+1\right)^{\frac{1}{1-q}+1}-1\right)}{\beta(q-2)},\ q \neq 2.$$

We have

$$lim_{x_{max}\to\infty}\frac{C_q\left(\left(\beta(q-1)x_{\max}+1\right)^{\frac{1}{1-q}+1}-1\right)}{\beta(q-2)} = \frac{C_q}{\beta(2-q)};$$

and

$$\frac{C_q\left(\left(\beta(q-1)x_{\max}+1\right)^{\frac{1}{1-q}+1}-1\right)}{\beta(q-2)} /.\, x_{\max} \to \frac{1}{\beta(1-q)} = \frac{C_q}{\beta(2-q)}.$$

So

$$C_q = \beta(2-q), \qquad \text{for } q \neq 2, \tag{B.11}$$

and

$$p(x) = \beta(2-q)(\beta(q-1)x+1)^{\frac{1}{1-q}};$$

finito.

**Recovering the Exponential)** In the limit $q \to 1$, we recover the exponential distribution:

$$p(x) = \beta e^{-\beta x}, \qquad x \in [0,\infty).$$

This is consistent with maximum entropy under a linear constraint for the classical Shannon/Boltzman-Gibbs entropy.

Part II

# THE LAW OF MEDIUM NUMBERS

# 7 | LIMIT DISTRIBUTIONS, A CONSOLIDATION*,†

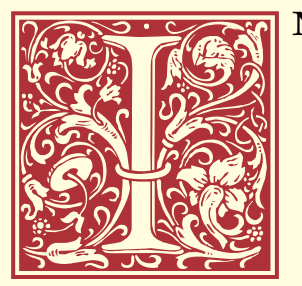

N THIS expository chapter we proceed to consolidate the literature on limit distributions seen from our purpose, with some shortcuts where indicated. After introducing the law of large numbers, we show the intuition behind the central limit theorem and illustrate how it varies preasymptotically across distributions. Then we discuss the law of large numbers as applied to higher moments. A more formal and deeper approach will be presented in the next chapter.

Both the law of large numbers and the central limit theorem are partial answers to a general problem: "What is the limiting behavior of a sum (or average) of random variables as the number of summands approaches infinity?". And our law of medium numbers (or preasymptotics) is: now what when the number of summands doesn't reach infinity?

## 7.1 REFRESHER: THE WEAK AND STRONG LLN

The standard presentation is as follows. Let $X_1, X_2, \ldots$ be an infinite sequence of independent and identically distributed (Lebesgue integrable) random variables with expected value $\mathbb{E}(X_n) = \mu$ (we will see further down one can somewhat relax the i.i.d. assumptions). For all $n$, the sample average $\overline{X}_n = \frac{1}{n}(X_1 + \cdots + X_n)$ converges to the expected value, $\overline{X}_n \rightarrow \mu$ ,for $n \rightarrow \infty$.

Finiteness of variance is not necessary (though of course the finite higher moments accelerate the convergence).

There are two modes of convergence: convergence in probability $\xrightarrow{P}$ (which implies convergence in distribution, though not always the reverse), and the stronger $\xrightarrow{a.s.}$ almost sure convergence (similar to pointwise convergence) (or almost every-

Discussion chapter (with some research).

where or almost always). Applied here the distinction corresponds to the weak and strong LLN respectively.

**The weak LLN** The weak law of large numbers (or Kinchin's law, or sometimes called Bernouilli's law) can be summarized as follows: the probability of a variation in excess of some threshold for the average becomes progressively smaller as the sequence progresses. In estimation theory, an estimator is called consistent if it thus converges in probability to the quantity being estimated.

$$\overline{X}_n \xrightarrow{P} \mu \text{ when } n \to \infty.$$

That is, for any positive number $\varepsilon$,

$$\lim_{n \to \infty} \mathbb{P}\left( |\overline{X}_n - \mu| > \varepsilon \right) = 0.$$

Note that standard proofs are based on Chebyshev's inequality: if $X$ has a finite non-zero variance $\sigma^2$. Then for any real number $k > 0$,

$$\Pr(|X - \mu| \geq k\sigma) \leq \frac{1}{k^2}.$$

**The strong LLN** The strong law of large numbers states that, as the number of summands $n$ goes to infinity, the probability that the average converges to the expectation equals 1.

$$\overline{X}_n \xrightarrow{\text{a.s.}} \mu \text{ when } n \to \infty.$$

That is,

$$\mathbb{P}\left( \lim_{n \to \infty} \overline{X}_n = \mu \right) = 1.$$

**Relaxations of i.i.d.** Now one can relax the identically distributed assumption under some conditions: Kolmogorov's proved that non identical distributions for the summands $X_i$ require for each summand the existence of a finite second moment.

As to independence, some weak dependence is allowed. Traditionally the conditions are, again, the usual finite variance 1) $\mathbb{V}(X_i) \leq c$ and some structure on the covariance matrix, 2) $\lim_{|i-j| \to +\infty} \text{Cov}(X_i, X_j) = 0$.

However it turns out 1) can be weakened to $\sum_{i=1}^{n} \mathbf{V}[X_i] = o(n^2)$, and 2) $|\text{Cov}(X_i, X_j)| \leq \varphi(|i - j|)$, where $\frac{1}{n} \sum_{i=1}^{n} \varphi(i) \to 0$. See Bernstein [22] and Kozlov [176] (in Russian).[2]

[2] Thanking "romanoved", a mysterious Russian speaking helper on Mathematics Stack Exchange.

**Our Interest** Our concern in this chapter and the next one is clearly to look at the "speed" of such convergence. Note that under the stronger assumption of i.i.d. we do not need variance to be finite, so we can focus on mean absolute deviation as a metric for divergence.

## 7.2 CENTRAL LIMIT IN ACTION

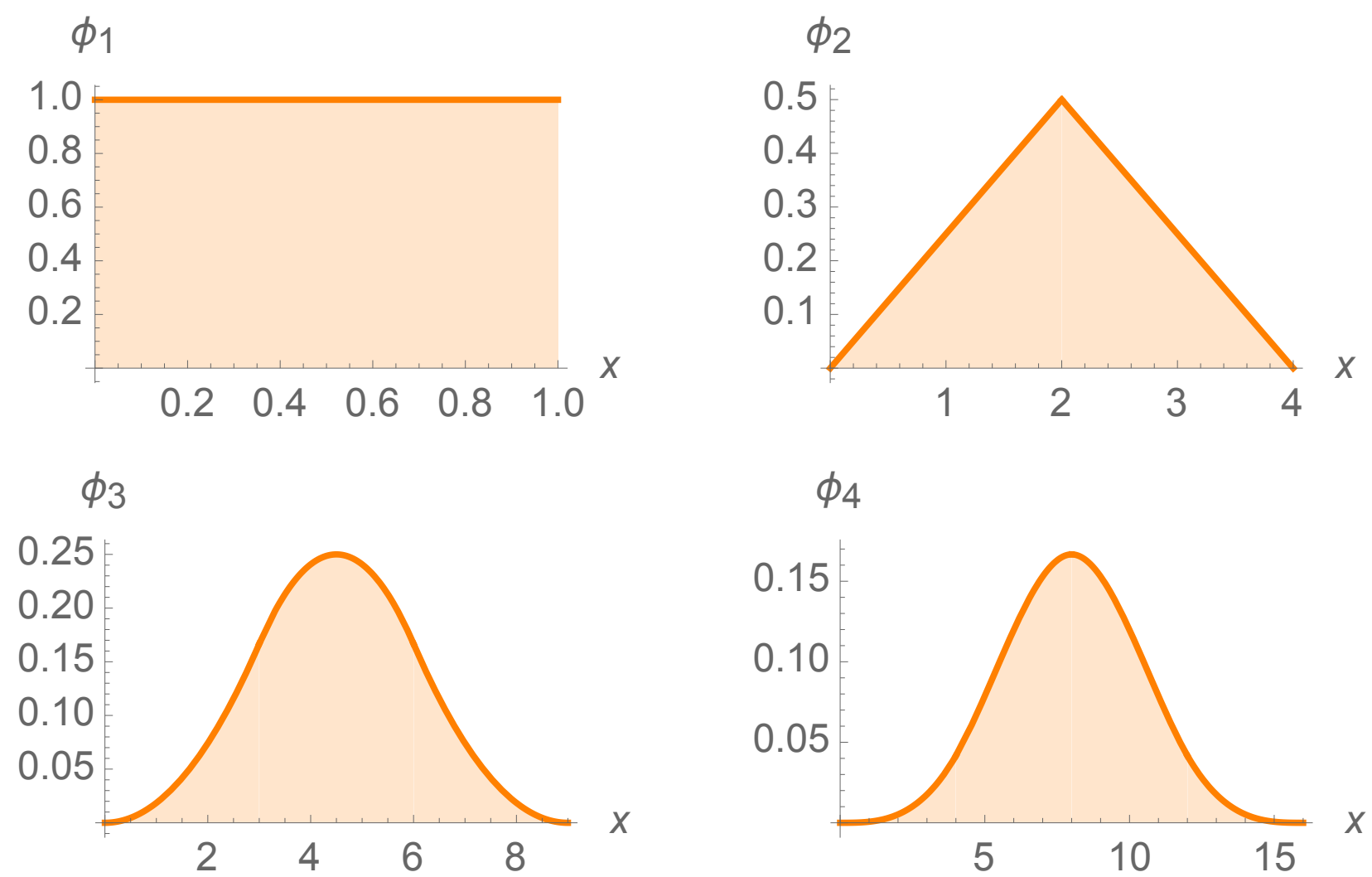

Figure 7.1: *The fastest CLT: the Uniform becomes Gaussian in a few steps. We have, successively, 1, 2, 3, and 4 summands. With 3 summands we see a well formed bell shape. (Note that we are plotting the unweighted sum, not the average.*

We will start with a simplification of the generalized central limit theorem (GCLT), as formulated by Paul Lévy (the traditional approaches to CLT as well as the technical backbone will be presented later):

### 7.2.1 The Stable Distribution

Using the same notation as above, let $X_1, \ldots, X_n$ be independent and identically distributed random variables. Consider their sum $S_n$. We have

$$\frac{Sn - a_n}{b_n} \xrightarrow{D} X_s, \tag{7.1}$$

where $X_s$ follows a stable distribution $\mathcal{S}$, $a_n$ and $b_n$ are norming constants, and, to repeat, $\xrightarrow{D}$ denotes convergence in distribution (the distribution of $X$ as $n \to \infty$). The properties of $\mathcal{S}$ will be more properly defined and explored in the next chapter. Take it for now that a random variable $X_s$ follows a stable (or $\alpha$-stable) distribution, symbolically $X_s \sim S(\alpha_s, \beta, \mu, \sigma)$, if its characteristic function $\chi(t) = \mathbb{E}(e^{itX_s})$ is of the form:

$$\chi(t) = e^{\left(i\mu t - |t\sigma|_s^{\alpha}\left(1 - i\beta \tan\left(\frac{\pi\alpha - s}{2}\right)\operatorname{sgn}(t)\right)\right)} \text{ when } \alpha_s \neq 1. \tag{7.2}$$

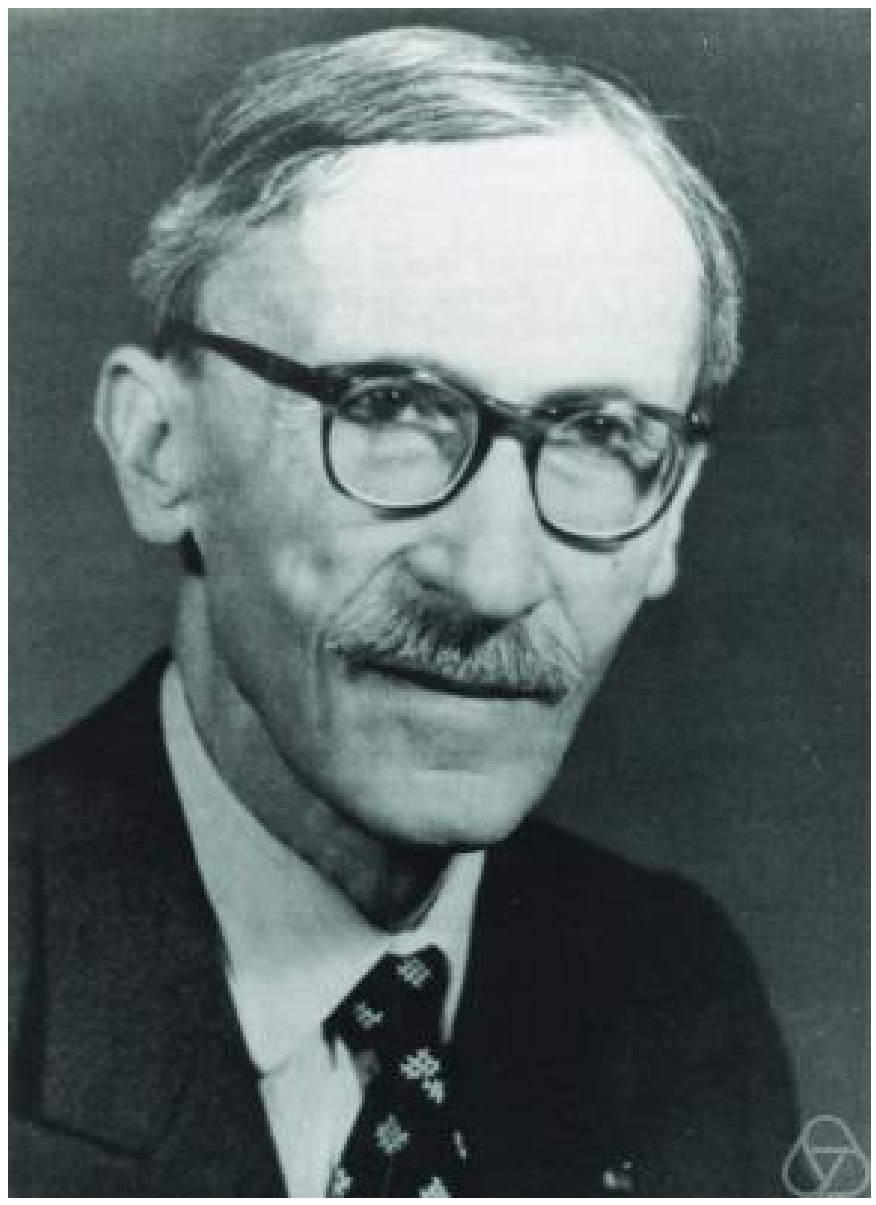

Figure 7.2: *Paul Lévy, 1886-1971, formulated the generalized central limit theorem.*

The constraints are $-1 \leq \beta \leq 1$ and $0 < \alpha_s \leq 2$.[3]

The designation stable distribution implies that the distribution (or class) is stable under summation: you sum up random variables following any the various distributions that are members of the class $\mathfrak{S}$ explained next chapter (actually the same distribution with different parametrizations of the characteristic function), and you stay within the same distribution. Intuitively, $\chi(t)^n$ is the same form as $\chi(t)$ , with $\mu \to n\mu$, and $\sigma \to n^{\frac{1}{\alpha}}\sigma$. The well known distributions in the class (or some people call it a "basin") are: the Gaussian, the Cauchy and the Lévy with $\alpha = 2, 1$, and $\frac{1}{2}$, respectively. Other distributions have no closed form density.[4]

### 7.2.2 The Law of Large Numbers for the Stable Distribution

Let us return to the law of large numbers.

---

3 We will try to use $\alpha_s \in (0, 2]$ to denote the exponent of the limiting and Platonic stable distribution and $\alpha_p \in (0, \infty)$ the corresponding Paretian (preasymptotic) equivalent but only in situations where there could be some ambiguity. Plain $\alpha$ should be understood in context.

4 Actually, there are ways to use special functions; for instance one discovered accidentally by the author: for the Stable $\mathcal{S}$ with standard parameters $\alpha = \frac{3}{2}, \beta = 1, \mu = 0, \sigma = 1$ , $PDF(x) = -\frac{\sqrt[3]{2}e^{\frac{x^3}{27}}\left(\sqrt[3]{3}x\text{Ai}\left(\frac{x^2}{3\, 2^{2/3}\sqrt[3]{3}}\right)+3\sqrt[3]{2}\text{Ai}'\left(\frac{x^2}{3\, 2^{2/3}\sqrt[3]{3}}\right)\right)}{3\, 3^{2/3}}$ used further down in the example on the limit distribution for Pareto sums.

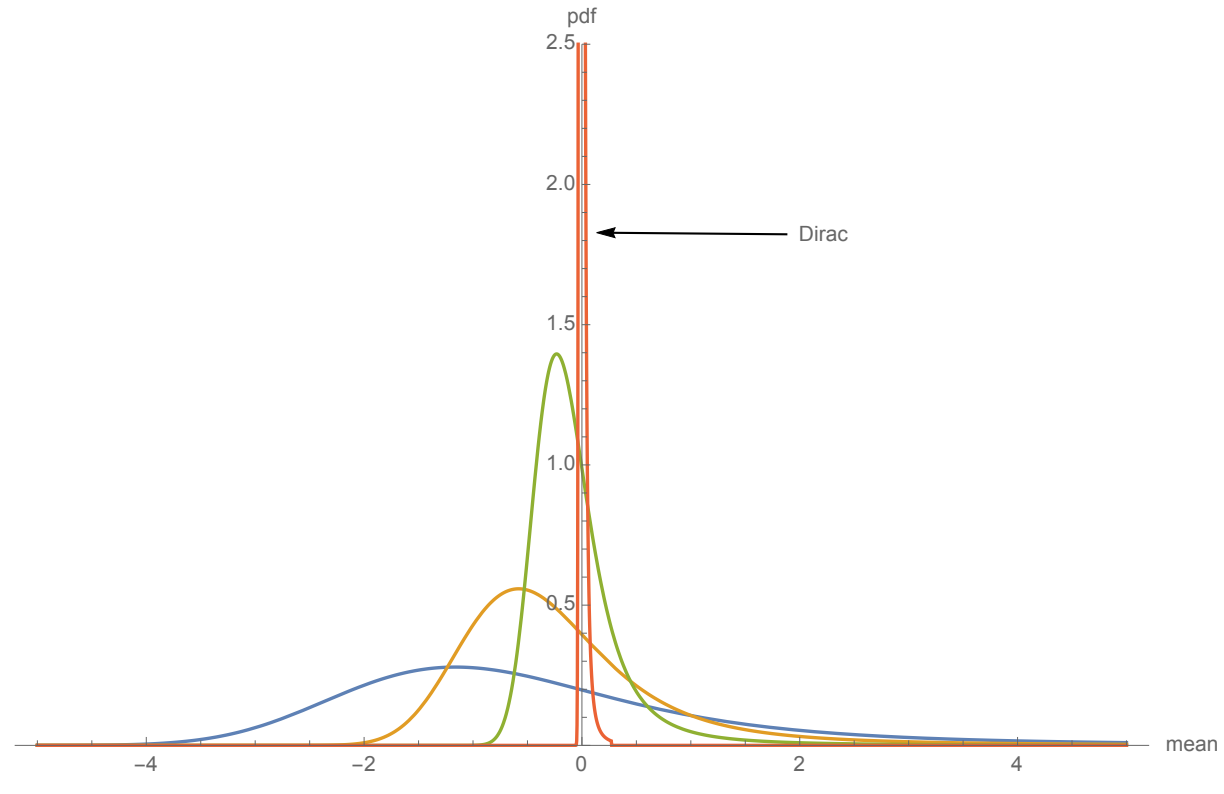

Figure 7.3: *The law of large numbers show a tightening distribution around the mean leading to degeneracy converging to a Dirac stick at the exact mean.*

By standard results, we can observe the law of large numbers at work for the stable distribution, as illustrated in Figure 7.3:

$$\lim_{n\to+\infty} \chi\left(\frac{t}{n}\right)^n = e^{i\mu t},\ 1 < \alpha_s \leq 2 \tag{7.3}$$

which is the characteristic functionof a Dirac delta at $\mu$, a degenerate distribution, since the Fourier transform $\mathcal{F}$ (here parametrized to be the inverse of the characteristic function) is:

$$\frac{1}{\sqrt{2\pi}}\mathcal{F}_t\left(e^{i\mu t}\right)(x) = \delta(\mu + x). \tag{7.4}$$

Further, we can observe the "real-time" operation for all $1 < n < +\infty$ in the following ways, as we will explore in the next sections.

## 7.3 SPEED OF CONVERGENCE OF CLT: VISUAL EXPLORATIONS

We note that if $X$ has a finite variance, the stable-distributed random variable $X_s$ will be Gaussian. But note that $X_s$ is a limiting construct as $n \to \infty$ and there are many, many complication with "how fast" we get there. Let us consider 4 cases that illustrate both the idea of CLT and the speed of it.

### 7.3.1 Fast Convergence: the Uniform Dist.

Consider a uniform distribution –the simplest of all. If its support is in $[0, 1]$, it will simply have a density of $\phi(x_1) = 1$ for $0 \leq x_1 \leq 1$ and integrates to 1. Now add another variable, $x_2$, identically distributed and independent. The sum $x_1 + x_2$ immediately changed in shape! Look at $\phi_2(.)$, the density of the sum in Figure 7.1. It is now a triangle. Add one variable and now consider the density $\phi_3$ of the distribution of $X_1 + X_2 + X_3$. It is already almost bell shaped, with $n = 3$ summands.

The uniform sum distribution

$$\phi_n(x) = \sum_{k=0}^{n} (-1)^k \begin{pmatrix} n \\ k \end{pmatrix} \left( \frac{x-L}{H-L} - k \right)^{n-1} \operatorname{sgn} \left( \frac{x-L}{H-L} - k \right) \text{ for } nL \leq x \leq nH$$

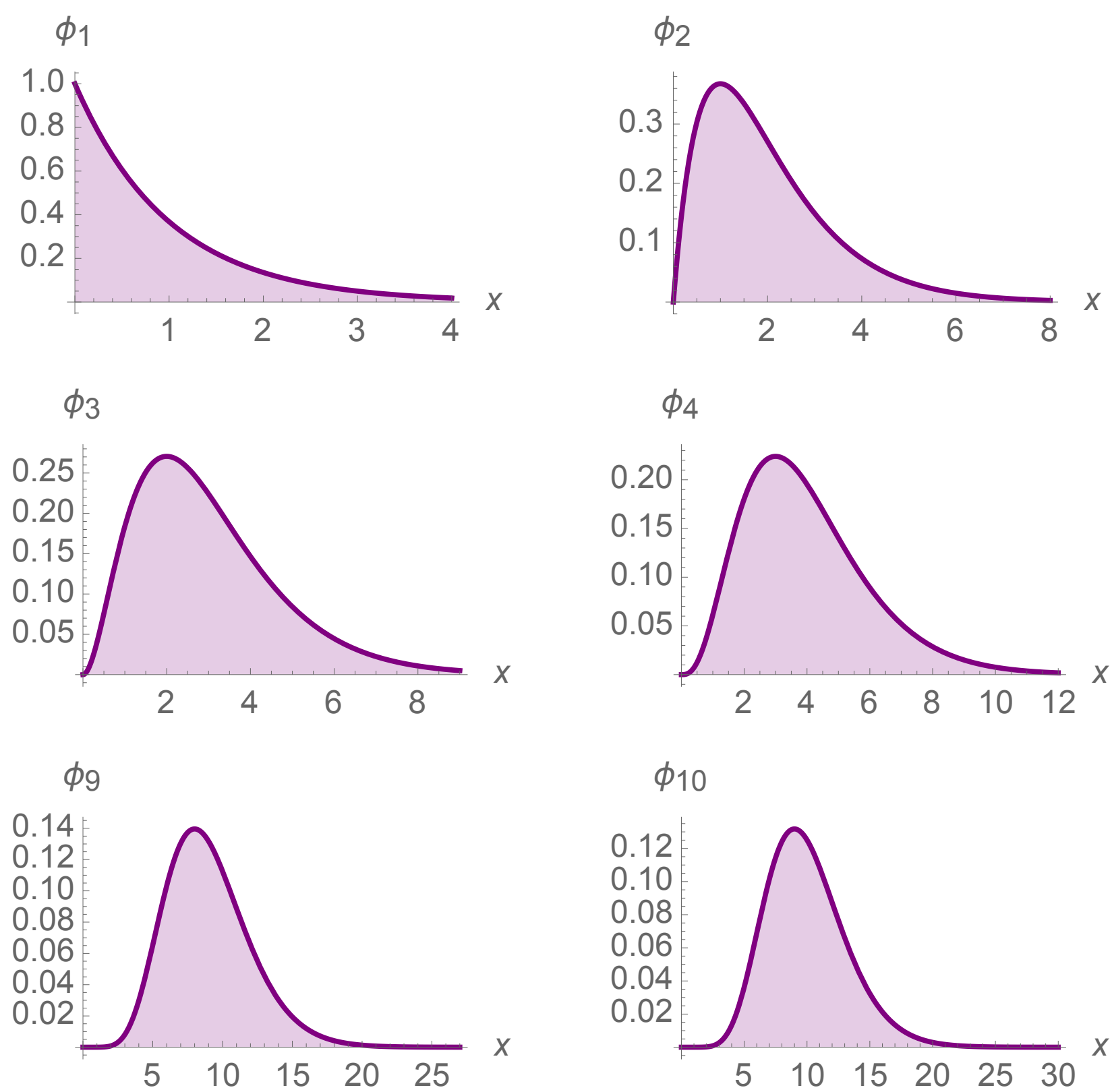

Figure 7.4: *The exponential distribution,$\phi$ indexed by the number of summands. Slower than the uniform, but good enough.*

### 7.3.2 Semi-slow convergence: the exponential

Let us consider a sum of exponential random variables.

We have for initial density

$$\phi_1(x) = \lambda e^{-\lambda x}, \; x \geq 0,$$

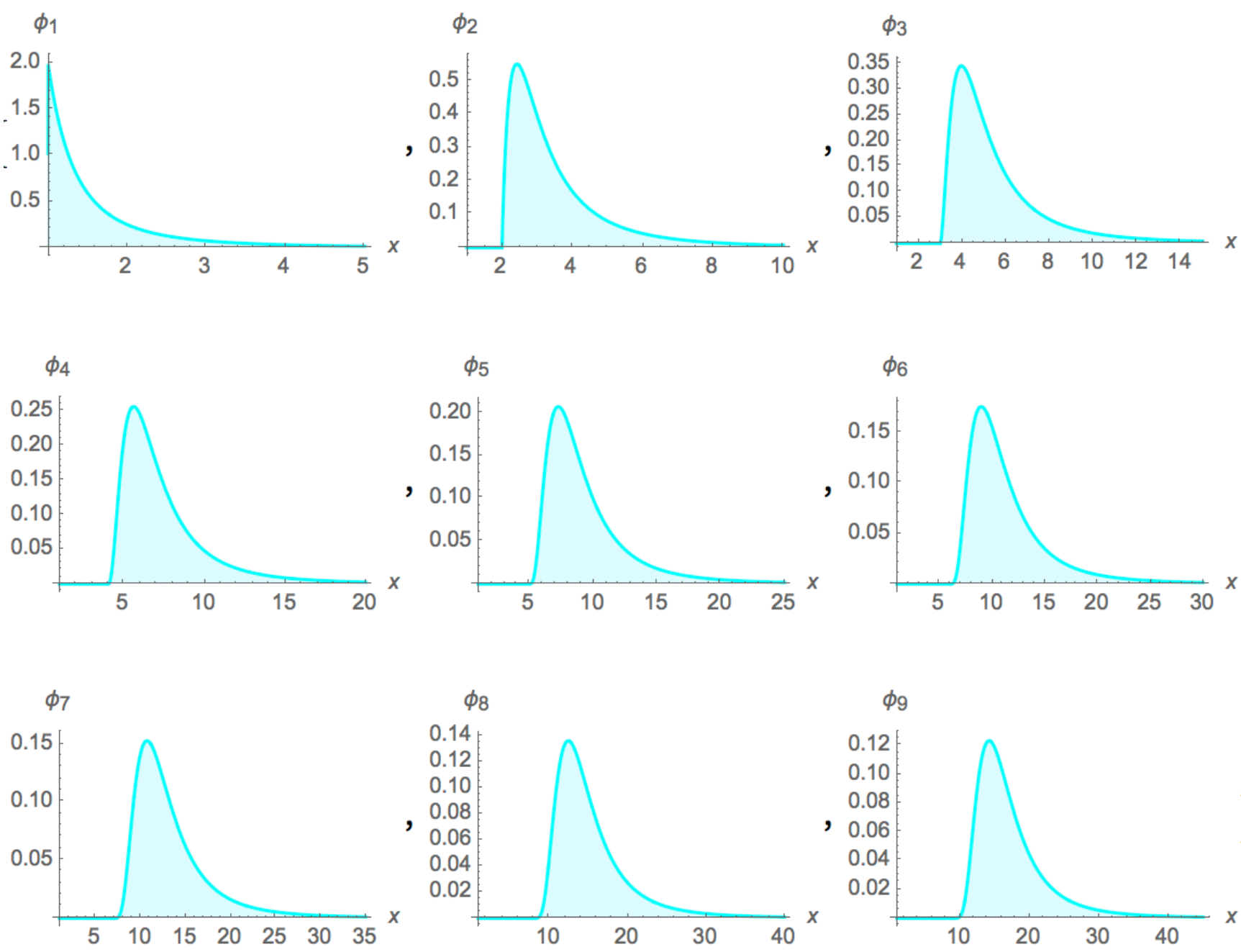

Figure 7.5: *The Pareto distribution. Doesn't want to lose its skewness, although in this case it should converge to the Gaussian... eventually.*

and for $n$ summands[5]

$$\phi_n(x) = \left(\frac{1}{\lambda}\right)^{-n} \frac{x^{n-1}e^{-\lambda x}}{\Gamma(n)}.$$

We have, replacing $x$ by $n/\lambda$ (and later in the illustrations in Fig. 7.4 $\lambda = 1$),

$$\frac{\left(\frac{1}{\lambda}\right)^{-n} x^{n-1} e^{\lambda(-x)}}{\Gamma(n)} \underset{n\to\infty}{\to} \frac{\lambda e^{-\frac{\lambda^2\left(x-\frac{n}{\lambda}\right)^2}{2n}}}{\sqrt{2\pi}\sqrt{n}},$$

which is the density of the normal distribution with mean $\frac{n}{\lambda}$ and variance $\frac{n}{\lambda^2}$.

We can see how we get more slowly to the Gaussian, as shown in Figure 7.4, mostly on account of its skewness. Getting to the Gaussian requires symmetry.

### 7.3.3 The slow Pareto

Consider the simplest Pareto distribution on $[1, \infty)$:

$$\phi_1(x) = 2x^{-3}$$

[5] We derive the density of sums either by convolving, easy in this case, or as we will see with the Pareto, via characteristic functions.

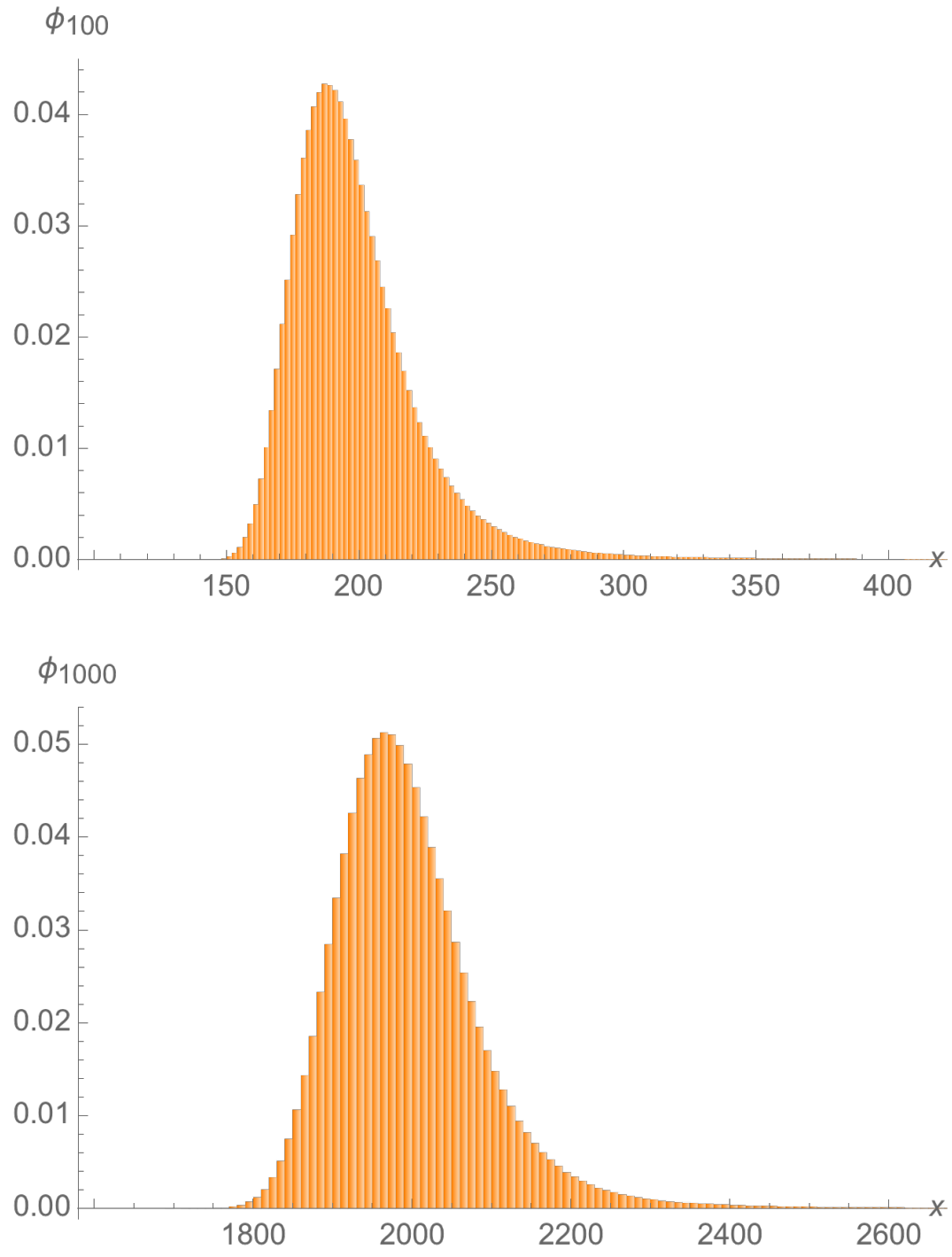

Figure 7.6: *The Pareto distribution, $\phi_{100}$ and $\phi_{1000}$, not much improvement towards Gaussianity, but an $\alpha = 2$ will eventually get you there if you are patient and have a long, very long, life.*

and inverting the characteristic function,

$$\phi_n(x) = \frac{1}{2\pi} \int_{-\infty}^{\infty} \exp(-itx)(2E_3(-it))^n \, dt, \; x \geq n$$

Where $E_{(.)}(.)$ is the exponential integral $E_n(z) = \int_1^{\infty} \frac{dte^{t(-z)}}{t^n}$. Clearly, the integration is done numerically (so far nobody has managed to pull out the distribution of a Pareto sum). It can be exponentially slow (up to 24 hours for $n = 50$ vs. 45 seconds for $n = 2$), so we have used Monte Carlo simulations for Figs. 7.3.1.

Recall from Eq. 7.1 that the convergence requires norming constants $a_n$ and $b_n$. From Uchaikin and Zolotarev [310], we have (narrowing the situation for $1 < \alpha_p \leq 2$):

$$\mathbb{P}(X > x) = cx^{-\alpha_p}$$

as $x \to \infty$ (assume here that $c$ is a constant we will present more formally the "slowly varying function" in the next chapter, and

$$\mathbb{P}(X < x) = d|x|^{-\alpha_p}$$

as $x \to \infty$. The norming constants become $a_n = n\ \mathbb{E}(X)$ for $\alpha_p > 1$ (for other cases, consult [310] as these are not likely to occur in practice), and

$$b_n = \begin{cases} \pi n^{1/\alpha_p} \left(2 \sin\left(\frac{\pi \alpha_p}{2}\right) \Gamma(\alpha_p)\right)^{-\frac{1}{\alpha_p}} (c+d)^{1/\alpha_p} & \text{for} \quad 1 < \alpha_p < 2 \\ \sqrt{c+d}\sqrt{n \log(n)} & \text{for} \quad \alpha_p = 2 \end{cases}. \tag{7.5}$$

And the symmetry parameter $\beta = \frac{c-d}{c+d}$. Clearly, the situation where the Paretian parameter $\alpha_p$ is greater than 2 leads to the Gaussian.

### 7.3.4 The half-cubic Pareto and its basin of convergence

Of interest is the case of $\alpha = \frac{3}{2}$. Unlike the situations where as in Figure 7.3.1, the distribution ends up slowly being symmetric. But, as we will cover in the next chapter, it is erroneous to conflate its properties with those of a stable. It is, in a sense, more fat-tailed.

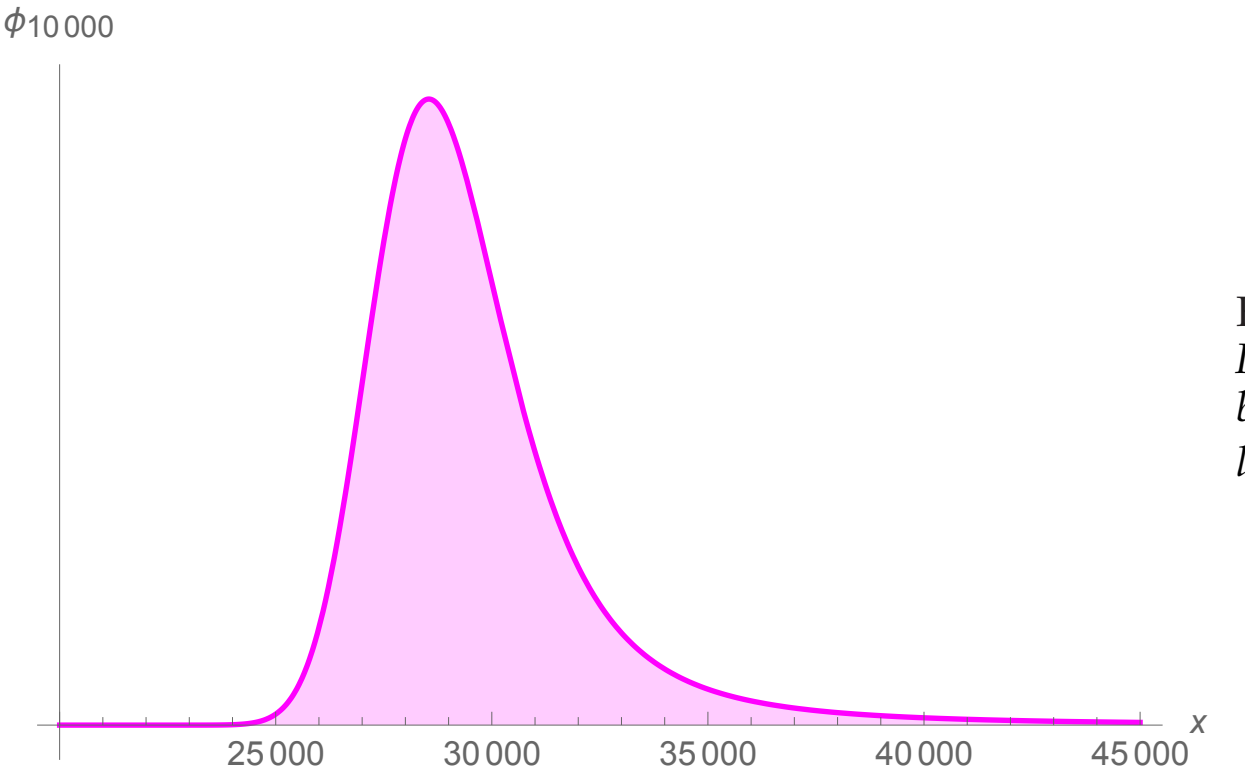

Figure 7.7: *The half-cubic Pareto distribution never becomes symmetric in real life. Here* $n = 10^4$

## 7.4 CUMULANTS AND CONVERGENCE

Since the Gaussian (as a basin of convergence) has skewness of 0 and (raw) kurtosis of 3, we can heuristically examine the convergence of these moments to establish the speed of the workings under CLT.

**Definition 7.1** (Excess p-cumulants)
*Let $\chi(\omega)$ be characteristic function of a given distribution, $n$ the number of summands (for independent random variables), $p$ the order of the moment. We define the ratio of cumulants for the corresponding $p^{th}$ moment:*

$$K_n^p \triangleq \frac{(-i)^p \partial^p \log(\chi(\omega)^n)}{(-\partial^2 \log(\chi(\omega)^n))^2}$$

*$K(n)$ is a metric of excess $p^{th}$ moment over that of a Gaussian, $p > 2$; in other words, $K_n^4 = 0$ denotes Gaussianity for $n$ independent summands.*

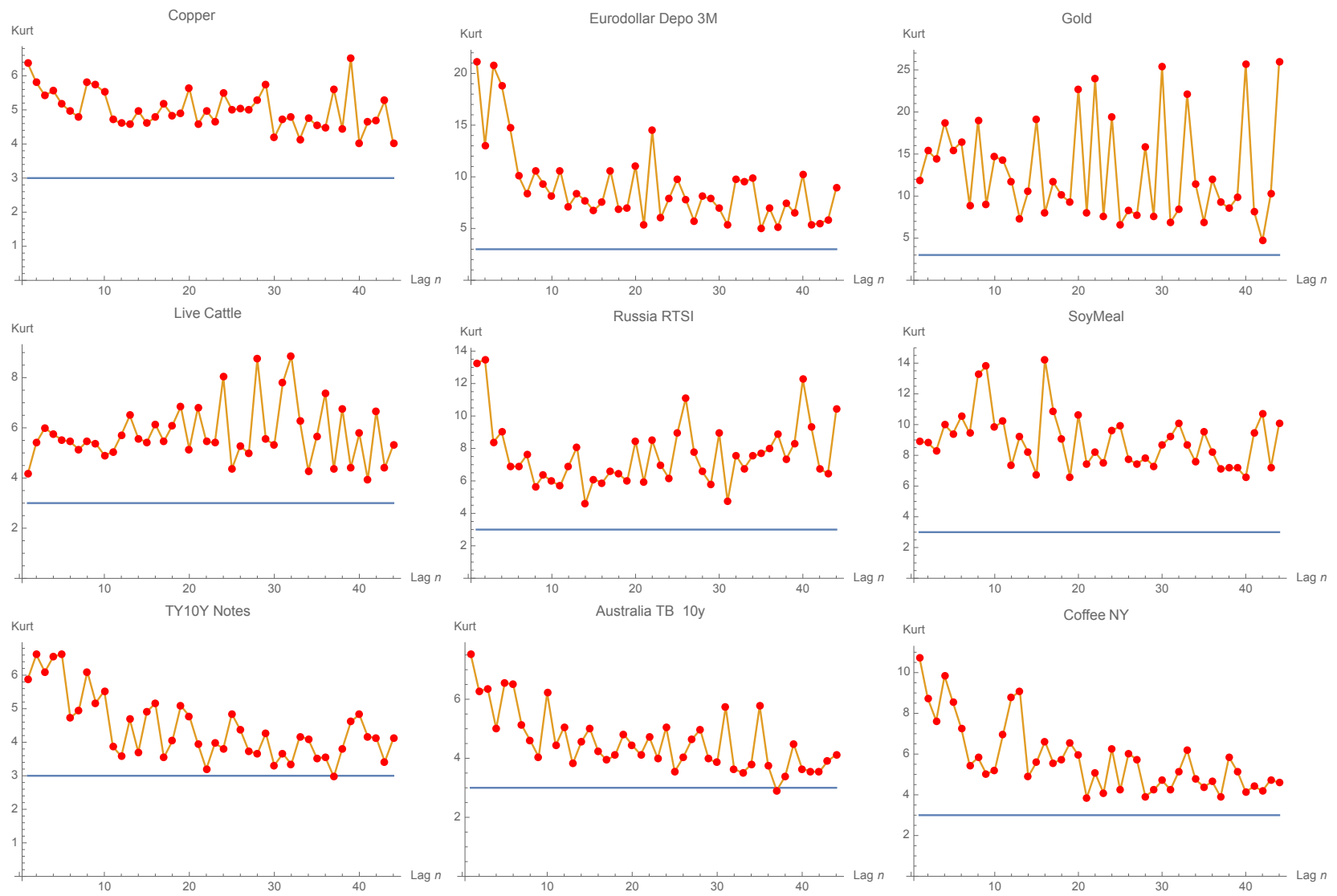

Figure 7.8: *Behavior of the* $4^{th}$ *moment under aggregation for a few financial securities deemed to converge to the Gaussian but in fact do not converge (backup data for [274]). There is no conceivable way to claim convergence to Gaussian for data sampled at a lower frequency.*

> **Remark 7.1**
>
> *We note that*
>
> $$\lim_{n\to\infty} K_N^p = 0$$
>
> *for all probability distributions outside the Power Law class.*
>
> *We also note that* $\lim_{p\to\infty} K_n^p$ *is finite for the thin-tailed class. In other words, we face a clear-cut basin of converging vs. diverging moments.*

For distributions outside the Power Law basin, $\forall p \in \mathbb{N}_{>2}$, $K_n^p$ decays at a rate $N^{p-2}$.

A sketch of the proof can be done using the stable distribution as the limiting basin and the nonderivability at order $p$ greater than its tail index, using Eq. 8.4.

Table 7.1 shows what happens to the cumulants $K(.)$ for $n$-summed variables.

We would expect a drop at a rate $\frac{1}{N^2}$ for stochastic volatility (gamma variance wlog). However, figure 10.2 shows the drop does not take place at any such speed. Visibly we are not in the basin. As seen in [274] there is an absence of convergence of kurtosis under summation across economic variables.

Table 7.1: *Table of Normalized Cumulants For Thin Tailed Distributions Speed of Convergence for N Independent Summands*

-

| **Distr.** | **Poisson** ($\lambda$) | **Expon.** ($\lambda$) | **Gamma** (a,b) | **Symmetric 2-state vol** $(\sigma_1, \sigma_2)$ | **Γ-Variance** $(a, b)$ |
|---|---|---|---|---|---|
| **K(2)** | 1 | 1 | 1 | 1 | 1 |
| **K(3)** | $\frac{1}{n\lambda}$ | $\frac{2\lambda}{n}$ | $\frac{2}{a\ b\ n}$ | 0 | 0 |
| **K(4)** | $\frac{1}{n\lambda^2}$ | $\frac{3!\lambda^2}{n}$ | $\frac{3!}{a^2\ b^2\ n}$ | $\frac{3(1-p)p}{n} \times \frac{\left(\sigma_1^2-\sigma_2^2\right)^2}{\left(p\sigma_1^2-(p-1)\sigma_2^2\right)^3}$ | $\frac{3b}{n}$ |

## 7.5 TECHNICAL REFRESHER: TRADITIONAL VERSIONS OF CLT

This is a refresher of the various approaches bundled under the designation CLT.

**The Standard (Lindeberg-Lévy) version of CLT** Suppose as before a sequence of i.i.d. random variables with $\mathbb{E}(X_i) = \mu$ and $\mathbb{V}(X_i) = \sigma^2 < +\infty$, and $\overline{X}_n$ the sample average for $n$. Then as $n$ approaches infinity, the sum of the random variables $\sqrt{n}(\overline{X}_n - \mu)$ converges in distribution to a Gaussian [24] [25]:

$$\sqrt{n}\left(\overline{X}_n - \mu\right) \xrightarrow{d} N\left(0, \sigma^2\right).$$

Convergence in distribution means that the CDF (cumulative distribution function) of $\sqrt{n}\left(\overline{X}_n - \mu\right)$ converges pointwise to the CDF of $\mathcal{N}(0, \sigma)$ for every real $z$,

$$\lim_{n \to \infty} \mathbb{P}\left(\sqrt{n}(\overline{X}_n - \mu) \leq z\right) = \lim_{n \to \infty} \mathbb{P}\left[\frac{\sqrt{n}(\overline{X}_n - \mu)}{\sigma} \leq \frac{z}{\sigma}\right] = \Phi\left(\frac{z}{\sigma}\right), \ \sigma > 0$$

where $\Phi(z)$ is the standard normal cdf evaluated at $z$. Note that the convergence is uniform in $z$ in the sense that

$$\lim_{n \to \infty} \sup_{z \in \mathbb{R}} \left| \mathbb{P}\left(\sqrt{n}(\overline{X}_n - \mu) \leq z\right) - \Phi\left(\frac{z}{\sigma}\right) \right| = 0,$$

where sup denotes the least upper bound, that is, the supremum of the set.

**Lyapunov's CLT** In Lyapunov's derivation, summands have to be independent, but not necessarily identically distributed. The theorem also requires that random variables $|X_i|$ have moments of some order $(2 + \delta)$, and that the rate of growth of these moments is limited by the Lyapunov condition, provided next.

The condition is as follows. Define

$$s_n^2 = \sum_{i=1}^{n} \sigma_i^2$$

If for some $\delta > 0$,

$$\lim_{n\to\infty} \frac{1}{s_n^{2+\delta}} \sum_{i=1}^{n} \mathbb{E}\left(|X_i - \mu_i|^{2+\delta}\right) = 0,$$

then a sum of $\frac{X_i - \mu_i}{s_i}$ converges in distribution to a standard normal random variable, as $n$ goes to infinity:

$$\frac{1}{s_n} \sum_{i=1}^{n} (X_i - \mu_i) \xrightarrow{D} N(0,1).$$

If a sequence of random variables satisfies Lyapunov's condition, then it also satisfies Lindeberg's condition that we cover next. The converse implication, however, does not hold.

**Lindeberg's condition** Lindeberg allows to reach CLT under weaker assumptions. With the same notations as earlier:

$$\lim_{n\to\infty} \frac{1}{s_n^2} \sum_{i=1}^{n} \mathbb{E}\left((X_i - \mu_i)^2 \cdot \mathbb{1}_{\{|X_i - \mu_i| > \varepsilon s_n\}}\right) = 0$$

for all $\varepsilon > 0$ , where $\mathbb{1}$ indicator function, then the random variable $Z_n = \frac{\sum_{i=1}^{n}(X_i - \mu_i)}{s_n}$ converges in distribution] to a Gaussian as $n \to \infty$.

Lindeberg's condition is sufficient, but not in general necessary except if the sequence under consideration satisfies:

$$\max_{1 \le k \le n} \frac{\sigma_i^2}{s_n^2} \to 0, \quad \text{as } n \to \infty,$$

then Lindeberg's condition is both sufficient and necessary, i.e. it holds if and only if the result of central limit theorem holds.

## 7.6 THE LAW OF LARGE NUMBERS FOR HIGHER MOMENTS

### 7.6.1 Higher Moments

A test of fat tailedness can be seen by applying the law of large number to higher moments and see how they converge. A visual examination of the behavior of the cumulative mean of the moment can be done in a similar way to the standard visual tests of LLN we saw in Chapter 3– except that it applies to $X^p$ (raw or centered) rather than $X$. We check the functioning of the law of large numbers by seeing if adding observations causes a reduction of the variability of the average (or its variance if it exists). Moments that do not exist will show occasional jumps –or, equivalently, large sub-samples will produce different averages. When moments exist, adding observations eventually prevents further jumps.

Another visual technique is to consider the contribution of the maximum observation to the total, and see how it behaves as $n$ grows larger. It is called the MS plot [135], "maximum to sum", and shown in Figure 7.9.

Table 7.2: *Kurtosis $K(t)$ for $t$ daily, 10-day, and 66-day windows for the random variables*

| | K(1) | K(10) | K(66) | Max Quartic | Years |
|---|---|---|---|---|---|
| Australian Dollar/USD | 6.3 | 3.8 | 2.9 | 0.12 | 22. |
| Australia TB 10y | 7.5 | 6.2 | 3.5 | 0.08 | 25. |
| Australia TB 3y | 7.5 | 5.4 | 4.2 | 0.06 | 21. |
| BeanOil | 5.5 | 7.0 | 4.9 | 0.11 | 47. |
| Bonds 30Y | 5.6 | 4.7 | 3.9 | 0.02 | 32. |
| Bovespa | 24.9 | 5.0 | 2.3 | 0.27 | 16. |
| British Pound/USD | 6.9 | 7.4 | 5.3 | 0.05 | 38. |
| CAC40 | 6.5 | 4.7 | 3.6 | 0.05 | 20. |
| Canadian Dollar | 7.4 | 4.1 | 3.9 | 0.06 | 38. |
| Cocoa NY | 4.9 | 4.0 | 5.2 | 0.04 | 47. |
| Coffee NY | 10.7 | 5.2 | 5.3 | 0.13 | 37. |
| Copper | 6.4 | 5.5 | 4.5 | 0.05 | 48. |
| Corn | 9.4 | 8.0 | 5.0 | 0.18 | 49. |
| Crude Oil | 29.0 | 4.7 | 5.1 | 0.79 | 26. |
| CT | 7.8 | 4.8 | 3.7 | 0.25 | 48. |
| DAX | 8.0 | 6.5 | 3.7 | 0.20 | 18. |
| Euro Bund | 4.9 | 3.2 | 3.3 | 0.06 | 18. |
| Euro Currency/DEM previously | 5.5 | 3.8 | 2.8 | 0.06 | 38. |
| Eurodollar Depo 1M | 41.5 | 28.0 | 6.0 | 0.31 | 19. |
| Eurodollar Depo 3M | 21.1 | 8.1 | 7.0 | 0.25 | 28. |
| FTSE | 15.2 | 27.4 | 6.5 | 0.54 | 25. |
| Gold | 11.9 | 14.5 | 16.6 | 0.04 | 35. |
| Heating Oil | 20.0 | 4.1 | 4.4 | 0.74 | 31. |
| Hogs | 4.5 | 4.6 | 4.8 | 0.05 | 43. |
| Jakarta Stock Index | 40.5 | 6.2 | 4.2 | 0.19 | 16. |
| Japanese Gov Bonds | 17.2 | 16.9 | 4.3 | 0.48 | 24. |
| Live Cattle | 4.2 | 4.9 | 5.6 | 0.04 | 44. |
| Nasdaq Index | 11.4 | 9.3 | 5.0 | 0.13 | 21. |
| Natural Gas | 6.0 | 3.9 | 3.8 | 0.06 | 19. |
| Nikkei | 52.6 | 4.0 | 2.9 | 0.72 | 23. |
| Notes 5Y | 5.1 | 3.2 | 2.5 | 0.06 | 21. |
| Russia RTSI | 13.3 | 6.0 | 7.3 | 0.13 | 17. |
| Short Sterling | 851.8 | 93.0 | 3.0 | 0.75 | 17. |
| Silver | 160.3 | 22.6 | 10.2 | 0.94 | 46. |
| Smallcap | 6.1 | 5.7 | 6.8 | 0.06 | 17. |
| SoyBeans | 7.1 | 8.8 | 6.7 | 0.17 | 47. |
| SoyMeal | 8.9 | 9.8 | 8.5 | 0.09 | 48. |
| Sp500 | 38.2 | 7.7 | 5.1 | 0.79 | 56. |
| Sugar #11 | 9.4 | 6.4 | 3.8 | 0.30 | 48. |
| SwissFranc | 5.1 | 3.8 | 2.6 | 0.05 | 38. |

Table 7.2: *(continued from previous page)*

| | K(1) | K(10) | K(66) | Max Quartic | Years |
|---|---|---|---|---|---|
| TY10Y Notes | 5.9 | 5.5 | 4.9 | 0.10 | 27. |
| Wheat | 5.6 | 6.0 | 6.9 | 0.02 | 49. |
| Yen/USD | 9.7 | 6.1 | 2.5 | 0.27 | 38. |

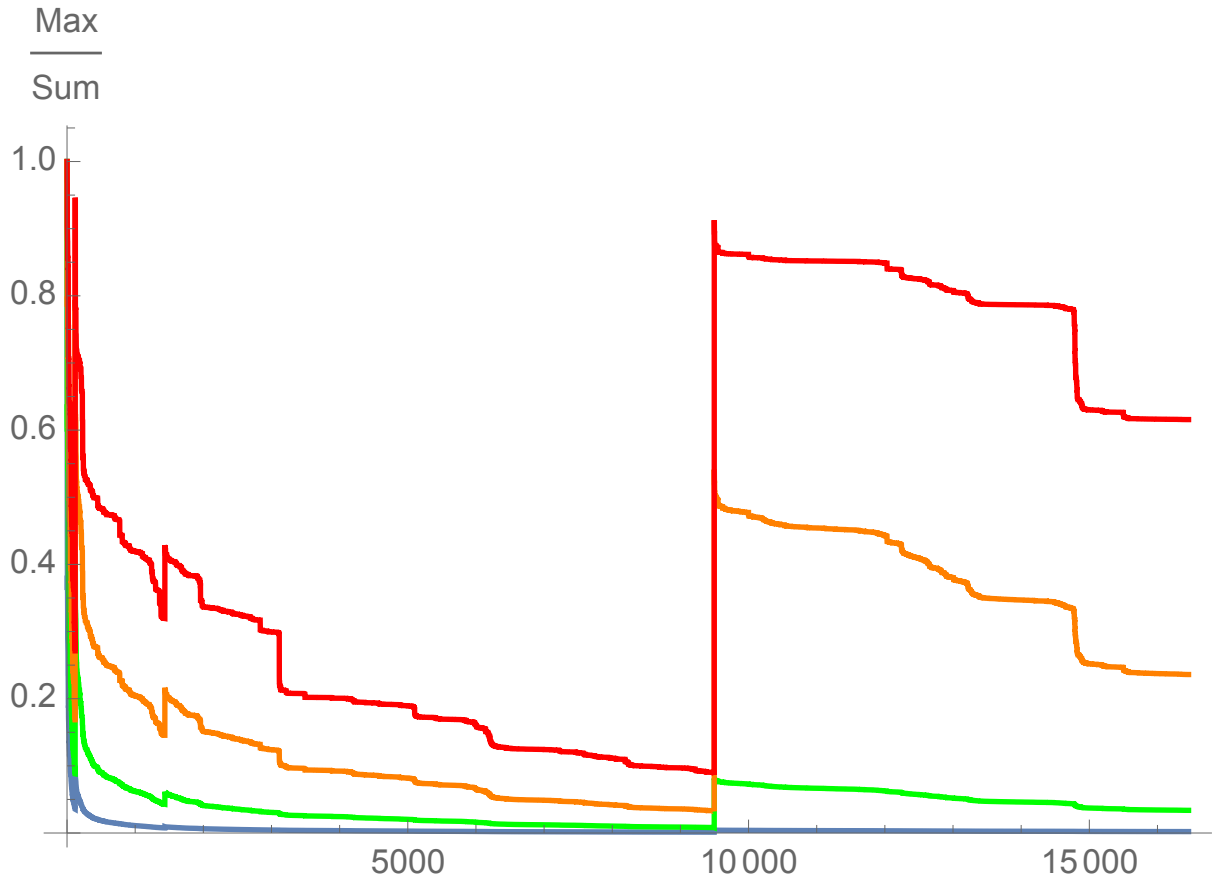

Figure 7.9: *MS Plot showing the behavior of cumulative moments $p = 1, 2, 3, 4$ for the SP500 over the 60 years ending in 2018. The MS plot (Maximum to sum) will be presented in 10.2.6.*

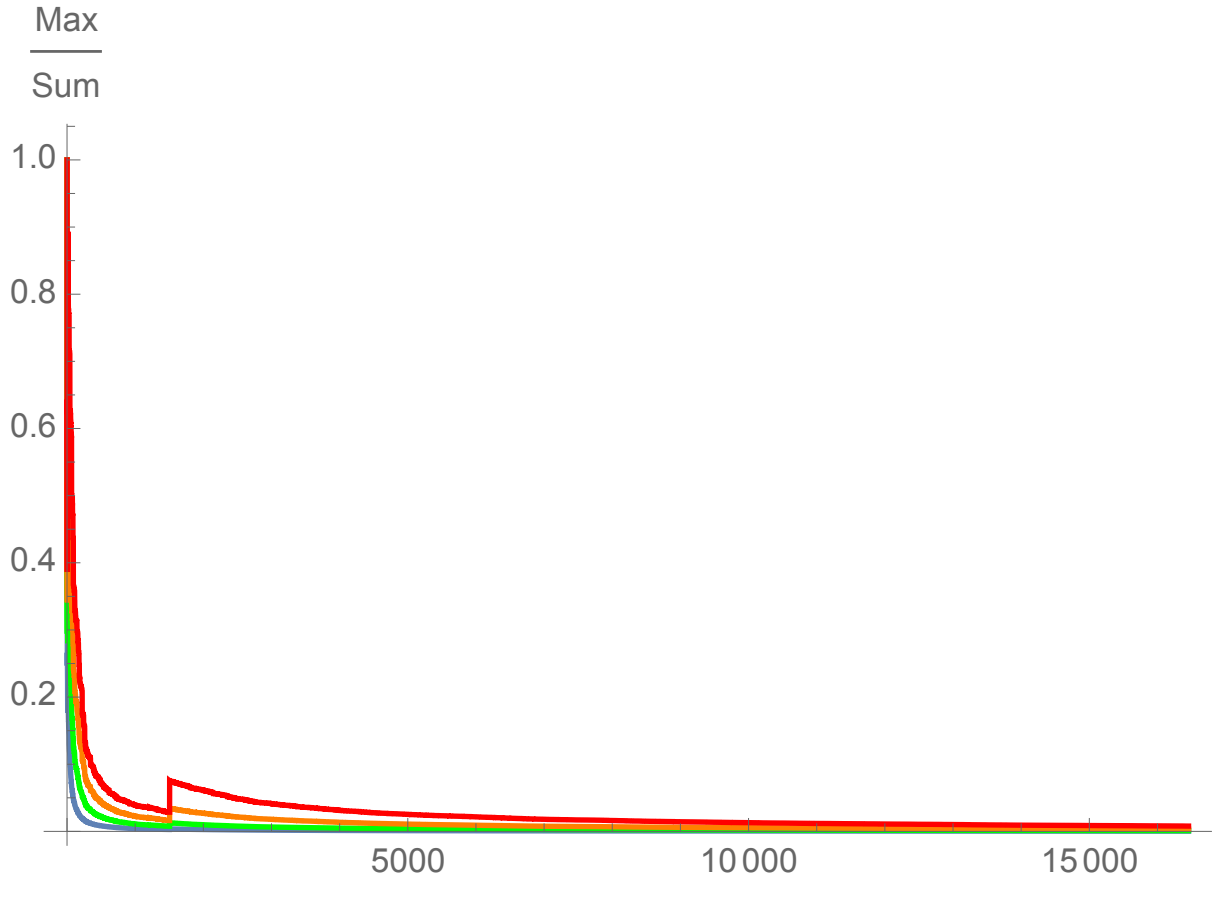

Figure 7.10: *Gaussian Control for the data in Figure 7.9.*

## 7.7 MEAN DEVIATION FOR A STABLE DISTRIBUTIONS

Let us prepare a result for the next chapter using the norm $L^1$ for situations of finite mean but infinite variance.[6] Clearly we have no way to measure the compression of the distribution around the mean within the norm $L^2$.

The error of a sum in the norm $L^1$ is as follows. Let $\theta(x)$ be the Heaviside function (whose value is zero for negative arguments and one for positive arguments). Since $\text{sgn}(x) = 2\theta(x) - 1$, its characteristic function will be:

$$\chi^{\text{sgn}(x)}(t) = \frac{2i}{t}. \tag{7.6}$$

Let $\chi^d(.)$ be the characteristic function of any nondegenerate distribution. Convoluting $\chi^{\text{sgn}(x)} * (\chi^d)^n$, we obtain the characteristic functionfor the positive variations for $n$ independent summands

$$\chi^m = \int_{-\infty}^{\infty} \chi^{\text{sgn}(x)}(t)\chi^d(u-t)^n \text{d}t.$$

In our case of mean absolute deviation being twice that of the positive values of $X$:

$$\chi(|S_n|) = (2i)\int_{-\infty}^{\infty} \frac{\chi(t-u)^n}{t}\, du,$$

which is the Hilbert transform of $\chi$ when $\int$ is taken in the p.v. sense (Pinelis, 2015)[226]. In our situation, given that all independents summands are copies from the same distribution, we can replace the product $\chi(t)^n$ with $\chi_s(t)$ which is the same characteristic function with $\sigma_s = n^{1/\alpha}\sigma$, $\beta$ remaining the same:

$$\mathbb{E}(|X|) = 2i\frac{\partial}{\partial u}\text{p.v.}\int_{-\infty}^{\infty} \frac{\chi_s(t-u)}{t}\, dt|_{t=0}. \tag{7.7}$$

Now, [226] the Hilbert transform $H$,

$$(Hf)(t) = \frac{2}{\pi i}\int_0^{\infty-} \chi_s(u+t) - \chi_s(u-t)\, dt$$

can be rewritten as

$$(Hf)(t) = -i\frac{\partial}{\partial u}\left(1+\chi_s(u)+\frac{1}{\pi i}\int_0^{\infty-} \chi_s(u+t)-\chi_s(u-t)-\chi_s(t)+\chi_s(-t)\,\frac{dt}{t}\right). \tag{7.8}$$

Consider the stable distribution defined in 7.2.1.

Deriving first inside the integral and using a change of variable, $z = \log(t)$,

$$\begin{aligned}
&\mathbb{E}|X|_{(\tilde{\alpha_s},\beta,\sigma_s,0)} = \\
&\int_{-\infty}^{\infty} 2i\alpha_s e^{-(\sigma_s e^z)^{\alpha_s}-z}\left(\sigma_s e^z\right)^{\alpha_s}\left(\beta\tan\left(\frac{\pi\alpha_s}{2}\right)\sin\left(\beta\tan\left(\frac{\pi\alpha_s}{2}\right)\left(\sigma_s e^z\right)^{\alpha_s}\right)\right. \\
&\left.+\cos\left(\beta\tan\left(\frac{\pi\alpha_s}{2}\right)\left(\sigma_s e^z\right)^{\alpha_s}\right)\right) dz
\end{aligned}$$

---

[6] We say, again by convention, *infinite* for the situation where the random variable, say $X^2$ (or the variance of any random variable), is one-tailed –bounded on one side– and undefined in situations where the variable is two-tailed, e.g. the infamous Cauchy.

which then integrates nicely to:

$$\mathbb{E}|X|_{(\tilde{\alpha_s},\beta,\sigma_s,0)} = \frac{\sigma_s}{2\pi}\Gamma\left(\frac{\alpha_s - 1}{\alpha_s}\right)\left(\left(1 + i\beta\tan\left(\frac{\pi\alpha_s}{2}\right)\right)^{1/\alpha_s} + \left(1 - i\beta\tan\left(\frac{\pi\alpha_s}{2}\right)\right)^{1/\alpha_s}\right). \tag{7.9}$$

## NEXT

The next chapter presents a central concept: how to work with the law of middle numbers? How can we translate between distributions?

# 8 HOW MUCH DATA DO YOU NEED? AN OPERATIONAL METRIC FOR FAT-TAILEDNESS‡

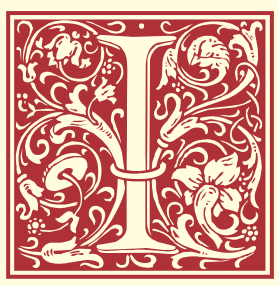

N THIS (RESEARCH) CHAPTER we discuss the laws of medium numbers. We present an operational metric for univariate unimodal probability distributions with finite first moment, in $[0, 1]$ where 0 is maximally thin-tailed (Gaussian) and 1 is maximally fat-tailed. It is based on "how much data one needs to make meaningful statements about a given dataset?"

Applications: Among others, it

- helps assess the sample size $n$ needed for statistical significance outside the Gaussian,
- helps measure the speed of convergence to the Gaussian (or stable basin),
- allows practical comparisons across classes of fat-tailed distributions,
- allows the assessment of the number of securities needed in portfolio construction to achieve a certain level of stability from diversification,
- helps understand some inconsistent attributes of the lognormal, pending on the parametrization of its variance.

The literature is rich for what concerns asymptotic behavior, but there is a large void for finite values of $n$, those needed for operational purposes.

Background : Conventional measures of fat-tailedness, namely 1) the tail index for the Power Law class, and 2) Kurtosis for finite moment distributions fail to apply to some distributions, and do not allow comparisons across classes and parametrization, that is between power laws outside the Levy-Stable basin, or

Research chapter.

The author owes the most to the focused comments by Michail Loulakis who, in addition, provided the rigorous derivations for the limits of the $\kappa$ for the Student T and lognormal distributions, as well as to the patience and wisdom of Spyros Makridakis. The paper was initially presented at *Extremes and Risks in Higher Dimensions*, Sept 12-16 2016, at the Lorentz Center, Leiden and at Jim Gatheral's Festschrift at the Courant Institute, in October 2017. The author thanks Jean-Philippe Bouchaud, John Einmahl, Pasquale Cirillo, and others. Laurens de Haan suggested changing the name of the metric from "gamma" to "kappa" to avoid confusion. Additional thanks to Colman Humphrey, Michael Lawler, Daniel Dufresne and others for discussions and insights with derivations.

power laws to distributions in other classes, or power laws for different number of summands. How can one compare a sum of 100 Student T distributed random variables with 3 degrees of freedom to one in a Levy-Stable or a Lognormal class? How can one compare a sum of 100 Student T with 3 degrees of freedom to a single Student T with 2 degrees of freedom?

We propose an operational and heuristic metric that allows us to compare $n$-summed independent variables under all distributions with finite first moment. The method is based on the rate of convergence of the law of large numbers for finite sums, $n$-summands specifically.

We get either explicit expressions or simulation results and bounds for the log-normal, exponential, Pareto, and the Student T distributions in their various calibrations –in addition to the general Pearson classes.

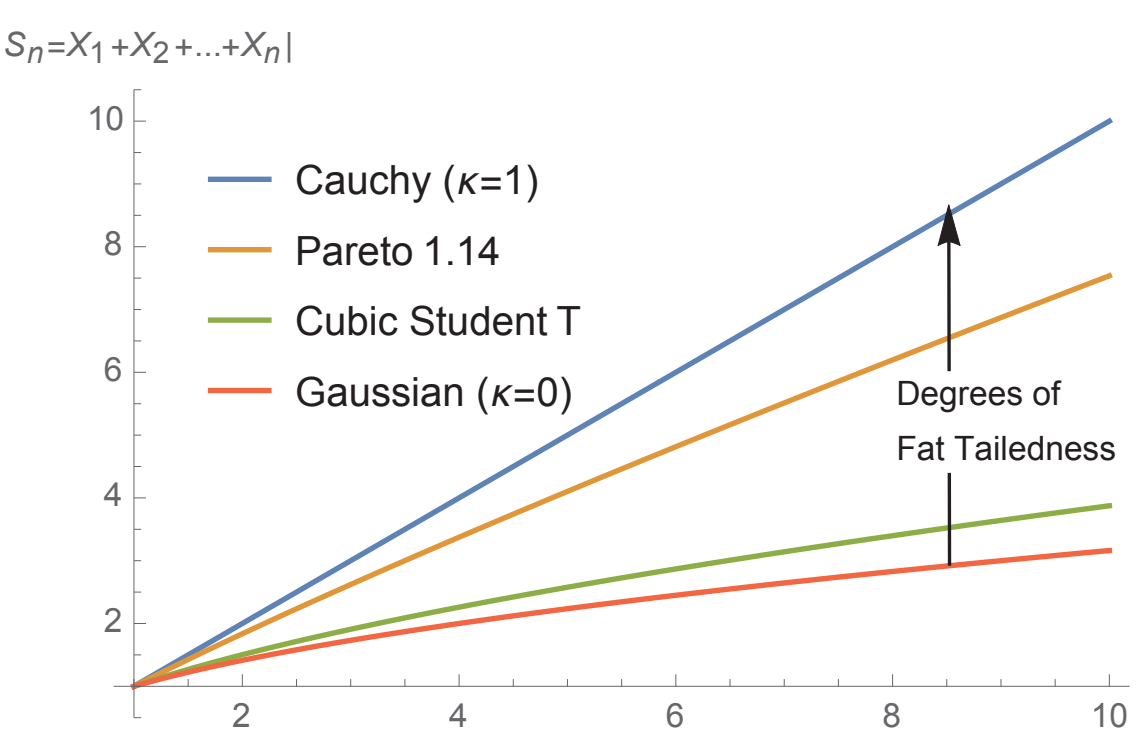

Figure 8.1: *The intuition of what $\kappa$ is measuring: how the mean deviation of the sum of identical copies of a r.v. $S_n = X_1 + X_2 + \ldots X_n$ grows as the sample increases and how we can compare preasymptotically distributions from different classes.*

## 8.1 INTRODUCTION AND DEFINITIONS

How can one compare a Pareto distribution with tail $\alpha = 2.1$ that is, with finite variance, to a Gaussian? Asymptotically, these distributions in the regular variation class with finite second moment, under summation, become Gaussian, but pre-asymptotically, we have no standard way of comparing them given that metrics that depend on higher moments, such as kurtosis, cannot be of help. Nor can we easily compare an infinite variance Pareto distribution to its limiting $\alpha$-Stable distribution (when both have the same tail index or tail exponent ). Likewise, how can one compare the "fat-tailedness" of, say a Student T with 3 degrees of freedom to that of a Levy-Stable with tail exponent of 1.95? Both distributions have a finite mean; of the two, only the first has a finite variance but, for a small number of summands, behaves more "fat-tailed" according to some operational criteria.

**Criterion for "fat-tailedness"** There are various ways to "define" Fat Tails and rank distributions according to each definition. In the narrow class of distributions having all moments finite, it is the kurtosis, which allows simple comparisons and measure departures from the Gaussian, which is used as a norm. For the Power Law class, it can be the tail exponent . One can also use extremal values, taking the probability of exceeding a maximum value, adjusted by the scale (as practiced in extreme value theory). For operational uses, practitioners' fat-tailedness is a degree of concentration, such as "how much of the statistical properties will be attributable to a single observation?", or, appropriately adjusted by the scale (or the mean dispersion), "how much is the total wealth of a country in the hands of the richest individual?"

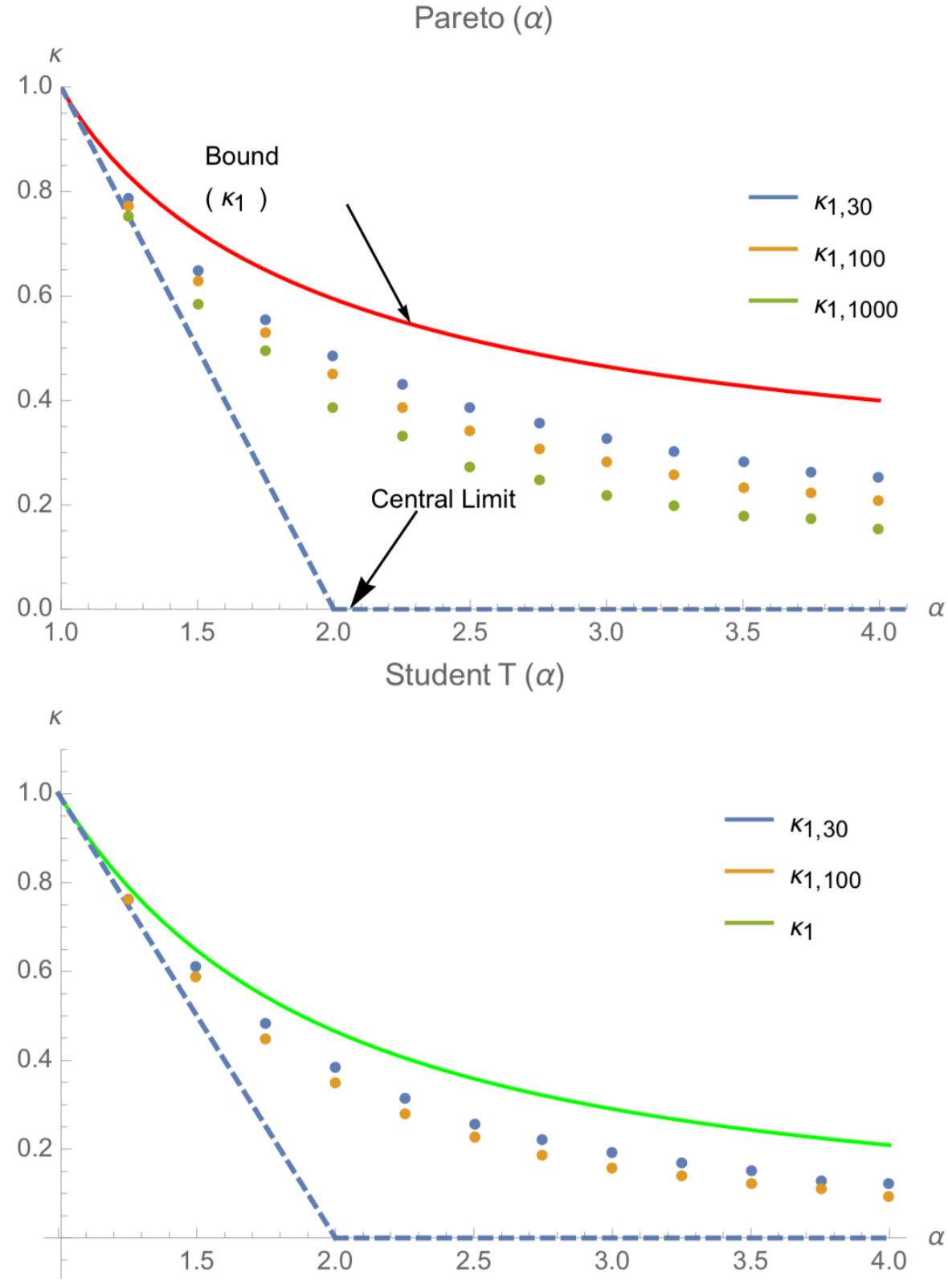

*Figure 8.2: Watching the effect of the Generalized Central Limit Theorem: Pareto and Student T Distribution, in the $\mathfrak{P}$ class, with $\alpha$ exponent, $\kappa$ converge to $2 - (\mathbb{1}_{\alpha<2}\alpha + \mathbb{1}_{\alpha\geq 2}2)$, or the Stable $\mathfrak{S}$ class. We observe how slow the convergence, even after 1000 summands. This discounts Mandelbrot's assertion that an infinite variance Pareto can be subsumed into a stable distribution.*

Here we use the following criterion for our purpose, which maps to the measure of concentration in the past paragraph: "How much will additional data (under such a probability distribution) help increase the stability of the observed mean". The purpose is not entirely statistical: it can equally mean: "How much will adding an additional security into my portfolio allocation (i.e., keeping the total constant) increase its stability?"

Our metric differs from the asymptotic measures (particularly ones used in extreme value theory) in the fact that it is fundamentally preasymptotic.

Real life, and real world realizations, are outside the asymptote.

**What does the metric do?** The metric we propose, $\kappa$ does the following:

- Allows comparison of $n$-summed variables of different distributions for a given number of summands , or same distribution for different $n$, and assess the preasymptotic properties of a given distributions.
- Provides a measure of the distance from the limiting distribution, namely the Lévy $\alpha$-Stable basin (of which the Gaussian is a special case).
- For statistical inference, allows assessing the "speed" of the law of large numbers, expressed in change of the mean absolute error around the average thanks to the increase of sample size $n$.
- Allows comparative assessment of the "fat-tailedness" of two different univariate distributions, when both have finite first moment.
- Allows us to know ahead of time how many runs we need for a Monte Carlo simulation.

**The state of statistical inference** The last point, the "speed", appears to have been ignored (see earlier comments in Chapter 3 about the 9,400 pages of the *Encyclopedia of Statistical Science* [175]). It is very rare to find a discussion about how long it takes to reach the asymptote, or how to deal with $n$ summands that are large but perhaps not sufficiently so for the so-called "normal approximation".

To repeat our motto, "statistics is never standard". This metric aims at showing *how standard is standard*, and measure the exact departure *from the standard* from the standpoint of statistical significance.

## 8.2 THE METRIC

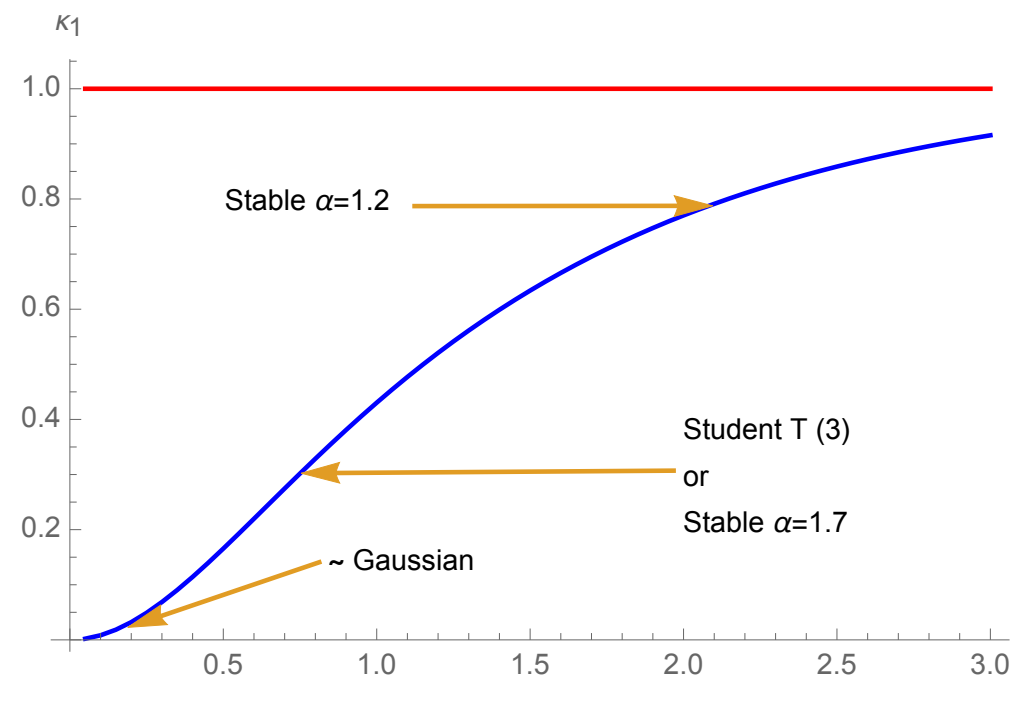

Figure 8.3: *The lognormal distribution behaves like a Gaussian for low values of $\sigma$, but becomes rapidly equivalent to a power law. This illustrates why, operationally, the debate on whether the distribution of wealth was lognormal (Gibrat) or Pareto (Zipf) doesn't carry much operational significance.*

**Definition 8.1** (the $\kappa$ metric)
*Let $X_1, \ldots, X_n$ be i.i.d. random variables with finite mean, that is $\mathbb{E}(X) < +\infty$. Let $S_n = X_1 + X_2 + \ldots + X_n$ be a partial sum. Let $\mathbb{M}(n) = \mathbb{E}(|S_n - \mathbb{E}(S_n)|)$ be the expected mean absolute deviation from the mean for n summands. Define the "rate" of convergence for n additional summands starting with $n_0$:*

$$\kappa_{n_0,n} = \min \left\{ \kappa_{n_0,n} : \frac{\mathbb{M}(n)}{\mathbb{M}(n_0)} = \left( \frac{n}{n_0} \right)^{\frac{1}{2-\kappa_{n_0,n}}}, n_0 = 1, 2, \ldots \right\},$$

Table 8.1: *Kappa for 2 summands*, $\kappa_1$.

| Distribution | $\kappa_1$ |
|---|---|
| Student T $(\alpha)$ | $2-\frac{2\log(2)}{2\log\left(\frac{2^{2-\alpha}\Gamma\left(\alpha-\frac{1}{2}\right)}{\Gamma\left(\frac{\alpha}{2}\right)^2}\right)+\log(\pi)}$ |
| Exponential/Gamma | $2-\frac{\log(2)}{2\log(2)-1}\approx .21$ |
| Pareto $(\alpha)$ | $2-\frac{\log(2)}{\log\left((\alpha-1)^{2-\alpha}\alpha^{\alpha-1}\int_0^{\frac{2}{\alpha-1}}-2\alpha^2(y+2)^{-2\alpha-1}\left(\frac{2}{\alpha-1}-y\right)\left(B_{\frac{1}{y+2}}(-\alpha,1-\alpha)-B_{\frac{y+1}{y+2}}(-\alpha,1-\alpha)\right)dy\right)}$ [3] |
| Normal $(\mu,\sigma)$ with switching variance $\sigma^2 a$ w.p $p$[4]. | $2-\frac{\log(2)}{\log\left(\frac{\sqrt{2}\left(\sqrt{\frac{ap}{p-1}+\sigma^2}+p\left(-2\sqrt{\frac{ap}{p-1}+\sigma^2}+p\left(\sqrt{\frac{ap}{p-1}+\sigma^2}-\sqrt{2a\left(\frac{1}{p-1}+2\right)+4\sigma^2}+\sqrt{a+\sigma^2}\right)+\sqrt{2a\left(\frac{1}{p-1}+2\right)+4\sigma^2}\right)\right)}{p\sqrt{a+\sigma^2}-(p-1)\sqrt{\frac{ap}{p-1}+\sigma^2}}\right)}$ |
| Lognormal $(\mu,\sigma)$ | $\approx 2-\frac{\log(2)}{\log\left(\frac{2\,\mathrm{erf}\left(\frac{\sqrt{\log\left(\frac{1}{2}\left(e^{\sigma^2}+1\right)\right)}}{2\sqrt{2}}\right)}{\mathrm{erf}\left(\frac{\sigma}{2\sqrt{2}}\right)}\right)}.$ |

Table 8.2: *Summary of main results*

| Distribution | $\kappa_n$ |
|---|---|
| Exponential/Gamma | Explicit |
| Lognormal $(\mu,\sigma)$ | No explicit $\kappa_n$ but explicit lower and higher bounds (low or high $\sigma$ or $n$). Approximated with Pearson IV for $\sigma$ in between. |
| Pareto $(\alpha)$ (Constant) | Explicit for $\kappa_2$ (lower bound for all $\alpha$). |
| Student T$(\alpha)$ (slowly varying function) | Explicit for $\kappa_1$ , $\alpha = 3$. |

$n > n_0 \geq 1$, *hence*

$$\kappa(n_0, n) = 2 - \frac{\log(n) - \log(n_0)}{\log\left(\frac{\mathbb{M}(n)}{\mathbb{M}(n_0)}\right)}. \tag{8.1}$$

Table 8.3: *Comparing Pareto to Student T (Same tail exponent $\alpha$)*

| $\alpha$ | Pareto | Pareto | Pareto | Student | Student | Student |
|---|---|---|---|---|---|---|
| | $\kappa_1$ | $\kappa_{1,30}$ | $\kappa_{1,100}$ | $\kappa_1$ | $\kappa_{1,30}$ | $\kappa_{1,100}$ |
| 1.25 | 0.829 | 0.787 | 0.771 | 0.792 | 0.765 | 0.756 |
| 1.5 | 0.724 | 0.65 | 0.631 | 0.647 | 0.609 | 0.587 |
| 1.75 | 0.65 | 0.556 | 0.53 | 0.543 | 0.483 | 0.451 |
| 2. | 0.594 | 0.484 | 0.449 | 0.465 | 0.387 | 0.352 |
| 2.25 | 0.551 | 0.431 | 0.388 | 0.406 | 0.316 | 0.282 |
| 2.5 | 0.517 | 0.386 | 0.341 | 0.359 | 0.256 | 0.227 |
| 2.75 | 0.488 | 0.356 | 0.307 | 0.321 | 0.224 | 0.189 |
| 3. | 0.465 | 0.3246 | 0.281 | 0.29 | 0.191 | 0.159 |
| 3.25 | 0.445 | 0.305 | 0.258 | 0.265 | 0.167 | 0.138 |
| 3.5 | 0.428 | 0.284 | 0.235 | 0.243 | 0.149 | 0.121 |
| 3.75 | 0.413 | 0.263 | 0.222 | 0.225 | 0.13 | 0.10 |
| 4. | 0.4 | 0.2532 | 0.211 | 0.209 | 0.126 | 0.093 |

*Further, for the baseline values $n = n_0 + 1$, we use the shorthand $\kappa_{n_0}$.*

We can also decompose $\kappa(n_0, n)$ in term of "local" intermediate ones similar to "local" interest rates, under the constraint.

$$\kappa(n_0, n) = 2 - \frac{\log(n) - \log(n_0)}{\sum_{i=0}^{n} \frac{\log(i+1) - \log(i)}{2 - \kappa(i, i+1)}}. \tag{8.2}$$

**Use of Mean Deviation** Note that we use for measure of dispersion around the mean the mean absolute deviation, to stay in norm $L^1$ in the absence of finite variance –actually, even in the presence of finite variance, under Power Law regimes, distributions deliver an unstable and uninformative second moment. Mean deviation proves far more robust there. (Mean absolute deviation can be shown to be more "efficient" except in the narrow case of kurtosis equals 3 (the Gaussian), see a longer discussion in [283]; for other advantages, see [220].)

## 8.3 STABLE BASIN OF CONVERGENCE AS BENCHMARK

**Definition 8.2** (the class $\mathfrak{P}$)
*The $\mathfrak{P}$ class of power laws (regular variation) is defined for r.v. X as follows:*

$$\mathfrak{P} = \{X : \mathbb{P}(X > x) \sim L(x)\, x^{-\alpha}\} \tag{8.3}$$

*where $\sim$ means that the limit of the ratio or rhs to lhs goes to 1 as $x \to \infty$. $L : [x_{\min}, +\infty) \to (0, +\infty)$ is a slowly varying function function, defined as $\lim_{x \to +\infty} \frac{L(kx)}{L(x)} = 1$ for any $k > 0$. The constant $\alpha > 0$.*

Next we define the domain of attraction of the sum of identically distributed variables, in our case with identical parameters.

**Definition 8.3**
*(stable $\mathfrak{S}$ class) A random variable X follows a stable (or α-stable) distribution, symbolically $X \sim S(\tilde{\alpha}, \beta, \mu, \sigma)$, if its characteristic function* $\chi(t) = \mathbb{E}(e^{itX})$ *is of the form:*

$$\chi(t) = \begin{cases} e^{\left(i\mu t - |t\sigma|^{\tilde{\alpha}}\left(1 - i\beta \tan\left(\frac{\pi\tilde{\alpha}}{2}\right) sgn(t)\right)\right)} & \tilde{\alpha} \neq 1 \\ e^{it\left(\frac{2\beta\sigma\log(\sigma)}{\pi} + \mu\right) - |t\sigma|\left(1 + \frac{2i\beta sgn(t)\log(|t\sigma|)}{\pi}\right)} & \tilde{\alpha} = 1 \end{cases}, \tag{8.4}$$

Next, we define the corresponding stable $\tilde{\alpha}$:

$$\tilde{\alpha} \triangleq \begin{cases} \alpha\, \mathbb{1}_{\alpha<2} + 2\, \mathbb{1}_{\alpha \geq 2} & \text{if X is in } \mathfrak{P} \\ 2 & \text{otherwise.} \end{cases} \tag{8.5}$$

Further discussions of the class $\mathfrak{S}$ are as follows.

### 8.3.1 Equivalence for Stable distributions

For all $n_0$ and $n \geq 1$ in the Stable $\mathfrak{S}$ class with $\tilde{\alpha} \geq 1$:

$$\kappa_{(n_0,n)} = 2 - \tilde{\alpha},$$

simply from the property that

$$\mathbb{M}(n) = n^{\frac{1}{\alpha}} \mathbb{M}(1) \tag{8.6}$$

This, simply shows that $\kappa_{n_0,n} = 0$ for the Gaussian.

The problem of the preasymptotics for $n$ summands reduces to:

- What is the property of the distribution for $n_0 = 1$ (or starting from a standard, off-the shelf distribution)?
- What is the property of the distribution for $n_0$ summands?
- How does $\kappa_n \to 2 - \tilde{\alpha}$ and at what rate?

### 8.3.2 Practical significance for sample sufficiency

**Confidence intervals**: As a simple heuristic, the higher $\kappa$, the more disproportionally insufficient the confidence interval. Any value of $\kappa$ above .15 effectively indicates a high degree of unreliability of the "normal approximation". One can immediately doubt the results of numerous research papers in fat-tailed domains.

Computations of the sort done Table 8.2 for instance allows us to compare various distributions under various parametriazation. (comparing various Pareto distributions to symmetric Student T and, of course the Gaussian which has a flat kappa of 0)

As we mentioned in the introduction, required sample size for statistical inference is driven by $n$, the number of summands. Yet the law of large numbers is often invoked in erroneous conditions; we need a rigorous sample size metric.

Many papers, when discussing financial matters, say [117] use finite variance as a binary classification for fat tailedness: power laws with a tail exponent greater than 2 are therefore classified as part of the "Gaussian basin", hence allowing the use of variance and other such metrics for financial applications. A much more natural boundary is finiteness of expectation for financial applications [275]. Our metric can thus be useful as follows:

Let $X_{g,1}, X_{g,2}, \ldots, X_{g,n_g}$ be a sequence of Gaussian variables with mean $\mu$ and scale $\sigma$. Let $X_{\nu,1}, X_{\nu,2}, \ldots, X_{\nu,n_\nu}$ be a sequence of some other variables scaled to be of the same $\mathbb{M}(1)$, namely $\mathbb{M}^\nu(1) = \mathbb{M}^g(1) = \sqrt{\frac{2}{\pi}}\sigma$. We would be looking for values of $n_\nu$ corresponding to a given $n_g$.

$\kappa_n$ is indicative of both the rate of convergence under the law of large number, and for $\kappa_n \to 0$, for rate of convergence of summands to the Gaussian under the central limit, as illustrated in Figure 8.2.

$$n_{\min} = \inf \left\{ n_\nu : \mathbb{E}\left( \left| \sum_{i=1}^{n_\nu} \frac{X_{\nu,i} - m_p}{n_\nu} \right| \right) \leq \mathbb{E}\left( \left| \sum_{i=1}^{n_g} \frac{X_{g,i} - m_g}{n_g} \right| \right), n_\nu > 0 \right\} \tag{8.7}$$

which can be computed using $\kappa_n = 0$ for the Gaussian and backing our from $\kappa_n$ for the target distribution with the simple approximation:

$$n_\nu = n_g^{-\frac{1}{\kappa_{1,n_g} - 1}} \approx n_g^{-\frac{1}{\kappa_1 - 1}}, n_g > 1 \tag{8.8}$$

The approximation is owed to the slowness of convergence. So for example, a Student T with 3 degrees of freedom ($\alpha = 3$) requires 120 observations to get the same drop in variance from averaging (hence confidence level) as the Gaussian with 30, that is 4 times as much. The one-tailed Pareto with the same tail exponent $\alpha = 3$ requires 543 observations to match a Gaussian sample of 30, 4.5 times more than the Student, which shows 1) finiteness of variance is not an indication of fat tailedness (in our statistical sense), 2) neither are tail exponent s good indicators 3) how the symmetric Student and the Pareto distribution are not equivalent because of the "bell-shapedness" of the Student (from the slowly varying function) that dampens variations in the center of the distribution.

We can also elicit quite counterintuitive results. From Eq. 8.8, the "Pareto 80/20" in the popular mind, which maps to a tail exponent around $\alpha \approx 1.14$, requires $> 10^9$ more observations than the Gaussian.

## 8.4 TECHNICAL CONSEQUENCES

### 8.4.1 Some Oddities With Asymmetric Distributions

The stable distribution, when skewed, has the same $\kappa$ index as a symmetric one (in other words, $\kappa$ is invariant to the $\beta$ parameter in Eq. 8.4, which conserves under summation). But a one-tailed simple Pareto distribution is fatter tailed (for our purpose here) than an equivalent symmetric one.

This is relevant because the stable is never really observed in practice and used as some limiting mathematical object, while the Pareto is more commonly seen. The point is not well grasped in the literature. Consider the following use of the substitution of a stable for a Pareto. In Uchaikin and Zolotarev [310]:

> Mandelbrot called attention to the fact that the use of the extremal stable distributions (corresponding to $\beta = 1$) to describe empirical principles was preferable to the use of the Zipf-Pareto distributions for a number of reasons. It can be seen from many publications, both theoretical and applied, that Mandelbrot's ideas receive more and more wide recognition of experts. In this way, the hope arises to confirm empirically established principles in the framework of mathematical models and, at the same time, to clear up the mechanism of the formation of these principles.

These are not the same animals, even for large number of summands.

### 8.4.2 Rate of Convergence of a Student T Distribution to the Gaussian Basin

We show in the appendix –thanks to the explicit derivation of $\kappa$ for the sum of students with $\alpha = 3$, the "cubic" commonly noticed in finance –that the rate of convergence of $\kappa$ to 0 under summation is $\frac{1}{\log(n)}$. This (and the semi-closed form for the density of an n-summed cubic Student) complements the result in Bouchaud and Potters [32] (see also [254]), which is as follows. Their approach is to separate the "Gaussian zone" where the density is approximated by that of a Gaussian, and a "Power Law zone" in the tails which retains the original distribution with Power Law decline. The "crossover" between the two moves right and left of the center at a rate of $\sqrt{n \log(n)}$ standard deviations) which is excruciatingly slow. Indeed, one can note that more summands fall at the center of the distribution, and fewer outside of it, hence the speed of convergence according to the central limit theorem will differ according to whether the density concerns the center or the tails.

Further investigations would concern the convergence of the Pareto to a Levy-Stable, which so far we only got numerically.

### 8.4.3 The Lognormal is Neither Thin Nor Fat Tailed

Naively, as we can see in Figure 8.2, at low values of the parameter $\sigma$, the lognormal behaves like a Gaussian, and, at high $\sigma$, it appears to have the behavior of a Cauchy of sorts (a one-tailed Cauchy, rather a stable distribution with $\alpha = 1$, $\beta = 1$), as $\kappa$ gets closer and closer to 1. This gives us an idea about some aspects of the debates as to whether some variable is Pareto or lognormally distributed, such as, say, the debates about wealth [191], [64], [65]. Indeed, such debates can be irrelevant to the real world. As P. Cirillo [53] observed, many cases of Paretianity are effectively lognormal situations with high variance; the practical statistical consequences, however, are smaller than imagined.

### 8.4.4 Can Kappa Be Negative?

Just as kurtosis for a mixed Gaussian (i.e., with stochastic mean, rather than stochastic volatility ) can dip below 3 (or become "negative" when one uses the convention of measuring kurtosis as excess over the Gaussian by adding 3 to the measure), the kappa metric can become negative when kurtosis is "negative". These situations require bimodality (i.e., a switching process between means under fixed variance, with modes far apart in terms of standard deviation). They do not appear to occur with unimodal distributions.

Details and derivations are presented in the appendix.

## 8.5 CONCLUSION AND CONSEQUENCES

To summarize, while the limit theorems (the law of large numbers and the central limit) are concerned with the behavior as $n \to +\infty$, we are interested in finite and exact $n$ both small and large.

We may draw a few operational consequences:

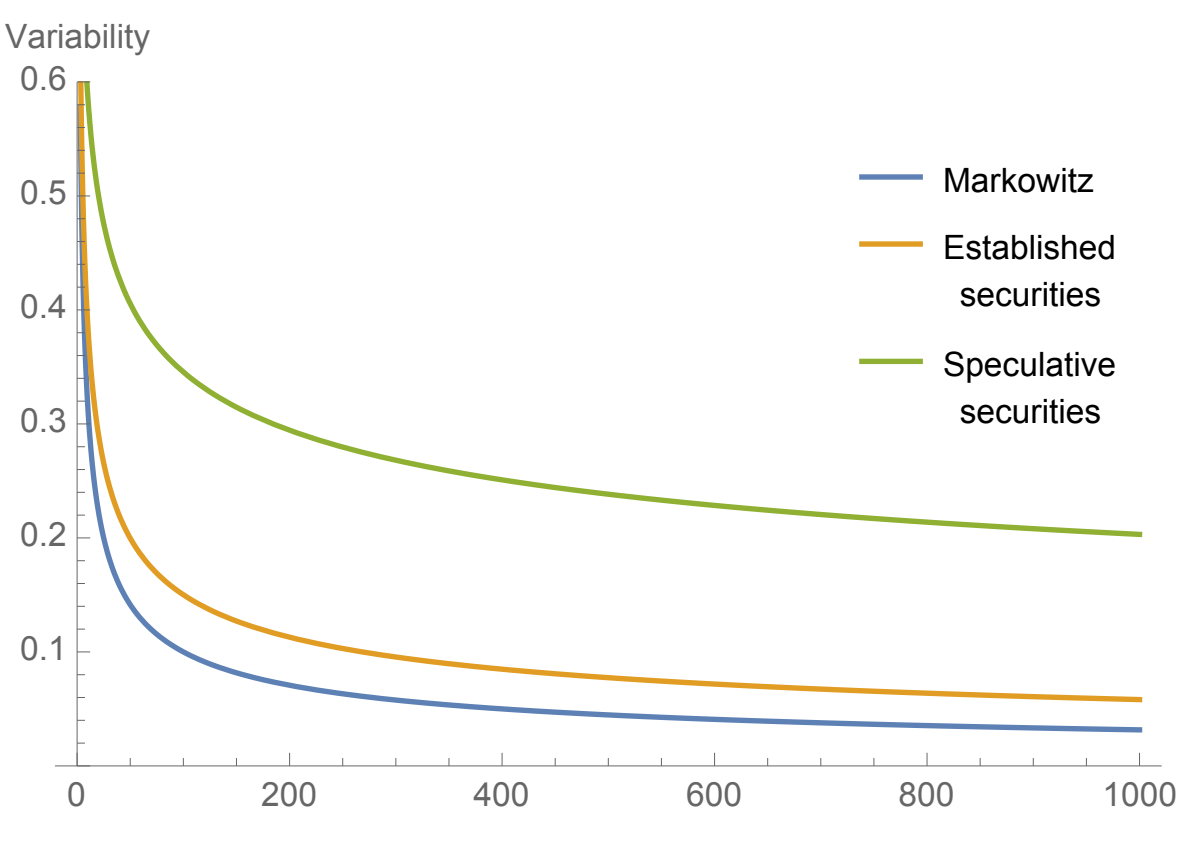

Figure 8.4: *In short, why the 1/n heuristic works: it takes many, many more securities to get the same risk reduction as via portfolio allocation according to the Markowitz. We assume to simplify that the securities are independent, which they are not, something that compounds the effect.*

### 8.5.1 Portfolio Pseudo-Stabilization

Our method can also naturally and immediately apply to portfolio construction and the effect of diversification since adding a security to a portfolio has the same "stabilizing" effect as adding an additional observation for the purpose of statistical significance. "How much data do you need?" translates into "How many securities do you need?". Clearly, the Markowicz allocation method in modern finance [195] (which seems to not be used by Markowitz himself for his own portfolio [208]) applies only for $\kappa$ near 0; people use convex heuristics, otherwise they will underestimate tail risks and "blow up" the way the famed portfolio-theory oriented hedge fund Long Term Management did in 1998 [282] [303].)

We mentioned earlier that a Pareto distribution close to the "80/20" requires up to $10^9$ more observations than a Gaussian; consider that the risk of a portfolio under such a distribution would be underestimated by at least 8 orders of magnitudes if one uses modern portfolio criteria. Following such a reasoning, one simply needs broader portfolios.

It has also been noted that there is practically no financial security that is not fatter tailed than the Gaussian, from the simple criterion of kurtosis [274], meaning Markowitz portfolio allocation is *never* the best solution. It happens that agents wisely apply a noisy approximation to the $\frac{1}{n}$ heuristic which has been classified as one of those biases by behavioral scientists but has in fact been debunked as false (a false bias is one in which, while the observed phenomenon is there, it does not constitute a "bias" in the bad sense of the word; rather it is the researcher who is mistaken owing to using the wrong tools instead of the decision-maker). This tendency to "overdiversify" has been deemed a departure from optimal investment behavior by Benartzi and Thaler [21], explained in [19] "when faced with $n$ options, divide assets evenly across the options. We have dubbed this heuristic the "$1/n$ rule."" However, broadening one's diversification is effectively as least as optimal as standard allocation(see critique by Windcliff and Boyle [320] and [74]). In short, an equally weighted portfolio outperforms the SP500 across a broad range range of metrics. But even the latter two papers didn't conceive of the full effect and properties of fat tails, which we can see here with some precision. Fig. 8.5 shows the effect for securities compared to Markowitz.

This false bias is one in many examples of policy makers "nudging" people into the wrong rationality [282] and driving them to increase their portfolio risk many folds.

A few more comments on financial portfolio risks. The SP500 has a $\kappa$ of around .2, but one needs to take into account that it is itself a basket of $n = 500$ securities, albeit unweighted and consisting of correlated members, overweighing stable stocks. Single stocks have kappas between .3 and .7, meaning a policy of "overdiversification" is a must.

Likewise the metric gives us some guidance in the treatment of data for forecasting, by establishing sample sufficiency, to state such matters as how many years of data do we need before stating whether climate conditions "have changed", see [189].

### 8.5.2 Other Aspects of Statistical Inference

So far we considered only univariate distributions. For higher dimensions, a potential area of investigation is an equivalent approach to the multivariate distribution of extreme fat tailed variables, the sampling of which is not captured by the Marchenko-Pastur (or Wishhart) distributions. As in our situation, adding variables doesn't easily remove noise from random matrices.

### 8.5.3 Final comment

As we keep saying, "statistics is never standard"; however there are heuristics methods to figure out where and by how much we depart from the standard.

## 8.6 APPENDIX, DERIVATIONS, AND PROOFS

We show here some derivations

### 8.6.1 Cubic Student T (Gaussian Basin)

The Student T with 3 degrees of freedom is of special interest in the literature owing to its prevalence in finance [117]. It is often mistakenly approximated to be Gaussian owing to the finiteness of its variance. Asymptotically, we end up with a Gaussian, but this doesn't tell us anything about the rate of convergence. Mandelbrot and Taleb [194] remarks that the cubic acts more like a power law in the distribution of the extremes, which we will elaborate here thanks to an explicit PDF for the sum.

Let $X$ be a random variable distributed with density $p(x)$:

$$p(x) = \frac{6\sqrt{3}}{\pi \left(x^2+3\right)^2}, x \in (-\infty, \infty) \tag{8.9}$$

**Proposition 8.1**
*Let $Y$ be a sum of $X_1, \ldots, X_n$, $n$ identical copies of $X$. Let $\mathbb{M}(n)$ be the mean absolute deviation from the mean for $n$ summands. The "rate" of convergence $\kappa_{1,n} = \left\{\kappa : \frac{\mathbb{M}(n)}{\mathbb{M}(1)} = n^{\frac{1}{2-\kappa}}\right\}$ is:*

$$\kappa_{1,n} = 2 - \frac{\log(n)}{\log\left(e^n n^{-n} \Gamma(n+1,n) - 1\right)} \tag{8.10}$$

where $\Gamma(.,.)$ is the incomplete gamma function $\Gamma(a,z) = \int_z^\infty dt t^{a-1} e^{-t}$.

Since the mean deviation $\mathbb{M}(n)$:

$$\mathbb{M}(n) = \begin{cases} \frac{2\sqrt{3}}{\pi} & \text{for } n = 1 \\ \frac{2\sqrt{3}}{\pi}\left(e^n n^{-n} \Gamma(n+1,n) - 1\right) & \text{for } n > 1 \end{cases} \tag{8.11}$$

The derivations are as follows. For the pdf and the MAD we followed different routes.

We have the characteristic functionfor $n$ summands:

$$\varphi(\omega) = (1 + \sqrt{3}|\omega|)^n \, e^{-n\sqrt{3}\,|\omega|}$$

The pdf of $Y$ is given by:

$$p(y) = \frac{1}{\pi} \int_0^\infty (1 + \sqrt{3}\,\omega)^n \, e^{-n\sqrt{3}\,\omega} \cos(\omega y)\, d\omega$$

After arduous integration we get the result in 8.11. Further, since the following result does not appear to be found in the literature, we have a side useful result: the PDF of $Y$ can be written as

$$p(y) = \frac{e^{n - \frac{iy}{\sqrt{3}}} \left( e^{\frac{2iy}{\sqrt{3}}} E_{-n}\left(n + \frac{iy}{\sqrt{3}}\right) + E_{-n}\left(n - \frac{iy}{\sqrt{3}}\right) \right)}{2\sqrt{3}\pi} \tag{8.12}$$

where $E_{(.)}(.)$ is the exponential integral $E_n z = \int_1^\infty \frac{e^{t(-z)}}{t^n} dt$.

Note the following identities (from the updating of Abramowitz and Stegun) [81]

$$n^{-n-1}\Gamma(n+1, n) = E_{-n}(n) = e^{-n} \frac{(n-1)!}{n^n} \sum_{m=0}^{n} \frac{n^m}{m!}$$

As to the asymptotics, we have the following result (proposed by Michail Loulakis): Reexpressing Eq. 8.11:

$$\mathbb{M}(n) = \frac{2\sqrt{3}n!}{\pi n^n} \sum_{m=0}^{n-1} \frac{n^m}{m!}$$

Further,

$$e^{-n} \sum_{m=0}^{n-1} \frac{n^m}{m!} = \frac{1}{2} + O\left(\frac{1}{\sqrt{n}}\right)$$

(From the behavior of the sum of Poisson variables as they converge to a Gaussian by the central limit theorem: $e^{-n} \sum_{m=0}^{n-1} \frac{n^m}{m!} = \mathbb{P}(X_n < n)$ where $X_n$ is a Poisson random variable with parameter $n$. Since the sum of $n$ independent Poisson random variables with parameter 1 is Poisson with parameter $n$, the Central Limit Theorem says the probability distribution of $Z_n = (X_n - n)/\sqrt{n}$ approaches a standard normal distribution. Thus $\mathbb{P}(X_n < n) = \mathbb{P}(Z_n < 0) \to 1/2$ as $n \to \infty$.[5] For another approach, see [210] for proof that $1 + \frac{n}{1!} + \frac{n^2}{2!} + \cdots + \frac{n^{n-1}}{(n-1)!} \sim \frac{e^n}{2}$.)

Using the property that $\lim_{n\to\infty} \frac{n!\exp(n)}{n^n\sqrt{n}} = \sqrt{2\pi}$, we get the following exact asymptotics:

$$\lim_{n\to\infty} \log(n)\kappa_{1,n} = \frac{\pi^2}{4}$$

thus $\kappa$ goes to 0 (i.e, the average becomes Gaussian) at speed $\frac{1}{\log(n)}$, which is excruciatingly slow. In other words, even with $10^6$ summands, the behavior cannot be

5 Robert Israel on Math Stack Exchange

summarized as that of a Gaussian, an intuition often expressed by B. Mandelbrot [194].

### 8.6.2 Lognormal Sums

From the behavior of its cumulants for $n$ summands, we can observe that a sum behaves likes a Gaussian when $\sigma$ is low, and as a lognormal when $\sigma$ is high –and in both cases we know explicitly $\kappa_n$.

The lognormal (parametrized with $\mu$ and $\sigma$) doesn't have an explicit characteristic function. But we can get cumulants $K_i$ of all orders $i$ by recursion and for our case of summed identical copies of r.v. $X_i$, $K_i^n = K_i(\sum_n X_i) = nK_i(X_1)$.

Cumulants:

$$
\begin{aligned}
K_1^n &= n e^{\mu+\frac{\sigma^2}{2}} \\
K_2^n &= n\left(e^{\sigma^2}-1\right) e^{2\mu+\sigma^2} \\
K_3^n &= n\left(e^{\sigma^2}-1\right)^2 \left(e^{\sigma^2}+2\right) e^{3\mu+\frac{3\sigma^2}{2}} \\
K_4^n &= \ldots
\end{aligned}
$$

Which allow us to compute: $\textit{Skewness} = \frac{\sqrt{e^{\sigma^2}-1}\left(e^{\sigma^2}+2\right) e^{\frac{1}{2}\left(2\mu+\sigma^2\right)-\mu-\frac{\sigma^2}{2}}}{\sqrt{n}}$ and $\textit{Kurtosis} = 3 + \frac{e^{2\sigma^2}\left(e^{\sigma^2}\left(e^{\sigma^2}+2\right)+3\right)-6}{n}$.

We can immediately prove from the cumulants/moments that:

$$
\lim_{n\to+\infty} \kappa_{1,n} = 0, \lim_{\sigma\to 0} \kappa_{1,n} = 0;
$$

and our bound on $\kappa$ becomes explicit:

Let $\kappa^*_{1,n}$ be the situation under which the sums of lognormal conserve the lognormal density, with the same first two moments. We have

$$
0 \leq \kappa^*_{1,n} \leq 1,
$$

$$
\kappa^*_{1,n} = 2 - \frac{\log(n)}{\log\left(\frac{n \operatorname{erf}\left(\frac{\sqrt{\log\left(\frac{n+e^{\sigma^2}-1}{n}\right)}}{2\sqrt{2}}\right)}{\operatorname{erf}\left(\frac{\sigma}{2\sqrt{2}}\right)}\right)}
$$

**Heuristic attempt** Among other heuristic approaches, we can see in two steps how 1) under high values of $\sigma$, $\kappa_{1,n} \to \kappa^*_{1,n}$, since the law of large numbers slows down, and 2) $\kappa^*_{1,n} \overset{\sigma\to\infty}{\to} 1$.

**Loulakis' Proof** Proving the upper bound, that for high variance $\kappa_{1,n}$ approaches 1 has been shown formally my Michail Loulakis[6] which we summarize as follows. We start with the identify $\mathbb{E}\left(|X-m|\right) = 2\int_m^\infty (x-m)f(x)dx = 2\int_m^\infty \bar{F}_X(t)dt$, where $f(.)$ is the density, $m$ is the mean, and $\bar{F}_X(.)$ is the survival function. Further, $\mathbb{M}(n) = 2\int_{nm}^\infty \bar{F}(x)dx$. Assume $\mu = \frac{1}{2}\sigma^2$, or $X = \exp\left(\sigma Z - \frac{\sigma^2}{2}\right)$ where Z is a standard normal variate. Let $S_n$ be the sum $X_1 + \ldots + X_n$; we get $\mathbb{M}(n) = 2\int_n^\infty \mathbb{P}(S_n > t)dt$. Using the property of subexponentiality ([229]), $\mathbb{P}(S_n > t) \geq \mathbb{P}(\max_{0<i\leq n}(X_i) > t) \geq n\mathbb{P}(X_1 > t) - \binom{n}{2}\mathbb{P}\left(X_1 > t\right)^2$. Now $\mathbb{P}\left(X_1 > t\right) \overset{\sigma\to\infty}{\to} 1$ and the second term to 0 (using Hölder's inequality).

Skipping steps, we get $\liminf_{\sigma\to\infty} \frac{\mathbb{M}(n)}{\mathbb{M}(1)} \geq n$, while at the same time we need to satisfy the bound $\frac{\mathbb{M}(n)}{\mathbb{M}(1)} \leq n$. So for $\sigma \to \infty$ , $\frac{\mathbb{M}(n)}{\mathbb{M}(1)} = n$, hence $\kappa_{1,n} \overset{\sigma\to\infty}{\to} 1$.

**Pearson Family Approach for Computation** For computational purposes, for the $\sigma$ parameter not too large (below $\approx .3$, we can use the Pearson family for computational convenience –although the lognormal does not belong to the Pearson class (the normal does, but we are close enough for computation). Intuitively, at low $\sigma$, the first four moments can be sufficient because of the absence of large deviations; not at higher $\sigma$ for which conserving the lognormal would be the right method.

The use of Pearson class is practiced in some fields such as information/communication theory, where there is a rich literature: for summation of lognormal variates see Nie and Chen, [212], and for Pearson IV, [45], [77].

The Pearson family is defined for an appropriately scaled density $f$ satisfying the following differential equation.

$$f'(x) = -\frac{(a_0 + a_1 x)}{b_0 + b_1 x + b_2 x^2} f(x) \tag{8.13}$$

We note that our parametrization of $a_0$, $b_2$, etc. determine the distribution within the Pearson class –which appears to be the Pearson IV. Finally we get an expression of mean deviation as a function of $n$, $\sigma$, and $\mu$.

Let $m$ be the mean. Diaconis et al [79] from an old trick by De Moivre, Suzuki [261] show that we can get explicit mean absolute deviation. Using, again, the identity $\mathbb{E}(|X-m|) = 2\int_m^\infty (x-m)f(x)\mathrm{d}x$ and integrating by parts,

$$\mathbb{E}(|X-m|) = \frac{2\left(b_0 + b_1 m + b_2 m^2\right)}{a_1 - 2b_2} f(m) \tag{8.14}$$

[6] Review of the paper version; Loulakis proposed a formal proof in place of the heuristic derivation.

We use cumulants of the n-summed lognormal to match the parameters. Setting $a_1 = 1$, and $m = \frac{b_1 - a_0}{1 - 2b_2}$, we get

$$
\begin{cases}
a_0 = \dfrac{e^{\mu+\frac{\sigma^2}{2}}\left(-12n^2+(3-10n)e^{4\sigma^2}+6(n-1)e^{\sigma^2}+12(n-1)e^{2\sigma^2}-(8n+1)e^{3\sigma^2}+3e^{5\sigma^2}+e^{6\sigma^2}+12\right)}{2\left(6(n-1)+e^{2\sigma^2}\left(e^{\sigma^2}\left(5e^{\sigma^2}+4\right)-3\right)\right)} \\
b_2 = \dfrac{e^{2\sigma^2}\left(e^{\sigma^2}-1\right)\left(2e^{\sigma^2}+3\right)}{2\left(6(n-1)+e^{2\sigma^2}\left(e^{\sigma^2}\left(5e^{\sigma^2}+4\right)-3\right)\right)} \\
b_1 = \dfrac{\left(e^{\sigma^2}-1\right)e^{\mu+\frac{\sigma^2}{2}}\left(e^{\sigma^2}\left(e^{\sigma^2}\left(e^{\sigma^2}\left(-4n+e^{\sigma^2}\left(e^{\sigma^2}+4\right)+7\right)-6n+6\right)+6(n-1)\right)+12(n-1)\right)}{2\left(6(n-1)+e^{2\sigma^2}\left(e^{\sigma^2}\left(5e^{\sigma^2}+4\right)-3\right)\right)} \\
b_0 = -\dfrac{n\left(e^{\sigma^2}-1\right)e^{2\left(\mu+\sigma^2\right)}\left(e^{\sigma^2}\left(-2(n-1)e^{\sigma^2}-3n+e^{3\sigma^2}+3\right)+6(n-1)\right)}{2\left(6(n-1)+e^{2\sigma^2}\left(e^{\sigma^2}\left(5e^{\sigma^2}+4\right)-3\right)\right)}
\end{cases}
$$

**Polynomial Expansions** Other methods, such as Gram-Charlier expansions, such as Schleher [248], Beaulieu,[17], proved less helpful to obtain $\kappa_n$. At high values of $\sigma$, the approximations become unstable as we include higher order Lhermite polynomials. See review in Dufresne [83] and [84].

### 8.6.3 Exponential

The exponential is the "entry level" fat tails, just at the border.

$$f(x) = \quad \lambda e^{-\lambda x}, \quad x \geq 0.$$

By convolution the sum $Z = X_1, X_2, \ldots X_n$ we get, by recursion, since

$$f(y) = \int_0^y f(x) f(y-x)\, dx = \lambda^2 y e^{-\lambda y} :$$

$$f_n(z) = \frac{\lambda^n z^{n-1} e^{-\lambda z}}{(n-1)!} \tag{8.15}$$

which is the gamma distribution; we get the mean deviation for $n$ summands:

$$\mathbb{M}(n) = \frac{2 e^{-n} n^n}{\lambda \Gamma(n)}, \tag{8.16}$$

hence:

$$\kappa_{1,n} = 2 - \frac{\log(n)}{n \log(n) - n - \log(\Gamma(n)) + 1} \tag{8.17}$$

We can see the asymptotic behavior is equally slow (similar to the student) although the exponential distribution is sitting at the cusp of subexponentiality:

$$\lim_{n \to \infty} \log(n) \kappa_{1,n} = 4 - 2\log(2\pi)$$

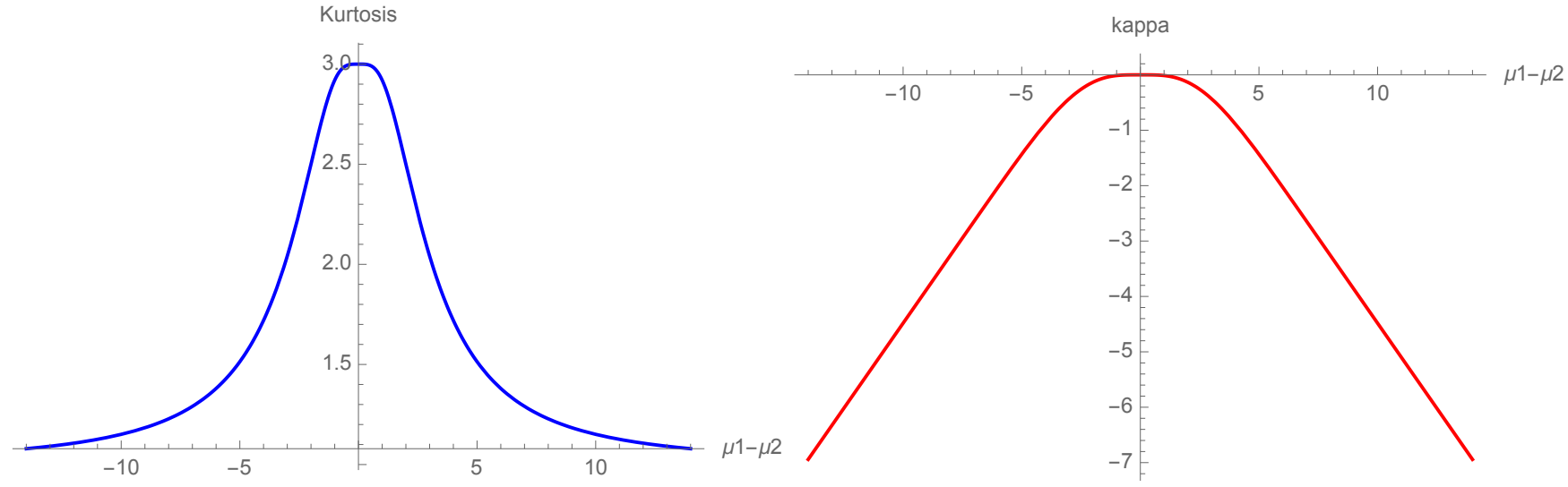

Figure 8.5: *Negative kurtosis from A.3 and corresponding kappa.*

### 8.6.4 Negative Kappa, Negative Kurtosis

Consider the simple case of a Gaussian with switching means and variance: with probability $\frac{1}{2}$, $X \sim \mathcal{N}(\mu_1, \sigma_1)$ and with probability $\frac{1}{2}$, $X \sim \mathcal{N}(\mu_2, \sigma_2)$.

These situations with thinner tails than the Gaussian are encountered with bimodal situations where $\mu_1$ and $\mu_2$ are separated; the effect becomes acute when they are separated by several standard deviations. Let d= $\mu_1 - \mu_2$ and $\sigma = \sigma_1 = \sigma_2$ (to achieve minimum kurtosis),

$$\kappa_1 = \frac{\log(4)}{\log(\pi) - 2\log\left(\frac{\sqrt{\pi}\, d e^{\frac{d^2}{4\sigma^2}} \operatorname{erf}\left(\frac{d}{2\sigma}\right) + 2\sqrt{\sigma^2}\, e^{\frac{d^2}{4\sigma^2}} + 2\sigma}{d e^{\frac{d^2}{4\sigma^2}} \operatorname{erf}\left(\frac{d}{2\sqrt{2}\sigma}\right) + 2\sqrt{\frac{2}{\pi}}\, \sigma e^{\frac{d^2}{8\sigma^2}}}\right)} + 2 \tag{8.18}$$

which we see is negative for wide values of $\mu_1 - \mu_2$.

NEXT

Next we consider some simple diagnostics for power laws with application to the SP500 . We show the differences between naive methods and those based on ML estimators that allow extrapolation into the tails.

# 9 EXTREME VALUES AND HIDDEN TAILS *,†

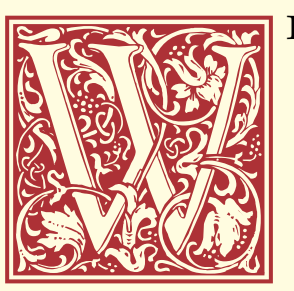
**HEN** the data is thick tailed, there is a hidden part of the distribution, not shown in past samples. Past extrema (maximum or minimum) is not a good predictor of future extrema – visibly records take place and past higher water mark is a naive estimation, what is referred to in Chapter 3 as the Lucretius fallacy, which as we saw can be paraphrased as: *the fool believes that the tallest river and tallest mountain there is equals the tallest ones he has personally seen.*

This chapter, after a brief introduction to extreme value theory, focuses on its application to thick tails. When the data is power law distributed, the maximum of $n$ observations follows a distribution easy to build from scratch. We show practically how the Fréchet distribution is, asymptotically, the maximum domain of attraction MDA of power law distributed variables.

More generally extreme value theory allows a rigorous approach to deal with extremes and the extrapolation past the sample maximum. We present some results on the "hidden mean", as it relates to a variety of fallacies in the risk management literature.

## 9.1 PRELIMINARY INTRODUCTION TO EVT

Let $X_1, \dots X_n$ be independent and distributed Pareto random variables with CDF $F(.)$ .

---

Exposition chapter with somme research.

Lucretius in *De Rerum Natura*:
Scilicet et fluvius qui visus maximus ei,
Qui non ante aliquem majorem vidit; et ingens
Arbor, homoque videtur, et omnia de genere omni
Maxima quae vidit quisque, haec ingentia fingit.

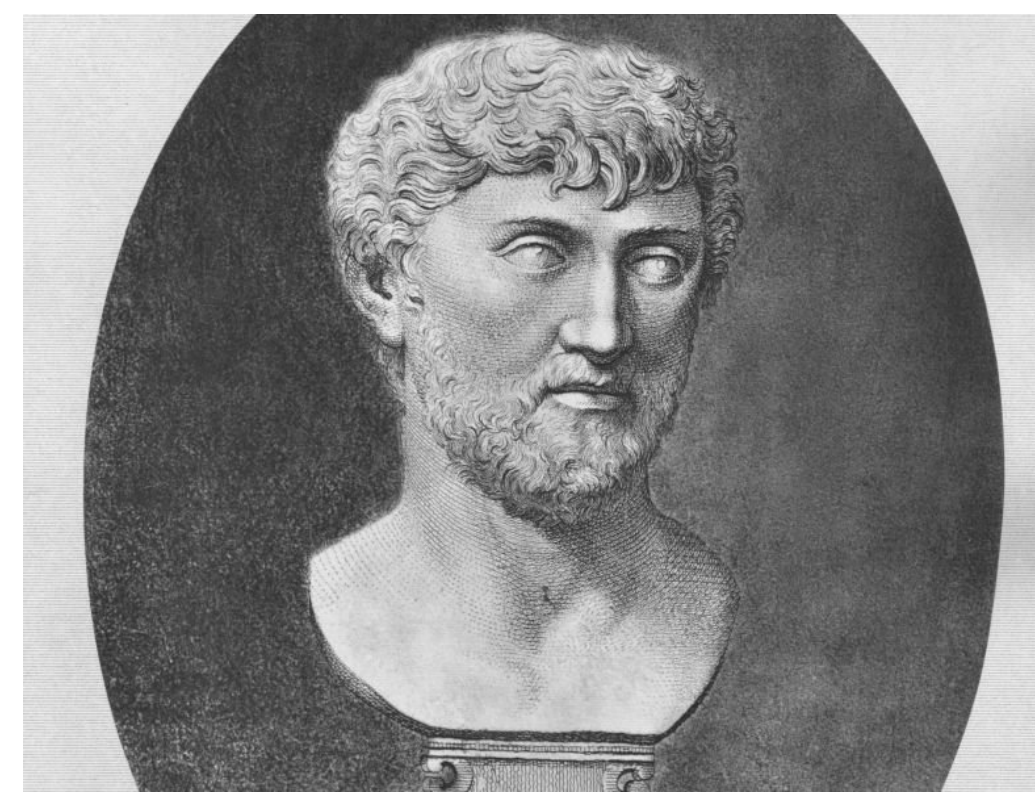

Figure 9.1: *The Roman philosophical poet Lucretius.*

We can get an exact distribution of the max (or minimum). The CDF of the maximum of the $n$ variables will be

$$\begin{aligned} \mathbb{P}\left(X_{max} \leq x\right) =& \mathbb{P}\left(X_1 \leq x, \ldots, X_n \leq x\right) = \mathbb{P}\left(X_1 \leq x\right) \\ & \cdots \mathbb{P}\left(X_n \leq x\right) = F(x)^n \end{aligned} \tag{9.1}$$

that is, the probability that all values of $x$ falling at or below $X_{max}$. The PDF is the first derivative : $\psi(x) = \frac{\partial F(x)^n}{\partial x}$.

The extreme value distribution concerns that of the maximum r.v., when $x \to x^*$, where $x^* = \sup\{x : F(x) < 1\}$ (the right "endpoint" of the distribution) is in the maximum domain of attraction, MDA [136]. In other words,

$$\max(X_1, \ldots X_n) \xrightarrow{P} x^*,$$

where $\xrightarrow{P}$ denotes convergence in probability. The central question becomes: what is the distribution of $x^*$? We said that we have the exact distribution, so as engineers we could be satisfied with the PDF from Eq. 9.1. As a matter of fact, we could get all test statistics from there, provided we have patience, computer power, and the will to investigate –it is the only way to deal with preasymptotics, that is "what happens when $n$ is small enough so $x$ is not quite $x^*$.

But it is quite useful for general statistical work to understand the general asymptotic structure.

The Fisher-Tippett-Gnedenko theorem (Embrechts et al. [97], de Haan and Ferreira [136]) states the following. If there exist sequences of "norming" constants $a_n > 0$ and $b_n \in \mathbb{R}$ such that

$$\mathbb{P}\left(\frac{M_n - b_n}{a_n} \leq x\right) \underset{n\to\infty}{\to} G(x), \tag{9.2}$$

then

$$G(x) \propto \exp\left(-(1+\xi x)^{-1/\xi}\right)$$

where $\xi$ is the extreme value index, and governs the tail behavior of the distribution. $G$ is called the (generalized) extreme value distribution, GED . The sub-families defined by $\xi = 0$, $\xi > 0$ and $\xi < 0$ correspond, respectively, to the Gumbel, Fréchet and Weibull families:

**Gumbel distribution (Type 1)** Here $\xi = 0$; rather $\lim_{\xi \to 0} \exp\left(-(\xi x + 1)^{-\frac{1}{\xi}}\right)$:

$$G(x) = \exp\left(-\exp\left(-\left(\frac{x - b_n}{a_n}\right)\right)\right) \text{ for } x \in \mathbb{R}.$$

when the distribution of $M_n$ has an exponential tail.

**Fréchet distribution(Type 2)** Here $\xi = \frac{1}{\alpha}$:

$$G(x) = \begin{cases} 0 & x \leq b_n \\ \exp\left(-\left(\frac{x-b_n}{a_n}\right)^{-\alpha}\right) & x > b_n. \end{cases}$$

when the distribution of $M_n$ has power law right tail, as we saw earlier. Note that $\alpha > 0$.

**Weibull distribution (Type 3)** Here $\xi = -\frac{1}{\alpha}$

$$G(x) = \begin{cases} \exp\left(-\left(-\left(\frac{x-b_n}{a_n}\right)\right)^{\alpha}\right) & x < b_n \\ 1 & x \geq b \end{cases}$$

when the distribution of $M_n$ has a finite support on the right (i.e., bounded maximum). Note here again that $\alpha > 0$.

### 9.1.1 How Any Power Law Tail Leads to Fréchet

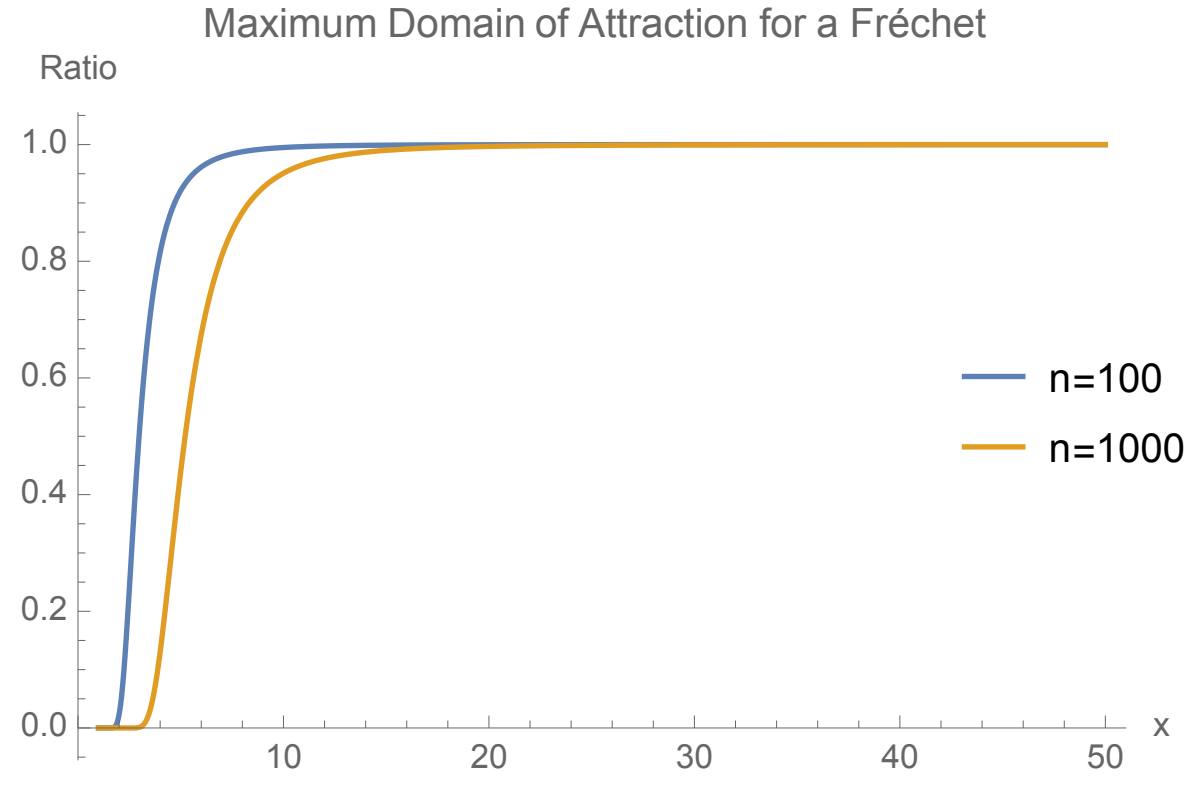

Figure 9.2: *Shows the ration of distributions of the CDF of the exact distribution over that of a Fréchet. We can visualize the acceptable level of approximation and see how x reaches the Maximum Domain of Attraction, MDA. Here $\alpha = 2$, $L = 1$. We note that the ratio for the PDF shows the same picture, unlike the Gaussian, as we will see further down.*

Let us proceed now like engineers rather than mathematicians, and consider two existing distributions, the Pareto and the Fréchet, and see how one can me made to converge to the other, in other words rederive the Fréchet from the asymptotic properties of power laws.

The reasoning we will follow next can be generalized to any Pareto-tailed variable considered above the point where slowly varying function satisfactorily approximates a constant –the "Karamata point".

The CDF of the Pareto with minimum value (and scale) $L$ and tail exponent $\alpha$:

$$F(x) = 1 - \left(\frac{L}{x}\right)^{\alpha},$$

so the PDF of the maximum of $n$ observations:

$$\psi(x) = \frac{\alpha n \left(\frac{L}{x}\right)^{\alpha} \left(1 - \left(\frac{L}{x}\right)^{\alpha}\right)^{n-1}}{x}. \tag{9.3}$$

The PDF of the Frechét:

$$\varphi(x) = \alpha \beta^{\alpha} x^{-\alpha-1} e^{\beta^{\alpha}\left(-x^{-\alpha}\right)}. \tag{9.4}$$

Let us now look for $x$ "very large" where the two functions equate, or $\psi(x^*) \to \varphi(x^*)$.

$$\lim_{x \to \infty} \frac{\psi(x)}{\varphi(x)} = n \left(\frac{1}{\beta}\right)^{\alpha} L^{\alpha}. \tag{9.5}$$

Accordingly, for $x$ deemed "large", we can use $\beta = L n^{1/\alpha}$. Equation 9.5 shows us how the tail $\alpha$ conserves across transformations of distribution:

**Property 9.1**

*The tail exponent of the maximum of i.i.d random variables is the same as that of the random variables themselves.*

Now, in practice, "where" do we approximate is shown in figure 9.2.

**Property 9.2**

*We get an exact asymptotic fitting for power law extrema.*

### 9.1.2 Gaussian Case

The Fréchet case is quite simple –power laws are usually simpler analytically, and we can get limiting parametrizations. For the Gaussian and other distributions, more involved derivations and approximations are required to fit the norming constants $a_n$ and $b_n$, usually entailing quantile functions. The seminal paper by Fisher and Tippet [112] warns us that "from the normal distribution, the limiting distribution is approached with extreme slowness" ( cited by Gasull et al. [119]).

In what follows we look for norming constants for a Gaussian, based on [140] and later developments.

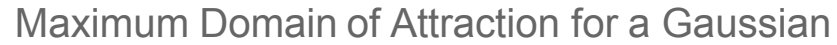

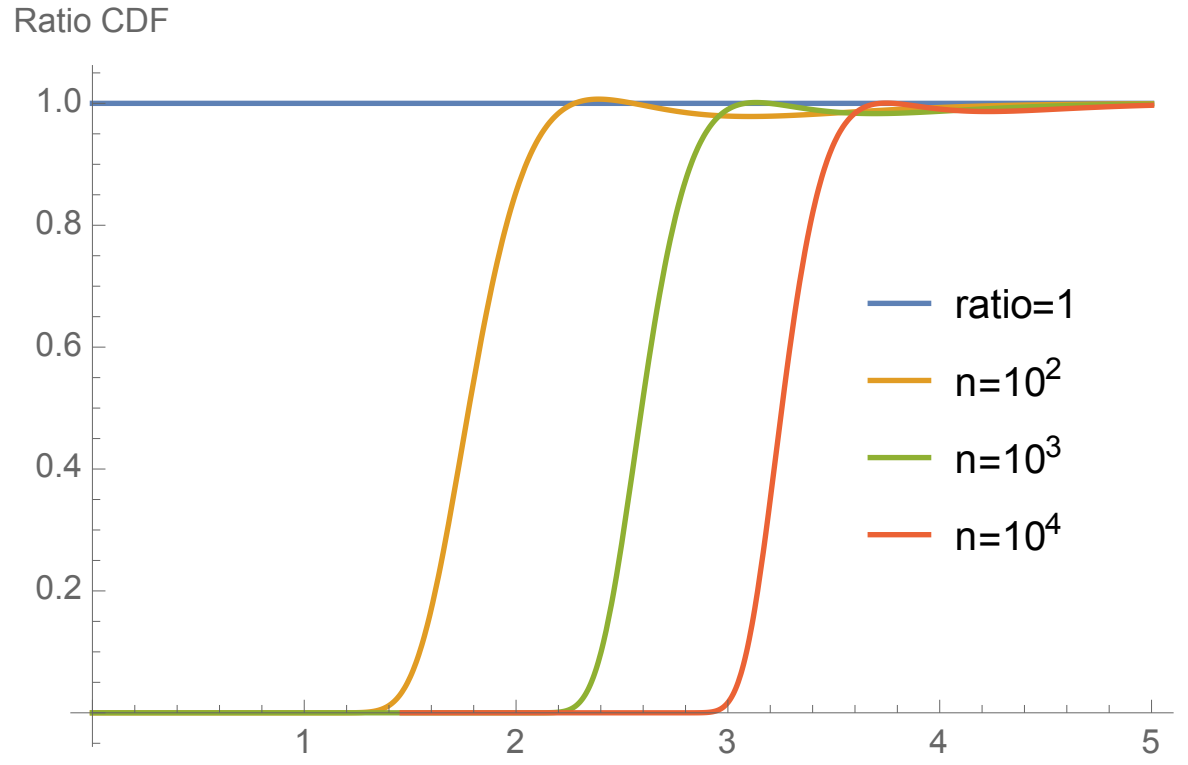

Figure 9.3: *The behavior of the Gaussian; it is hard to get a good parametrization, unlike with power laws. The y axis shows the ratio for the CDF of thee exact maximum distribution for n variables over that of the parametrized EVT.*

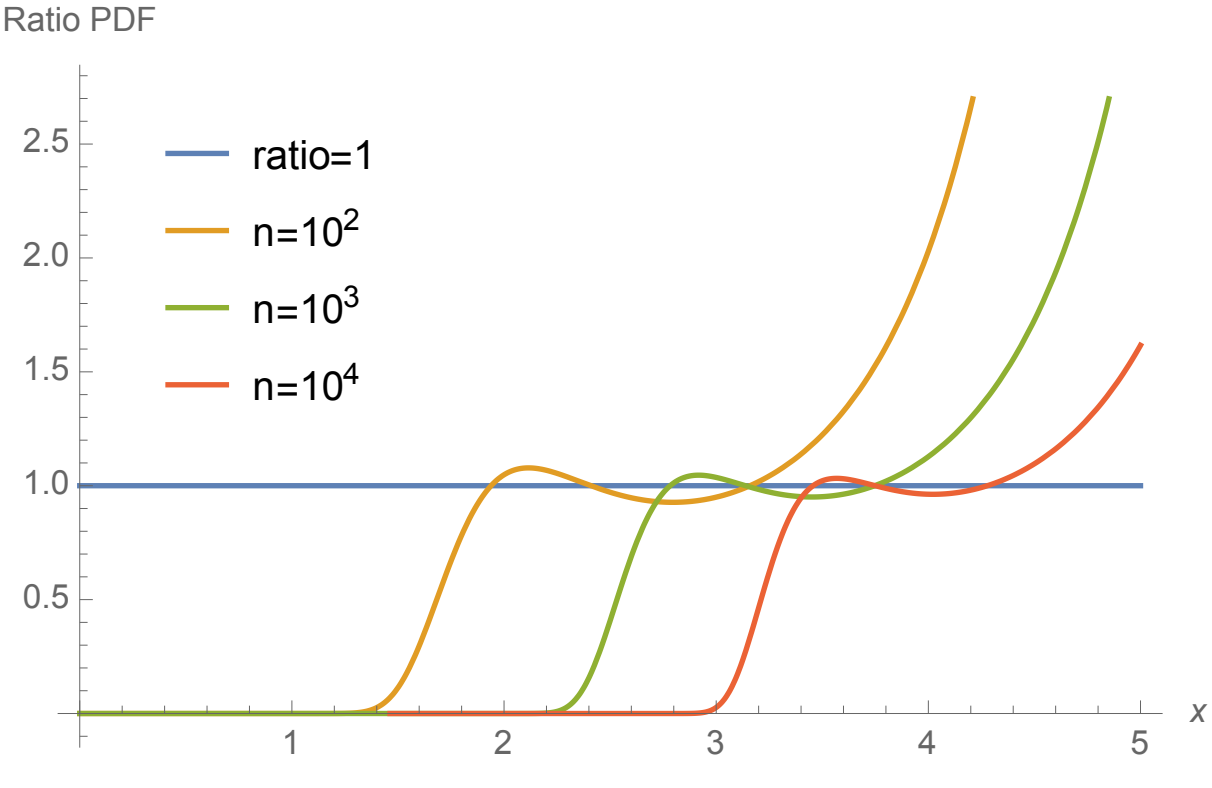

Figure 9.4: *The same as figure 9.3 but using PDF. It is not possible to obtain a good approximation in the tails.*

Consider $M_n = a_n x + b_n$ in Eq. 9.2. We assume then that $M_n$ follows the Extreme Value Distribution EVT (the CDF is $e^{-e^x}$, the mirror distribution of the Gumbel for minima, obtained by transforming the distribution of $-M_n$ where $\frac{M_n - b_n}{a_n}$ follows a Gumbel with CDF $1 - e^{-e^x}$.) [3] The parametrized CDF for $M_n$ is $e^{-e^{-\frac{x-b_n}{a_n}}}$.

An easy shortcut comes from the following approximation[4]: $a_n = \frac{b_n}{b_n^2+1}$ and $b_n = -\sqrt{2}\,\text{erfc}^{-1}\left(2\left(1-\frac{1}{n}\right)\right)$, where $\text{erfc}^{-1}$ is the inverse complementary error function.

3 The convention we follow considers the Gumbel for minima only, with the properly parametrized EVT for the maxima.

4 Embrechts et al [97] proposes $a_n = \frac{1}{\sqrt{2\log(n)}}$, $b_n = \sqrt{2\log(n)} - \frac{\log(\log(n)) + \log(4\pi)}{2\sqrt{2\log(n)}}$, the second term for $b_n$ only needed for large values of $n$. The approximation is of order $\sqrt{\log(n)}$.

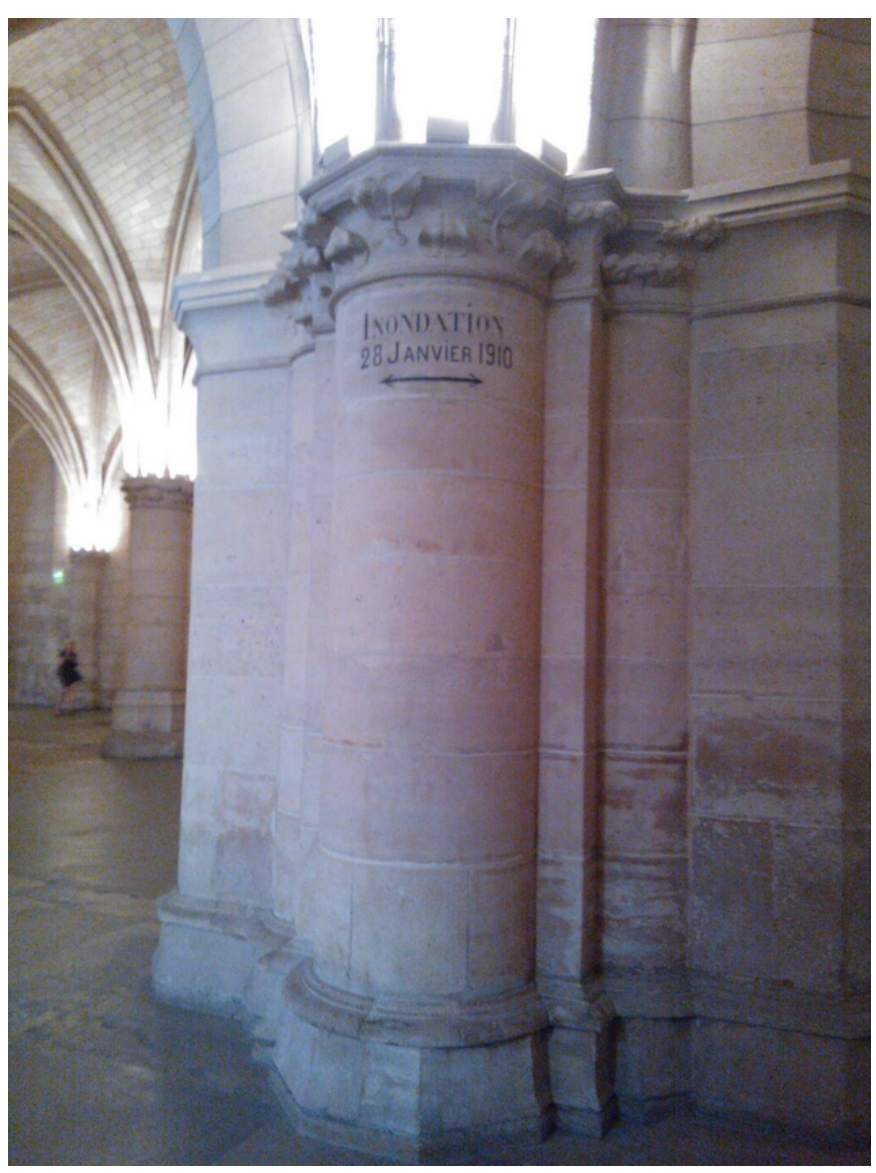

Figure 9.5: *The high watermark: the level of flooding in Paris in 1910 as a maxima. Clearly one has to consider that such record will be topped some day in the future and proper risk management consists in "how much" more than such a level one should seek protection. We have been repeating the Lucretius fallacy forever.*

**Property 9.3**

*For tail risk and properties, it is vastly preferable to work with the exact distribution for the Gaussian, namely for n variables, we have the exact distribution of the maximum from the CDF of the Standard Gaussian $F^{(g)}$:*

$$\frac{\partial F^{(g)}(K)}{\partial K} = \frac{e^{-\frac{K^2}{2}} 2^{\frac{1}{2}-n} n \, \text{erfc}\left(-\frac{K}{\sqrt{2}}\right)^{n-1}}{\sqrt{\pi}}, \tag{9.6}$$

*where erfc is the complementary error function.*

### 9.1.3 The Picklands-Balkema-de Haan Theorem

The conditional excess distribution function is the equivalent in density to the "Lindy" conditional expectation of excess deviation [136, 223], –we will make use of it in Chapter 18.

Consider an unknown distribution function $F$ of a random variable $X$; we are interested in estimating the conditional distribution function $F_u$ of the variable $X$ above a certain threshold $u$, defined as

$$F_u(y) = \mathbb{P}(X - u \leq y | X > u) = \frac{F(u+y) - F(u)}{1 - F(u)} \tag{9.7}$$

for $0 \leq y \leq x^* - u$, where $x^*$ is the finite or infinite right endpoint of the underlying distribution F. Then there exists a measurable function $\sigma(u)$ such that

$$\lim_{u \to x^*} \sup_{0 \leq x < x^* - u} \left| F_u(x) - G_{\xi,\sigma(u)}(x) \right| = 0 \tag{9.8}$$

and vice versa where $G_{\xi,\sigma(u)}(x)$ is the generalized Pareto distribution (GPD) :

$$G_{\xi,\sigma}(x) = \begin{cases} 1 - (1 + \xi x/\sigma)^{-1/\xi} & \text{if } \xi \neq 0 \\ 1 - \exp(-x/\sigma) & \text{if } \xi = 0 \end{cases} \tag{9.9}$$

If $\xi > 0$, $G_{.,.}$ is a Pareto distribution. If $\xi = 0$, $G_{.,.}$ (as we saw above) is a an exponential distribution. If $\xi = -1$, $G_{.,.}$ is uniform.

The theorem allows us to do some data inference by isolating exceedances. More on it in our discussion of wars and trends of violence in Chapter 18.

## 9.2 THE INVISIBLE TAIL FOR A POWER LAW

Consider $K_n$ the maximum of a sample of $n$ independent identically distributed variables in the power law class; $K_n = \max(X_1, X_2, \ldots, X_n)$. Let $\phi(.)$ be the density of the underlying distribution. We can decompose the moments in two parts, with the "hidden" moment above $K_0$, as shown in Fig 9.6:

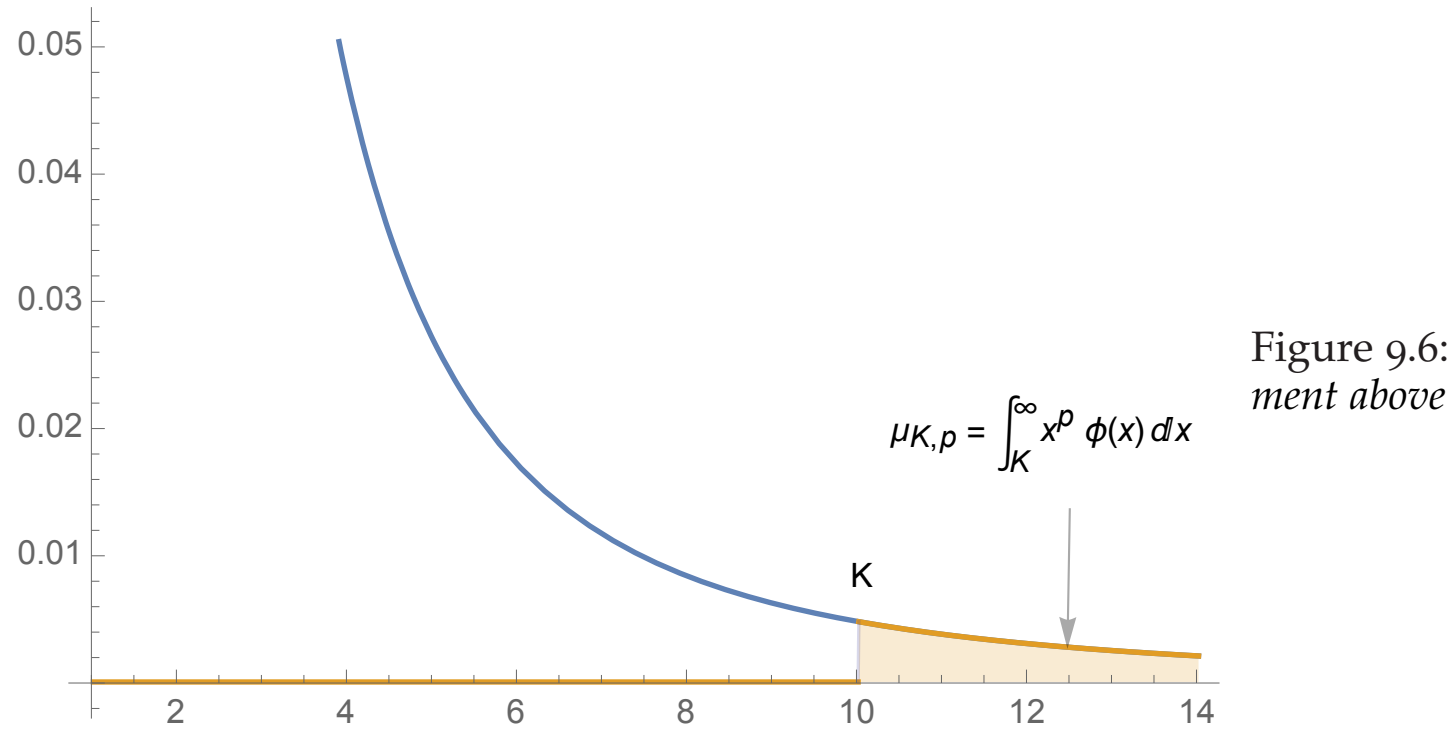

Figure 9.6: *The $p^{th}$ moment above K*

$$\mathbb{E}(X^p) = \underbrace{\int_L^{K_n} x^p \phi(x)\, dx}_{\mu_{0,p}} + \underbrace{\int_{K_n}^{\infty} x^p \phi(x)\, dx}_{\mu_{K,p}}$$

where $\mu_0$ is the visible part of the distribution and $\mu_n$ the hidden one.

We can also consider using $\phi_e$ as the empirical distribution by normalizing. Since:

$$\underbrace{\left( \int_L^{K_n} \phi_e(x) dx - \int_{K_n}^{\infty} \phi(x)\, dx \right)}_{\text{Corrected}} + \int_{K_n}^{\infty} \phi(x)\, dx = 1, \tag{9.10}$$

we can use the Radon-Nikodym derivative

$$\mathbb{E}(X^p) = \int_L^{K_n} x^p \frac{\partial \mu(x)}{\partial \mu_e(x)} \phi_e(x) dx + \int_{K_n}^{\infty} x^p \phi(x)\, dx. \tag{9.11}$$

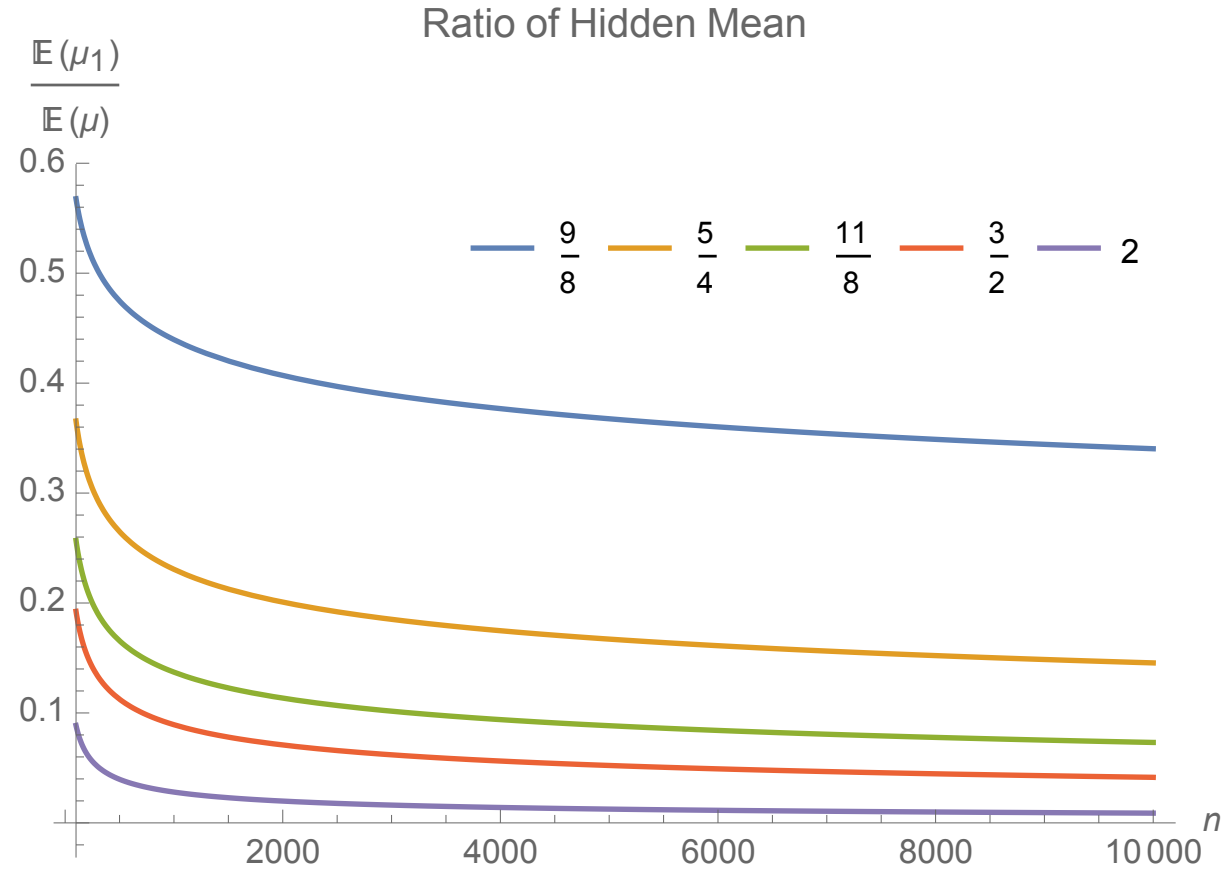

Figure 9.7: *Proportion of the hidden mean in relation to the total mean, for different parametrizations of the tail exponent $\alpha$.*

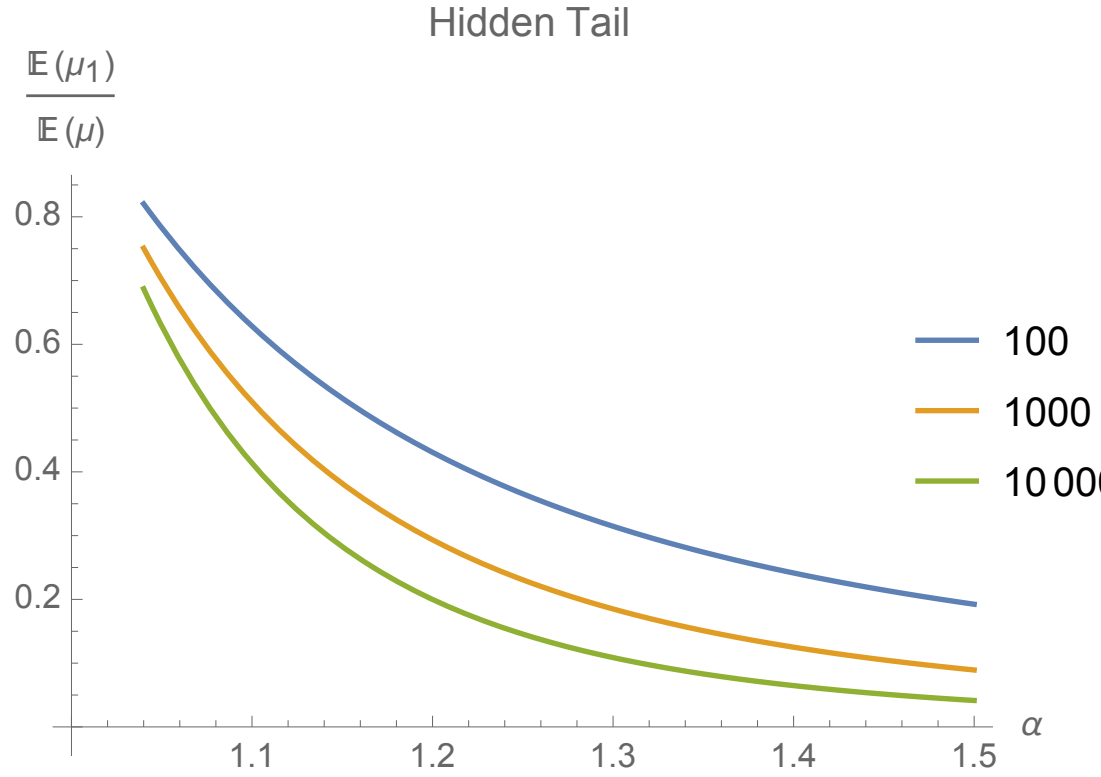

Figure 9.8: *Proportion of the hidden mean in relation to the total mean, for different sample sizes $n$.*

**Proposition 9.1**
*Let $K^*$ be point where the survival function of the random variable X can be satisfactorily approximated by a constant, that is $\mathbb{P}(X > x) \approx L^{-\alpha} x^{-\alpha}$.*

*Under the assumptions that $K > K^*$, the distribution for the hidden moment, $\mu_{K,p}$, for $n$ observation has for density $g_{(.,.,.)}(.)$:*

$$g_{n,p,\alpha}(z) = nL^{\frac{\alpha p}{p-\alpha}} \left(z - \frac{pz}{\alpha}\right)^{\frac{p}{\alpha - p}} \exp\left(n\left(-L^{\frac{\alpha p}{p-\alpha}}\right)\left(z - \frac{pz}{\alpha}\right)^{-\frac{\alpha}{p-\alpha}}\right) \tag{9.12}$$

*for $z \geq 0$, $p > \alpha$, and $L > 0$.*

The expectation of the $p^{th}$ moment above $K$, with $K > L > 0$ can be derived as

$$\mathbb{E}(\mu_{K,p}) = \frac{\alpha\left(L^p - L^\alpha K^{p-\alpha}\right)}{\alpha - p}. \tag{9.13}$$

We note that the distribution of the sample survival function (that is, $p = 0$) is an exponential distribution with PDF:

$$g_{n,0,\alpha}(z) = n e^{-nz} \tag{9.14}$$

which we can see depends only on $n$. Exceedance probability for an empirical distribution does not depend on the fatness of the tails.

To get the mean, we just need to get the integral with a stochastic lower bound $K > K_{min}$:

$$\int_{K_{min}}^{\infty} \left( \underbrace{\int_{K_n}^{\infty} x^p \phi(x)\, dx}_{\mu_{K,p}} \right) f_K(K) dK.$$

For the full distribution $g_{n,p,\alpha}(z)$, let us decompose the mean of a Pareto with scale $L$, so $K_{min} = L$.

By standard transformation, a change of variable, $K \sim \mathcal{F}(\alpha, L n^{\frac{1}{\alpha}})$ a Fréchet distribution with PDF: $f_K(K) = \alpha n K^{-\alpha-1} L^\alpha e^{n\left(-\left(\frac{L}{K}\right)^\alpha\right)}$, we get the required result.

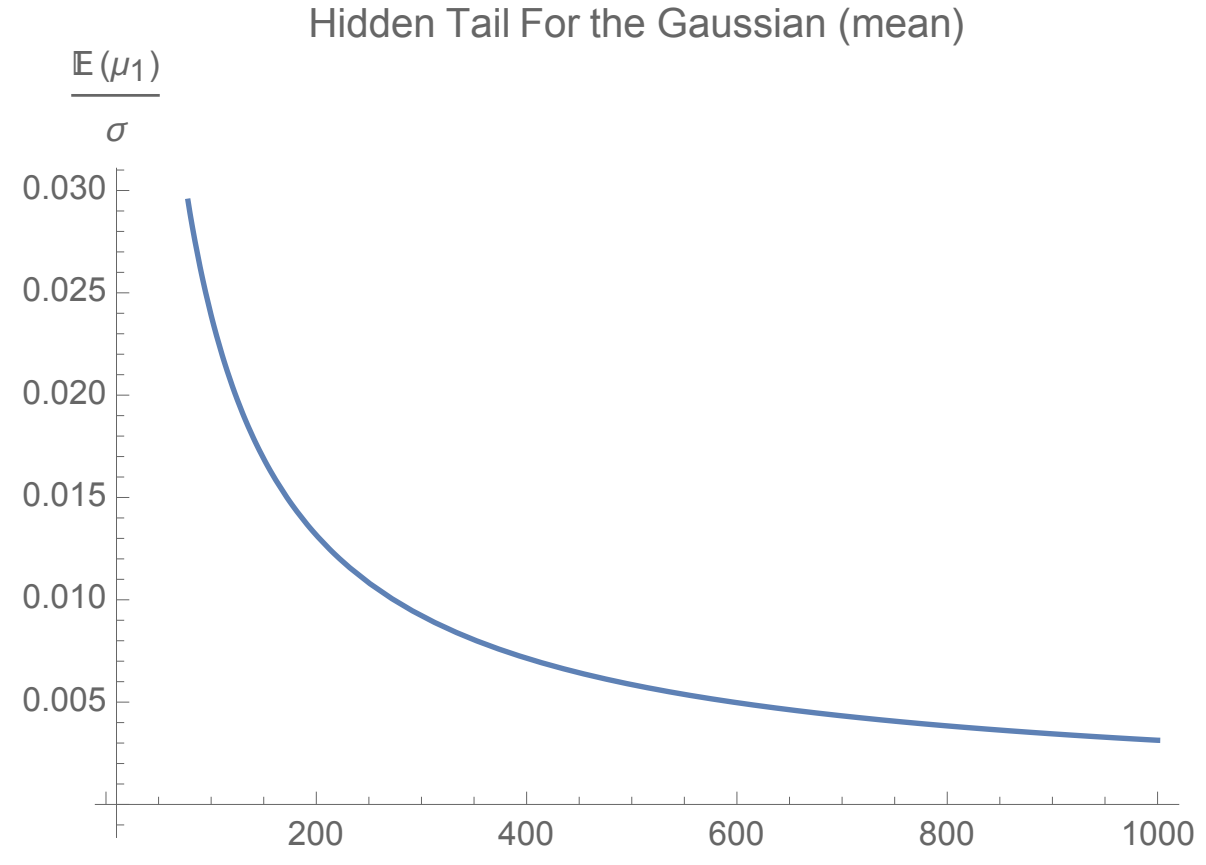

Figure 9.9: *Proportion of the hidden mean in relation to the standard deviation, for different values of $n$.*

### 9.2.1 Comparison with the Normal Distribution

For a Gaussian with PDF $\phi^{(g)}(.)$ indexed by $(g)$, $\mu_K^{(g)} = \int_K^\infty \phi^{(g)}(x)dx = \frac{2^{\frac{p}{2}-1}\Gamma\left(\frac{p+1}{2}, \frac{K^2}{2}\right)}{\sqrt{\pi}}$. As we saw earlier, without going through the Gumbel (rather EVT or "mirror-Gumbel"), it is preferable to the exact distribution of the maximum from the CDF of the Standard Gaussian $F^{(g)}$:

$$\frac{\partial F^{(g)}(K)}{\partial K} = \frac{e^{-\frac{K^2}{2}} 2^{\frac{1}{2}-n} n \operatorname{erfc}\left(-\frac{K}{\sqrt{2}}\right)^{n-1}}{\sqrt{\pi}},$$

where ertc is the complementary error function

For $p = 0$, the expectation of the "invisible tail" $\approx \frac{1}{n}$.

$$\int_0^{\infty} \frac{e^{-\frac{K^2}{2}} 2^{-n-\frac{1}{2}} n \Gamma\left(\frac{1}{2}, \frac{K^2}{2}\right) \left(\operatorname{erf}\left(\frac{K}{\sqrt{2}}\right) + 1\right)^{n-1}}{\pi} dK = \frac{1-2^{-n}}{n+1}.$$

## 9.3 APPENDIX: THE EMPIRICAL DISTRIBUTION IS NOT EMPIRICAL

Crash Beliefs From Investor Surveys

William N. Goetzmann — *Yale School of Management, Yale University*

Dasol Kim — *Weatherhead School of Management, Case Western Reserve University*

Robert J. Shiller — *Yale University*

Draft: March 19, 2016

*Please do not quote without permission*

Abstract: Historical data suggest that the base rate for a severe, single-day stock market crash is relatively low. Surveys of individual and institutional investors, conducted regularly over a 26 year period in the United States, show that they assess the probability to be much higher. We examine the factors that influence investor responses and test the role of media influence. We find evidence consistent with an availability bias. Recent market declines and adverse market events made salient by the financial press are associated with higher subjective crash probabilities. Non-market-related, rare disasters are also associated with higher subjective crash probabilities.

Keywords: Crash Beliefs, Availability Bias, Investor Surveys

JEL: G00, G11, G23, E03, G02

Figure 9.10: *The base rate fallacy, revisited —or, rather in the other direction. The "base rate" is an empirical evaluation that bases itself on the worst past observations, an error identified in [273] as the fallacy identified by the Roman poet Lucrecius in De rerum natura of thinking the tallest future mountain equals the tallest on has previously seen. Quoted without permission after warning the author.*

There is a prevalent confusion about the nonparametric empirical distribution based on the following powerful property: as $n$ grows, the errors around the empirical histogram for cumulative frequencies are Gaussian *regardless of the base distribution*, even if the true distribution is fat-tailed (assuming infinite support). For the CDF (or survival functions) are both uniform on $[0, 1]$, and, further, by the Donsker theorem, the sequence $\sqrt{n}\,(F_n(x) - F(x))$ ($F_n$ is the observed CDF or survival function for $n$ summands, $F$ the true CDF or survival function) converges in distribution to a Normal Distribution with mean 0 and variance $F(x)\,(1 - F(x))$ (one may find even stronger forms of convergence via the Glivenko– Cantelli theorem).

Owing to this remarkable property, one may mistakenly assume that the effect of tails of the distribution converge in the same manner independently of the distribution. Further, and what contributes to the confusion, the variance, $F(x)(1 - F(x))$ for both empirical CDF and survival function, drops at the extremes –though not its corresponding payoff.

In truth, and that is a property of extremes, the error effectively increases in the tails if one multiplies by the deviation that corresponds to the probability.

For the U.S. stock market indices, while the first method is deemed to be ludicrous, using the second method leads to an underestimation of the payoff in the tails of between 5 and 70 times, as can be shown in Figure 9.11. The topic is revisited again in Chapter 11 with our discussion of the difference between binary and continuous payoffs, and the conflation between probability and real world payoffs when said payoffs are from a fat tailed distribution.

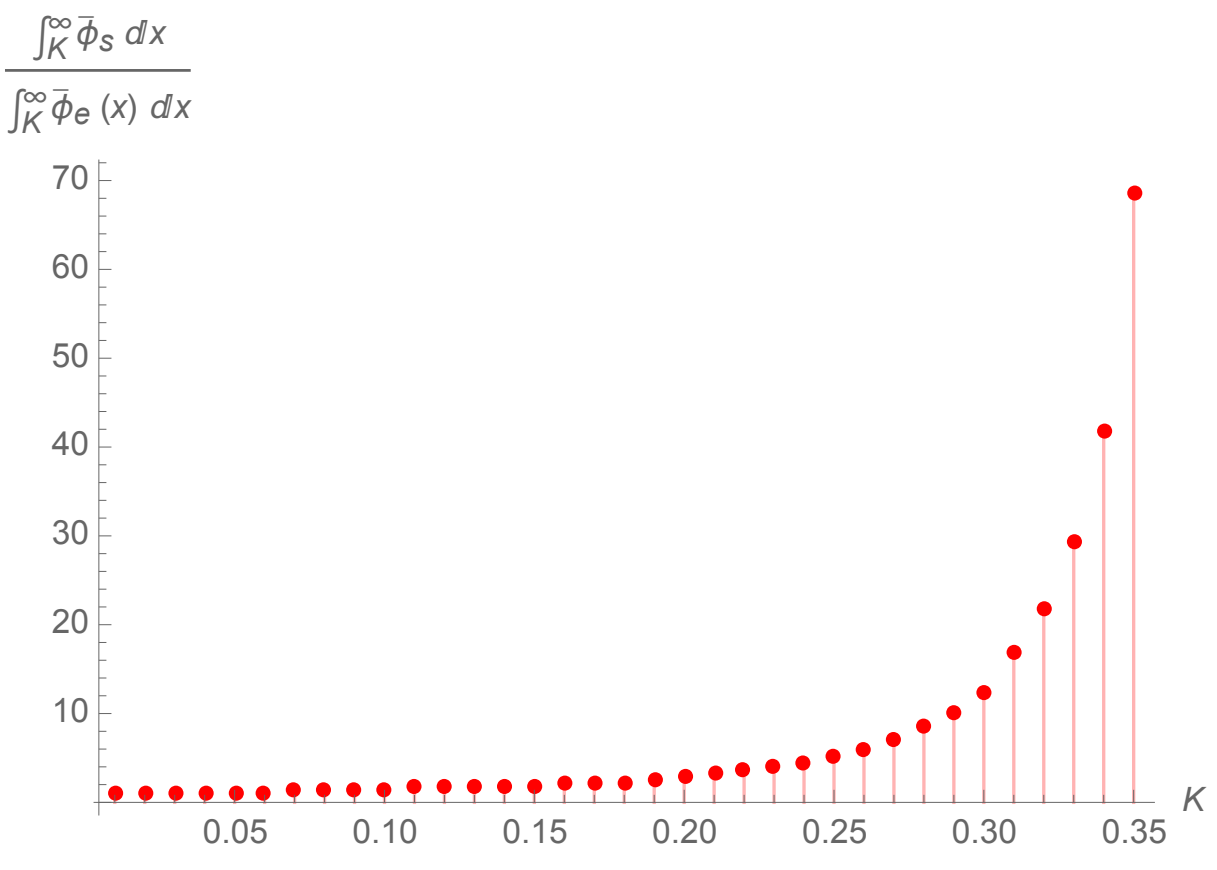

Figure 9.11: *This figure shows the relative value of tail CVar-style measure compared to that from the (smoothed) empirical distribution. The deep tail is underestimated up to 70 times by current methods, even those deemed "empirical".*

# C | GROWTH RATE & OUTCOME: NOT IN THE SAME DISTRIBUTION CLASS

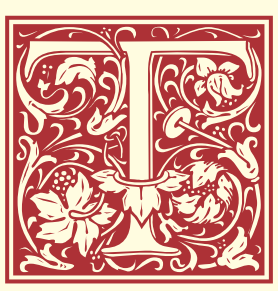

HE AUTHOR AND Pasquale Cirillo showed that fatalities from pandemics follow power laws with a tail exponent patently lower than 1. This means that all the information resides in the tail. So unless one has some real reason to ignore general and unconditional statistics (of the style "this one is different"), one should not base risk management decisions on the behavior of the expected average or some point estimate.

The following paradox arose: $X_t$ the number of fatalities between period $t_0$ and $t$ is Paretian with undefined mean. However its exponential growth rate is not! It is going to be thin tailed, exponentially distributed or so.

Cirillo and Taleb (2020) [57] (CT) showed via extreme value theory that Pandemics have a tail $\alpha < 1$ when seem in $X_T$, the number of fatalities at some date $T$ in the future, with survival function $\mathbb{P}(X > x) = L(x)x^{-\alpha}$. Assume to simplify that, with a minimum value $L$, $L(x) \sim L$ so we get the survival function

$$\mathbb{P}(X > x) = Lx^{-\alpha}. \tag{C.1}$$

## C.1 THE PUZZLE

Consider the usual model,

$$X_t = X_0 e^{r(t-t_0)}, \tag{C.2}$$

where

$$r = \frac{1}{(t-t_0)} \int_{t_0}^{t} r_s ds \tag{C.3}$$

and $r_s$ is the instantaneous rate. Normalize the distribution to $L = 1$. We can thus prove the following (under the assumption above that $X_t$ has survival function in Eq. 15.13):

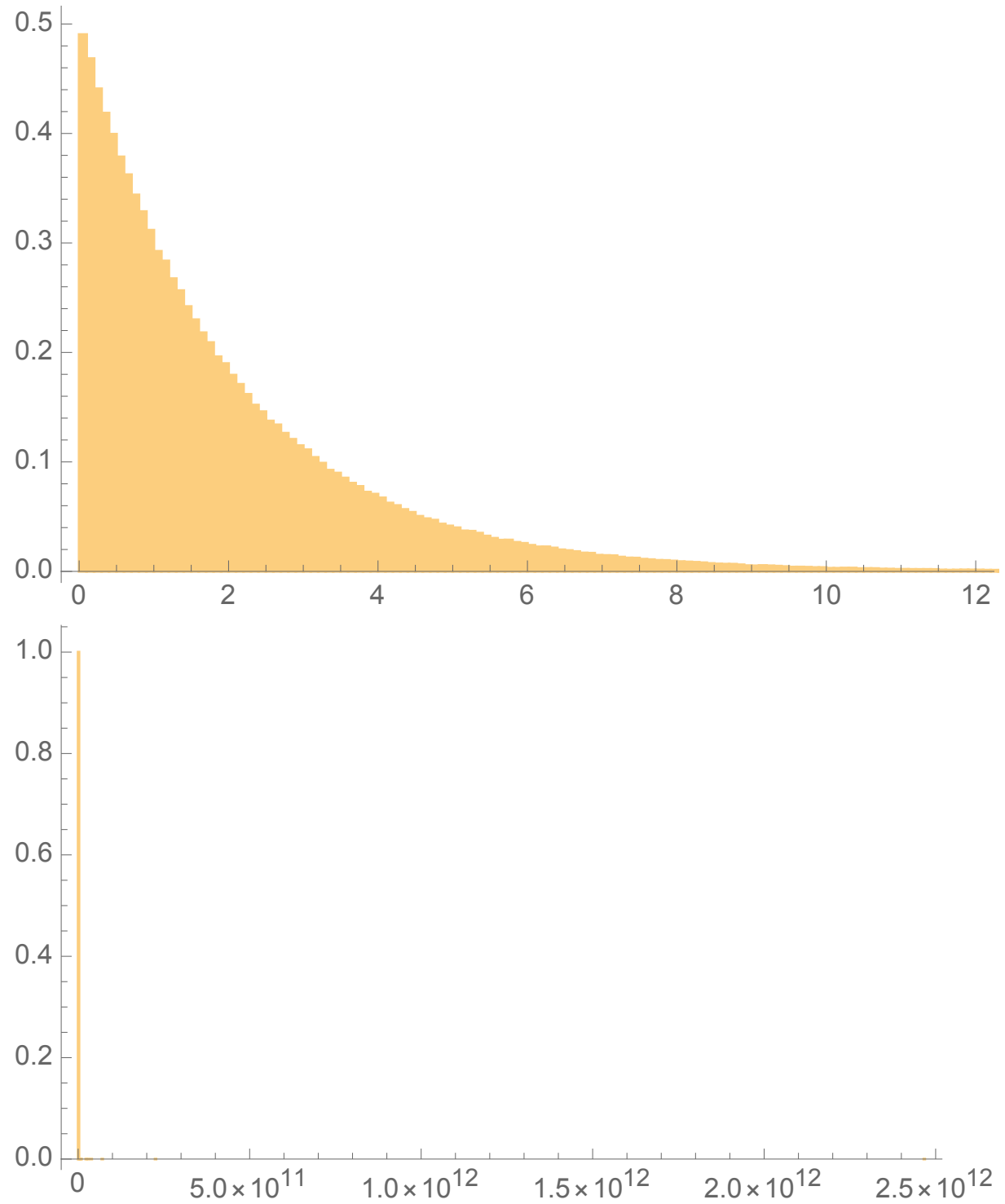

Figure C.1: *Above, a histogram of* $10^6$ *realisations of* $r$*, from an exponential distribution with parameter* $\lambda = \frac{1}{2}$*. Below, the histogram of* $X = e^r$*. We can see the difference between the two distributions. The sample kurtosis are 9 and* $10^6$ *respectively (in fact it is theoretically infinite for the second); all values for the latter are dominated by a single large deviation.*

**Theorem 1**

*If* $r$ *has support in* $(-\infty, \infty)$*, then its PDF* $\varphi$ *for the scaled rate* $\rho = r(t - t_0)$ *can be parametrized as*

$$\varphi(\rho) = \begin{cases} \frac{e^{-\frac{\rho}{b}}}{2b} & \rho \geq 0 \\ \frac{e^{-\frac{-\rho}{b}}}{2b} & \textit{otherwise} \end{cases}$$

*where* $b = \frac{1}{\alpha}$.

*If* $r$ *has support in* $(0, \infty)$*, then its PDF* $\varphi$

$$\varphi(\rho) = \begin{cases} \alpha e^{\alpha(-\rho)} & \rho \geq 0 \\ 0 & \textit{otherwise} \end{cases}$$

What we have here is versions of the exponential or double exponential distribution (Laplace).

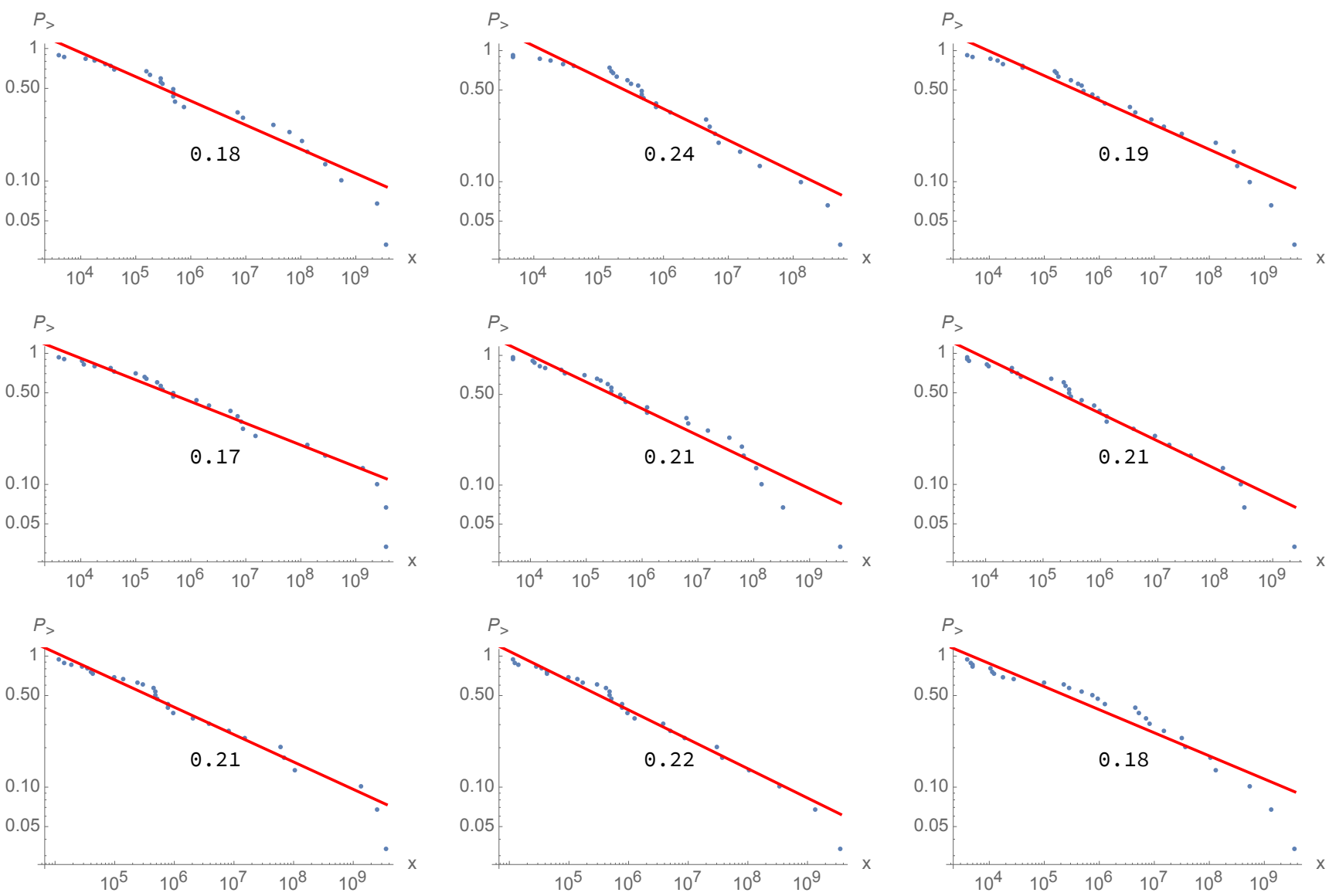

Figure C.2: *We take the 60 largest pandemics and randomly subselect half. We normalize the data by today's population. The Paretian properties (and parametrization) are robust to these perturbations. EVT provides slightly higher tail exponent, but firmly below one. This about the lowest tail exponent the authors have ever seen in their careers.*

**Remark C.1**

*Implication: One cannot naively translate properties between the rate of growth r and $X_T$ because errors in r could be small (but nonzero) for r but explosive in translation owing to the exponentiation.*

The reverse is also true: if $r$ follows an exponential distribution then $X_T$ must be Pareto distributed as in Eq. 15.13.

The sketch of the derivation is as follows, via change of variables. Let $r$ follow a distribution with density $\phi$, with support $(a, b)$; under some standard conditions, $u = g(r)$ follows a new distribution with density

$$\psi(u) = \frac{\phi\left(g^{(-1)}(u)\right)}{g'\left(g^{(-1)}(u)\right)},$$

and support $[g(a), g(b)]$.

## C.2 PANDEMICS ARE REALLY FAT TAILED

Figure C.2 shows how we get a power law with a low $\alpha$ no matter what random subsample from the data we select. We used in [57] extreme value theory but the graphs show the preliminary analysis (not in paper). This is the lowest tail exponent we have ever seen anywhere. The implication is that epidemiology studies need to be used for research but policy making must be done using EVT or simply relying on precautionary principles –that is, to cut the cancer when it is cheap to do so.[1]

[1] A gross error is the reliance on single point forecast for policy –in fact as we show in chapter 11, it is always wrong to use the forecast of the survival function –to gauge forecasting ability thinking it "how science is done" –outside binary bets.

# D THE LARGE DEVIATION PRINCIPLE, IN BRIEF

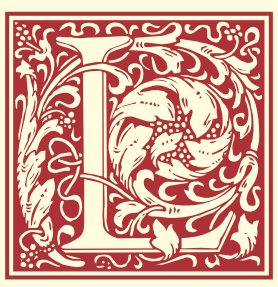

LET US RETURN to the Cramer bound with a rapid exposition of the surrounding literature. The idea behind the tall vs. rich outliers in 3.1 is that under some conditions, their tail probabilities decay exponentially. Such a property that is central in risk management –as we mentioned earlier, the catastrophe principle explains that for diversification to be effective, such exponential decay is necessary.

The large deviation principle helps us understand such a tail behavior.It also helps us figure out why things do not blow-up under thin-tailedness –but, more significantly, why they could under fat tails, or where the Cramèr condition is not satisfied [138].

Let $M_N$ be the mean of a sequence of realizations (identically distributed) of $N$ random variables. For large $N$, consider the tail probability:

$$\mathbb{P}(M_N > x) \approx e^{-N I(x)},$$

where $I(.)$ is the Cramer (or rate) function (Varadhan [314], Denbo and Zeitouni [71]). If we know the distribution of $X$, then, by Legendre transformation, $I(x) = \sup_{\theta>0} \left(\theta x - \lambda(\theta)\right)$, where $\lambda(\theta) = \log \mathbb{E}\left(e^{\theta(X)}\right)$ is the cumulant generating function.

The behavior of the function $\theta(x)$ informs us on the contribution of a single event to the overall payoff. (It connects us to the *Cramer condition* which requires existence of exponential moments).

A special case for Bernoulli variables is the Chernoff Bound, which provides tight bounds for that class of discrete variables.

## SIMPLE CASE: CHERNOFF BOUND

A binary payoff is subjected to very tight bounds. Let $(X_i)_{1<i\leq n}$ be a sequence of independent Bernouilli trials taking values in $\{0,1\}$, with $\mathbb{P}(X = 1) = p$ and $\mathbb{P}(X = 0) = 1 - p$. Consider the sum $S_n = \sum_{1<i\leq n} X_i$. with expectation $\mathbb{E}(S_n)$= $np = \mu$. Taking $\delta$ as a "distance from the mean", the Chernoff bounds gives:
For any $\delta > 0$

$$\mathbb{P}\left(S \geq (1+\delta)\mu\right) \leq \left(\frac{e^{\delta}}{(1+\delta)^{1+\delta}}\right)^{\mu}$$

and for $0 < \delta \leq 1$

$$\mathbb{P}\left(S \geq (1+\delta)\mu\right) \leq 2e^{-\frac{\mu\delta^2}{3}}$$

Let us compute the probability of coin flips $n$ of having 50% higher than the true mean, with p= $\frac{1}{2}$ and $\mu = \frac{n}{2}$: $\mathbb{P}\left(S \geq \left(\frac{3}{2}\right)\frac{n}{2}\right) \leq 2e^{-\frac{\mu\delta^2}{3}} = e^{-n/24}$, which for $n = 1000$ happens every 1 in $1.24 \times 10^{18}$.

**Proof** The Markov bound gives: $\mathbb{P}(X \geq c) \leq \frac{\mathbb{E}(X)}{c}$, but allows us to substitute $X$ with a positive function $g(x)$, hence $\mathbb{P}(g(x) \geq g(c)) \leq \frac{\mathbb{E}(g(X))}{g(c)}$. We will use this property in what follows, with $g(X) = e^{\omega X}$.

Now consider $(1+\delta)$, with $\delta > 0$, as a "distance from the mean", hence, with $\omega > 0$,

$$\mathbb{P}\left(S_n \geq (1+\delta)\mu\right) = \mathbb{P}\left(e^{\omega S_n} \geq e^{\omega(1+\delta)\mu}\right) \leq e^{-\omega(1+\delta)\mu}\mathbb{E}(e^{\omega S_n}) \tag{D.1}$$

Now $\mathbb{E}(e^{\omega S_n}) = \mathbb{E}(e^{\omega \Sigma(X_i)}) = \mathbb{E}(e^{\omega X_i})^n$, by independence of the stopping time, becomes $\left(\mathbb{E}(e^{\omega X})\right)^n$.

We have $\mathbb{E}(e^{\omega X}) = 1 - p + pe^{\omega}$. Since $1 + x \leq e^x$,

$$\mathbb{E}(e^{\omega S_n}) \leq e^{\mu(e^{\omega a}-1)}$$

Substituting in D.1, we get:

$$\mathbb{P}\left(e^{\omega S_n} \geq e^{\omega(1+\delta)\mu}\right) \leq e^{-\omega(1+\delta)\mu}e^{\mu(e^{\omega}-1)} \tag{D.2}$$

We tighten the bounds by playing with values of $\omega$ that minimize the right side. $\omega^* = \left\{\omega : \frac{\partial e^{\mu\left(e^{\omega}-1\right)-(\delta+1)\mu\omega}}{\partial\omega} = 0\right\}$ yields $\omega^* = \log(1+\delta)$.

Which recovers the bound: $e^{\delta\mu}(\delta+1)^{(-\delta-1)\mu}$.

An extension of Chernoff bounds was made by Hoeffding [152] who broadened it to bounded independent random variables, but not necessarily Bernouilli..

# E | CALIBRATING UNDER PARETIANITY

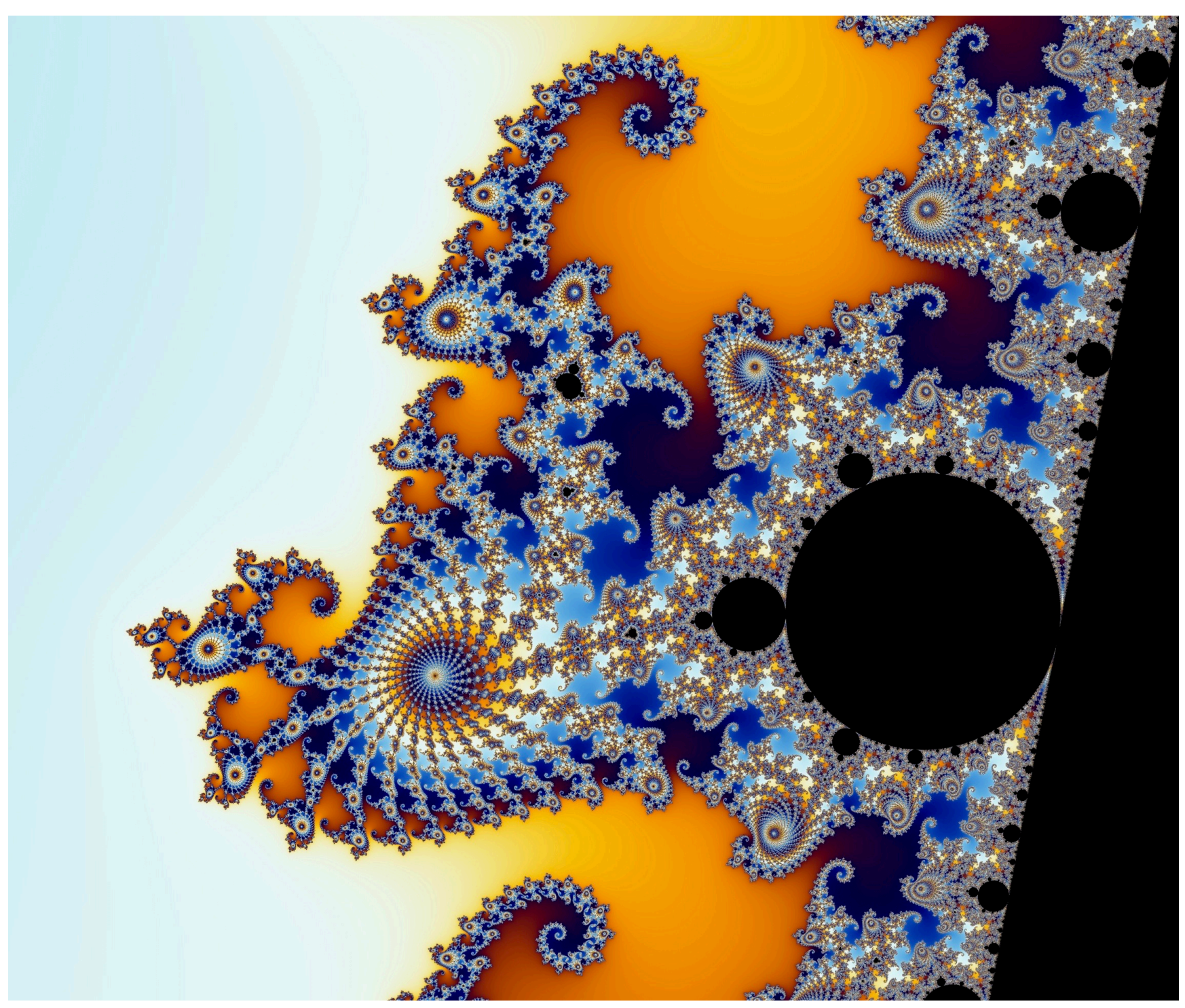

Figure E.1: *The great Benoit Mandelbrot linked fractal geometry to statistical distributions via self-affinity at all scales. When asked to explain his work, he said: "rugosité", meaning"roughness" –it took him fifty years to realize that was his specialty. (Seahorse Created by Wolfgang Beyer, Wikipedia Commons.)*

We start with a refresher:

**Definition E.1** (Power Law Class $\mathfrak{P}$)
*The r.v. $X \in \mathbb{R}$ belongs to $\mathfrak{P}$, the class of slowly varying functions (a.k.a. Paretiantail or*

*power law-tailed) if its survival function (for the variable taken in absolute value) decays asymptotically at a fixed exponent* $\alpha$*, or* $\alpha'$*, that is*

$$\mathbb{P}(X > x) = L(x)\, x^{-\alpha} \tag{E.1}$$

*(right tail) or*

$$\mathbb{P}(-X > x) = L(x)\, x^{-\alpha'} \tag{E.2}$$

*(left tail)*
*where* $\alpha, \alpha' > 0$ *and* $L : (0, \infty) \to (0, \infty)$ *is a slowly varying function,defined as*

$$\lim_{x \to \infty} \frac{L(kx)}{L(x)} = 1$$

*for all* $k > 0$*.*

The happy result is that the parameter $\alpha$ obeys an inverse gamma distribution that converges rapidly to a Gaussian and does not require a large $n$ to get a good estimate. This is illustrated in Figure E.2, where we can see the difference in fit.

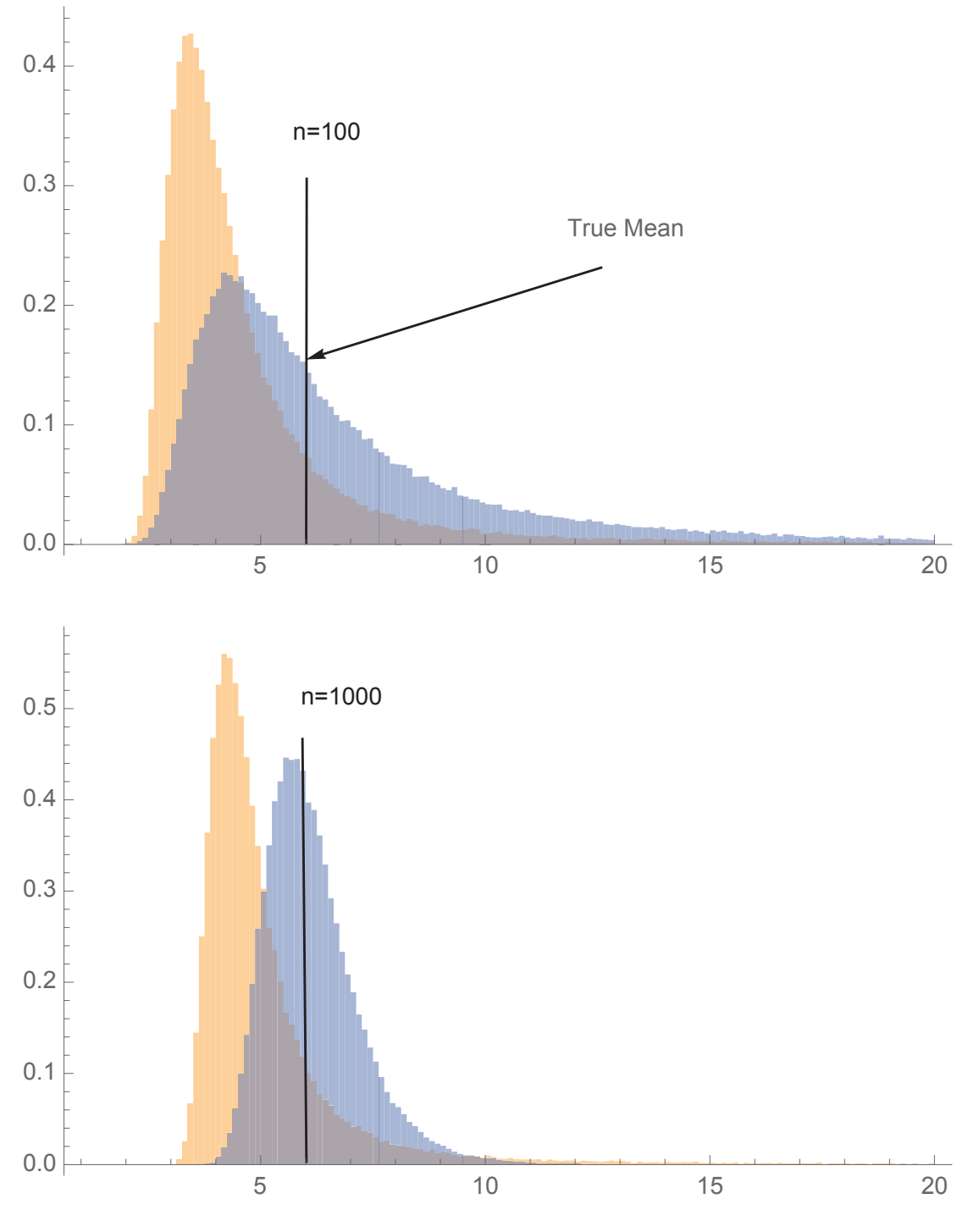

Figure E.2: *Monte Carlo Simulation* ($10^5$) *of a comparison of sample mean (Methods 1 and 2) vs maximum likelihood mean estimations (Method 3) for a Pareto Distribution with* $\alpha = 1.2$ *(yellow and blue respectively), for* $n = 100, 1000$*. We can see how the MLE tracks the distribution more reliably. We can also observe the bias as Methods 1 and 2 underestimate the sample mean in the presence of skewness in the data. We need* $10^7$ *more data in order to get the same error rate.*

As we saw, there is a problem with the so-called finite variance power laws: finiteness of variance doesn't help as we saw in Chapter 8.

## E.1 DISTRIBUTION OF THE SAMPLE TAIL EXPONENT

Consider the standard Pareto distribution for a random variable $X$ with PDF:

$$\phi_X(x) = \alpha L^\alpha x^{-\alpha-1}, x > L \tag{E.3}$$

Assume $L = 1$ by scaling.

The likelihood function is $\mathcal{L} = \prod_{i=1}^n \alpha x_i^{-\alpha-1}$. Maximizing the Log of the likelihood function (assuming we set the minimum value) $\log(\mathcal{L}) = n(\log(\alpha) + \alpha \log(L)) - (\alpha+1)\sum_{i=1}^n \log(x_i)$ yields: $\hat{\alpha} = \frac{n}{\sum_{i=1}^n \log(x_i)}$. Now consider $l = -\frac{\sum_{i=1}^n \log X_i}{n}$. Using the characteristic function to get the distribution of the average logarithm yield:

$$\psi(t)^n = \left( \int_1^\infty f(x) \exp\left(\frac{it \log(x)}{n}\right) dx \right)^n = \left(\frac{\alpha n}{\alpha n - it}\right)^n$$

which is the characteristic function of the gamma distribution $(n, \frac{1}{\alpha n})$. A standard result is that $\hat{\alpha}' \triangleq \frac{1}{l}$ will follow the inverse gamma distribution with density:

$$\phi_{\hat{\alpha}}(a) = \frac{e^{-\frac{\alpha n}{\hat{\alpha}}} \left(\frac{\alpha n}{\hat{\alpha}}\right)^n}{\hat{\alpha}\Gamma(n)}, a > 0$$

.

**Debiasing** Since $\mathbb{E}(\hat{\alpha}) = \frac{n}{n-1}\alpha$ we elect another –unbiased– random variable $\hat{\alpha}' = \frac{n-1}{n}\hat{\alpha}$ which, after scaling, will have for distribution $\phi_{\hat{\alpha}'}(a) = \frac{e^{\frac{\alpha - \alpha n}{a}} \left(\frac{\alpha(n-1)}{a}\right)^{n+1}}{\alpha \Gamma(n+1)}$.

**Truncating for $\alpha > 1$** Given that values of $\alpha \leq 1$ lead to absence of mean we restrict the distribution to values greater than $1+\epsilon$, $\epsilon > 0$. Our sampling now applies to lower-truncated values of the estimator, those strictly greater than 1, with a cut point $\epsilon > 0$, that is, $\sum \frac{n-1}{\log(x_i)} > 1+\epsilon$, or $\mathbb{E}(\hat{\alpha}|_{\hat{\alpha}>1+\epsilon})$: $\phi_{\hat{\alpha}''}(a) = \frac{\phi_{\hat{\alpha}'}(a)}{\int_{1+\epsilon}^\infty \phi_{\hat{\alpha}'}(a)\, \mathrm{d}a}$, hence the distribution of the values of the exponent conditional of it being greater than 1 becomes:

$$\phi_{\hat{\alpha}''}(a) = \frac{e^{\frac{\alpha n^2}{a - a n}} \left(\frac{\alpha n^2}{a(n-1)}\right)^n}{a\left(\Gamma(n) - \Gamma\left(n, \frac{n^2 \alpha}{(n-1)(\epsilon+1)}\right)\right)}, a \geq 1+\epsilon \tag{E.4}$$

So as we can see in Figure E.2, the "plug-in" mean via the tail $\alpha$ might be a good approach under one-tailed Paretianity.

# 10 "IT IS WHAT IT IS": DIAGNOSING THE SP500†

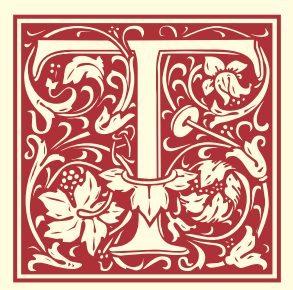

HIS IS A DIAGNOSTICS tour of the properties of the SP500 index in its history. We engage in a battery of tests and check what statistical picture emerges. Clearly, its returns are power law distributed (with some added complications, such as an asymmetry between upside and downside) which, again, invalidates common methods of analysis. We look, among other things to:

- The behavior of Kurtosis under aggregation (as we lengthen the observation window )
- The behavior of the conditional expectation $\mathbb{E}(X|_{X>K})$ for various values of $K$.
- The maximum-to-sum plot (MS Plot).
- Drawdowns (that is, maximum excursions over a time window)
- Extremes and records to see if extremes are independent.

These diagnostics allow us to confirm that an entire class of analyses in $L2$ such as modern portfolio theory, factor analysis, GARCH, conditional variance, or stochastic volatility are methodologically (and practically) invalid.

## 10.1 PARETIANITY AND MOMENTS

**The problem** As we said in the Prologue, switching from thin-tailed to fat-tailed is not just *changing the color of the dress*. The finance and economic rent seekers hold the message "we know it is fat tailed" but then fail to grasp the consequences on many things such as the slowness of the law of large numbers and the failure of sample means or higher moments to be sufficient statistic (

---

This is largely a graphical chapter made to be read from the figures more than from the text as the arguments largely repose on the absence of convergence in the graphs.

as well as the ergodicity effect, among others). Likewise it leads to a bevy of uninformative analytics in the investment industry.

Paretianity is clearly defined by the absence of some higher moment, exhibited by lack of convergence under LLN.

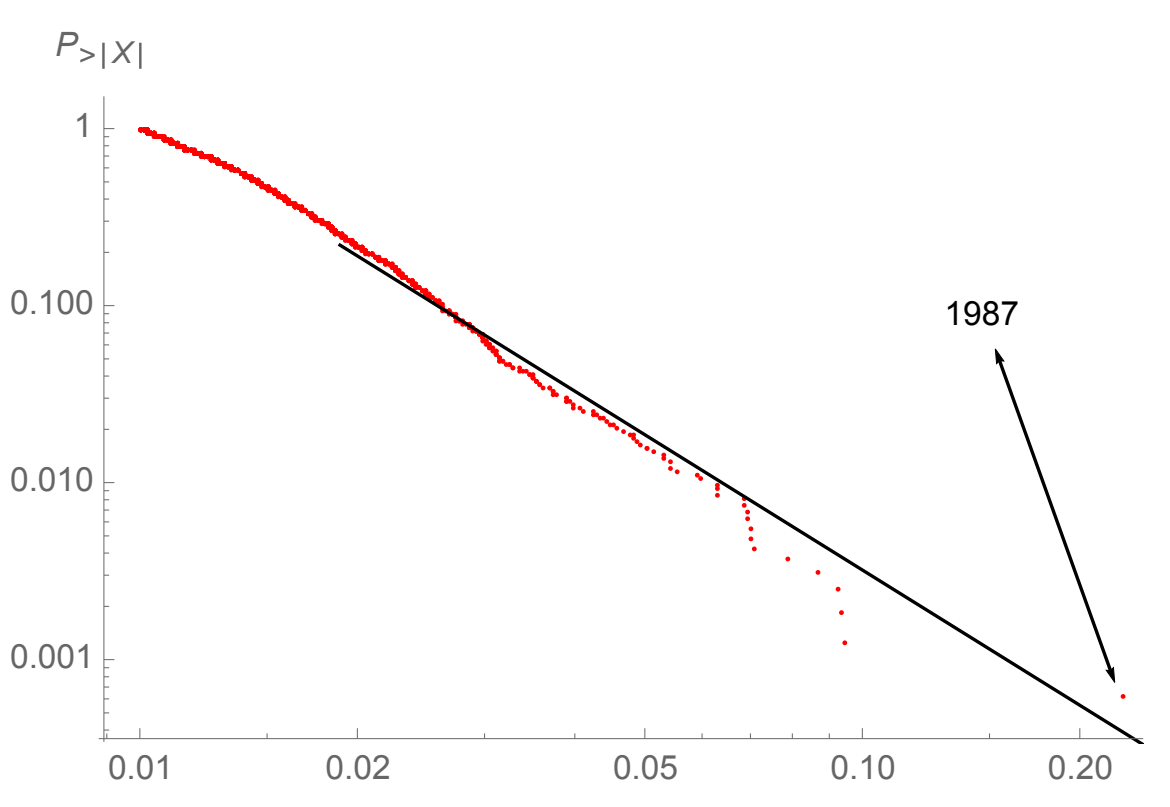

Figure 10.1: *Visual Identification of Paretianity on a standard log-log plot with (absolute) returns on the horizontal axis and the survival function on the vertical one. If one removes the data point corresponding to the crash of 1987, a lognormal would perhaps work, or some fat tailed mixed distribution outside the power law class. For we can see the survival function becoming vertical, indicative of an infinite asymptotic tail exponent. But as the saying goes, all one needs is a single event...*

**Remark 10.1**

*Given that:*

*1) the regularly varying class has no higher moments than $\alpha$, more precisely,*

- *if $p > \alpha$, $\mathbb{E}(X^p) = \infty$ if $p$ is even or the distribution has one-tailed support and*
- *$\mathbb{E}(X^p)$ is undefined if $p$ is odd and the distribution has two-tailed support,*

*and*

*2) distributions outside the regularly varying class have all moments $\forall p \in \mathbb{N}^+$, $\mathbb{E}(X^p) < \infty$.*

*$\exists p \in \mathbb{N}^+$ s.t. $\mathbb{E}(X^p)$ is either undefined or infinite $\Leftrightarrow X \in \mathfrak{P}$.*

Next we examine ways to detect "infinite" moments. Much confusion attends the notion of infinite moments and its identification since by definition sample moments are finite and measurable under the counting measure. We will rely on the nonconvergence of moments. Let $\|\mathbf{X}\|_p$ be the weighted $p$-norm

$$\|\mathbf{X}\|_p \triangleq \left( \frac{1}{n} \sum_{i=1}^{n} |x_i|^p \right)^{1/p},$$

we have the property of power laws:

$$\mathbb{E}(X^p) \nless \infty \Leftrightarrow \|\mathbf{x}\|_p \text{ is not convergent.}$$

**Question** How does belonging to the class of Power Law tails (with $\alpha \leq 4$) cancel much of the methods in $L2$?

Section 5.10 shows the distribution of the mean deviation of the second moment for a finite variance power law. Simply, even if the fourth moment does not exist, under infinite higher moments, the second moment of the variance has itself infinite variance, and we fall in the sampling problems seen before: just as with a power law of $\alpha$ close to 1 (though slightly above it), the mean exists but will never be observed, in a situation of infinite third moment, the observed second moment will fail to be informative as it will almost never converge to its value.

## 10.2 CONVERGENCE TESTS

Convergence laws can help us *exclude* some classes of probability distributions.

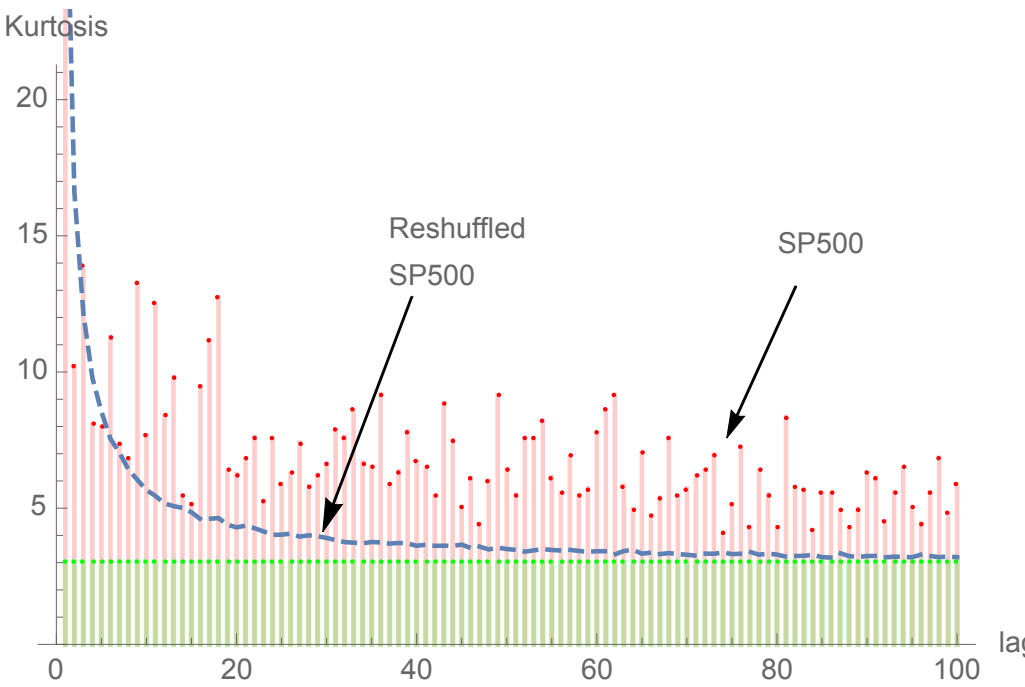

Figure 10.2: *Visual convergence diagnostics for the kurtosis of the SP500 over the past 17000 observations. We compute the kurtosis at different lags for the raw SP500 and reshuffled data. While the $4^{th}$ norm is not convergent for raw data, it is clearly so for the reshuffled series. We can thus assume that the "fat tailedness" is attributable to the temporal structure of the data, particularly the clustering of its volatility. See Table 7.1 for the expected drop at speed $1/n$ for thin-tailed distributions.*

### 10.2.1 Test 1: Kurtosis under Aggregation

If Kurtosis existed, it would end up converging to that of a Gaussian as one lengthens the time window. So we test for the computations of returns over longer and longer lags, as we can see in Fig 10.2.

**Result** The verdict as shown in Figure 10.2 is that the one-month kurtosis is not lower than the daily kurtosis and, as we add data, no drop in kurtosis is observed. Further we would expect a drop $\sim n^{-1}$. This allows us to safely eliminate numerous classes, which includes stochastic volatility in its simple formulations such as gamma variance. Next we will get into the technicals of the point and the strength of the evidence.

A typical misunderstanding is as follows. In a note "What can Taleb learn from Markowitz" [307], Jack L. Treynor, one of the founders of portfolio theory, defended the field with the argument that the data may be fat tailed "short term" but in something called the "long term" things become Gaussian. Sorry, it is not so. (We add the ergodic problem that blurs, if not eliminate, the distinction between long term and short term).

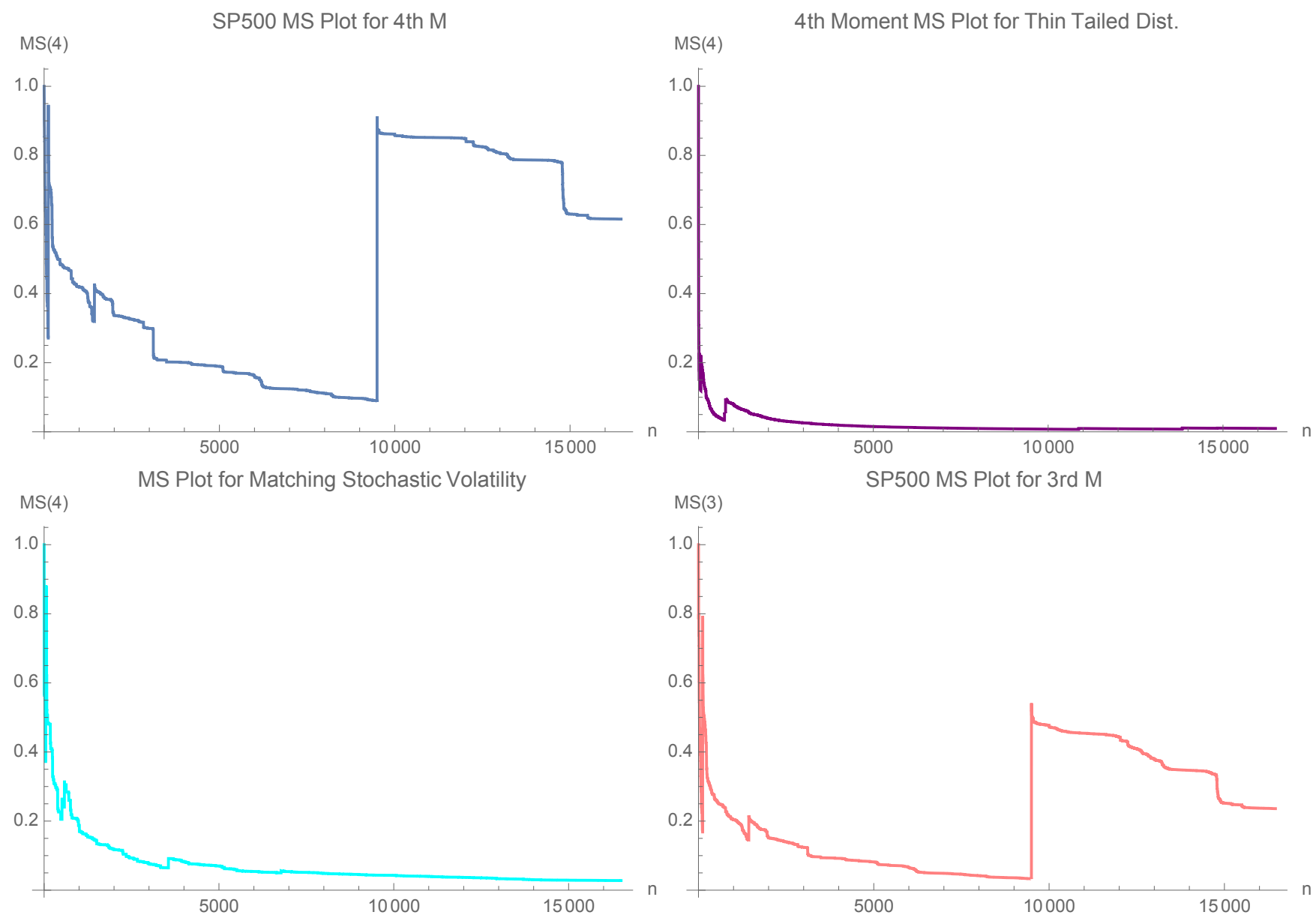

Figure 10.3: *MS Plot (or "law of large numbers for p moments") for p = 4 for the SP500 compared to p = 4 for a Gaussian and stochastic volatility for a matching Kurtosis ( 30) over the entire period. Convergence, if any, does not take place in any reasonable time. MS Plot for moment p = 3 for the SP500 compared to p = 4 for a Gaussian. We can safely say that the* $4^{th}$ *moment is infinite and the* $3^{rd}$ *one is indeterminate*

The reason is that, simply we cannot possibly talk about "Gaussian" if kurtosis is infinite, even when lower moments exist. Further, for $\alpha \approx 3$, Central limit operates very slowly, requires $n$ of the order of $10^6$ to become acceptable, not what we have in the history of markets. [31]

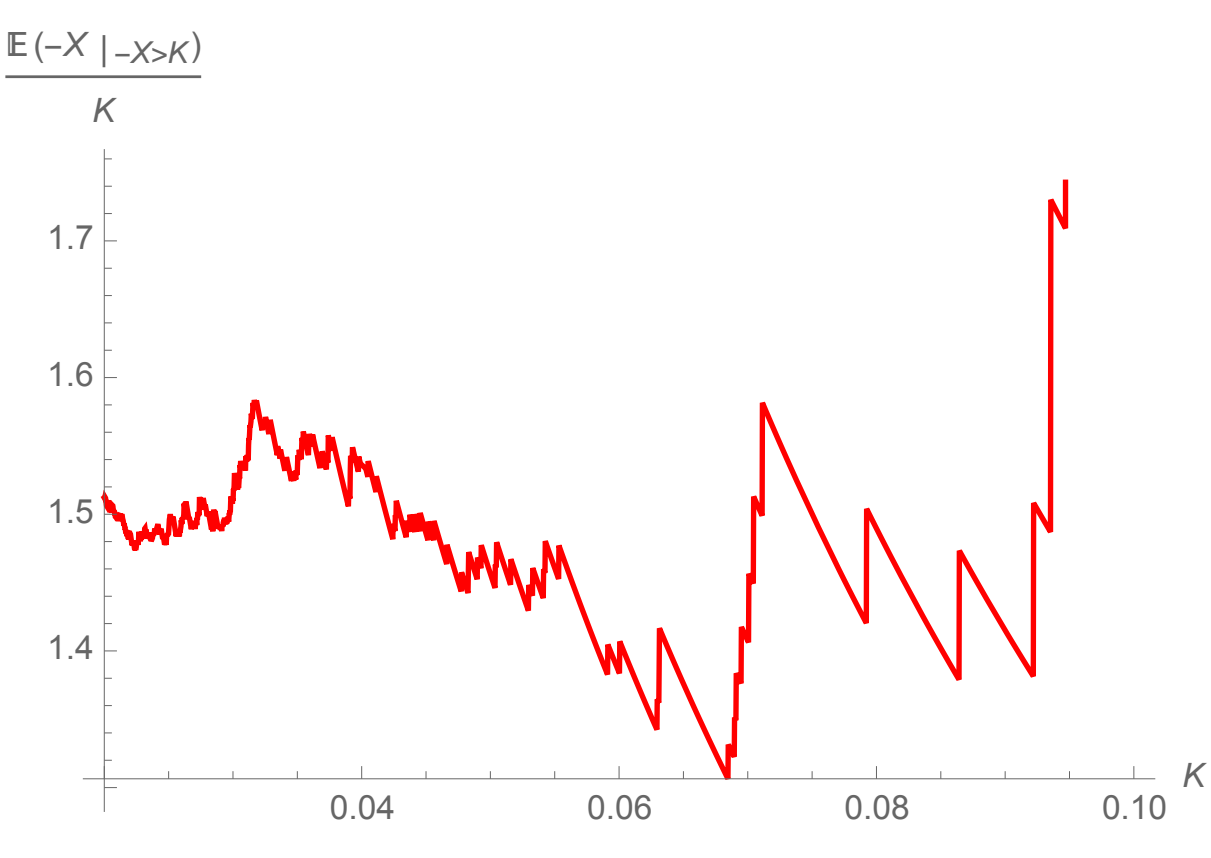

Figure 10.4: *The "Lindy test" or Condexp, using the conditional expectation below K as K varies as test of scalability. As we move K, the measure should drop.*

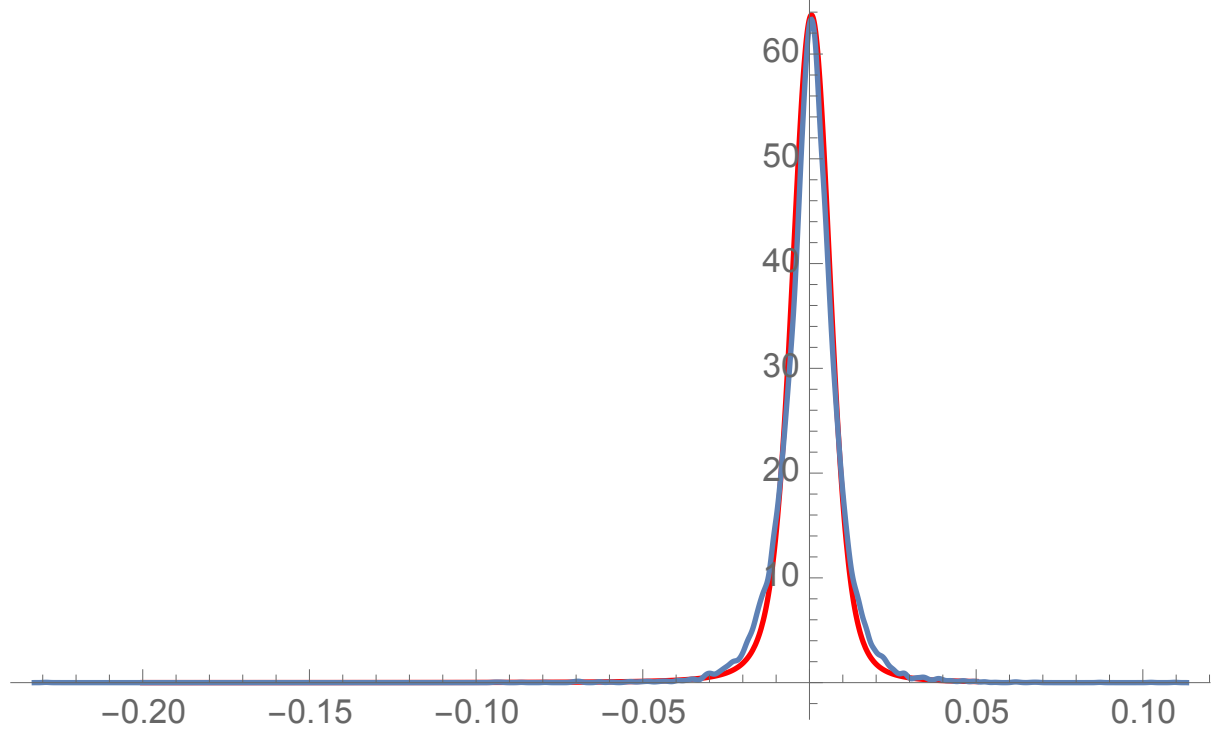

Figure 10.5: *The empirical distribution could conceivably fit a Lévy stable distribution with $\alpha_l = 1.62$.*

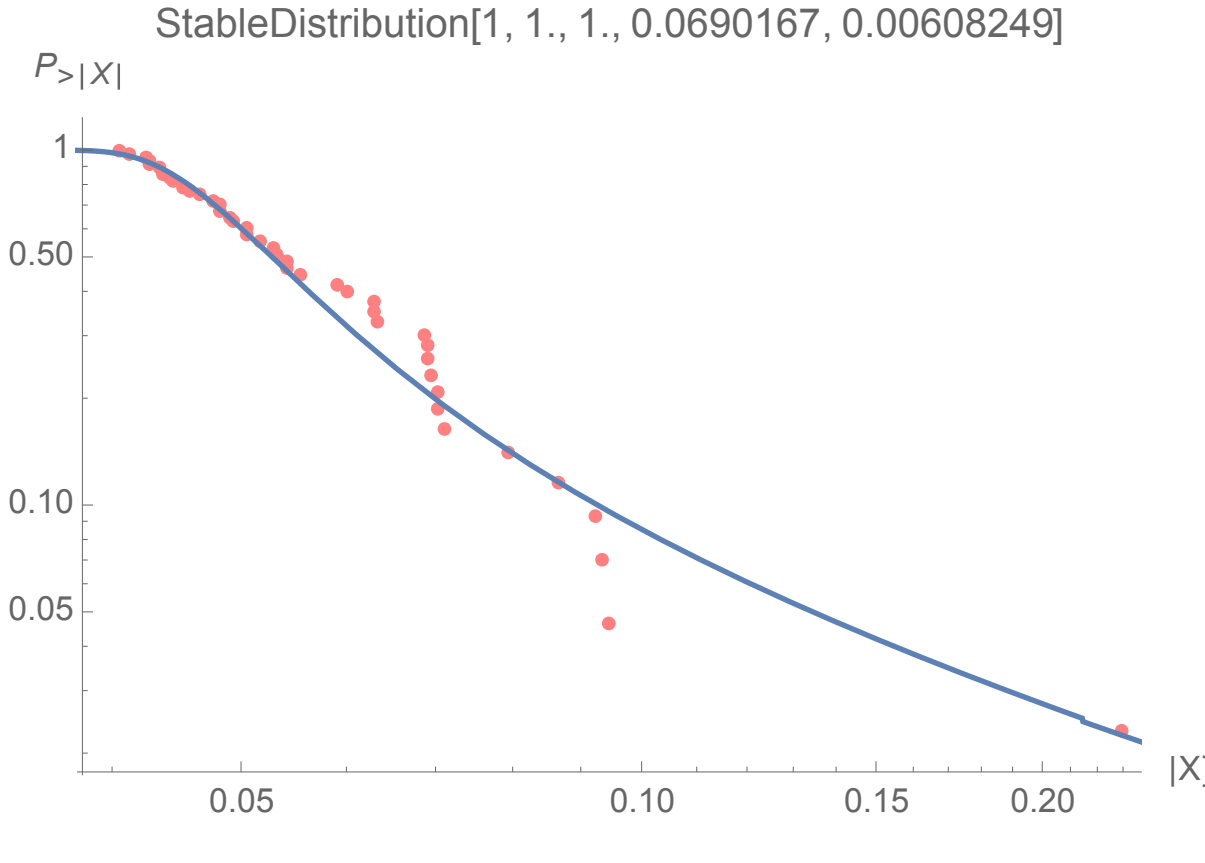

Figure 10.6: *The tails can even possibly fit an infinite mean stable distribution with $\alpha_l = 1$.*

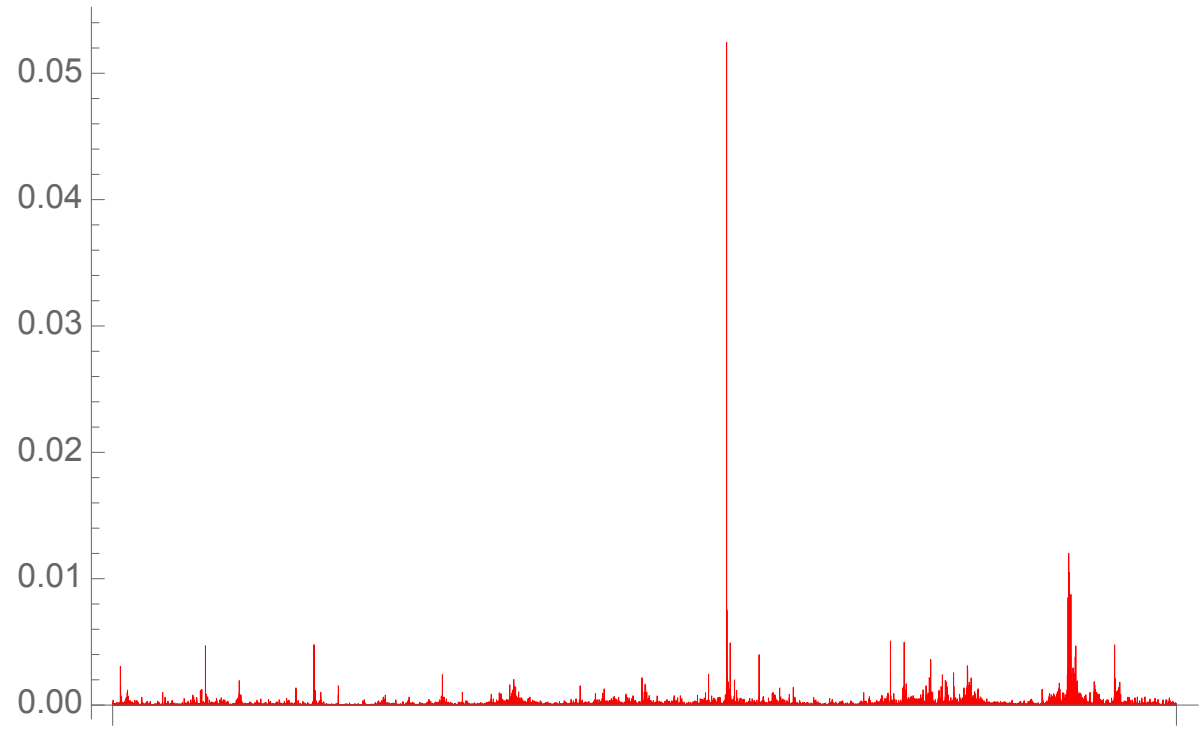

Figure 10.7: *SP500 squared returns for 16500 observations. No GARCH(1,1) can produce such jaggedness or what the great Benoit Mandelbrot called "rugosité".*

### 10.2.2 Maximum Drawdowns

For a time series for asset $S$ taken over $(t_0, t_0 + \Delta t, t_0 + n\Delta t)$, we are interested in the behavior of

$$\delta\left(t_0, t, \Delta t\right) = \text{Min}\left(S_{i\Delta t+t_0} - \left(\text{Min} S_{j\Delta t+t_0}\right)_{j=i+1}^{n}\right)_{i=0}^{n} \qquad (10.1)$$

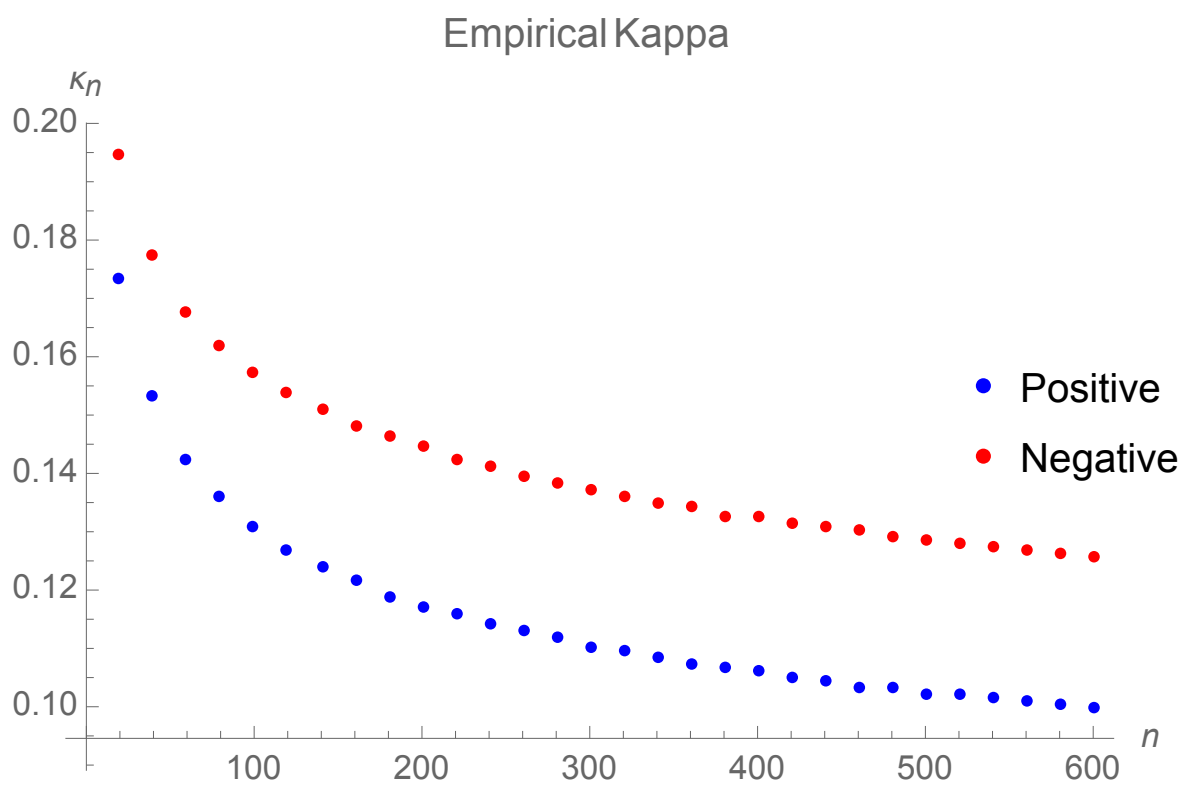

Figure 10.8: *kappa-n estimated empirically.*

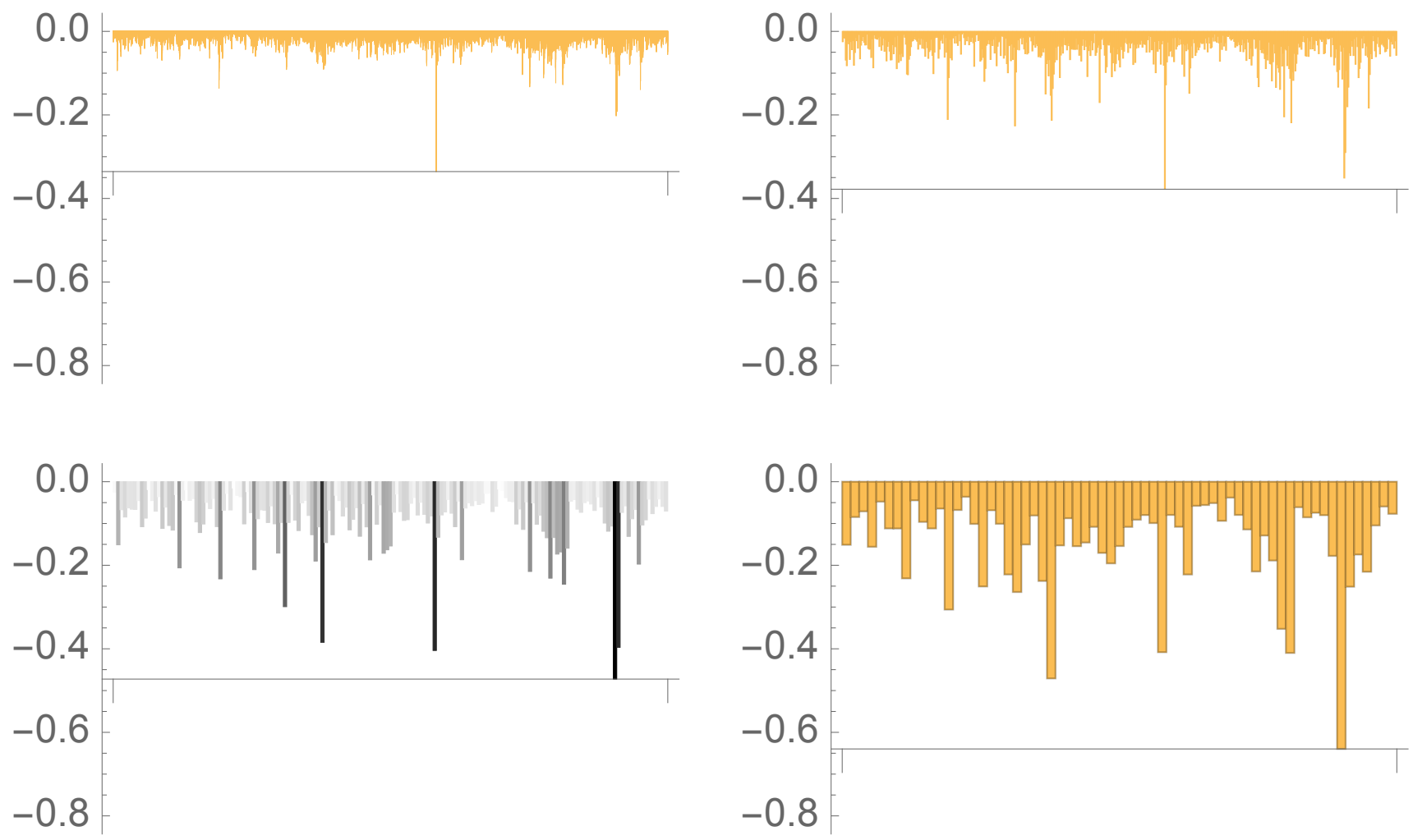

Figure 10.9: *Drawdowns for windows* $n = 5$, 30, 100, *and* 252 *days, respectively. Maximum drawdowns are excursions mapped in Eq. 10.1. We use here the log of the minimum of S over a window of n days following a given S.*

We can consider the relative drawdown by using the log of that minimum, as we do with returns. The window for the drawdown can be $n = 5, 100, 252$ days. As seen in Figure 10.10, drawdowns are Paretian.

### 10.2.3 Empirical Kappa

From our kappa equation in Chapter 8:

$$\kappa(n_0, n) = 2 - \frac{\log(n) - \log(n_0)}{\log\left(\frac{\mathbb{M}(n)}{\mathbb{M}(n_0)}\right)}. \tag{10.2}$$

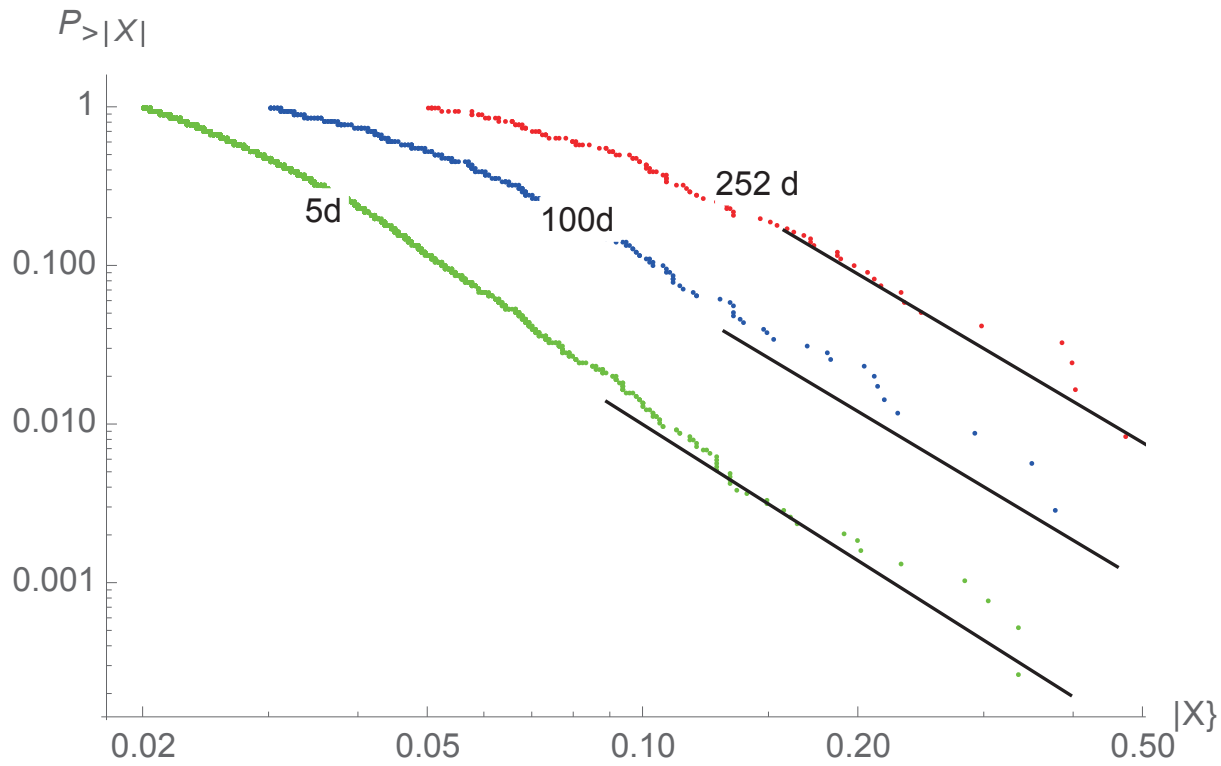

Figure 10.10: *Paretianity of Drawdowns and Scale*

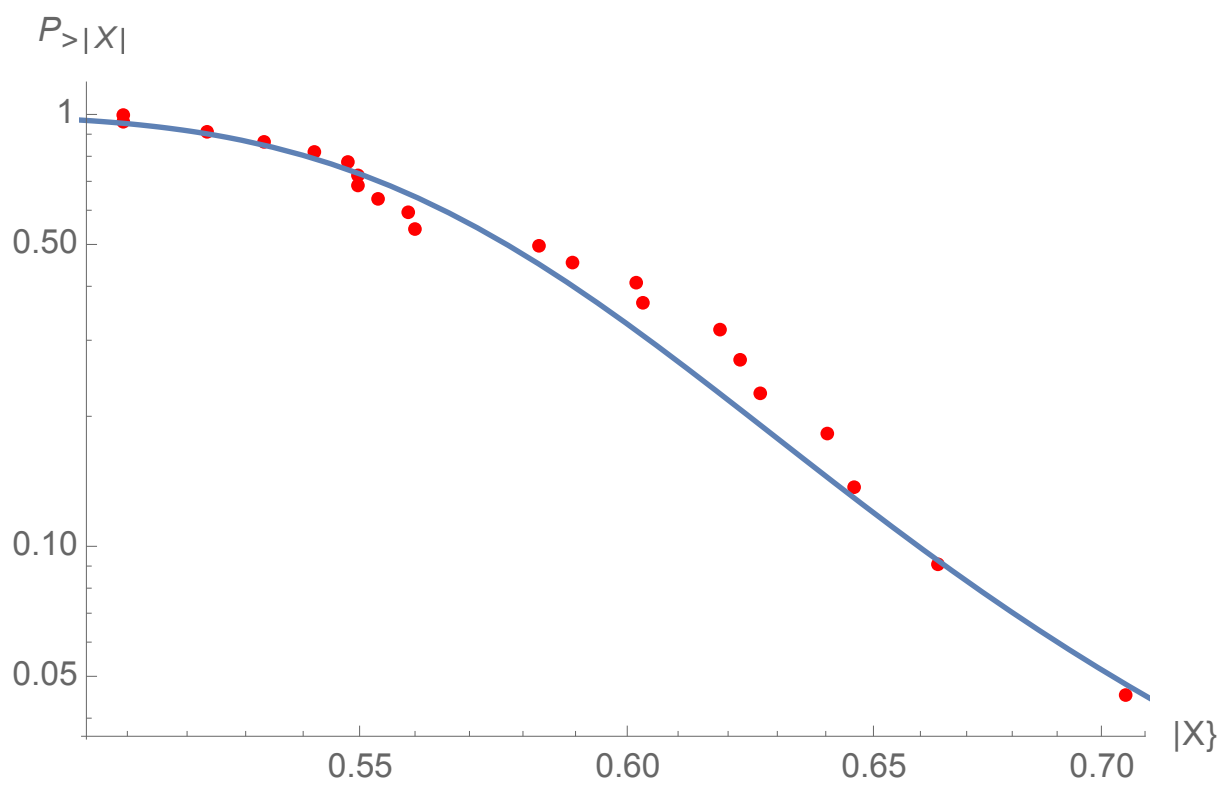

Figure 10.11: *Fitting a Stable Distribution to drawdowns*

with shortcut $\kappa_n = \kappa(1, n)$. We estimate empirically via bootstrapping and can effectively see how it maps to that of a power law – with $\alpha < 3$ for the negative returns.

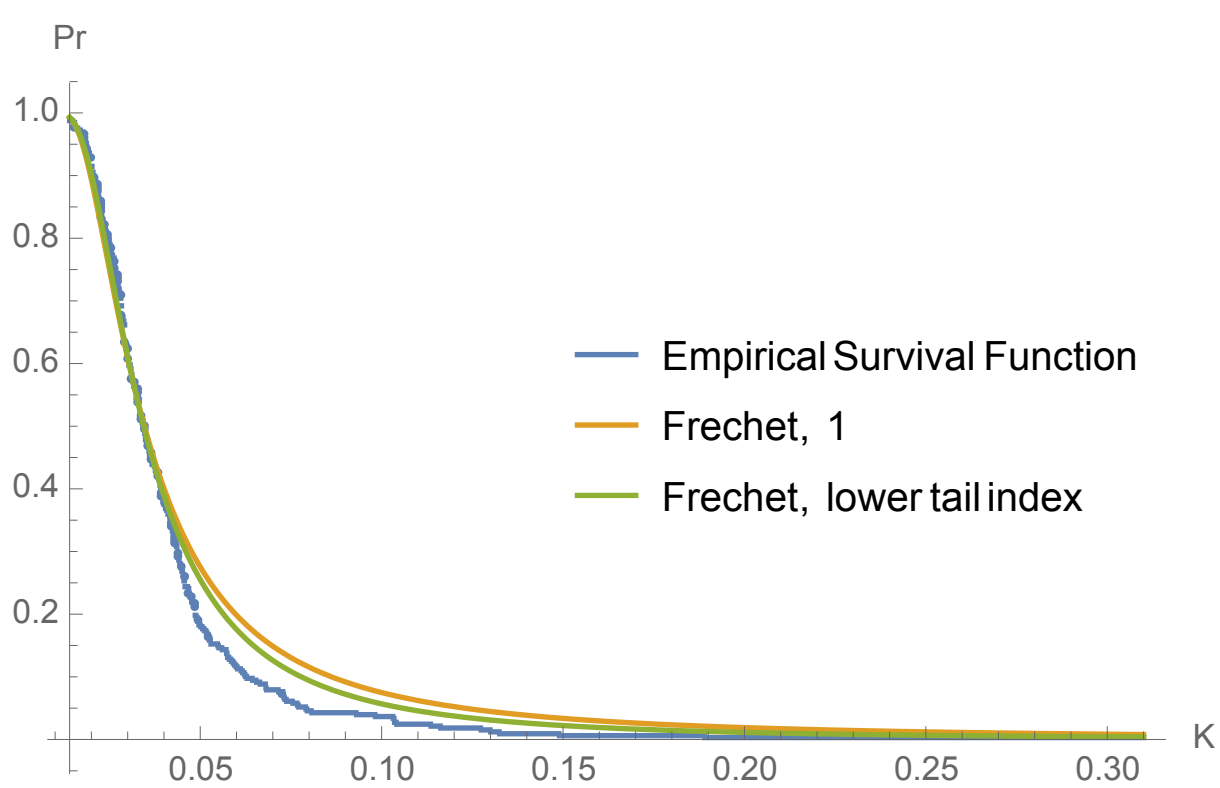

Figure 10.12: *Correcting the empirical distribution function with a Frechet for the SP500*

### 10.2.4 Test 2: Excess Conditional Expectation

**Result:** The verdict from this test is that, as we can see in Figure 10.4, that the conditional expectation of $X$ (and $-X$), conditional on $X$ is greater than some arbitrary value $K$, remains proportional to $K$.

**Definition 10.1**
*Let $K$ be in $\mathbb{R}^+$, the relative excess conditional expectation:*

$$\varphi_K^+ \triangleq \frac{\mathbb{E}(X)|_{X>K}}{K},$$

$$\varphi_K^- \triangleq \frac{\mathbb{E}(-X)|_{X>K}}{K}.$$

We have

$$\lim_{K\to\infty} \varphi_K = 0,$$

for distributions outside the power law basin, and

$$\lim_{K\to\infty} \varphi_K/K = \frac{\alpha}{1-\alpha}$$

for distribution satisfying the fat tailed class. Note the van der Wijk's law [53], [274].

Figure 10.4 shows the following: the conditional expectation does not drop for large values, which is incompatible with non-Paretian distributions.

### 10.2.5 Test 3- Instability of $4^{th}$ moment

A main argument in [274] is that in 50 years of SP500 observations, a single one represents >80 % of the Kurtosis. Similar effect are seen with other socioeconomic variables, such as gold, oil, silver other stock markets, soft commodities. Such sample dependence of the kurtosis means that the fourth moment does not have the stability, that is, does not exist.

### 10.2.6 Test 4: MS Plot

An additional approach to detect if $\mathbb{E}(X^p)$ exists consists in examining convergence according to the law of large numbers (or, rather, absence of), by looking the behavior of higher moments in a given sample. One convenient approach is the Maximum-to-Sum plot, or MS plot as shown in Figure 19.1. The MS Plot relies on a consequence of the law of large numbers [95] when it comes to the maximum of a variable. For a sequence $X_1, X_2, ..., X_n$ of non-negative i.i.d. random variables, if for $p = 1, 2, 3, \ldots$, $\mathbb{E}[X^p] < \infty$, then

$$R_n^p = M_n^p / S_n^p \to^{a.s.} 0$$

as $n \to \infty$, where $S_n^p = \sum_{i=1}^{n} X_i^p$ is the partial sum, and $M_n^p = \max(X_1^p, ..., X_n^p)$ the partial maximum. (Note that we can have $X$ the absolute value of the random variable in case the r.v. can be negative to allow the approach to apply to odd moments.)

We show by comparison the MS plot for a Gaussian and that for a Student T with a tail exponent of 3. We observe that the SP500 show the typical characteristics of a steep power law, as in 16,000 observations (50 years) it does not appear to drop to the point of allowing the functioning of the law of large numbers.

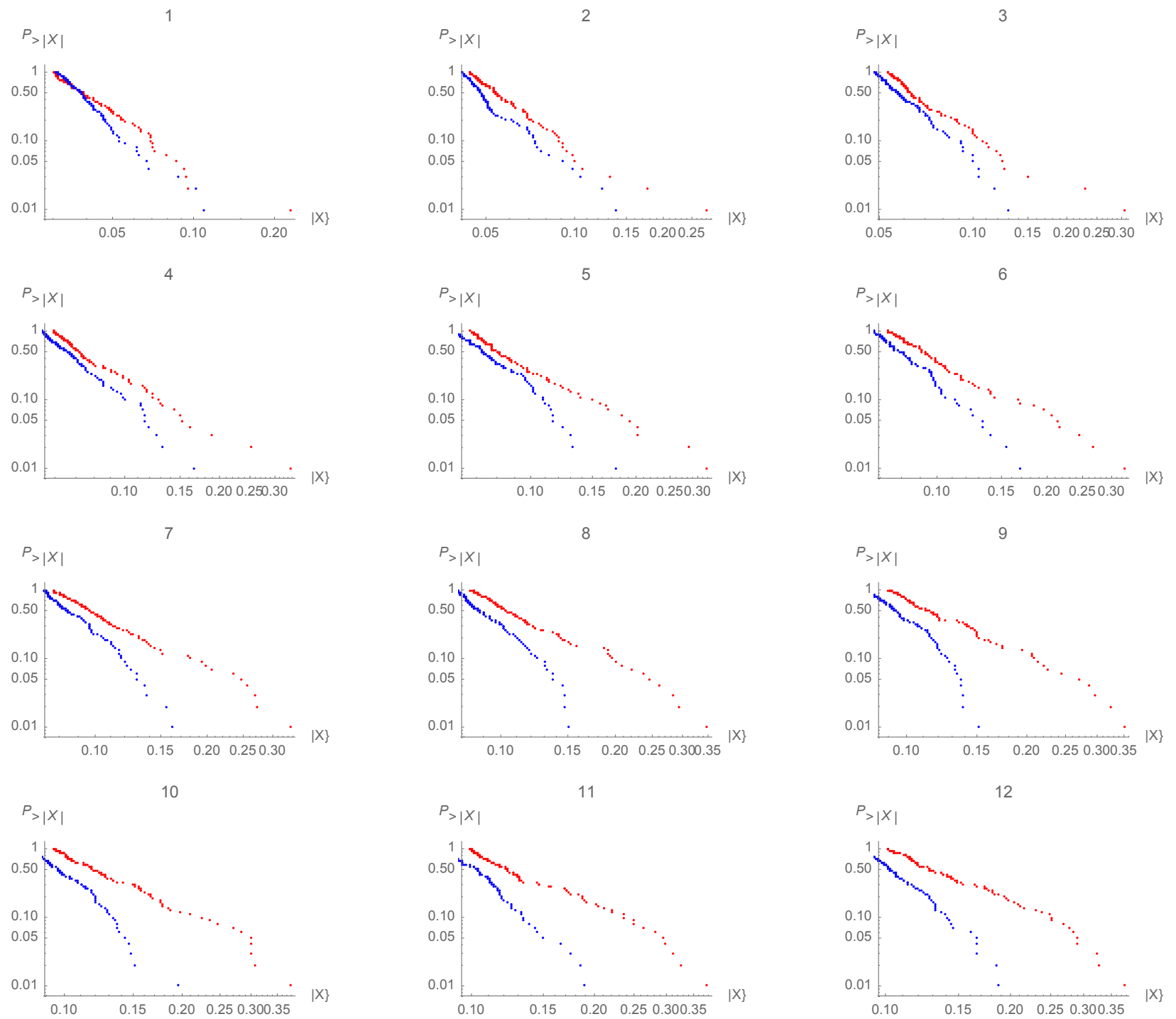

Figure 10.13: *We separate positive and negative logarithmic returns and use overlapping cumulative returns from 1 up to 15. Clearly the negative returns appear to follow a Power Law while the Paretianity of the right one is more questionable.*

### 10.2.7 Records and Extrema

The Gumbel record methods is as follows (Embrechts et al [97]). Let $X_1, X_2, \ldots$ be a discrete time series, with a maximum at period $t \geq 2$, $M_t = \max(X_1, X_2, \ldots, X_t)$, we have the record counter $N_{1,t}$ for $n$ data points.

$$N_{1,t} = 1 + \sum_{k=2}^{t} \mathbb{1}_{X_t > M_{t-1}}. \tag{10.3}$$

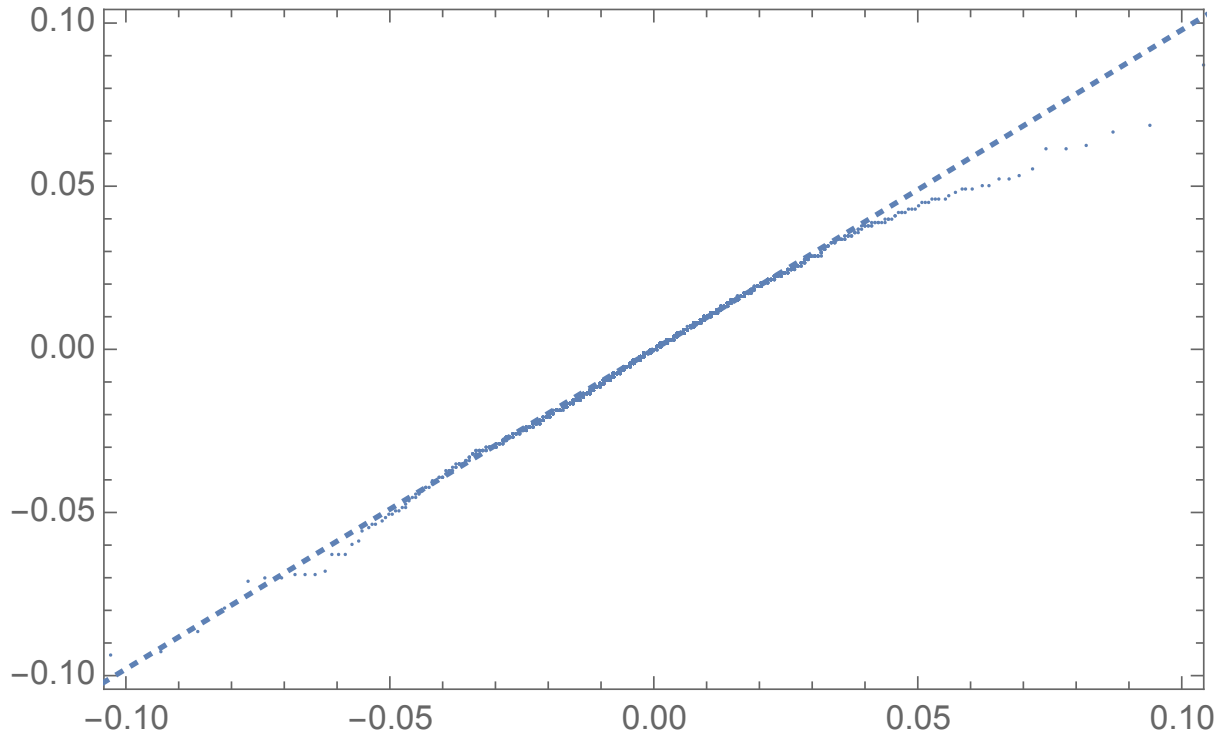

Figure 10.14: *QQ Plot comparing the Student T to the empirical distribution of the SP500: the left tail fits, not the right tail.*

Regardless of the underlying distribution, the expectation $\mathbb{E}(N_t)$ is the Harmonic Number $H_t$, and the variance $H_t - H_t^2$, where $H_t = \sum_{i=1}^{t} \frac{1}{i^r}$. We note that the harmonic number is concave and very slow in growth, logarithmic, as it can be approximated with $log(n) + \gamma$, where $\gamma$ is the Euler Mascheroni constant. The approximation is such that $\frac{1}{2(t+1)} \leq H_t - log(t) - \gamma \leq \frac{1}{2t}$ (*Wolfram Mathworld* [317]).

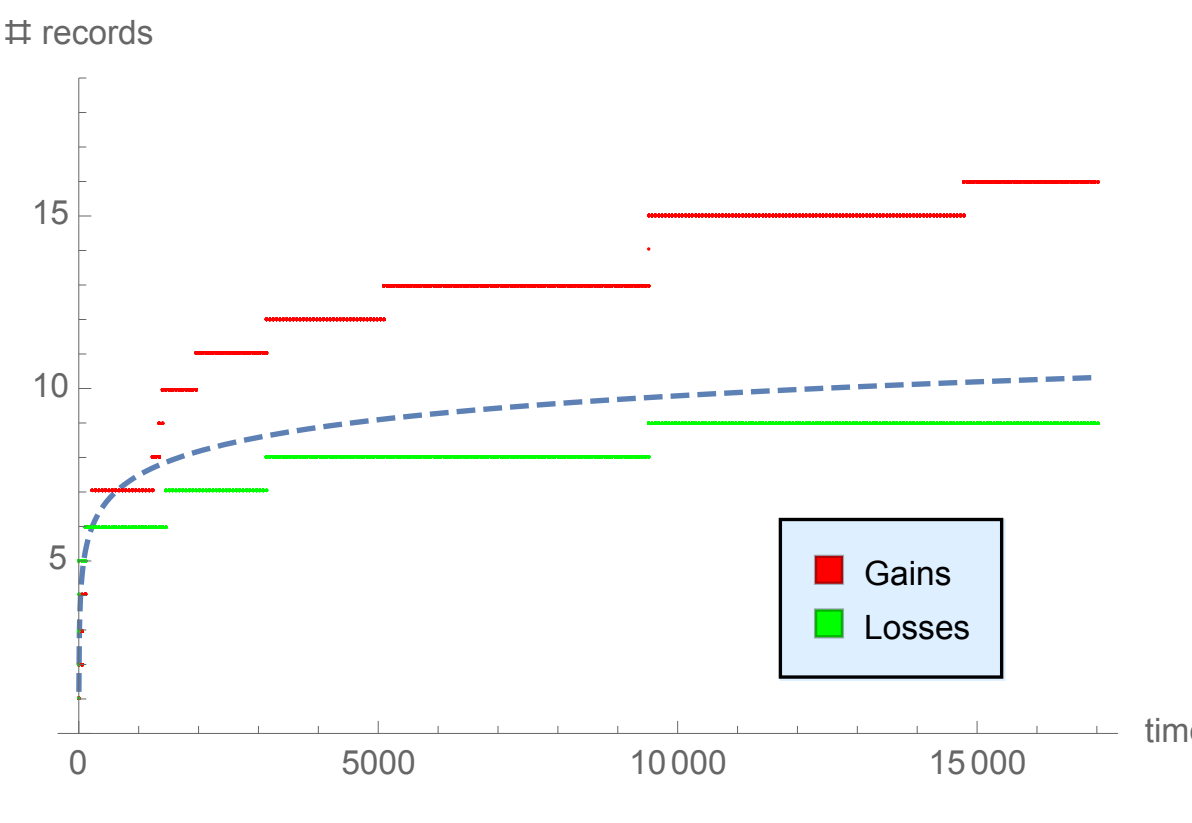

Figure 10.15: *The record test shows independence for extremes of negative returns, dependence for positive ones. The number of records for independent observations grows over time at the harmonic number $H(t)$ (dashed line), ≈ logarithmic but here appears to grow $>$ 2.5 standard deviations faster for positive returns, hence we cannot assume independence for extremal gains. The test does not make claims about dependence outside extremes.*

> **Remark 10.2**
>
> *The Gumbel test of independence above is sufficient condition for the convergence of extreme negative values of the log-returns of the SP500 to the Maximum Domain of Attraction (MDA) of the extreme value distribution.*

**Entire series** We reshuffled the SP500 (i.e. bootstrapped with replacement, using a sample size equal to the original ≈ 17000 points, with $10^3$ repeats) and ran records across all of them. As shown in Figures 10.18 and 10.17, the mean was 10.4 (approximated by the harmonic number, with a corresponding standard deviation.) The survival function S(.) of $N_{1.7\times10^4} = 16$, $S(16) = \frac{1}{40}$ which allows us to consider the independence of positive extrema implausible.

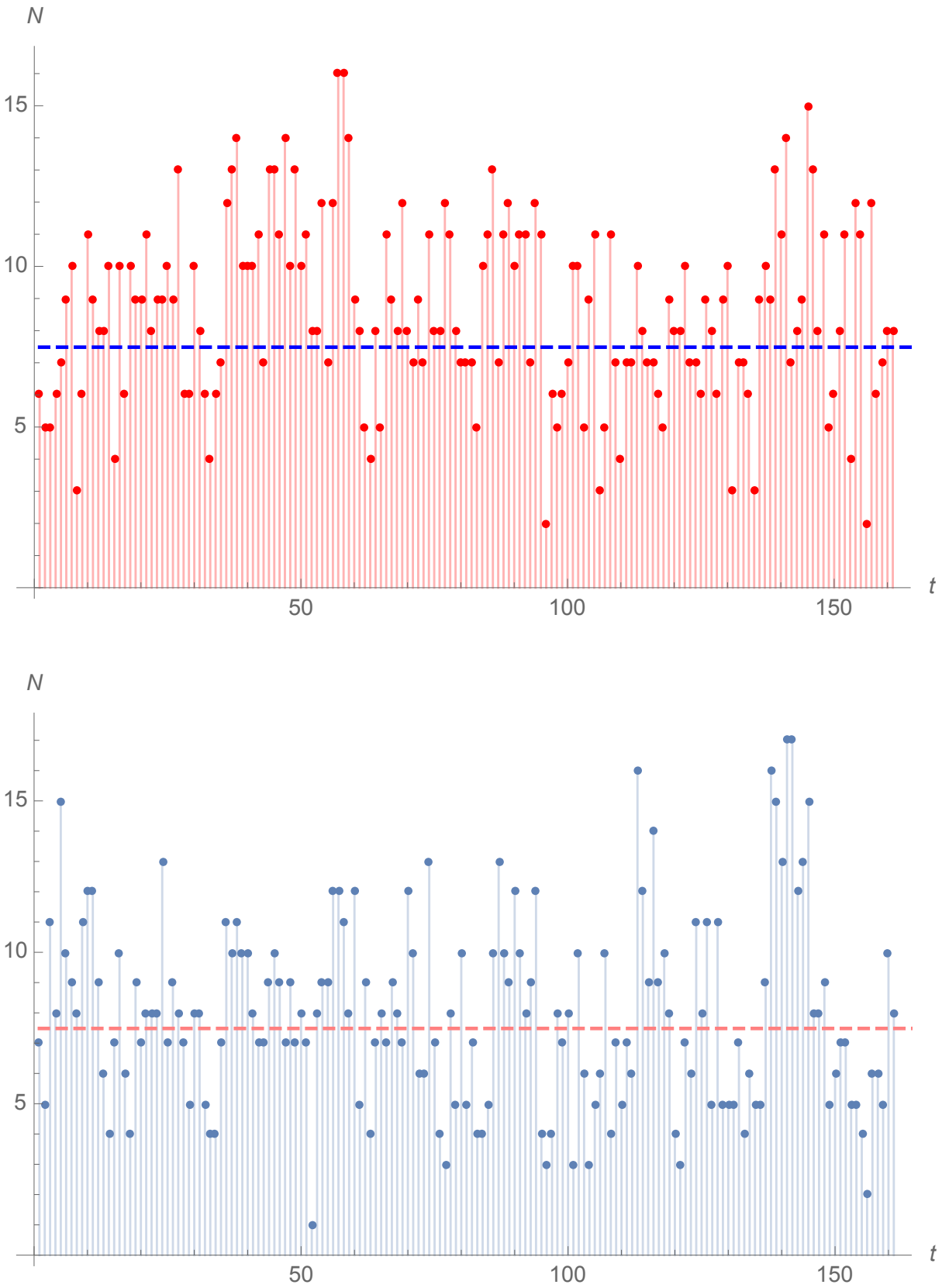

Figure 10.16: *Running shorter period,* $t = 1000$ *days of overlapping observations for the records of maxima(top) and minima (bottom), compared to the expected Harmonic number* $H(1000)$.

On the other hand the negative extrema (9 counts) show realizations close to what is expected (10.3), diverting by $\frac{1}{2}$ a s.t.d. from expected, enough to justify a failure to reject independence.

**Subrecords** If instead of taking the data as one block over the entire period, we broke the period into sub-periods, we get (because of the concavity of the measure and Jensen's inequality), $N_{t_1+\delta, t_1+\Delta+\delta}$, we obtain $T/\delta$ observations. We took $\Delta = 10^3$ and $\delta = 10^2$, thus getting 170 subperiods for the $T \approx 17 \times 10^3$ days. The picture as shown in Figure 10.16 cannot reject independence for both positive and negative deviations.

**Conclusion for subrecords** We can at least apply EVT methods for negative observations.

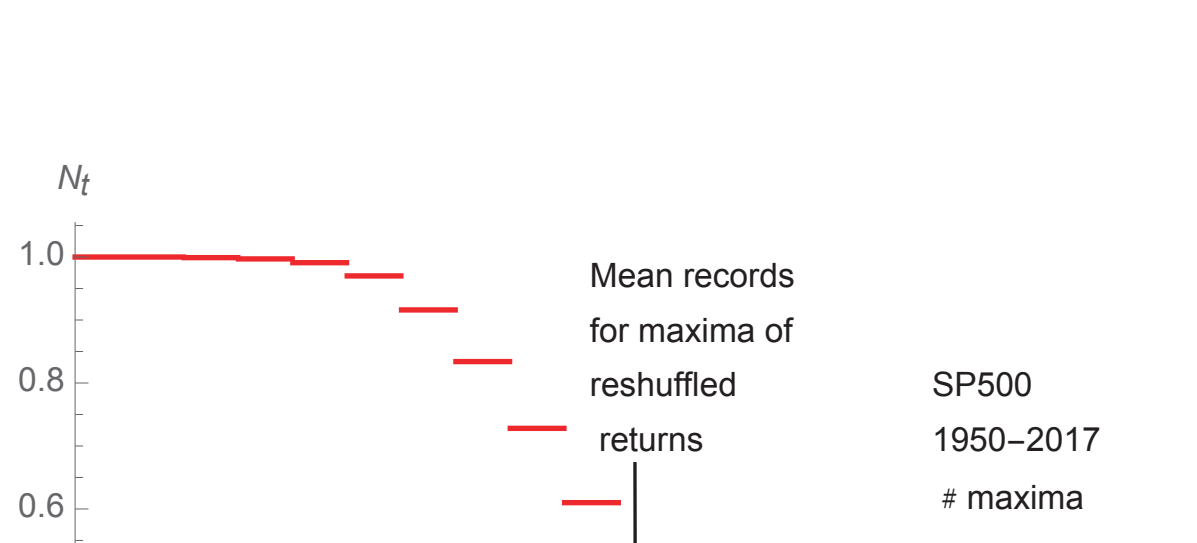

Figure 10.17: *The survival function of the records of positive maxima for the resampled SP500 ($10^3$ times) by keeping all returns but reshuffling them, thus removing the temporal structure. The mass above 16 (observed number of maxima records for SP500 over the period) is* $\frac{1}{40}$.

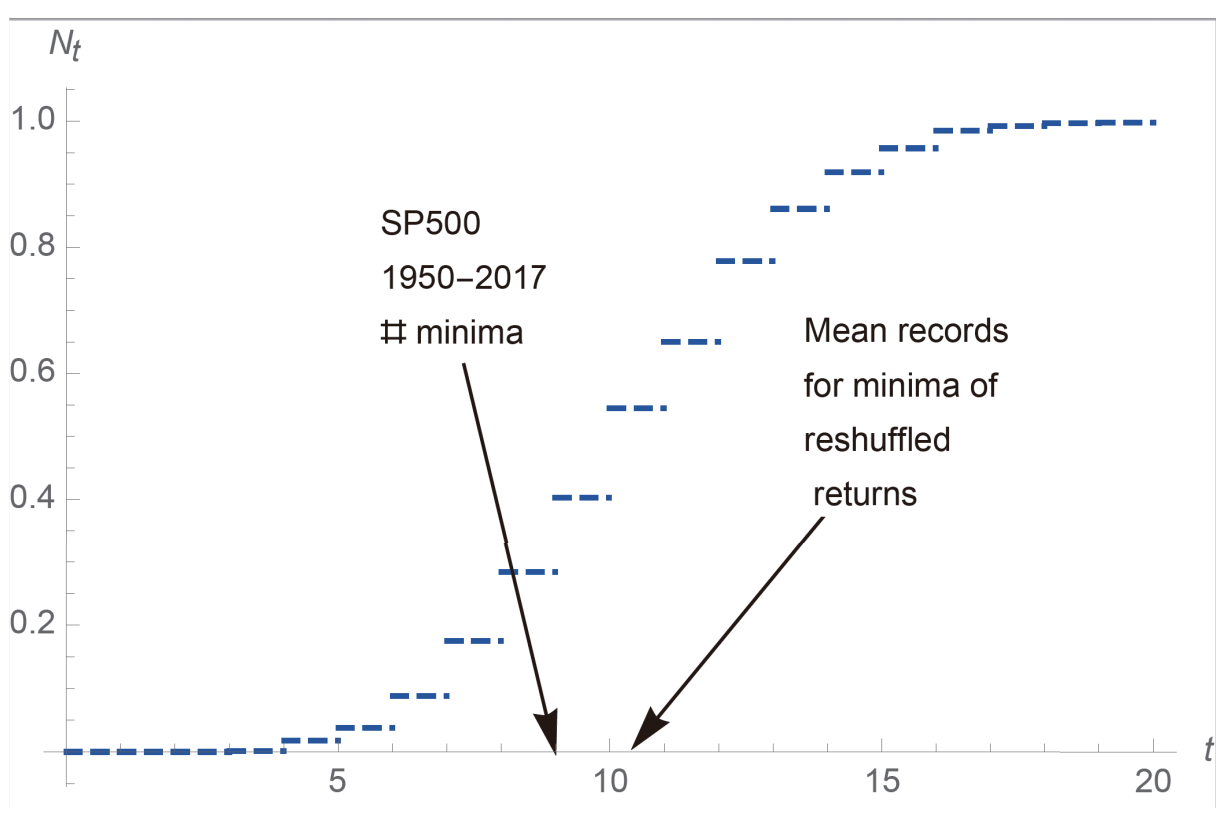

Figure 10.18: *The CDF of the records of negative extrema for the resampled SP500 ($10^3$ times) reshuffled as above. The mass above 9 (observed number of minima records for SP500 over the period) is* $\frac{2}{5}$.

### 10.2.8 Asymmetry right-left tail

We note an asymmetry as seen in Figure 10.13, with the left tail considerably thicker than the right one. It may be a nightmare for modelers looking for some precise process, but not necessarily to people interested in risk, and option trading.

## 10.3 conclusion: it is what it is

This chapter allowed us to explore a simple topic: returns on the SP500 index (which represents the bulk of the U.S. stock market capitalization) are, simply power law distributed –by Wittgenstein's ruler, it is irresponsible to model them in any other manner. Standard methods such as Modern Portfolio Theory (MPT) or "base rate crash" verbalisms (claims that people overestimate the probabilities of tail events) are totally bogus –we are talking of $> 70,000$ papers and entire cohorts of research, not counting about $10^6$ papers in general economics with results depending on "variance" and "correlation". You need to live with the fact that these metrics are bogus. As the ancients used to say, *dura lex sed lex*, or in more modern mafia terms:

*It is what it is.*

# F | THE PROBLEM WITH ECONOMETRICS

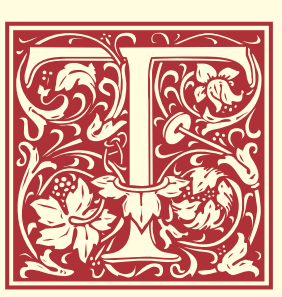

HERE IS SOMETHING WRONG with econometrics, as almost all papers don't replicate in the real world. Two reliability tests in Chapter 10, one about parametric methods the other about robust statistics, show that there must be something rotten in econometric methods, fundamentally wrong, and that the methods are not dependable enough to be of use in anything remotely related to risky decisions. Practitioners keep spinning inconsistent *ad hoc* statements to explain failures. This is a brief nontechnical exposition from the results in [274].

With economic variables one single observation in 10,000, that is, one single day in 40 years, can explain the bulk of the "kurtosis", the finite-moment standard measure of "fat tails", that is, both a measure how much the distribution under consideration departs from the standard Gaussian, or the role of remote events in determining the total properties. For the U.S. stock market, a single day, the crash of 1987, determined 80% of the kurtosis for the period between 1952 and 2008. The same problem is found with interest and exchange rates, commodities, and other variables. Redoing the study at different periods with different variables shows a total instability to the kurtosis. The problem is not just that the data had "fat tails", something people knew but sort of wanted to forget; it was that we would never be able to determine "how fat" the tails were within standard methods. Never.[1]

[1] Macroeconomic variables, such as U.S. weekly jobless claims have traditionally appeared to be tractable inside the (ugly and drab) buildings housing economic departments. They ended up breaking the models with a bang. Jobless claims experienced "unexpected" jumps with Covid 19 (the coronavirus) described of "thirty standard deviations": the kurtosis (of the log changes) rose from 8 to $> 550$ *after a single observation* in April 2020. Almost all in-sample higher moments are attributable to a one data point, and the higher the moment the higher such an effect –hence one must accept that there are *no* higher moments, and no informative lower moment, and the variable must be power law distributed.

Such a role for the tail cancels the entire history of macroeconomic modeling, as well as policies based on the conclusion of economists using Mediocristan-derived metrics. While economists in the citation-rings may not be aware of their fraudulent behavior, others are not missing the point. At the time of writing people are starting to realize that the fatter the tails, the more policies should be based on the expected extrema, using extreme value theory (EVT), and the differences between Gaussian and power law models are even starker for the extremes.

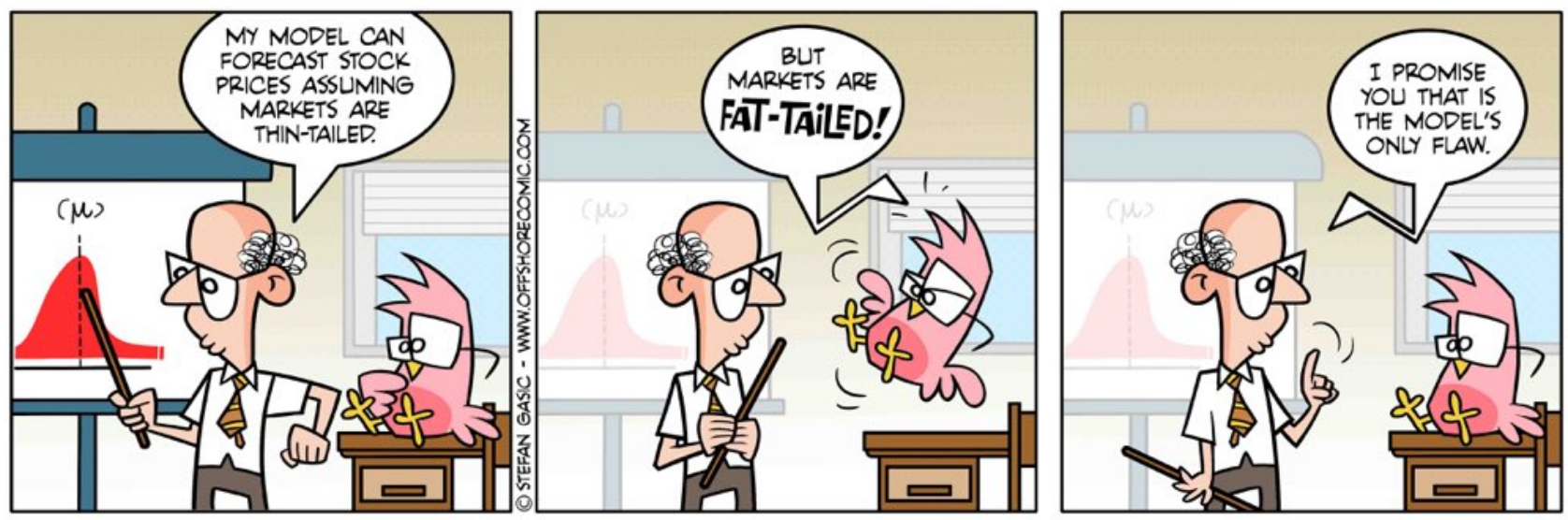

Figure F.1: *Credit: Stefan Gasic*

The implication is that those tools used in economics that are *based on squaring variables* (more technically, the $\mathcal{L}^2$ norm), such as standard deviation, variance, correlation, regression, the kind of stuff you find in textbooks, are not valid *scientifically* (except in some rare cases where the variable is bounded). The so-called "p values" you find in studies have no meaning with economic and financial variables. Even the more sophisticated techniques of stochastic calculus used in mathematical finance do not work in economics except in selected pockets.

## F.1 PERFORMANCE OF STANDARD PARAMETRIC RISK ESTIMATORS

The results of most papers in economics based on these standard statistical methods are thus not expected to replicate, and they effectively don't. Further, these tools invite foolish risk taking. Neither do alternative techniques yield reliable measures of rare events, except that we can tell if a remote event is underpriced, without assigning an exact value.

From [274]), using log returns, $X_t \triangleq \log\left(\frac{P(t)}{P(t-i\Delta t)}\right)$. Consider the $n$-sample maximum quartic observation $\text{Max}(X_{t-i\Delta t}{}^4)_{i=0}^n$. Let $Q(n)$ be the contribution of the maximum quartic variations over $n$ samples and frequency $\Delta t$.

$$Q(n) := \frac{\text{Max}\left(X_{t-i\Delta t}^4\right)_{i=0}^n}{\sum_{i=0}^n X_{t-i\Delta t}^4}.$$

Note that for our purposes, where we use central or noncentral kurtosis makes no difference –results are nearly identical.

For a Gaussian (i.e., the distribution of the square of a Chi-square distributed variable) show $Q\left(10^4\right)$ the maximum contribution should be around .008 ± .0028. Visibly we can see that the observed distribution of the 4$^{\text{th}}$ moment has the property

$$\mathbb{P}\left(X > \max(x_i^4)_{i \leq 2 \leq n}\right) \approx \mathbb{P}\left(X > \sum_{i=1}^n x_i^4\right).$$

Table F.1: *Maximum contribution to the fourth moment from a single daily observation*

| Security | Max Q | Years. |
|---|---|---|
| Silver | 0.94 | 46. |
| SP500 | 0.79 | 56. |
| CrudeOil | 0.79 | 26. |
| Short Sterling | 0.75 | 17. |
| Heating Oil | 0.74 | 31. |
| Nikkei | 0.72 | 23. |
| FTSE | 0.54 | 25. |
| JGB | 0.48 | 24. |
| Eurodollar Depo 1M | 0.31 | 19. |
| Sugar #11 | 0.3 | 48. |
| Yen | 0.27 | 38. |
| Bovespa | 0.27 | 16. |
| Eurodollar Depo 3M | 0.25 | 28. |
| CT | 0.25 | 48. |
| DAX | 0.2 | 18. |

Recall that, naively, the fourth moment expresses the stability of the second moment. And the second moment expresses the stability of the measure across samples.

Note that taking the snapshot at a different period would show extremes coming from other variables while these variables showing high maximma for the kurtosis, would drop, a mere result of the instability of the measure across series and time.

**Description of the dataset** All tradable macro markets data available as of August 2008, with "tradable" meaning actual closing prices corresponding to transactions (stemming from markets not bureaucratic evaluations, includes interest rates, currencies, equity indices).

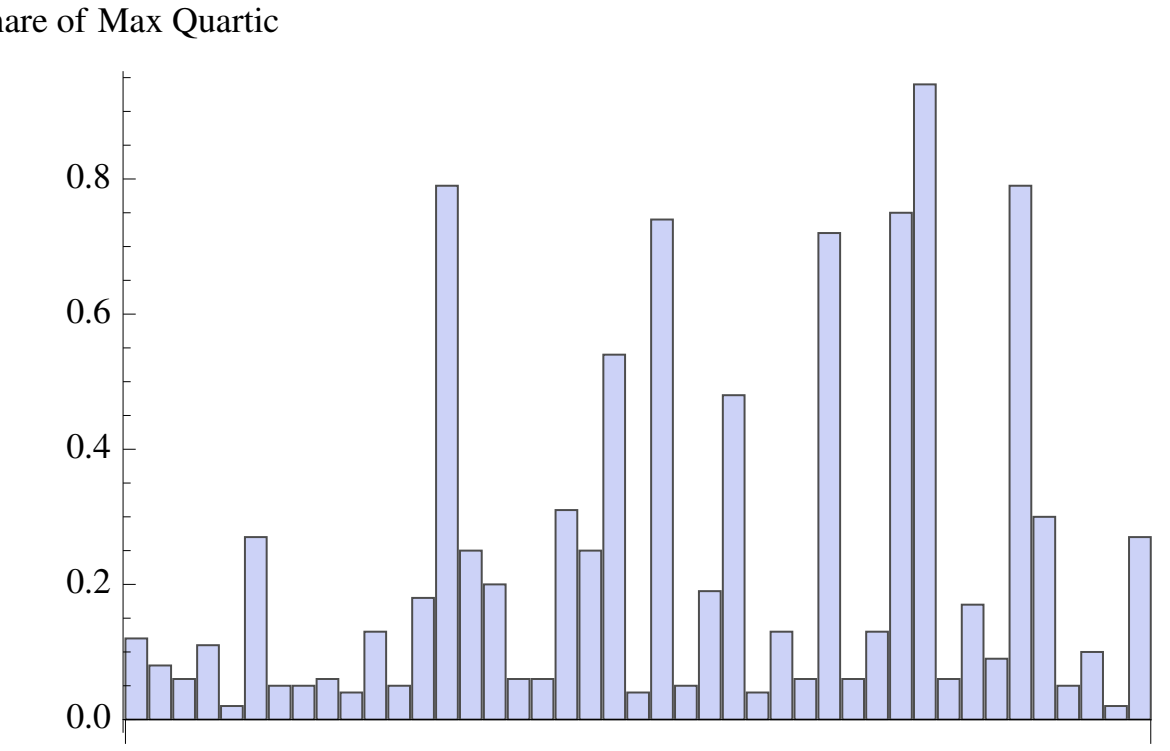

Figure F.2: *Max quartic across securities in Table F.1.*

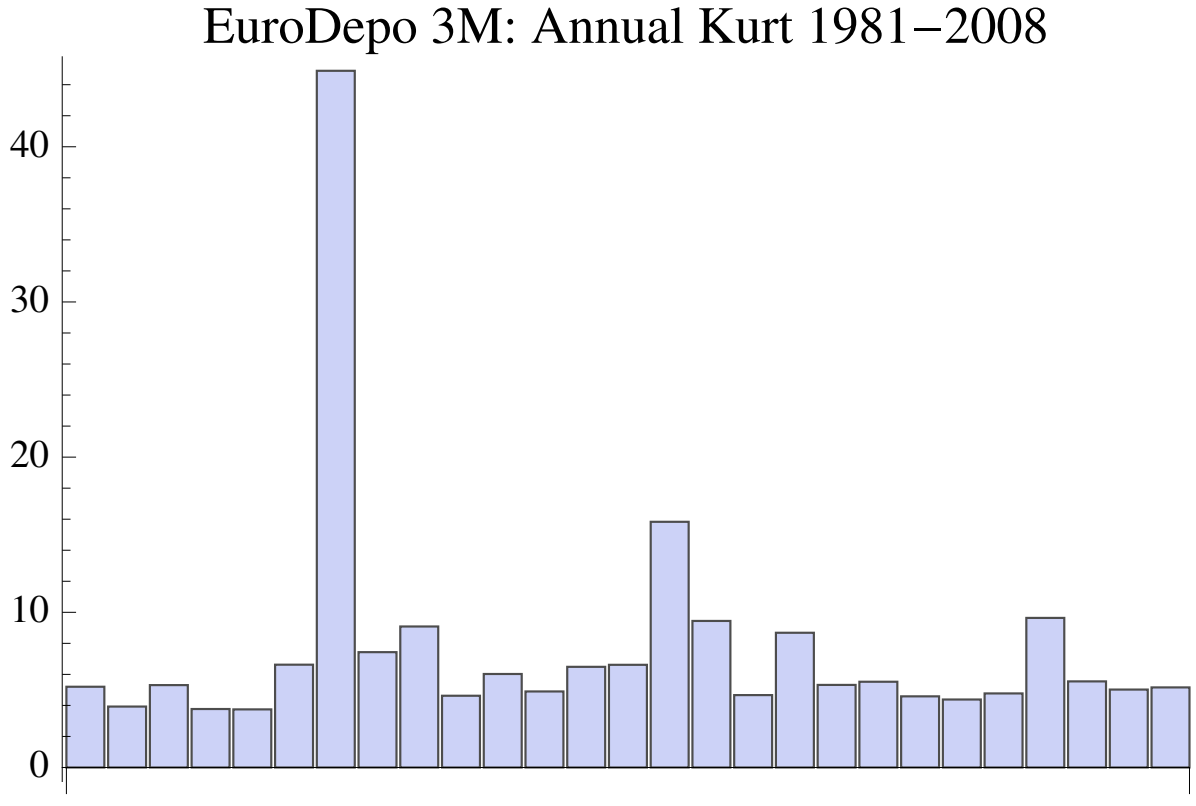

Figure F.3: *Kurtosis across nonoverlapping periods for Eurodeposits.*

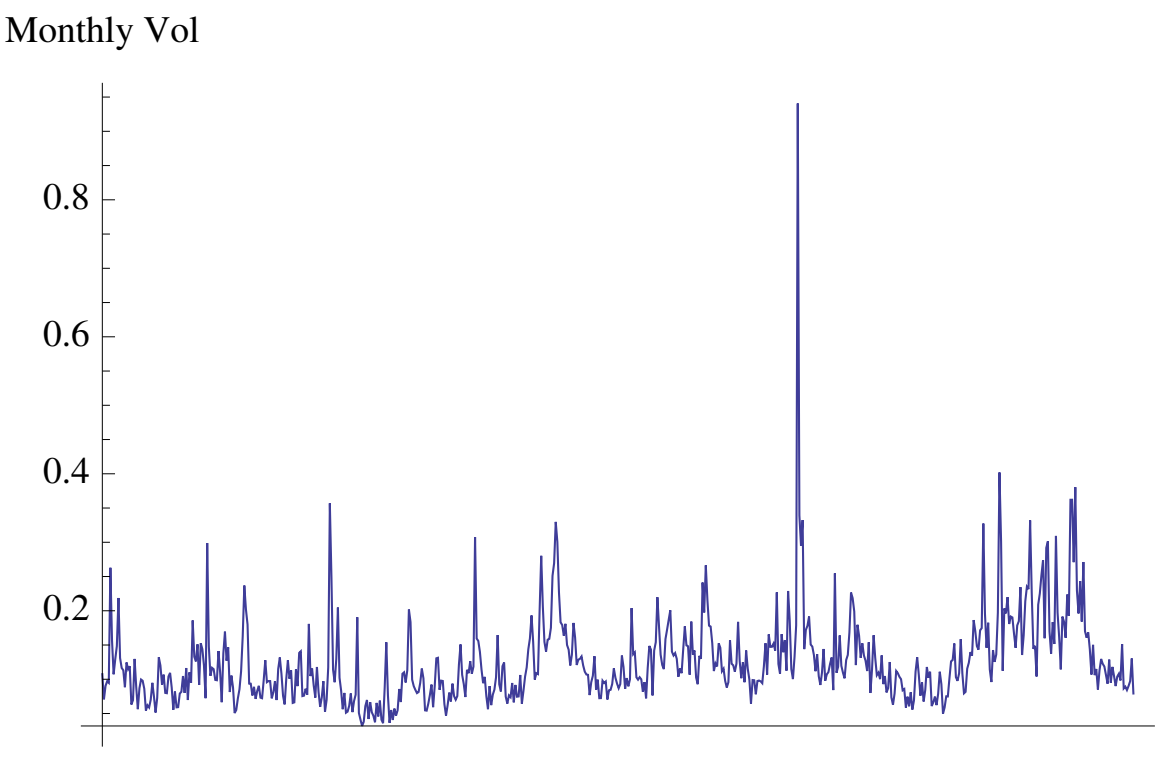

Figure F.4: *Monthly delivered volatility in the SP500 (as measured by standard deviations). The only structure it seems to have comes from the fact that it is bounded at 0. This is standard.*

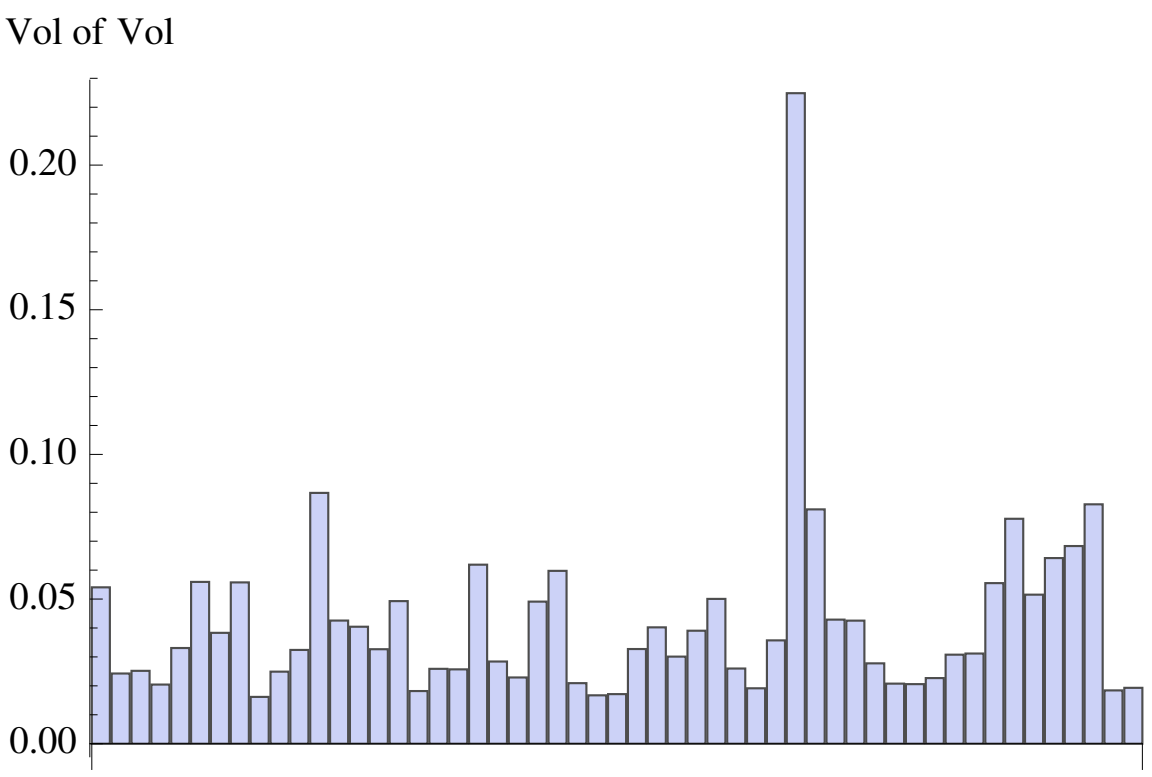

Figure F.5: *Montly volatility of volatility from the same dataset in Table F.1, predictably unstable.*

## F.2 PERFORMANCE OF STANDARD NONPARAMETRIC RISK ESTIMATORS

Does the past resemble the future in the tails? The following tests are nonparametric, that is entirely based on empirical probability distributions.

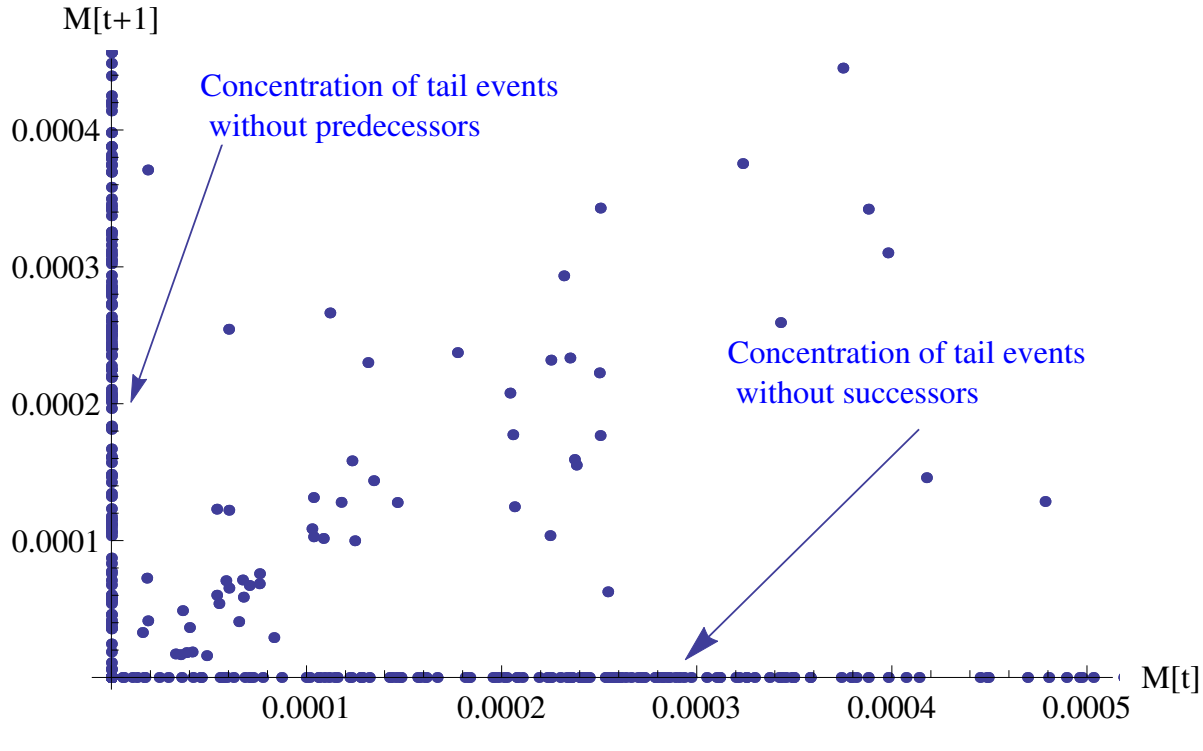

Figure F.6: *Comparing one absolute deviation M[t] and the subsequent one M[t+1] over a certain threshold (here 4% in stocks); illustrated how large deviations have no (or few) predecessors, and no (or few) successors– over the past 50 years of data.*

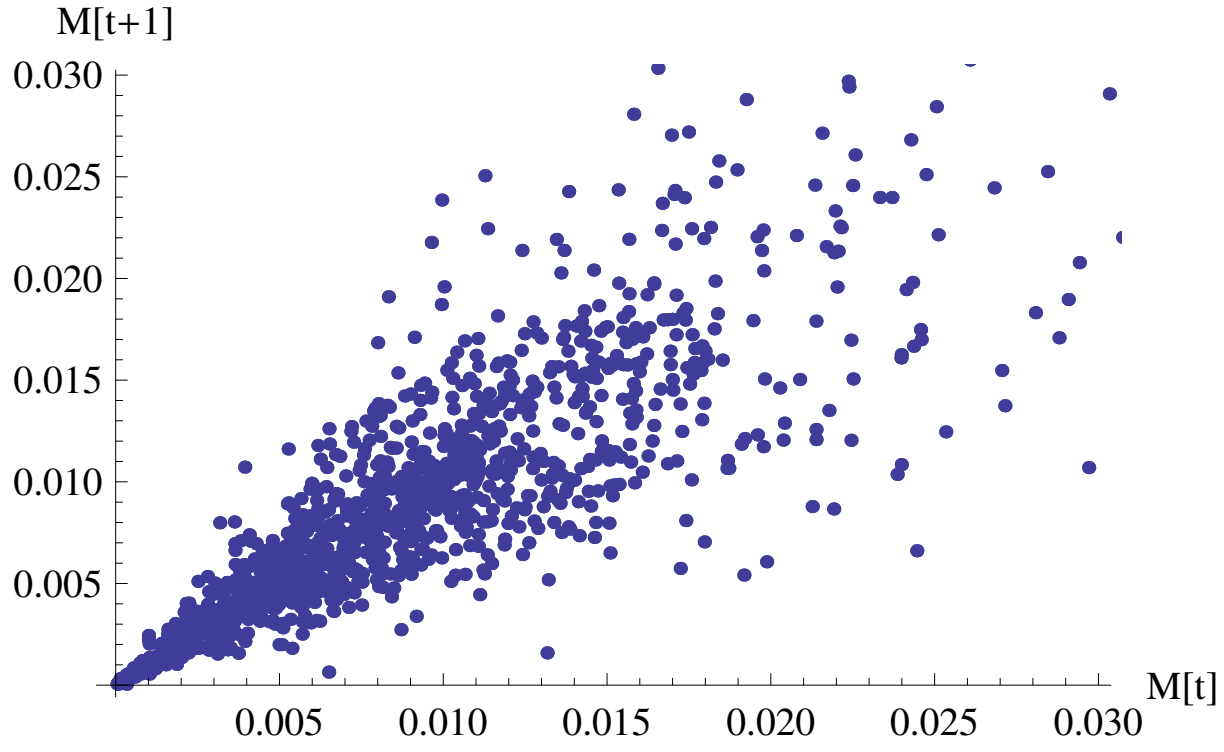

Figure F.7: *The "regular" is predictive of the regular, that is mean deviation. Comparing one absolute deviation M[t] and the subsequent one M[t+1] for macroeconomic data.*

So far we stayed in dimension 1. When we look at higher dimensional properties, such as covariance matrices, things get worse. We will return to the point with the treatment of model error in mean-variance optimization.

When $x_t$ are now in $\mathbb{R}^N$, the problems of sensitivity to changes in the covariance matrix makes the empirically observed moments and conditional moments extremely unstable. Tail events for a vector are vastly more difficult to calibrate, and increase in dimensions.

**The Responses so far by members of the economics/econometrics establishment** No answer as to why they still use STD, regressions, GARCH , value-at-risk and similar methods.

**Peso problem** Benoit Mandelbrot used to insist that one can fit anything with Poisson jumps. This is similar to the idea that one can always perfectly fit $n$ data points with a polynomial with $n-1$ parameters. If you need to change your parameters, it's not a power law.

Many researchers invoke "outliers" or "peso problem"[2] as acknowledging fat tails (or the role of the tails for the distribution), yet ignore them analytically (outside of Poisson models that are not possible to calibrate except after the fact: conventional Poisson jumps are thin-tailed). Our approach here is exactly the opposite: do not push outliers under the rug, rather build everything around them. In other words, just like the FAA and the FDA who deal with safety by focusing on catastrophe avoidance, we will throw away the ordinary under the rug and retain extremes as the sole sound approach to risk management. And this extends beyond safety since much of the analytics and policies that can be destroyed by tail events are inapplicable.

[2] The peso problem is a discovery of an outlier in money supply, became a name for outliers and unexplained behavior in econometrics.

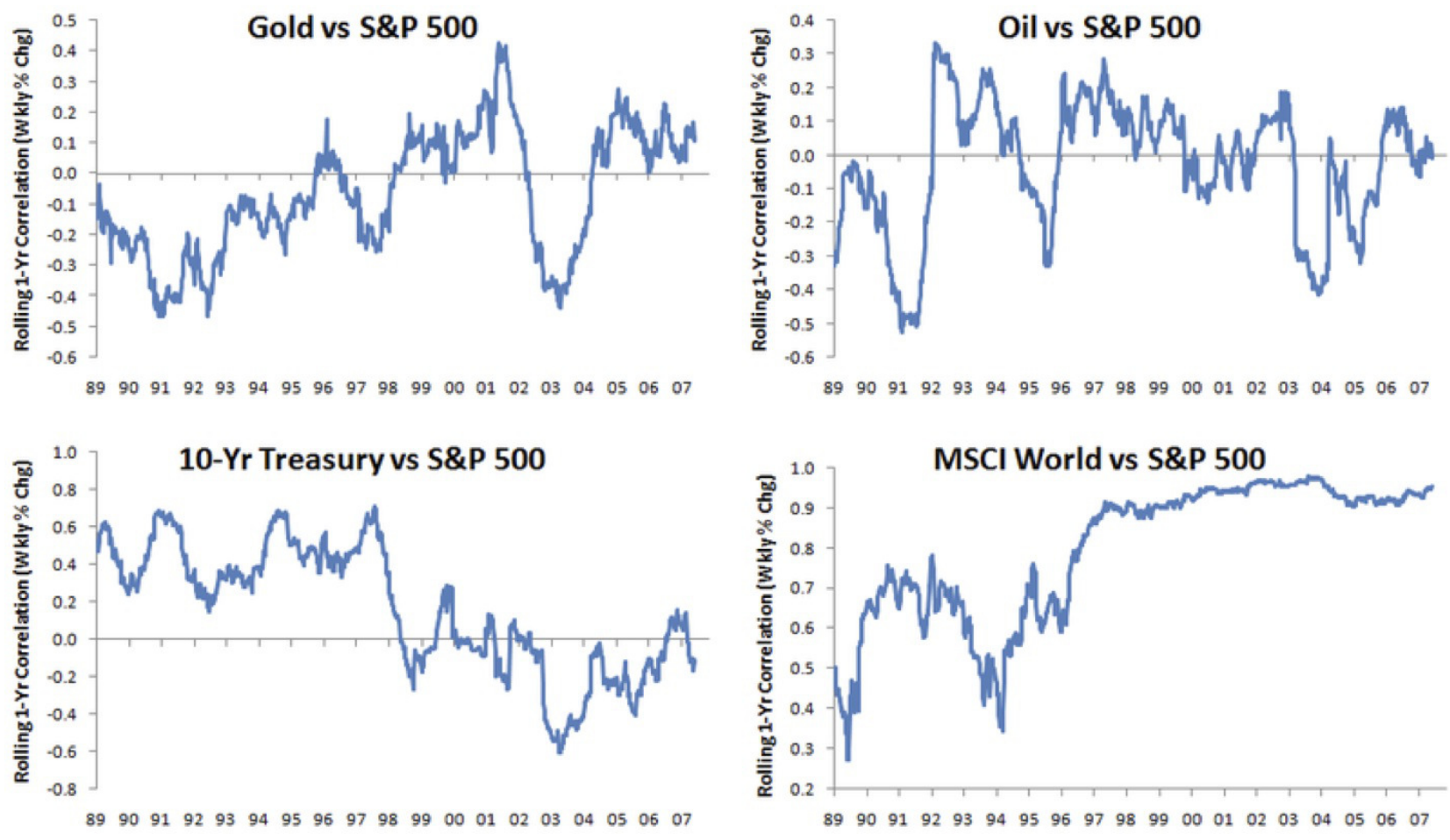

Figure F.8: *Correlations are also problematic, which flows from the instability of single variances and the effect of multiplication of the values of random variables. Under such stochasticity of correlations it makes no sense, no sense whatsoever, to use covariance-based methods such as portfolio theory.*

**Peso problem confusion about the Black Swan problem** :

> "(...) "Black Swans" (Taleb, 2007). These cultural icons refer to disasters that occur so infrequently that they are virtually impossible to analyze using standard statistical inference. However, we find this perspective less than helpful because it suggests a state of hopeless ignorance in which we resign ourselves to being buffeted and battered by the unknowable."
>
> Andrew Lo, who obviously did not bother to read the book he was citing.

**Lack of skin in the game.** Indeed one wonders why econometric methods keep being used while being wrong, so shockingly wrong, how "University" researchers (adults) can partake of such acts of artistry. Basically these capture the ordinary and mask higher order effects. Since blowups are not frequent, these events do not show in data and the researcher looks smart most of the time while being fundamentally wrong. At the source, researchers, "quant" risk manager, and academic economist do not have skin in the game so they are not hurt by wrong risk measures: other people are hurt by them. And the artistry should continue perpetually so long as people are allowed to harm others with impunity. (More in Taleb and Sandis [297], Taleb [282] ).

# G | MACHINE LEARNING CONSIDERATIONS

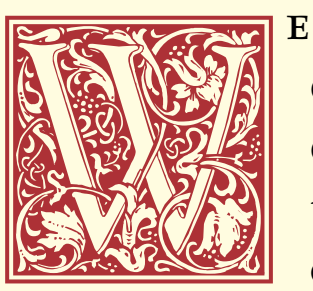E HAVE learned from option trading that you can express any one-dimensional function as a weighted linear combination of call or put options –smoothed by adding time value to the option. An option becomes a building block. A payoff constructed via option is more precisely as follows $S = \sum_i^n \omega_i C(K_i, t_i)$, $i = 1, 2, \ldots, n$, where $C$ is the call price (or, rather, valuation), $\omega$ is a weight $K$ is the strike price, and $t$ the time to expiration of the option. A European call $C$ delivers $max(S - K, 0)$ at expiration $t$. [a]

Neural networks and nonlinear regression, the predecessors of machine learning, on the other hand, focused on the Heaviside step function, again smoothed to produce a sigmoid type "S" curve. A collection of different sigmoids would fit *in sample*.

[a] This appears to be an independent discovery by traders of the universal approximation theorem, initially for sigmoid functions, which are discussed further down (Cybenko [63]).

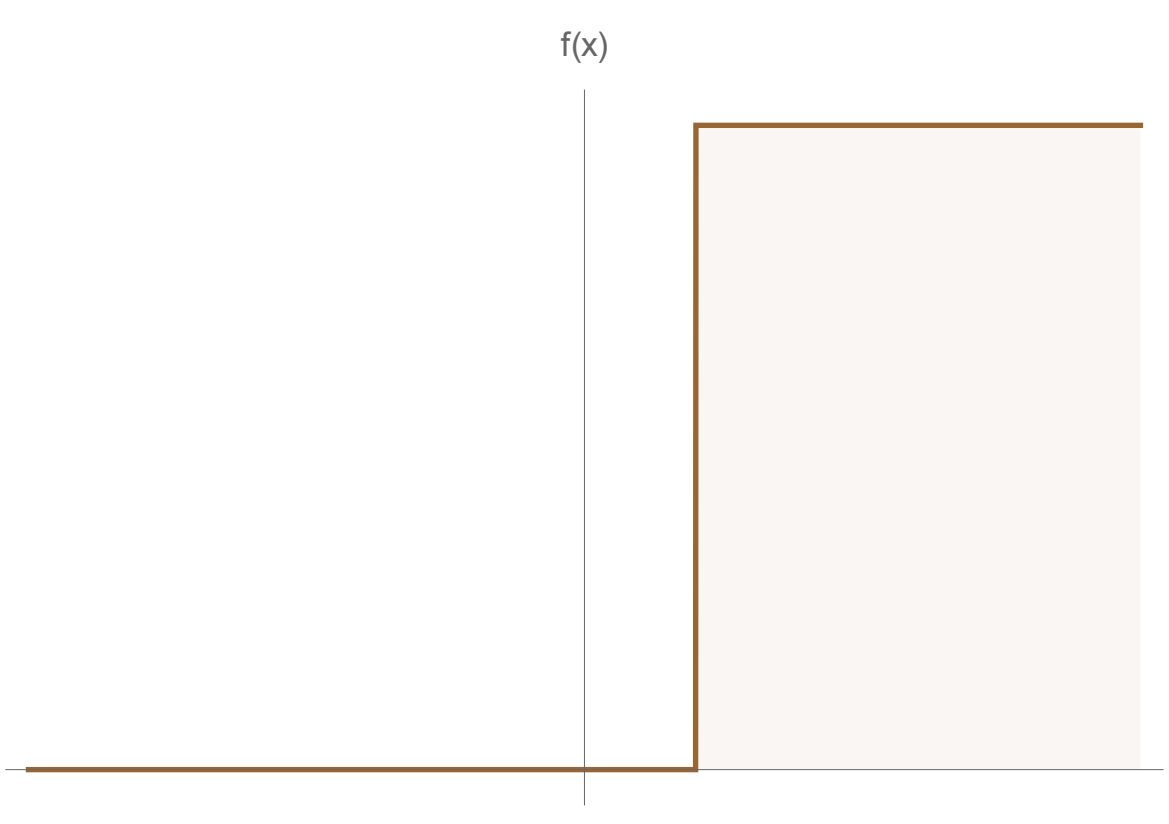

Figure G.1: *The heaviside $\theta$ function: note that it is the payoff of the "binary option" and can be decomposed as* $\lim_{\Delta K \to 0} \frac{C(K) - C(K+\Delta K)}{\delta K}$.

So this discussion is about ...fattailedness and how the different building blocks can accommodate them. Statistical machine learning switched to "ReLu" or "ramp" functions that act exactly like call options rather than an aggregation of "S" curves.

Researchers then discovered that it allows better handling of out of sample tail events (since there are by definition no unexpected tail events in sample) owing to the latter's extrapolation properties.

What is a sigmoid? Consider a payoff function as shown in G.7 that can be expressed with formula $S : (-\infty, \infty) \to (0,1)$, $S(x) = \frac{1}{2} \tanh\left(\frac{\kappa x}{\pi}\right) + \frac{1}{2}\Big)$, or, more precisely, a three parameter function $S_i : (-\infty, \infty) \to (0, a_1)$ $S_i(x) = \frac{a_i}{e^{(c_i - b_i x)} + 1}$. It can also be the cumulative normal distribution, $\mathcal{N}(\mu, \sigma)$ where $\sigma$ controls the smoothness (it then becomes the Heaviside of Fig. G.7 at the limit of $\sigma \to 0$). The (bounded) sigmoid is the smoothing using parameters of the Heaviside function.

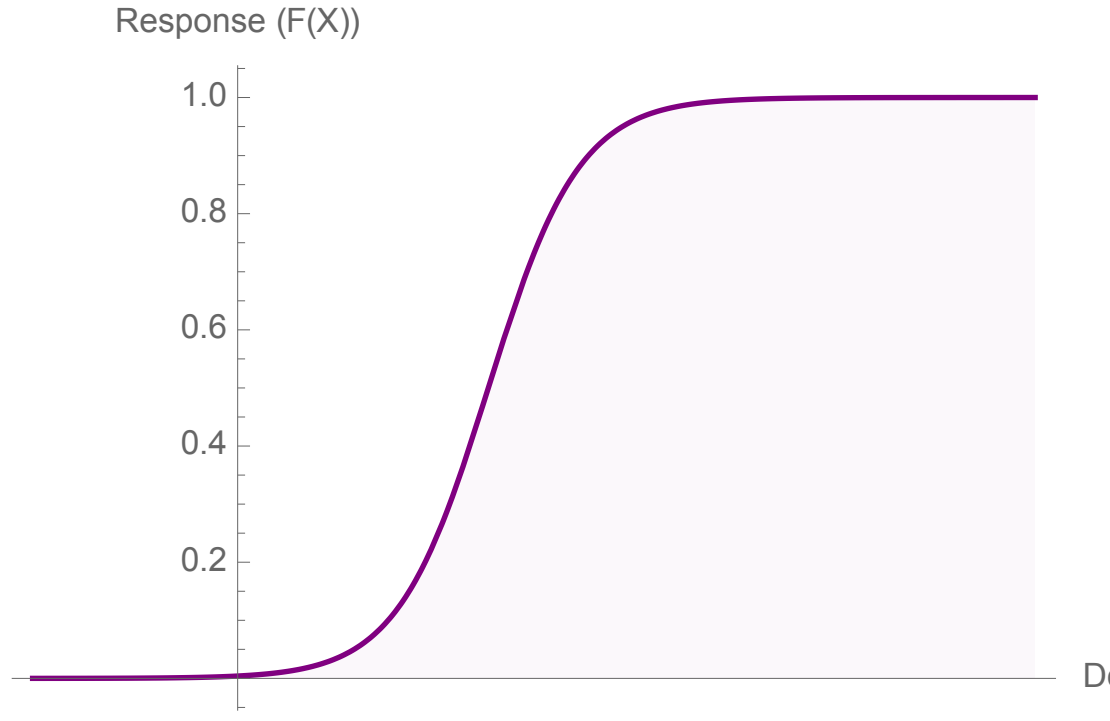

Figure G.2: *The sigmoid function ; note that it is bounded to both the left and right sides owing to saturation: it looks like a smoothed Heaviside $\theta$.*

We can build composite "S" functions with $n$ summands $\chi^n(x) = \sum_i^n \omega_i S_i(x)$ as in G.3. But:

**Remark G.1**

*For $\chi^n(x) \in [0, \infty) \vee [-\infty, 0) \vee (-\infty, \infty)$, we must have $n \to \infty$.*

We need an infinity of summands for an unbounded function. So wherever the "empirical distribution" will be maxed, the last observation will match the flat part of the sig. For the definition of an empirical distribution see 3.4.

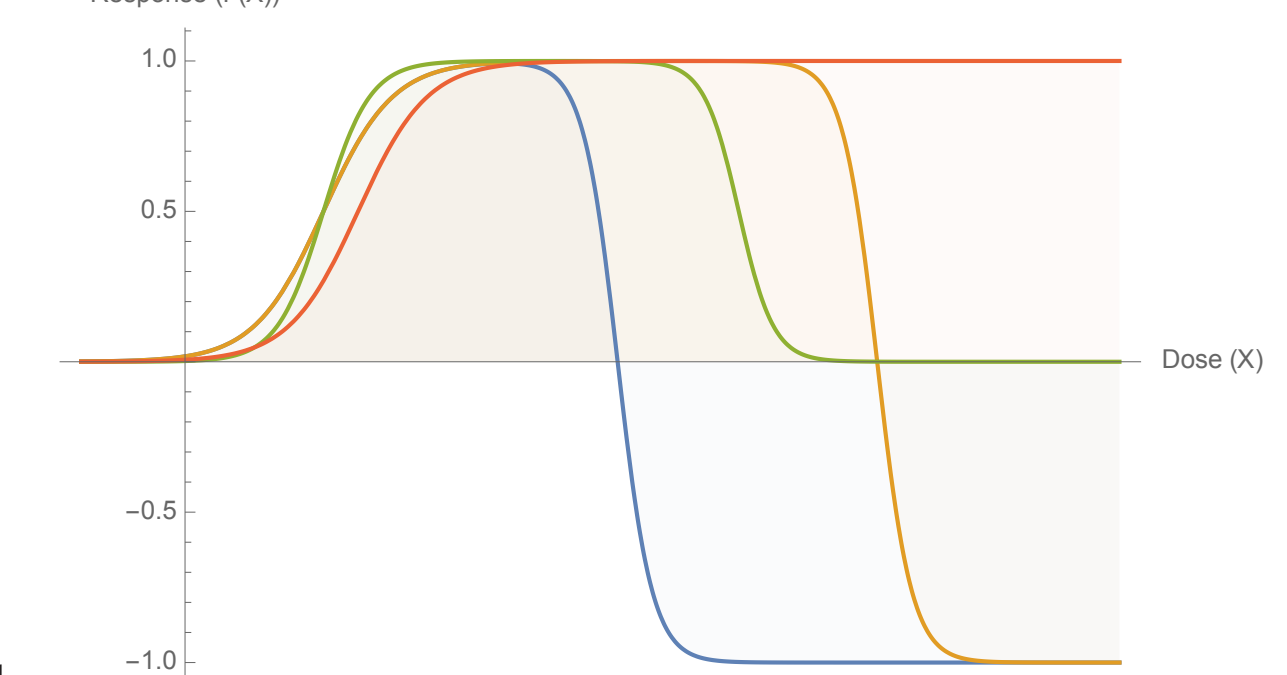

Figure G.3: *A sum of sigmoids will always be bounded, so one needs an infinite sum to replicate an "open" payoff, one that is not subjected to saturation.*

the

Now let us consider option payoffs. Fig.G.4 shows the payoff of a regular option at expiration –the definition of which which matches a Rectifier Linear Unit (ReLu) in machine learning. Now Fig. G.5 shows the following function: consider a function $\rho : (-\infty, \infty) \to [k, \infty)$, with $K \in \mathbb{R}$:

$$\rho(x, K, p) = k + \frac{\log\left(e^{p(x-K)} + 1\right)}{p}. \tag{G.1}$$

We can sum the function as $\sum_i = 1^n \rho(x, K_i, p_i)$ to fit a nonlinear function, which in fact replicates what we did with call options –the parameters $p_i$ allow to smooth time value.

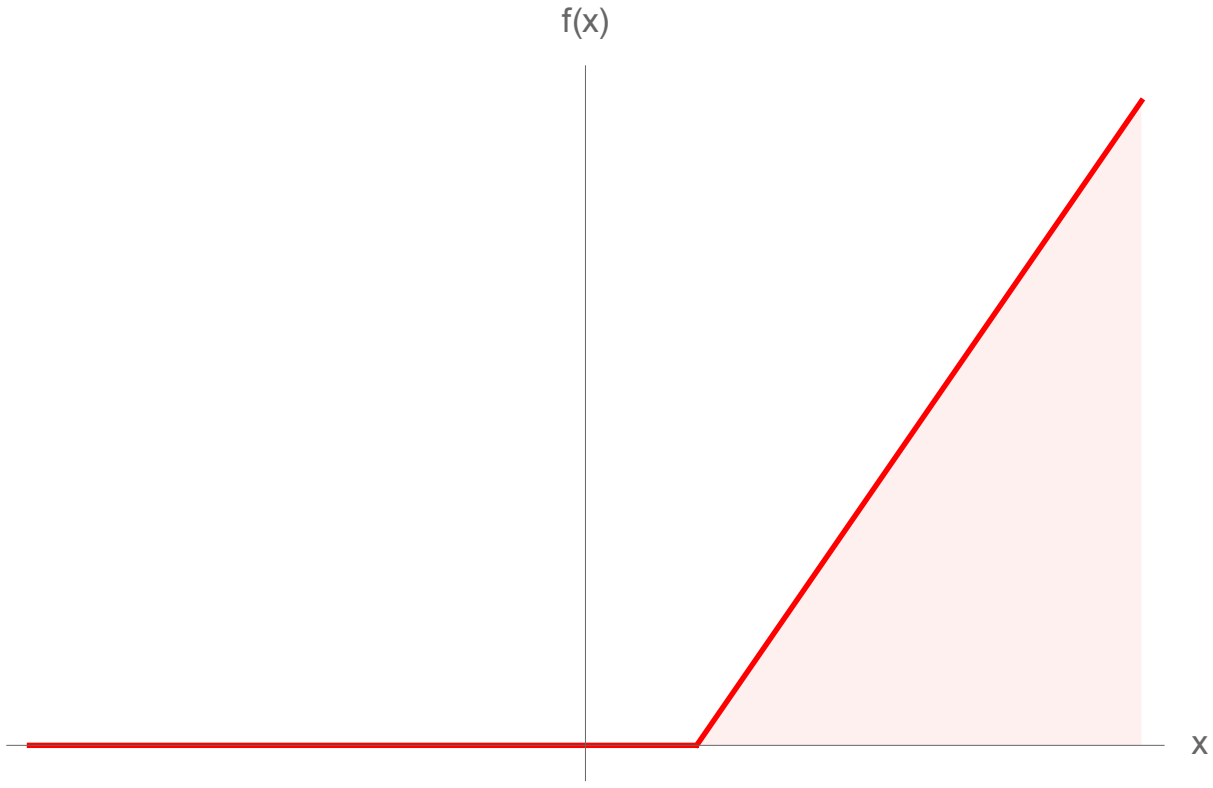

Figure G.4: *An option payoff at expiration, open on the right.*

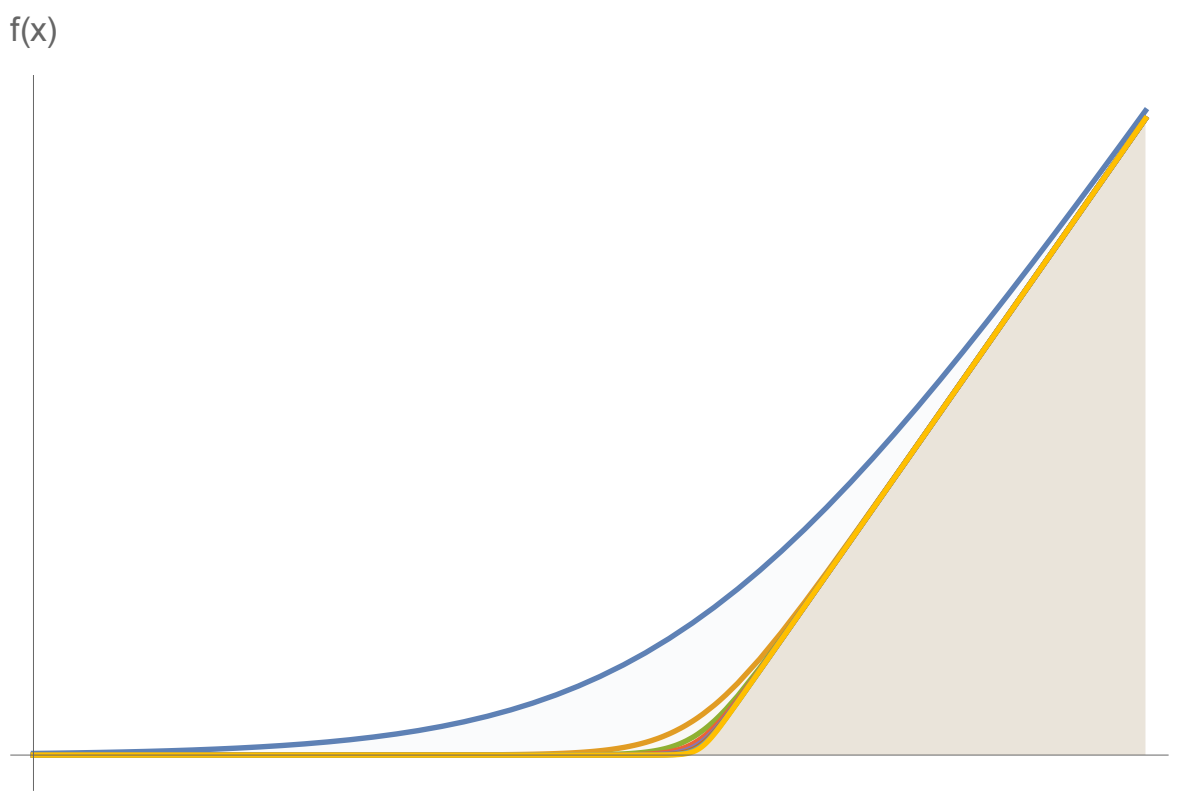

Figure G.5: $\rho$ *function, from Eq. 11.18 , with* $k = 0$. *We calibrate and smooth the payoff with different values of* $p$.

### G.0.1 Calibration via Angles

From figure G.6 we can see that, in the equation, $S = \sum_i^n \omega_i \; C(K_i, t_i)$, the $\omega_i$ corresponds to the arc tangent of the angle made –if positive (as illustrated in figure G.7), or the negative of the arctan of the supplementary angle.

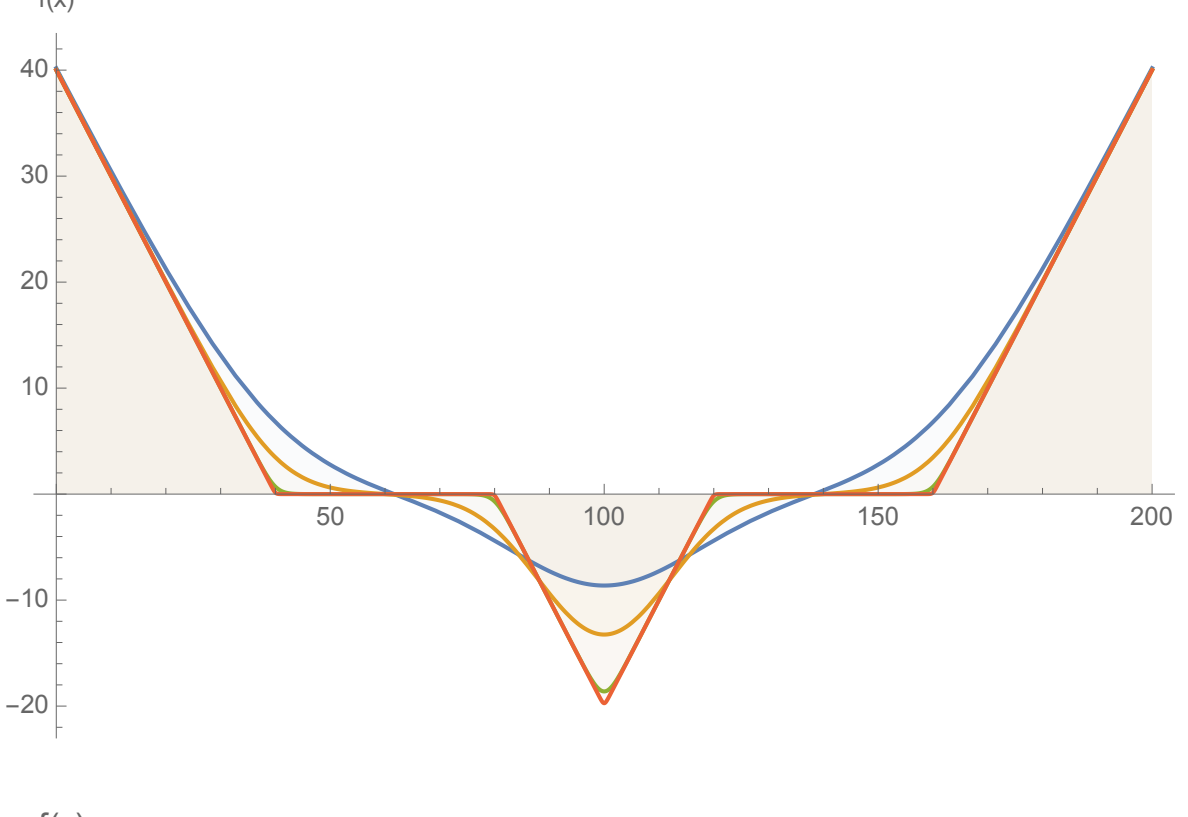

Figure G.6: *A butterfly (built via a sum of options/ReLu, not sigmoids), with open tails on both sides and flipping first and second derivatives. This example is particularly potent as it has no verbalistic correspondence but can be understood by option traders and machine learning.*

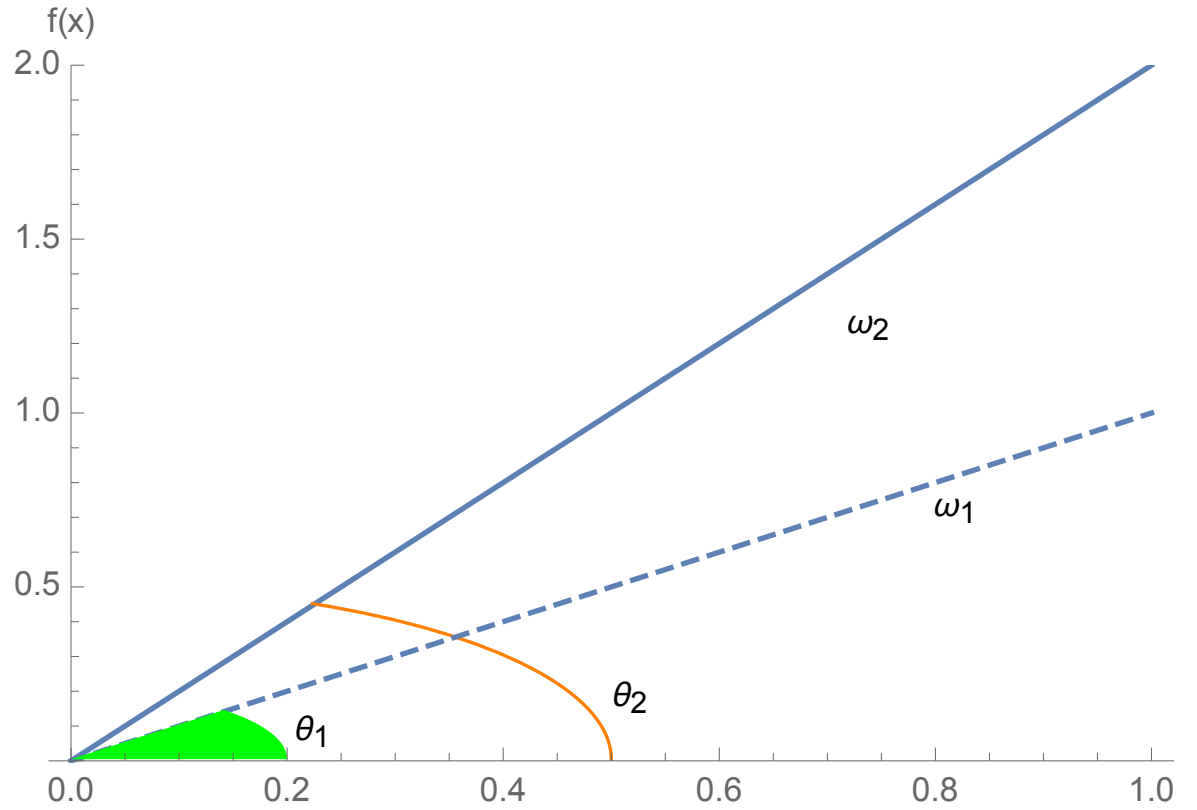

Figure G.7: *How* $\omega = \arctan \theta$. *By fitting angles we can translate a nonlinear function into its option summation.*

**Summary**

We can express all nonlinear univariate functions using a weighted sum of call options of different strikes, which in machine learning applications maps to the tails better than a sum of sigmoids (themselves a net of a long and a short options of neighboring strikes). We can get the weights implicitly using the angles of the functions relative to Cartesian coordinates.

Part III

# PREDICTIONS, FORECASTING, AND UNCERTAINTY

# 11 PROBABILITY CALIBRATION UNDER FAT TAILS ‡

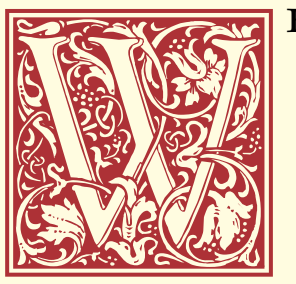

HAT DO BINARY (or probabilistic) forecasting abilities have to do with performance? We map the difference between (univariate) binary predictions or "beliefs" (expressed as a specific "event" will happen/will not happen) and real-world continuous payoffs (numerical benefits or harm from an event) and show the effect of their conflation and mischaracterization in the decision-science literature.

The effects are:

**A) Spuriousness of psychological research** particularly those documenting that humans overestimate tail probabilities and rare events, or that they overreact to fears of market crashes, ecological calamities, etc. Many perceived "biases" are just mischaracterizations by psychologists. There is also a misuse of Hayekian arguments in promoting prediction markets.

**B) Being a "good forecaster" in binary space doesn't lead to having a good performance**, and vice versa, especially under nonlinearities. A binary forecasting record is likely to be a reverse indicator under some classes of distributions. Deeper uncertainty or more complicated and realistic probability distribution worsen the conflation .

**C) Machine Learning:** Some nonlinear payoff functions, while not lending themselves to verbalistic expressions and "forecasts", are well captured by ML or expressed in option contracts.

**D) M Competitions Methods:** The score for the M4-M5 competitions appear to be closer to real world variables than the Brier score.

The appendix shows the mathematical properties and exact distribution of the various payoffs, along with an exact distribution for the Brier score helpful for significance testing and sample sufficiency.

---

Research chapter.

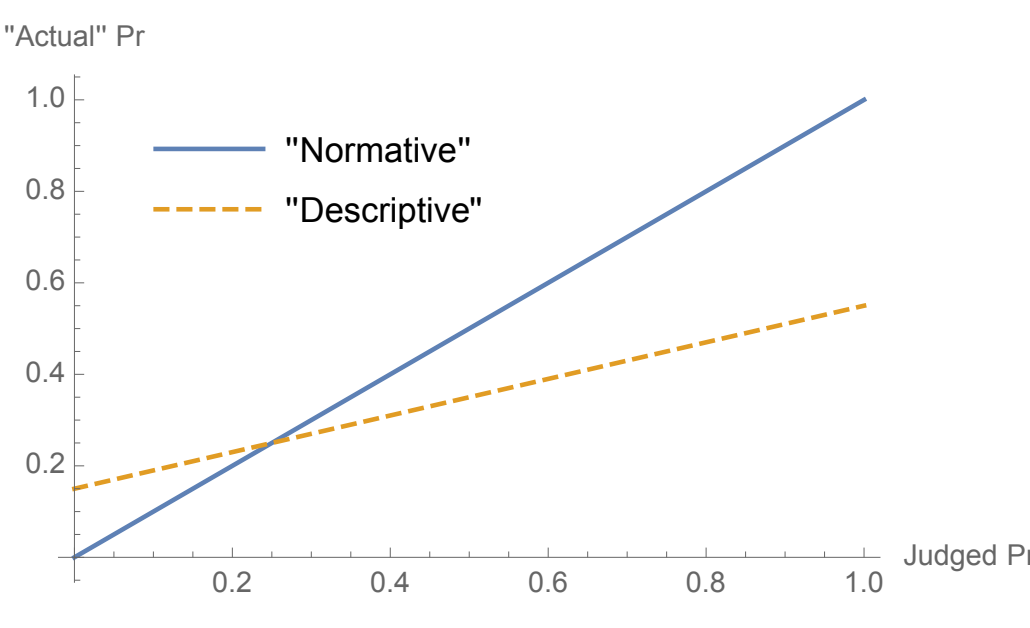

Figure 11.1: *"Typical patterns" as stated and described in [16], a representative claim in psychology of decision making that people overestimate small probability events. The central findings are in 1977 and 1978 [181] and [182]. We note that to the left, in the estimation part, 1) events such as floods, tornados, botulism, mostly patently thick tailed variables, matters of severe consequences that agents might have incorporated in the probability, 2) these probabilities are subjected to estimation error that, when endogenised, increase the estimation.*

## 11.1 CONTINUOUS VS. DISCRETE PAYOFFS: DEFINITIONS & COMMENTS

**Example 11.1** ("One does not eat beliefs and (binary) forecasts")
*In the first volume of the Incerto ( Fooled by Randomness, 2001 [272]), the narrator, a trader, is asked by the manager "do you predict that the market is going up or down?" "Up", he answered, with confidence. Then the boss got angry when, looking at the firm's exposures, he discovered that the narrator was short the market, i.e., would benefit from the market going down.*

*The trader had difficulties conveying the idea that there was no contradiction, as someone could hold the (binary) belief that the market had a higher probability of going up than down, but that, should it go down, there is a very small probability that it could go down considerably, hence a short position had a positive expected return and the rational response was to engage in a short exposure. "You do not eat forecasts, but P/L" (or "one does not monetize forecasts") goes the saying among traders.*

If exposures and beliefs do not go in the same direction, it is because beliefs are verbalistic reductions that contract a higher dimensional object into a single dimension. To express the manager's error in terms of decision-making research, there can be a conflation in something as elementary as the notion of a binary event (related to the zeroth moment) or the *probability* of an event and *expected payoff* from it (related to the first moment and, when nonlinear, to all higher moments) as the payoff functions of the two can be similar in some circumstances and different in others.

**Commentary 11.1**
*In short, probabilistic calibration requires estimations of the zeroth moment while the real world requires all moments (outside of gambling bets or artificial environments such as psychological experiments where payoffs are necessarily truncated), and it is a central property of thick tails that higher moments are explosive (even "infinite") and count more and more.*

### 11.1.1 Away from the Verbalistic

While the trader story is mathematically trivial (though the mistake is committed a bit too often), more serious gaps are present in decision making and risk management, particularly when the payoff function is more complicated, or nonlinear (and related to higher moments). So once we map the contracts or exposures mathematically, rather than focus on words and verbal descriptions, some serious distributional issues arise.

**Definition 11.1** ( Event)
*A (real-valued) random variable $X\colon \Omega \to \mathbb{R}$ defined on the probability space $(\Omega, \mathcal{F}, P)$ is a function $X(\omega)$ of the outcome $\omega \in \Omega$. An event is a measurable subset (countable or not) of $\Omega$, measurable meaning that it can be defined through the value(s) of one of several random variable(s).*

**Definition 11.2** (Binary forecast/payoff)
*A binary forecast (belief, or payoff) is a random variable taking two values*

$$X : \Omega \to \{X_1, X_2\},$$

*with realizations $X_1, X_2 \in \mathbb{R}$.*

In other words, it lives in the binary set (say $\{0, 1\}$, $\{-1, 1\}$, etc.), i.e., the specified event will or will not take place and, if there is a payoff, such payoff will be mapped into two finite numbers (a fixed sum if the event happened, another one if it didn't). Unless otherwise specified, in this discussion we default to the $\{0, 1\}$ set.

Example of situations in the real world where the payoff is binary:

- Casino gambling, lotteries , coin flips, "ludic" environments, or binary options paying a fixed sum if, say, the stock market falls below a certain point and nothing otherwise –deemed a form of gambling[2].
- Elections where the outcome is binary (e.g., referenda, U.S. Presidential Elections), though not the economic effect of the result of the election.[3]
- Medical prognoses for a single patient entailing survival or cure over a specified duration, though not the duration itself as variable, or disease-specific survival expressed in time, or conditional life expectancy. Also exclude anything related to epidemiology.
- Whether a given person who has an online profile will buy or not a unit or more of a specific product at a given time (not the quantity or units).

**Commentary 11.2** (A binary belief is equivalent to a payoff)
*A binary "belief" should map to an economic payoff (under some scaling or normalization*

[2] Retail binary options are typically used for gambling and have been banned in many jurisdictions, such as, for instance, by the European Securities and Markets Authority (ESMA), www.esma.europa.eu, as well as the United States where it is considered another form of internet gambling, triggering a complaint by a collection of decision scientists, see Arrow et al. [5]. We consider such banning as justified since bets have practically no economic value, compared to financial markets that are widely open to the public, where natural exposures can be properly offset.

[3] Note the absence of spontaneously forming gambling markets with binary payoffs for continuous variables. The exception might have been binary options but these did not remain in fashion for very long, from the experiences of the author, for a period between 1993 and 1998, largely motivated by tax gimmicks.

*necessarily to constitute a probability), an insight owed to De Finetti [68] who held that a "belief" and a "prediction" (when they are concerned with two distinct outcomes) map into the equivalent of the expectation of a binary random variable and bets with a payoff in $\{0,1\}$. An "opinion" becomes a choice price for a gamble, and one at which one is equally willing to buy or sell. Inconsistent opinions therefore would lead to a violation of arbitrage rules, such as the "Dutch book", where a combination of mispriced bets can guarantee a future loss.*

**Definition 11.3** (Real world open continuous payoff)

$$X : \Omega \to [a,\infty) \vee (-\infty, b] \vee (-\infty,\infty).$$

*A continuous payoff "lives" in an interval, not a finite set. It corresponds to an unbounded random variable either doubly unbounded or semi-bounded, with the bound on one side (one tailed variable).*

**Caveat** We are limiting for the purposes of our study the consideration to binary vs. continuous and open-ended (i.e., no compact support). Many discrete payoffs are subsumed into the continuous class using standard arguments of approximation. We are also omitting triplets, that is, payoffs in, say $\{-1,0,3\}$, as these obey the properties of binaries (and can be constructed using a sum of binaries). Further, many variable with a floor and a remote ceiling (hence, formally with compact support), such as the number of victims or a catastrophe, are analytically and practically treated as if they were open-ended [55].

Example of situations in the real world where the payoff is continuous:

- Wars casualties, calamities due to earthquake, medical bills, etc.
- Magnitude of a market crash, severity of a recession, rate of inflation
- Income from a strategy
- Sales and profitability of a new product
- In general, anything covered by an insurance contract

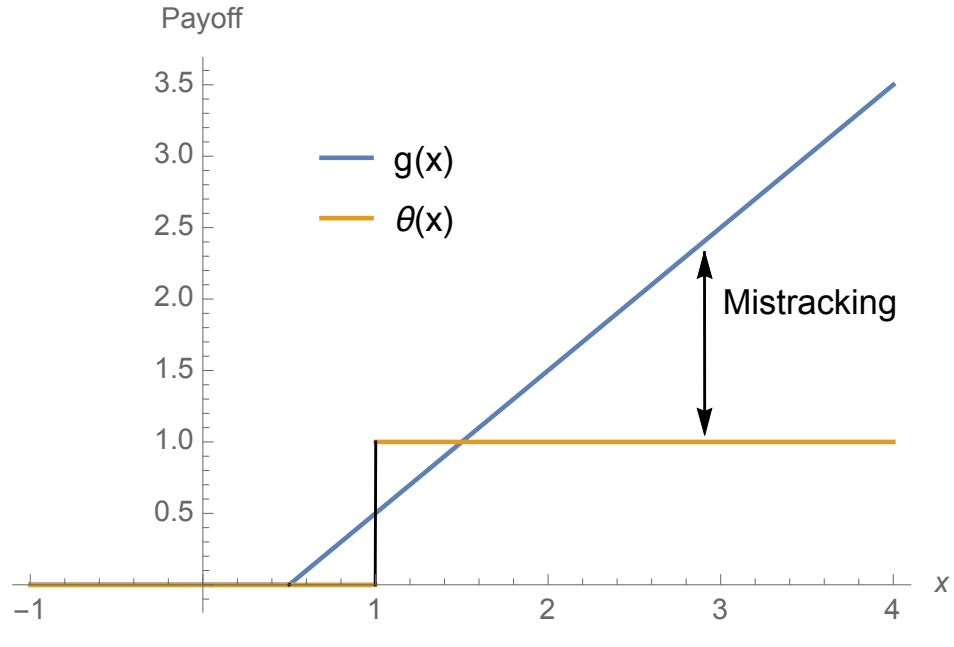

Figure 11.2: *Comparing the payoff of a binary bet (The Heaviside $\theta(.)$) to a continuous open-ended exposure $g(x)$. Visibly there is no way to match the (mathematical) derivatives for any form of hedging.*

Most natural and socio-economic variables are continuous and their statistical distribution does not have a compact support in the sense that we do not have a handle of an exact upper bound.

**Example 11.2**

*Predictive analytics in binary space $\{0,1\}$ can be successful in forecasting if, from his*

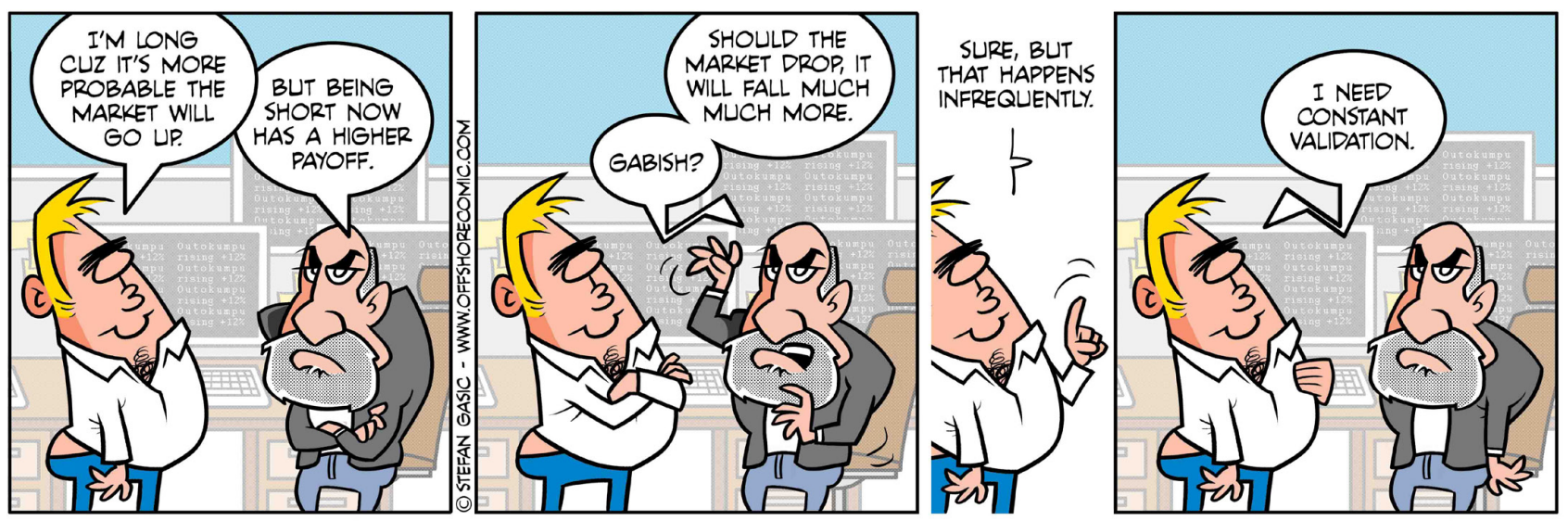

Figure 11.3: *Conflating probability and expected return is deeply entrenched in psychology and finance. Credit: Stefan Gasic.*

*online activity, online consumer Iannis Papadopoulos will purchase a certain item, say a wedding ring, based solely on computation of the probability. But the probability of the "success" for a potential new product might be –as with the trader's story– misleading. Given that company sales are typically thick tailed, a very low probability of success might still be satisfactory to make a decision. Consider venture capital or option trading –an out of the money option can often be attractive yet may have less than 1 in 1000 probability of ever paying off.*

*More significantly, the tracking error for probability guesses will not map to that of the performance.* $\lambda^{(M_4)}$ *would.*

This difference is well known by option traders as there are financial derivative contracts called "binaries" that pay in the binary set $\{0, 1\}$ (say if the underlying asset $S$, say, exceeds a strike price $K$), while others called "vanilla" that pay in $[0, \infty)$, i.e. $\max(S - K, 0)$ (or, worse, in $(-\infty, 0)$ for the seller can now be exposed to bankruptcy owing to the unbounded exposure). The considerable mathematical and economic difference between the two has been discussed and is the subject of *Dynamic Hedging: Managing Vanilla and Exotic Options* [271]. Given that the former are bets paying a fixed amount and the latter have full payoff, one cannot be properly replicated (or hedged) using another, especially under fat tails and parametric uncertainty –meaning performance in one does not translate to performance into the other. While this knowledge is well known in mathematical finance it doesn't seem to have been passed on to the decision-theory literature.

**Commentary 11.3** (Derivatives theory)
*Our approach here is inspired from derivatives (or option) theory and practice where there are different types of derivative contracts, 1) those with binary payoffs (that pay a fixed sum if an event happens) and 2) "vanilla" ones (standard options with continuous payoffs). It is practically impossible to hedge one with another [271]. Furthermore a bet with a strike price K and a call option with same strike K, with K in the tails of the distribution, almost always have their valuations react in opposite ways when one increases the kurtosis of the distribution, (while preserving the first three moments) or, in an example further down in the lognormal environment, when one increases uncertainty via the scale of the distribution.*

**Commentary 11.4** (Term sheets)
*Note that, thanks to "term sheets" that are necessary both legally and mathematically,*

*financial derivatives practive provides precise legalistic mapping of payoffs in a way to make their mathematical, statistical, and economic differences salient.*

There has been a tension between prediction markets and real financial markets. As we can show here, prediction markets may be useful for gamblers, but they cannot hedge economic exposures.

The mathematics of the difference and the impossibility of hedging can be shown in the following. Let $X$ be a random variable in $\mathbb{R}$, we have the payoff of the bet or the prediction $\theta_K : \mathbb{R} \to \{0, 1\}$,

$$\theta_K(x) = \begin{cases} 1, & x \geq K \\ 0 & \text{otherwise,} \end{cases} \tag{11.1}$$

and $g : \mathbb{R} \to \mathbb{R}$ that of the natural exposure. Since $\frac{\partial}{\partial x}\theta_K(x)$ is a Dirac delta function at $K$, $\delta(K)$ and $\frac{\partial)}{\partial x} g_k(x)$ is at least once differentiable for $x \geq K$ (or constant in case the exposure is globally linear or, like an option, piecewise linear above $K$), matching derivatives for the purposes of offsetting variations is not a possible strategy.[4] The point is illustrated in Fig 11.2.

### 11.1.2 There is no defined "collapse", "disaster", or "success" under fat tails

The fact that an "event" has some uncertainty around its magnitude carries some mathematical consequences. Some verbalistic papers in 2019 still commit the fallacy of binarizing an event in $[0, \infty)$: A recent paper on calibration of beliefs says "...if a person claims that the United States is on the verge of an economic collapse or that a climate disaster is imminent..." An economic "collapse" or a climate "disaster" must not be expressed as an event in $\{0, 1\}$ when in the real world it can take many values. For that, a characteristic scale is required. In fact under fat tails, there is no "typical" collapse or disaster, owing to the absence of characteristic scale, hence verbal binary predictions or beliefs cannot be used as gauges.

We present the difference between thin tailed and fat tailed domains as follows.

**Definition 11.4** (Characteristic scale)
*Let $X$ be a random variable that lives in either $(0, \infty)$ or $(-\infty, \infty)$ and $\mathbb{E}$ the expectation operator under "real world" (physical) distribution. By classical results [97]:*

$$\lim_{K\to\infty} \frac{1}{K}\mathbb{E}(X|_{X>K}) = \lambda, \tag{11.2}$$

- *If $\lambda = 1$, $X$ is said to be in the thin tailed class $\mathcal{D}_1$ and has a characteristic scale*
- *If $\lambda > 1$, $X$ is said to be in the fat tailed regular variation class $\mathcal{D}_2$ and has no characteristic scale*
- *If*

$$\lim_{K\to\infty} \mathbb{E}(X|_{X>K}) - K = \mu$$

4 To replicate an open-ended continuous payoff with binaries, one needs an infinite series of bets, which cancels the entire idea of a prediction market by transforming it into a financial market. Distributions with compact support always have finite moments, not the case of those on the real line.

*where $\mu > 0$, then X is in the borderline exponential class*

The point can be made clear as follows. One cannot have a binary contract that adequately hedges someone against a "collapse", given that one cannot know in advance the size of the collapse or how much the face value or such contract needs to be. On the other hand, an insurance contract or option with continuous payoff would provide a satisfactory hedge. Another way to view it: reducing these events to verbalistic "collapse", "disaster" is equivalent to a health insurance payout of a lump sum if one is "very ill" –regardless of the nature and gravity of the illness – and 0 otherwise.

And it is highly flawed to separate payoff and probability in the integral of expected payoff.[5] Some experiments of the type shown in Figure 11 ask agents what is their estimates of deaths from botulism or some such disease: agents are blamed for misunderstanding the probability. This is rather a problem with the experiment: people do not necessarily separate probabilities from payoffs.

## 11.2 SPURIOUS OVERESTIMATION OF TAIL PROBABILITY IN PSYCHOLOGY

**Definition 11.5** (Substitution of integral)
*Let $K \in \mathbb{R}^+$ be a threshold, $f(.)$ a density function and $p_K \in [0,1]$ the probability of exceeding it, and $g(x)$ an impact function. Let $I_1$ be the expected payoff above K:*

$$I_1 = \int_K^\infty g(x) f(x) \, \mathrm{d}x,$$

*and Let $I_2$ be the impact at K multiplied by the probability of exceeding K:*

$$I_2 = g(K) \int_K^\infty f(x) \, dx = g(K) p_K.$$

*The substitution comes from conflating $I_1$ and $I_2$, which becomes an identity if and only if $g(.)$ is constant above K (say $g(x) = \theta_K(x)$, the Heaviside theta function). For $g(.)$ a variable function with positive first derivative, $I_1$ can be close to $I_2$ only under thin-tailed distributions, not under the fat tailed ones.*[6]

For the discussions and examples in this section assume $g(x) = x$ as we will consider the more advanced nonlinear case in Section 11.5.

---

5 Practically all economic and informational variables have been shown since the 1960s to belong to the $\mathcal{D}_2$ class, or at least the intermediate subexponential class (which includes the lognormal), [117, 191, 192, 193, 272], along with social variables such as size of cities, words in languages, connections in networks, size of firms, incomes for firms, macroeconomic data, monetary data, victims from interstate conflicts and civil wars[55, 232], operational risk, damage from earthquakes, tsunamis, hurricanes and other natural calamities, income inequality [44], etc. Which leaves us with the more rational question: where are Gaussian variables? These appear to be at best one order of magnitude fewer in decisions entailing formal predictions.

6 This can also explain, as we will see in Chapter 11 that binary bets can never represent "skin in the game" under fat tailed distributions.

**Theorem 2: Convergence of $\frac{I_1}{I_2}$**

*If X is in the thin tailed class $\mathcal{D}_1$ as described in 11.2,*

$$\lim_{K\to\infty} \frac{I_1}{I_2} = 1 \tag{11.3}$$

*If X is in the regular variation class $\mathcal{D}_2$,*

$$\lim_{K\to\infty} \frac{I_1}{I_2} = \lambda > 1. \tag{11.4}$$

*Proof.* From Eq. 11.2. Further comments:

### 11.2.1 Thin tails

By our very definition of a thin tailed distribution (more generally any distribution outside the subexponential class, indexed by $(g)$), where $f^{(g)}(.)$ is the PDF:

$$\lim_{K\to\infty} \frac{\int_K^\infty x f^{(g)}(x)\, dx}{K \int_K^\infty f^{(g)}(x)\, dx} = \frac{I_1}{I_2} = 1. \tag{11.5}$$

Special case of a Gaussian: Let $g(.)$ be the PDF of predominantly used Gaussian distribution (centered and normalized),

$$\int_K^\infty x g(x)\, dx = \frac{e^{-\frac{K^2}{2}}}{\sqrt{2\pi}} \tag{11.6}$$

and $K_p = \frac{1}{2}\text{erfc}\left(\frac{K}{\sqrt{2}}\right)$, where erfc is the complementary error function, and $K_p$ is the threshold corresponding to the probability $p$.

We note that $K_p \frac{I_1}{I_2}$ corresponds to the inverse Mills ratio used in insurance.

### 11.2.2 Fat tails

For all distributions in the regular variation class, defined by their tail survival function: for $K$ large,

$$\mathbb{P}(X > K) \approx L K^{-\alpha},\ \alpha > 1,$$

where $L > 0$ and $f^{(p)}$ is the PDF of a member of that class:

$$\lim_{K_p\to\infty} \frac{\int_K^\infty x f^{(p)}(x)\, dx}{K \int_{K_p}^\infty f^{(p)}(x)\, dx} = \frac{\alpha}{\alpha - 1} > 1. \tag{11.7}$$

□

### 11.2.3 Conflations

**Conflation of $I_1$ and $I_2$** In numerous experiments, which include the prospect theory paper by Kahneman and Tversky (1978) [165], it has been repeatedly established that agents overestimate small probabilities in experiments where the odds are shown to them, and when the outcome corresponds to a single payoff. The well known Kahneman-Tversky result proved robust, but interpretations make erroneous claims from it. Practically all the subsequent literature, relies on $I_2$ and conflates it with $I_1$, what this author has called *the ludic fallacy* in *The Black Swan* [272], as games are necessarily truncating a dimension from reality. The psychological results might be robust, in the sense that they replicate when repeated in the exact similar conditions, but all the claims outside these conditions and extensions to real risks will be an exceedingly dubious generalization –given that our exposures in the real world rarely map to $I_2$. Furthermore, one can overestimate the probability yet underestimate the expected payoff.

**Stickiness of the conflation** The misinterpretation is still made four decades after Kahneman-Tversky (1979). In a review of behavioral economics, with emphasis on miscaculation of probability, Barberis (2003) [15] treats $I_1 = I_2$. And Arrow et al. [5], a long list of decision scientists pleading for deregulation of the betting markets also misrepresented the fitness of these binary forecasts to the real world (particularly in the presence of real financial markets).

Another stringent –and dangerous –example is the "default VaR" (Value at risk) which is explicitly given as $I_2$ , i.e. default probability $x(1 -$ expected recovery rate$)$, which can be quite different from the actual loss expectation in case of default. Finance presents erroneous approximations of CVaR[7], and the approximation is the risk-management flaw that may have caused the crisis of 2008 [294].

The fallacious argument is that they compute the recovery rate as the expected value of collateral, without conditioning by the default event. The expected value of the collateral conditionally to a default is often far less then its unconditional expectation. In 2007, after a massive series of foreclosures, the value of most collaterals dropped to about 1/3 of its expected value!

**Misunderstanding of Hayek's knowledge arguments** "Hayekian" arguments for the consolidation of beliefs via prices does not lead to prediction markets as discussed in such pieces as [34], or Sunstein's [260]: prices exist in financial and commercial markets; prices are not binary bets. For Hayek [148] consolidation of knowledge is done via prices and *arbitrageurs* (his words)–and arbitrageurs trade products, services, and financial securities, not binary bets.

---

7 The mathematical expression of the Value at Risk, VaR, for a random variable $X$ with distribution function $F$ and threshold $\alpha \in [0, 1]$

$$\text{VaR}_\alpha(X) = -\inf \{x \in \mathbb{R} : F_X(x) > \alpha\},$$

and the corresponding CVar

$$\text{ES}_\alpha(X) = \mathbb{E}\left(-X \,|_{X \leq -\text{VaR}_\alpha(X)}\right)$$

Table 11.1: *Gaussian pseudo-overestimation*

| p | $K_p$ | $\int_{K_p}^{\infty} x f(x) dx$ | $K_p \int_{K_p}^{\infty} f(x) dx$ | $p^*$ | $\frac{p^*}{p}$ |
|---|---|---|---|---|---|
| $\frac{1}{10}$ | 1.28 | $1.75 \times 10^{-1}$ | $1.28 \times 10^{-1}$ | $1.36 \times 10^{-1}$ | 1.36 |
| $\frac{1}{100}$ | 2.32 | $2.66 \times 10^{-2}$ | $2.32 \times 10^{-2}$ | $1.14 \times 10^{-2}$ | 1.14 |
| $\frac{1}{1000}$ | 3.09 | $3.36 \times 10^{-3}$ | $3.09 \times 10^{-3}$ | $1.08 \times 10^{-3}$ | 1.08 |
| $\frac{1}{10000}$ | 3.71 | $3.95 \times 10^{-4}$ | $3.71 \times 10^{-4}$ | $1.06 \times 10^{-4}$ | 1.06 |

Table 11.2: *Paretian pseudo-overestimation*

| p | $K_p$ | $\int_{K_p}^{\infty} x f(x) dx$ | $K_p \int_{K_p}^{\infty} f(x) dx$ | $p^*$ | $\frac{p^*}{p}$ |
|---|---|---|---|---|---|
| $\frac{1}{10}$ | 8.1 | 8.92 | 0.811 | 1.1 (sic) | 11. |
| $\frac{1}{100}$ | 65.7 | 7.23 | 0.65 | 0.11 | 11. |
| $\frac{1}{1000}$ | 533 | 5.87 | 0.53 | 0.011 | 11. |
| $\frac{1}{10000}$ | 4328 | 4.76 | 0.43 | 0.0011 | 11. |

**Definition 11.6** (Corrected probability in binarized experiments)
*Let $p^*$ be the equivalent probability to make $I_1 = I_2$ and eliminate the effect of the error, so*

$$p^* = \{p : I_1 = I_2 = K\}$$

Now let' s solve for $K_p$ "in the tails", working with a probability $p$. For the Gaussian, $K_p = \sqrt{2}\text{erfc}^{-1}(2p)$; for the Paretian tailed distribution, $K_p = p^{-1/\alpha}$.

Hence, for a Paretian distribution, the ratio of real continuous probability to the binary one

$$\frac{p^*}{p} = \frac{\alpha}{1-\alpha},$$

which can allow in absurd cases $p^*$ to exceed 1 when the distribution is grossly misspecified.

Tables 11.1 and 11.2 show, for a probability level $p$, the corresponding tail level $K_p$, such as

$$K_p = \{\inf K : \mathbb{P}(X > K) > p\},$$

and the corresponding adjusted probability $p^*$ that de-binarize the event [8][9]– probabilities here need to be in the bottom half, i.e., $p < .5$. Note that we are operating under the mild case of known probability distributions, as it gets worse under parametric uncertainty.[10]

The most commonly known distribution among the public, the "Pareto 80/20" (based on Pareto discovering that 20 percent of the people in Italy owned 80 percent of the land), maps to a tail index $\alpha = 1.16$, so the adjusted probability is $> 7$ times the naive one.

[8] The analysis is invariant to whether we use the right or left tail .By convention, finance uses negative value for losses, whereas other areas of risk management express the negative of the random variance, hence focus on the right tail.

[9] $K_p$ is equivalent to the Value at Risk $VaR_p$ in finance, where $p$ is the probability of loss.

[10] Note the van der Wijk's law, see Cirillo [53]: $\frac{I_1}{I_2}$ is related to what is called in finance the expected shortfall for $K_p$.

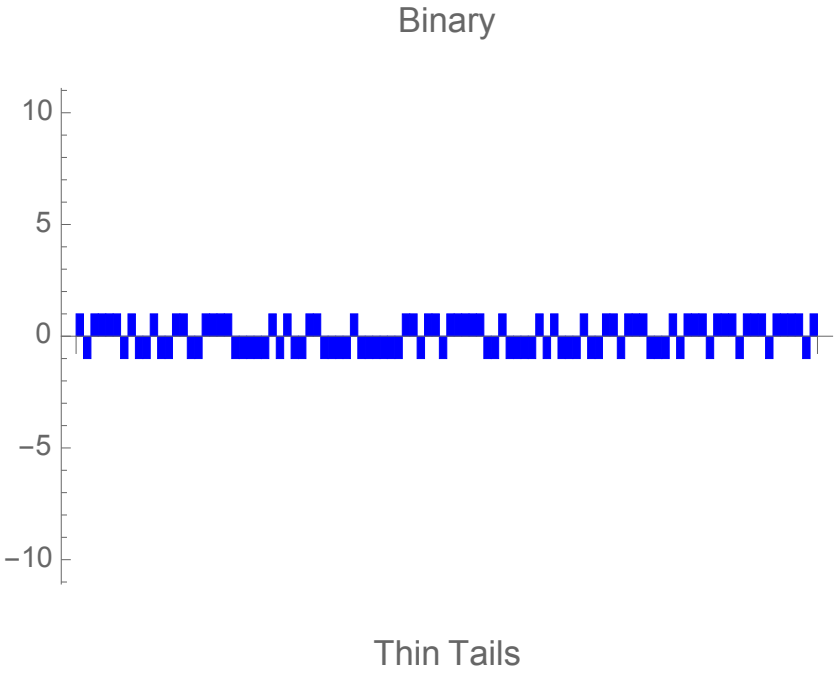

Figure 11.4: *Comparing the three payoffs under two distributions –the binary has the same profile regardless of whether the distribution is thin or fat tailed. The first two subfigures are to scale, the third (representing the Pareto 80/20 with* $\alpha = 1.16$ *requires multiplying the scale by two orders of magnitude.*

**Example of probability and expected payoff reacting in opposite direction under increase in uncertainty** An example showing how, under a skewed distribution, the binary and the expectation reacting in opposite directions is as follows. Consider the risk-neutral lognormal distribution $\mathcal{L}(X_0 - \frac{1}{\sigma^2}, \sigma)$ with PDF $f_L(.)$, mean $X_0$ and variance $\left(e^{\sigma^2} - 1\right) X_0^2$. We can increase its uncertainty with the parameter $\sigma$. We have the expectation of a contract above $X_0$, $\mathbb{E}_{>X_0}$:

$$\mathbb{E}_{>X_0} = \int_{X_0}^{\infty} x f_L(x)\, \mathrm{d}x = \frac{1}{2} X_0 \left(1 + \operatorname{erf}\left(\frac{\sigma}{2\sqrt{2}}\right)\right)$$

and the probability of exceeding $X_0$,

$$\mathbb{P}(X > X_0) = \frac{1}{2}\left(1 - \operatorname{erf}\left(\frac{\sigma}{2\sqrt{2}}\right)\right),$$

where erf is the error function. As $\sigma$ rises $\text{erf}\left(\frac{\sigma}{2\sqrt{2}}\right) \to 1$, with $\mathbb{E}_{>X_0} \to X_0$ and $\mathbb{P}(X > X_0) \to 0$. This example is well known by option traders (see *Dynamic Hedging* [271]) as the binary option struck at $X_0$ goes to 0 while the standard call of the same strike rises considerably to reach the level of the asset –regardless of strike. This is typically the case with venture capital: the riskier the project, the less likely it is to succeed but the more rewarding in case of success. So, the expectation can go to $+\infty$ while to probability of success goes to 0.

### 11.2.4 Distributional Uncertainty

**Remark 11.1: Distributional uncertainty**

*Owing to Jensen's inequality, the discrepancy $(I_1 - I_2)$ increases under parameter uncertainty, expressed in higher kurtosis, via stochasticity of $\sigma$ the scale of the thin-tailed distribution, or that of $\alpha$ the tail index of the Paretian one.*

*Proof.* First, the Gaussian world. We consider the effect of $I_1 - I_2 = \int_K^\infty x f^{(g)}(x) - \int_K^\infty f^{(g)}(x)$ under stochastic volatility, i.e. the parameter from increase of volatility. Let $\sigma$ be the scale of the Gaussian, with $K$ constant:

$$\frac{\partial^2 (\int_K^\infty x f^{(g)}(x) dx)}{\partial \sigma^2} - \frac{\partial^2 (\int_K^\infty f^{(g)}(x) dx)}{\partial \sigma^2} = \frac{e^{-\frac{K^2}{2\sigma^2}} \left((K-1)K^3 - (K-2)K\sigma^2\right)}{\sqrt{2\pi}\sigma^5}, \tag{11.8}$$

which is positive for all values of $K > 0$ (given that $K^4 - K^3 - K^2 + 2K > 0$ for $K$ positive).

Second, consider the sensitivity of the ratio $\frac{I_1}{I_2}$ to parameter uncertainty for $\alpha$ in the Paretian case (for which we can get a streamlined expression compared to the difference). For $\alpha > 1$ (the condition for a finite mean):

$$\frac{\partial^2 \left(\int_K^\infty x f^{(p)}(x) dx / \int_K^\infty f^{(p)}(x) dx\right)}{\partial \alpha^2} = \frac{2K}{(\alpha - 1)^3}, \tag{11.9}$$

which is positive and increases markedly at lower values of $\alpha$, meaning the fatter the tails, the worse the uncertainty about the expected payoff and the larger the difference between $I_1$ and $I_2$.

□

## 11.3 CALIBRATION AND MISCALIBRATION

The psychology literature also examines the "calibration" of probabilistic assessment –an evaluation of how close someone providing odds of events turns out to be on average (under some operation of the law of large number deemed satisfactory) [181], [171], see Fig. 3.13 (as we saw in Chapter 3). The methods, for the reasons we have shown here, are highly flawed except in narrow circumstances of purely binary payoffs (such as those entailing a "win/lose" outcome) –and generalizing from these payoffs is either not possible or produces misleading results. Accordingly, Fig. 11 makes little sense empirically.

At the core, calibration metrics such as the Brier score are always thin-tailed, when the variable under measurement is fat-tailed, which worsens the tractability.

To use again the saying "You do not eat forecasts", most businesses have severely skewed payoffs, so being calibrated in probability is meaningless.

**Remark 11.2: Distributional differences**

*Binary forecasts and calibration metrics via the Brier score belong to the thin-tailed class.*

We will show proofs next.

## 11.4 SCORING METRICS

Table 11.3: *Scoring Metrics for Performance Evaluation*

| Metric | Name | Fitness to reality |
|---|---|---|
| $P^{(r)}(T)$ | Cumulative P/L | Adapted to real world distributions, particularly under a survival filter |
| $P^{(p)}(n)$ | Tally of Bets | Misrepresents the performance under fat tails, works only for binary bets and/or thin tailed domains. |
| $\lambda(n)$ | Brier Score | Misrepresents performance precision under fat tails, ignores higher moments. |
| $\lambda_n^{(M4)}$ | M4 Score | Represents precision not exactly real world performance but maps to real distribution of underlying variables. |
| $\lambda_n^{(M5)}$ | Proposed M5 Score | Represents both precision and survival conditions by predicting extrema of time series. |
| $g(.)$ | Machine learning nonlinear payoff function (not a metric) | Expresses exposures without verbalism and reflects true economic or other P/L. Resembles financial derivatives term sheets. |

This section, summarized in Table 11.3, compares the probability distributions of the various metrics used to measure performance, either by explicit formulation or by linking it to a certain probability class. Clearly one may be mismeasuring performance if the random variable is in the wrong probability class. Different underlying distributions will require a different number of sample sizes owing to the differences in the way the law of numbers operates across distributions. A series of binary forecasts will converge very rapidly to a thin-tailed Gaussian even if the underlying distribution is fat-tailed, but an economic P/L tracking performance for someone with a real exposure will require a considerably larger sample size if, say, the underlying is Pareto distributed [281].

We start by precise expressions for the four possible ones:

1. Real world performance under conditions of survival, or, in other words, P/L or a quantitative cumulative score.
2. A tally of bets, the naive sum of how often a person's binary prediction is correct
3. De Finetti's Brier score $\lambda(B)_n$
4. The M4 score $\lambda_n^{M4}$ for $n$ observations used in the M4 competition, and its prosed sequel M5.

**P/L in Payoff Space (under survival condition)** The "P/L" is short for the natural profit and loss index, that is, a cumulative account of performance. Let $X_i$ be realizations of an unidimensional generic random variable $X$ with support in $\mathbb{R}$ and $t = 1, 2, \ldots n$. Real world payoffs $P_r(.)$ are expressed in a simplified way as

$$P_r(n) = P(0) + \sum_{k \leq N} g(x_t), \tag{11.10}$$

where $g_t : \mathbb{R} \rightarrow \mathbb{R}$ is a measurable function representing the payoff; $g$ may be path dependent (to accommodate a survival condition), that is, it is a function of the preceding period $\tau < t$ or on the cumulative sum $\sum_{\tau \leq t} g(x_\tau)$ to introduce an absorbing barrier, say, bankruptcy avoidance, in which case we write:

$$P^{(r)}(T) = P^{(r)}(0) + \sum_{t \leq n} \mathbb{1}_{(\sum_{\tau < t} g(x_\tau) > b)}\, g(x_t), \tag{11.11}$$

where $b$ is is any arbitrary number in $\mathbb{R}$ that we call the survival mark and $\mathbb{1}_{(.)}$ an indicator function $\in \{0, 1\}$.

The last condition from the indicator function in Eq. 11.11 is meant to handle ergodicity or lack of it [272].

**Commentary 11.5**
*P/L tautologically corresponds to the real world distribution, with an absorbing barrier at the survival condition.*

**Frequency Space,** The standard psychology literature has two approaches.

**A-When tallying forecasts as a counter**

$$P^{(p)}(n) = \frac{1}{n} \sum_{i \leq n} \mathbb{1}_{X_t \in \chi}, \tag{11.12}$$

where $\mathbb{1}_{X_t \in \chi} \in \{0,1\}$ is an indicator that the random variable $x \in \chi_t$ in in the "forecast range", and $T$ the total number of such forecasting events. where $f_t \in [0,1]$ is the probability announced by the forecaster for event $t$

**B-When dealing with a score (calibration method)** in the absence of a visible net performance, researchers produce some more advanced metric or score to measure calibration. We select below the gold standard", De Finetti's Brier score(DeFinetti, [69]). It is favored since it doesn't allow arbitrage and requires perfect probabilistic calibration: someone betting than an event has a probability 1 of occurring will get a perfect score only if the event occurs all the time.

$$\lambda_n^{(B)} = \frac{1}{n} \sum_{t \leq n} (f_t - \mathbb{1}_{X_t \in \chi})^2, \tag{11.13}$$

which needs to be minimized for a perfect probability assessor.

**Applications: M4 and M5 Competitions** The M series (Makridakis [188]) evaluate forecasters using various methods to predict a point estimate (along with a range of possible values). The last competition in 2018, M4, largely relied on a series of scores, $\lambda^{M4_j}$, which works well in situations where one has to forecast the first moment of the distribution and the dispersion around it.

**Definition 11.7** (The M4 first moment forecasting scores)
*The M4 competition precision score (Makridakis et al. [188]) judges competitors on the following metrics indexed by* $j = 1, 2$

$$\lambda_n^{(M4)_j} = \frac{1}{n} \sum_i^n \frac{\left| X_{f_i} - X_{r_i} \right|}{s_j} \tag{11.14}$$

*where* $s_1 = \frac{1}{2}\left( |X_{f_i}| + |X_{r_i}| \right)$ *and* $s_2$ *is (usually) the raw mean absolute deviation for the observations available up to period i (i.e., the mean absolute error from either "naive" forecasting or that from in sample tests),* $X_{f_i}$ *is the forecast for variable i as a point estimate,* $X_{r_i}$ *is the realized variable and n the number of experiments under scrutiny.*

In other word, it is an application of the Mean Absolute Scaled Error (MASE) and the symmetric Mean Absolute Percentage Error (sMAPE) [155].

The suggested M5 score (expected for 2020) adds the forecasts of extrema of the variables under considerations and repeats the same tests as the one for raw variables in Definition 11.7.

### 11.4.1 Deriving Distributions

**Distribution of** $P^{(p)}(n)$

**Remark 11.3**

*The tally of binary forecast* $P^{(p)}(n)$ *is asymptotically normal with mean* $p$ *and standard deviation* $\sqrt{\frac{1}{n}(p-p^2)}$ *regardless of the distribution class of the random variable* $X$.

The results are quite standard, but see appendix for the re-derivations.

### Distribution of the Brier Score $\lambda_n$

**Theorem 3**

*Regardless of the distribution of the random variable* $X$*, without even assuming independence of* $(f_1 - \mathbb{1}_{A1}), \ldots, (f_n - \mathbb{1}_{An})$*, for* $n < +\infty$*, the score* $\lambda_n$ *has all moments of order* $q$*,* $\mathbb{E}(\lambda_n^q) < +\infty$.

*Proof.* For all $i$, $(f_i - \mathbb{1}_{Ai})^2 \leq 1$ . □

We can get actually closer to a full distribution of the score across independent betting policies. Assume binary predictions $f_i$ are independent and follow a beta distribution $\mathcal{B}(a,b)$ (which approximates or includes all unimodal distributions in $[0,1]$ (plus a Bernoulli via two Dirac functions), and let $p$ be the rate of success $p = \mathbb{E}\left(\mathbb{1}_{Ai}\right)$, the characteristic function of $\lambda_n$ for $n$ evaluations of the Brier score is

$$\begin{aligned}\varphi_n(t) = \pi^{n/2} \Bigg( & 2^{-a-b+1}\Gamma(a+b) \\ & \left( p\, {}_2\tilde{F}_2\left(\frac{b+1}{2}, \frac{b}{2}; \frac{a+b}{2}, \frac{1}{2}(a+b+1); \frac{it}{n}\right)\right. \\ & \left.\left. - (p-1)\, {}_2\tilde{F}_2\left(\frac{a+1}{2}, \frac{a}{2}; \frac{a+b}{2}, \frac{1}{2}(a+b+1); \frac{it}{n}\right)\right)\right) . \end{aligned} \tag{11.15}$$

Here ${}_2\tilde{F}_2$ is the generalized hypergeometric function regularized ${}_2\tilde{F}_2(.,.;.,.;.) = \frac{{}_2F_2(a;b;z)}{(\Gamma(b_1)\ldots\Gamma(b_q))}$ and ${}_pF_q(a;b;z)$ has series expansion $\sum_{k=0}^{\infty} \frac{(a_1)_k \ldots (a_p)_k}{(b_1)_k \ldots (b_p)_k} z^k / k!$, were $(a)_{(.)}$ is the Pochhammer symbol.

Hence we can prove the following: under the conditions of independence of the summands stated above,

$$\lambda_n \xrightarrow{D} \mathcal{N}\left(\mu, \sigma_n\right) \tag{11.16}$$

where $\mathcal{N}$ denotes the Gaussian distribution with for first argument the mean and for second argument the standard deviation.

The proof and parametrization of $\mu$ and $\sigma_n$ is in the appendix.

**Distribution of the economic P/L or quantitative measure $P_r$**

**Remark 11.4**
*Conditional on survival to time $T$, the distribution of the quantitative measure $P^{(r)}(T)$ will follow the distribution of the underlying variable $g(x)$.*

The discussion is straightforward if there is no absorbing barrier (i.e., no survival condition).

**Distribution of the M4 score** The distribution of an absolute deviation is in the same probability class as the variable itself. Thee Brier score is in the norm L2 and is based on the second moment (which always exists) as De Finetti has shown that it is more efficient to just a probability in square deviations. However for nonbinaries, it is vastly more efficient under fat tails to rely on absolute deviations, even when the second moment exists [285].

## 11.5 non-verbalistic payoff functions/machine learning

Earlier examples focused on simple payoff functions, with some cases where the conflation $I_1$ and $I_2$ can be benign (under the condition of being in a thin tailed environment). However

**Inseparability of probability under nonlinear payoff function** Now when we introduce a payoff function $g(.)$ that is nonlinear, that is that the economic or other quantifiable response to the random variable $X$ varies with the levels of $X$, the discrepancy becomes greater and the conflation worse.

**Commentary 11.6** (Probability as an integration kernel)
*Probability is just a kernel inside an integral or a summation, not a real thing on its own. The economic world is about quantitative payoffs.*

**Remark 11.5: Inseparability of probability**

*Let $F : \mathcal{A} \to [0,1]$ be a probability distribution (with derivative $f$) and $g : \mathbb{R} \to \mathbb{R}$ a measurable function, the "payoff"". Clearly, for $\mathcal{A}'$ a subset of $\mathcal{A}$:*

$$\begin{aligned} \int_{\mathcal{A}'} g(x) \mathrm{d}F(x) &= \int_{\mathcal{A}'} f(x) g(x) \mathrm{d}x \\ &\neq \int_{\mathcal{A}'} f(x) \mathrm{d}x \; g\left( \int_{\mathcal{A}'} \mathrm{d}x \right) \end{aligned}$$

*In discrete terms, with $\pi(.)$ a probability mass function:*

$$\begin{aligned} \sum_{x \in \mathcal{A}'} \pi(x) g(x) &\neq \sum_{x \in \mathcal{A}'} \pi(x) \; g\left( \frac{1}{n} \sum_{x \in \mathcal{A}'} x \right) \\ &= \text{probability of event } \times \text{payoff of average event} \end{aligned} \tag{11.17}$$

*Proof.* Immediate by Jensen's inequality. □

In other words, the probability of an event is an expected payoff only when, as we saw earlier, $g(x)$ is a Heaviside theta function.

Next we focus on functions tractable mathematically or legally but not reliable verbalistically via "beliefs" or "predictions".

**Misunderstanding $g$** Figure 11.5 showing the mishedging story of Morgan Stanley is illustrative of verbalistic notions such as "collapse" mis-expressed in nonlinear exposures. In 2007 the Wall Street firm Morgan Stanley decided to "hedge" against a real estate "collapse", before the market in real estate started declining. The problem is that they didn't realize that "collapse" could take many values, some worse than they expected, and set themselves up to benefit if there were a mild decline, but lose much if there is a larger one. They ended up right in predicting the crisis, but lose \$10 billion from the "hedge".

Figure G.6 shows a more complicated payoff, dubbed a "butterfly"

**The function $g$ and machine learning** We note that $g$ maps to various machine learning functions that produce exhaustive nonlinearities via the universal universal approximation theorem (Cybenko [63]), or the generalized option payoff decompositions (see *Dynamic Hedging* [271]).

Consider the function $\rho : (-\infty, \infty) \to [K, \infty)$, with $K$, the r.v. $X \in \mathbb{R}$:

$$\rho_{K,p}(x) = k + \frac{\log\left( e^{p(x-K)} + 1 \right)}{p} \tag{11.18}$$

We can express all nonlinear payoff functions $g$ as, with the weighting $\omega_i \in \mathbb{R}$:

$$g(x) = \sum_i \omega_i \, \rho_{K_i,p}(x) \tag{11.19}$$

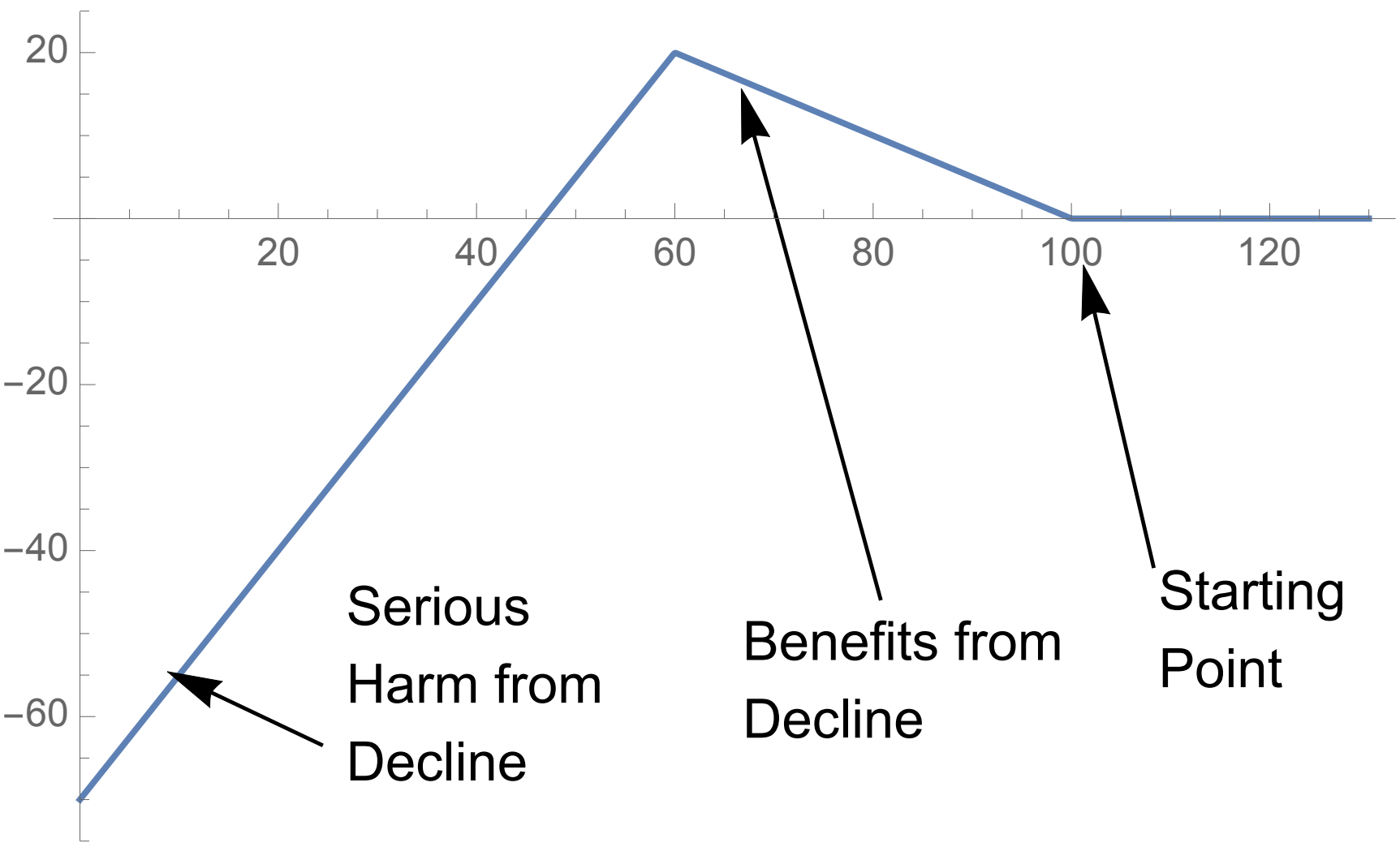

Figure 11.5: *The Morgan Stanley Story: an example of an elementary nonlinear payoff that cannot be described verbalistically. This exposure is called in derivatives traders jargon a "Christmas Tree", achieved by purchasing a put with strike K and selling a put with lower strike $K-\Delta_1$ and another with even lower strike $K-\Delta_2$, with $\Delta_2 \geq \Delta_1 \geq 0$.*

by some similarity, $\rho_{K,p}(x)$ maps to the value a call price with strike $K$ and time $t$ to expiration normalized to 1, all rates set at 0, with sole other parameter $\sigma$ the standard deviation of the underlying.

We note that the expectation of $g(.)$ is the sum of expectation of the ReLu functions:

$$\mathbb{E}\left(g(x)\right) = \sum_i \omega_i \; \mathbb{E}\left(\rho_{K_i,p}(x)\right) \tag{11.20}$$

The variance and other higher order statistical measurements are harder to obtain in closed or simple form.

**Commentary 11.7**
*Risk management is about changing the payoff function $g(.)$ rather than making "good forecasts".*

We note than $\lambda$ is not a metric but a target to which one can apply various metrics.

**Survival**

Decision making is sequential. Accordingly, miscalibration may be a good idea if it reduces the odds of being absorbed. See the appendix of *Skin in the Game* [272], which shows the difference between ensemble probability and time probability. The expectation of the sum of $n$ gamblers over a given day is different from that of a single gambler over $n$ days, owing to the conditioning.

In that sense, measuring the performance of an agent who will eventually go bust (with probability one) is meaningless.[11]

## 11.6 CONCLUSION:

Finally, that in the real world, it is the net performance (economic or other) that counts, and making "calibration" mistakes where it doesn't matter or can be helpful should be encouraged, not penalized. The bias variance argument is well known in machine learning [145] as means to increase performance, in discussions of rationality (see *Skin in the Game* [272]) as a necessary mechanism for survival, and a very helpful psychological adaptation (Brighton and Gigerenzer [37] show a potent argument that if it is a bias, it is a pretty useful one.) If a mistake doesn't cost you anything –or helps you survive or improve your outcomes– it is clearly not a mistake. And if it costs you something, and has been present in society for a long time, consider that there may be hidden evolutionary advantages to these types of mistakes –of the following sort: **mistaking a bear for a stone** is worse than **mistaking a stone for a bear**.

We have shown that, in risk management, one should never operate in probability space.

## 11.7 APPENDIX: PROOFS AND DERIVATIONS

### 11.7.1 Distribution of Binary Tally $P^{(p)}(n)$

We are dealing with an average of Bernoulli random variables, with well known results but worth redoing. The characteristic function of a Bernoulli distribution with parameter $p$ is $\psi(t) = 1 - p + e^{(It)}p$. We are concerned with the $N$-summed cumulant generating function $\psi'(\omega) = \log \psi(\frac{\omega}{N})^N$. We have $\kappa(p)$ the cumulant of order $p$:

$$\kappa(p) = -i^p \frac{\partial^p \psi'}{\partial t^p}\bigg|_{t\to 0}$$

So: $\kappa(1) = p$, $\kappa(2) = \frac{(1-p)p}{N}$, $\kappa(3) = \frac{(p-1)p(2p-1)}{N^2}$, $\kappa(4) = \frac{(1-p)p(6(p-1)p+1)}{N^3}$, which proves that $P^{(p)}(N)$ converges by the law of large numbers at speed $\sqrt{N}$, and by the central limit theorem arrives to the Gaussian at a rate of $\frac{1}{N}$, (since from the cumulants above, its kurtosis = $3 - \frac{6(p-1)p+1}{n(p-1)p}$).

### 11.7.2 Distribution of the Brier Score

[11] The M5 competition is expected to correct for that by making "predictors" predict the minimum (or maximum) in a time series.

**Base probability $f$** First, we consider the distribution of $f$ the base probability. We use a beta distribution that covers both the conditional and unconditional case (it is a matter of parametrization of $a$ and $b$ in Eq. 11.15).

**Distribution of the probability** Let us refresh a standard result behind nonparametric discussions and tests, dating from Kolmogorov [174] to show the rationale behind the claim that the probability distribution of probability (sic) is robust –in other words the distribution of the probability of $X$ doesn't depend on the distribution of $X$, ([80] [171]).

The probability integral transform is as follows. Let $X$ have a continuous distribution for which the cumulative distribution function (CDF) is $F_X$. Then –in the absence of additional information –the random variable $U$ defined as $U = F_X(X)$ is uniform between 0 and 1. The proof is as follows: For $t \in [0, 1]$,

$$\mathbb{P}(Y \leq u) = P(F_X(X) \leq u) = \mathbb{P}(X \leq F_X^{-1}(u)) = F_X(F_X^{-1}(u)) = u \quad (11.21)$$

which is the cumulative distribution function of the uniform. This is the case regardless of the probability distribution of $X$.

Clearly we are dealing with 1) $f$ beta distributed (either as a special case the uniform distribution when purely random, as derived above, or a beta distribution when one has some accuracy, for which the uniform is a special case), and 2) $\mathbb{1}_{At}$ a Bernoulli variable with probability $p$.

Let us consider the general case. Let $g_{a,b}$ be the PDF of the Beta:

$$g_{a,b}(x) = \frac{x^{a-1}(1-x)^{b-1}}{B(a,b)}, \; 0 < x < 1$$

The results, a bit unwieldy but controllable:

$$\mu = \frac{\left(a^2(-(p-1)) - ap + a + b(b+1)p\right)\Gamma(a+b)}{\Gamma(a+b+2)}$$

$$\sigma_n^2 = -\frac{1}{n(a+b)^2(a+b+1)^2}\left(a^2(p-1)+a(p-1)-b(b+1)p\right)^2 + \frac{1}{(a+b+2)(a+b+3)}(a + b)(a+b+1)(p(a-b)(a+b+3)(a(a+3)+(b+1)(b+2)) - a(a+1)(a+2)(a+3))$$

We can further verify that the Brier score has thinner tails than the Gaussian as its kurtosis is lower than 3.

*Proof.* We start with $y_j = (f - \mathbb{1}_{Aj})$, the difference between a continuous Beta distributed random variable and a discrete Bernoulli one, both indexed by $j$. The characteristic function of $y_j$, $\Psi_f^{(y)} = \left(1 + p\left(-1 + e^{-it}\right)\right) {}_1F_1(a; a+b; it)$ where ${}_1F_1(.;.;.)$ is the Kummer confluent hypergeometric function ${}_1F_1(a;b;z) = \sum_{k=0}^{\infty} \frac{a_k \frac{z^k}{k}!}{b_k}$.

From here we get the characteristic function for $y_j^2 = (f_j - \mathbb{1}_{Aj})^2$

$$\Psi^{(y^2)}(t) = \sqrt{\pi}2^{-a-b+1}\Gamma(a+b)\left(p\,{}_2\tilde{F}_2\left(\frac{b+1}{2},\frac{b}{2};\frac{a+b}{2},\frac{1}{2}(a+b+1);it\right) - (p-1)\,{}_2\tilde{F}_2\left(\frac{a+1}{2},\frac{a}{2};\frac{a+b}{2},\frac{1}{2}(a+b+1);it\right)\right) \tag{11.22}$$

where ${}_2\tilde{F}_2$ is the generalized hypergeometric function regularized ${}_2\tilde{F}_2(.,.;.,.;.) = \frac{{}_2F_2(a;b;z)}{(\Gamma(b_1)\ldots\Gamma(b_q))}$ and ${}_pF_q(a;b;z)$ has series expansion $\sum_{k=0}^{\infty}\frac{(a_1)_k\ldots(a_p)_k}{(b_1)_k\ldots(b_p)_k}z^k/k!$, were $(a)_{(.)}$ is the Pochhammer symbol.

We can proceed to prove directly from there the convergence in distribution for the average $\frac{1}{n}\sum_i^n y_i^2$:

$$\lim_{n\to\infty}\Psi_{y^2}(t/n)^n = \exp\left(-\frac{it(p(a-b)(a+b+1)-a(a+1))}{(a+b)(a+b+1)}\right) \tag{11.23}$$

which is that of a degenerate Gaussian (Dirac) with location parameter $\frac{p(b-a)+\frac{a(a+1)}{a+b+1}}{a+b}$.

We can finally assess the speed of convergence, the rate at which higher moments map to those of a Gaussian distribution: consider the behavior of the $4^{th}$ cumulant $\kappa_4 = -i\frac{\partial^4 \log \Psi_{.}(.)}{\partial t^4}|_{t\to 0}$:

1) in the maximum entropy case of $a = b = 1$:

$$\kappa_4|_{a=1,b=1} = -\frac{6}{7n}$$

regardless of $p$.

2) In the maximum variance case, using l'Hôpital:

$$\lim_{\substack{a\to 0\\ b\to 0}}\kappa_4 = -\frac{6(p-1)p+1}{n(p-1)p}$$

Se we have $\frac{\kappa_4}{\kappa_2^2}\underset{n\to\infty}{\to} 0$ at rate $n^{-1}$. □

Further, we can extract its probability density function of the Brier score for $N = 1$: for $0 < z < 1$,

$$p(z) = \frac{\Gamma(a+b)\left((p-1)z^{a/2}\left(1-\sqrt{z}\right)^b - p\left(1-\sqrt{z}\right)^a z^{b/2}\right)}{2\left(\sqrt{z}-1\right)z\Gamma(a)\Gamma(b)}. \tag{11.24}$$

# 12 ON SINGLE POINT FORECASTS FOR FAT-TAILED VARIABLES

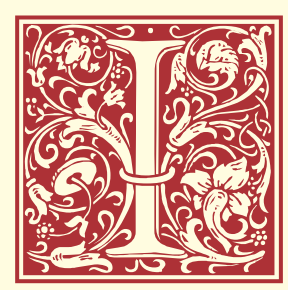
N THIS (RESEARCH) CHAPTER we show consequences of fat tails on pandemics in general, and the conditions and reactions of the last one in particular. The authors fought a few general misunderstandings of the dynamics of pandemics trying to be squeezed into naive "science is forecasting" frameworks. The following statements ensue:

## 12.1 MAIN STATEMENTS

(i)– Forecasting single variables in fat-tailed domains is in violation of both common sense and probability theory.

(ii)– Pandemics are extremely fat-tailed events, with potentially destructive tail risk.

(iii)– Science is not about making single points predictions but about understanding properties (which can *sometimes* be tested by single point estimates and predictions).

(iv)– Sound risk management is concerned with extremes, tails and their properties, and not with averages or the bulk of a distribution.

(v)– Naive fortune-cookie evidentiary methods fail to work under both risk management and fat tails, because the absence of evidence can play a large role in the properties.

## 12.2 COMMENTARY

**Both forecasters and their critics are wrong** At the onset of the COVID-19 pandemic, many researcher groups and agencies produced single point "forecasts"

Research chapter with Yaneer Bar-Yam and Pasquale Cirillo.

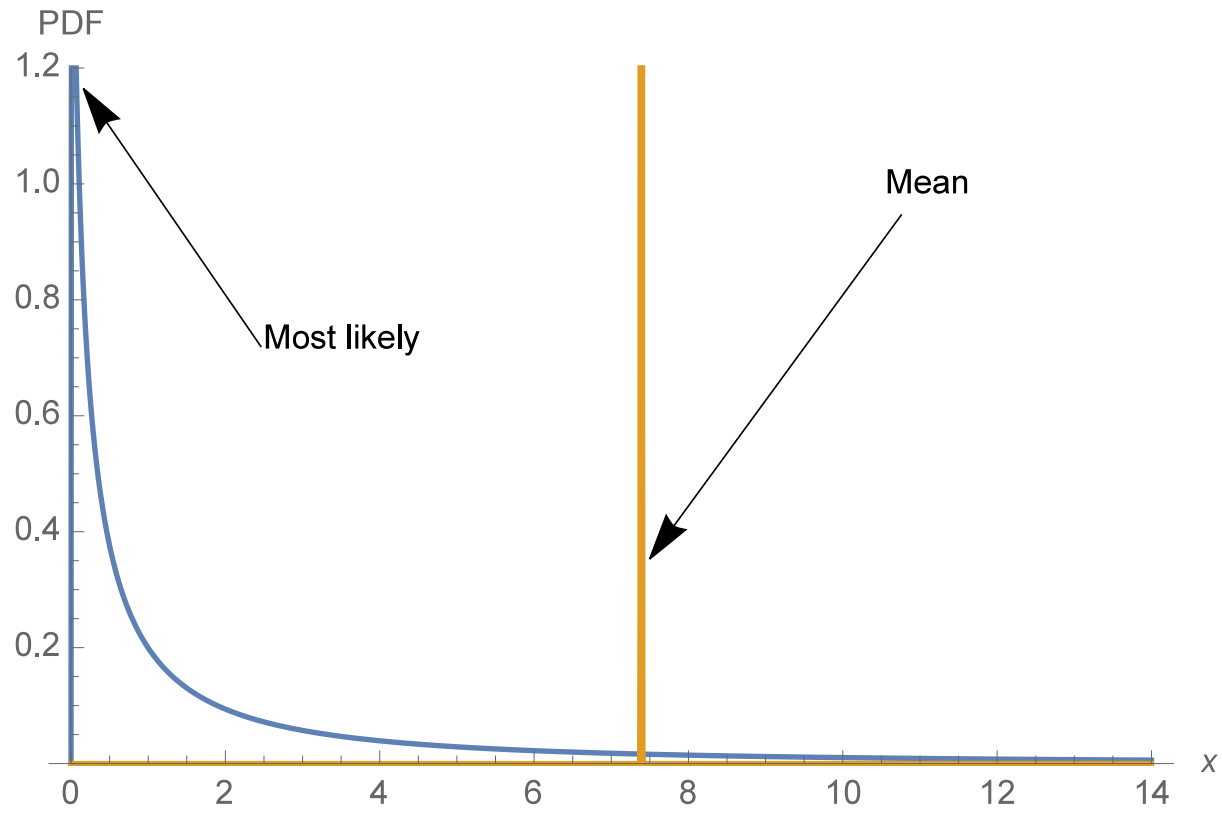

Figure 12.1: *A high variance Lognormal distributions. 85% of observations fall below the mean. Half the observations fall below 13% of the mean (here 1). The lognormal has milder tails than the Pareto which has been shown to represent pandemics.*

for the pandemic—most relied on the compartmental SIR model, sometimes supplemented with cellular automata or agent based models assuming various social and behavioral assumptions. The prevailing idea is that *producing a numerical estimate* is how science is done, and how science-informed decision-making ought to be done: bean counters producing precise numbers. Always within a narrowly considered set of options identified by the researchers.

Well, no. That's not how "science is done", at least in this domain, and that's not how informed decision-making out to be done. Furthermore, subsequently, many criticized the predictions because these did not play out (no surprise there). This is also wrong. Both forecasters (who missed) and their critics were wrong –and the forecasters would have been wrong even if they got the prediction right.

**Statistical attributes of pandemics** Using tools from extreme value theory (EVT), Cirillo and Taleb [? ] determined that pandemics are *patently* fat tailed (with a tail exponent patently in the heaviest class[2]: $\alpha < 1$ ) –a fact that was well known (and communicated by Benoit Mandelbrot) but not formally investigated. Pandemics do represent existential risk. The implication is that much of what takes place in the body of the distribution is just noise. One must not forecast, discuss, or theorize from noise. But it does not mean we can ignore the –very tractable – attributes as in fact they were shown to have remarkably stable extreme value properties.

> **Remark 12.1**
>
> *Random variables with unstable (and uninformative) sample moments may still have stable tail properties centrally useful for inference and risk taking.*[a].
>
> [a] This is the central problem with the misunderstanding of *The Black Swan*: some events have stable and well known properties yet do not lend themselves to prediction

[2] Simple characterization of the power law class by the survival function: for $X$ in the tails, $\mathbb{P}(X > x) = Kx^{-\alpha}$, where $K$ is a constant.

**Fortune cookie evidentiary methods** At about the same time Ioannidis (2020) [158] made statements to the effect that one should wait for "more evidence" before acting with respect to the pandemic, claiming that "we are making decisions without reliable data".

Firstly, there seems to be a huge probabilistic confusion. We do not need more evidence under fat tailed distributions –it is there in the properties themselves (properties for which we have ample *evidence*) and these clearly represent risk that must be killed in the egg (when it is still cheap to do so).

Secondly, unreliable data –or any source of uncertainty – should make us follow the most paranoid route. The entire idea of the author's *Incerto2001* is that more uncertainty in a system makes precautionary decisions very easy to make (if I am *uncertain* about the skills of the pilot, I get off the plane).

More generally, evidence follows, does not precede, rare impactful events and waiting for the accident before putting the seat belt on, or evidence of fire before buying insurance would make the perpetrator exit the gene pool. As the Latin sayings: *cineri nunc medicina datur* (one does not give remedies to the dead) or *post bellum auxilium* ([one must not] send troups after the battle).

**Remark 12.2: Fundamental Risk Asymmetry**

*For matters of survival, particularly when systemic, under such classes as multiplicative pandemics, we require "evidence of no harm" rather than "evidence of harm".*

## 12.3 more technical commentary

**LLN a nd Evidence** In order to leave the domain of ancient divination (or modern anecdote) and enter proper empirical science, forecasting must abide by both evidentiary and probabilistic rigor. Thus any forecasting activity requires the working of the law of large number (LLN), in other words, some convergence at a known rate so one can establish some significance given $n$ observations. This is well known and established... except that many are not aware that, with the theory remaining exactly the same, the story changes under fat tails.

If one claims fitness or nonfitness of a forecasting ability based on a single observation $n = 1$, she or he would be deemed to be making an unscientific claim. For fat tailed variables that "$n = 1$" error can be made with $n = 10^6$. In the case of pandemics, $n = \infty$ is still anecdotal.

**Remark 12.3**

*Random variables in the power law class with tail exponent* $\alpha \leq 1$ *are, simply, not forecastable. They do not obey the LLN. But we can still understand their properties.*

As a matter of fact, owing to preasymptotic properties, a heuristic is to consider variables with up to $\alpha <= \frac{5}{2}$ as not forecastable– the mean will be too unstable and requires way too much data for it to be possible to do so in reasonable time. It takes $10^{14}$ observations for a "Pareto 80/20" (the most commonly referred to probability distribution, that is with $\alpha \approx 1.13$) for the average thus obtained to emulate the significance of a Gaussian with only 30 observations.

Assuming significance with a low $n$ in relation to the properties of the variable is an insult to everything we have learned since Bernoulli, or perhaps even Cardano.

**Science is about understanding properties, not forecasting single outcomes.** Figures 12.1 and 15.13 show the extent of the problem of forecasting under fat tails. Most of the information is away from the center of the distribution.

In some situations of fast-acting LLN, as in physics, properties can be revealed by single predictive experiments. But it is a fallacy to assume from that that single predictive experiments must be able to test any theory –though an $n = 1$ tail event with low conditional probability can falsify a theory.

Sometimes, as has been shown in the IJF by the author (just off the press, [265]), the forecaster may find a single variable that is forecastable, say the survival function which is uninformative for risk (or any practical purpose) . In fact, tail survival functions have errors, for $n$ observations of $o(\frac{1}{n})$, even when tail moments are not tractable, which is why many predict binary outcomes –as with the "superforecasters" masquerade. In fact the paper shows how the more intractable the higher moments of the variable, the more tractable the survival function. Metrics such as the Brier score are well adapted to binary survival functions, though not to the corresponding random variables. In finance and insurance, for instance, one never uses survival functions for risk management or hedging, only expected shortfalls –binary functions are reserved to (illegal) gambling. See paper for ample details.

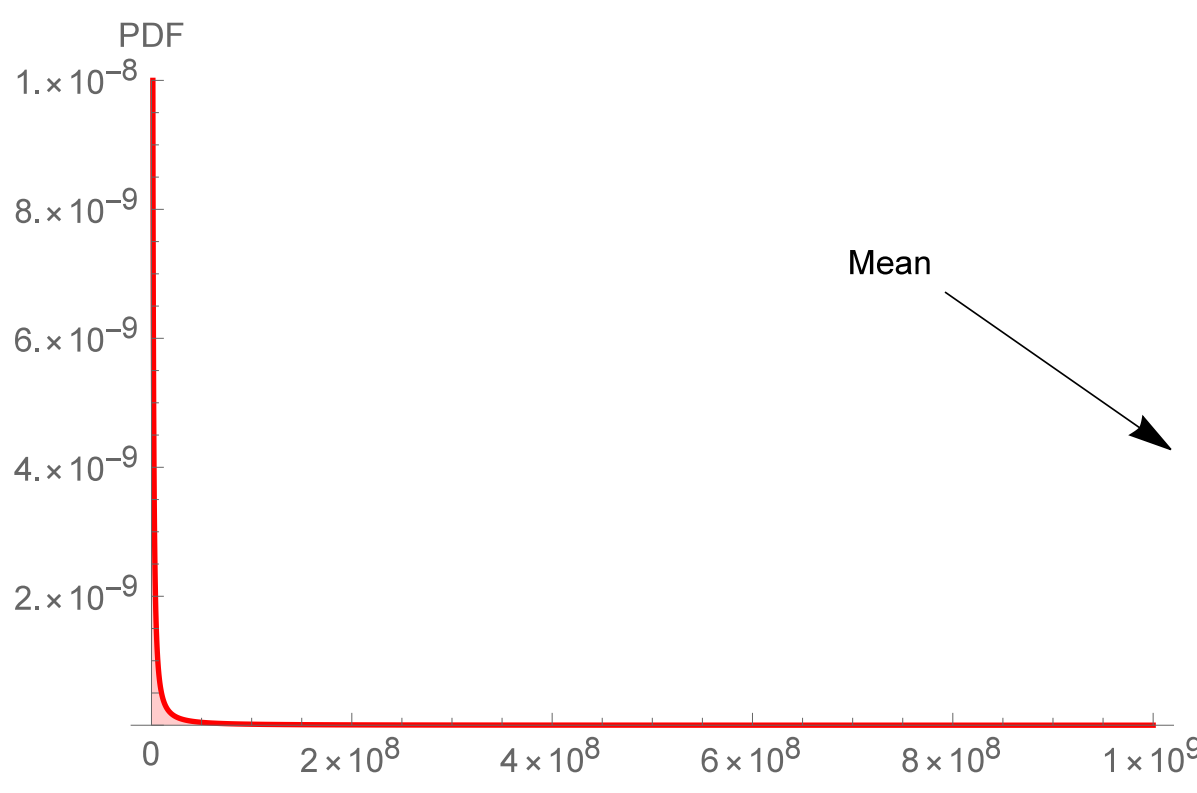

Figure 12.2: *A Pareto distribution with a tail similar to that of the pandemics. Makes no sense to forecast a single point. The "mean" is so far away you almost never observe it. You need to forecast things other than the mean. And most of the density is where there is noise.*

**Never cross a river that's 4 feet deep on average** As we saw, risk management (or policy making) focuses on tail properties not the body of probability distributions. For instance, Holland has a policy to calibrate their dams not on average height of sea levels but on the properties of the maxima –all it takes is one mistake to cause a disaster.

**Science is not about safety** Finally, science is a procedure to update knowledge; it can be wrong provided it produces interesting discussions that lead to more discoveries. But real life is not an experiment. If we used a p-value of .01 or other method of statistical comfort for airplane safety, no pilot would be alive. For matters that have systemic effects and/or entail survival, the asymmetry is even more pronounced.

—

**Forecasts can result in adjustments** If forecasts lead to adjustments and these responses then one can no longer judge these forecasts on their accuracy.

By various mechanisms, including what has been known as Goodhart's law [48], a forecast cab become a target that is gamed by participants –see also the Lucas critique applying the point more generally to dynamical systems. In that sense a forecast can be a warning of the style "if you don't act these are the costs".

More generally any game theoretical framework has an interplay of information and expectation that cancels forecasts.

## 12.4 FURTHER COMMENTARY SPECIFIC TO PANDEMICS

(i)– The reason to act in response to a pandemic is that we care about the consequences to human life and suffering.

(ii)– Actions are based upon anticipation of consequences, and consequences are based upon actions. This is not a loop it is a cause and consequence relationship.

(iii)– Forecasting without contingency on decisions is denying the existence of independent (control) variable, e.g. treatment of a disease, response to a hurricane.

(iv)– Multiplicative / cascading processes are "chaotic"—fundamentally unpredictable (sensitive to initial conditions, to incorrect assumptions, to stochastic details in space and time, and importantly to changes in human behavior or other actions taken)—and give rise to fat tailed distributions.

For reasons that remain to be identified, many scientific studies of COVID fail to consider directly the impact of actions on the suffering and death that would arise and to directly connect action with consequences. This is hard to understand because the entire purpose of pandemic response is to minimize the harmful consequences. When they did consider harm in terms of numbers affected, they failed to directly describe the contingency relationship, providing instead an estimated prediction based upon hypothesized conditions, these include conditions that are inherently difficult to know such as social actions of populations, and are outside the expertise of the practitioners, typically epidemiologists trained in biology (not that social science domains of science are capable of representing those behaviors, but still). Moreover the outcomes by definition of a multiplicative process are highly sensitive to those actions guaranteeing that point predictions are unreliable and present an incorrect view of the role of policy decisions in changing the outcomes.

Thus, every statement that does not describe directly the conditions under which consequences happened, should not be given any credence. For example, there is a difference of about two weeks in similar response efforts in Greece and Spain, resulting in two orders of magnitude higher numbers of cases in Spain than in Greece. Similar differences are observable in multiple countries distinguishing those that have already been able to extinguish the impact of the virus and those that continue to suffer therefrom. The impact of actions is demonstrably large, and this is not surprising given the sensitivity of a multiplicative process to everything from initial conditions to actions taken.

Science in this context has a role to play in risk management, warning of the high consequences that are possible, and recognizing that the very sensitivity to conditions means that the difference between right and wrong actions is the difference between life and death. While uncertainty drives the initial phases of response efforts, over time progressive understanding refines understanding to focus attention on those actions that are the most impactful on the harm. Continued recognition of the remaining uncertainty, such as longer term consequences of the disease and the uncertainty of whether or at what time vaccinations may be available is essential.

For COVID at this time, the key is not about uncertainty but about the manifest harm that occurs without strong action. The difference between countries that acted early and effectively compared to those that did not is sufficient, though many other available data support this conclusion.

Indeed, far too few acted based upon projections of potential harm. This is apparent because of the distinction between countries that acted early and those who did not. For many countries that experienced the outbreak, including Italy, Spain, Germany, and the US, actions were taken based upon definite knowledge of current harm that was already devastating in terms of disease and death. Absent such impacts, it is apparent from policies adopted that no action would have been taken. It is important to recognize that countries do not shut down much of their economic activity without manifest evidence of severe harm as consequences of inaction. It is possible, but not convincingly demonstrated, that the motivation for actions also included anticipations based upon understanding the likelihood of continuing increases, i.e. if it is so bad now and it is growing it will be much worse soon. More clearly, in those countries where action did not occur early in the outbreak, the existing conditions were untenable. Thus, this decision making was not about statistical uncertainty of hypotheticals but about the harm experienced.

And yet, it is surely illogical from even a simple evidence argument to say that if there is an asteroid headed for earth, we should wait for it to arrive to see what the impact will be. We might counter that there were asteroids in the past that had devastating impacts, and besides we can calculate the physics. The logical fallacy runs deeper: "We didn't see this particular asteroid yet" misses the very nature of the power of science to generalize and the power of actions to change the outcome of events. The avoidance of harm asymmetry is apparent even if someone were to argue that perhaps the asteroid is actually made of cotton. Logically true but not actionable information.

Similarly, if we had a hurricane headed for Florida, a statement that "We haven't seen this hurricane yet, perhaps it won't be like other hurricanes." misses the essential role of science. And if science predicts that it will be devastating and people take action boarding up windows, and evacuating, a claim that someone might make of "look it wasn't so devastating", is where we might find a lunatic conspiracy fringe, not a scientific discourse.

Similarly, someone with this view might say "We don't need heart surgery, look how few people die from heart attacks." Or perhaps "Don't worry about jumping off a cliff, previous evidence might not apply to you." The very opposite of evidence in science, and in real life.

Fortunately, while it took some countries too long to act, when they acted the impact was greatly reduced over what it could be. All of the harm experienced was due to delayed action in the face of manifest risk and adequate existing knowledge. People waited till they were hit before they acted even though they could have seen the impact ahead of time. Today, still or again, some areas of the world are not doing what is needed to stop the outbreak.

When any scientist does not anchor their analysis in compassion for those who have been and will be affected, their analysis can be dismissed. This is not an academic game of logic and demonstration of cleverness. For those who have separated themselves too long from it, more experience of the real world may be needed, because policy is ultimately anchored in the world and not in mathematical models. It is the relationship between the models and the real world that is meaningful. If there are those who don't want to contract the disease themselves in order to experience its consequences first hand, they can at minimum talk to those who have been severely sick, and to the spouses and families of those who have been sick or died. They can't talk to the dead though reading their words or encountering the narrative of their lives and contributions lost, including those due to relationships with others, may be helpful. This is essential to inform scientific analysis for policy.

These failures are manifest in the starting and continuing nature of science during the outbreak and its failure to effectively inform policy.

### REMARKS SPECIFIC TO IOANNIDIS ET AL.

**Systemic risks vs individual risks** A fundamental problem, in both [156] and [157], lies in ignoring scaling: systemic risks do not resemble (even qualitatively) individual risks. The macro and the micro-properties of contagious events, given their infective multiplicative nature, don't map directly onto one another.

Ioannidis et al. [156] write: "the average daily risk of dying from coronavirus for a person <65 years old is equivalent to the risk of dying driving a distance of 13 to 101 miles by car per day during that COVID-19 fatality season in 17 of the 24 hotbeds (...) For many hotbeds, the risk of death is in the same level roughly as dying from a car accident during daily commute."

Even if Ioannidis et al.'s computation were tp hold true (it does not) for one individual, conditionally on an excess of $10^3$ of such individuals dying, the probability that the cause of death is COVID-19 and not a car accident converges to 1. When you die of a contagious disease, people around you are at risk of contagion, and they can then infect other people, in a cascading effect. If you die of a car accident, or because you fall from a ladder, the problem of contagion does not hold, or just holds locally (those sitting in the cars crashing), but it cannot spread. It is quite elementary: car accidents are not contagious, while COVID-19 is. You cannot conflate the two objects: one is additive in the aggregate, the other is multiplicative. In [266] [This volume], it has been shown that this is a severe error, leading to macroscopic blunders.[3]

**Remark 12.4: Scaling of probability**

*Under multiplicative effects the risks for a collective do not scale up from the risks of an individual. Trivially, systemic risks can be extreme, where the individual ones are low, or vice-versa.*

**Trade-offs and Ergodicity** One could say: panic saves lives, but at what economic price? Let us put aside ethical arguments, and answer it, ignoring for a moment the value of human life.

The fact is that some classes of (systemic) risks require being killed in the egg, also from an economic point of view. The good news is that there are not so many–but pandemics as we said fall squarely within the category.

The "dismal" theorem [318] mentioned earlier tells us that it is an error to use trade-off analysis under existential risk. There have been many proofs of similar arguments on grounds of ergodicity, well-known by insurance companies since Cramèr: simply, you cannot use naive B-school costs-benefit analyses for Russian roulette, because of the presence of an absorption barrier [263]. But one should not blame Ioannidis et al. [156] for this error in reasoning: it has been shown to be unfortunately prevalent in the decision-science literature [219].

**Remark 12.5: Ruin Problems**

*Traditional cost-benefit analysis fails to apply to situations where statistical averages are unreliable, if not invalid.*

Moreover, it is not correct to assume, more or less implicitly, that a disease brings no or little costs, while mitigation is burdensome. There are indeed severe nonlinearities at play.

First of all, risk is beyond the simple and direct disease-specific mortality rate. In fact, letting the disease run above a certain threshold would compound its effect

3 Note that this is also a typical example of "size fallacy," in which different risky events are compared just on the basis of their probabilities of occurrence, without caring about their different nature [142].

(in an explosive manner), because of the saturation of services, causing for example the displacement of other patients, many in potentially critical conditions; something that we have seen happening in the Region of Lombardy in Italy, in New York City, and elsewhere for several weeks during the spring of 2020 [253]. Furthermore, for survivors the illness itself represents a large economic drain, be it only from lost working hours, not counting the costs of hospitalization. And for every severe infection, there is an unspecified number of morbidities, with unknown (but definitely larger than zero) additional mortality and long term costs for the health system [2, 253], as it has been the case for other diseases like SARS [211].[4]

**Remark 12.6: False Dichotomies**

*One should not treat the economy and the disease as separate independent items, particularly by viewing a naive trade-off between economic costs and pandemic mitigation.*

Moreover, never underestimate consumers' (nonlinear) behavior. When risks are visible (and a pandemic definitely is), people tend to modify their behavior, rationally or not, also switching to alternatives, with nonlinear effects on the businesses concerned [238]. This is the reason why the airline industry in the U.S. manages to have fewer than 1 fatal crash in $25 \times 10^6$ flights (and aims at an even more favorable ratio). One may claim that it is irrational to spend so much of our resources mitigating plane crashes, but airline companies know that, in case of fewer checks and efforts, consumers would then probably switch to other companies, if not directly to other types of transportation.

Take the hospitality industry. Unless there is once again comfort on the part of the public, restaurants and hotels will be unprofitable. The rule of thumb in NYC is that a drop of 15% in revenues is sufficient to make a restaurant shutter permanently; there has been a large drop in restaurant attendance in Sweden where the state did not enforce lockdowns, owing to a high rate of voluntary self-isolation [167].

The United States (and many other countries worldwide) have spent trillions of dollars on sophisticated weaponry in the past decades, to counter *uncertain* threats. It would be a good idea to question these expenditures first, before doubting the spending to stave off certain pandemics.

Likewise, it would be a good idea to question first the excessive burden on Western economies, particularly the U.S., of measures taken to ensure workplace and transportation safety which, we saw, are driven by the legal system and the tort mechanisms.

[4] **Geronticide:** This discussion does not even cover the ethical discussion of trade-offs and their inapplicability in some domains, perhaps the most central discussion. At what price will you kill your parents/grandparents? A million dollars? Ten million? A billion? Furthermore, the fact that older people are more vulnerable to the disease brings considerations of geronticide (senicide): one misses that the silver rule [263] command treating older generations under a moral liability, as one wishes to be treated by the next generation. Letting the disease run through older generations violates the interdicts on geronticide and intergenerational obligations. The fact that your parents did not sacrifice their own parents creates an obligation to not sacrifice them; your children will spare you in turn, under the same rule.

**Remark 12.7: Domain Dependence**

*It is not rational to worry about pandemic costs (extremely fat-tailed exposure), while not also questioning other sizable insurance-style expenditures for transportation and workers safety.*

It is therefore incorrect to claim that it is the authorities' response to the pandemic that caused unemployment in the transportation or hospitality industries. As a matter of fact, the arguments proposed by two of the authors last January 2020, were aimed at lowering the economic effect of the pandemic: prevention is orders of magnitude cheaper than the cure–recall that *sed prior est sanitas quam sit curatio morbi.*

We note that many comments of the type "the pandemic has caused *only* 640K fatalities" (as of July 25, 2020) simply ignore the fact that, in practically every location subject to the pandemic, there has been local or governmental action to mitigate it–we do not consider the counterfactual of "what if" because it is not visible.

**Remark 12.8: Economic Fragility**

*The argument in [263] is that we live in an over-optimized environment, in which a slight drop in sales or a change in consumer preferences may cause wild interlocking industry collapses. This nonlinearity is similar to "large a movie theater with a very small door at the times of fire."*

*It is more cogent to blame the over-optimized economic structure than the general reaction to the disease.*

**Early Mitigation and Economic History** We note here that while the Great Plague took place in the fourteenth century, quarantines were enforced five centuries later as economies understood they could not afford recurrences. Between the Habsburg and the Ottoman Empire, there were *lazarettos* along the border, and every active Mediterranean port enforced quarantines for travelers along the expanded silk road, while pilgrim routes were subjected to similar measures. For instance, in the 1830s, in *the Count of Monte Christo*, a traveler from Paris to Ioanina (where Prof. Ioannidis was previously located), had to spend four days in quarantine to get there, while there was no particular threat of disease. In fact, the novelist was underestimating, for historian records show mandatory nine days for ordinary travelers and fifteen days for merchants according to [235, 247]. Economies adapted to early mitigation throughout the centuries preceding our era. Furthermore, the Ottoman Empire has ready *lazaretos* for additional quarantining along specified stations at the first signs of a pandemic.

Mitigation has another effect: to delay and temporize, while we can understand the properties of the disease. While initial treatments are under high opacity, later treatments allow for gains of collective experience[5] .

[5] Aside from considerations of geronticide, when the costs of the Swedish experiment are finally told, one of the factors will be the early loss of life when later (current) medical practice would have saved them even before a vaccine.

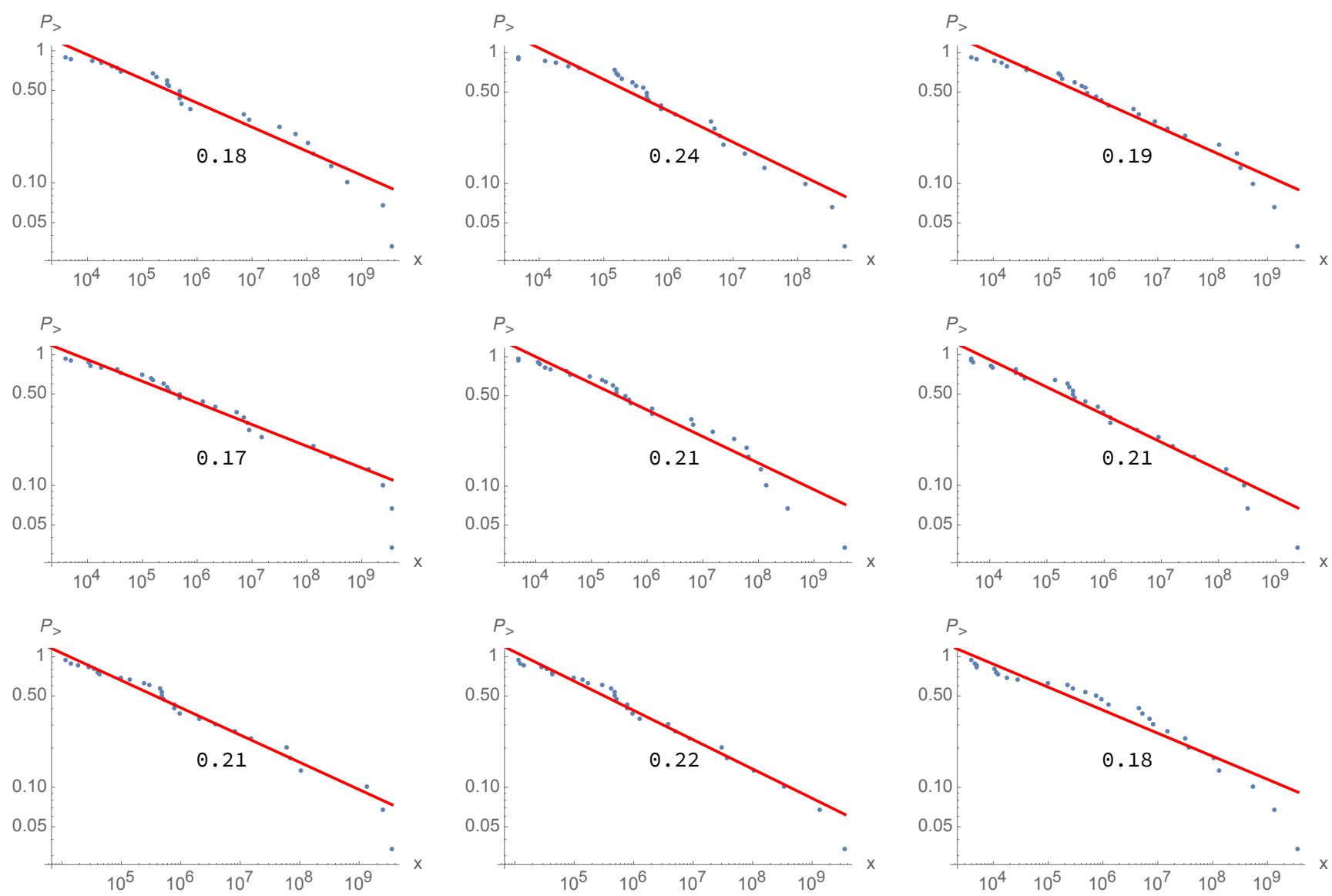

Figure 12.3: *Zipf plots (log-log plots of the empirical survival function* $P_>$*) for nine random selections of 30 out of the 72 pandemics in [52]. The number in the center represents the a naive (OLS) estimate of the tail parameter* $\alpha$*, readable as the absolute slope of the red negative line. The values of* $\alpha$ *appear to be stable notwithstanding the sampling, signalling the robustness of the approach and the inconsistency of the "selection bias" critique. The values are also in line with the more rigorous EVT-based findings of [52].*

**Selection Bias and Class of Events** In [157], the authors erroneously maintain that choosing tail events as done by [52] is "selection bias". Actually, the standard technique there used is the exact opposite of selection bias: in EVT, one purposely focuses on extremes to derive properties that influence the outcomes, especially from a risk management point of view. One could more reasonably argue that the data in [52] do not contain all the extremes, but, by jackknifing and bootstrapping the data, the authors actually show the robustness of their results to variations and holes in historical observations: the tail index $\alpha$ is consistently lower than 1. In Figure 12.3 a simple illustration is given, showing that one can be quite radical in dealing with the uncertainty in pandemic fatalities, and still find out that the findings of [52] hold true.

When the authors in [157] state that "Tens of millions of outbreaks with a couple deaths must have happened throughout time," to support their selection bias claim against [52], they seem to overlook the fact that the analysis deals with pandemics and not with a single sternutation. The class of events under considerations in [52] is precisely defined as "pandemics with fatalities in excess of 1$K$," and their dataset likely contains most (if not all) of them. Worrying about many missing observations in the left tail of the distribution of pandemic deaths is thus misplaced.

**Conditional information** One may be entitled to ask: as we get to know the disease, do the tails get thinner? Early in the game one must rely on conditional information, but as our knowledge of the disease progresses, shouldn't we be allowed to ignore tails?

Alas, no. The scale of the pandemic might change, but the tail properties will remain invariant. Furthermore, there is an additional paradox. If one does not take the pandemic seriously, it will likely run wild (particularly under the connectivity of the modern world, several orders of magnitude higher than in the past [3]). And diseases mutate, increasing or decreasing in both lethality and contagiousness. The argument would therefore resemble the following: "we have not observed many plane crashes lately, let's relax our safety measures".

Finally, we conclude this section with an encouraging point: fat tails do not make the world more complicated and do not cause frivolous worries, to the contrary. Understanding them actually reduces costs of reaction because they tell us what to target–and when to do so. Because network models tend to follow certain patterns to generate large tail events, wisdom in action is to kill the exponential growth in the egg via three central measures 1) reducing super-spreader events; 2) monitoring and reducing mobility for those coming from far-away places (via quarantines); 3) looking for cheap measures with large payoffs in terms of the reduction of the multiplicative effects (e.g. face masks[6]). Anything that "demultiplies the multiplicative" helps [264]. The trader lore taught by generations of operators says "if you must panic, panic early." The Ottoman Empire integrated Byzantine knowledge accumulated since at least the Plague of Justinian; it is sad to see ancient cultures more risk-conscious, better learners from history, and economically more effective than modern governments. Indeed, had it not been for such a collective ancestral risk-awareness and an innate understanding of asymmetry, we doubt that many of us would be here today.

[6] Most of the trillions spent could have been saved if authorities understood the double nonlinearities in face masks: 1) the compounding effect of both parties having protection, 2) the nonlinearity of the dose response with disproportional drop in the probability of infection from a reduction in viral load **[?]**

# 13 THE PROBABILITY CONFLATION

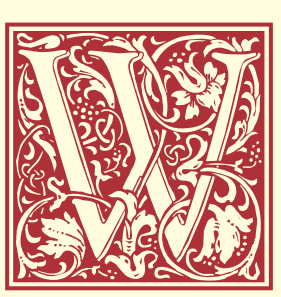
E RESPOND to criticism by Tetlock et al. (2022) showing 1) how expert judgment fails to reflect tail risk, 2) the lack of compatibility between forecasting tournaments and tail risk assessment methods (such as extreme value theory). More importantly, we communicate a new result showing a greater gap between the properties of tail expectation and those of the corresponding probability.

This constitutes a nice simplification of the previous three chapters on verbalism in probability.

## 13.1 SUMMARY OF THE CRITICISM

## 13.2 THE FAT TAILS PROBLEM

Tetlock et al. (2022) [299], in their criticism of claims by a paper titled "On single point forecasts for fat-tailed variables" in this journal (Taleb et al., 2022 [289]) insist that discriminating between a binary probability and a continuous distribution is a false dichotomy, that binary probabilities derived from expert forecasting tournaments can provide information on tail risk, in addition to some claims about a collaboration with the first author and a "challenge".

We apologize for not answering most of their points as these are already amply covered in two papers in this very journal, which includes the one they are criticizing, [289] and the more formal [287]. Alas "probability" is a mathematical concept that does not easily accommodate verbal discussions and requires a formal treatment, which necessitates precise definitions.

Research chapter, written with the collaboration of Ronald Richman, Marcos Carreira, and James Sharpe, thanks to the online discussion group on Uncertainty.

At the gist of what we referred to as "the so-called masquerade problem" is the following conflation, we simplify from [287] using a continuous distribution for ease of exposition:

Let $K \in \mathbb{R}^+$ be a threshold, $f(.)$ a density function for a random variable $X \in \mathbb{R}^+$ , $P_K = \mathbb{P}(X > K) \in [0,1]$ the probability of exceeding it, and $g(x)$: $\mathbb{R}^+ \to \mathbb{R}$, an impact function. Let $G_K$ be a partial expectation of $g(.)$ for the function of realizations of $X$ above $K$:

$$G_K = \int_K^\infty g(x) f(x)\, \mathrm{d}x,$$

and for clarity lets write the survival function (that is, the complementary cumulative distribution function, CDF) at $K$:

$$P_K = \int_K^\infty f(x)\, dx$$

The error comes from conflating the properties of $G_K$ which those of $P_K$, often associating $P_K$ with some constant representing the presumed impact associated with the threshold $K$.

The intuition of the difference can be shown as follows: assuming $g(x) = x$, for $X$ a random variable with finite first moment, we have, focusing on the positive domain, generalizing the Tail Probability Expectation Formula,

$$G_K = \underbrace{K\, P_K}_{\text{Prob times impact at threshold}} + \underbrace{\int_K^\infty P_x\, dx}_{\text{additional term}}, \qquad (13.1)$$

with a second term that can dominate the first, particularly under heavy tailed distributions.

As explained in the two referenced papers, $P_K$ *as a random variable*, being bounded between 0 and 1, necessarily has thin tails, with finite mean, variance, and all moments. By the probability integral transform, its unconditional probability distribution is the Uniform $\mathcal{U}(0,1)$ –and its conditional is usually treated, particularly in the Bayesian literature, as a Beta (a special case of which is the Uniform) – the Beta distribution can accommodate practically all shapes of double bounded unimodal distributions. A sum of such bets becomes rapidly Gaussian.

**Remark 13.1: Moments**

*$P_K$ corresponds to the zeroeth moment and is always thin-tailed. $G_K$ maps to higher moments, and up to the infinite moment (i.e. the extremum) dealt with in extreme value theory.*[a]

[a] Consider the expectation of the $p^{th}$ moment $\mathbb{E}(x^p)$. $P_K$ corresponds to $p = 0$ and the expected maximum to $\lim_{p\to\infty} \mathbb{E}(x^p)$. In general, risk management concerns extrema, another point of divergence with Tetlock et al.

As discussed extensively in Taleb (1997) reflecting the author's experiences as a derivatives market-maker [271], one fails to "hedge" the other in practice (and, of course, in theory). For instance, a rise in skewness of the distribution will tend to increase one side ($G_K$) while decreasing the other ($P_K$): simply, the number of realizations above $K$ drops, but their impact becomes larger.[2]

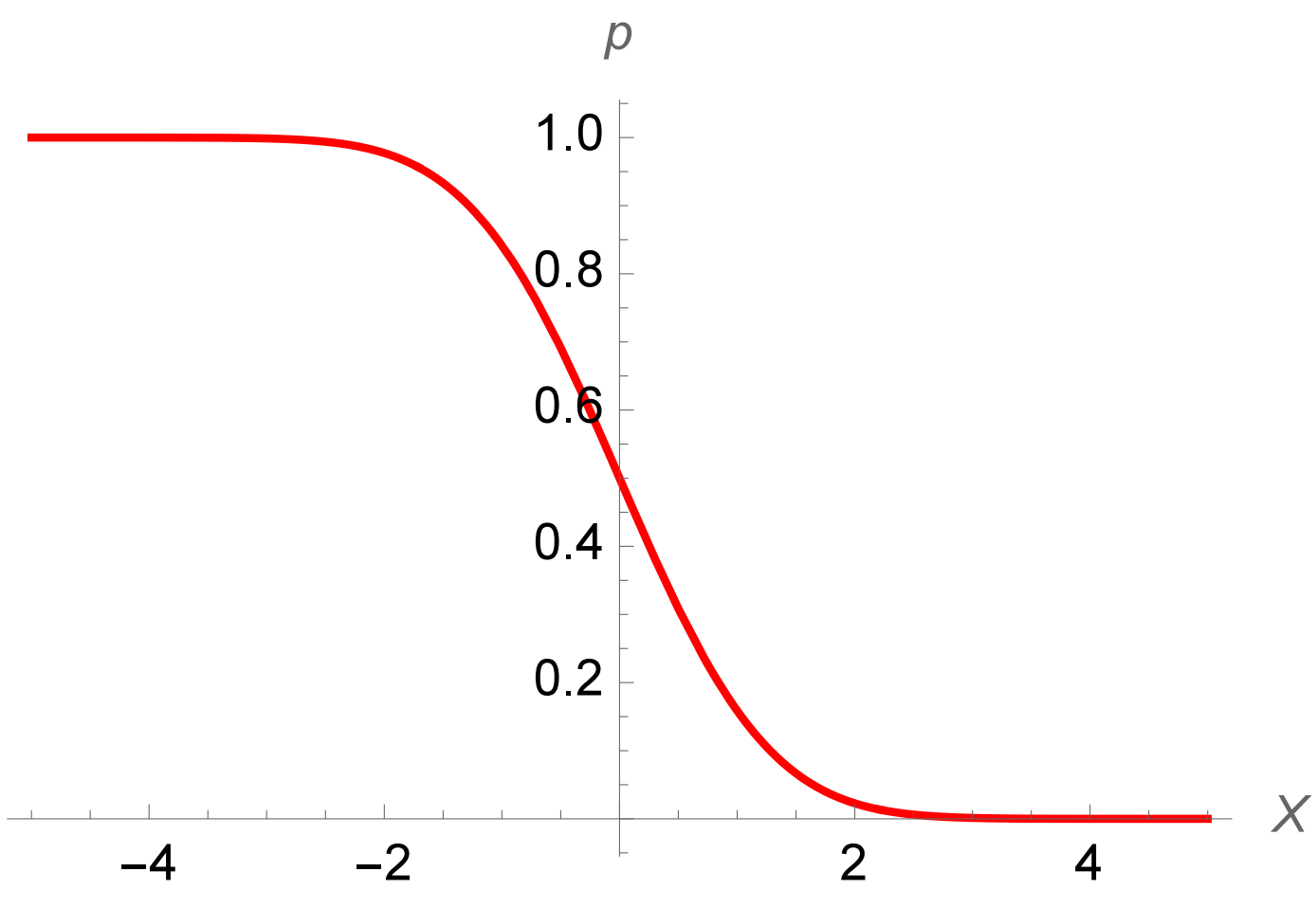

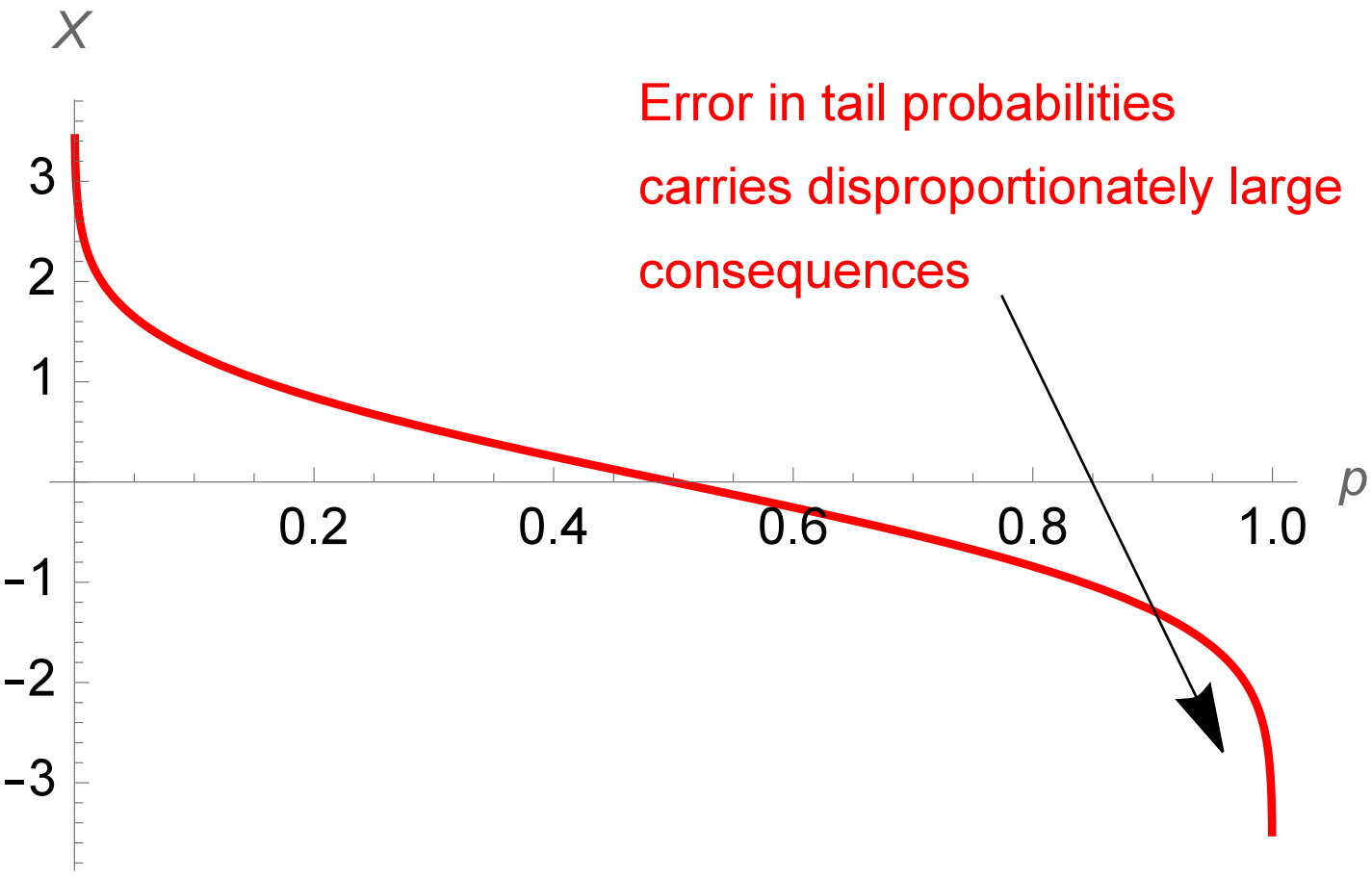

Figure 13.1: *The inverse survival function (complementary quantile function) corresponding to a complementary CDF ($P_K$) is unbounded while the complementary CDF is bounded; it is extremely concave for tail probabilities, and compounds the estimation errors on p. Simply the transformation reverses the signs of the second derivative and compounds it. We show how errors on $P_K$ translate in larger and larger values for K, possibly infinite.*

[2] Tetlock et al states that Taleb and Tetlock (2013) claims that the two methods are "complementary". Our understanding of the latter paper is that says the exact opposite: there is no such complementarity

And if one does not hedge the other, then being "good at predicting $P_K$" provides *no* information on $G_K$.

**Remark 13.2: Probability Classes**

*If $P_K$ and $G_K$ are not in the same probability class, that is, while $P_K$ is always thin tailed, if $G_K$ is not thin-tailed, then one cannot be a practical proxy for the other.*

The field of extreme value theory was designed to deal with the issue. Basically, "probability" is not a tangible object like a tomato; it is, mathematically, a kernel inside an integral (or a summation), inseparable from other integrands, and one should avoid drowning it with verbalism. This point applies no matter the probability interpretation (Bayesian, frequentist, propensional, or subjectivist). Furthermore, science is not about precise measurements of exceedance probability, but understanding properties in a comprehensive and useful ways. As explained in the paper criticized by Tetlock et al (2022), one does not handle pandemics via forecasting tournaments and reliance on champions for single point forecasts, but by getting full distributional properties, particularly the shape in the tails. Decisions must be informed by the shape of the total distribution, and some understanding of the dynamics involved in generating such a distribution (multiplicative effects cause thick tails while additive ones tend to be benign stochastic outcomes).

Just as warning is not predicting, understanding distribution classes and tail properties is not forecasting. Furthermore, the language of "false positive", while useful in medicine and similar applications based on signal, in not useful in risk and insurance based on distributional considerations.

**Remark 13.3: Gambling**

*It is worth noting that binary options on financial instruments (that is, the trading of $P_K$ or $1 - P_K$) proved of little economic value and are not considered an investment in the U.S. and the European Union; they are banned by most corresponding regulators, as they are considered gambling devices. The European Securities and Markets Authority (ESMA) have disallowed retail dealing with binary options. These binaries were also traded at Lloyds, until banned by U.K. legislation with the 1909 Marine Insurance Act.*

**Note on "dichotomies":** Tetlock et al (2022) seem to mix the "dichotomy" between binary and full payoffs with another distinction, that probability estimates help to flag tail risk. Our representation can accommodate both with the function $g(.)$ which as mentioned earlier can reflect the infinite moment, i.e. the extremum.

fat tailed variables, which is what this discussion about "masquerade" is about. This is the reason why the first author (Taleb) refused to be involved in the IARPA forecasting exercise.

## 13.3 A NEW RESULT: VALUE AT RISK FAILS UNDER ERROR PROPAGATION

At the Global Uncertainty Reading Group discussion around Tetlock et al (2022), on Dec 1, 2022, a new useful result emerged, which we find worth communicating[3].

**Remark 13.4: Events are not defined**

*A well known problem with heavy tails is that, at the core, in that class of distributions, "events" are not defined verbally: a "war" can have 200 or a million casualties, so it does not have a quantitative meaning. But setting a precise threshold no longer maps to a precise probability under an error rate, and, vice versa.*

As illustrated in Fig. 13.1, The error in the evaluation of the probability $P_K$ can translate into an explosive value for the corresponding $K$, and the more fat tailed the distribution, the more explosive such corresponding value.

There is no space for a general proof, so we shall provide one for any distribution that ends (for large values) with Paretian tails, which is the standard case. Let us assume the probability $P_K$ follows a Beta Distribution $\beta(a, b)$ (both parameters $> 0$ and as we mentioned this fits the unconditional uniform with $a = b = 1$).

$$f_{P_K}(p) = \frac{p^{a-1}(1-p)^{b-1}}{B(a,b)}, \tag{13.2}$$

$0 < p < 1$, where $(B.,.)$ is the standard Beta function.

The mean and variance will be $M_{P_K} = \frac{a}{a+b}$, $V_{P_K} = \frac{ab}{(a+b)^2(a+b+1)}$.

Now assume the underlying distribution for $X$ where $X > K$, lies in the strong Pareto basin, meaning $P(X > K) = L^{\alpha} K^{-\alpha}$, where $\alpha$ is the tail index an $L$ a scaling constant –this is general for large values of $X$ under all fat-tailed distributions.

The inverse complementary CDF (quantile function) can be expressed as: $K = L\left(P_K\right)^{-1/\alpha}$ if $0 \leq P_K \leq 1$.

If probability taken as a random variable $P_K$, that is, $1 - CDF$ (the cumulative distribution function), follows a Beta distribution $\beta(a, b)$, then $K$, the inverse complementary CDF has for density:

$$f_K(k) = \frac{\alpha\, K^{-a\alpha-1} L^{a\alpha}\left(1 - K^{-\alpha} L^{\alpha}\right)^{b-1}}{B(a,b)}, \tag{13.3}$$

with mean

$$M_K = \frac{L\,\Gamma\left(a - \frac{1}{\alpha}\right)\Gamma(a+b)}{\Gamma(a)\Gamma\left(a+b-\frac{1}{\alpha}\right)},$$

[3] This result can be useful in financial risk management, particulaly the mapping between "VaR" (Value at Risk, maps to $K$ for a set probability $P_K$ of losses above that threshold) and expected shortfall, "CVaR" (maps to $G_K$, that is, includes the impact of losses).

and variance

$$V_K = \frac{L^2\,\Gamma(a+b)\left(\frac{\Gamma(a)\Gamma\left(a-\frac{2}{\alpha}\right)}{\Gamma\left(a+b-\frac{2}{\alpha}\right)} - \frac{\Gamma\left(a-\frac{1}{\alpha}\right)^2\Gamma(a+b)}{\Gamma\left(a+b-\frac{1}{\alpha}\right)^2}\right)}{\Gamma(a)^2}.$$

The proof is done via the standard Jacobian Method for the transformation of probability distributions.

As we can see the first moment exists only if $\alpha > \frac{1}{a}$ and $\alpha > \frac{1}{a+b}$; the second moment exists only if $\alpha > \frac{2}{a}$ and $\alpha > \frac{2}{a+b}$; more generally the $n^{th}$ moment exists only if $\alpha > \frac{n}{a}$ and $\alpha > \frac{n}{a+b}$.

**Remark 13.5: Error Propagation**

*While the error on the probability can be small and controlled, the error on the corresponding quantity under consideration can be infinite.*

We note that the tail index $\alpha$ for pandemics (as addressed at length in the paper criticized by Tetlock et al. (2022)) is well below 1. The same applies to wars, which means that when it comes to conflicts, forecasts for tail events are not compatible with probability theory.

## 13.4 A RATHER UNSCIENTIFIC CHALLENGE

> "We challenge Taleb et al. (2022) to be equally transparent about the performance of their tail-risk hedging strategies -a controversial topic (Brown, 2020)."

We are surprised to see such a remark coming from professional evidence-based researchers: requesting the single track record of a tail hedging program as a backup for a claim about the mathematical inadequacy of using a binary forecast for, say, Covid 19 or similar events under fat tailed distributions. Said tail-hedging strategy consists in capturing the *difference* between idiosyncratically selected option prices in the market and subsequent market jumps. (Incidentally Professor Tetlock has appeared to conflate tail events –which take place in the tails of any distribution – and the fat-tailedness attributes of statistical distributions). And, on top, such request is made to the author of an entire book, *Fooled by Randomness* about the futility of such claims. To repeat the famous disparagement by the economist Jagdish Bhagwati of the claim by the speculator George Soros that he "falsified the random walk" [23], we find it highly unscientific to use a single track record to make any general claim – this bizarre demand on the part of professional researchers is no different from anecdotal claims ($n = 1$) used by medical charlatans. In addition, trading records are not like points in soccer games, particularly when they can hide tail risks.[4]

[4] Since Tetlock et al (2022) is uncritically citing a web *opinion* article by Aaron Brown, we would like to debunk the claim in it that the performance record is *not* available: not being a retail product, it has been continuously available to what the Security and Exchange Commission (S.E.C.) defines as *qualified* investors, not Twitter activists, and Mr. Brown (who by his own disclosure had a severe conflict of interest) violated, willingly or unwillingly journalistic standards. For it turned out that

So we prefer the more robust challenge which, under these circumstances, becomes fair. We believe that academia is about search for knowledge and understanding the world, not a commercial enterprise. The same with societal risk management, which is about the public good and does not issue precise point forecasts (recall that Taleb et al.(2022), as its title indicates, is against single point prediction in some domains). So the burden is on forecasters to forecast. And we fail to get how a practical project with remarkable forecasting skills could work for government and not the private sector. The superforecasting project is just about such betting. So, as much as we would have preferred to voice the unspoken question, here we are obligated to put the old adage as "if they claim to be so good at forecasting, and their forecasts are actionable and related to reality, why aren't they so rich?" –in other words, why do they depend on taxpayer funds and, possibly, tax deductible (that is, charitable) contributions to finance such forecasts?

## 13.5 conclusion: some more evidence required

Finally, can Tetlock et al. prove that better estimates of $P_K$ can provide *real* benefits to decision makers?

In addition to the problems in the financial domain mentioned above, we completely fail to see the link in insurance –and in event risk in general. In our experience as risk and insurance practitioners, decision makers are usually not well equipped to deal with probabilistic information (compounding the difficulty in translating probabilistic information into practical effects)[5]. For instance, in a military context, if we refine the estimate of a South China sea conflict from e.g. 15% to 17%, can Tetlock et al prove that this makes a difference in any practical situation?

Also, one is allowed to wonder why the superforecasting project is not applied to sports and election forecasts where 1) $P_K$ applies, 2) compatible with probability theory, 3) provides repeatable tests with overly abundant data and, centrally, 4) is "bankable" (that is, translatable into dollars and cents)?

We conclude with the following recommendation: in future work, it would be helpful if Tetlock et al. provided more rigorous backing of their claims about the link between $P_K$ and $G_K$. For, as it stands, we see neither theoretical nor practical benefits to that "superforecasting" enterprise.

## 13.6 proofs

Let $X$ be a non-negative random variable and $p$ its probability density function:

Brown never did the fact checking, and never asked to see the *audited* returns. We also note that only dimensionless returns are to be compared.

[5] There is the other problem that payoffs are in in expectation space not in frequency space. For instance hedge funds with the best track records turned out to be the most vulnerable to tail events in 2008, see [286].

$$\int_K^{\infty} x\, p(x)\, dx = \int_K^{\infty} \mathbb{P}(X > x)\, dx + K\, \mathbb{P}(X > K) \tag{13.4}$$

*Proof.* We write $x = \int_K^{\infty} \mathbb{1}_{x>u} du + K$.

$$\int_K^{\infty} x\, p(x)\, dx = \int_K^{\infty} \left( \int_K^{\infty} \mathbb{1}_{x>u} du + K \right) p(x)\, dx \tag{13.5}$$

$$= \int_K^{\infty} \int_K^{\infty} \mathbb{1}_{x>u} du\, p(x)\, dx + K\, \mathbb{P}(X > K) \tag{13.6}$$

$$\overset{\text{Fubini}}{=} \int_K^{\infty} \underbrace{\int_K^{\infty} \mathbb{1}_{x>u}\, p(x)\, dx}_{\mathbb{P}(X>u)}\, du + K\, \mathbb{P}(X > K) \tag{13.7}$$

□

# 14 ELECTION PREDICTIONS AS MARTINGALES: AN ARBITRAGE APPROACH‡

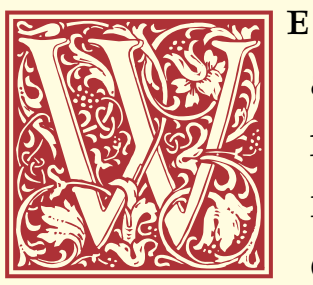E EXAMINE the effect of uncertainty on binary outcomes, with application to elections. A standard result in quantitative finance is that when the volatility of the underlying security increases, arbitrage pressures push the corresponding binary option to trade closer to 50%, and become less variable over the remaining time to expiration. Counterintuitively, the higher the uncertainty of the underlying security, the lower the volatility of the binary option. This effect should hold in all domains where a binary price is produced – yet we observe severe violations of these principles in many areas where binary forecasts are made, in particular those concerning the U.S. presidential election of 2016. We observe stark errors among political scientists and forecasters, for instance with 1) assessors giving the candidate D. Trump between 0.1% and 3% chances of success , 2) jumps in the revisions of forecasts from 48% to 15%, both made while invoking uncertainty.

Conventionally, the quality of election forecasting has been assessed statically by De Finetti's method, which consists in minimizing the Brier score, a metric of divergence from the final outcome (the standard for tracking the accuracy of probability assessors across domains, from elections to weather). No intertemporal evaluations of changes in estimates appear to have been imposed outside the

---

‡ Research chapter.

The author thanks Dhruv Madeka and Raphael Douady for detailed and extensive discussions of the paper as well as thorough auditing of the proofs across the various iterations, and, worse, the numerous changes of notation. Peter Carr helped with discussions on the properties of a bounded martingale and the transformations. I thank David Shimko,Andrew Lesniewski, and Andrew Papanicolaou for comments. I thankArthur Breitman for guidance with the literature for numerical approximations of the various logistic-normal integrals. I thank participants of the Tandon School of Engineering and Bloomberg Quantitative Finance Seminars. I also thank Bruno Dupire, MikeLawler, the Editors-In-Chief of Quantitative Finance, and various friendly people on social media. DhruvMadeka, then at Bloomberg, while working on a similar problem, independently came up with the same relationships between the volatility of an estimate and its bounds and the same arbitrage bounds. All errors are mine.

quantitative finance practice and literature. Yet De Finetti's own principle is that a probability should be treated like a two-way "choice" price, which is thus violated by conventional practice.

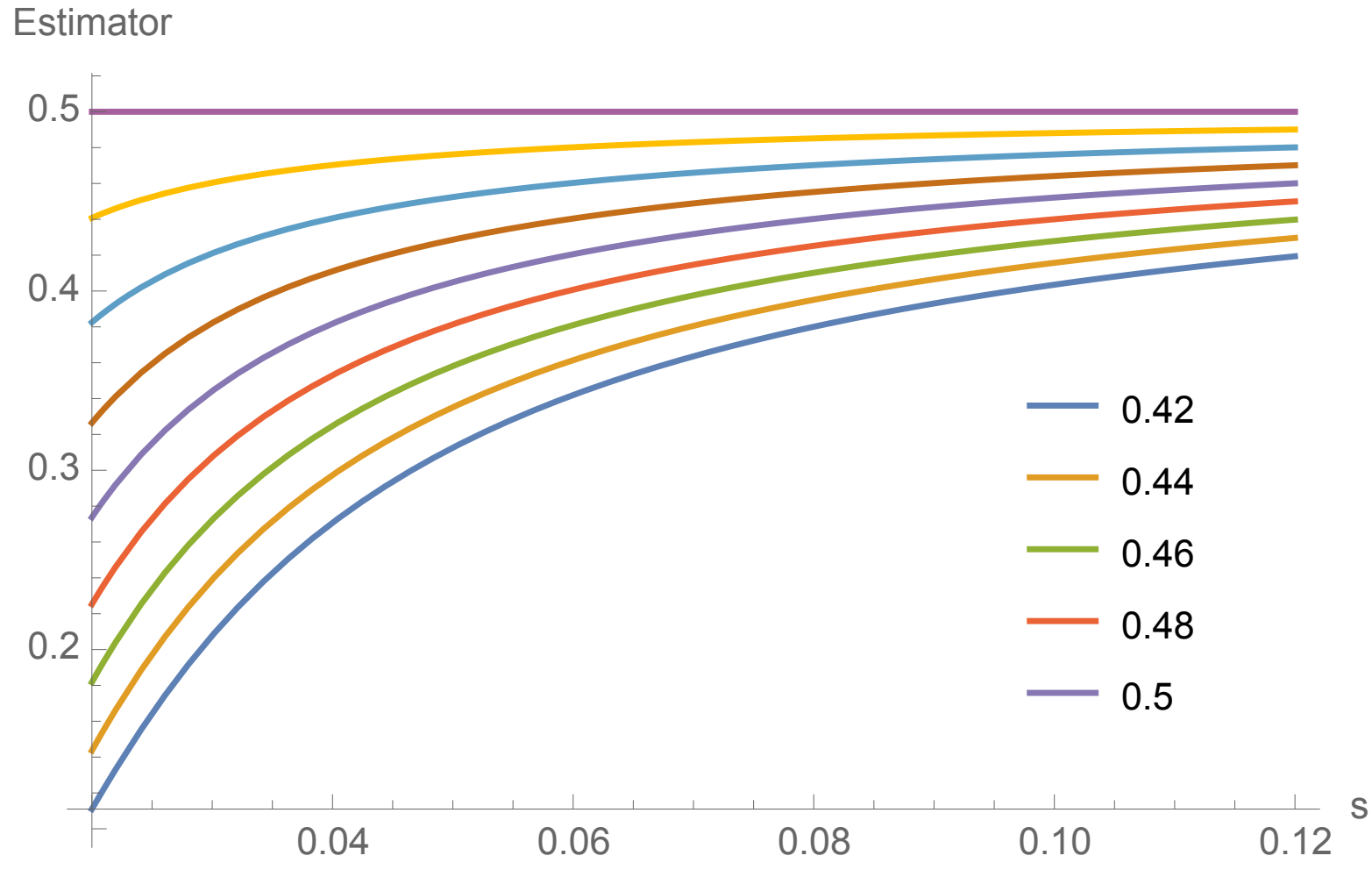

Figure 14.1: *Election arbitrage "estimation" (i.e., valuation) at different expected proportional votes $Y \in [0, 1]$, with s the expected volatility of Y between present and election results. We can see that under higher uncertainty, the estimation of the result gets closer to 0.5, and becomes insensitive to estimated electoral margin.*

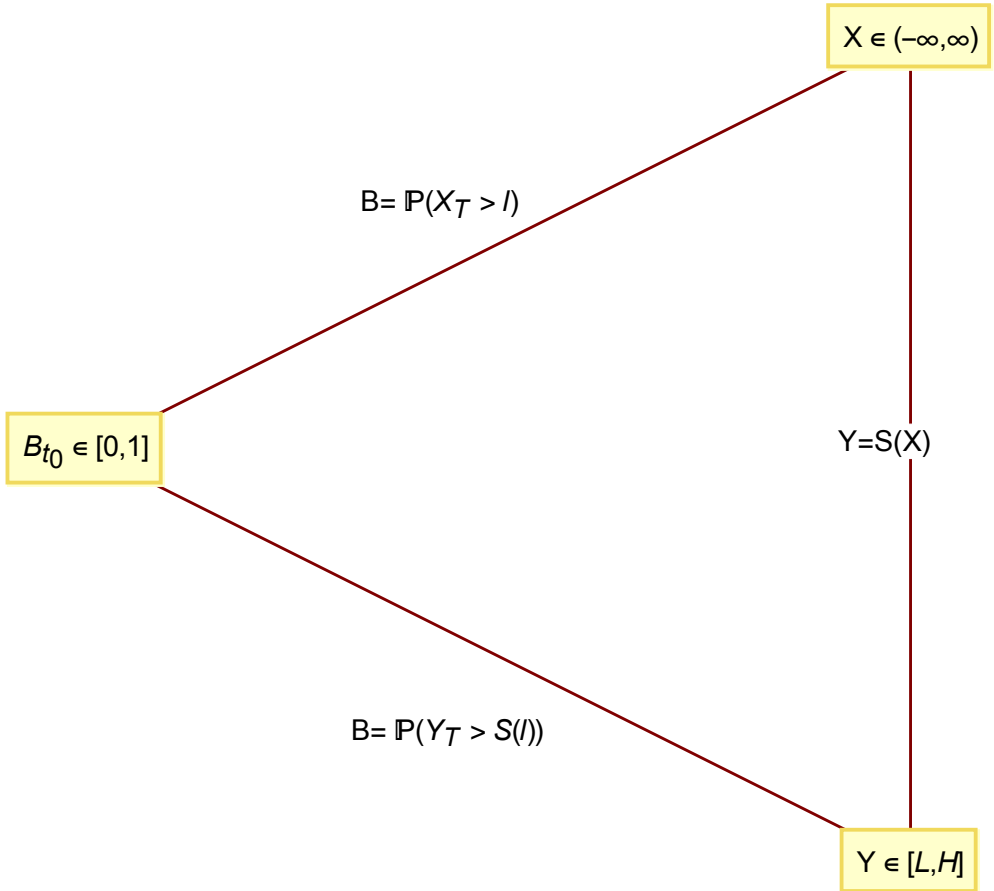

Figure 14.2: *X is an open non observable random variable (a shadow variable of sorts) on ℝ, Y, its mapping into "votes" or "electoral votes" via a sigmoidal function S(.), which maps one-to-one, and the binary as the expected value of either using the proper corresponding distribution.*

In this chapter we take a dynamic, continuous-time approach based on the principles of quantitative finance and argue that a probabilistic estimate of an election outcome by a given "assessor" needs be treated like a tradable price, that is, as a binary option value subjected to arbitrage boundaries (particularly since binary options are actually used in betting markets). Future revised estimates need to be compatible with martingale pricing, otherwise intertemporal arbitrage is created, by "buying" and "selling" from the assessor.

A mathematical complication arises as we move to continuous time and apply the standard martingale approach: namely that as a probability forecast, the underlying security lives in $[0,1]$. Our approach is to create a dual (or "shadow") martingale process $Y$, in an interval $[L,H]$ from an arithmetic Brownian motion, $X$ in $(-\infty,\infty)$ and price elections accordingly. The dual process $Y$ can for example represent the numerical votes needed for success. A complication is that, because of the transformation from $X$ to $Y$, if $Y$ is a martingale, $X$ cannot be a martingale (and vice-versa).

The process for $Y$ allows us to build an arbitrage relationship between the volatility of a probability estimate and that of the underlying variable, e.g. the vote number. Thus we are able to show that when there is a high uncertainty about the final outcome, 1) indeed, the arbitrage value of the forecast (as a binary option) gets closer to 50% and 2) the estimate should not undergo large changes even if polls or other bases show significant variations.[3]

The pricing links are between 1) the binary option value (that is, the forecast probability), 2) the estimation of $Y$ and 3) the volatility of the estimation of $Y$ over the remaining time to expiration (see Figures 14.1 and 14.2 ).

### 14.0.1 Main results

For convenience, we start with our notation.

#### Notation

[3] A central property of our model is that it prevents $B(.)$ from varying more than the estimated $Y$: in a two candidate contest, it will be capped (floored) at $Y$ if lower (higher) than .5. In practice, we can observe probabilities of winning of 98% vs. 02% from a narrower spread of estimated votes of 47% vs. 53%; our approach prevents, under high uncertainty, the probabilities from diverging away from the estimated votes. But it remains conservative enough to not give a higher proportion.

| | |
|---|---|
| $Y_0$ | the observed estimated proportion of votes expressed in $[0,1]$ at time $t_0$. These can be either popular or electoral votes, so long as one treats them with consistency. |
| $T$ | period when the irrevocable final election outcome $Y_T$ is revealed, or expiration. |
| $t_0$ | present evaluation period, hence $T - t_0$ is the time until the final election, expressed in years. |
| $s$ | annualized volatility of $Y$, or uncertainty attending outcomes for $Y$ in the remaining time until expiration. We assume $s$ is constant without any loss of generality –but it could be time dependent. |
| $B(.)$ | "forecast probability", or estimated continuous-time arbitrage evaluation of the election results, establishing arbitrage bounds between $B(.)$, $Y_0$ and the volatility $s$. |

**Main results**

$$B(Y_0, \sigma, t_0, T) = \frac{1}{2}\text{erfc}\left(\frac{l - \text{erf}^{-1}(2Y_0 - 1)e^{\sigma^2(T-t_0)}}{\sqrt{e^{2\sigma^2(T-t_0)} - 1}}\right), \tag{14.1}$$

where

$$\sigma \approx \frac{\sqrt{\log\left(2\pi s^2 e^{2\text{erf}^{-1}(2Y_0-1)^2} + 1\right)}}{\sqrt{2}\sqrt{T - t_0}}, \tag{14.2}$$

$l$ is the threshold needed (defaults to .5), and erfc(.) is the standard complementary error function, 1-erf(.), with $\text{erf}(z) = \frac{2}{\sqrt{\pi}}\int_0^z e^{-t^2}dt$. □

We find it appropriate here to answer the usual comment by statisticians and people operating outside of mathematical finance: "why not simply use a Beta-style distribution for $Y$?". The answer is that 1) the main purpose of the paper is establishing (arbitrage-free) time consistency in binary forecasts, and 2) we are not aware of a continuous time stochastic process that accommodates a beta distribution or a similarly bounded conventional one.

### 14.0.2 Organization

The remaining parts of the paper are organized as follows. First, we show the process for $Y$ and the needed transformations from a specific Brownian motion. Second, we derive the arbitrage relationship used to obtain equation (14.1). Finally, we discuss De Finetti's approach and show how a martingale valuation relates to minimizing the conventional standard in the forecasting industry, namely the Brier Score.

**A comment on absence of closed form solutions for $\sigma$** We note that for $Y$ we lack a closed form solution for the integral reflecting the total variation:

$\int_{t_0}^{T} \frac{\sigma}{\sqrt{\pi}} e^{-\text{erf}^{-1}(2y_s-1)^2} ds$, though the corresponding one for $X$ is computable. Accordingly, we have relied on propagation of uncertainty methods to obtain a closed form solution for the probability density of $Y$, though not explicitly its moments as the logistic normal integral does not lend itself to simple expansions [228].

**Time slice distributions for $X$ and $Y$** The time slice distribution is the probability density function of $Y$ from time $t$, that is the one-period representation, starting at $t$ with $y_0 = \frac{1}{2} + \frac{1}{2}\text{erf}(x_0)$. Inversely, for $X$ given $y_0$, the corresponding $x_0$, $X$ may be found to be normally distributed for the period $T - t_0$ with

$$\mathbb{E}(X,T) = X_0 e^{\sigma^2(T-t_0)},$$

$$\mathbb{V}(X,T) = \frac{e^{2\sigma^2(T-t_0)} - 1}{2}$$

and a kurtosis of 3. By probability transformation we obtain $\varphi$, the corresponding distribution of $Y$ with initial value $y_0$ is given by

$$\varphi(y; y_0, T) = \frac{1}{\sqrt{e^{2\sigma^2(t-t_0)} - 1}} \exp\left\{ \text{erf}^{-1}(2y-1)^2 - \frac{1}{2}\left(\coth\left(\sigma^2 t\right) - 1\right)\left(\text{erf}^{-1}(2y-1) - \text{erf}^{-1}(2y_0-1)e^{\sigma^2(t-t_0)}\right)^2 \right\} \tag{14.3}$$

and we have $\mathbb{E}(Y_t) = Y_0$.

As to the variance, $\mathbb{E}(Y^2)$, as mentioned above, does not lend itself to a closed-form solution derived from $\varphi(.)$, nor from the stochastic integral; but it can be easily estimated from the closed form distribution of $X$ using methods of propagation of uncertainty for the first two moments (the delta method).

Since the variance of a function $f$ of a finite moment random variable $X$ can be approximated as $V(f(X)) = f'(\mathbb{E}(X))^2 V(X)$:

$$\left.\frac{\partial S^{-1}(y)}{\partial y}\right|_{y=Y_0} s^2 \approx \frac{e^{2\sigma^2(T-t_0)} - 1}{2}$$

$$s \approx \sqrt{\frac{e^{-2\text{erf}^{-1}(2Y_0-1)^2}\left(e^{2\sigma^2(T-t_0)} - 1\right)}{2\pi}}. \tag{14.4}$$

Likewise for calculations in the opposite direction, we find

$$\sigma \approx \frac{\sqrt{\log\left(2\pi s^2 e^{2\text{erf}^{-1}(2Y_0-1)^2} + 1\right)}}{\sqrt{2}\sqrt{T-t_0}},$$

which is (14.2) in the presentation of the main result.

Note that expansions including higher moments do not bring a material increase in precision – although $s$ is highly nonlinear around the center, the range of values for the volatility of the total or, say, the electoral college is too low to affect higher order terms in a significant way, in addition to the boundedness of the sigmoid-style transformations.

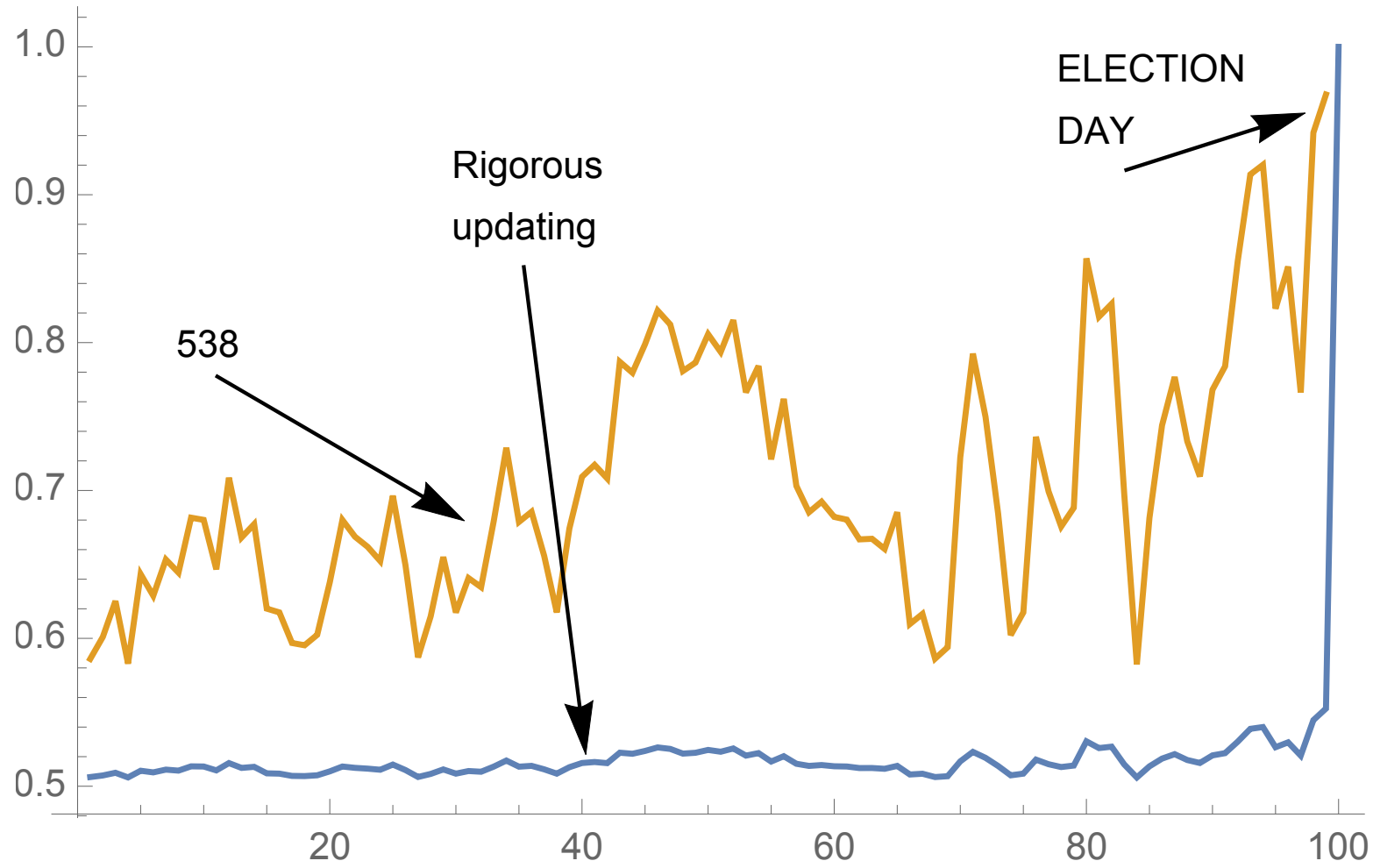

Figure 14.3: *Theoretical approach (top) vs practice (bottom). Shows how the estimation process cannot be in sync with the volatility of the estimation of (electoral or other) votes as it violates arbitrage boundaries.*

### 14.0.3 A Discussion on Risk Neutrality

We apply risk neutral valuation, for lack of conviction regarding another way, as a default option. Although $Y$ may not necessarily be tradable, adding a risk premium for the process involved in determining the arbitrage valuation would necessarily imply a negative one for the other candidate(s), which is hard to justify. Further, option values or binary bets, need to satisfy a no Dutch Book argument (the De Finetti form of no-arbitrage) (see [115]), i.e. properly priced binary options interpreted as probability forecasts give no betting "edge" in all outcomes without loss. Finally, any departure from risk neutrality would degrade the Brier score (about which, below) as it would represent a diversion from the final forecast.

Also note the absence of the assumptions of financing rate usually present in financial discussions.

## 14.1 THE BACHELIER-STYLE VALUATION

Let $F(.)$ be a function of a variable $X$ satisfying

$$dX_t = \sigma^2 X_t dt + \sigma \, dW_t. \tag{14.5}$$

We wish to show that $X$ has a simple Bachelier option price $B(.)$. The idea of no arbitrage is that a continuously made forecast must itself be a martingale.

Applying Itô's Lemma to $F \triangleq B$ for $X$ satisfying (14.5) yields

$$dF = \left[\sigma^2 X \frac{\partial F}{\partial X} + \frac{1}{2}\sigma^2 \frac{\partial^2 F}{\partial X^2} + \frac{\partial F}{\partial t}\right] dt + \sigma \frac{F}{X} dW$$

so that, since $\frac{\partial F}{\partial t} \triangleq 0$, $F$ must satisfy the partial differential equation

$$\frac{1}{2}\sigma^2 \frac{\partial^2 F}{\partial X^2} + \sigma^2 X \frac{\partial F}{\partial X} + \frac{\partial F}{\partial t} = 0, \tag{14.6}$$

which is the driftless condition that makes $B$ a martingale.

For a binary (call) option, we have for terminal conditions $B(X,t) \triangleq F, F_T = \theta(x-l)$, where $\theta(.)$ is the Heaviside theta function and $l$ is the threshold:

$$\theta(x) := \begin{cases} 1, & x \geq l \\ 0, & x < l \end{cases}$$

with initial condition $x_0$ at time $t_0$ and terminal condition at $T$ given by:

$$\frac{1}{2}\text{erfc}\left(\frac{x_0 e^{\sigma^2 t} - l}{\sqrt{e^{2\sigma^2 t} - 1}}\right)$$

which is, simply, the survival function of the Normal distribution parametrized under the process for $X$.

Likewise we note from the earlier argument of one-to one (one can use Borel set arguments ) that

$$\theta(y) := \begin{cases} 1, & y \geq S(l) \\ 0, & y < S(l), \end{cases}$$

so we can price the alternative process $B(Y,t) = \mathbb{P}(Y > \frac{1}{2})$ (or any other similarly obtained threshold $l$, by pricing

$$B(Y_0, t_0) = \mathbb{P}(x > S^{-1}(l)).$$

The pricing from the proportion of votes is given by:

$$B(Y_0, \sigma, t_0, T) = \frac{1}{2}\text{erfc}\left(\frac{l - \text{erf}^{-1}(2Y_0 - 1)e^{\sigma^2(T-t_0)}}{\sqrt{e^{2\sigma^2(T-t_0)} - 1}}\right),$$

the main equation (14.1), which can also be expressed less conveniently as

$$B(y_0, \sigma, t_0, T) = \frac{1}{\sqrt{e^{2\sigma^2 t} - 1}} \int_l^1 \exp\Bigg(\text{erf}^{-1}(2y-1)^2 \\ - \frac{1}{2}\left(\coth\left(\sigma^2 t\right) - 1\right)\left(\text{erf}^{-1}(2y-1) - \text{erf}^{-1}(2y_0 - 1)e^{\sigma^2 t}\right)^2\Bigg) dy$$

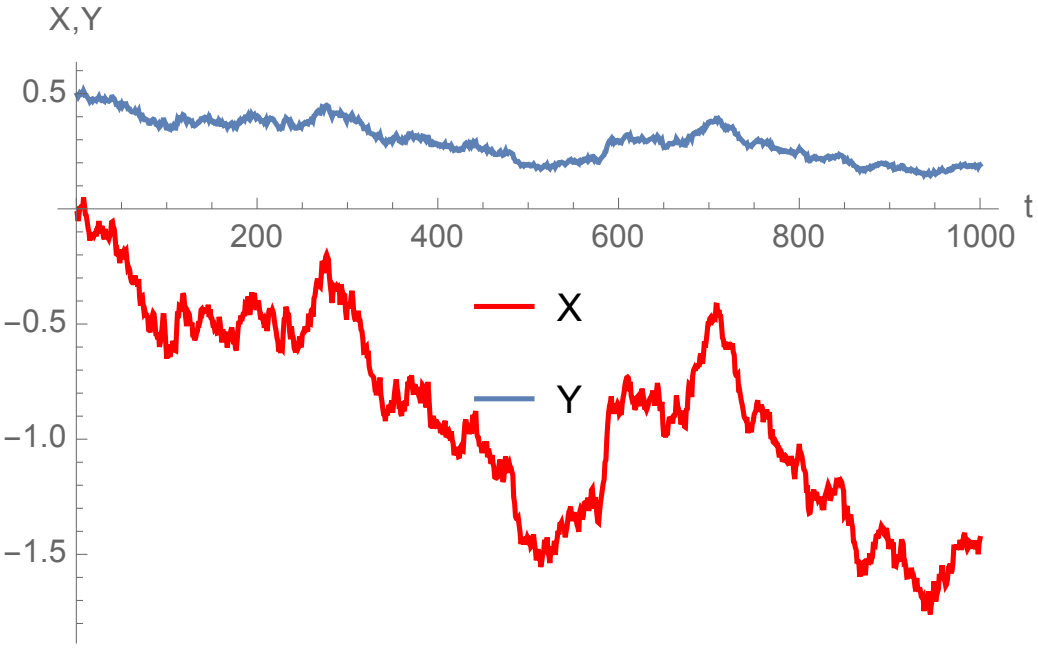

Figure 14.4: *Process and Dual Process*

## 14.2 BOUNDED DUAL MARTINGALE PROCESS

$Y_T$ is the terminal value of a process on election day. It lives in $[0,1]$ but can be generalized to the broader $[L,H]$, $L,H \in [0,\infty)$. The threshold for a given candidate to win is fixed at $l$. $Y$ can correspond to raw votes, electoral votes, or any other metric. We assume that $Y_t$ is an intermediate realization of the process at $t$, either produced synthetically from polls (corrected estimates) or other such systems.

Next, we create, for an unbounded arithmetic stochastic process, a bounded "dual" stochastic process using a sigmoidal transformation. It can be helpful to map processes such as a bounded electoral process to a Brownian motion, or to map a bounded payoff to an unbounded one, see Figure 14.2.

**Proposition 14.1**
*Under sigmoidal style transformations $S: x \mapsto y, \mathbb{R} \to [0,1]$ of the form a) $\frac{1}{2}+\frac{1}{2}\text{erf}(x)$, or b) $\frac{1}{1+\exp(-x)}$, if $X$ is a martingale, $Y$ is only a martingale for $Y_0=\frac{1}{2}$, and if $Y$ is a martingale, $X$ is only a martingale for $X_0=0$ .*

*Proof.* The proof is sketched as follows. From Itô's lemma, the drift term for $dX_t$ becomes 1) $\sigma^2 X(t)$, or 2) $\frac{1}{2}\sigma^2 \text{Tanh}\left(\frac{X(t)}{2}\right)$, where $\sigma$ denotes the volatility, respectively with transformations of the forms a) of $X_t$ and b) of $X_t$ under a martingale for $Y$. The drift for $dY_t$ becomes: 1) $\frac{\sigma^2 e^{-\text{erf}^{-1}(2Y-1)^2}\text{erf}^{-1}(2Y-1)}{\sqrt{\pi}}$ or 2) $\frac{1}{2}\sigma^2 Y(Y-1)(2Y-1)$ under a martingale for $X$. □

We therefore select the case of $Y$ being a martingale and present the details of the transformation a). The properties of the process have been developed by Carr [39]. Let $X$ be the arithmetic Brownian motion (14.5), with $X$-dependent drift and constant scale $\sigma$:

$$dX_t = \sigma^2 X_t dt + \sigma dW_t, \quad 0 < t < T < +\infty.$$

We note that this has similarities with the Ornstein-Uhlenbeck process normally written $dX_t = \theta(\mu - X_t)dt + \sigma dW$, except that we have $\mu = 0$ and violate the rules by using a negative mean reversion coefficient, rather more adequately described as "mean repelling", $\theta = -\sigma^2$.

We map from $X \in (-\infty, \infty)$ to its dual process $Y$ as follows. With $S : \mathbb{R} \to [0,1]$, $Y = S(x)$,

$$S(x) = \frac{1}{2} + \frac{1}{2}\operatorname{erf}(x)$$

the dual process (by unique transformation since $S$ is one to one, becomes, for $y \triangleq S(x)$, using Itô's lemma (since $S(.)$ is twice differentiable and $\partial S/\partial t = 0$):

$$dS = \left(\frac{1}{2}\sigma^2 \frac{\partial^2 S}{\partial x^2} + X\sigma^2 \frac{\partial S}{\partial x}\right) \mathrm{d}t + \sigma \frac{\partial S}{\partial x} dW$$

which with zero drift can be written as a process

$$dY_t = s(Y) dW_t,$$

for all $t > \tau$, $\mathbb{E}(Y_t | Y_\tau) = Y_\tau$. and scale

$$s(Y) = \frac{\sigma}{\sqrt{\pi}} e^{-\operatorname{erf}^{-1}(2y-1)^2}$$

which as we can see in Figure 14.5, $s(y)$ can be approximated by the quadratic function $y(1-y)$ times a constant.

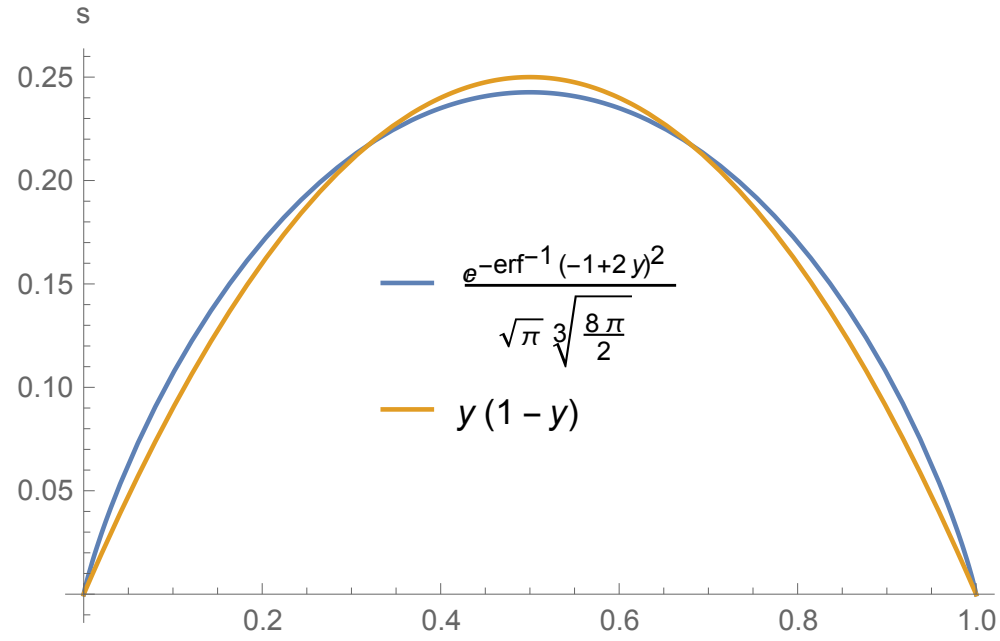

Figure 14.5: *The instantaneous volatility of $Y$ as a function of the level of $Y$ for two different methods of transformations of $X$, which appear to not be substantially different. We compare to the quadratic form $y - y^2$ scaled by a constant $\frac{1}{\sqrt[3]{\frac{8\pi}{2}}}$. The volatility declines as we move away from $\frac{1}{2}$ and collapses at the edges, thus maintaining $Y$ in $(0,1)$. For simplicity we assumed $\sigma = t = 1$.*

We can recover equation (14.5) by inverting, namely $S^{-1}(y) = \operatorname{erf}^{-1}(2y-1)$, and again applying Itô's Lemma. As a consequence of gauge invariance option prices are identical whether priced on $X$ or $Y$, even if one process has a drift while the other is a martingale. In other words, one may apply one's estimation to the electoral threshold, or to the more complicated $X$ with the same results. And, to summarize our method, pricing an option on $X$ is familiar, as it is exactly a Bachelier-style option price.

## 14.3 RELATION TO DE FINETTI'S PROBABILITY ASSESSOR

This section provides a brief background for the conventional approach to probability assessment. The great De Finetti [69] has shown that the "assessment" of the "probability" of the realization of a random variable in $\{0,1\}$ requires a nonlinear loss function – which makes his definition of *probabilistic assessment* differ from that of the P/L of a trader engaging in binary bets.

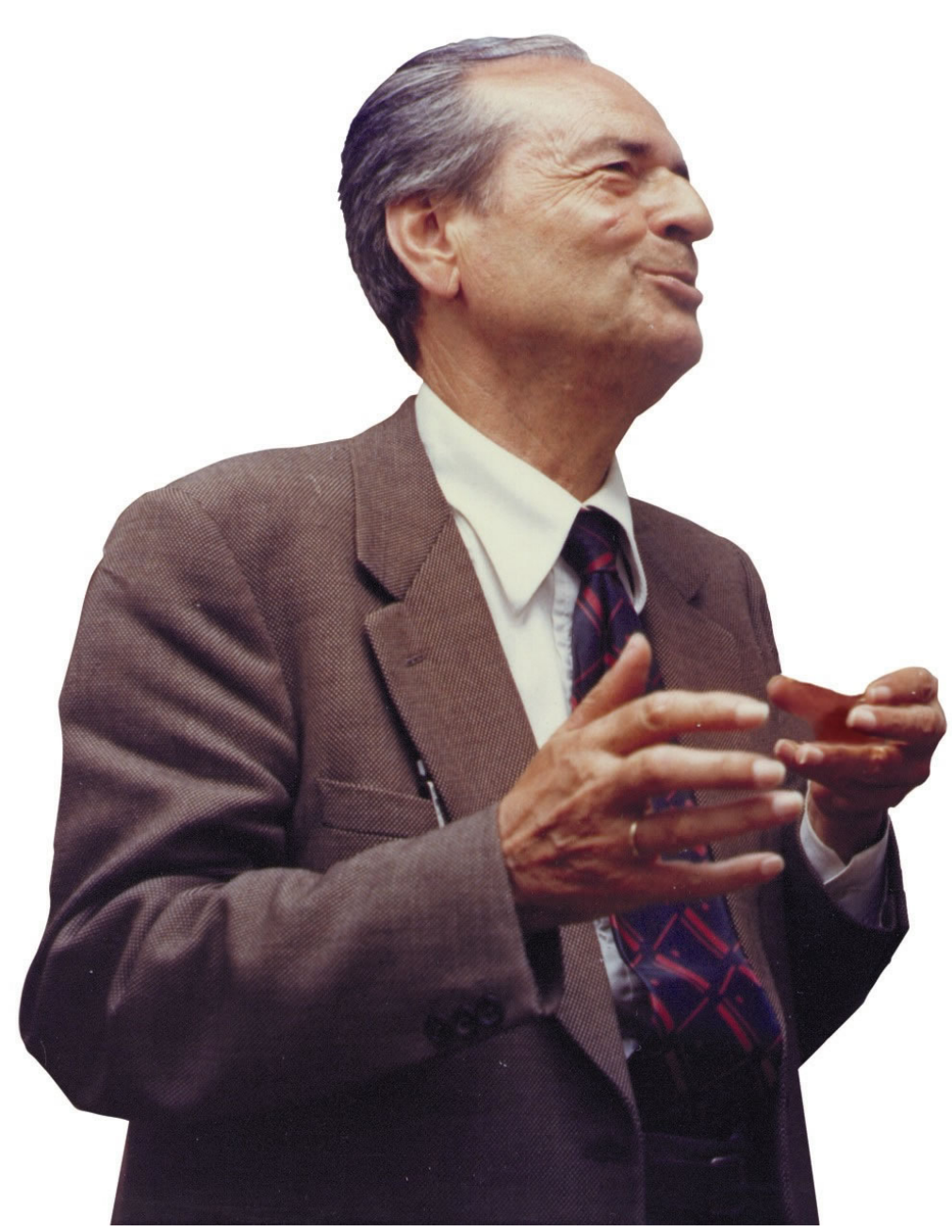

Figure 14.6: *Bruno de Finetti (1906-1985). A probabilist, philosopher, and insurance mathematician, he formulated the Brier score for probabilistic assessment which we show is compatible dynamically with a martingale. Source: DeFinetti.org*

Assume that a betting agent in an $n$-repeated two period model, $t_0$ and $t_1$, produces a strategy $\mathfrak{S}$ of bets $b_{0,i} \in [0,1]$ indexed by $i = 1, 2, \ldots, n$, with the realization of the binary r.v. $\mathbb{1}_{t_1,i}$. If we take the absolute variation of his P/L over $n$ bets, it will be

$$L_1(\mathfrak{S}) = \frac{1}{n} \sum_{i=1}^{n} \left| \mathbb{1}_{t_1,i} - b_{t_0,i} \right| .$$

For example, assume that $\mathbb{E}(\mathbb{1}_{t_1}) = \frac{1}{2}$. Betting on the probability, here $\frac{1}{2}$, produces a loss of $\frac{1}{2}$ in expectation, which is the same as betting either 0 or 1 – hence not favoring the agent to bet on the exact probability.

If we work with the same random variable and non-time-varying probabilities, the $L^1$ metric would be appropriate:

$$L_1(\mathfrak{S}) = \frac{1}{n} \left| \mathbb{1}_{t_1,i} - \sum_{i=1}^{n} b_{t_0,i} \right| .$$

De Finetti proposed a "Brier score" type function, a quadratic loss function in $\mathcal{L}^2$:

$$L_2(\mathfrak{S}) = \frac{1}{n} \sum_{i=1}^{n} (\mathbb{1}_{t_1,i} - b_{t_0,i})^2,$$

the minimum of which is reached for $b_{t_0,i} = \mathbb{E}(\mathbb{1}_{t_1})$.

In our world of continuous time derivative valuation, where, in place of a two period lattice model, we are interested, for the same final outcome at $t_1$, in the stochastic process $b_t$, $t_0 \geq t \geq t_1$, the arbitrage "value" of a bet on a binary out-come needs to match the expectation, hence, again, we map to the Brier score – *by an arbitrage argument*. Although there is no quadratic loss function involved, the fact that the bet is a function of a martingale, which is required to be itself a martingale, i.e. that the conditional expectation remains invariant to time, does not allow an arbitrage to take place. A "high" price can be "shorted" by the arbitrageur, a "low" price can be "bought", and so on repeatedly. The consistency between bets at period $t$ and other periods $t + \Delta t$ enforces the probabilistic discipline. In other words, someone can "buy" from the forecaster then "sell" back to him, generating a positive expected "return" if the forecaster is out of line with martingale valuation.

As to the current practice by forecasters, although some election forecasters appear to be aware of the need to minimize their Brier score, the idea that the revisions of estimates should also be subjected to martingale valuation is not well established.

## 14.4 CONCLUSION AND COMMENTS

As can be seen in Figure 14.1, a binary option reveals more about uncertainty than about the true estimation, a result well known to traders, see [271].

In the presence of more than 2 candidates, the process can be generalized with the following heuristic approximation. Establish the stochastic process for $Y_{1,t}$, and just as $Y_{1,t}$ is a process in $[0,1]$, $Y_{2,t}$ is a process $\in (Y_{1,t}, 1]$, with $Y_{3,t}$ the residual $1 - Y_{2,t} - Y_{1,t}$, and more generally $Y_{n-1,t} \in (Y_{n_2,t}, 1]$ and $Y_{n,t}$ is the residual $Y_n = 1 - \sum_{i=1}^{n-1} Y_{i,t}$. For $n$ candidates, the n$^{th}$ is the residual.

## ADDENDUM: ALL ROADS LEAD TO QUANTITATIVE FINANCE

**Background** *Aubrey Clayton sent a letter to the editor complaining about the previous piece on grounds of "errors" in the above methodology. The author answered, with Dhruv Madeka, not quite to Clayton, rather to express the usefulness of quantitative finance methods in life.*

We are happy to respond to Clayton's (non-reviewed) letter, in spite of its confusions, as it will give us the opportunity to address more fundamental misunderstandings of the role of quantitative finance in general, and arbitrage pricing in particular, and proudly show how "all roads lead to quantitative finance", that is, that arbitrage approaches are universal and applicable to all manner of binary forecasting. It also allows the second author to comment from his paper, Madeka (2017)[187], which independently and simultaneously obtained similar results to Taleb (2018)[280].

### Incorrect claims

The claims:

> Taleb's criticism of popular forecast probabilities, specifically the election forecasts of FiveThirtyEight...

and

> He [Taleb] claims this means the FiveThirtyEight forecasts must have "violate[d] arbitrage boundaries.

are factually incorrect.

There is no mention of FiveThirtyEight in [280], and Clayton must be confusing scientific papers with Twitter debates. The paper is an attempt at addressing elections in a rigorous manner, not journalistic discussion, and only mentions the 2016 election in one illustrative sentence.[4]

Let us however continue probing Clayton's other assertions, in spite of his confusion and the nature of the letter.

### Incorrect arbitrage valuation

Clayton's claims either an error ("First, one of the "standard results" of quantitative finance that his election forecast assessments rely on is false", he initially writes), or, as he confusingly retracts, something "only partially true". Again, let us set aside that Taleb(2018)[280] makes no "assessment" of FiveThirtyEight's record and outline his reasoning.

Clayton considers three periods, $t_0 = 0$, an intermediate period $t$ and a terminal one $T$, with $t_0 \leq t < T$. Clayton shows a special case of the distribution of the forward probability, seen at $t_0$, for time starting at $t = \frac{T}{2}$ and ending at $T$. It is a uniform distribution for that specific time period. In fact under his construction, using the probability integral transform, one can show that the probabilities follow what resembles a symmetric beta distribution with parameters $a$ and $b$, and with $a = b$. When $t = \frac{T}{2}$, we have $a = b = 1$ (hence the uniform distribution). Before $T/2$ it has a $\cap$ shape, with Dirac at $t = t_0$. Beyond $T/2$ it has a $\cup$ shape, ending with two Dirac sticks at 0 and 1 (like a Bernoulli) when $t$ is close to $T$ (and close to an arcsine distribution with $a = b = \frac{1}{2}$ somewhere in between).

Clayton's construction is indeed misleading, since he analyzes the distribution of the price at time $t$ with the filtration at time $t_0$, particularly when discussing arbitrage pricing and arbitrage pressures. Agents value options between $t$ and $T$ at time $t$ (not period $t_0$), with an underlying price: under such constraint, the binary option automatically converges towards $\frac{1}{2}$ as $\sigma \to \infty$, and that for *any* value of the underlying price, no matter how far away from the strike price (or threshold). The $\sigma$ here is never past realized, only future unrealized volatility. This can be seen within the framework presented in Taleb (2018) [280] but also by taking any binary option pricing model. A price is not a probability (less even a probability distribution), but an expectation. Simply, as arbitrage operators, we look at *future* volatility given information about the underlying when pricing a binary option, not the distribution of probability itself in the unconditional abstract.

---

[4] Incidentally, the problem with FiveThirtyEight isn't changing probabilities from .55 to .85 within a 5 months period, it is performing abrupt changes within a much shorter timespan –and *that* was discussed in Madeka (2017)[187].

At infinite $\sigma$, it becomes all noise, and such a level of noise drowns all signals.

Another way to view the pull of uncertainty towards $\frac{1}{2}$ is in using information theory and the notion of maximum entropy under deep uncertainty : the entropy ($I$) of a Bernoulli distribution with probabilities $p$ and $(1-p)$, $I = -((1-p)\log(1-p) + p\log(p))$ is maximal at $\frac{1}{2}$.

To beat a $\frac{1}{2}$ pricing one needs to have enough information to beat the noise. As we will see in the next section, it is not easy.

### Arbitrage matters

Another result from quantitative finance that puts bounds on the volatility of forecasting is as follows. Since election forecasts can be interpreted as a European binary option, we can exploit the fact that the price process of this option is bounded between 0 and 1 to make claims about the volatility of the price itself.

Essentially, if the price of the binary option varies too much, a simple trading strategy of buying low and selling high is guaranteed to produce a profit[5]. The argument can be summed up by noting that if we consider an arithmetic brownian motion that's bounded between $[L, H]$:

$$dB_t = \sigma dW_t \tag{14.7}$$

The stochastic integral $2\int^T (B_0 - B_t)dB_t = \sigma^2 T - (B_T - B_0)^2$ can be replicated at zero cost, indicating that the value of $B_T$ is bounded by the maximum value of the square difference on the right hand side of the equation. That is, a forecaster who produces excessively volatile probabilities – if he or she is willing to trade on such a forecast (i.e. they have skin in the game) – can be arbitraged by following a strategy that sells (proportionally) when the forecast is too high and buys (proportionally) when the forecast is too low.

To conclude, any numerical probabilistic forecasting should be treated like a choice price —De Finetti's intuition is that forecasts should have skin in the game. Under these conditions, binary forecasting belongs to the rules of arbitrage and derivative pricing, well mapped in quantitative finance. Using a quantitative finance approach to produce binary forecasts does not prevent Bayesian methods (Taleb(2018) does not say probabilities should be $\frac{1}{2}$, only that there is a headwind towards that level owing to arbitrage pressures and constraints on how variable a forecast can be). It is just that there is one price that counts at the end, 1 or 0, which puts a structure on the updating.[6]

---

5 We take this result from Bruno Dupire's notes for his continuous time finance class at NYU's Courant Institute, particularly his final exam for the Spring of 2019.

6 Another way to see it, from outside our quantitative finance models: consider a standard probabilistic score. Let $X_1, \ldots, X_n$ be random variables in $[0, 1$ and a $B_T$ a constant $B_T \in \{0, 1\}$, we have the $\lambda$ score

$$\lambda_n = \frac{1}{n}\sum_{i=1}^{n}(x_i - B_T)^2,$$

which needs to be minimized (on a single outcome $B_T$). For any given $B_T$ and an average forecast $\overline{x} = \sum_{i=1}^{n} x_i$, the minimum value of $\lambda_n$ is reached for $x_1 = \ldots = x_n$. To beat a Dirac forecast $x_1 = \ldots =$

The reason Clayton might have trouble with quantitative finance could be that probabilities and underlying polls may not be martingales in real life; traded probabilities (hence real forecasts) must be martingales. Which is why in Taleb (2018)[280] the process for the polls (which can be vague and nontradable) needs to be transformed into a process for probability in $[0,1]$.

## ACKNOWLEDGMENTS

Raphael Douady, students at NYU's Tandon School of Engineering, participants at the Bloomberg Quantitative Finance Seminar in New York.

$x_n = \frac{1}{2}$ for which $\lambda = \frac{1}{4}$ with a high variance strategy, one needs to have 75% accuracy. (Note that a uniform forecast has a score of $\frac{1}{3}$.) This shows us the trade-off between volatility and signal.

# Part IV

# INEQUALITY ESTIMATORS UNDER FAT TAILS

# 15 | GINI ESTIMATION UNDER INFINITE VARIANCE ‡

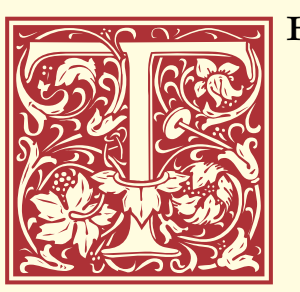HIS CHAPTER is about the problems related to the estimation of the Gini index in presence of a fat-tailed data generating process, i.e. one in the stable distribution class with finite mean but infinite variance (i.e. with tail index $\alpha \in (1, 2)$). We show that, in such a case, the Gini coefficient cannot be reliably estimated using conventional nonparametric methods, because of a downward bias that emerges under fat tails. This has important implications for the ongoing discussion about economic inequality.

We start by discussing how the nonparametric estimator of the Gini index undergoes a phase transition in the symmetry structure of its asymptotic distribution, as the data distribution shifts from the domain of attraction of a light-tailed distribution to that of a fat-tailed one, especially in the case of infinite variance. We also show how the nonparametric Gini bias increases with lower values of $\alpha$. We then prove that maximum likelihood estimation outperforms nonparametric methods, requiring a much smaller sample size to reach efficiency.

Finally, for fat-tailed data, we provide a simple correction mechanism to the small sample bias of the nonparametric estimator based on the distance between the mode and the mean of its asymptotic distribution.

## 15.1 INTRODUCTION

Wealth inequality studies represent a field of economics, statistics and econophysics exposed to fat-tailed data generating processes, often with infinite variance [43, 172]. This is not at all surprising if we recall that the prototype of fat-tailed distributions, the Pareto, has been proposed for the first time to model household incomes [217]. However, the fat-tailedness of data can be problem-

Research chapter.
(With A. Fontanari and P. Cirillo), coauthors

atic in the context of wealth studies, as the property of efficiency (and, partially, consistency) does not necessarily hold for many estimators of inequality and concentration [97, 172].

The scope of this work is to show how fat tails affect the estimation of one of the most celebrated measures of economic inequality, the Gini index [91, 128, 172], often used (and abused) in the econophysics and economics literature as the main tool for describing the distribution and the concentration of wealth around the world [43, 224**?** ].

The literature concerning the estimation of the Gini index is wide and comprehensive (e.g. [91, 262] for a review), however, strangely enough, almost no attention has been paid to its behavior in presence of fat tails, and this is curious if we consider that: 1) fat tails are ubiquitous in the empirical distributions of income and wealth [172, 224], and 2) the Gini index itself can be seen as a measure of variability and fat-tailedness [89, 92, 93, 113].

The standard method for the estimation of the Gini index is nonparametric: one computes the index from the empirical distribution of the available data using Equation (15.5) below. But, as we show in this paper, this estimator suffers from a downward bias when we deal with fat-tailed observations. Therefore our goal is to close this gap by deriving the limiting distribution of the nonparametric Gini estimator in presence of fat tails, and propose possible strategies to reduce the bias. We show how the maximum likelihood approach, despite the risk of model misspecification, needs much fewer observations to reach efficiency when compared to a nonparametric one.[2]

Our results are relevant to the discussion about wealth inequality, recently rekindled by Thomas Piketty in [224], as the estimation of the Gini index under fat tails and infinite variance may cause several economic analyses to be unreliable, if not markedly wrong. Why should one trust a biased estimator?

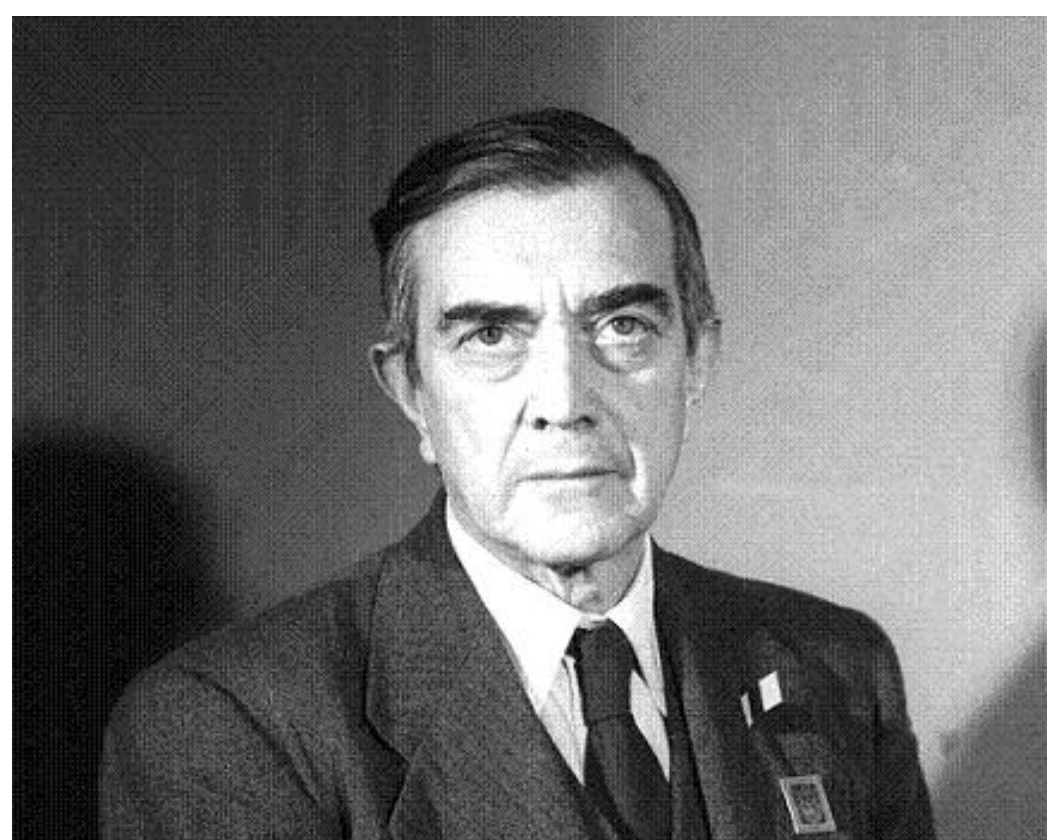

Figure 15.1: *The Italian statistician Corrado Gini, 1884-1965. source: Bocconi.*

[2] A similar bias also affects the nonparametric measurement of quantile contributions, i.e. those of the type “the top 1% owns x% of the total wealth" [292]. This paper extends the problem to the more widespread Gini coefficient, and goes deeper by making links with the limit theorems.

By fat-tailed data we indicate those data generated by a positive random variable $X$ with cumulative distribution function (c.d.f.) $F(x)$, which is regularly-varying of order $\alpha$ [162], that is, for $\bar{F}(x) := 1 - F(x)$, one has

$$\lim_{x\to\infty} x^{\alpha}\bar{F}(x) = L(x), \tag{15.1}$$

where $L(x)$ is a slowly-varying function such that $\lim_{x\to\infty} \frac{L(cx)}{L(x)} = 1$ with $c > 0$, and where $\alpha > 0$ is called the tail exponent .

Regularly-varying distributions define a large class of random variables whose properties have been extensively studied in the context of extreme value theory [97, 136], when dealing with the probabilistic behavior of maxima and minima. As pointed out in [53], regularly-varying and fat-tailed are indeed synonyms. It is known that, if $X_1, ..., X_n$ are i.i.d. observations with a c.d.f. $F(x)$ in the regularly-varying class, as defined in Equation (15.1), then their data generating process falls into the maximum domain of attraction of a Fréchet distribution with parameter $\rho$, in symbols $X \in MDA(\Phi(\rho))$[136]. This means that, for the partial maximum $M_n = \max(X_1, ..., X_n)$, one has

$$P\left(a_n^{-1}\left(M_n - b_n\right) \leq x\right) \xrightarrow{d} \Phi(\rho) = e^{-x^{-\rho}}, \qquad \rho > 0, \tag{15.2}$$

with $a_n > 0$ and $b_n \in \mathbb{R}$ two normalizing constants. Clearly, the connection between the regularly-varying coefficient $\alpha$ and the Fréchet distribution parameter $\rho$ is given by: $\alpha = \frac{1}{\rho}$ [97].
The Fréchet distribution is one of the limiting distributions for maxima in extreme value theory, together with the Gumbel and the Weibull; it represents the fat-tailed and unbounded limiting case [136]. The relationship between regularly-varying random variables and the Fréchet class thus allows us to deal with a very large family of random variables (and empirical data), and allows us to show how the Gini index is highly influenced by maxima, i.e. extreme wealth, as clearly suggested by intuition [113, 172], especially under infinite variance. Again, this recommends some caution when discussing economic inequality under fat tails.

It is worth remembering that the existence (finiteness) of the moments for a fat-tailed random variable $X$ depends on the tail exponent $\alpha$, in fact

$$\begin{aligned} E(X^{\delta}) &< \infty \text{ if } \delta \leq \alpha, \\ E(X^{\delta}) &= \infty \text{ if } \delta > \alpha. \end{aligned} \tag{15.3}$$

In this work, we restrict our focus on data generating processes with finite mean and infinite variance, therefore, according to Equation (15.3), on the class of regularly-varying distributions with tail index $\alpha \in (1, 2)$.

Table 15.1 and Figure 15.2 present numerically and graphically our story, already suggesting its conclusion, on the basis of artificial observations sampled from a Pareto distribution (Equation (15.13) below) with tail parameter $\alpha$ equal to 1.1.

Table 15.1 compares the nonparametric Gini index of Equation (15.5) with the maximum likelihood (ML) tail-based one of Section 15.3. For the different sample sizes in Table 15.1, we have generated $10^8$ samples, averaging the estimators via Monte Carlo. As the first column shows, the convergence of the nonparametric estimator to the true Gini value ($g = 0.8333$) is extremely slow and monotonically increasing; this suggests an issue not only in the tail structure of the distribution of the nonparametric estimator but also in its symmetry.

Figure 15.2 provides some numerical evidence that the limiting distribution of the nonparametric Gini index loses its properties of normality and symmetry [107], shifting towards a skewed and fatter-tailed limit, when data are characterized by an infinite variance. As we prove in Section 15.2, when the data generating process is in the domain of attraction of a fat-tailed distribution, the asymptotic distribution of the Gini index becomes a skewed-to-the-right $\alpha$-stable law. This change of behavior is responsible of the downward bias of the nonparametric Gini under fat tails. However, the knowledge of the new limit allows us to propose a correction for the nonparametric estimator, improving its quality, and thus reducing the risk of badly estimating wealth inequality, with all the possible consequences in terms of economic and social policies [172, 224].

Table 15.1: *Comparison of the Nonparametric (NonPar) and the Maximum Likelihood (ML) Gini estimators, using Paretian data with tail $\alpha = 1.1$ (finite mean, infinite variance) and different sample sizes. Number of Monte Carlo simulations: $10^8$. Error Ratio:* $\frac{\sum_{i=1}^{n} |obs_i - (Nonpar)|}{\sum_{i=1}^{n} |obs_i - (ML)|}$.

| $n$ | Nonpar | | ML | | Error Ratio[3] |
|---|---|---|---|---|---|
| *(number of obs.)* | *Mean* | *Bias* | *Mean* | *Bias* | |
| $10^3$ | 0.711 | -0.122 | 0.8333 | 0 | 1.4 |
| $10^4$ | 0.750 | -0.083 | 0.8333 | 0 | 3 |
| $10^5$ | 0.775 | -0.058 | 0.8333 | 0 | 6.6 |
| $10^6$ | 0.790 | -0.043 | 0.8333 | 0 | 156 |
| $10^7$ | 0.802 | -0.031 | 0.8333 | 0 | $10^5$+ |

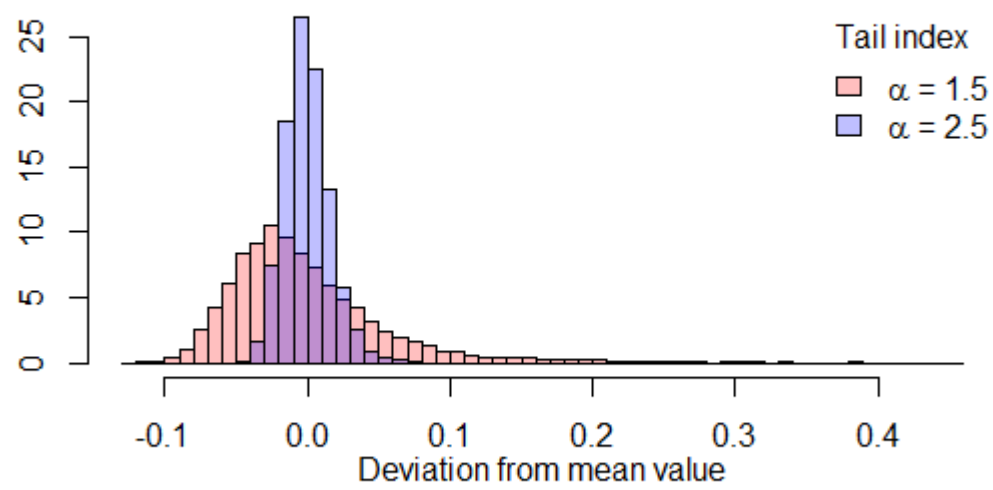

Figure 15.2: *Histograms for the Gini nonparametric estimators for two Paretian (type I) distributions with different tail indices, with finite and infinite variance (plots have been centered to ease comparison). Sample size: $10^3$. Number of samples: $10^2$ for each distribution.*

The rest of the paper is organized as follows. In Section 15.2 we derive the asymptotic distribution of the sample Gini index when data possess an infinite variance. In Section 15.3 we deal with the maximum likelihood estimator; in Section 15.4 we provide an illustration with Paretian observations; in Section 15.5 we propose a simple correction based on the mode-mean distance of the asymptotic distribution of the nonparametric estimator, to take care of its small-sample bias. Section 15.6 closes the paper. A technical Appendix contains the longer proofs of the main results in the work.

## 15.2 ASYMPTOTICS OF THE NONPARAMETRIC ESTIMATOR UNDER INFINITE VARIANCE

We now derive the asymptotic distribution for the nonparametric estimator of the Gini index when the data generating process is fat-tailed with finite mean but infinite variance.

The so-called stochastic representation of the Gini $g$ is

$$g = \frac{1}{2} \frac{\mathbb{E}\left(|X' - X''|\right)}{\mu} \in [0, 1], \tag{15.4}$$

where $X'$ and $X''$ are i.i.d. copies of a random variable $X$ with c.d.f. $F(x) \in [c, \infty)$, $c > 0$, and with finite mean $\mathbb{E}(X) = \mu$. The quantity $\mathbb{E}\left(|X' - X''|\right)$ is known as the "Gini Mean Difference" (GMD) [262]. For later convenience we also define $g = \frac{\theta}{\mu}$ with $\theta = \frac{\mathbb{E}(|X'-X''|)}{2}$.

The Gini index of a random variable $X$ is thus the mean expected deviation between any two independent realizations of $X$, scaled by twice the mean [94].

The most common nonparametric estimator of the Gini index for a sample $X_1, ..., X_n$ is defined as

$$G^{NP}(X_n) = \frac{\sum_{1 \le i < j \le n} |X_i - X_j|}{(n-1) \sum_{i=1}^{n} X_i}, \tag{15.5}$$

which can also be expressed as

$$G^{NP}(X_n) = \frac{\sum_{i=1}^{n} (2(\frac{i-1}{n-1} - 1) X_{(i)}}{\sum_{i=1}^{n} X_{(i)}} = \frac{\frac{1}{n} \sum_{i=1}^{n} Z_{(i)}}{\frac{1}{n} \sum_{i=1}^{n} X_i}, \tag{15.6}$$

where $X_{(1)}, X_{(2)}, ..., X_{(n)}$ are the ordered statistics of $X_1, ..., X_n$, such that: $X_{(1)} < X_{(2)} < ... < X_{(n)}$ and $Z_{(i)} = 2\left(\frac{i-1}{n-1} - 1\right) X_{(i)}$. The asymptotic normality of the estimator in Equation (15.6) under the hypothesis of finite variance for the data generating process is known [172, 262]. The result directly follows from the properties of the U-statistics and the L-estimators involved in Equation (15.6)

A standard methodology to prove the limiting distribution of the estimator in Equation (15.6), and more in general of a linear combination of order statistics, is to show that, in the limit for $n \to \infty$, the sequence of order statistics can be approximated by a sequence of i.i.d random variables [67, 180]. However, this usually requires some sort of $L^2$ integrability of the data generating process, something we are not assuming here.

Lemma 15.1 (proved in the Appendix) shows how to deal with the case of sequences of order statistics generated by fat-tailed $L^1$-only integrable random variables.

**Lemma 15.1**
*Consider the following sequence* $R_n = \frac{1}{n} \sum_{i=1}^{n} (\frac{i}{n} - U_{(i)}) F^{-1}(U_{(i)})$ *where* $U_{(i)}$ *are the order* *statistics of a uniformly distributed i.i.d random sample. Assume that* $F^{-1}(U) \in L^1$. *Then the following results hold:*

$$R_n \xrightarrow{L^1} 0, \tag{15.7}$$

*and*

$$\frac{n^{\frac{\alpha-1}{\alpha}}}{L_0(n)} R_n \xrightarrow{L^1} 0, \tag{15.8}$$

*with* $\alpha \in (1,2)$ *and* $L_0(n)$ *a slowly-varying function.*

### 15.2.1 A Quick Recap on $\alpha$-Stable Random Variables

We here introduce some notation for $\alpha$-stable distributions, as we need them to study the asymptotic limit of the Gini index.

A random variable $X$ follows an $\alpha$-stable distribution, in symbols $X \sim S(\alpha, \beta, \gamma, \delta)$, if its characteristic functionis

$$E(e^{itX}) = \begin{cases} e^{-\gamma^\alpha |t|^\alpha (1 - i\beta \operatorname{sign}(t)) \tan(\frac{\pi\alpha}{2}) + i\delta t} & \alpha \neq 1 \\ e^{-\gamma |t| (1 + i\beta \frac{2}{\pi} \operatorname{sign}(t)) \ln |t| + i\delta t} & \alpha = 1 \end{cases},$$

where $\alpha \in (0,2)$ governs the tail, $\beta \in [-1,1]$ is the skewness, $\gamma \in \mathbb{R}^+$ is the scale parameter, and $\delta \in \mathbb{R}$ is the location one. This is known as the $S1$ parametrization of $\alpha$-stable distributions [213, 246].

Interestingly, there is a correspondence between the $\alpha$ parameter of an $\alpha$-stable random variable, and the $\alpha$ of a regularly-varying random variable as per Equation (15.1): as shown in [107, 213], a regularly-varying random variable of order $\alpha$ is $\alpha$-stable, with the same tail coefficient. This is why we do not make any distinction in the use of the $\alpha$ here. Since we aim at dealing with distributions characterized by finite mean but infinite variance, we restrict our focus to $\alpha \in (1,2)$, as the two $\alpha$'s coincide.

Recall that, for $\alpha \in (1,2]$, the expected value of an $\alpha$-stable random variable $X$ is equal to the location parameter $\delta$, i.e. $\mathbb{E}(X) = \delta$. For more details, we refer to [213, 246].

The standardized $\alpha$-stable random variable is expressed as

$$S_{\alpha,\beta} \sim S(\alpha, \beta, 1, 0). \tag{15.9}$$

We note that $\alpha$-stable distributions are a subclass of infinitely divisible distributions. Thanks to their closure under convolution, they can be used to describe the limiting behavior of (rescaled) partials sums, $S_n = \sum_{i=1}^n X_i$, in the General central limit theorem (GCLT) setting [107]. For $\alpha = 2$ we obtain the normal distribution as a special case, which is the limit distribution for the classical CLTs, under the hypothesis of finite variance.

In what follows we indicate that a random variable is in the domain of attraction of an $\alpha$-stable distribution, by writing $X \in DA(S_\alpha)$. Just observe that this condition for the limit of partial sums is equivalent to the one given in Equation (15.2) for the limit of partial maxima [97, 107].

### 15.2.2 The $\alpha$-Stable Asymptotic Limit of the Gini Index

Consider a sample $X_1, ..., X_n$ of i.i.d. observations with a continuous c.d.f. $F(x)$ in the regularly-varying class, as defined in Equation (15.1), with tail index $\alpha \in (1, 2)$. The data generating process for the sample is in the domain of attraction of a Fréchet distribution with $\rho \in (\frac{1}{2}, 1)$, given that $\rho = \frac{1}{\alpha}$.

For the asymptotic distribution of the Gini index estimator, as presented in Equation (15.6), when the data generating process is characterized by an infinite variance, we can make use of the following two theorems: Theorem 1 deals with the limiting distribution of the Gini Mean Difference (the numerator in Equation (15.6)), while Theorem 2 extends the result to the complete Gini index. Proofs for both theorems are in the Appendix.

**Theorem 1**
*Consider a sequence $(X_i)_{1 \leq i \leq n}$ of i.i.d random variables from a distribution $X$ on $[c, +\infty)$ with $c > 0$, such that $X$ is in the domain of attraction of an $\alpha$-stable random variable, $X \in DA(S_\alpha)$, with $\alpha \in (1, 2)$. Then the sample Gini mean deviation (GMD) $\frac{\sum_{i=1}^n Z_{(i)}}{n}$ satisfies the following limit in distribution:*

$$\frac{n^{\frac{\alpha-1}{\alpha}}}{L_0(n)} \left( \frac{1}{n} \sum_{i=1}^{n} Z_{(i)} - \theta \right) \xrightarrow{d} S_{\alpha,1}, \tag{15.10}$$

*where $Z_i = (2F(X_i) - 1)X_i$, $\mathbb{E}(Z_i) = \theta$, $L_0(n)$ is a slowly-varying function such that Equation (15.37) holds (see the Appendix), and $S_{\alpha,1}$ is a right-skewed standardized $\alpha$-stable random variable defined as in Equation (15.9).*

*Moreover the statistic $\frac{1}{n}\sum_{i=1}^n Z_{(i)}$ is an asymptotically consistent estimator for the GMD, i.e. $\frac{1}{n}\sum_{i=1}^n Z_{(i)} \xrightarrow{P} \theta$.*

Note that Theorem 1 could be restated in terms of the maximum domain of attraction $MDA(\Phi(\rho))$ as defined in Equation (15.2).

**Theorem 2**
*Given the same assumptions of Theorem 1, the estimated Gini index $G^{NP}(X_n) = \frac{\sum_{i=1}^n Z_{(i)}}{\sum_{i=1}^n X_i}$ satisfies the following limit in distribution*

$$\frac{n^{\frac{\alpha-1}{\alpha}}}{L_0(n)} \left( G^{NP}(X_n) - \frac{\theta}{\mu} \right) \xrightarrow{d} Q, \tag{15.11}$$

*where $\mathbb{E}(Z_i) = \theta$, $\mathbb{E}(X_i) = \mu$, $L_0(n)$ is the same slowly-varying function defined in Theorem 1 and $Q$ is a right-skewed $\alpha$-stable random variable $S(\alpha, 1, \frac{1}{\mu}, 0)$.*

*Furthermore the statistic $\frac{\sum_{i=1}^n Z_{(i)}}{\sum_{i=1}^n X_i}$ is an asymptotically consistent estimator for the Gini index, i.e. $\frac{\sum_{i=1}^n Z_{(i)}}{\sum_{i=1}^n X_i} \xrightarrow{P} \frac{\theta}{\mu} = g$.*

In the case of fat tails with $\alpha \in (1, 2)$, Theorem 2 tells us that the asymptotic distribution of the Gini estimator is always right-skewed notwithstanding the distribution of the underlying data generating process. Therefore heavily fat-tailed data not only induce a fatter-tailed limit for the Gini estimator, but they also change the shape of the limit law, which definitely moves away from the usual symmetric Gaussian. As a consequence, the Gini estimator, whose asymptotic consistency is still guaranteed [180], will approach its true value more slowly, and from below. Some evidence of this was already given in Table 15.1.

## 15.3 THE MAXIMUM LIKELIHOOD ESTIMATOR

Theorem 2 indicates that the usual nonparametric estimator for the Gini index is not the best option when dealing with infinite-variance distributions, due to the skewness and the fatness of its asymptotic limit. The aim is to find estimators that still preserve their asymptotic normality under fat tails, which is not possible with nonparametric methods, as they all fall into the $\alpha$-stable Central Limit Theorem case [97, 107]. Hence the solution is to use parametric techniques.

Theorem 3 shows how, once a parametric family for the data generating process has been identified, it is possible to estimate the Gini index via MLE. The resulting estimator is not just asymptotically normal, but also asymptotically efficient.

In Theorem 3 we deal with random variables $X$ whose distribution belongs to the large and flexible exponential family [250], i.e. whose density can be represented as

$$f_\theta(x) = h(x)e^{(\eta(\theta)T(x)-A(\theta))},$$

with $\theta \in \mathbb{R}$, and where $T(x)$, $\eta(\theta)$, $h(x)$, $A(\theta)$ are known functions.

**Theorem 3**
*Let $X \sim F_\theta$ such that $F_\theta$ is a distribution belonging to the exponential family. Then the Gini index obtained by plugging-in the maximum likelihood estimator of $\theta$, $G^{ML}(X_n)_\theta$, is asymptotically normal and efficient. Namely:*

$$\sqrt{n}\left(G^{ML}\left(X_n\right)_\theta - g_\theta\right) \xrightarrow{D} \mathcal{N}\left(0, g_\theta'^2 I^{-1}(\theta)\right), \tag{15.12}$$

*where $g_\theta' = \frac{dg_\theta}{d\theta}$ and $I(\theta)$ is the Fisher Information.*

$$\sqrt{n}\left(G^{ML}\left(X_n\right)_\theta - g_\theta\right) \xrightarrow{D} \mathcal{N}\left(0, g_\theta'^2 I^{-1}(\theta)\right),$$

*Proof.* The result follows easily from the asymptotic efficiency of the maximum likelihood estimators of the exponential family, and the invariance principle of MLE. In particular, the validity of the invariance principle for the Gini index is granted by the continuity and the monotonicity of $g_\theta$ with respect to $\theta$. The asymptotic variance is then obtained by application of the delta-method [250]. □

## 15.4 A PARETIAN ILLUSTRATION

We provide an illustration of the obtained results using some artificial fat-tailed data. We choose a Pareto I [217], with density

$$f(x) = \alpha c^\alpha x^{-\alpha-1}, x \geq c. \tag{15.13}$$

It is easy to verify that the corresponding survival function $\bar{F}(x)$ belongs to the regularly-varying class with tail parameter $\alpha$ and slowly-varying function $L(x) = c^{\alpha}$. We can therefore apply the results of Section 15.2 to obtain the following corollaries.

**Corollary 15.1**
*Let $X_1, ..., X_n$ be a sequence of i.i.d. observations with Pareto distribution with tail parameter $\alpha \in (1,2)$. The nonparametric Gini estimator is characterized by the following limit:*

$$D_n^{NP} = G^{NP}(X_n) - g \sim S\left(\alpha, 1, \frac{C_\alpha^{-\frac{1}{\alpha}}}{n^{\frac{\alpha-1}{\alpha}}} \frac{(\alpha-1)}{\alpha}, 0\right). \tag{15.14}$$

*Proof.* Without loss of generality we can assume $c = 1$ in Equation (15.13). The results is a mere application of Theorem 2, remembering that a Pareto distribution is in the domain of attraction of $\alpha$-stable random variables with slowly-varying function $L(x) = 1$. The sequence $c_n$ to satisfy Equation (15.37) becomes $c_n = n^{\frac{1}{\alpha}} C_\alpha^{-\frac{1}{\alpha}}$, therefore we have $L_0(n) = C_\alpha^{-\frac{1}{\alpha}}$, which is independent of $n$. Additionally the mean of the distribution is also a function of $\alpha$, that is $\mu = \frac{\alpha}{\alpha-1}$. □

**Corollary 15.2**
*Let the sample $X_1, ..., X_n$ be distributed as in Corollary 15.1, let $G_\theta^{ML}$ be the maximum likelihood estimator for the Gini index as defined in Theorem 3. Then the MLE Gini estimator, rescaled by its true mean $g$, has the following limit:*

$$D_n^{ML} = G_\alpha^{ML}(X_n) - g \sim N\left(0, \frac{4\alpha^2}{n(2\alpha-1)^4}\right), \tag{15.15}$$

*where $N$ indicates a Gaussian.*

*Proof.* The functional form of the maximum likelihood estimator for the Gini index is known to be $G_\theta^{ML} = \frac{1}{2\alpha^{ML}-1}$ [172]. The result then follows from the fact that the Pareto distribution (with known minimum value $x_m$) belongs to an exponential family and therefore satisfies the regularity conditions necessary for the asymptotic normality and efficiency of the maximum likelihood estimator. Also notice that the Fisher information for a Pareto distribution is $\frac{1}{\alpha^2}$. □

Now that we have worked out both asymptotic distributions, we can compare the quality of the convergence for both the MLE and the nonparametric case when dealing with Paretian data, which we use as the prototype for the more general class of fat-tailed observations.

In particular, we can approximate the distribution of the deviations of the estimator from the true value $g$ of the Gini index for finite sample sizes, by using Equations (15.14) and (15.15).

Figure 15.3 shows how the deviations around the mean of the two different types of estimators are distributed and how these distributions change as the number of observations increases. In particular, to facilitate the comparison between the maximum likelihood and the nonparametric estimators, we fixed the number of observation in the MLE case, while letting them vary in the nonparametric one. We perform this study for different types of tail indices to show how large the impact is on the consistency of the estimator. It is worth noticing that, as the tail index decreases towards 1 (the threshold value for a infinite mean), the mode of the distribution of the nonparametric estimator moves farther away from the mean of the distribution (centered on 0 by definition, given that we are dealing with deviations from the mean). This effect is responsible for the small sample bias observed in applications. Such a phenomenon is not present in the MLE case, thanks to the the normality of the limit for every value of the tail parameter.

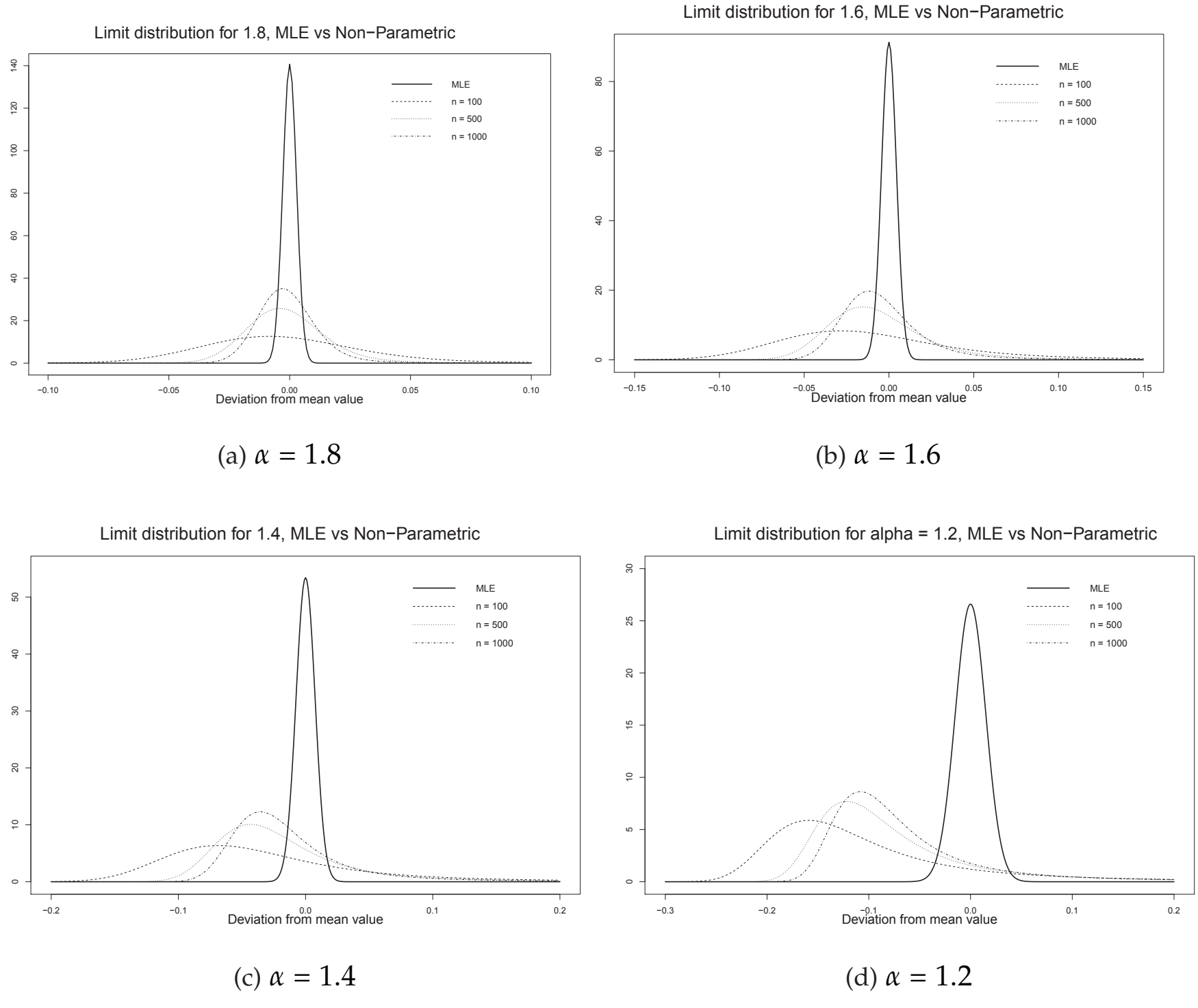

(a) $\alpha = 1.8$ (b) $\alpha = 1.6$

(c) $\alpha = 1.4$ (d) $\alpha = 1.2$

Figure 15.3: *Comparisons between the maximum likelihood and the nonparametric asymptotic distributions for different values of the tail index* $\alpha$*. The number of observations for MLE is fixed to* $n = 100$*. Note that, even if all distributions have mean zero, the mode of the distributions of the nonparametric estimator is different from zero, because of the skewness.*

We can make our argument more rigorous by assessing the number of observations $\tilde{n}$ needed for the nonparametric estimator to be as good as the MLE one, under different tail scenarios. Let's consider the likelihood-ratio-type function

$$r(c, n) = \frac{P_S(|D_n^{NP}| > c)}{P_N(|D_{100}^{ML}| > c)}, \tag{15.16}$$

where $P_S(|D_n^{NP}| > c)$ and $P_N(|D_{100}^{ML}| > c)$ are the probabilities ($\alpha$-stable and Gaussian respectively) of the centered estimators in the nonparametric, and in the MLE cases, of exceeding the thresholds $\pm c$, as per Equations (15.15) and (15.14). In the nonparametric case the number of observations $n$ is allowed to change, while in the MLE case it is fixed to 100. We then look for the value $\tilde{n}$ such that $r(c, \tilde{n}) = 1$ for fixed $c$.

Table 15.2 displays the results for different thresholds $c$ and tail parameters $\alpha$. In particular, we can see how the MLE estimator outperforms the nonparametric one, which requires a much larger number of observations to obtain the same tail probability of the MLE with $n$ fixed to 100. For example, we need at least $80 \times 10^6$ observations for the nonparametric estimator to obtain the same probability of exceeding the $\pm 0.02$ threshold of the MLE one, when $\alpha = 1.2$.

Table 15.2: *The number of observations $\tilde{n}$ needed for the nonparametric estimator to match the tail probabilities, for different threshold values $c$ and different values of the tail index $\alpha$, of the maximum likelihood estimator with fixed $n = 100$.*

| | *Threshold c as per Equation (15.16):* | | | |
|---|---|---|---|---|
| $\alpha$ | 0.005 | 0.01 | 0.015 | 0.02 |
| 1.8 | $27 \times 10^3$ | $12 \times 10^5$ | $12 \times 10^6$ | $63 \times 10^5$ |
| 1.5 | $21 \times 10^4$ | $21 \times 10^4$ | $46 \times 10^5$ | $81 \times 10^7$ |
| 1.2 | $33 \times 10^8$ | $67 \times 10^7$ | $20 \times 10^7$ | $80 \times 10^6$ |

Interestingly, the number of observations needed to match the tail probabilities in Equation (15.16) does not vary uniformly with the threshold. This is expected, since as the threshold goes to infinity or to zero, the tail probabilities remain the same for every value of $n$. Therefore, given the unimodality of the limit distributions, we expect that there will be a threshold maximizing the number of observations needed to match the tail probabilities, while for all the other levels the number of observations will be smaller.

We conclude that, when in presence of fat-tailed data with infinite variance, a plug-in MLE based estimator should be preferred over the nonparametric one.

## 15.5 SMALL SAMPLE CORRECTION

Theorem 2 can be also used to provide a correction for the bias of the nonparametric estimator for small sample sizes. The key idea is to recognize that, for unimodal distributions, most observations come from around the mode. In symmetric distributions the mode and the mean coincide, thus most observations will be close to the mean value as well, not so for skewed distributions: for right-skewed continuous unimodal distributions the mode is lower than the mean. Therefore, given that the asymptotic distribution of the nonparametric Gini index is right-skewed, we expect that the observed value of the Gini index will be usually lower than the true one (placed at the mean level). We can quantify this difference (i.e.

the bias) by looking at the distance between the mode and the mean, and once this distance is known, we can correct our Gini estimate by adding it back[4].

Formally, we aim to derive a corrected nonparametric estimator $G^C(X_n)$ such that

$$G^C(X_n) = G^{NP}(X_n) + ||m(G^{NP}(X_n)) - \mathbb{E}(G^{NP}(X_n))||, \quad (15.17)$$

where $||m(G^{NP}(X_n)) - \mathbb{E}(G^{NP}(X_n))||$ is the distance between the mode $m$ and the mean of the distribution of the nonparametric Gini estimator $G^{NP}(X_n)$.

Performing the type of correction described in Equation (15.17) is equivalent to shifting the distribution of $G^{NP}(X_n)$ in order to place its mode on the true value of the Gini index.

Ideally, we would like to measure this mode-mean distance

$$||m(G^{NP}(X_n)) - \mathbb{E}(G^{NP}(X_n))||$$

on the exact distribution of the Gini index to get the most accurate correction. However, the finite distribution is not always easily derivable as it requires assumptions on the parametric structure of the data generating process (which, in most cases, is unknown for fat-tailed data [172]). We therefore propose to use the limiting distribution for the nonparametric Gini obtained in Section 15.2 to approximate the finite sample distribution, and to estimate the mode-mean distance with it. This procedure allows for more freedom in the modeling assumptions and potentially decreases the number of parameters to be estimated, given that the limiting distribution only depends on the tail index and the mean of the data, which can be usually assumed to be a function of the tail index itself, as in the Paretian case where $\mu = \frac{\alpha}{\alpha-1}$.

By exploiting the location-scale property of $\alpha$-stable distributions and Equation (15.11), we approximate the distribution of $G^{NP}(X_n)$ for finite samples by

$$G^{NP}(X_n) \sim S\left(\alpha, 1, \gamma(n), g\right), \quad (15.18)$$

where $\gamma(n) = \frac{1}{n^{\frac{\alpha-1}{\alpha}}} \frac{L_0(n)}{\mu}$ is the scale parameter of the limiting distribution.

As a consequence, thanks to the linearity of the mode for $\alpha$-stable distributions, we have

$$||m(G^{NP}(X_n)) - \mathbb{E}(G^{NP}(X_n))|| \approx ||m(\alpha, \gamma(n)) + g - g|| = ||m(\alpha, \gamma(n))||,$$

where $m(\alpha, \gamma(n))$ is the mode function of an $\alpha$-stable distribution with zero mean.

The implication is that, in order to obtain the correction term, knowledge of the true Gini index is not necessary, given that $m(\alpha, \gamma(n))$ does not depend on $g$. We then estimate the correction term as

$$\hat{m}(\alpha, \gamma(n)) = \arg\max_x s(x), \quad (15.19)$$

where $s(x)$ is the numerical density of the associated $\alpha$-stable distribution in Equation (15.18), but centered on 0. This comes from the fact that, for $\alpha$-stable distri-butions, the mode is not available in closed form, but it can be easily computed numerically [213], using the unimodality of the law.

---

4 Another idea, which we have tested in writing the paper, is to use the distance between the median and the mean; the performances are comparable.

The corrected nonparametric estimator is thus

$$G^C(X_n) = G^{NP}(X_n) + \hat{m}(\alpha, \gamma(n)), \tag{15.20}$$

whose asymptotic distribution is

$$G^C(X_n) \sim S\left(\alpha, 1, \gamma(n), g + \hat{m}(\alpha, \gamma(n))\right). \tag{15.21}$$

Note that the correction term $\hat{m}\left(\alpha, \gamma(n)\right)$ is a function of the tail index $\alpha$ and is connected to the sample size $n$ by the scale parameter $\gamma(n)$ of the associated limiting distribution. It is important to point out that $\hat{m}(\alpha, \gamma(n))$ is decreasing in $n$, and that $\lim_{n\to\infty} \hat{m}(\alpha, \gamma(n)) \to 0$. This happens because, as $n$ increases, the distribution described in Equation (15.18) becomes more and more centered around its mean value, shrinking to zero the distance between the mode and the mean. This ensures the asymptotic equivalence of the corrected estimator and the nonparametric one. Just observe that

$$\begin{aligned}\lim_{n\to\infty} |G(X_n)^C - G^{NP}(X_n)| &= \lim_{n\to\infty} |G^{NP}(X_n) + \hat{m}(\alpha, \gamma(n)) - G^{NP}(X_n)| \\ &= \lim_{n\to\infty} |\hat{m}(\alpha, \gamma(n))| \to 0.\end{aligned}$$

Naturally, thanks to the correction, $G^C(X_n)$ will always behave better in small samples. Consider also that, from Equation (15.21), the distribution of the corrected estimator has now for mean $g + \hat{m}(\alpha, \gamma(n))$, which converges to the true Gini $g$ as $n \to \infty$.

From a theoretical point of view, the quality of this correction depends on the distance between the exact distribution of $G^{NP}(X_n)$ and its $\alpha$-stable limit; the closer the two are to each other, the better the approximation. However, given that, in most cases, the exact distribution of $G^{NP}(X_n)$ is unknown, it is not possible to give more details.

From what we have written so far, it is clear that the correction term depends on the tail index of the data, and possibly also on their mean. These parameters, if not assumed to be known a priori, must be estimated. Therefore the additional uncertainty due to the estimation will reflect also on the quality of the correction.

We conclude this Section with the discussion of the effect of the correction procedure with a simple example. In a Monte Carlo experiment, we simulate 1000 Paretian samples of increasing size, from $n = 10$ to $n = 2000$, and for each sample size we compute both the original nonparametric estimator $G^{NP}(X_n)$ and the corrected $G^C(X_n)$. We repeat the experiment for different $\alpha$'s. Figure 15.4 presents the results.

It is clear that the corrected estimators always perform better than the uncorrected ones in terms of absolute deviation from the true Gini value. In particular, our numerical experiment shows that for small sample sizes with $n \leq 1000$ the gain is quite remarkable for all the different values of $\alpha \in (1, 2)$. However, as expected, the difference between the estimators decreases with the sample size, as the correction term decreases both in $n$ and in the tail index $\alpha$. Notice that, when the tail index equals 2, we obtain the symmetric Gaussian distribution and the two estimators coincide, given that, thanks to the finiteness of the variance, the nonparametric estimator is no longer biased.

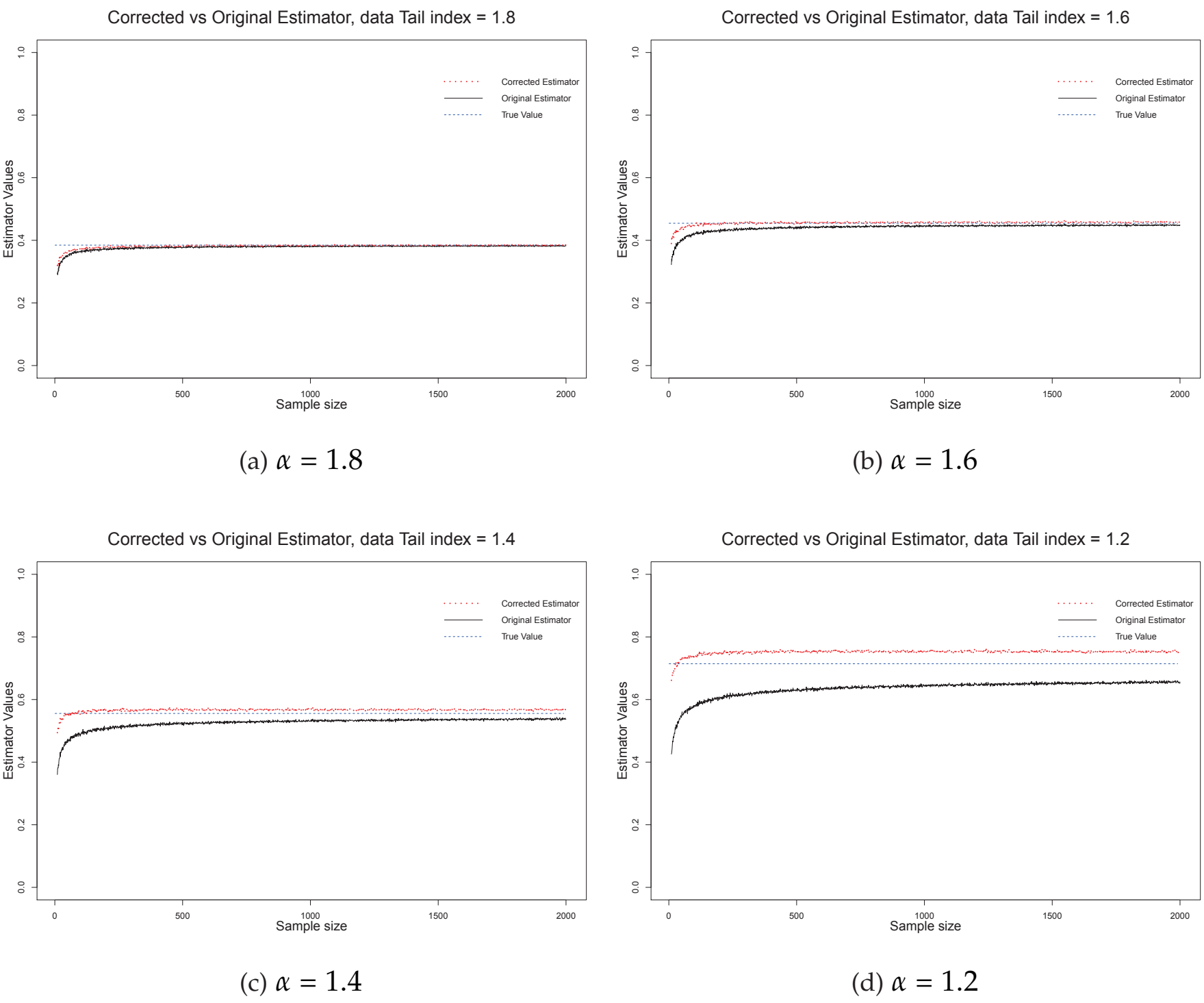

(a) $\alpha = 1.8$ (b) $\alpha = 1.6$

(c) $\alpha = 1.4$ (d) $\alpha = 1.2$

Figure 15.4: *Comparisons between the corrected nonparametric estimator (in red, the one on top) and the usual nonparametric estimator (in black, the one below). For small sample sizes the corrected one clearly improves the quality of the estimation.*

## 15.6 CONCLUSIONS

In this chapter we address the issue of the asymptotic behavior of the nonparametric estimator of the Gini index in presence of a distribution with infinite variance, an issue that has been curiously ignored by the literature. The central mistake in the nonparametric methods largely used is to believe that asymptotic consistency translates into equivalent pre-asymptotic properties.

We show that a parametric approach provides better asymptotic results thanks to the properties of maximum likelihood estimation. Hence we strongly suggest that, if the collected data are suspected to be fat-tailed, parametric methods should be preferred.

In situations where a fully parametric approach cannot be used, we propose a simple correction mechanism for the nonparametric estimator based on the distance between the mode and the mean of its asymptotic distribution. Even if the correction works nicely, we suggest caution in its use owing to additional uncertainty from the estimation of the correction term.

## TECHNICAL APPENDIX

**Proof of Lemma 15.1**

Let $U = F(X)$ be the standard uniformly distributed integral probability transform of the random variable $X$. For the order statistics, we then have [67]: $X_{(i)} \stackrel{a.s.}{=} F^{-1}(U_{(i)})$. Hence

$$R_n = \frac{1}{n}\sum_{i=1}^{n}(i/n - U_{(i)})F^{-1}(U_{(i)}). \tag{15.22}$$

Now by definition of empirical c.d.f it follows that

$$R_n = \frac{1}{n}\sum_{i=1}^{n}(F_n(U_{(i)}) - U_{(i)})F^{-1}(U_{(i)}), \tag{15.23}$$

where $F_n(u) = \frac{1}{n}\sum_{i=1}^{n} 1_{U_i \leq u}$ is the empirical c.d.f of uniformly distributed random variables.

To show that $R_n \xrightarrow{L^1} 0$, we are going to impose an upper bound that goes to zero. First we notice that

$$\mathbb{E}|R_n| \leq \frac{1}{n}\sum_{i=1}^{n}\mathbb{E}|(F_n(U_{(i)}) - U_{(i)})F^{-1}(U_{(i)})|. \tag{15.24}$$

To build a bound for the right-hand side (r.h.s) of (15.24), we can exploit the fact that, while $F^{-1}(U_{(i)})$ might be just $L^1$-integrable, $F_n(U_{(i)}) - U_{(i)}$ is $L^\infty$ integrable, therefore we can use Hölder's inequality with $q = \infty$ and $p = 1$. It follows that

$$\frac{1}{n}\sum_{i=1}^{n}\mathbb{E}|(F_n(U_{(i)}) - U_{(i)})F^{-1}(U_{(i)})| \leq \frac{1}{n}\sum_{i=1}^{n}\mathbb{E}\sup_{U_{(i)}}|(F_n(U_{(i)}) - U_{(i)})|\mathbb{E}|F^{-1}(U_{(i)})|. \tag{15.25}$$

Then, thanks to the Cauchy-Schwarz inequality, we get

$$\begin{aligned} &\frac{1}{n}\sum_{i=1}^{n}\mathbb{E}\sup_{U_{(i)}}|(F_n(U_{(i)}) - U_{(i)})|\mathbb{E}|F^{-1}(U_{(i)})| \\ &\leq \left(\frac{1}{n}\sum_{i=1}^{n}(\mathbb{E}\sup_{U_{(i)}}|(F_n(U_{(i)}) - U_{(i)})|)^2 \frac{1}{n}\sum_{i=1}^{n}(\mathbb{E}(F^{-1}(U_{(i)})))^2\right)^{\frac{1}{2}}. \end{aligned} \tag{15.26}$$

Now, first recall that $\sum_{i=1}^{n} F^{-1}(U_{(i)}) \overset{a.s.}{=} \sum_{i=1}^{n} F^{-1}(U_i)$ with $U_i$, $i = 1, ..., n$, being an i.i.d sequence, then notice that $\mathbb{E}(F^{-1}(U_i)) = \mu$, so that the second term of Equation (15.26) becomes

$$\mu \left( \frac{1}{n} \sum_{i=1}^{n} (\mathbb{E} \sup_{U_{(i)}} |(F_n(U_{(i)}) - U_{(i)})|)^2 \right)^{\frac{1}{2}}. \tag{15.27}$$

The final step is to show that Equation (15.27) goes to zero as $n \to \infty$.

We know that $F_n$ is the empirical c.d.f of uniform random variables. Using the triangular inequality the inner term of Equation (15.27) can be bounded as

$$\begin{aligned} &\frac{1}{n} \sum_{i=1}^{n} (\mathbb{E} \sup_{U_{(i)}} |(F_n(U_{(i)}) - U_{(i)})|)^2 \\ &\leq \frac{1}{n} \sum_{i=1}^{n} (\mathbb{E} \sup_{U_{(i)}} |(F_n(U_{(i)}) - F(U_{(i)}))|)^2 + \frac{1}{n} \sum_{i=1}^{n} (\mathbb{E} \sup_{U_{(i)}} |(F(U_{(i)}) - U_{(i)})|)^2. \end{aligned} \tag{15.28}$$

Since we are dealing with uniforms, we known that $F(U) = u$, and the second term in the r.h.s of (15.28) vanishes.

We can then bound $\mathbb{E}(\sup_{U_{(i)}} |(F_n(U_{(i)}) - F(U_{(i)})|)$ using the Vapnik-Chervonenkis (VC) inequality, a uniform bound for empirical processes [33, 66, 312], getting

$$\mathbb{E} \sup_{U_{(i)}} |(F_n(U_{(i)}) - F(U_{(i)})| \leq \sqrt{\frac{\log(n+1) + \log(2)}{n}}. \tag{15.29}$$

Combining Equation (15.29) with Equation (15.27) we obtain

$$\mu \left( \frac{1}{n} \sum_{i=1}^{n} (\mathbb{E} \sup_{U_{(i)}} |(F_n(U_{(i)}) - U_{(i)})|)^2 \right)^{\frac{1}{2}} \leq \mu \sqrt{\frac{\log(n+1) + \log(2)}{n}}, \tag{15.30}$$

which goes to zero as $n \to \infty$, thus proving the first claim.

For the second claim, it is sufficient to observe that the r.h.s of (15.30) still goes to zero when multiplied by $\frac{n^{\frac{\alpha-1}{\alpha}}}{L_0(n)}$ if $\alpha \in (1, 2)$.

**Proof of Theorem 1**

The first part of the proof consists in showing that we can rewrite Equation (15.10) as a function of i.i.d random variables in place of order statistics, to be able to apply a Central Limit Theorem (CLT) argument.

Let's start by considering the sequence

$$\frac{1}{n} \sum_{i=1}^{n} Z_{(i)} = \frac{1}{n} \sum_{i=1}^{n} \left( 2 \frac{i-1}{n-1} - 1 \right) F^{-1}(U_{(i)}). \tag{15.31}$$

Using the integral probability transform $X \stackrel{d}{=} F^{-1}(U)$ with $U$ standard uniform, and adding and removing $\frac{1}{n}\sum_{i=1}^{n}\left(2U_{(i)}-1\right)F^{-1}(U_{(i)})$, the r.h.s. in Equation (15.31) can be rewritten as

$$\frac{1}{n}\sum_{i=1}^{n} Z_{(i)} = \frac{1}{n}\sum_{i=1}^{n}(2U_{(i)}-1)F^{-1}(U_{(i)}) + \frac{1}{n}\sum_{i=1}^{n} 2\left(\frac{i-1}{n-1}-U_{(i)}\right)F^{-1}(U_{(i)}). \quad (15.32)$$

Then, by using the properties of order statistics [67] we obtain the following almost sure equivalence

$$\frac{1}{n}\sum_{i=1}^{n} Z_{(i)} \stackrel{a.s.}{=} \frac{1}{n}\sum_{i=1}^{n}(2U_{i}-1)F^{-1}(U_{i}) + \frac{1}{n}\sum_{i=1}^{n} 2\left(\frac{i-1}{n-1}-U_{(i)}\right)F^{-1}(U_{(i)}). \quad (15.33)$$

Note that the first term in the r.h.s of (15.33) is a function of i.i.d random variables as desired, while the second term is just a reminder, therefore

$$\frac{1}{n}\sum_{i=1}^{n} Z_{(i)} \stackrel{a.s.}{=} \frac{1}{n}\sum_{i=1}^{n} Z_i + R_n,$$

with $Z_i = (2U_i - 1)F^{-1}(U_i)$ and $R_n = \frac{1}{n}\sum_{i=1}^{n}(2(\frac{i-1}{n-1} - U_{(i)}))F^{-1}(U_{(i)})$.

Given Equation (15.10) and exploiting the decomposition given in (15.33) we can rewrite our claim as

$$\frac{n^{\frac{\alpha-1}{\alpha}}}{L_0(n)}\left(\frac{1}{n}\sum_{i=1}^{n} Z_{(i)} - \theta\right) = \frac{n^{\frac{\alpha-1}{\alpha}}}{L_0(n)}\left(\frac{1}{n}\sum_{i=1}^{n} Z_i - \theta\right) + \frac{n^{\frac{\alpha-1}{\alpha}}}{L_0(n)} R_n. \quad (15.34)$$

From the second claim of the Lemma 15.1 and Slutsky Theorem, the convergence in Equation (15.10) can be proven by looking at the behavior of the sequence

$$\frac{n^{\frac{\alpha-1}{\alpha}}}{L_0(n)}\left(\frac{1}{n}\sum_{i=1}^{n} Z_i - \theta\right), \quad (15.35)$$

where $Z_i = (2U_i - 1)F^{-1}(U_i) = (2F(X_i) - 1)X_i$. This reduces to proving that $Z_i$ is in the fat tails domain of attraction.

Recall that by assumption $X \in DA(S_\alpha)$ with $\alpha \in (1,2)$. This assumption enables us to use a particular type of CLT argument for the convergence of the sum of fat-tailed random variables. However, we first need to prove that $Z \in DA(S_\alpha)$ as well, that is $P(|Z| > z) \sim L(z)z^{-\alpha}$, with $\alpha \in (1,2)$ and $L(z)$ slowly-varying.

Notice that

$$P(|\tilde{Z}| > z) \leq P(|Z| > z) \leq P(2X > z),$$

where $\tilde{Z} = (2U - 1)X$ and $U \perp X$. The first bound holds because of the positive dependence between $X$ and $F(X)$ and it can be proven rigorously by noting that $2UX \leq 2F(X)X$ by the so-called re-arrangement inequality [143]. The upper bound conversely is trivial.

Using the properties of slowly-varying functions, we have $P(2X > z) \sim 2^\alpha L(z)z^{-\alpha}$. To show that $\tilde{Z} \in DA(S_\alpha)$, we use the Breiman's Theorem, which ensure the sta-bility of the $\alpha$-stable class under product, as long as the second random variable is not too fat-tailed [322].

To apply the Theorem we re-write $P(|\tilde{Z}| > z)$ as

$$P(|\tilde{Z}| > z) = P(\tilde{Z} > z) + P(-\tilde{Z} > z) = P(\tilde{U}X > z) + P(-\tilde{U}X > z),$$

where $\tilde{U}$ is a standard uniform with $\tilde{U} \perp X$.

We focus on $P(\tilde{U}X > z)$ since the procedure is the same for $P(-\tilde{U}X > z)$. We have

$$P(\tilde{U}X > z) = P(\tilde{U}X > z|\tilde{U} > 0)P(\tilde{U} > 0) + P(\tilde{U}X > z|\tilde{U} \leq 0)P(\tilde{U} \leq 0),$$

for $z \to +\infty$.

Now, we have that $P(\tilde{U}X > z|\tilde{U} \leq 0) \to 0$, while, by applying Breiman's Theorem, $P(\tilde{U}X > z|\tilde{U} > 0)$ becomes

$$P(\tilde{U}X > z|\tilde{U} > 0) \to E(\tilde{U}^{\alpha}|U > 0)P(X > z)P(U > 0).$$

Therefore

$$P(|\tilde{Z}| > z) \to \frac{1}{2}E(\tilde{U}^{\alpha}|U > 0)P(X > z) + \frac{1}{2}E((-\tilde{U})^{\alpha}|U \leq 0)P(X > z).$$

From this

$$\begin{aligned} P(|\tilde{Z}| > z) &\to \frac{1}{2}P(X > z)[E(\tilde{U})^{\alpha}|U > 0) + E((-\tilde{U}^{\alpha}|U \leq 0)] \\ &= \frac{2^{\alpha}}{1-\alpha}P(X > z) \sim \frac{2^{\alpha}}{1-\alpha}L(z)z^{-\alpha}. \end{aligned}$$

We can then conclude that, by the squeezing Theorem [107],

$$P(|Z| > z) \sim L(z)z^{-\alpha},$$

as $z \to \infty$. Therefore $Z \in DA(S_{\alpha})$.

We are now ready to invoke the Generalized Central Limit Theorem (GCLT) [97] for the sequence $Z_i$, i.e.

$$nc_n^{-1}\left(\frac{1}{n}\sum_{i=1}^{n} Z_i - \mathbb{E}(Z_i)\right) \xrightarrow{d} S_{\alpha,\beta}. \tag{15.36}$$

with $\mathbb{E}(Z_i) = \theta$, $S_{\alpha,\beta}$ a standardized $\alpha$-stable random variable, and where $c_n$ is a sequence which must satisfy

$$\lim_{n\to\infty} \frac{nL(c_n)}{c_n^{\alpha}} = \frac{\Gamma(2-\alpha)|\cos(\frac{\pi\alpha}{2})|}{\alpha - 1} = C_{\alpha}. \tag{15.37}$$

Notice that $c_n$ can be represented as $c_n = n^{\frac{1}{\alpha}}L_0(n)$, where $L_0(n)$ is another slowly-varying function possibly different from $L(n)$.

The skewness parameter $\beta$ is such that

$$\frac{P(Z > z)}{P(|Z| > z)} \to \frac{1+\beta}{2}.$$

Recalling that, by construction, $Z \in [-c, +\infty)$, the above expression reduces to

$$\frac{P(Z > z)}{P(Z > z) + P(-Z > z)} \to \frac{P(Z > z)}{P(Z > z)} = 1 \to \frac{1+\beta}{2}, \tag{15.38}$$

therefore $\beta = 1$. This, combined with Equation (15.34), the result for the reminder $R_n$ of Lemma 15.1 and Slutsky Theorem, allows us to conclude that the same weak limits holds for the ordered sequence of $Z_{(i)}$ in Equation (15.10) as well.

**Proof of Theorem 2**

The first step of the proof is to show that the ordered sequence $\frac{\sum_{i=1}^n Z_{(i)}}{\sum_{i=1}^n X_i}$, characterizing the Gini index, is equivalent in distribution to the i.i.d sequence $\frac{\sum_{i=1}^n Z_i}{\sum_{i=1}^n X_i}$. In order to prove this, it is sufficient to apply the factorization in Equation (15.33) to Equation (15.11), getting

$$\frac{n^{\frac{\alpha-1}{\alpha}}}{L_0(n)} \left( \frac{\sum_{i=1}^n Z_i}{\sum_{i=1}^n X_i} - \frac{\theta}{\mu} \right) + \frac{n^{\frac{\alpha-1}{\alpha}}}{L_0(n)} R_n \frac{n}{\sum_{i=1}^n X_i}. \tag{15.39}$$

By Lemma 15.1 and the application of the continuous mapping and Slutsky Theorems, the second term in Equation (15.39) goes to zero at least in probability. Therefore to prove the claim it is sufficient to derive a weak limit for the following sequence

$$n^{\frac{\alpha-1}{\alpha}} \frac{1}{L_0(n)} \left( \frac{\sum_{i=1}^n Z_i}{\sum_{i=1}^n X_i} - \frac{\theta}{\mu} \right). \tag{15.40}$$

Expanding Equation (15.40) and recalling that $Z_i = (2F(X_i) - 1)X_i$, we get

$$\frac{n^{\frac{\alpha-1}{\alpha}}}{L_0(n)} \frac{n}{\sum_{i=1}^n X_i} \left( \frac{1}{n} \sum_{i=1}^n X_i \left( 2F(X_i) - 1 - \frac{\theta}{\mu} \right) \right). \tag{15.41}$$

The term $\frac{n}{\sum_{i=1}^n X_i}$ in Equation (15.41) converges in probability to $\frac{1}{\mu}$ by an application of the continuous mapping Theorem, and the fact that we are dealing with positive random variables $X$. Hence it will contribute to the final limit via Slutsky Theorem.

We first start by focusing on the study of the limit law of the term

$$\frac{n^{\frac{\alpha-1}{\alpha}}}{L_0(n)} \frac{1}{n} \sum_{i=1}^n X_i \left( 2F(X_i) - 1 - \frac{\theta}{\mu} \right). \tag{15.42}$$

Set $\hat{Z}_i = X_i(2F(X_i) - 1 - \frac{\theta}{\mu})$ and note that $\mathbb{E}(\hat{Z}_i) = 0$, since $\mathbb{E}(Z_i) = \theta$ and $\mathbb{E}(X_i) = \mu$.

In order to apply a GCLT argument to characterize the limit distribution of the sequence $\frac{n^{\frac{\alpha-1}{\alpha}}}{L_0(n)}\frac{1}{n}\sum_{i=1}^{n}\hat{Z}_i$ we need to prove that $\hat{Z} \in DA(S_\alpha)$. If so then we can apply GCLT to

$$\frac{n^{\frac{\alpha-1}{\alpha}}}{L_0(n)}\left(\frac{\sum_{i=1}^{n}\hat{Z}_i}{n} - \mathbb{E}(\hat{Z}_i)\right). \tag{15.43}$$

Note that, since $\mathbb{E}(\hat{Z}_i) = 0$, Equation (15.43) equals Equation (15.42).

To prove that $\hat{Z} \in DA(S_\alpha)$, remember that $\hat{Z}_i = X_i(2F(X_i) - 1 - \frac{\theta}{\mu})$ is just $Z_i = X_i(2F(X_i) - 1)$ shifted by $\frac{\theta}{\mu}$. Therefore the same argument used in Theorem 1 for $Z$ applies here to show that $\hat{Z} \in DA(S_\alpha)$. In particular we can point out that $\hat{Z}$ and $Z$ (therefore also $X$) share the same $\alpha$ and slowly-varying function $L(n)$.

Notice that by assumption $X \in [c, \infty)$ with $c > 0$ and we are dealing with continuous distributions, therefore $\hat{Z} \in [-c(1 + \frac{\theta}{\mu}), \infty)$. As a consequence the left tail of $\hat{Z}$ does not contribute to changing the limit skewness parameter $\beta$, which remains equal to 1 (as for $Z$) by an application of Equation (15.38).

Therefore, by applying the GCLT we finally get

$$n^{\frac{\alpha-1}{\alpha}}\frac{1}{L_0(n)}\left(\frac{\sum_{i=1}^{n}Z_i}{\sum_{i=1}^{n}X_i} - \frac{\theta}{\mu}\right) \xrightarrow{d} \frac{1}{\mu}S(\alpha, 1, 1, 0). \tag{15.44}$$

We conclude the proof by noting that, as proven in Equation (15.39), the weak limit of the Gini index is characterized by the i.i.d sequence of $\frac{\sum_{i=1}^{n}Z_i}{\sum_{i=1}^{n}X_i}$ rather than the ordered one, and that an $\alpha$-stable random variable is closed under scaling by a constant [246].

# 16 ON THE SUPER-ADDITIVITY AND ESTIMATION BIASES OF QUANTILE CONTRIBUTIONS ‡

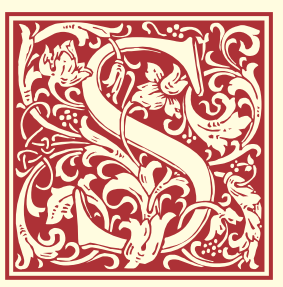

SAMPLE MEASURES[a] of top centile contributions to the total (concentration) are downward biased, unstable estimators, extremely sensitive to sample size and concave in accounting for large deviations. It makes them particularly unfit in domains with Power Law tails, especially for low values of the exponent. These estimators can vary over time and increase with the population size, as shown in this article, thus providing the illusion of structural changes in concentration. They are also inconsistent under aggregation and mixing distributions, as the weighted average of concentration measures for $A$ and $B$ will tend to be lower than that from $A \cup B$. In addition, it can be shown that under such thick tails, increases in the total sum need to be accompanied by increased sample size of the concentration measurement. We examine the estimation superadditivity and bias under homogeneous and mixed distributions.

[a] With R. Douady

## 16.1 INTRODUCTION

Vilfredo Pareto noticed that 80% of the land in Italy belonged to 20% of the population, and vice-versa, thus both giving birth to the power law class of distributions and the popular saying 80/20. The self-similarity at the core of the property of power laws [191] and [192] allows us to recurse and reapply the 80/20 to the remaining 20%, and so forth until one obtains the result that the top percent of the population will own about 53% of the total wealth.

It looks like such a measure of concentration can be seriously biased, depending on how it is measured, so it is very likely that the true ratio of concentration of

Research chapter.

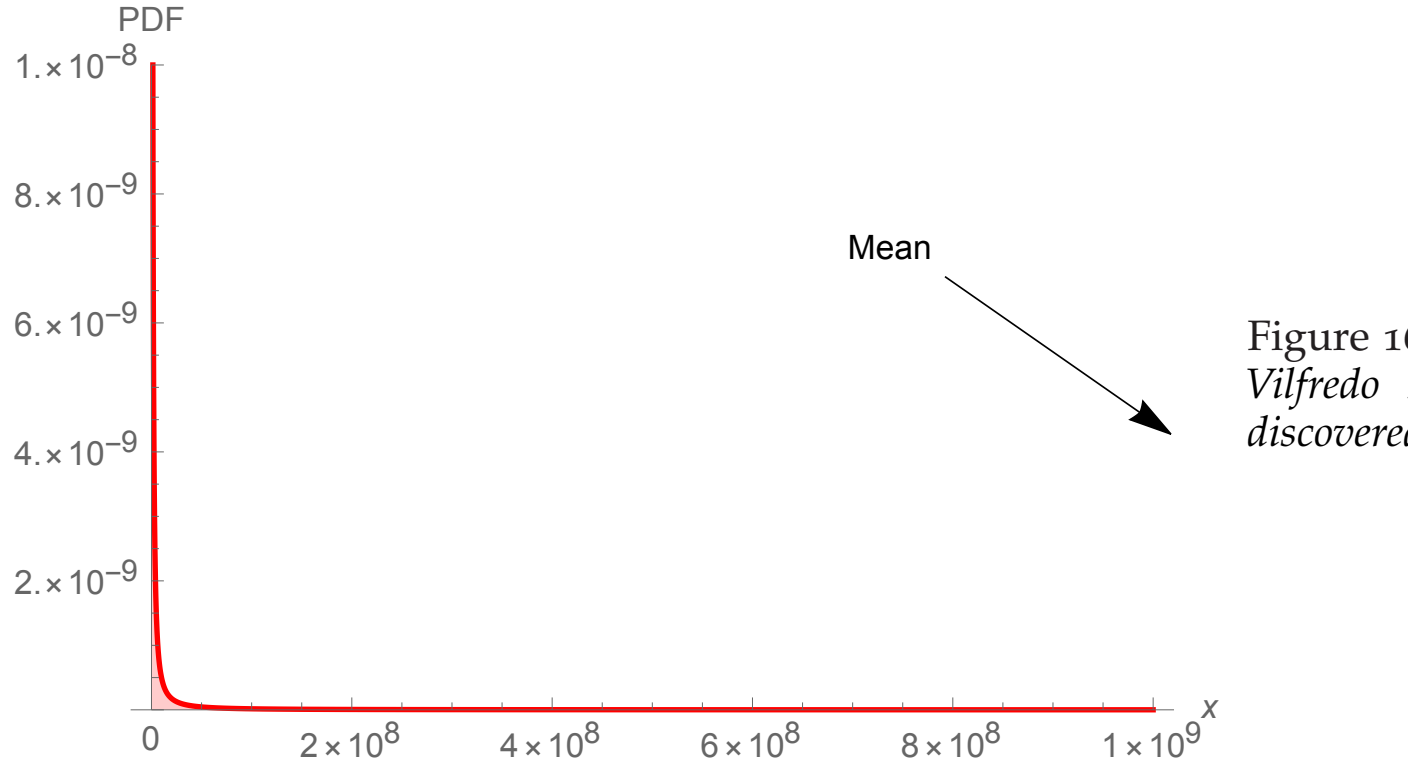

Figure 16.1: *The young Vilfredo Pareto, before he discovered power laws.*

what Pareto observed, that is, the share of the top percentile, was closer to 70%, hence changes year-on-year would drift higher to converge to such a level from larger sample. In fact, as we will show in this discussion, for, say wealth, more complete samples resulting from technological progress, and also larger population and economic growth will make such a measure converge by increasing over time, for no other reason than expansion in sample space or aggregate value.

The core of the problem is that, for the class one-tailed fat-tailed random variables, that is, bounded on the left and unbounded on the right, where the random variable $X \in [x_{\min}, \infty)$, the in-sample quantile contribution is a biased estimator of the true value of the actual quantile contribution.

Let us define the *quantile contribution*

$$\kappa_q = q \frac{\mathbb{E}[X|X > h(q)]}{\mathbb{E}[X]}$$

where $h(q) = \inf\{h \in [x_{min}, +\infty), \mathbb{P}(X > h) \leq q\}$ is the exceedance threshold for the probability $q$.

For a given sample $(X_k)_{1 \leq k \leq n}$, its "natural" estimator $\widehat{\kappa}_q \equiv \frac{q^{th}\text{percentile}}{total}$, used in most academic studies, can be expressed, as

$$\widehat{\kappa}_q \equiv \frac{\sum_{i=1}^n \mathbb{1}_{X_i > \hat{h}(q)} X_i}{\sum_{i=1}^n X_i}$$

where $\hat{h}(q)$ is the estimated exceedance threshold for the probability $q$ :

$$\hat{h}(q) = \inf\{h : \frac{1}{n} \sum_{i=1}^n \mathbb{1}_{x > h} \leq q\}$$

We shall see that the observed variable $\widehat{\kappa}_q$ is a downward biased estimator of the true ratio $\kappa_q$, the one that would hold out of sample, and such bias is in proportion to the fatness of tails and, for very thick tailed distributions, remains significant, even for very large samples.

## 16.2 ESTIMATION FOR UNMIXED PARETO-TAILED DISTRIBUTIONS

Let $X$ be a random variable belonging to the class of distributions with a "power law" right tail, that is:

$$\mathbb{P}(X > x) = L(x)\, x^{-\alpha} \tag{16.1}$$

where $L : [x_{\min}, +\infty) \to (0, +\infty)$ is a slowly varying function, defined as $\lim_{x\to+\infty} \frac{L(kx)}{L(x)} = 1$ for any $k > 0$.

There is little difference for small exceedance quantiles (<50%) between the various possible distributions such as Student's t, Lévy $\alpha$-stable, Dagum,[64],[65] Singh-Maddala distribution [252], or straight Pareto.

For exponents $1 \leq \alpha \leq 2$, as observed in [281] (Chapter 8 in this book), the law of large numbers operates, though *extremely* slowly. The problem is acute for $\alpha$ around, but strictly above 1 and severe, as it diverges, for $\alpha = 1$.

### 16.2.1 Bias and Convergence

**Simple Pareto Distribution** Let us first consider $\phi_\alpha(x)$ the density of a $\alpha$-Pareto distribution bounded from below by $x_{\min} > 0$, in other words:

$$\phi_\alpha(x) = \alpha x_{\min}^{\alpha} x^{-\alpha-1} \mathbb{1}_{x \geq x_{\min}},$$

kummer and $\mathbb{P}(X > x) = \left(\frac{x_{\min}}{x}\right)^{\alpha}$. Under these assumptions, the cutpoint of exceedance is $h(q) = x_{\min}\, q^{-1/\alpha}$ and we have:

$$\kappa_q = \frac{\int_{h(q)}^{\infty} x\, \phi(x) dx}{\int_{x_{min}}^{\infty} x\, \phi(x) dx} = \left(\frac{h(q)}{x_{\min}}\right)^{1-\alpha} = q^{\frac{\alpha-1}{\alpha}} \tag{16.2}$$

If the distribution of $X$ is $\alpha$-Pareto only beyond a cut-point $x_{\text{cut}}$, which we assume to be below $h(q)$, so that we have $\mathbb{P}(X > x) = \left(\frac{\lambda}{x}\right)^{\alpha}$ for some $\lambda > 0$, then we still have $h(q) = \lambda q^{-1/\alpha}$ and

$$\kappa_q = \frac{\alpha}{\alpha - 1} \frac{\lambda}{\mathbb{E}\,[X]} q^{\frac{\alpha-1}{\alpha}}$$

The estimation of $\kappa_q$ hence requires that of the exponent $\alpha$ as well as that of the scaling parameter $\lambda$, or at least its ratio to the expectation of $X$.

Table 16.1 shows the bias of $\widehat{\kappa}_q$ as an estimator of $\kappa_q$ in the case of an $\alpha$-Pareto distribution for $\alpha = 1.1$, a value chosen to be compatible with practical economic measures, such as the wealth distribution in the world or in a particular country, including developped ones.[2] In such a case, the estimator is extemely sensitive to *"small"* samples, *"small"* meaning in practice $10^8$. We ran up to a trillion simulations across varieties of sample sizes. While $\kappa_{0.01} \approx 0.657933$, even a sample size of 100 million remains severely biased as seen in the table.

Naturally the bias is rapidly (and nonlinearly) reduced for $\alpha$ further away from 1, and becomes weak in the neighborhood of 2 for a constant $\alpha$, though not under a mixture distribution for $\alpha$, as we shall se later. It is also weaker outside the top 1% centile, hence this discussion focuses on the famed "one percent" and on low values of the $\alpha$ exponent.

---

[2] This value, which is lower than the estimated exponents one can find in the literature – around 2 – is, following [102], a lower estimate which cannot be excluded from the observations.

Table 16.1: *Biases of Estimator of* $\kappa = 0.657933$ *From* $10^{12}$ *Monte Carlo Realizations*

| $\widehat{\kappa}(n)$ | Mean | Median | STD across MC runs |
|---|---|---|---|
| $\widehat{\kappa}(10^3)$ | 0.405235 | 0.367698 | 0.160244 |
| $\widehat{\kappa}(10^4)$ | 0.485916 | 0.458449 | 0.117917 |
| $\widehat{\kappa}(10^5)$ | 0.539028 | 0.516415 | 0.0931362 |
| $\widehat{\kappa}(10^6)$ | 0.581384 | 0.555997 | 0.0853593 |
| $\widehat{\kappa}(10^7)$ | 0.591506 | 0.575262 | 0.0601528 |
| $\widehat{\kappa}(10^8)$ | 0.606513 | 0.593667 | 0.0461397 |

In view of these results and of a number of tests we have performed around them, we can conjecture that the bias $\kappa_q - \widehat{\kappa}_q(n)$ is "of the order of" $c(\alpha, q)n^{-b(q)(\alpha-1)}$ where constants $b(q)$ and $c(\alpha, q)$ need to be evaluated. Simulations suggest that $b(q) = 1$, whatever the value of $\alpha$ and $q$, but the rather slow convergence of the estimator and of its standard deviation to 0 makes precise estimation difficult.

**General Case** In the general case, let us fix the threshold $h$ and define:

$$\kappa_h = \mathbb{P}(X > h)\frac{\mathbb{E}[X|X > h]}{\mathbb{E}[X]} = \frac{\mathbb{E}[X\mathbb{1}_{X>h}]}{\mathbb{E}[X]}$$

so that we have $\kappa_q = \kappa_{h(q)}$. We also define the $n$-sample estimator:

$$\widehat{\kappa}_h \equiv \frac{\sum_{i=1}^{n} \mathbb{1}_{X_i>h} X_i}{\sum_{i=1}^{n} X_i}$$

where $X_i$ are $n$ independent copies of $X$. The intuition behind the estimation bias of $\kappa_q$ by $\widehat{\kappa}_q$ lies in a difference of concavity of the concentration measure with respect to an innovation (a new sample value), whether it falls below or above the threshold. Let $A_h(n) = \sum_{i=1}^{n} \mathbb{1}_{X_i>h} X_i$ and $S(n) = \sum_{i=1}^{n} X_i$, so that $\widehat{\kappa}_h(n) = \dfrac{A_h(n)}{S(n)}$ and assume a frozen threshold $h$. If a new sample value $X_{n+1} < h$ then the new value is $\widehat{\kappa}_h(n+1) = \dfrac{A_h(n)}{S(n) + X_{n+1}}$. The value is convex in $X_{n+1}$ so that uncertainty on $X_{n+1}$ increases its expectation. At variance, if the new sample value $X_{n+1} > h$, the new value $\widehat{\kappa}_h(n+1) \approx \frac{A_h(n)+X_{n+1}-h}{S(n)+X_{n+1}-h} = 1 - \frac{S(n)-A_h(n)}{S(n)+X_{n+1}-h}$, which is now concave in $X_{n+1}$, so that uncertainty on $X_{n+1}$ reduces its value. The competition between these two opposite effects is in favor of the latter, because of a higher concavity with respect to the variable, and also of a higher variability (whatever its measurement) of the variable conditionally to being above the threshold than to being below. The fatter the right tail of the distribution, the stronger the effect. Overall, we find that $\mathbb{E}\left[\widehat{\kappa}_h(n)\right] \leq \dfrac{\mathbb{E}\left[A_h(n)\right]}{\mathbb{E}\left[S(n)\right]} = \kappa_h$ (note that unfreezing the threshold $\hat{h}(q)$ also tends to reduce the concentration measure estimate, adding to the effect, when introducing one extra sample because of a slight increase in the expected value of the estimator $\hat{h}(q)$, although this effect is rather negligible). We have in fact the following:

**Proposition 16.1**

*Let $\boldsymbol{X} = (X)_{i=1}^n$ a random sample of size $n > \frac{1}{q}$, $Y = X_{n+1}$ an extra single random observation, and define: $\widehat{\kappa}_h(\boldsymbol{X} \sqcup Y) = \dfrac{\sum_{i=1}^n \mathbb{1}_{X_i > h} X_i + \mathbb{1}_{Y > h} Y}{\sum_{i=1}^n X_i + Y}$. We remark that, whenever $Y > h$, one has:*

$$\frac{\partial^2 \widehat{\kappa}_h(\boldsymbol{X} \sqcup Y)}{\partial Y^2} \leq 0.$$

*This inequality is still valid with $\widehat{\kappa}_q$ as the value $\hat{h}(q, \boldsymbol{X} \sqcup Y)$ doesn't depend on the particular value of $Y > \hat{h}(q, \boldsymbol{X})$.*

We face a different situation from the common small sample effect resulting from high impact from the rare observation in the tails that are less likely to show up in small samples, a bias which goes away by repetition of sample runs. The concavity of the estimator constitutes a upper bound for the measurement in finite $n$, clipping large deviations, which leads to problems of aggregation as we will state below in Theorem 1.

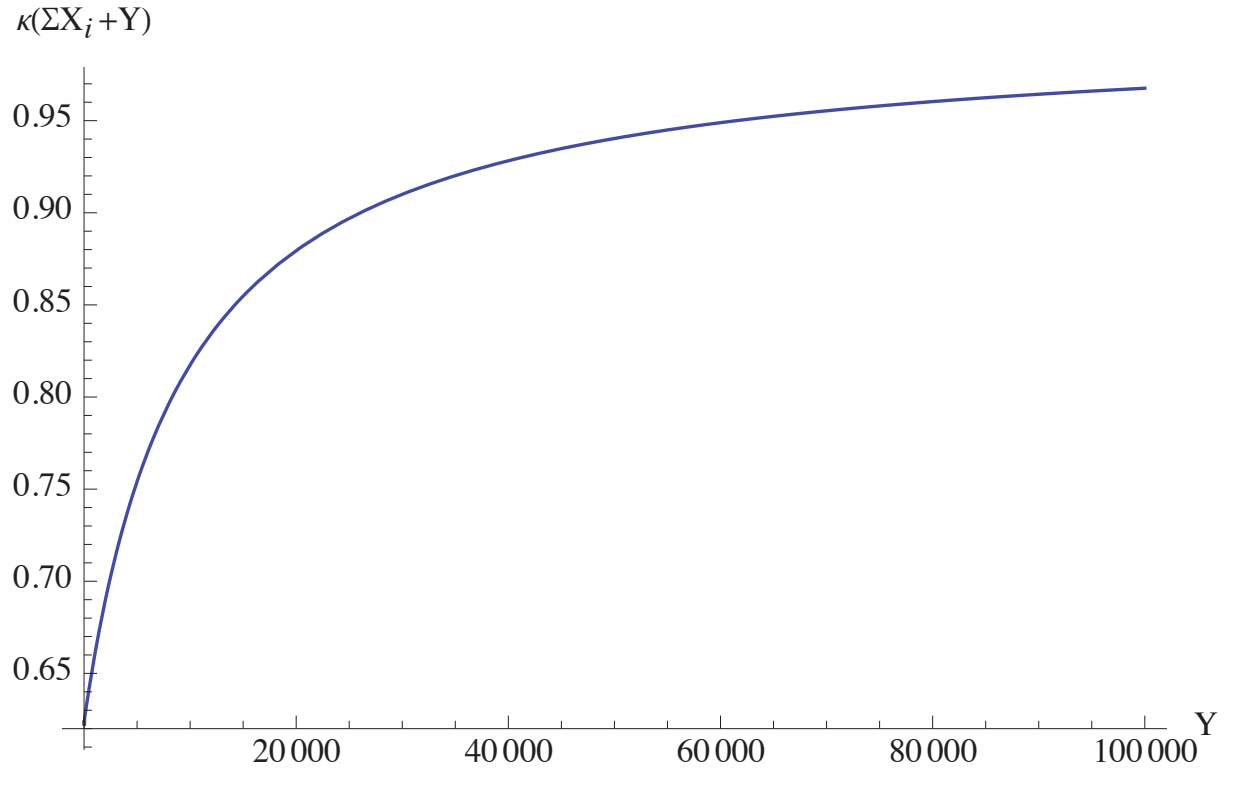

Figure 16.2: *Effect of additional observations on $\kappa$*

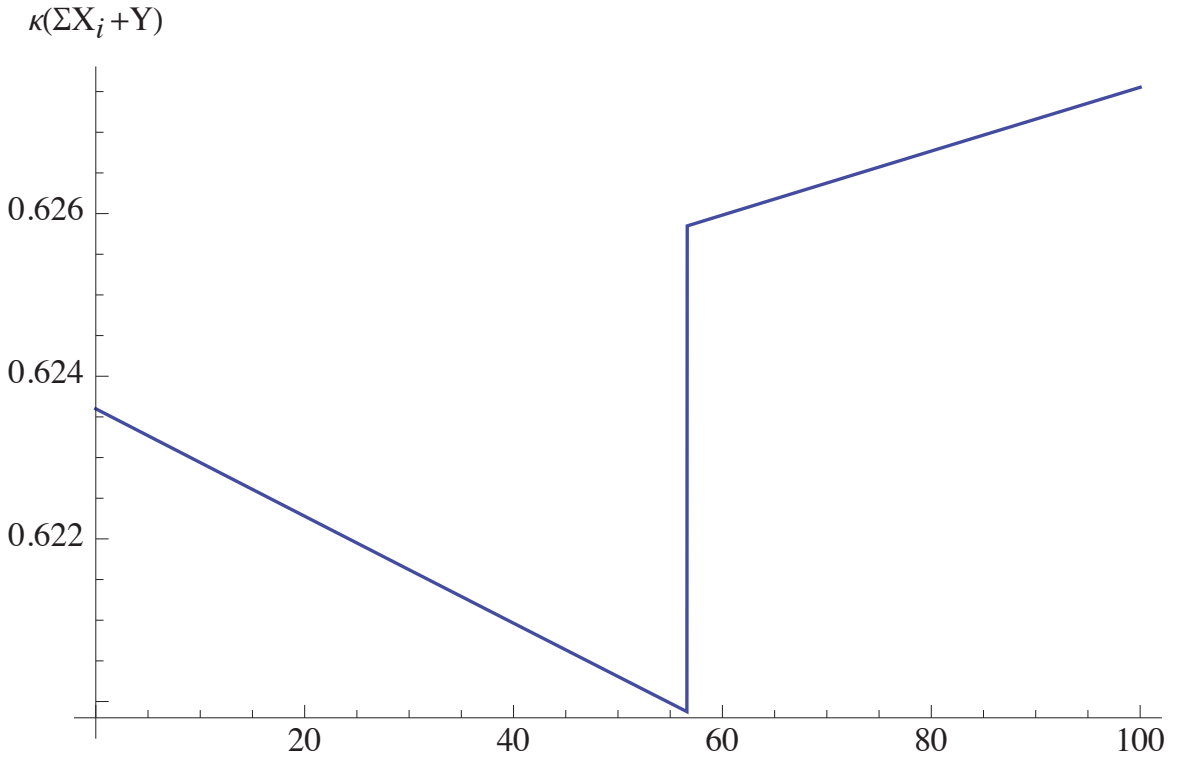

Figure 16.3: *Effect of additional observations on $\kappa$, we can see convexity on both sides of $h$ except for values of no effect to the left of $h$, an area of order $1/n$*

In practice, even in very large sample, the contribution of very large rare events to $\kappa_q$ slows down the convergence of the sample estimator to the true value. For a better, unbiased estimate, one would need to use a different path: first estimating the distribution parameters $\left(\hat{\alpha}, \hat{\lambda}\right)$ and only then, estimating the theoretical tail contribution $\kappa_q(\hat{\alpha}, \hat{\lambda})$. Falk [102] observes that, even with a proper estimator of $\alpha$ and $\lambda$, the convergence is extremely slow, namely of the order of $n^{-\delta}/\ln n$, where the exponent $\delta$ depends on $\alpha$ and on the tolerance of the actual distribution vs. a theoretical Pareto, measured by the Hellinger distance. In particular, $\delta \to 0$ as $\alpha \to 1$, making the convergence really slow for low values of $\alpha$.

## 16.3 AN INEQUALITY ABOUT AGGREGATING INEQUALITY

For the estimation of the mean of a fat-tailed r.v. $(X)_i^j$, in $m$ sub-samples of size $n_i$ each for a total of $n = \sum_{i=1}^m n_i$, the allocation of the total number of observations $n$ between $i$ and $j$ does not matter so long as the total $n$ is unchanged. Here the allocation of $n$ samples between $m$ sub-samples does matter because of the concavity of $\kappa$.[3] Next we prove that global concentration as measured by $\widehat{\kappa}_q$ on a broad set of data will appear higher than local concentration, so aggregating European data, for instance, would give a $\widehat{\kappa}_q$ higher than the average measure of concentration across countries – an *"inequality about inequality"*. In other words, we claim that the estimation bias when using $\widehat{\kappa}_q(n)$ is even increased when dividing the sample into sub-samples and taking the weighted average of the measured values $\widehat{\kappa}_q(n_i)$.

**Theorem 4**

*Partition the $n$ data into $m$ sub-samples $N = N_1 \cup \ldots \cup N_m$ of respective sizes $n_1, \ldots, n_m$, with $\sum_{i=1}^m n_i = n$, and let $S_1, \ldots, S_m$ be the sum of variables over each sub-sample, and $S = \sum_{i=1}^m S_i$ be that over the whole sample. Then we have:*

$$\mathbb{E}\left[\widehat{\kappa}_q(N)\right] \geq \sum_{i=1}^m \mathbb{E}\left[\frac{S_i}{S}\right] \mathbb{E}\left[\widehat{\kappa}_q(N_i)\right]$$

*If we further assume that the distribution of variables $X_j$ is the same in all the sub-samples. Then we have:*

$$\mathbb{E}\left[\widehat{\kappa}_q(N)\right] \geq \sum_{i=1}^m \frac{n_i}{n} \mathbb{E}\left[\widehat{\kappa}_q(N_i)\right]$$

In other words, averaging concentration measures of sub-samples, weighted by the total sum of each subsample, produces a downward biased estimate of the concentration measure of the full sample.

*Proof.* An elementary induction reduces the question to the case of two sub-samples. Let $q \in (0,1)$ and $(X_1, \ldots, X_m)$ and $\left(X_1', \ldots, X_n'\right)$ be two samples of positive i.i.d. random variables, the $X_i$'s having distributions $p(dx)$ and the $X_j'$'s having distribution $p'(dx')$. For simplicity, we assume that both $qm$ and $qn$ are integers. We set $S = \sum_{i=1}^{m} X_i$ and $S' = \sum_{i=1}^{n} X_i'$. We define $A = \sum_{i=1}^{mq} X_{[i]}$ where $X_{[i]}$ is the $i$-th largest value of $(X_1, \ldots, X_m)$, and $A' = \sum_{i=1}^{mq} X_{[i]}'$ where $X_{[i]}'$ is the $i$-th largest value of $\left(X_1', \ldots, X_n'\right)$. We also set $S'' = S + S'$ and $A'' = \sum_{i=1}^{(m+n)q} X_{[i]}''$ where $X_{[i]}''$ is the $i$-th largest value of the joint sample $(X_1, \ldots, X_m, X_1', \ldots, X_n')$.

[3] The same concavity – and general bias – applies when the distribution is lognormal, and is exacerbated by high variance.

The $q$-concentration measure for the samples $\boldsymbol{X} = (X_1, ..., X_m)$, $\boldsymbol{X}' = (X_1', ..., X_n')$ and $\boldsymbol{X}'' = (X_1, \ldots, X_m, X_1', \ldots, X_n')$ are:

$$\kappa = \frac{A}{S} \qquad \kappa' = \frac{A'}{S'} \qquad \kappa'' = \frac{A''}{S''}$$

We must prove that the following inequality holds for expected concentration measures:

$$\mathbb{E}\left[\kappa''\right] \geq \mathbb{E}\left[\frac{S}{S''}\right] \mathbb{E}\left[\kappa\right] + \mathbb{E}\left[\frac{S'}{S''}\right] \mathbb{E}\left[\kappa'\right]$$

We observe that:

$$A = \max_{\substack{J \subset \{1,...,m\} \\ |J| = \theta m}} \sum_{i \in J} X_i$$

and, similarly

$$A' = \max_{J' \subset \{1,...,n\}, |J'| = qn} \sum_{i \in J'} X_i'$$

and

$$A'' = \max_{J'' \subset \{1,...,m+n\}, |J''| = q(m+n)} \sum_{i \in J''} X_i,$$

where we have denoted $X_{m+i} = X_i'$ for $i = 1 \ldots n$. If $J \subset \{1, ..., m\}$, $|J| = \theta m$ and $J' \subset \{m+1, ..., m+n\}$, $|J'| = qn$, then $J'' = J \cup J'$ has cardinal $m+n$, hence $A + A' = \sum_{i \in J''} X_i \leq A''$, whatever the particular sample. Therefore $\kappa'' \geq \frac{S}{S''}\kappa + \frac{S'}{S''}\kappa'$ and we have:

$$\mathbb{E}\left[\kappa''\right] \geq \mathbb{E}\left[\frac{S}{S''}\kappa\right] + \mathbb{E}\left[\frac{S'}{S''}\kappa'\right]$$

Let us now show that:

$$\mathbb{E}\left[\frac{S}{S''}\kappa\right] = \mathbb{E}\left[\frac{A}{S''}\right] \geq \mathbb{E}\left[\frac{S}{S''}\right] \mathbb{E}\left[\frac{A}{S}\right]$$

If this is the case, then we identically get for $\kappa'$ :

$$\mathbb{E}\left[\frac{S'}{S''}\kappa'\right] = \mathbb{E}\left[\frac{A'}{S''}\right] \geq \mathbb{E}\left[\frac{S'}{S''}\right] \mathbb{E}\left[\frac{A'}{S'}\right]$$

hence we will have:

$$\mathbb{E}\left[\kappa''\right] \geq \mathbb{E}\left[\frac{S}{S''}\right] \mathbb{E}\left[\kappa\right] + \mathbb{E}\left[\frac{S'}{S''}\right] \mathbb{E}\left[\kappa'\right]$$

Let $T = X_{[mq]}$ be the cut-off point (where $[mq]$ is the integer part of $mq$), so that $A = \sum_{i=1}^{m} X_i \mathbb{1}_{X_i \geq T}$ and let $B = S - A = \sum_{i=1}^{m} X_i \mathbb{1}_{X_i < T}$. Conditionally to $T$, $A$ and $B$ are independent: $A$ is a sum if $m\theta$ samples constarined to being above $T$, while $B$ is the sum of $m(1-\theta)$ independent samples constrained to being below $T$. They are also independent of $S'$. Let $p_A(t, da)$ and $p_B(t, db)$ be the distribution of $A$ and $B$ respectively, given $T = t$. We recall that $p'(ds')$ is the distribution of $S'$ and denote $q(dt)$ that of $T$. We have:

$$\mathbb{E}\left[\frac{S}{S''}\kappa\right] = \iint \frac{a+b}{a+b+s'} \frac{a}{a+b} p_A(t, da)\, p_B(t, db)\, q(dt)\, p'(ds')$$

For given $b$, $t$ and $s'$, $a \to \frac{a+b}{a+b+s'}$ and $a \to \frac{a}{a+b}$ are two increasing functions of the same variable $a$, hence conditionally to $T$, $B$ and $S'$, we have:

$$\begin{aligned}\mathbb{E}\left[\frac{S}{S''}\kappa \,\middle|\, T, B, S'\right] &= \mathbb{E}\left[\frac{A}{A+B+S'} \,\middle|\, T, B, S'\right] \\ &\geq \mathbb{E}\left[\frac{A+B}{A+B+S'} \,\middle|\, T, B, S'\right] \mathbb{E}\left[\frac{A}{A+B} \,\middle|\, T, B, S'\right]\end{aligned}$$

This inequality being valid for any values of $T$, $B$ and $S'$, it is valid for the unconditional expectation, and we have:

$$\mathbb{E}\left[\frac{S}{S''}\kappa\right] \geq \mathbb{E}\left[\frac{S}{S''}\right] \mathbb{E}\left[\frac{A}{S}\right]$$

If the two samples have the same distribution, then we have:

$$\mathbb{E}\left[\kappa''\right] \geq \frac{m}{m+n}\mathbb{E}\left[\kappa\right] + \frac{n}{m+n}\mathbb{E}\left[\kappa'\right]$$

Indeed, in this case, we observe that $\mathbb{E}\left[\frac{S}{S''}\right] = \frac{m}{m+n}$. Indeed $S = \sum_{i=1}^{m} X_i$ and the $X_i$ are identically distributed, hence $\mathbb{E}\left[\frac{S}{S''}\right] = m\mathbb{E}\left[\frac{X}{S''}\right]$. But we also have $\mathbb{E}\left[\frac{S''}{S''}\right] = 1 = (m+n)\mathbb{E}\left[\frac{X}{S''}\right]$ therefore $\mathbb{E}\left[\frac{X}{S''}\right] = \frac{1}{m+n}$. Similarly, $\mathbb{E}\left[\frac{S'}{S''}\right] = \frac{n}{m+n}$, yielding the result.

This ends the proof of the theorem. □

Let $X$ be a positive random variable and $h \in (0,1)$. We remind the theoretical $h$-concentration measure, defined as:

$$\kappa_h = \frac{\mathbb{P}(X > h)\mathbb{E}\left[X \,|X > h\right]}{\mathbb{E}\left[X\right]}$$

whereas the $n$-sample $\theta$-concentration measure is $\widehat{\kappa}_h(n) = \frac{A(n)}{S(n)}$, where $A(n)$ and $S(n)$ are defined as above for an $n$-sample $\boldsymbol{X} = (X_1, \ldots, X_n)$ of i.i.d. variables with the same distribution as $X$.

**Theorem 5**

*For any $n \in \mathbb{N}$, we have:*

$$\mathbb{E}\left[\widehat{\kappa}_h(n)\right] < \kappa_h$$

*and*

$$\lim_{n \to +\infty} \widehat{\kappa}_h(n) = \kappa_h \quad \textit{a.s. and in probability}$$

*Proof.* The above corrolary shows that the sequence $n\mathbb{E}\left[\widehat{\kappa}_h(n)\right]$ is super-additive, hence $\mathbb{E}\left[\widehat{\kappa}_h(n)\right]$ is an increasing sequence. Moreover, thanks to the law of large numbers, $\frac{1}{n}S(n)$ converges almost surely and in probability to $\mathbb{E}\left[X\right]$ and $\frac{1}{n}A(n)$ converges almost surely and in probability to $\mathbb{E}\left[X\mathbb{1}_{X>h}\right] = \mathbb{P}(X > h)\mathbb{E}\left[X \,|X > h\right]$, hence their ratio also converges almost surely to $\kappa_h$. On the other hand, this ratio is bounded by 1. Lebesgue dominated convergence theorem concludes the argument about the convergence in probability. □

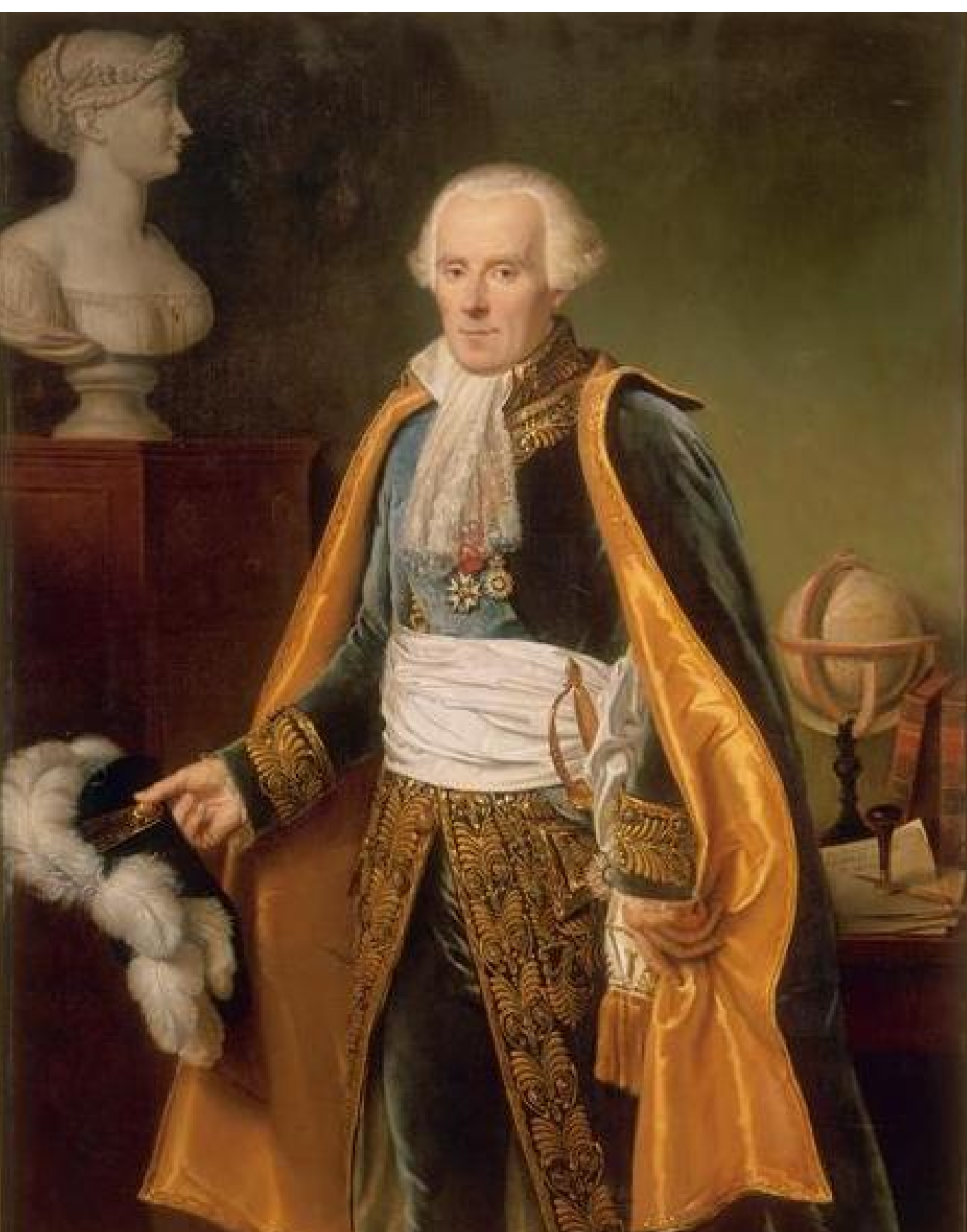

Figure 16.4: *Pierre Simon, Marquis de Laplace. He got his name on a distribution and a few results, but was behind both the Cauchy and the Gaussian distributions (see the Stigler law of eponymy [258]). Posthumous portrait by Jean-Baptiste Paulin Guérin, 1838.*

## 16.4 MIXED DISTRIBUTIONS FOR THE TAIL EXPONENT

Consider now a random variable $X$, the distribution of which $p(dx)$ is a mixture of parametric distributions with different values of the parameter: $p(dx) = \sum_{i=1}^{m} \omega_i p_{\alpha_i}(dx)$. A typical $n$-sample of $X$ can be made of $n_i = \omega_i n$ samples of $X_{\alpha_i}$ with distribution $p_{\alpha_i}$. The above theorem shows that, in this case, we have:

$$\mathbb{E}\left[\widehat{\kappa}_q(n, X)\right] \geq \sum_{i=1}^{m} \mathbb{E}\left[\frac{S(\omega_i n, X_{\alpha_i})}{S(n, X)}\right] \mathbb{E}\left[\widehat{\kappa}_q(\omega_i n, X_{\alpha_i})\right]$$

When $n \to +\infty$, each ratio $\dfrac{S(\omega_i n, X_{\alpha_i})}{S(n, X)}$ converges almost surely to $\omega_i$ respectively, therefore we have the following convexity inequality:

$$\kappa_q(X) \geq \sum_{i=1}^{m} \omega_i \kappa_q(X_{\alpha_i})$$

The case of Pareto distribution is particularly interesting. Here, the parameter $\alpha$ represents the tail exponent of the distribution. If we normalize expectations to 1, the cdf of $X_\alpha$ is $F_\alpha(x) = 1 - \left(\frac{x}{x_{\min}}\right)^{-\alpha}$ and we have:

$$\kappa_q(X_\alpha) = q^{\frac{\alpha-1}{\alpha}}$$

and

$$\frac{d^2}{d\alpha^2} \kappa_q(X_\alpha) = q^{\frac{\alpha-1}{\alpha}} \frac{(\log q)^2}{\alpha^3} > 0$$

Hence $\kappa_q(X_\alpha)$ is a convex function of $\alpha$ and we can write:

$$\kappa_q(X) \geq \sum_{i=1}^{m} \omega_i \kappa_q(X_{\alpha_i}) \geq \kappa_q(X_{\bar{\alpha}})$$

where $\bar{\alpha} = \sum_{i=1}^{m} \omega_i \alpha$.

Suppose now that $X$ is a positive random variable with unknown distribution, except that its tail decays like a power low with unknown exponent. An unbiased estimation of the exponent, with necessarily some amount of uncertainty (i.e., a distribution of possible true values around some average), would lead to a downward biased estimate of $\kappa_q$.

Because the concentration measure only depends on the tail of the distribution, this inequality also applies in the case of a mixture of distributions with a power decay, as in Equation 27.1:

$$\mathbb{P}(X > x) = \sum_{j=1}^{N} \omega_i L_i(x) x^{-\alpha_j} \tag{16.3}$$

The slightest uncertainty about the exponent increases the concentration index. One can get an actual estimate of this bias by considering an average $\bar{\alpha} > 1$ and two surrounding values $\alpha^+ = \alpha + \delta$ and $\alpha^- = \alpha - \delta$. The convexity inequaly writes as follows:

$$\kappa_q(\bar{\alpha}) = q^{1-\frac{1}{\bar{\alpha}}} < \frac{1}{2}\left(q^{1-\frac{1}{\alpha+\delta}} + q^{1-\frac{1}{\alpha-\delta}}\right)$$

So in practice, an estimated $\bar{\alpha}$ of around $3/2$, sometimes called the "half-cubic" exponent, would produce similar results as value of $\alpha$ much closer ro 1, as we used in the previous section. Simply $\kappa_q(\alpha)$ is convex, and dominated by the second order effect $\frac{\ln(q)q^{1-\frac{1}{\alpha+\delta}}(\ln(q)-2(\alpha+\delta))}{(\alpha+\delta)^4}$, an effect that is exacerbated at lower values of $\alpha$.

To show how unreliable the measures of inequality concentration from quantiles, consider that a standard error of 0.3 in the measurement of $\alpha$ causes $\kappa_q(\alpha)$ to rise by 0.25.

## 16.5 A LARGER TOTAL SUM IS ACCOMPANIED BY INCREASES IN $\widehat{\kappa}_q$

There is a large dependence between the estimator $\widehat{\kappa}_q$ and the sum $S = \sum_{j=1}^{n} X_j$ : conditional on an increase in $\widehat{\kappa}_q$ the expected sum is larger. Indeed, as shown in theorem 4, $\widehat{\kappa}_q$ and $S$ are positively correlated.

For the case in which the random variables under concern are wealth, we observe as in Figure 16.5 such conditional increase; in other words, since the distribution is of the class of thick tails under consideration, the maximum is of the same order as the sum, additional wealth means more measured inequality. Under such dynamics, is quite absurd to assume that additional wealth will arise from the bottom or even the middle. (The same argument can be applied to wars, pandemics, size or companies, etc.)

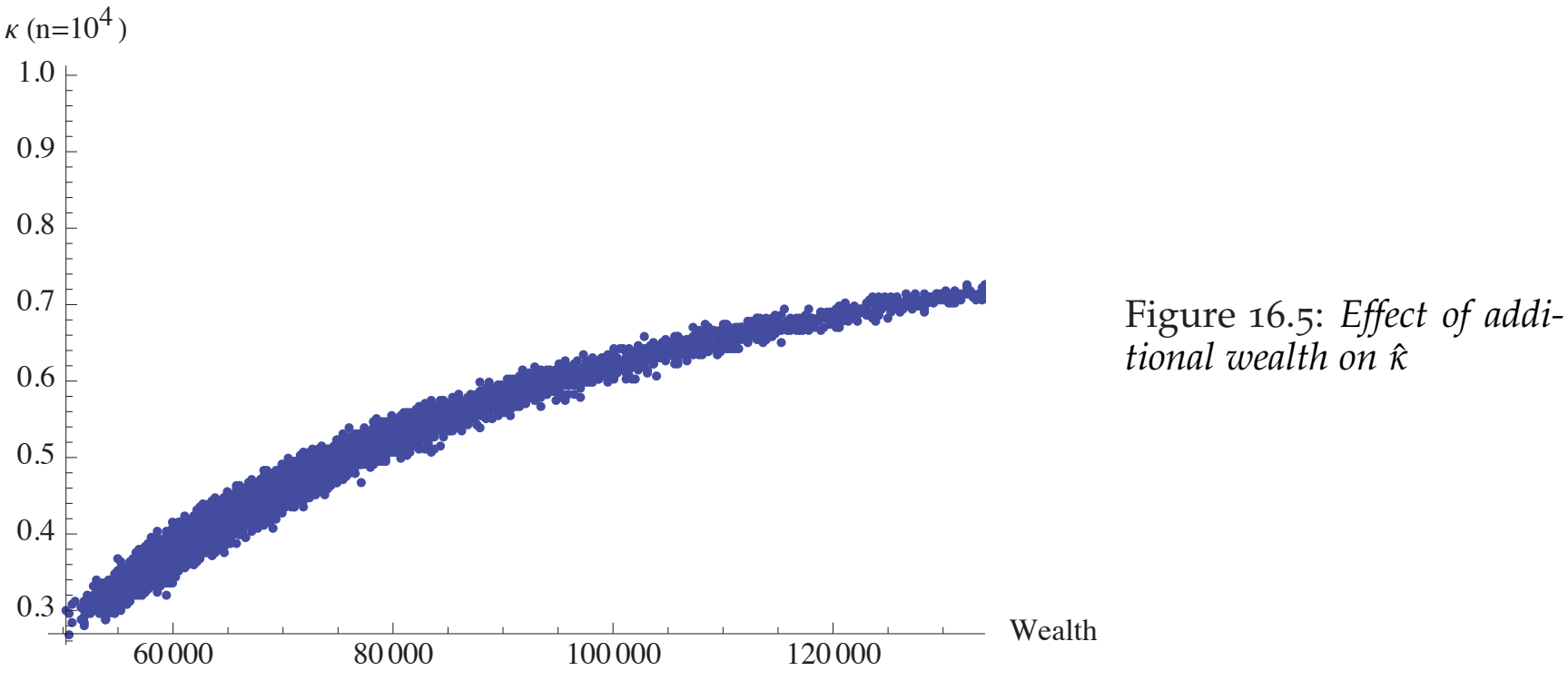

Figure 16.5: *Effect of additional wealth on $\hat{\kappa}$*

## 16.6 CONCLUSION AND PROPER ESTIMATION OF CONCENTRATION

Concentration can be high at the level of the generator, but in small units or subsections we will observe a lower $\kappa_q$. So examining times series, we can easily get a historical illusion of rise in, say, wealth concentration when it has been there all along at the level of the process; and an expansion in the size of the unit measured can be part of the explanation.[4]

Even the estimation of $\alpha$ can be biased in some domains where one does not see the entire picture: in the presence of uncertainty about the "true" $\alpha$, it can be shown that, unlike other parameters, the one to use is not the probability-weighted exponents (the standard average) but rather the minimum across a section of exponents.

One must not perform analyses of year-on-year changes in $\widehat{\kappa}_q$ without adjustment. It did not escape our attention that some theories are built based on claims of such "increase" in inequality, as in [224], without taking into account the true nature of $\kappa_q$, and promulgating theories about the "variation" of inequality without reference to the stochasticity of the estimation $-$ and the lack of consistency of $\kappa_q$ across time and sub-units. What is worse, rejection of such theories also ignored the size effect, by countering with data of a different sample size, effectively making the dialogue on inequality uninformational statistically.[5]

The mistake appears to be commonly made in common inference about fat-tailed data in the literature. The very methodology of using concentration and changes in concentration is highly questionable. For instance, in the thesis by Steven Pinker [227] that the world is becoming less violent, we note a fallacious inference about the concentration of damage from wars from a $\widehat{\kappa}_q$ with minutely small population in relation to the fat-tailedness.[6] Owing to the fat-tailedness of war casualties and consequences of violent conflicts, an adjustment would rapidly invalidate such claims that violence from war has statistically experienced a decline.

### 16.6.1 Robust methods and use of exhaustive data

We often face argument of the type "the method of measuring concentration from quantile contributions $\hat{\kappa}$ is robust and based on a complete set of data". Robust methods, alas, tend to fail with fat-tailed data, see Chapter 8. But, in addition, the problem here is worse: even if such "robust" methods were deemed unbiased, a method of direct centile estimation is still linked to a static and specific population and does not aggregage. Accordingly, such techniques do not allow us to make statistical claims or scientific statements about the true properties which should necessarily carry out of sample.

---

[4] Accumulated wealth is typically thicker tailed than income, see [117].

[5] *Financial Times*, May 23, 2014 "Piketty findings undercut by errors" by Chris Giles.

[6] Using Richardson's data, [227]: "(Wars) followed an 80:2 rule: almost eighty percent of the deaths were caused by *two percent* (his emph.) of the wars". So it appears that both Pinker and the literature cited for the quantitative properties of violent conflicts are using a flawed methodology, one that produces a severe bias, as the centile estimation has extremely large biases with fat-tailed wars. Furthermore claims about the mean become spurious at low exponents.

Take an insurance (or, better, reinsurance) company. The "accounting" profits in a year in which there were few claims do not reflect on the "economic" status of the company and it is futile to make statements on the concentration of losses per insured event based on a single year sample. The "accounting" profits are not used to predict variations year-on-year, rather the exposure to tail (and other) events, analyses that take into account the stochastic nature of the performance. This difference between "accounting" (deterministic) and "economic" (stochastic) values matters for policy making, particularly under thick tails. The same with wars: we do not estimate the severity of a (future) risk based on past in-sample historical data.

### 16.6.2 How Should We Measure Concentration?

Practitioners of risk managers now tend to compute CVaR and other metrics, methods that are extrapolative and nonconcave, such as the information from the $\alpha$ exponent, taking the one closer to the lower bound of the range of exponents, as we saw in our extension to Theorem 2 and rederiving the corresponding $\kappa$, or, more rigorously, integrating the functions of $\alpha$ across the various possible states. Such methods of adjustment are less biased and do not get mixed up with problems of aggregation –they are similar to the "stochastic volatility" methods in mathematical finance that consist in adjustments to option prices by adding a "smile" to the standard deviation, in proportion to the variability of the parameter representing volatility and the errors in its measurement. Here it would be "stochastic alpha" or "stochastic tail exponent "[7] By extrapolative, we mean the built-in extension of the tail in the measurement by taking into account realizations outside the sample path that are in excess of the extrema observed.[8] [9]

ACKNOWLEDGMENT

The late Benoît Mandelbrot, Branko Milanovic, Dominique Guéguan, Felix Salmon, Bruno Dupire, the late Marc Yor, Albert Shiryaev, an anonymous referee, the staff at Luciano Restaurant in Brooklyn and Naya in Manhattan.

7 Also note that, in addition to the centile estimation problem, some authors such as [225] when dealing with censored data, use Pareto interpolation for unsufficient information about the tails (based on tail parameter), filling-in the bracket with conditional average bracket contribution, which is not the same thing as using full power law extension; such a method retains a significant bias.

8 Even using a lognormal distribution, by fitting the scale parameter, works to some extent as a rise of the standard deviation extrapolates probability mass into the right tail.

9 We also note that the theorems would also apply to Poisson jumps, but we focus on the powerlaw case in the application, as the methods for fitting Poisson jumps are interpolative and have proved to be easier to fit in-sample than out of sample.

# Part V

# SHADOW MOMENTS PAPERS

# 17 SHADOW MOMENTS OF APPARENTLY INFINITE-MEAN PHENOMENA ‡

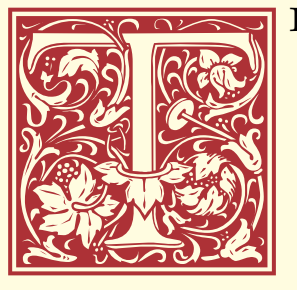

HIS CHAPTER proposes an approach to compute the conditional moments of fat-tailed phenomena that, only looking at data, could be mistakenly considered as having infinite mean. This type of problems manifests itself when a random variable $Y$ has a heavy-tailed distribution with an extremely wide yet bounded support.

We introduce the concept of dual distribution, by means of a log-transformation that smoothly removes the upper bound. The tail of the dual distribution can then be studied using extreme value theory, without making excessive parametric assumptions, and the estimates one obtains can be used to study the original distribution and compute its moments by reverting the transformation.

The central difference between our approach and a simple truncation is in the smoothness of the transformation between the original and the dual distribution, allowing use of extreme value theory.

War casualties, operational risk, environment blight, complex networks and many other econophysics phenomena are possible fields of application.

## 17.1 INTRODUCTION

Consider a heavy-tailed random variable $Y$ with finite support $[L, H]$. W.l.o.g. set $L >> 0$ for the lower bound, while for upper one $H$, assume that its value is remarkably large, yet finite. It is so large that the probability of observing values in its vicinity is extremely small, so that in data we tend to find observations only below a certain $M << H < \infty$.

Figure 17.1 gives a graphical representation of the problem. For our random variable $Y$ with remote upper bound $H$ the real tail is represented by the con-tinuous line. However, if we only observe values up to $M << H$, and - willing or not - we ignore the existence of $H$, which is unlikely to be seen, we could be inclined to believe the the tail is the dotted one, the apparent one. The two tails are indeed essentially indistinguishable for most cases, as the divergence is only evident when we approach $H$.

‡ Research chapter, with P. Cirillo.

Now assume we want to study the tail of $Y$ and, since it is fat-tailed and despite $H < \infty$, we take it to belong to the so-called Fréchet class[2]. In extreme value theory [95], a distribution $F$ of a random variable $Y$ is said to be in the Fréchet class if $\bar{F}(y) = 1 - F(y) = y^{-\alpha}L(y)$, where $L(y)$ is a slowly varying function . In other terms, the Fréchet class is the class of all distributions whose right tail behaves as a power law.

Looking at the data, we could be led to believe that the right tail is the dotted line in Figure 17.1, and our estimation of $\alpha$ shows it be smaller than 1. Given the properties of power laws, this means that $E[Y]$ is not finite (as all the other higher moments). This also implies that the sample mean is essentially useless for making inference, in addition to any considerations about robustness [197]. But if $H$ is finite, this cannot be true: all the moments of a random variable with bounded support are finite.

A solution to this situation could be to fit a parametric model, which allows for fat tails and bounded support, such as for example a truncated Pareto [1]. But what happens if $Y$ only shows a Paretian behavior in the upper tail, and not for the whole distribution? Should we fit a mixture model?

In the next section we propose a simple general solution, which does not rely on strong parametric assumptions.

## 17.2 THE DUAL DISTRIBUTION

Instead of altering the tails of the distribution we find it more convenient to transform the data and rely on distributions with well known properties. In Figure 17.1, the real and the apparent tails are indistinguishable to a great extent. We can use this fact to our advantage, by transforming $Y$ to remove its upper bound $H$, so that the new random variable $Z$ - the dual random variable - has the same tail as the apparent tail. We can then estimate the shape parameter $\alpha$ of the tail of $Z$ and come back to $Y$ to compute its moments or, to be more exact, to compute its excess moments, the conditional moments above a given threshold, view that we will just extract the information from the tail of $Z$.

Take $Y$ with support $[L, H]$, and define the function

$$\varphi(Y) = L - H \log\left(\frac{H - Y}{H - L}\right). \tag{17.1}$$

We can verify that $\varphi$ is "smooth": $\varphi \in C^{\infty}$, $\varphi^{-1}(\infty) = H$, and $\varphi^{-1}(L) = \varphi(L) = L$. Then $Z = \varphi(Y)$ defines a new random variable with lower bound $L$ and an infinite upper bound. Notice that the transformation induced by $\varphi(\cdot)$ does not depend on any of the parameters of the distribution of $Y$.

[2] Note that treating $Y$ as belonging to the Fréchet class is a mistake. If a random variable has a finite upper bound, it cannot belong to the Fréchet class, but rather to the Weibull class [136].

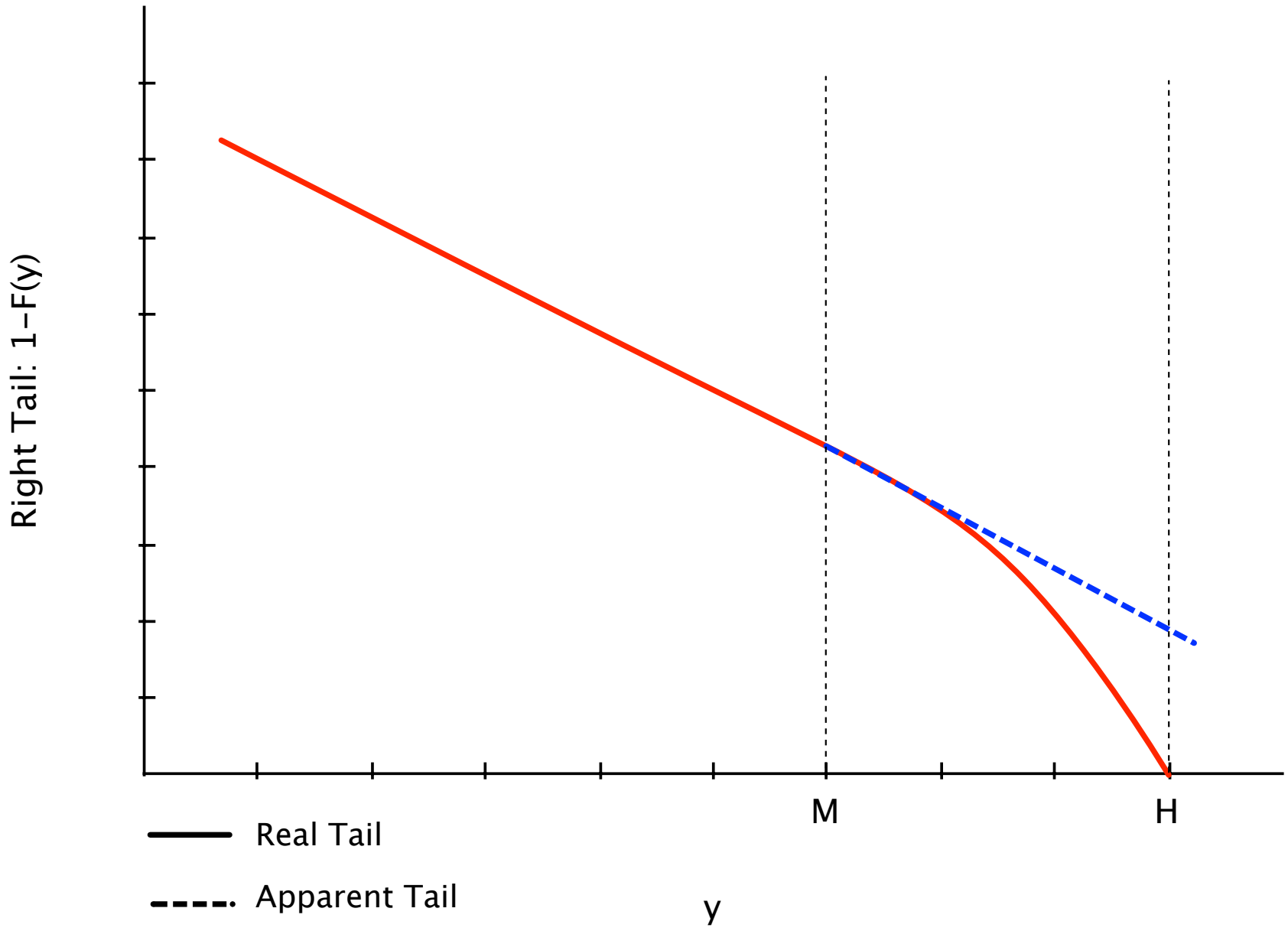

Figure 17.1: *Graphical representation of what may happen if one ignores the existence of the finite upper bound H, since only M is observed.*

By construction, $z = \varphi(y) \approx y$ for very large values of $H$. This means that for a very large upper bound, unlikely to be touched, the results we get for the tail of $Y$ and $Z = \varphi(Y)$ are essentially the same, until we do not reach $H$. But while $Y$ is bounded, $Z$ is not. Therefore we can safely model the unbounded dual distribution of $Z$ as belonging to the Fréchet class, study its tail, and then come back to $Y$ and its moments, which under the dual distribution of $Z$ could not exist.[3]

The tail of $Z$ can be studied in different ways, see for instance [95] and [103]. Our suggestions is to rely on the so-called de Pickands, Balkema and de Haan's Theorem [136]. This theorem allows us to focus on the right tail of a distribution, without caring too much about what happens below a given threshold threshold $u$. In our case $u \geq L$.

Consider a random variable $Z$ with distribution function $G$, and call $G_u$ the conditional df of $Z$ above a given threshold $u$. We can then define the r.v. $W$, representing the rescaled excesses of $Z$ over the threshold $u$, so that

$$G_u(w) = P(Z - u \leq w | Z > u) = \frac{G(u+w) - G(u)}{1 - G(u)},$$

for $0 \leq w \leq z_G - u$, where $z_G$ is the right endpoint of $G$.

[3] Note that the use of logarithmic transformation is quite natural in the context of utility.

Pickands, Balkema and de Haan have showed that for a large class of distribution functions $G$, and a large $u$, $G_u$ can be approximated by a Generalized Pareto distribution, i.e. $G_u(w) \to GPD(w;\xi,\sigma)$, as $u \to \infty$ where

$$GPD(w;\xi,\sigma) = \begin{cases} 1-(1+\xi\frac{w}{\sigma})^{-1/\xi} & if\, \xi \neq 0 \\ 1-e^{-\frac{w}{\sigma}} & if\, \xi = 0 \end{cases}, \; w \geq 0. \tag{17.2}$$

The parameter $\xi$, known as the shape parameter, and corresponding to $1/\alpha$, governs the fatness of the tails, and thus the existence of moments. The moment of order $p$ of a Generalized Pareto distributed random variable only exists if and only if $\xi < 1/p$, or $\alpha > p$ [95]. Both $\xi$ and $\sigma$ can be estimated using MLE or the method of moments [136].[4]

## 17.3 BACK TO $y$: THE SHADOW MEAN (OR POPULATION MEAN)

With $f$ and $g$, we indicate the densities of $Y$ and $Z$.

We know that $Z = \varphi(Y)$, so that $Y = \varphi^{-1}(Z) = (L-H)e^{\frac{L-Z}{H}} + H$.

Now, let's assume we found $u = L^* \geq L$, such that $G_u(w) \approx \text{GPD}(w;\xi,\sigma)$. This implies that the tail of $Y$, above the same value $L^*$ that we find for $Z$, can be obtained from the tail of $Z$, i.e. $G_u$.

First we have

$$\int_{L^*}^{\infty} g(z)\,\mathrm{d}z = \int_{L^*}^{\varphi^{-1}(\infty)} f(y)\,\mathrm{d}y. \tag{17.3}$$

And we know that

$$g(z;\xi,\sigma) = \frac{1}{\sigma}\left(1+\frac{\xi z}{\sigma}\right)^{-\frac{1}{\xi}-1}, \qquad z \in [L^*,\infty). \tag{17.4}$$

Setting $\alpha = \xi^{-1}$, we get

$$f(y;\alpha,\sigma) = \frac{H\left(1+\frac{H(\log(H-L)-\log(H-y))}{\alpha\sigma}\right)^{-\alpha-1}}{\sigma(H-y)}, \; y \in [L^*,H], \tag{17.5}$$

or, in terms of distribution function,

$$F(y;\alpha,\sigma) = 1-\left(1+\frac{H(\log(H-L)-\log(H-y))}{\alpha\sigma}\right)^{-\alpha}. \tag{17.6}$$

Clearly, given that $\varphi$ is a one-to-one transformation, the parameters of $f$ and $g$ obtained by maximum likelihood methods will be the same —the likelihood functions of $f$ and $g$ differ by a scaling constant.

4 There are alternative methods to face finite (or concave) upper bounds, i.e., the use of tempered power laws (with exponential dampening)[230] or stretched exponentials [178]; while being of the same nature as our exercise, these methods do not allow for immediate applications of extreme value theory or similar methods for parametrization.

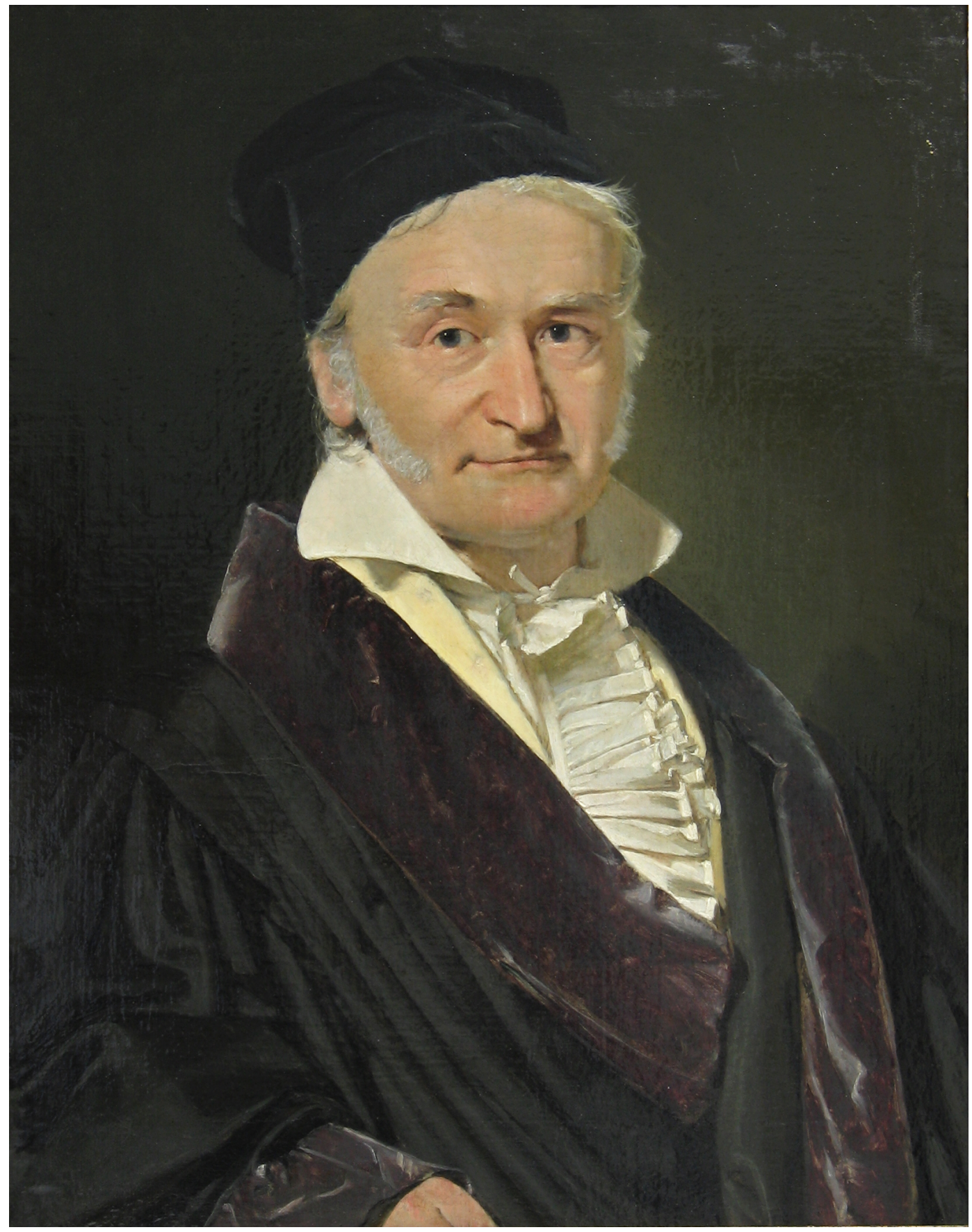

Figure 17.2: *C.F. Gauss, painted by Christian Albrecht Jensen. Gauss has his name on the distribution, generally attributed to Laplace.*

We can derive the shadow mean[5] of $Y$, conditionally on $Y > L^*$, as

$$E[Y|Y > L^*] = \int_{L^*}^{H} y\, f(y;\alpha,\sigma)\,\mathrm{d}y, \tag{17.7}$$

obtaining

$$E[Y|Z > L^*] = (H - L^*)e^{\frac{\alpha\sigma}{H}}\left(\frac{\alpha\sigma}{H}\right)^{\alpha}\Gamma\left(1-\alpha, \frac{\alpha\sigma}{H}\right) + L^*. \tag{17.8}$$

The conditional mean of $Y$ above $L^* \geq L$ can then be estimated by simply plugging in the estimates $\hat{\alpha}$ and $\hat{\sigma}$, as resulting from the GPD approximation of the tail of $Z$. It is worth noticing that if $L^* = L$, then $E[Y|Y > L^*] = E[Y]$, i.e. the conditional mean of $Y$ above $Y$ is exactly the mean of $Y$.

Naturally, in a similar way, we can obtain the other moments, even if we may need numerical methods to compute them.

5 We call the population average –as opposed to the sample one – "shadow", as it is not immediately visible from the data.

Our method can be used in general, but it is particularly useful when, from data, the tail of $Y$ appears so fat that no single moment is finite, as it is often the case when dealing with operational risk losses, the degree distribution of large complex networks, or other econophysical phenomena.

For example, assume that for $Z$ we have $\xi > 1$. Then both $E[Z|Z > L^*]$ and $E[Z]$ are not finite[6]. Figure 17.1 tells us that we might be inclined to assume that also $E[Y]$ is infinite - and this is what the data are likely to tell us if we estimate $\hat{\xi}$ from the tail[7] of $Y$. But this cannot be true because $H < \infty$, and even for $\xi > 1$ we can compute the expected value $E[Y|Z > L^*]$ using equation (19.2).

### Value-at-Risk and Expected Shortfall

Thanks to equation (17.6), we can compute by inversion the quantile function of $Y$ when $Y \geq L^*$, that is

$$Q(p; \alpha, \sigma, H, L) = e^{-\gamma(p)} \left( L^* e^{\frac{\alpha\sigma}{H}} + H e^{\gamma(p)} - H e^{\frac{\alpha\sigma}{H}} \right), \tag{17.9}$$

where $\gamma(p) = \frac{\alpha\sigma(1-p)^{-1/\alpha}}{H}$ and $p \in [0, 1]$. Again, this quantile function is conditional on $Y$ being larger than $L^*$.

From equation (17.9), we can easily compute the Value-at-Risk ($VaR$) of $Y|Y \geq L^*$ for whatever confidence level. For example, the 95% $VaR$ of $Y$, if $Y$ represents operational losses over a 1-year time horizon, is simply $VaR^Y_{0.95} = Q(0.95; \alpha, \sigma, H, L)$.

Another quantity we might be interested in when dealing with the tail risk of $Y$ is the so-called expected shortfall ($ES$), that is $E[Y|Y > u \geq L^*]$. This is nothing more than a generalization of equation (19.2).

We can obtain the expected shortfall by first computing the mean excess function of $Y|Y \geq L^*$, defined as

$$e_u(Y) = E[Y - u|Y > u] = \frac{\int_u^\infty (u - y) f(y; \alpha, \sigma) \mathrm{d}y}{1 - F(u)},$$

for $y \geq u \geq L^*$. Using equation (17.5), we get

$$\begin{aligned} e_u(Y) \quad &= \quad (H - L) e^{\frac{\alpha\sigma}{H}} \left( \frac{\alpha\sigma}{H} \right)^\alpha \left( \frac{H \log\left(\frac{H-L}{H-u}\right)}{\alpha\sigma} + 1 \right)^\alpha \times \\ & \qquad \Gamma\left(1 - \alpha, \frac{\alpha\sigma}{H} + \log\left(\frac{H - L}{H - u}\right)\right). \end{aligned} \tag{17.10}$$

The Expected Shortfall is then simply computed as

$$E[Y|Y > u \geq L^*] = e_u(Y) + u.$$

6 Remember that for a GPD random variable $Z$, $E[Z^p] < \infty$ iff $\xi < 1/p$.

7 Because of the similarities between $1 - F(y)$ and $1 - G(z)$, at least up until $M$, the GPD approximation will give two statistically undistinguishable estimates of $\xi$ for both tails [95].

As in finance and risk management, ES and VaR can be combined. For example we could be interested in computing the 95% ES of $Y$ when $Y \geq L^*$. This is simply given by $VaR^Y_{0.95} + e_{VaR^Y_{0.95}}(Y)$.

## 17.4 COMPARISON TO OTHER METHODS

There are three ways to go about explicitly cutting a Paretian distribution in the tails (not counting methods to stretch or "temper" the distribution).

1) The first one consists in hard truncation, i.e. in setting a single endpoint for the distribution and normalizing. For instance the distribution would be normalized between $L$ and $H$, distributing the excess mass across all points.

2) The second one would assume that $H$ is an absorbing barrier, that all the realizations of the random variable in excess of $H$ would be compressed into a Dirac delta function at $H$ –as practiced in derivative models. In that case the distribution would have the same density as a regular Pareto except at point $H$.

3) The third is the one presented here.

The same problem has cropped up in quantitative finance over the use of truncated normal (to correct for Bachelier's use of a straight Gaussian) vs. logarithmic transformation (Sprenkle, 1961 [256]), with the standard model opting for logarithmic transformation and the associated one-tailed lognormal distribution. Aside from the additivity of log-returns and other such benefits, the models do not produce a "cliff", that is an abrupt change in density below or above, with the instability associated with risk measurements on non-smooth function.

As to the use of extreme value theory, Breilant et al. (2014)[**?** ] go on to truncate the distribution by having an excess in the tails with the transformation $Y^{-\alpha} \to (Y^{-\alpha} - H^{-\alpha})$ and apply EVT to the result. Given that the transformation includes the estimated parameter, a new MLE for the parameter $\alpha$ is required. We find issues with such a non-smooth transformation. The same problem occurs as with financial asset models, particularly the presence an abrupt "cliff" below which there is a density, and above which there is none. The effect is that the expectation obtained in such a way will be higher than ours, particularly at values of $\alpha < 1$, as seen in Figure 17.3.

We can demonstrate the last point as follows. Assume we observe distribution is Pareto that is in fact truncated but treat it as a Pareto. The density is $f(x) = \frac{1}{\sigma}\left(\frac{x-L}{\alpha\sigma}+1\right)^{-\alpha-1}, x \in [L, \infty)$. The truncation gives $g(x) = \frac{\left(\frac{x-L}{\alpha\sigma}+1\right)^{-\alpha-1}}{\sigma(1-\alpha^\alpha\sigma^\alpha(\alpha\sigma+H-L)^{-\alpha})}, x \in [L, H]$.

Moments of order $p$ of the truncated Pareto (i.e. what is seen from realizations of the process), $M(p)$ are:

$$
\begin{aligned}
M(p) = &\alpha e^{-i\pi p}(\alpha\sigma)^\alpha(\alpha\sigma - L)^{p-\alpha} \\
&\frac{\left(B_{\frac{H}{L-\alpha\sigma}}(p+1, -\alpha) - B_{\frac{L}{L-\alpha\sigma}}(p+1, -\alpha)\right)}{\left(\frac{\alpha\sigma}{\alpha\sigma+H-L}\right)^\alpha - 1}
\end{aligned}
\tag{17.11}
$$

where $B(.,.)$ is the Euler Beta function, $B(a,b) = \frac{\Gamma(a)\Gamma(b)}{\Gamma(a+b)} = \int_0^1 t^{a-1}(1-t)^{b-1}\,dt$.

We end up with $r(H,\alpha)$, the ratio of the mean of the soft truncated to that of the truncated Pareto.

$$r(H,\alpha) = e^{-\frac{\alpha}{H}} \left(\frac{\alpha}{H}\right)^{\alpha} \left(\frac{\alpha}{\alpha+H}\right)^{-\alpha} \left(\frac{\alpha+H}{\alpha}\right)^{-\alpha} \frac{\left(-\left(\frac{\alpha+H}{\alpha}\right)^{\alpha} + H + 1\right)}{(\alpha-1)\left(\left(\frac{\alpha}{H}\right)^{\alpha} - \left(\frac{\alpha+H}{H}\right)^{\alpha}\right) E_{\alpha}\left(\frac{\alpha}{H}\right)} \tag{17.12}$$

where $E_{\alpha}\left(\frac{\alpha}{H}\right)$ is the exponential integral $e_{\alpha} z = \int_1^{\infty} \frac{e^{t(-\alpha)}}{t^n}\,dt$.

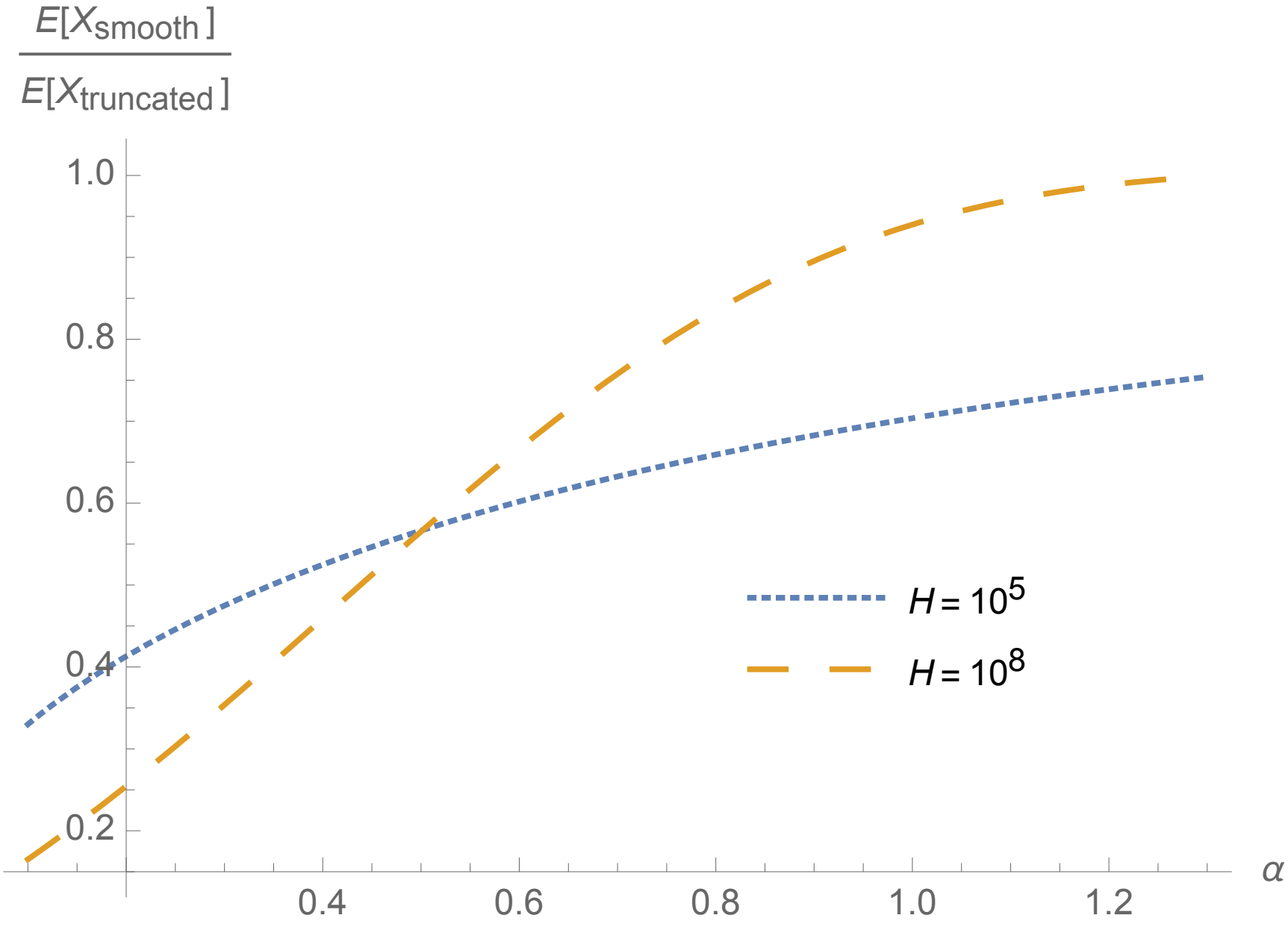

Figure 17.3: *Ratio of the expectation of smooth transformation to truncated.*

## 17.5 APPLICATIONS

**Operational risk** The losses for a firm are bounded by the capitalization, with well-known maximum losses.

**Capped reinsurance contracts** Reinsurance contracts almost always have caps (i.e., a maximum claim); but a reinsurer can have many such contracts on the same source of risk and the addition of the contract pushes the upper bound in such a way as to cause larger potential cumulative harm.

**Violence** While wars are extremely fat-tailed, the maximum effect from any such event cannot exceed the world's population.

**Credit risk** A loan has a finite maximum loss, in a way similar to reinsurance contracts.

**City size** While cities have been shown to be Zipf distributed, the size of a given city cannot exceed that of the world's population.

**Environmental harm** While these variables are exceedingly fat-tailed, the risk is confined by the size of the planet (or the continent on which they take place) as a firm upper bound.

**Complex networks** The number of connections is finite.

**Company size** The sales of a company is bound by the GDP.

**Earthquakes** The maximum harm from an earthquake is bound by the energy.

**Hydrology** The maximum level of a flood can be determined.

# 18 ON THE TAIL RISK OF VIOLENT CONFLICT (WITH P. CIRILLO)‡

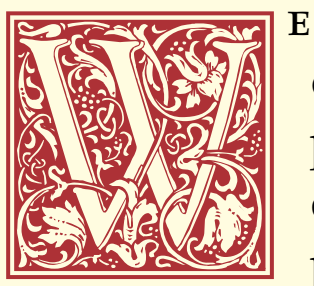

WE EXAMINE all possible statistical pictures of violent conflicts over common era history with a focus on dealing with incompleteness and unreliability of data. We apply methods from extreme value theory on log-transformed data to remove compact support, then, owing to the boundedness of maximum casualties, retransform the data and derive expected means. We find the estimated mean likely to be at least three times larger than the sample mean, meaning severe underestimation of the severity of conflicts from naive observation. We check for robustness by sampling between high and low estimates and jackknifing the data. We study inter-arrival times between tail events and find (first-order) memorylessless of events. The statistical pictures obtained are at variance with the claims about "long peace".

## 18.1 INTRODUCTION/SUMMARY

This study is as much about new statistical methodologies with thick tailed (and unreliable data), as well as bounded random variables with local Power Law behavior, as it is about the properties of violence.[2]

Violence is much more severe than it seems from conventional analyses and the prevailing "long peace" theory which claims that violence has declined. Adapting methods from extreme value theory, and adjusting for errors in reporting of conflicts and historical estimates of casualties, we look at the various statistical pictures of violent conflicts, with focus for the parametrization on those with

Research chapter.

2 Acknowledgments: Captain Mark Weisenborn engaged in the thankless and gruesome task of compiling the data, checking across sources and linking each conflict to a narrative on Wikipedia (see Appendix 1). We also benefited from generous help on social networks where we put data for scrutiny, as well as advice from historians thanked in the same appendix. We also thank the late Benoit Mandelbrot for insights on the tail properties of wars and conflicts, as well as Yaneer Bar-Yam, Raphael Douady...

Figure 18.1: *Values of the tail exponent $\alpha$ from Hill estimator obtained across 100,000 different rescaled casualty numbers uniformly selected between low and high estimates of conflict. The exponent is slightly (but not meaningfully) different from the Maximum Likelihood for all data as we focus on top 100 deviations.*

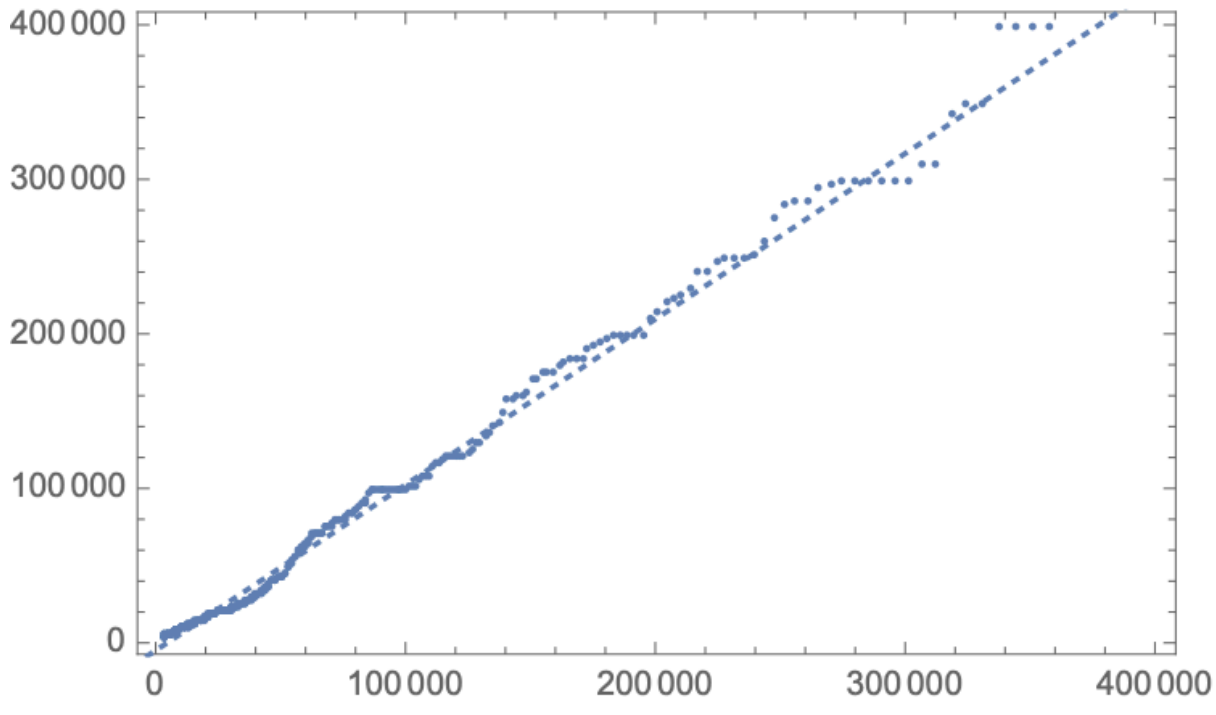

Figure 18.2: *Q-Q plot of the rescaled data in the near-tail plotted against a Pareto II-Lomax Style distribution.*

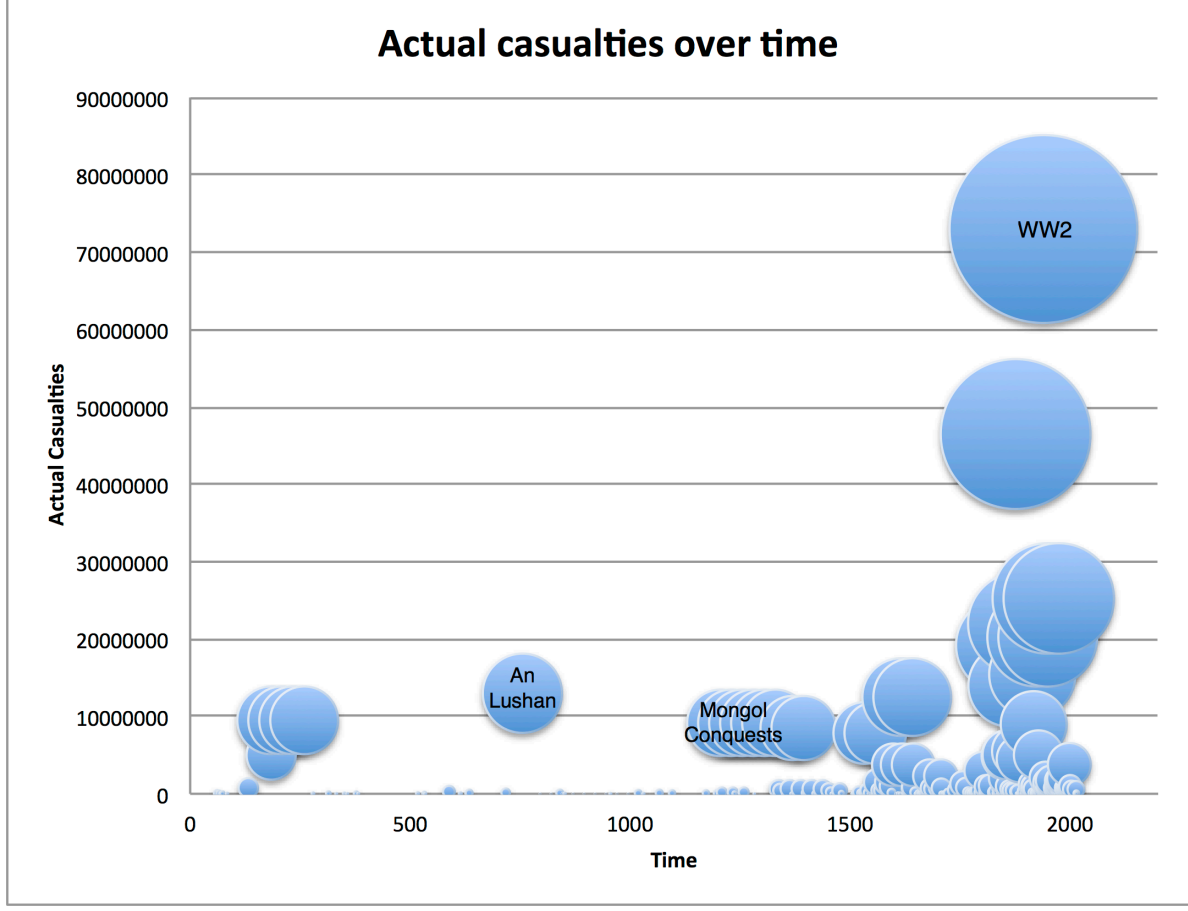

Figure 18.3: *Death toll from "named conflicts" over time. Conflicts lasting more than 25 years are disaggregated into two or more conflicts, each one lasting 25 years.*

more than 50k victims (in equivalent ratio of today's population, which would correspond to $\approx$ 5k in the $18^{th}$ C.). Contrary to current discussions, *all* statistical pictures thus obtained show that 1) the risk of violent conflict has not been decreasing, but is rather underestimated by techniques relying on naive year-on-year changes in the mean, or using sample mean as an estimator of the true mean of an extremely fat-tailed phenomenon; 2) armed conflicts have memoryless inter-arrival times, thus incompatible with the idea of a time trend. Our analysis uses 1) raw data, as recorded and estimated by historians; 2) a naive transformation, used by certain historians and sociologists, which rescales past conflicts and casualties with respect to the actual population; 3) more importantly, a log transformation to account for the fact that the number of casualties in a conflict cannot be larger than the world population. (This is similar to the transformation of data into log-returns in mathematical finance in order to use distributions with support on the real line.)

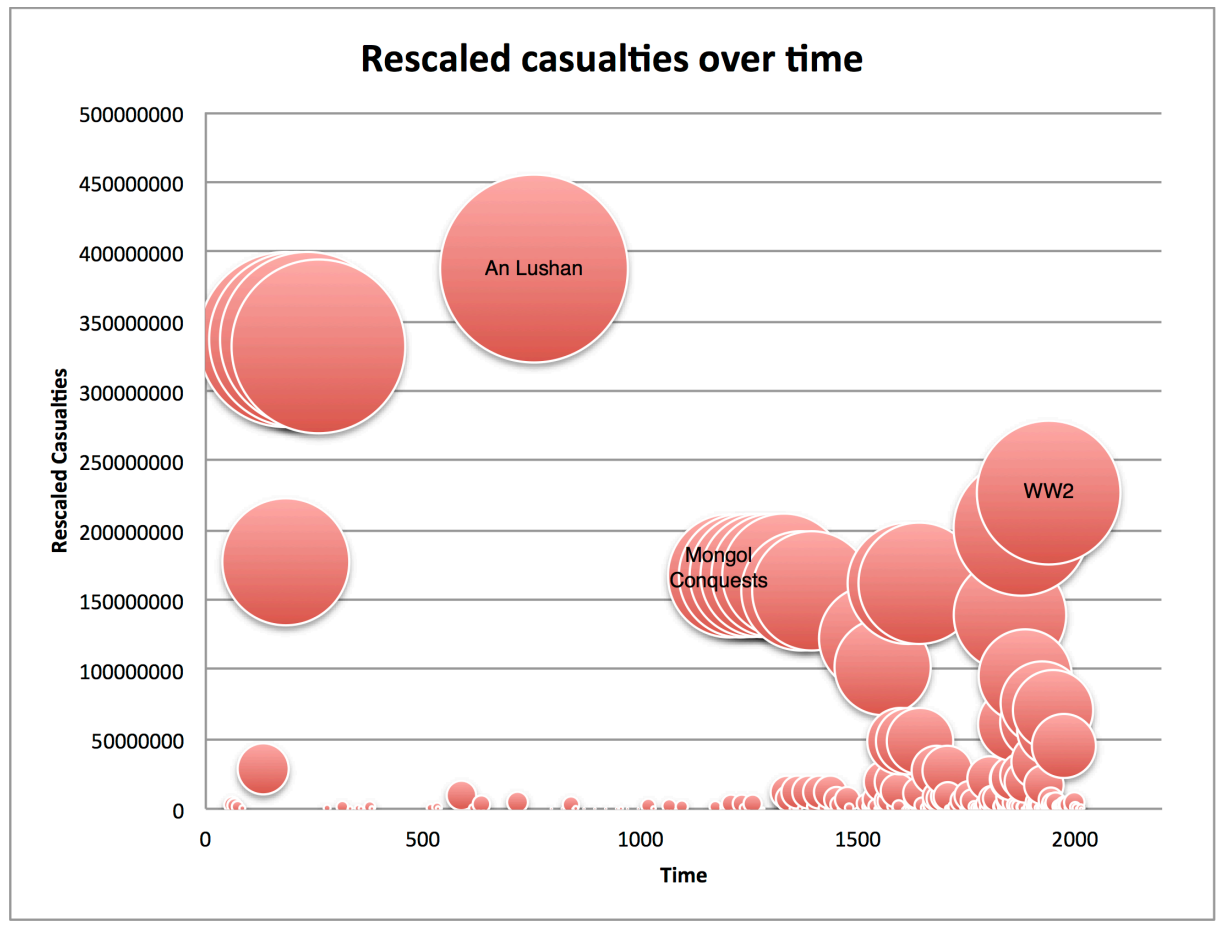

Figure 18.4: *Rescaled death toll of armed conflict and regimes over time. Data are rescaled w.r.t. today's world population. Conflicts lasting more than 25 years are disaggregated into two or more conflicts, each one lasting 25 years.*

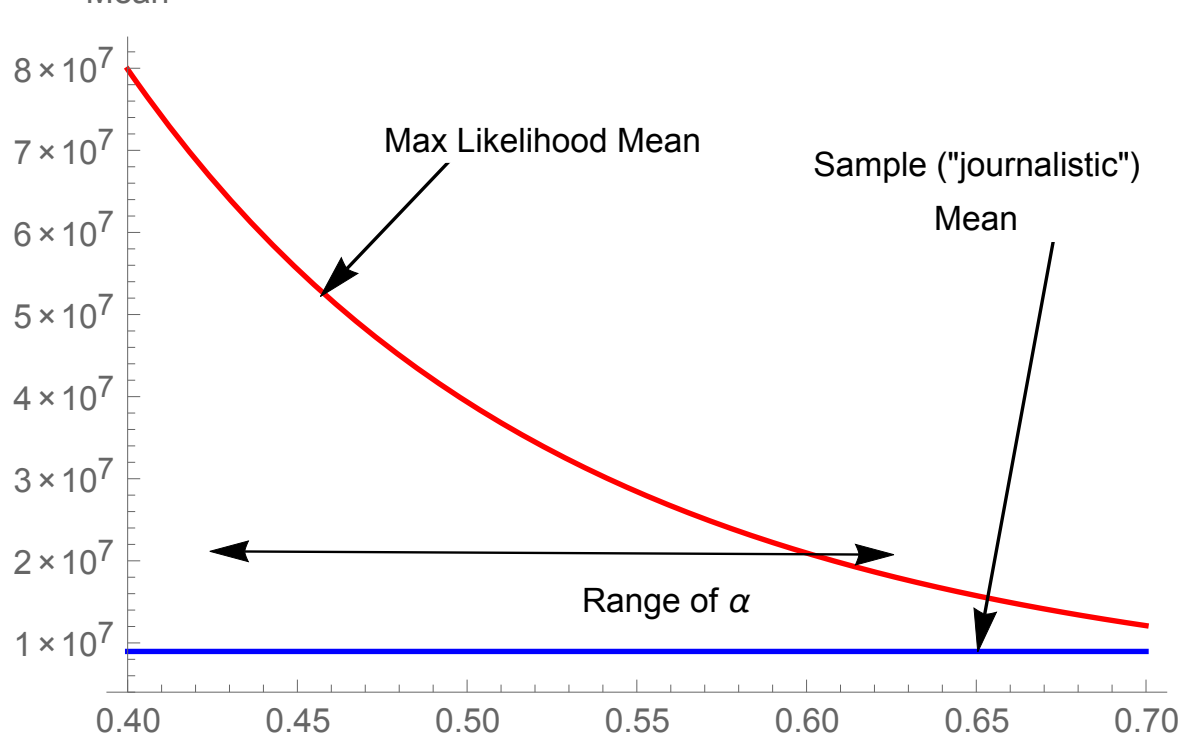

Figure 18.5: *Observed "journalistic" mean compared to MLE mean (derived from rescaling back the data to compact support) for different values of $\alpha$ (hence for permutations of the pair $(\sigma_\alpha, \alpha)$). The "range of $\alpha$ is the one we get from possible variations of the data from bootstrap and reliability simulations.*

All in all, among the different classes of data (raw and rescaled), we observe that 1) casualties are Power Law distributed.[3] In the case of log-rescaled data we observe $.4 \leq \alpha \leq .7$, thus indicating an extremely fat-tailed phenomenon with an undefined mean (a result that is robustly obtained); 2) the inter-arrival times of conflicts above the 50k threshold follow a homogeneous Poisson process, indicating no particular trend, and therefore contradicting a popular narrative about the decline of violence; 3) the true mean to be expected in the future, and the most compatible with the data, though highly stochastic, is $\approx 3\times$ higher than past mean.

[3] Many earlier studies have found Paretianity in data, [? ],[42]. Our study, aside from the use of extreme value techniques, reliability bootstraps, and compact support transformations, varies in both calibrations and interpretation.

Further, we explain: 1) how the mean (in terms of expected casualties) is severely underestimated by conventional data analyses as the observed mean is not an estimator of true mean (unlike the tail exponent that provides a picture with smaller noise); 2) how misconceptions arise from the deceiving lengthy (and volatile) inter-arrival times between large conflicts.

To remedy the inaccuracies of historical numerical assessments, we provide a standard bootstrap analysis of our estimates, in addition to Monte Carlo checks for unreliability of wars and absence of events from currently recorded history.

## 18.2 SUMMARY STATISTICAL DISCUSSION

### 18.2.1 Results

**Paretian tails** Peak-Over-Threshold methods show (both raw and rescaled variables) exhibit strong Paretiantail behavior, with survival probability $\mathbb{P}(X > x) = \lambda(x)x^{-\alpha}$, where $\lambda : [L, +\infty) \to (0, +\infty)$ is a slowly varying function, defined as $\lim_{x\to+\infty} \frac{\lambda(kx)}{\lambda(x)} = 1$ for any $k > 0$.

We parametrize G(.), a Generalized Pareto Distribution (GPD) , see Table 18.4, $G(x) = 1 - (1 + \xi y/\beta)^{-1/\xi}$ , with $\xi \approx 1.88, \pm.14$ for rescaled data which corresponds to a tail $\alpha = \frac{1}{\xi} = .53, \pm.04$.

**Memorylessness of onset of conflicts** Tables 18.2 and 18.3 show inter-arrival times, meaning one can wait more than a hundred years for an event such as WWII without changing one's expectation. There is no visible autocorrelation, no statistically detectable temporal structure (i.e. we cannot see the imprint of a self-exciting process), see Figure 18.8.

**Full distribution(s)** Rescaled data fits a Lomax-Style distribution with same tail as obtained by POT, with strong goodness of fit. For events with casualties $> L = 10K, 25K, 50K, etc.$ we fit different Pareto II (Lomax) distributions with corresponding tail $\alpha$ (fit from GPD), with scale $\sigma = 84,360$, i.e., with density $\frac{\alpha\left(\frac{-L+\sigma+x}{\sigma}\right)^{-\alpha-1}}{\sigma}, x \geq L$.

We also consider a wider array of statistical "pictures" from pairs $\alpha, \sigma_\alpha$ across the data from potential alternative values of $\alpha$, with recalibration of maximum likelihood $\sigma$, see Figure 18.5.

**Difference between sample mean and maximum likelihood mean** : Table 18.1 shows the true mean using the parametrization of the Pareto distribution above and inverting the transformation back to compact support. "True" or maximum likelihood, or "statistical" mean is between 3 and 4 times observed mean.

This means the "journalistic" observation of the mean, aside from the conceptual mistake of relying on sample mean, underestimates the true mean by at least 3 times and higher future observations would not allow the conlusion that violence has "risen".

Table 18.1: *Sample means and estimated maximum likelihood mean across minimum values L – Rescaled data.*

| L | Sample Mean | ML Mean | Ratio |
|---|---|---|---|
| 10K | $9.079 \times 10^6$ | $3.11 \times 10^7$ | 3.43 |
| 25K | $9.82 \times 10^6$ | $3.62 \times 10^7$ | 3.69 |
| 50K | $1.12 \times 10^7$ | $4.11 \times 10^7$ | 3.67 |
| 100K | $1.34 \times 10^7$ | $4.74 \times 10^7$ | 3.53 |
| 200K | $1.66 \times 10^7$ | $6.31 \times 10^7$ | 3.79 |
| 500K | $2.48 \times 10^7$ | $8.26 \times 10^7$ | 3.31 |

### 18.2.2 Conclusion

History as seen from tail analysis is far more risky, and conflicts far more violent than acknowledged by naive observation of behavior of averages in historical time series.

Table 18.2: *Average inter-arrival times and their mean absolute deviation for events with more than 1, 2, 5 and 10 million casualties, using actual estimates.*

| Threshold | Average | MAD |
|---|---|---|
| 1 | 26.71 | 31.66 |
| 2 | 42.19 | 47.31 |
| 5 | 57.74 | 68.60 |
| 10 | 101.58 | 144.47 |

Table 18.3: *Average inter-arrival times and their mean absolute deviation for events with more than 1, 2, 5, 10, 20, and 50 million casualties, using rescaled amounts.*

| Threshold | Average | MAD |
|---|---|---|
| 1 | 11.27 | 12.59 |
| 2 | 16.84 | 18.13 |
| 5 | 26.31 | 27.29 |
| 10 | 37.39 | 41.30 |
| 20 | 48.47 | 52.14 |
| 50 | 67.88 | 78.57 |

Table 18.4: *Estimates (and standard errors) of the Generalized Pareto Distribution parameters for casualties over a 50k threshold. For both actual and rescaled casualties, we also provide the number of events lying above the threshold (the total number of events in our data is 99).*

| **Data** | **Nr. Excesses** | $\xi$ | $\beta$ |
|---|---|---|---|
| Raw Data | 307 | 1.5886 (0.1467) | 3.6254 (0.8191) |
| Naive Rescaling | 524 | 1.8718 (0.1259) | 14.3254 (2.1111) |
| Log-rescaling | 524 | 1.8717 (0.1277) | 14.3261 (2.1422) |

## 18.3 METHODOLOGICAL DISCUSSION

### 18.3.1 Rescaling Method

We remove the compact support to be able to use power laws as follows (see earlier chapters). Using $X_t$ as the r.v. for number of incidences from conflict at times $t$, consider first a naive rescaling of $X'_t = \frac{X_t}{H_t}$, where $H_t$ is the total human population at period $t$. See appendix for methods of estimation of $H_t$.

Next, with today's maximum population $H$ and $L$ the naively rescaled minimum for our definition of conflict, we introduce a smooth rescaling function $\varphi : [L, H] \to [L, \infty)$ satisfying:

i $\varphi$ is "smooth": $\varphi \in C^\infty$,

ii $\varphi^{-1}(\infty) = H$,

iii $\varphi^{-1}(L) = \varphi(L) = L$.

In particular, we choose:

$$\varphi(x) = L - H \log\left(\frac{H - x}{H - L}\right). \tag{18.1}$$

We can perform appropriate analytics on $x_r = \varphi(x)$ given that it is unbounded, and properly fit Power Law exponents. Then we can rescale back for the properties of $X$. Also notice that the $\varphi(x) \approx x$ for very large values of $H$. This means that for a very large upper bound, the results we will get for $x$ and $\varphi(x)$ will be essentially the same. The big difference is only from a philosophical/methodological point of view, in the sense that we remove the upper bound (unlikely to be reached).

In what follows we will use the naively rescaled casualties as input for the $\varphi(\cdot)$ function.

We pick $H = P_{t_0}$ for the exercise.

The distribution of $x$ can be rederived as follows from the distribution of $x_r$:

$$\int_L^\infty f(x_r)\, \mathrm{d}x_r = \int_L^{\varphi^{-1}(\infty)} g(x)\, \mathrm{d}x, \tag{18.2}$$

where $\varphi^{-1}(u) = (L - H)e^{\frac{L-u}{H}} + H$

In this case, from the Pareto-Lomax selected:

$$f(x_r) = \frac{\alpha \left(\frac{-L+\sigma+x_r}{\sigma}\right)^{-\alpha-1}}{\sigma}, x_r \in [L, \infty) \tag{18.3}$$

$$g(x) = \frac{\alpha H \left(\frac{\sigma - H \log\left(\frac{H-x}{H-L}\right)}{\sigma}\right)^{-\alpha-1}}{\sigma(H-x)}, x \in [L, H],$$

which verifies $\int_L^H x\, g(x)\, \mathrm{d}x = 1$. Hence the expectation

$$\mathbb{E}_g(x; L, H, \sigma, \alpha) = \int_L^H x\, g(x)\, \mathrm{d}x, \tag{18.4}$$

$$\mathbb{E}_g(X; L, H, \sigma, \alpha) = \alpha H \left(\frac{1}{\alpha} - \frac{(H-L)e^{\sigma/H} E_{\alpha+1}\left(\frac{\sigma}{H}\right)}{H}\right) \tag{18.5}$$

where $E_.(.)$ is the exponential integral $E_n z = \int_1^\infty \frac{e^{t(-z)}}{t^n} \mathrm{d}t$.

Note that we rely on the invariance property:

**Remark 18.1**

*If $\hat{\theta}$ is the maximum likelihood estimator (MLE) of $\theta$, then for an absolutely continuous function $\phi$, $\phi(\hat{\theta})$ is the MLE estimator of $\phi(\theta)$.*

For further details see [250].

### 18.3.2 Expectation by Conditioning (less rigorous)

We would be replacing a smooth function in $C^\infty$ by a Heaviside step function, that is the indicator function $\mathbb{1} : \mathbb{R} \to \{0, 1\}$, written as $\mathbb{1}_{X \in [L,H]}$:

$$\mathbb{E}(\mathbb{1}_{X \in [L,H]}) = \frac{\int_L^H x\, f(x)\, \mathrm{d}x}{\int_L^H f(x)\, \mathrm{d}x}$$

which for the Pareto Lomax becomes:

$$\mathbb{E}(\mathbb{1}_{X \in [L,H]}) = \frac{\frac{\alpha\sigma^\alpha (H-L)}{\sigma^\alpha - (H-L+\sigma)^\alpha} + (\alpha - 1)L + \sigma}{\alpha - 1} \tag{18.6}$$

### 18.3.3 Reliability of Data and Effect on Tail Estimates

Data from violence is largely anecdotal, spreading via citations, often based on some vague estimate, without anyone's ability to verify the assessments using period sources. An event that took place in the seventh century, such as the an Lushan rebellion, is "estimated" to have killed 26 million people, with no precise or reliable methodology to allow us to trust the number. The independence war of Algeria has various estimates, some from France, others from the rebels, and nothing scientifically or professionally obtained.

As said earlier, in this chapter, we use different data: raw data, naively rescaled data w.r.t. the current world population, and log-rescaled data to avoid the theoretical problem of the upper bound.

For some observations, together with the estimated number of casualties, as resulting from historical sources, we also have a lower and upper bound available. Let $X_t$ be the number of casualties in a given conflict at time $t$. In principle, we can define triplets like

- $\{X_t, X_t^l, X_t^u\}$ for the actual estimates (raw data), where $X_t^l$ and $X_t^u$ represent the lower and upper bound, if available.
- $\{Y_t = X_t \frac{P_{20015}}{P_t}, Y_t^l = X_t^l \frac{P_{20015}}{P_t}, Y_t^u = X_t^u \frac{P_{20015}}{P_t}\}$ for the naively rescaled data, where $P_{2015}$ is the world population in 2015 and $P_t$ is the population at time $t = 1, ..., 2014$.
- $\{Z_t = \varphi(Y_t), Z_t^l = \varphi(Y_t^l), Z_t^u = \varphi(Y_t^u)\}$ for the log-rescaled data.

To prevent possible criticism about the use of middle estimates, when bounds are present, we have decided to use the following Monte Carlo procedure (for more details [234]), obtaining no significant different in the estimates of all the quantities of interest (like the tail exponent $\alpha = 1/\xi$):

1. For each event $X$ for which bounds are present, we have assumed casualties to be uniformly distributed between the lower and the upper bound, i.e. $X \sim U(X^l, X^u)$. The choice of the uniform distribution is to keep things simple. All other bounded distributions would in fact generate the same results in the limit, thanks to the central limit theorem.
2. We have then generated a large number of Monte Carlo replications, and in each replication we have assigned a random value to each event $X$ according to $U(X^l, X^u)$.
3. For each replication we have computed the statistics of interest, typically the tail exponent, obtaining values that we have later averaged.

This procedure has shown that the precision of estimates does not affect the tail of the distribution of casualties, as the tail exponent is rather stable.

For those events for which no bound is given, the options were to use them as they are, or to perturb them by creating fictitious bounds around them (and then treat them as the other bounded ones in the Monte Carlo replications). We have chosen the second approach.

The above also applies to $Y_t$ and $Z_t$.

Note that the tail $\alpha$ derived from an average is different from an average alpha across different estimates, which is the reason we perform the various analyses across estimates.

**Technical comment** These simulations are largely looking for a "stochastic alpha" bias from errors and unreliability of data (Chapter 21). With a sample size of $n$, a parameter $\hat{\theta}_m$ will be the average parameter obtained across a large number of Monte Carlo runs. Let $X_i$ be a given Monte Carlo simulated vector indexed by $i$ and $X_\mu$ is the middle estimate between high and low bounds. Since, with $\frac{1}{m}\sum_{\leq m}\|X_j\|_1 = \|X_\mu\|_1$ across Monte Carlo runs but $\forall j\,, \|X_j\|_1 \neq \|X_\mu\|_1$, $\widehat{\theta}_m = \frac{1}{m}\sum_{\leq m}\widehat{\theta}(X_j) \neq \widehat{\theta}(X_\mu)$. For instance, consider the maximum likelihood estimation of a Paretian tail, $\widehat{\alpha}(X_i) \triangleq n\left(\sum_{1\leq i\leq n}\log\left(\frac{x_i}{L}\right)\right)^{-1}$. With $\Delta \geq x_m$, define

$$\widehat{\alpha}(X_i \sqcup \Delta) \triangleq \frac{1}{2}\left(\frac{n}{\sum_{i=1}^{n}\log\left(\frac{x_i}{L}\right) - \log\left(\frac{\Delta}{L}\right)} + \frac{n}{\sum_{i=1}^{n}\log\left(\frac{x_i}{L}\right) + \log\left(\frac{\Delta}{L}\right)}\right),$$

which, owing to the concavity of the logarithmic function, gives the inequality

$$\forall \Delta \geq x_m, \widehat{\alpha}(X_i \sqcup \Delta) \geq \widehat{\alpha}(X_i).$$

### 18.3.4 Definition of An "Event"

"Named" conflicts are an arbitrary designation that, often, does not make sense statistically: a conflict can have two or more names; two or more conflicts can have the same name, and we found no satisfactory hierarchy between war and conflict. For uniformity, we treat events as the shorter of event or its disaggregation into units with a maximum duration of 25 years each. Accordingly, we treat Mongolian wars, which lasted more than a century and a quarter, as more than a single event. It makes little sense otherwise as it would be the equivalent of treating the period from the Franco-Prussian war to WW II as "German(ic) wars", rather than multiple events because these wars had individual names in contemporary sources. Effectively the main sources such as the *Encyclopedia of War* [222] list numerous conflicts in place of "Mongol Invasions" –the more sophisticated the historians in a given area, the more likely they are to break conflicts into different "named" events and, depending on historians, Mongolian wars range between 12 and 55 conflicts.

What controversy about the definition of a "name" can be, once again, solved by bootstrapping. Our conclusion, incidentally, is invariant to the bundling or unbundling of the Mongolian wars.

Further, in the absence of a clearly defined protocol in historical studies, it has been hard to disentangle direct death from wars and those from less direct effects on populations (say blocades, famine). For instance the First Jewish War has confused historians as an estimated 30K death came from the war, and a considerably higher (between 350K and the number 1M according to Josephus) from the famine or civilian casualties.

### 18.3.5 Missing Events

We can assume that there are numerous wars that are not part of our sample, even if we doubt that such events are in the "tails" of the distribution, given that large conflicts are more likely to be reported by historians. Further, we also assume that their occurrence is random across the data (in the sense that they do not have an effect on clustering).

But we are aware of a bias from differential in both accuracy and reporting across time: events are more likely to be recorded in modern times than in the past. Raising the minimum value $L$ the number of such "missed" events and their impact are likely to drop rapidly. Indeed, as a robustness check, raising the bar to a minimum $L = 500K$ does not change our analysis.

A simple jackknife procedure, performed by randomly removing a proportion of events from the sample and repeating analyses, shows us the dependence of our analysis on missing events, dependence that we have found to be insignificant, when focusing on the tail of the distribution of casualties. In other words, given that we are dealing with extremes, if removing 30% of events and checking the effects on parameters produce no divergence from initial results, then we do not need to worry of having missed 30% of events, as missing events are not likely to cause thinning of the tails.[4]

### 18.3.6 Survivorship Bias

We did not take into account of the survivorship biases in the analysis, assuming it to be negligible before 1960, as the probability of a conflict affecting all of mankind was negligible. Such probability (and risk) became considerably higher since, especially because of nuclear and other mass destruction weapons.

## 18.4 DATA ANALYSIS

Figures 18.3 and 18.4 graphically represent our data: the number of casualties over time. Figure 18.3 refers to the estimated actual number of victims, while Figure 18.4 shows the rescaled amounts, obtained by rescaling the past observation with respect to the world population in 2015 (around 7.2 billion people)[5]. Figures 18.3 might suggest an increase in the death toll of armed conflicts over time, thus supporting the idea that war violence has increased. Figure 18.4, conversely, seems to suggest a decrease in the (rescaled) number of victims, especially in the last hundred years, and possibly in violence as well. In what follows we show that both interpretations are surely naive, because they do not take into consideration the fact that we are dealing with extreme events.

---

[4] The opposite is not true, which is at the core of the Black Swan asymmetry: such procedure does not remedy the missing of tail, "Black Swan" events from the record. A single "Black Swan" event can considerably fatten the tail. In this case the tail is fat enough and no missing information seems able to make it thinner.

[5] Notice that, in equation (19.1), for $H = 7.2$ billion, $\varphi(x) \approx x$. Therefore Figure 18.4 is also representative for log-rescaled data.

### 18.4.1 Peaks Over Threshold

Given the fat-tailed nature of the data, which can be easily observed with some basic graphical tools like histograms on the logs and QQplots (Figure 18.6 shows the QQplot of actual casualties against an exponential distribution: the clear concavity is a signal of fat-tailed distribution), it seems appropriate to use a well-known method of extreme value theory to model war casualties over time: the Peaks-over-Threshold or POT [95].

According to the POT method, excesses of an i.i.d. sequence over a high threshold $u$ (that we have to identify) occur at the times of a homogeneous Poisson process, while the excesses themselves can be modeled with a Generalized Pareto Distribution (GPD). Arrival times and excesses are assumed to be independent of each other.

In our case, assuming the independence of the war events does not seem a strong assumption, given the time and space separation among them. Regarding the other assumptions, on the contrary, we have to check them.

We start by identifying the threshold $u$ above which the GPD approximation may hold. Different heuristic tools can be used for this purpose, from Zipf plot to mean excess function plots, where one looks for the linearity which is typical of fat-tailed phenomena [53, 97]. Figure 19.4 shows the mean excess function plot for actual casualties[6]: an upward trend is clearly present, already starting with a threshold equal to 5k victims. For the goodness of fit, it might be appropriate to choose a slightly larger threshold, like $u = 50k$[7].

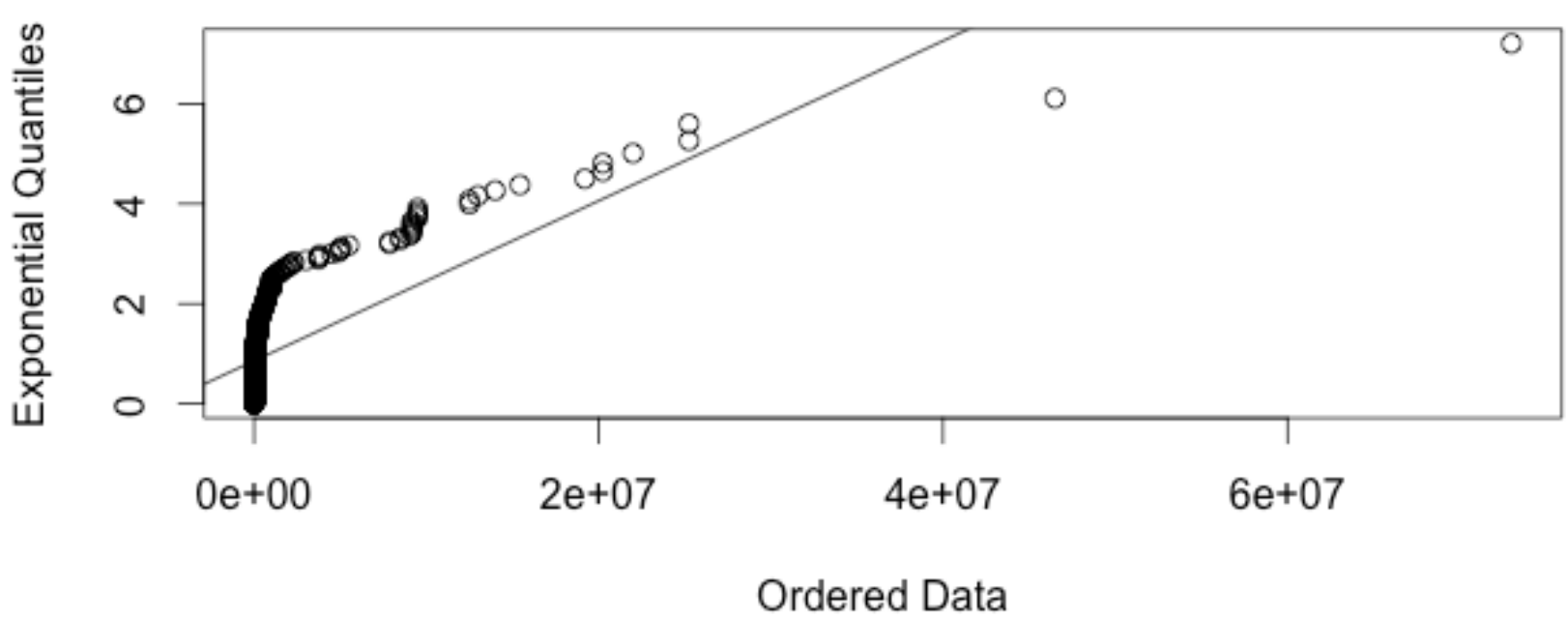

Figure 18.6: *QQplot of actual casualties against standard exponential quantile. The concave curvature of data points is a clear signal of heavy tails.*

[6] Similar results hold for the rescaled amounts (naive and log). For the sake of brevity we always show plots for one of the two variables, unless a major difference is observed.

[7] This idea has also been supported by subsequent goodness-of-fit tests.

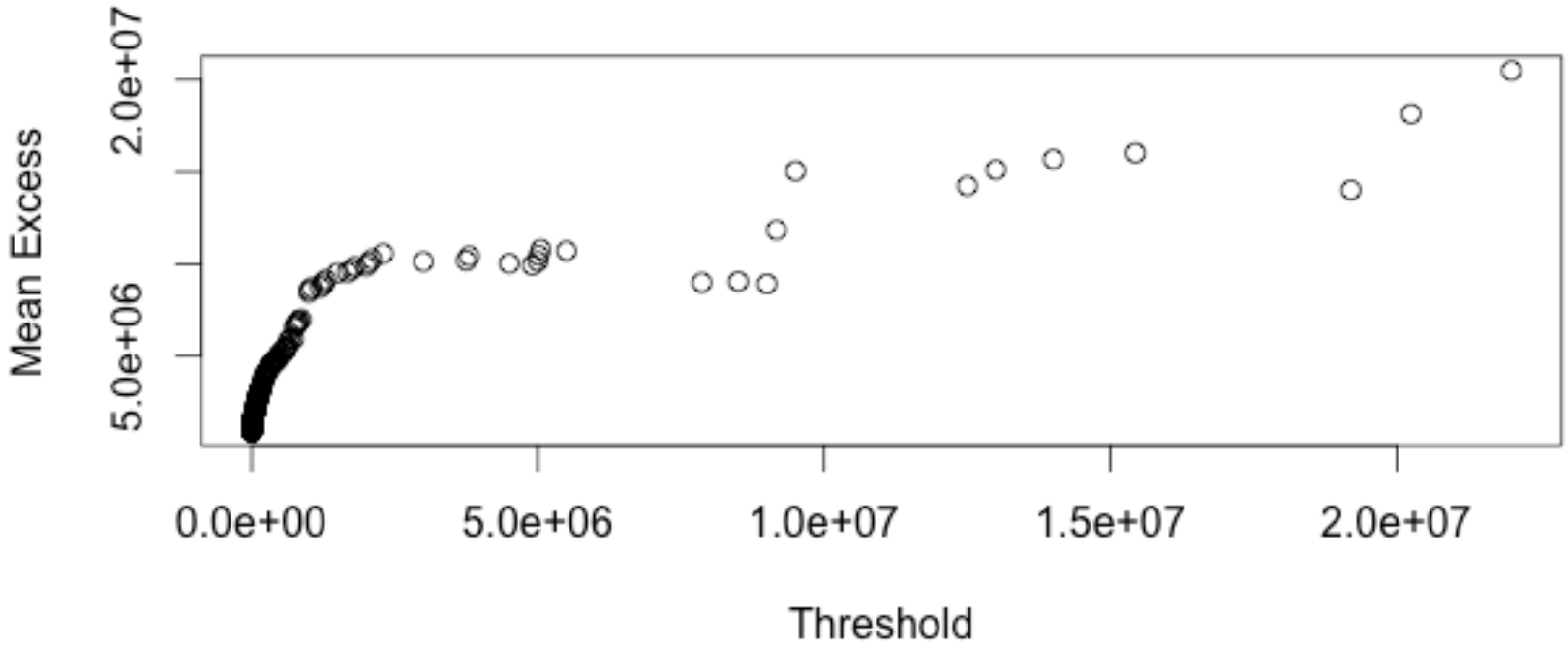

Figure 18.7: *Mean excess function plot (MEPLOT) for actual casualties. An upward trend - almost linear in the first part of the graph - is present, suggesting the presence of a fat right tail. The variability of the mean excess function for higher thresholds is due to the small number of observation exceeding those thresholds and should not be taken into consideration.*

### 18.4.2 Gaps in Series and Autocorrelation

To check whether events over time occur according to a homogeneous Poisson process, a basic assumption of the POT method, we can look at the distribution of the inter-arrival times or gaps, which should be exponential. Gaps should also show no autocorrelation.

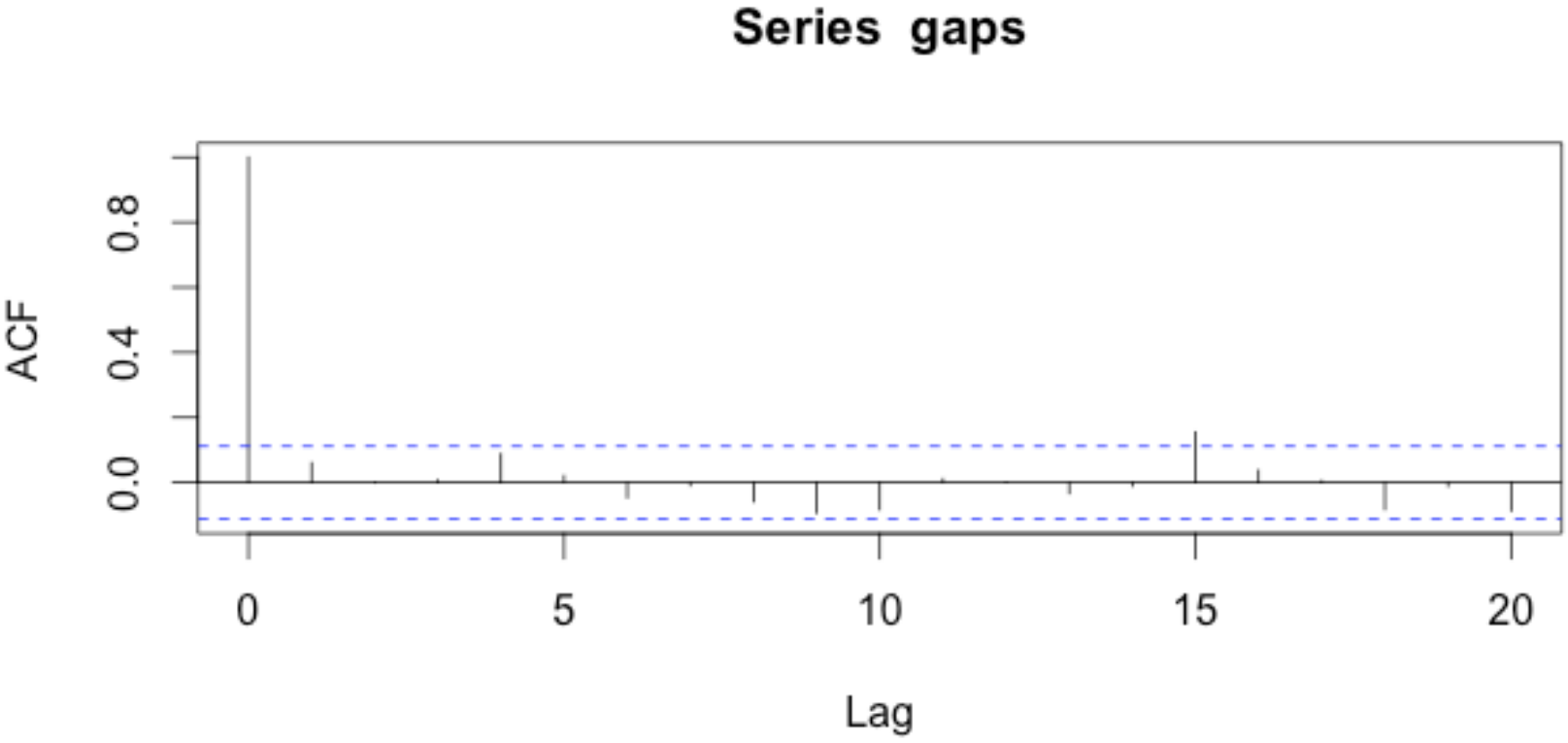

Figure 18.8: *ACF plot of gaps for actual casualties, no significant autocorrelation is visible.*

Figure 18.8 clearly shows the absence of autocorrelation. The plausibility of an exponential distribution for the inter-arrival times can be positively checked using both heuristic and analytic tools. Here we omit the positive results for brevity.

However, in order to provide some extra useful information, in Tables 18.2 and 18.3 we provide some basic statistics about the inter-arrival times for very catastrophic events in terms of casualties[8]. The simple evidence there contained should already be sufficient to underline how unreliable can be the statement that war violence has been decreasing over time. For an events with more than 10 million victims, if we refer to actual estimates, the average time delay is 101.58 years, with a mean absolute deviation of 144.47 years[9]. This means that it is totally plausible that in the last few years we have not observed such a large event. It could simply happen tomorrow or some time in the future. This also means that every trend extrapolation makes no great sense for this type of extreme events. Finally, we have to consider that an event as large as WW2 happened only once in 2014 years, if we deal with actual casualties (for rescaled casualties we can consider the An Lushan rebellion); in this case the possible waiting time is even longer.

### 18.4.3 Tail Analysis

Given that the POT assumptions about the Poisson process seem to be confirmed by data, it is finally the time to fit a Generalized Pareto Distribution to the exceedances.

Consider a random variable $X$ with df $F$, and call $F_u$ the conditional df of $X$ above a given threshold $u$. We can then define a r.v. $Y$, representing the rescaled excesses of $X$ over the threshold $u$, getting [95]

$$F_u(y) = \mathbb{P}(X - u \leq y | X > u) = \frac{F(u+y) - F(u)}{1 - F(u)}$$

for $0 \leq y \leq x_F - u$, where $x_F$ is the right endpoint of the underlying distribution $F$. Pickands [223], Balkema and de Haan [11], [12] and [13] showed that for a large class of underlying distribution functions $F$ (following in the so-called domain of attraction of the GEV distribution [95]), and a large $u$, $F_u$ can be approximated by a Generalized Pareto distribution: $F_u(y) \to G(y)$, as $u \to \infty$ where

$$G(y) = \begin{cases} 1 - (1 + \xi y/\beta)^{-1/\xi} & if\ \xi \neq 0 \\ 1 - e^{-y/\beta} & if\ \xi = 0. \end{cases} \tag{18.7}$$

It can be shown that the GPD distribution is a distribution interpolating between the exponential distribution (for $\xi = 0$) and a class of Pareto distributions. We refer to [95] for more details.

The parameters in (18.7) can be estimated using methods like maximum likelihood or probability weighted moments [95]. The goodness of fit can then be tested using bootstrap-based tests [316].

Table 18.4 contains our mle estimates for actual and rescaled casualties above a 50k victims threshold. This threshold is in fact the one providing the best compromise between goodness of fit and a sufficient number of observation, so that standard errors are reliable. The actual and both the rescaled data show two different sets of estimates, but their interpretation is strongly consistent. For this reason we just focus on actual casualties for the discussion.

[8] Table 18.2 does not show the average delay for events with 20M (50M) or more casualties. This is due to the limited amount of these observations in actual, non-rescaled data. In particular, all the events with more than 20 million victims have occurred during the last 150 years, and the average inter-arrival time is below 20 years. Are we really living in more peaceful world?

[9] In case of rescaled amounts, inter-arrival times are shorter, but the interpretation is the same.

The parameter $\xi$ is the most important for us: it is the parameter governing the fatness of the right tail. A $\xi$ greater than 1 (we have 1.5886) signifies that no moment is defined for our Generalized Pareto: a very fat-tailed situation. Naturally, in the sample, we can compute all the moments we are interested in, but from a theoretical point of view they are completely unreliable and their interpretation is extremely flawed (a very common error though). According to our fitting, very catastrophic events are not at all improbable. It is worth noticing that the estimates is significant, given that its standard error is 0.1467.

Figures 18.9 and 18.10 compare our fittings to actual data. In both figures it is possible to see the goodness of the GPD fitting for most of the observations above the 50k victims threshold. Some problems arise for the very large events, like WW2 and the An Lushan rebellion [10]. In this case it appears that our fitting expects larger events to have happened. This is a well-known problem for extreme data [95]. The very large event could just be behind the corner.

Similarly, events with 5 to 10 million victims (not at all minor ones!) seem to be slightly more frequent than what is expected by our GPD fitting. This is another signal of the extreme character of war casualties, which does not allow for the extrapolation of simplistic trends.

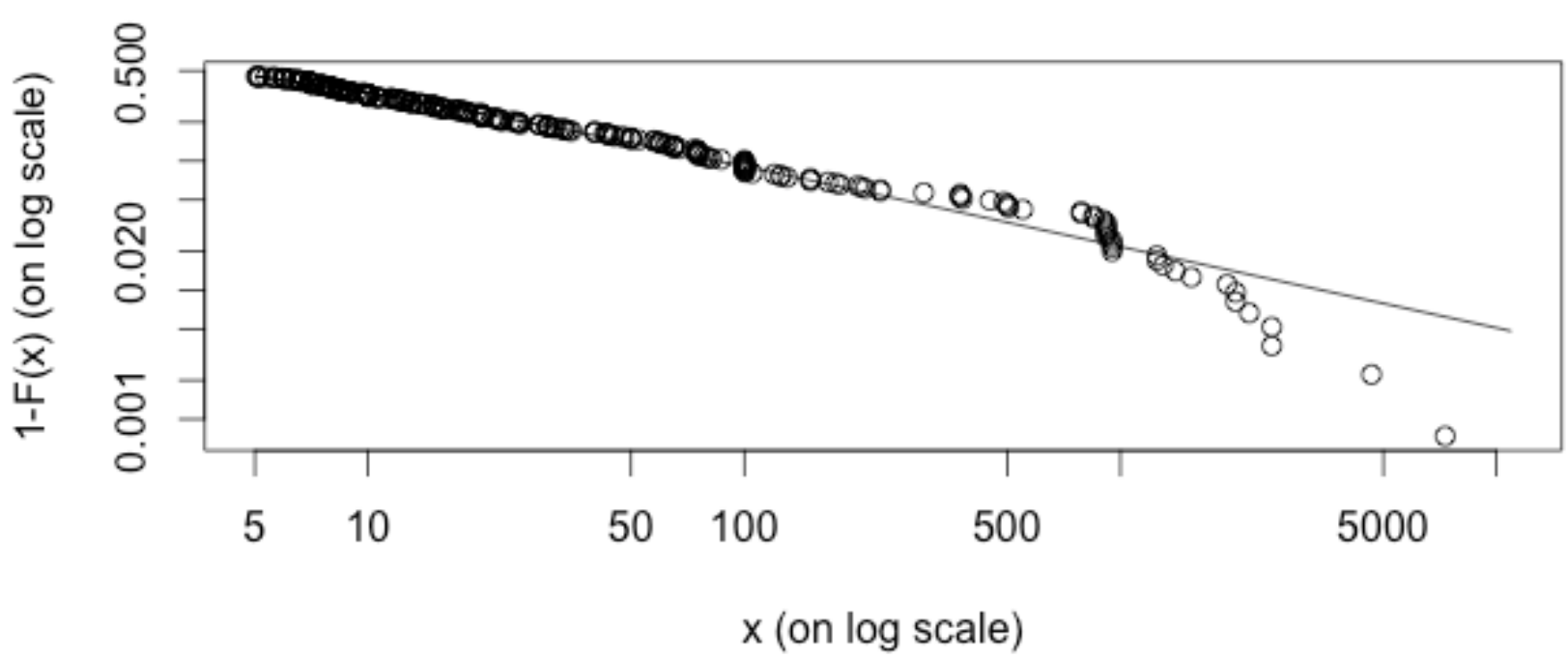

Figure 18.9: *GPD tail fitting to actual casualties' data (in 10k). Parameters as per Table 18.4, first line.*

[10] If we remove the two largest events from the data, the GPD hypothesis cannot be rejected at the 5% significance level.

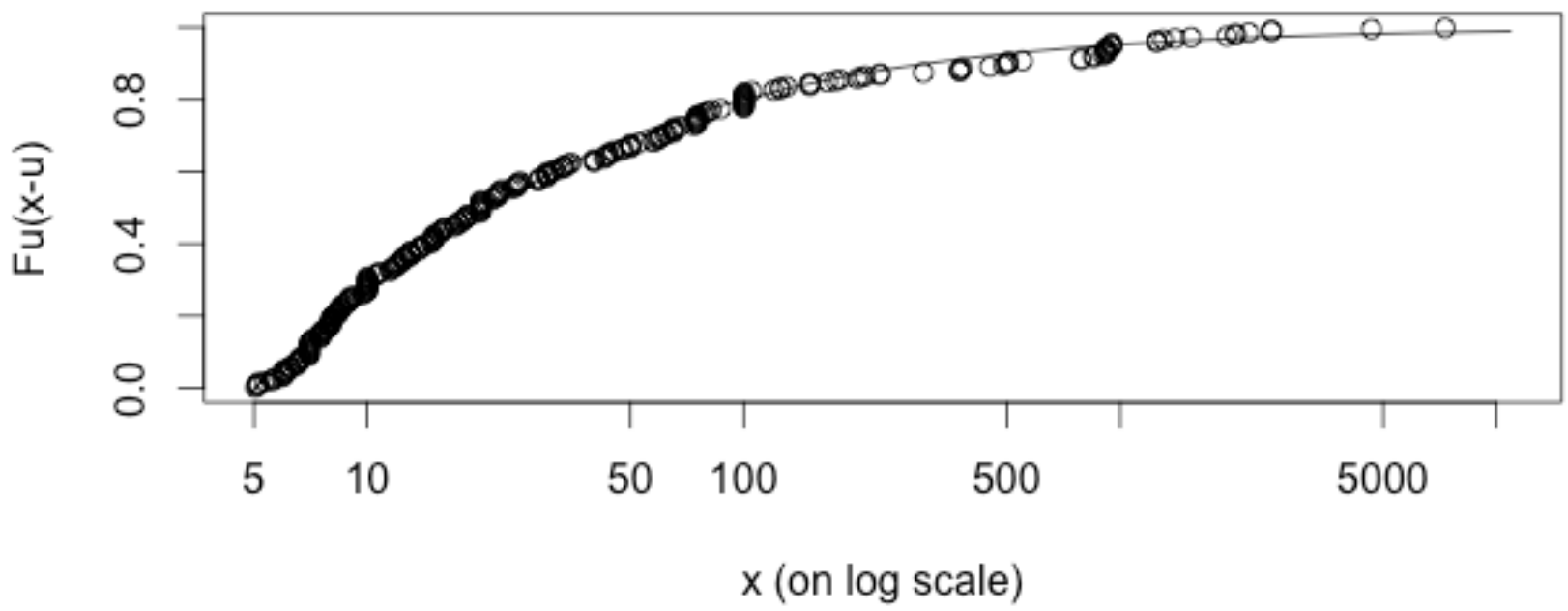

Figure 18.10: *GPD cumulative distribution fitting to actual casualties' data (in 10k). Parameters as per Table 18.4, first line.*

### 18.4.4 An Alternative View on Maxima

Another method is the block-maxima approach of extreme value theory. In this approach data are divided into blocks, and within each block only the maximum value is taken into consideration. The Fisher-Tippet theorem [95] then guarantees that the normalized maxima converge in distribution to a Generalized Extreme Value Distribution, or GEV.

$$GEV(x;\xi) = \begin{cases} \exp\left(-(1+\xi x)^{-\frac{1}{\xi}}\right) & \xi \neq 0 \\ \exp\left(-\exp(-x)\right) & \xi = 0 \end{cases}, \quad 1+\xi x > 0$$

This distribution is naturally related to the GPD, and we refer to [95] for more details.

If we divide our data into 100-year blocks, we obtain 21 observation (the last block is the residual one from 2001 to 2014). Maximum likelihood estimations give a $\xi$ larger than 2, indicating that we are in the so-called Fréchet maximum domain of attraction, compatible with very heavy-tailed phenomena. A value of $\xi$ greater than 2 under the GEV distribution further confirms the idea of the absence of moments, a clear signal of a very heavy right tail.

### 18.4.5 Full Data Analysis

Naturally, being aware of limitations, we can try to fit all our data, while for casualties in excess of 10000, we fit the Pareto Distribution from Equation 18.3 with $\alpha \approx 0.53$ throughout. The goodness of fit for the "near tail" (L=10K) can be see in Figure 18.2. Similar results to Figure 18.2 are seen for different values in table below, all with the same goodness of fit.

| $L$ | $\sigma$ |
|---|---|
| $10K$ | $84,260$ |
| $25K$ | $899,953$ |
| $50K$ | $116,794$ |
| $100K$ | $172,733$ |
| $200K$ | $232,358$ |
| $500K$ | $598,292$ |

The different possible values of the mean in Equation 18.4 can be calculated across different set values of $\alpha$, with one single degree of freedom: the corresponding $\sigma$ is a MLE estimate using such $\alpha$ as fixed: for a sample size $n$, and $x_i$ the observations higher than $L$, $\sigma_\alpha = \left\{\sigma : \frac{\alpha n}{\sigma} - (\alpha+1)\sum_{i=1}^{n} \frac{1}{x_i - L + \sigma} = 0, \sigma > 0\right\}$.

The sample average for $L = 10K$ is $9.12 \times 10^6$, across 100K simulations, with the spread in values showed in Figure 18.15.

The "true" mean from Equation 18.4 yields $3.1 * 10^7$ , and we repeated for $L$ =10K, 20K, 50K, 100K, 200K, and 500K, finding ratios of true estimated mean to observed safely between 3 and 4., see Table 18.1. Notice that this value for the mean of $\approx 3.5$ times the observed sample mean is only a general guideline, since, being stochastic, does not reveal any precise information other than prevent us from taking the naive mean estimation seriously.

For under fat tails, the mean derived from estimates of $\alpha$ is more rigorous and has a smaller error, since the estimate of $\alpha$ is asymptotically Gaussian while the average of a power law, even when it exists, is considerably more stochastic. See the discussion on "slowness of the law of large numbers" in 8 in connection with the point.

We get the mean by truncation for L=10K a bit lower, under equation 18.6; around $1.8835 \times 10^7$.

We finally note that, for values of $L$ considered, 96 % of conflicts with more than 10,000 victims are below the mean: where $m$ is the mean,

$$\mathbb{P}(X < m) = 1 - \left(1 - \frac{H \log\left(\alpha e^{\sigma/H} E_{\alpha+1}\left(\frac{\sigma}{H}\right)\right)}{\sigma}\right)^{-\alpha}.$$

## 18.5 ADDITIONAL ROBUSTNESS AND RELIABILITY TESTS

### 18.5.1 Bootstrap for the GPD

In order to check our sensitivity to the quality/precision of our data, we have decided to perform some bootstrap analysis. For both raw data and the rescaled ones we have generated 100K new samples by randomly selecting 90% of the observations, with replacement. Figures 18.11, 18.12 and 18.13 show the stability of our $\xi$ estimates. In particular $\xi > 0$ in all samples, indicating the extreme fat-tailedness of the number of victims in armed conflicts. The $\xi$ estimates in Table 18.4 appear to be good approximations for our GPD real shape parameters, notwithstanding imprecisions and missing observations in the data.

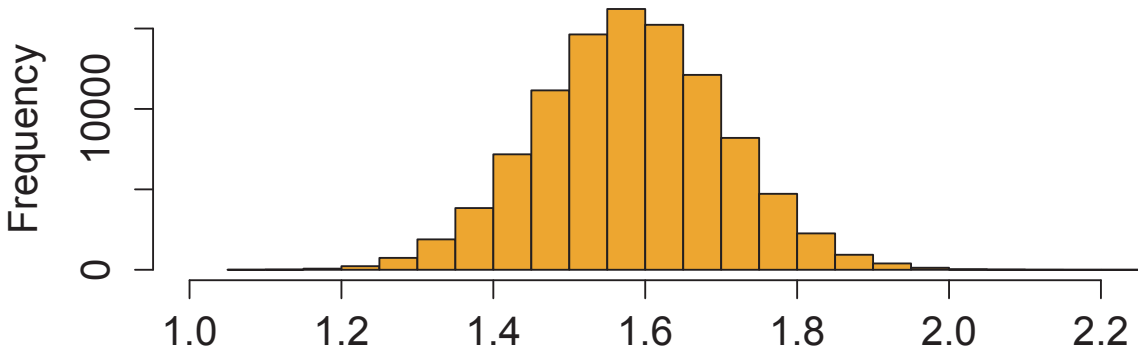

*Figure 18.11: $\xi$ parameter's distribution over 100K bootstrap samples for actual data. Each sample is randomly selected with replacement using 90% of the original observations.*

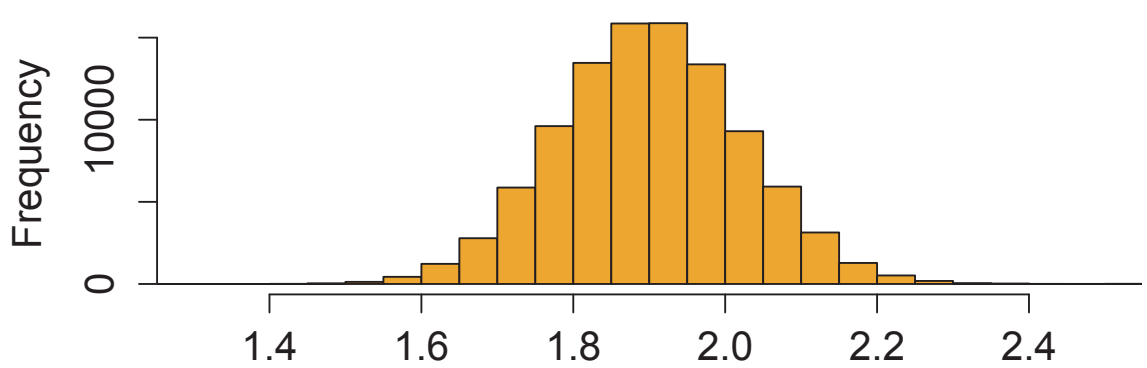

*Figure 18.12: $\xi$ parameter's distribution over 100K bootstrap samples for naively rescaled data. Each sample is randomly selected with replacement using 90% of the original observations.*

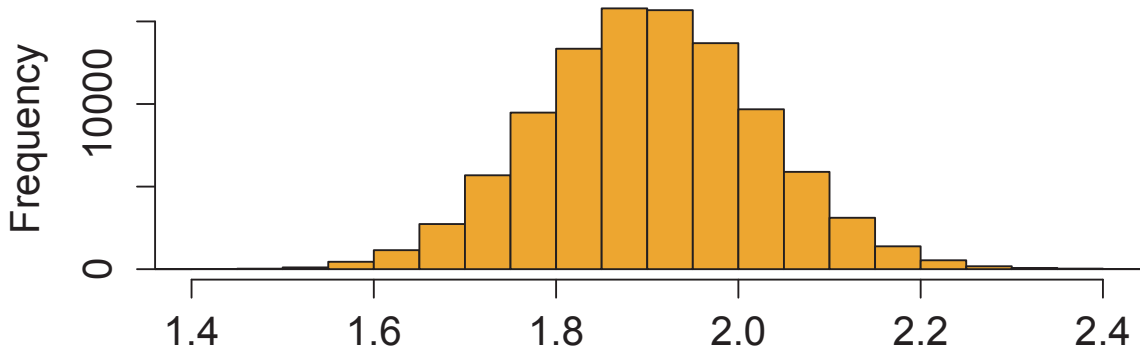

*Figure 18.13: $\xi$ parameter's distribution over 100K bootstrap samples for log-rescaled data. Each sample is randomly selected with replacement using 90% of the original observations.*

### 18.5.2 Perturbation Across Bounds of Estimates

We performed analyses for the "near tail" using the Monte Carlo techniques discussed in section 18.3.3. We look at second order "p-values", that is the sensitivity of the p-values across different estimates in Figure 18.14 –practically all results meet the same statistical significance and goodness of fit.

In addition, we look at values of both the sample means and the alpha-derived MLE mean across permutations, see Figures 18.15 and 18.16.

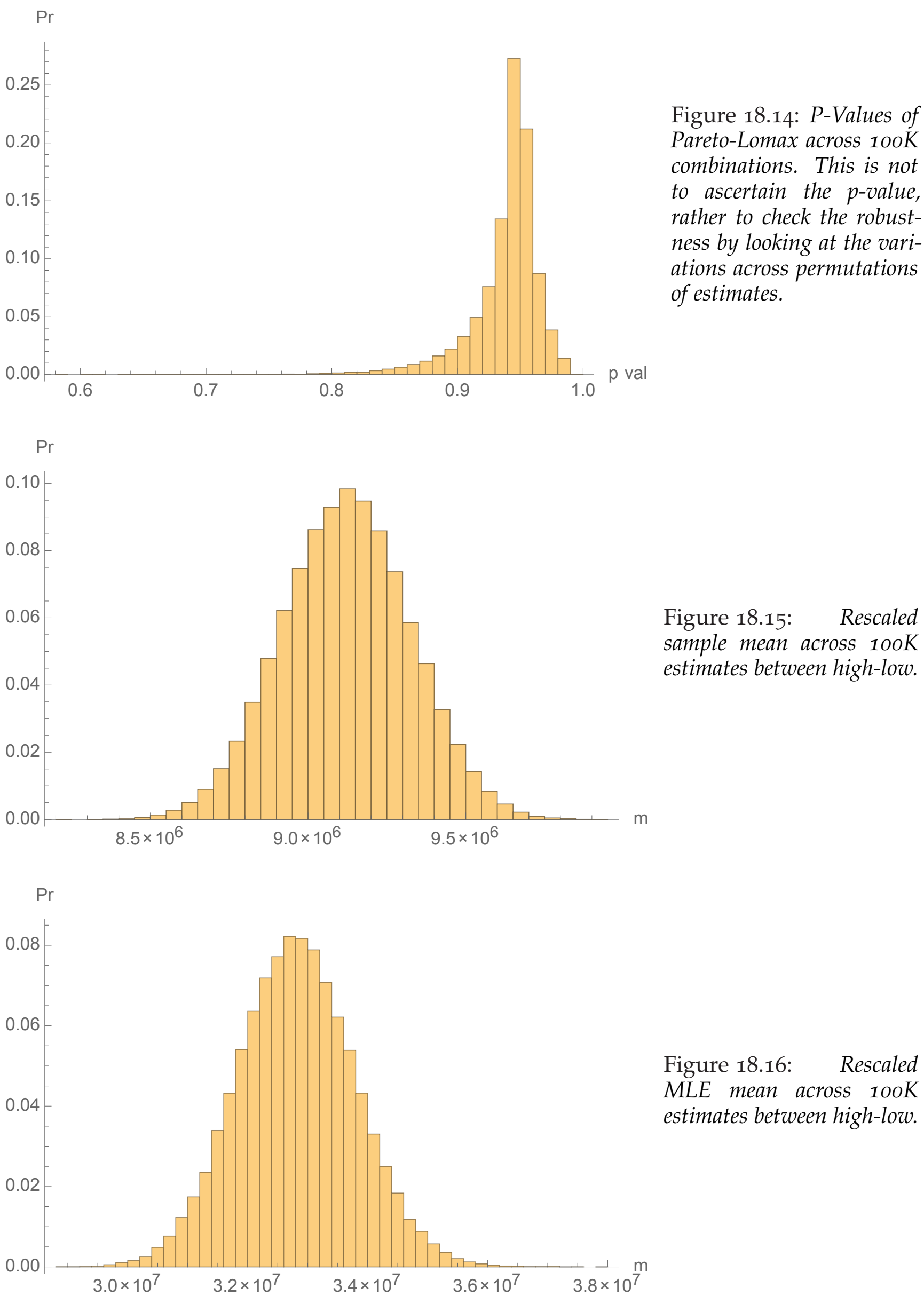

Figure 18.14: *P-Values of Pareto-Lomax across 100K combinations. This is not to ascertain the p-value, rather to check the robustness by looking at the variations across permutations of estimates.*

Figure 18.15: *Rescaled sample mean across 100K estimates between high-low.*

Figure 18.16: *Rescaled MLE mean across 100K estimates between high-low.*

## 18.6 CONCLUSION: IS THE WORLD MORE UNSAFE THAN IT SEEMS?

To put our conclusion in the simplest of terms: the occurrence of events that would raise the average violence by a multiple of 3 would not cause us to rewrite this chapter, nor to change the parameters calibrated within.

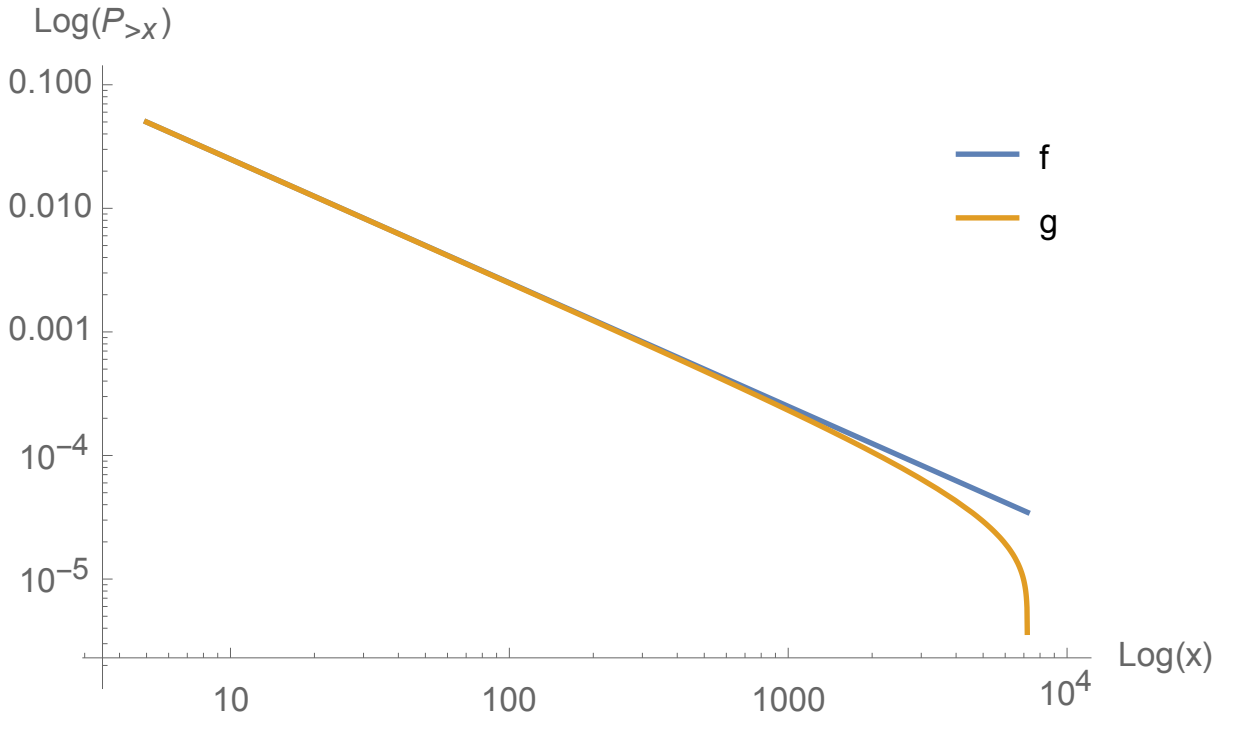

Figure 18.17: *Loglogplot comparison of $f$ and $g$, showing a pasting-boundary style capping around $H$.*

- Indeed, from statistical analysis alone, the world is more unsafe than casually examined numbers. Violence is underestimated by journalistic nonstatistical looks at the mean and lack of understanding of the stochasticity of under inter-arrival times.
- The transformation into compact support allowed us to perform the analyses and gauge such underestimation which , if noisy, gives us an idea of the underestimation and its bounds.
- In other words, a large event and even a rise in observed mean violence would not be inconsistent with statistical properties, meaning it would justify a "nothing has changed" reaction.
- We avoided discussions of homicide since we limited $L$ to values $> 10,000$, but its rate doesn't appear to have a particular bearing on the tails. It could be a drop in the bucket. It obeys different dynamics. We may have observed lower rate of homicide in societies but most risks of death come from violent conflict. (Casualties from homicide by rescaling from the rate 70 per $100k$, gets us $5.04 \times 10^6$ casualties per annum at today's population. A drop to minimum levels stays below the difference between errors on the mean of violence from conflicts with higher than 10,000 casualties.)
- We ignored survivorship bias in the data analysis (that is, the fact that *had the world been more violent, we wouldn't be here to talk about it*). Adding it would increase the risk. The presence of tail effects today makes further analysis require taking it into account. Since 1960, a single conflict –which almost happened– has the ability to reach the max casualties, something we did not have before. (We can rewrite the model with one of fragmentation of the world, constituted of "separate" isolated $n$ independent random variables $X_i$, each with a maximum value $H_i$, with the total $\sum_n \omega_i H_i = H$, with all $w_i > 0$, $\sum_n \omega_i = 1$. In that case the maximum (that is worst conflict) could require the joint probabilities that all $X_1, X_2, \cdots X_n$ are near their maximum value, which, under subexponentiality, is an event of much lower probability than having a single variable reach its maximum.)[11]

[11] How long do we have to wait before making a *scientific* pronouncement about the drop in the incidence of wars of a certain magnitude? Simply, because inter-arrival time follows a memoryless exponential distribution, roughly the survival function of a deviation of three times the mean is $e^{-3} \approx .05$.

## 18.7 ACKNOWLEDGMENTS

The data was compiled by Captain Mark Weisenborn. We thank Ben Kiernan for comments on East Asian conflicts.

It means wait for three times as long as the mean inter-arrival times before saying something scientific. For large wars such as WW1 and WW2, wait 300 years. *It is what it is*.

# H | WHAT ARE THE CHANCES OF A THIRD WORLD WAR?[*,†]

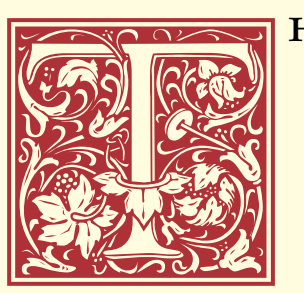
HIS IS FROM AN ARTICLE that is part of the debate with public intellectuals who claim that violence have dropped "from data", without realizing that science is hard; significance requires further data under fat tails and more careful examination. Our response (by the author and P. Cirillo) provides a way to summarize the main problem with naive empiricism under fat tails.

In a recent issue of *Significance* Mr. Peter McIntyre asked what the chances are that World War III will occur this century. Prof. Michael Spagat wrote that nobody knows, nobody can really answer–and we totally agree with him on this. Then he adds that "a really huge war is possible but, in my view, extremely unlikely." To support his statement, Prof. Spagat relies partly on the popular science work of Prof. Steven Pinker, expressed in The Better Angels of our Nature and journalistic venues. Prof. Pinker claims that the world has experienced a long-term decline in violence, suggesting a structural change in the level of belligerence of humanity.

It is unfortunate that Prof. Spagat, in his answer, refers to our paper (this volume, Chapter 18 ), which is part of a more ambitious project we are working on related to fat-tailed variables.

What characterizes fat tailed variables? They have their properties (such as the mean) dominated by extreme events, those "in the tails". The most popularly known version is the "Pareto 80/20".

We show that, simply, data do not support the idea of a structural change in human belligerence. So Prof. Spagat's first error is to misread our claim: we are making neither pessimistic nor optimistic declarations: we just believe that statisticians should abide by the foundations of statistical theory and avoid telling data what to say.

Let us go back to first principles.

---

Discussion chapter.

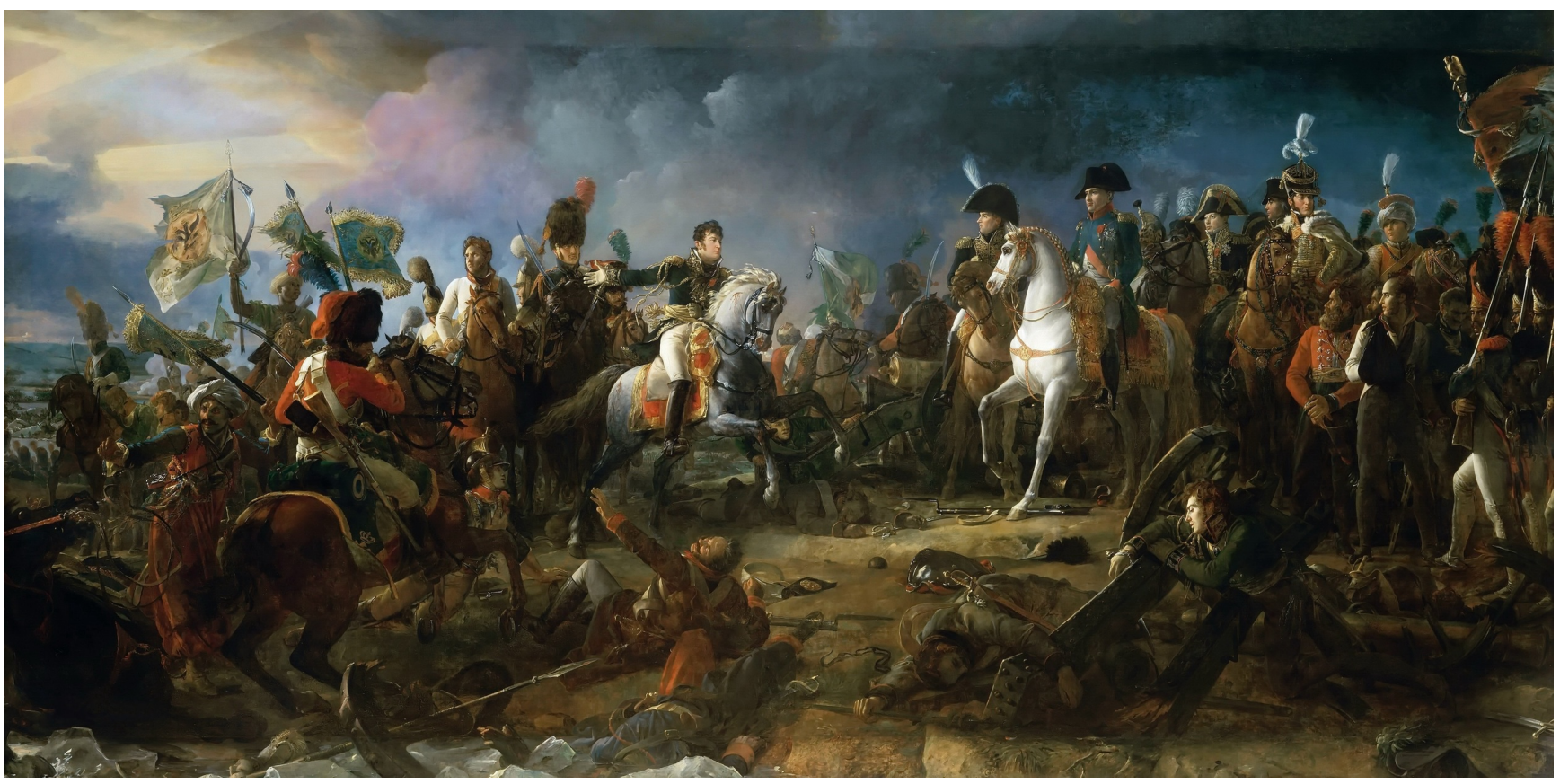

Figure H.1: *After Napoleon, there was a lull in Europe. Until nationalism came to change the story.*

**Foundational Principles**

Fundamentally, statistics is about ensuring people do not build scientific theories from hot air, that is without significant departure from random. Otherwise, it is patently "fooled by randomness".

Further, for fat tailed variables, the conventional mechanism of the law of large numbers is considerably slower and significance requires more data and longer periods. Ironically, there are claims that can be done on little data: inference is asymmetric under fat-tailed domains. ***We require more data to assert that there are no Black Swans than to assert that there are Black Swans*** hence we would need much more data to claim a drop in violence than to claim a rise in it.

Finally, statements that are not deemed statistically significant –and shown to be so –should never be used to construct scientific theories.

These foundational principles are often missed because, typically, social scientists' statistical training is limited to mechanistic tools from thin tailed domains [2]. In physics, one can often claim evidence from small data sets, bypassing standard statistical methodologies, simply because the variance for these variables is low. The higher the variance, the more data one needs to make statistical claims. For fat-tails, the variance is typically high and underestimated in past data.

The second –more serious –error Spagat and Pinker made is to believe that tail events and the mean are somehow different animals, not realizing that the mean includes these tail events.

> ***For fat-tailed variables, the mean is almost entirely determined by extremes. If you are uncertain about the tails, then you are uncertain about the mean.***

It is thus incoherent to say that violence has dropped but maybe not the risk of tail events; it would be like saying that someone is "extremely virtuous except during the school shooting episode when he killed 30 students".

### Robustness

Our study tried to draw the most robust statistical picture of violence, relying on methods from extreme value theory and statistical methods adapted to fat tails. We also put robustness checks to deal with the imperfection of data collected some thousand years ago: our results need to hold even if a third (or more) of the data were wrong.

### Inter-arrival times

We show that the inter-arrival times among major conflicts are extremely long, and consistent with a homogenous Poisson process: therefore no specific trend can be established: we as humans can not be deemed as less belligerent than usual. For a conflict generating at least 10 million casualties, an event less bloody than WW1 or WW2, the waiting time is on average 136 years, with a mean absolute deviation of 267 (or 52 years and 61 deviations for data rescaled to today's population). The seventy years of what is called the "Long Peace" are clearly not enough to state much about the possibility of WW3 in the near future.

### Underestimation of the mean

We also found that the average violence observed in the past underestimates the true statistical average by at least half. Why? Consider that about 90-97% of the observations fall below the mean, which requires some corrections with the help of extreme value theory. (Under extreme fat tails, the statistical mean can be closer to the past maximum observation than sample average.)

### A common mistake

Similar mistakes have been made in the past. In 1860, one H.T. Buckle[2] used the same unstatistical reasoning as Pinker and Spagat.

> That this barbarous pursuit is, in the progress of society, steadily declining, must be evident, even to the most hasty reader of European history. If we compare one country with another, we shall find that for a very long period wars have been becoming less frequent; and now so clearly is the movement marked, that, until the late commencement of hostilities, we had remained at peace for nearly forty years: a circumstance unparalleled (...) The question arises, as to what share our moral feelings have had in bringing about this great improvement.

Moral feelings or not, the century following Mr. Buckle's prose turned out to be the most murderous in human history.

We conclude by saying that we find it fitting –and are honored –to expose fundamental statistical mistakes in a journal called *Significance*, as the problem is precisely about significance and conveying notions of statistical rigor to the general public.

[2] Buckle, H.T. (1858) *History of Civilization in England, Vol. 1*, London: John W. Parker and Son.

# 19 | TAIL RISK OF CONTAGIOUS DISEASES

Applying a modification of Extreme value Theory (thanks to a dual distribution technique by the authors in [51]) on data over the past 2,500 years, we show that pandemics are extremely fat-tailed in terms of fatalities, with a marked potentially existential risk for humanity.

Such a macro property should invite the use of Extreme Value Theory (EVT) rather than naive interpolations and expected averages for risk management purposes. An implication is that potential tail risk overrides conclusions on decisions derived from compartmental epidemiological models and similar approaches.

## 19.1 INTRODUCTION AND POLICY IMPLICATIONS

We examine the distribution of fatalities from major pandemics in history (spanning about 2,500 years), and build a statistical picture of their tail properties. Using tools from Extreme Value Theory (EVT), we show for that the distribution of the victims of infectious diseases is extremely fat-tailed, more than what one could be led to believe from the outset[1].

A non-negative continuous random variable $X$ is fat-tailed, in the regular variation class, if its survival function $S(x) = P(X \geq x)$ decays as a power law $x^{-\frac{1}{\xi}}$, the more we move into the tail[2], that is for $x$ growing towards the right endpoint of $X$. The parameter $\xi$ is known as the tail parameter, and it governs the fatness of the tail (the larger $\xi$ the fatter the tail) and the existence of moments ($E[X^p] < \infty$ if and only if $\xi < 1/p$). In some literature, e.g. [58], the tail index is re-parametrized as $\alpha = 1/\xi$, and its interpretation is naturally reversed.

[1] In this comment we do not discuss the possible generating mechanisms behind these fat tails, a topic of separate research. Networks analysis, e.g. [3], proposes mechanisms for the spreading of contagion and the existence of super spreaders, a plausible joint cause of fat tails. Likewise simple automata processes can lead to high uncertainty of outcomes owing to "computational irreducibility" [? ].

[2] More technically, a non-negative continuous random variable $X$ has a fat-tailed distribution (in the maximum domain of attraction of the Fréchet distribution), if its survival function is regularly varying, i.e. $S(x) = L(x)x^{-\frac{1}{\xi}}$, where $L(x)$ is a slowly varying function, such that $\lim_{x\to\infty} \frac{L(cx)}{L(x)} = 1$ for $c > 0$ [70, 96].

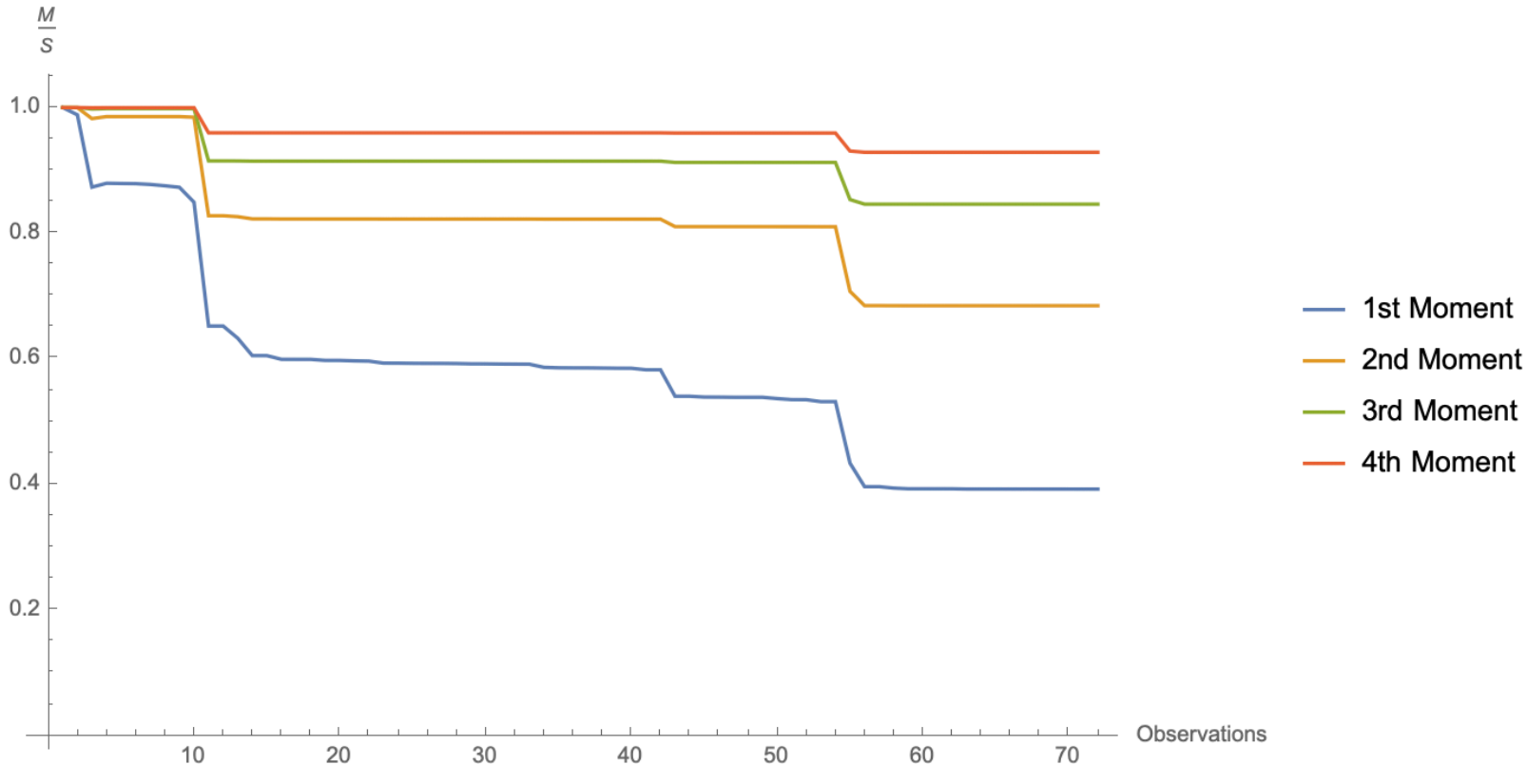

Figure 19.1: *Maximum to Sum plot (MS plot) of the average death numbers in pandemic events in history, as per Table 1.*

While it is known that fat tails represent a common–yet often ignored [267] in modeling–regularity in many fields of science and knowledge [58], for the best of our knowledge, only war casualties and operational risk losses show a behavior [50, 51, 209] as erratic and wild as the one we observe for pandemic fatalities.

The core of the problem is shown in Figure 19.1, with the Maximum-to-Sum plot [97] of the number of pandemic fatalities in history (data in Table 1). Such a plot relies on a simple consequence of the law of large numbers: for a sequence $X_1, X_2, ..., X_n$ of non-negative i.i.d. random variables, if $E[X^p] < \infty$ for $p = 1, 2, 3...$, then $R_n^p = M_n^p / S_n^p \to^{a.s.} 0$ as $n \to \infty$, where $S_n^p = \sum_{i=1}^{n} X_i^p$ is the partial sum of order $p$, and $M_n^p = \max(X_1^p, ..., X_n^p)$ the corresponding partial maximum. Figure 19.1 clearly shows that no finite moment is likely to exist for the number of victims in pandemics, as the $R_n$ ratio does not converge to 0 for $p = 1, 2, 3, 4$, no matter how many data points we use. Such a behavior hints that the victims distribution has such a fat right tail that not even the first theoretical moment is finite. We are looking at a phenomenon for which observed quantities such as the naive sample average and standard deviation are therefore meaningless for inference.

However, Figure 19.1 (or a naive use of EVT) does not imply that pandemic risk is actually infinite and there is nothing we can do or model. Using the methodology we developed to study war casualties [51, 268], we are in fact able to extract useful information from the data, quantifying the large yet finite risk of pandemic diseases. The method provides in fact rough estimates for quantities not immediately observable in the data.

**The tail wags the dog effect**

Centrally, the more fat-tailed the distribution, the more "the tail wags the dog", that is, the more statistical information resides in the extremes and the less in the "bulk" (that is the events of high frequency), where it becomes almost noise. This makes EVT the most effective approach, and our sample of extremes very highly sufficient and informative for risk management purposes[3].

The fat-tailedness of the distribution of pandemic fatalities has the following policy implications, useful in the wake of the Covid-19 pandemic.

First, it should be evident that one cannot compare fatalities from multiplicative infectious diseases (fat-tailed, like a Pareto) to those from car accidents, heart attacks or falls from ladders (thin-tailed, like a Gaussian). Yet this is a common (and costly) error in policy making, and in both the decision science and the journalistic literature[4]. Some research papers even criticise people's "paranoïa" with respect to pandemics, not understanding that such a paranoïa is merely responsible (and realistic) risk management in front of potentially destructive events [267]. The main problem is that those articles–often relied upon for policy making –consistently use the wrong thin-tailed distributions, underestimating tail risk, so that every conservative or preventative reaction is bound to be considered an overreaction.

Second, epidemiological models like the SIR [149] differential equations, sometimes supplemented with simulative experiments like [109], while useful for scientific discussions for the bulk of the distributions of infections and deaths, or to understand the dynamics of events after they happened, should never be used for precautionary risk management, which should focus on maxima and tail exposures instead. It is highly unrigorous to use naive (and reassuring) statistics, like the expected average outcome of compartmental models, or one or more point estimates, as a motivation for policies. Owing to the compounding effect of parameters' uncertainty, the "tail wagging the dog" effect easily invalidates both point estimates and scenario analyses[5].

EVT is the natural candidate to handle pandemics. It was born to cope with maxima [101], and it evolved to deal with tail risk in a robust way, even with a limited number of observations and the uncertainty associated with it [97]. In the Netherlands, for example, EVT was used to get a handle on the distribution of the maxima–not the average!–of sea levels in order to build dams and dykes high and strong enough for the safety of citizens [70].

Finally, EVT-based risk management is compatible with the (non-naïve) precautionary principle of [159], which should be the leading driver for policy decisions under jointly systemic and extreme risks.

---

3 Since the law of large numbers works slowly under fat tails, the bulk becomes increasingly dominated by noise, and averages and higher moments–even when they exist–become uninformative and unreliable, while extremes are rich in information [267].

4 Sadly, this mistake is sometimes made by professional statisticians as well. Thin tailed (discrete) variables are subjected to Chernov bounds, unlike fat-tailed ones [267].

5 The current Covid-19 pandemic is generating a lot of research, and finally some scholars are looking at the impact of parameters' uncertainty on the scenarios generated by epidemiological models, e.g. [82].

## 19.2 DATA AND DESCRIPTIVE STATISTICS

We investigate the distribution of deaths from the major epidemic and pandemic diseases of history, from 429 BC until now. The data are available in Table 1, together with their sources, and only refer to events with more than 1K estimated victims, for a total of 72 observations. As a consequence, potentially high-risk diseases, like the Middle East Respiratory Syndrome (MERS), do not appear in our collection[6]. All diseases whose end year is 2020 are to be taken as still occurring worldwide, as for the running COVID-19 pandemic.

Three estimates of the reported cumulative death toll have been used: minimum, average and maximum. When the three numbers coincide in Table 1, our sources simply do not provide intervals for the estimates. Since we are well aware of the volatility and possible unreliability of historical data [249, 268], in Section 19.4 we deal with such an issue by perturbing and omitting observations.

In order to compare fatalities with respect to the coeval population (that is, the relative impact of pandemics), column *Rescaled* of Table 1 provides the rescaled version of column *Avg Est*, using the information in column *Population*[7] [130, 131, 205]. For example, the Antonine plague of 165-180 killed an average of 7.5M people, that is to say 3.7% of the coeval world population of 202M people. Using today's population, such a number would correspond to about 283M deaths, a terrible hecatomb, killing more people than WW2.

For space considerations, we restrict our attention to the actual average estimates in Table 1, but all our findings and conclusions hold true for the lower, the upper and the rescaled estimates as well[8].

Figure 19.2 shows the histogram of the actual average numbers of deaths in the 72 large contagious events. The distributions appears highly skewed and possibly fat-tailed. The numbers are as follows: the sample average is 4.9M, while the median is 76K, compatibly with the skewness observable in Figure 19.2. The 90% quantile is 6.5M and the 99% quantile is 137.5M. The sample standard deviation is 19M.

Using common graphical tools for fat tails [97], in Figure 19.3 we show the log log plot (also known as Zipf plot) of the empirical survival functions for the average victims over the diverse contagious events. In such a plot possible fat tails can be identified in the presence of a linearly decreasing behavior of the plotted curve. To improve interpretability a naive linear fit is also proposed. Figure 19.3 suggests the presence of fat tails.

The Zipf plot shows a necessary but not sufficient condition for fat-tails [49]. Therefore, in Figure 19.4 we complement the analysis with a mean excess function plot, or meplot. If a random variable $X$ is possibly fat-tailed, its mean excess function $e_X(u) = E[X - u | X \geq u]$ should grow linearly in the threshold $u$, at least above a certain value identifying the actual power law tail [97]. In a meplot, where the empirical $e_X(u)$ is plotted against the different values of $u$, one thus looks for some (more or less) linearly increasing trend, as the one we observe in Figure 19.4.

[6] Up to the present, MERS has killed 858 people as reported in https://www.who.int/emergencies/mers-cov/en. For SARS the death toll is between 774 and 916 victims until now https://www.nytimes.com/2003/10/05/world/taiwan-revises-data-on-sars-total-toll-drops.html.

[7] Population estimates are by definitions estimates, and different sources can give different results (most of the times differences are minor), especially for the past. However our methodology is robust to this type of variability, as we stress later in the paper.

[8] The differences in the estimates do not change the main message: we are dealing with an extremely erratic phenomenon, characterised by very fat tails.

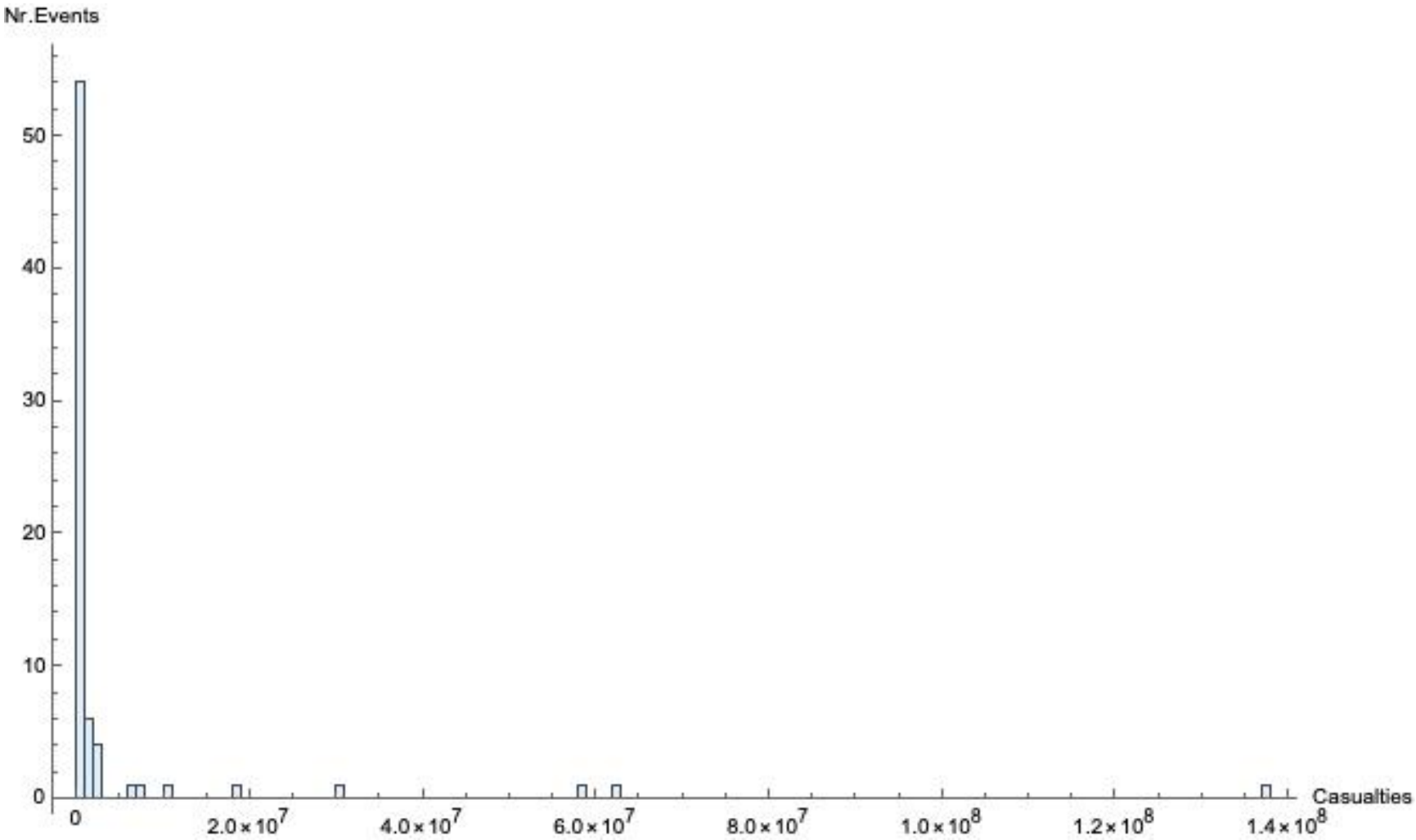

Figure 19.2: *Histogram of the average number of deaths in the 72 contagious diseases of Table 1.*

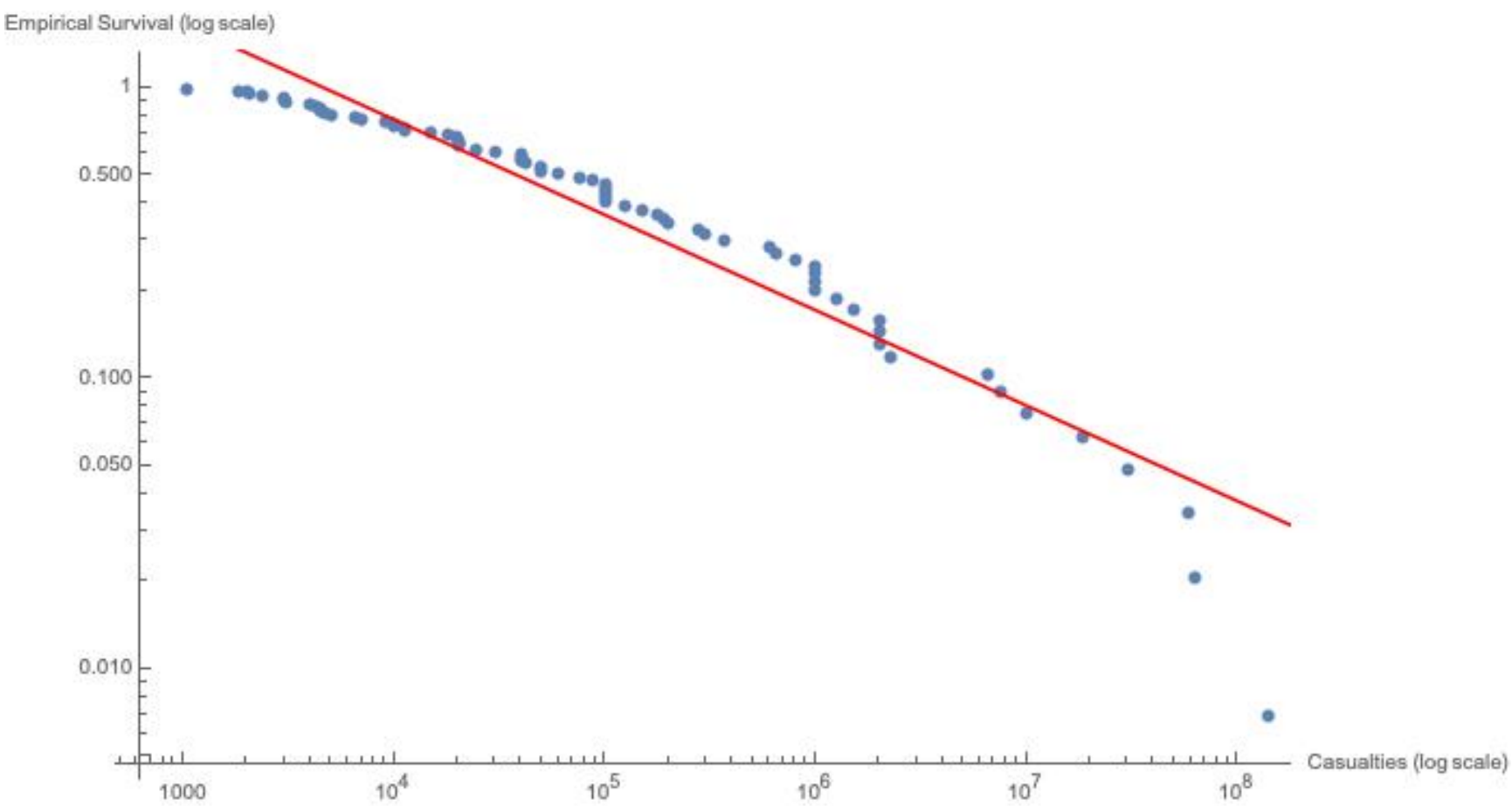

Figure 19.3: *Log log plot of the empirical survival function (Zipf plot) of the actual average death numbers in Table 1. The red line represents a naive linear fit of the decaying tail.*

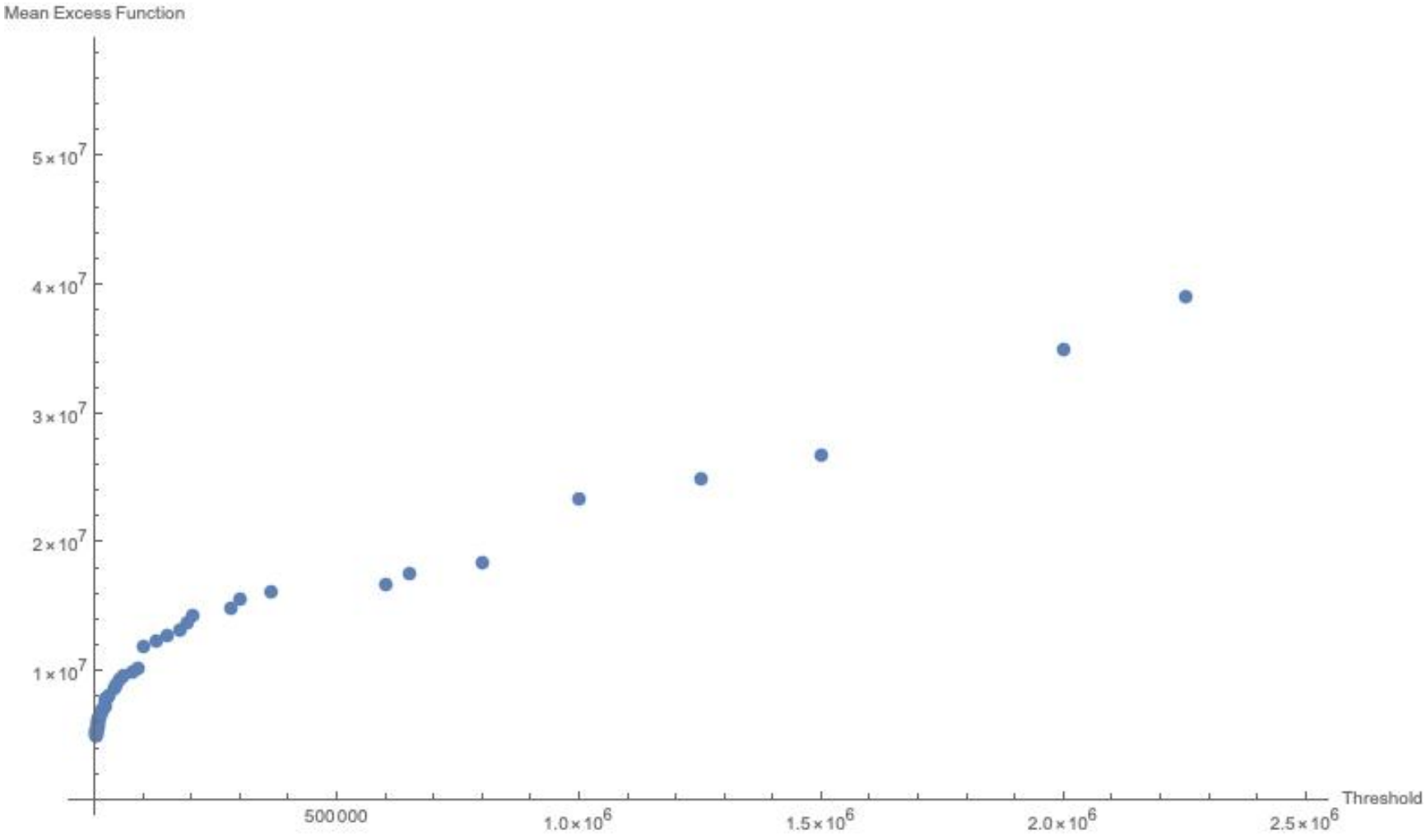

Figure 19.4: *Mean excess function plot (meplot) of the average death numbers in Table 1. The plot excludes 3 points on the top right corner, consistently with the suggestions in [97] about the exclusion of the more volatile observations.*

A useful tool for the analysis of tails–when one suspects them to be fat–is the nonparametric Hill estimator [97]. For a collection $X_1, ..., X_n$, let $X_{n,n} \leq ... \leq X_{1,n}$ be the corresponding order statistics. Then we can estimate the tail parameter $\xi$ as

$$\hat{\xi} = \frac{1}{k} \sum_{i=1}^{k} \log(X_{i,n}) - \log(X_{k,n}), \qquad 2 \leq k \leq n.$$

In Figure 19.5, $\hat{\xi}$ is plotted against different values of $k$, creating the so-called Hill plot [97]. The plot suggests $\xi > 1$, in line with Figure 19.1, further supporting the evidence of infinite moments.

Other graphical tools could be used and they would all confirm the point: we are in the presence of fat tails in the distribution of the victims of pandemic diseases. Even more, a distribution with possibly no finite moment.

### The dual distribution

As we observed for war casualties [51], the non-existence of moments for the distribution of pandemic victims is questionable. Since the distribution of victims is naturally bounded by the coeval world population, no disease can kill more people than those living on the planet at a given time time. We are indeed looking at an *apparently* infinite-mean phenomenon, like in the case of war casualties [51, 268] and operational risk [50].

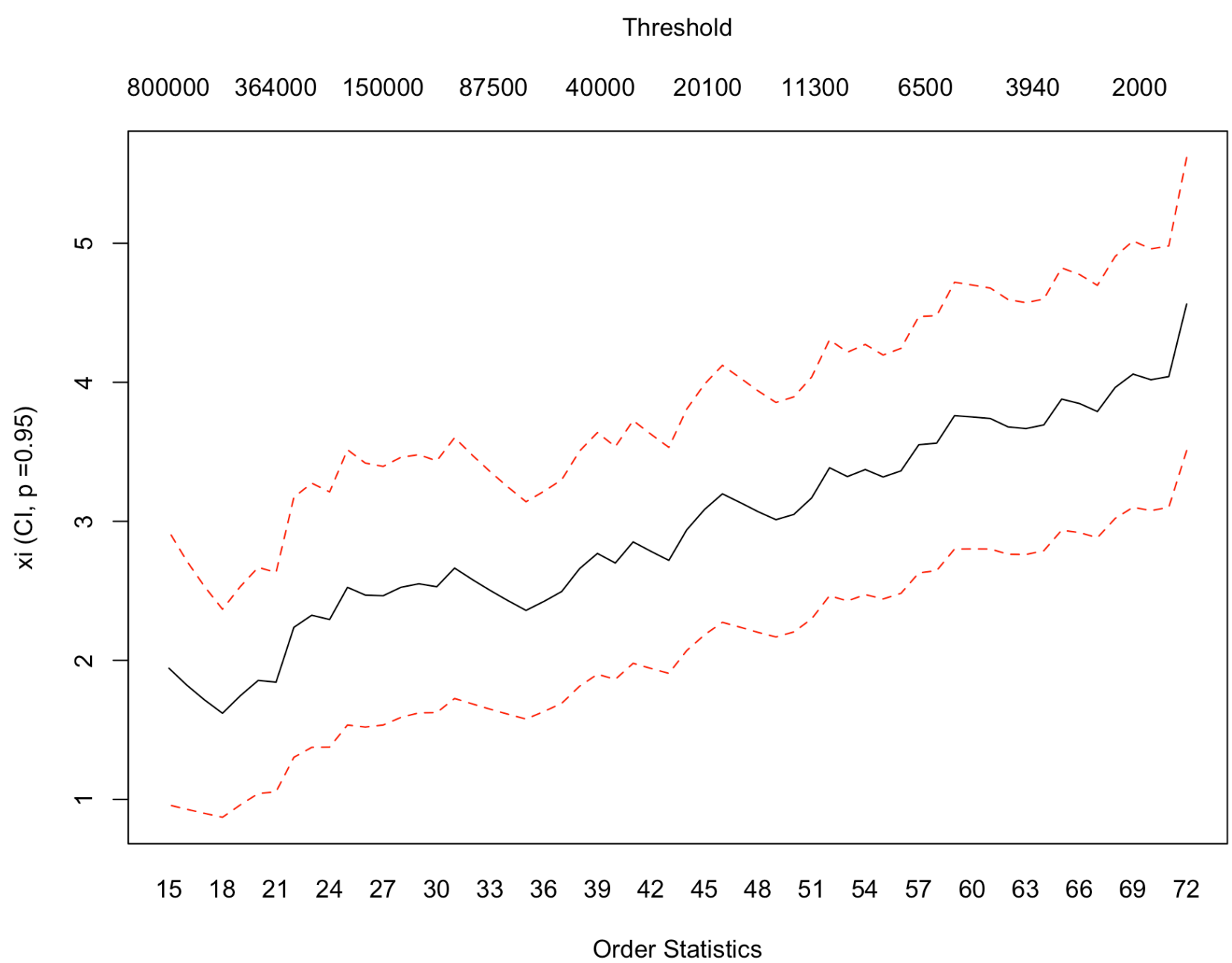

Figure 19.5: *Hill plot of the average death numbers in Table 1, with 95% confidence intervals. Clearly* $\xi > 1$, *suggesting the non-existence of moments.*

Let $[L, H]$ be the support of the distribution of pandemic victims today, with $L >> 0$ to ignore small events not officially definable as pandemic [215]. For what concerns $H$, its value cannot be larger than the world population, i.e. 7.7 billion people in 2020[9]. Evidently $H$ is so large that the probability of observing values in its vicinity is in practice zero, and one always finds observations below a given $M << H < \infty$ (something like 150M deaths using actual data). Thus one could be fooled by data into ignoring $H$ and taking it as infinite, up to the point of believing in an infinite mean phenomenon, as Figure 19.1 suggests. However notice that a finite upper bound $H$–no matter how large it is–is not compatible with infinite moments, hence Figure 19.1 risks to be dangerously misleading.

In Figure 19.6, the real tail of the random variable $Y$ with remote upper bound $H$ is represented by the dashed line. If one only observes values up to $M << H$, and more or less consciously ignores the existence of $H$, one could be fooled by the data into believing that the tail is actually the continuous one, the so-called apparent tail [50]. The tails are indeed indistinguishable for most cases, virtually in all finite samples, as the divergence is only clear in the vicinity of $H$. A bounded tail with very large upper limit is therefore mistakenly taken for an unbounded one, and no model will be able to see the difference, even if epistemologically we are in two extremely different situations. This is the typical case in which critical reasoning, and the a priori analysis of the characteristics of the phenomenon under scrutiny, should precede any instinctive and uncritical fitting of the data.

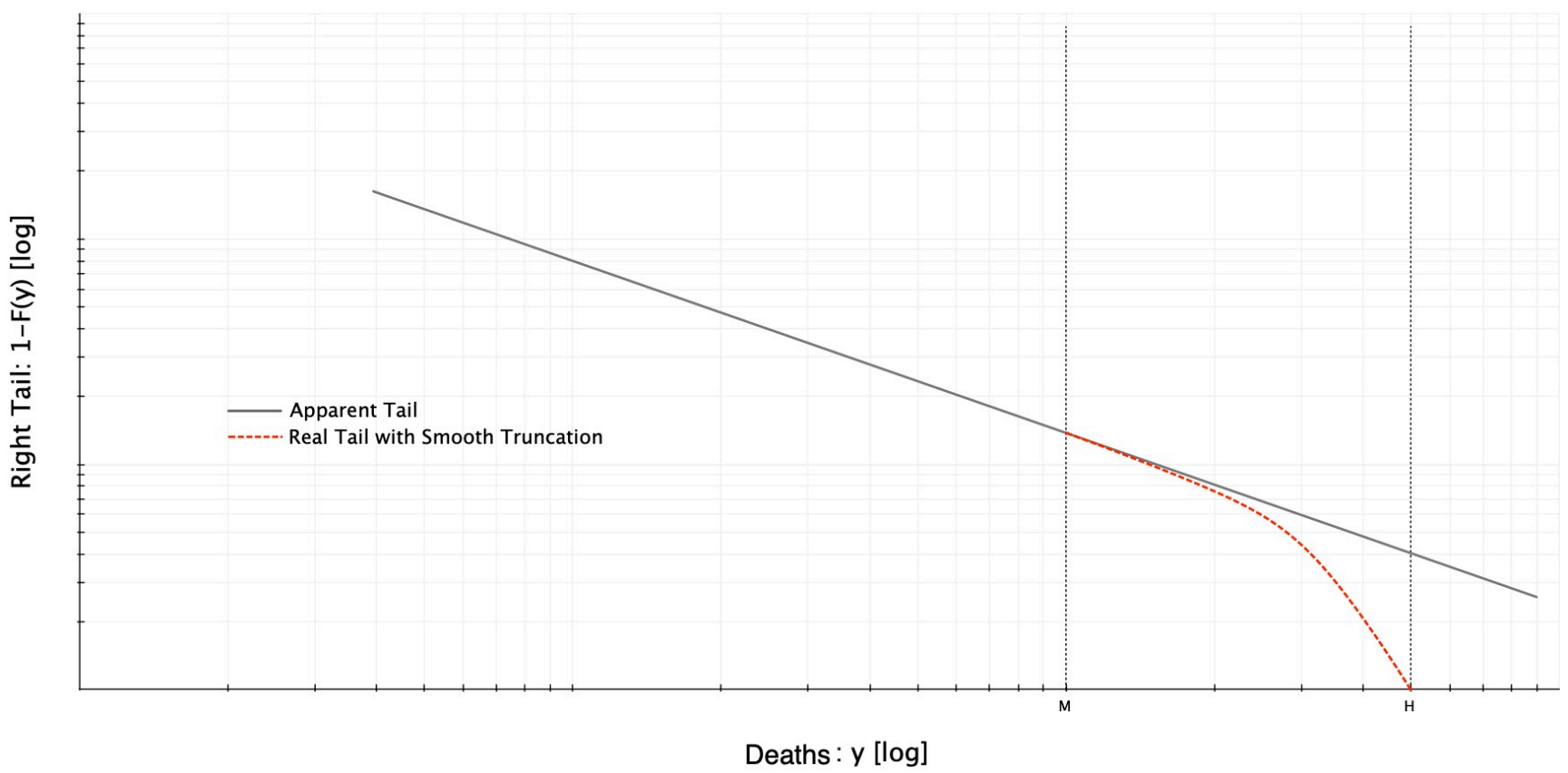

Figure 19.6: *Graphical representation (log-log plot) of what may happen if one ignores the existence of the finite upper bound $H$, since only $M$ is observed.*

A solution is the approach of [50, 51], which introduces the concept of dual data via a special log-transformation [10]. The basic idea is to find a way of matching naive extrapolations (apparently infinite moments) with correct modelling.

9 Today's world population [205] can be safely taken as the upper bound also for the past.

10 Other log-transformations have been proposed in the literature, but they are all meant to thin the tails, without actually taking care of the upper bound problem: the number of victims can still be Let $L$ and $H$ be respectively the finite lower and upper bounds of a random variable $Y$, and define the function

$$\varphi(Y) = L - H \log\left(\frac{H-Y}{H-L}\right). \tag{19.1}$$

We can easily check that

1. $\varphi \in C^\infty$,
2. $\varphi^{-1}(\infty) = H$,
3. $\varphi^{-1}(L) = \varphi(L) = L$.

Then $Z = \varphi(Y)$ defines a new random variable with lower bound $L$ and an infinite upper bound. Notice that the transformation induced by $\varphi(\cdot)$ does not depend on any of the parameters of the distribution of $Y$, and that $\varphi(\cdot)$ is monotone. From now on, we call the distributions of $Y$ and $Z$, respectively the real and the dual distribution. It is easy to verify that for values smaller than $M << H$, $Y$ and $Z$ are in practice undistinguishable (and do are their quantiles [50]).

As per [50, 51], we take the observations in the column "Avg Est" of Table 1, our $Y$'s, and transform them into their dual $Z$'s. We then study the actually unbounded duals using EVT (see Section 19.5), to find out that the naive observation of infinite moments can makes sense in such a framework (but not for the bounded world population!). Finally, by reverting to the real distribution, we compute the so-called *shadow* means [50] of pandemics, equal to

$$E[Y] = (H-L)e^{\frac{\frac{1}{\xi}\sigma}{H}}\left(\frac{\sigma}{H\xi}\right)^{\frac{1}{\xi}}\Gamma\left(1-\frac{1}{\xi},\frac{\sigma}{H\xi}\right)+L, \tag{19.2}$$

where $\Gamma(\cdot,\cdot)$ is the gamma function.

Notice that the random quantity $Y$ is defined above $L$, therefore its expectation corresponds to a tail expectation with respect to the random variable $Z$, an expected shortfall in the financial jargon, being only valid in the tail above $\mu$ [51]. All moments of the random variable $Y$ are called shadow moments in [50], as they are not immediately visible from the data, but from plug-in estimation.

## 19.3 THE DUAL TAIL VIA EVT AND THE SHADOW MEAN

Take the dual random variable $Z$ whose distribution function $G$ is unknown, and let $z_G = \sup\{z \in \mathbb{R} : G(z) < 1\}$ be its right-end point, which can be finite or infinite. Given a threshold $u < z_G$, we can define the exceedance distribution of $Z$ as

$$G_u(z) = P(Z \le z | Z > u) = \frac{G(z) - G(u)}{1 - G(u)}, \tag{19.3}$$

for $z \ge u$.

infinite. The rationale behind those transformations is given by the observation that if X is a random variable whose distribution function is in the domain of attraction of a FrÃl'chet, the family of fat-tailed distributions, then $\log(X)$ is in the domain of attraction of a Gumbel, the more reassuring family of normals and exponentials [97].

For a large class of distributions $G$, and high thresholds $u \to z_G$, $G_u$ can be approximated by a Generalized Pareto distribution (GPD) [70], i.e.

$$G_u(z) \approx GPD(z; \xi, \beta, u) = \begin{cases} 1 - (1 + \xi \frac{z-u}{\beta})^{-1/\xi} & \xi \neq 0 \\ 1 - e^{-\frac{z-u}{\beta}} & \xi = 0 \end{cases}, \tag{19.4}$$

where $z \geq u$ for $\xi \geq 0$, $u \leq z \leq u - \beta/\xi$ for $\xi < 0$, $u \in \mathbb{R}$, $\xi \in \mathbb{R}$ and $\beta > 0$.

Let us just consider $\xi > 0$, being $\xi = 0$ not relevant for fat tails. From equation (19.3), we see that $G(z) = (1 - G(u))G_u(z) + G(u)$, hence we obtain

$$\begin{aligned} G(z) &\approx (1 - G(u))GPD(z; \xi, \beta, u) + G(u) \\ &= 1 - \bar{G}(u) \left(1 + \xi \frac{z-u}{\beta}\right)^{-1/\xi}, \end{aligned}$$

with $\bar{G}(x) = 1 - G(x)$. The tail of $Z$ is therefore

$$\bar{G}(z) = \bar{G}(u) \left(1 + \xi \frac{z-u}{\beta}\right)^{-1/\xi}. \tag{19.5}$$

Equation (19.5) is called the tail estimator of $G(z)$ for $z \geq u$. Given that $G$ is in principle unknown, one usually substitutes $G(u)$ with its empirical estimator $n_u/n$, where $n$ is the total number of observations in the sample, and $n_u$ is the number of exceedances above $u$.

Equation (19.5) then changes into

$$\bar{G}(z) = \frac{n_u}{n} \left(1 + \xi \frac{z-u}{\beta}\right)^{-1/\xi} \approx 1 - GPD(z^*; \xi, \sigma, \mu), \tag{19.6}$$

where $\sigma = \beta \left(\frac{n_u}{n}\right)^{\xi}$, $\mu = u - \frac{\beta}{\xi}\left(1 - \left(\frac{n_u}{n}\right)^{\xi}\right)$, and $z^* \geq \mu$ is an auxiliary random variable. Both $\sigma$ and $\mu$ can be estimated semi-parametrically, starting from the estimates of $\xi$ and $\beta$ in equation (19.4). If $\xi > -1/2$, the preferred estimation method is maximum likelihood [70], while for $\xi \leq -1/2$ other approaches are better used [97]. For both the exceedances distribution and the recovered tail, the parameter $\xi$ is the same, and it also coincides with the tail parameter we have used to define fat tails[11].

One can thus study the tail of $Z$ without caring too much about the rest of the distribution, i.e. the part below $u$. All in all, the most destructive risks come from the right tail, and not from the first quantiles or even the bulk of the distribution. The identification of the correct $u$ is a relevant question in extreme value statistics [70, 96]. One can rely on heuristic graphical tools [49], like the Zipf plot and the meplot we have seen before, or on statistical tests for extreme value conditions [101] and GPD goodness-of-fit [6].

What is important to stress–once again–is that the GPD fit needs to be performed on the dual quantities, to be statistically and epistemologically correct. One could in fact work with the raw observation directly, without the logarithmic transformation of Equation (19.1), surely ending up with $\xi > 1$, in line with Figures 19.1 and 19.5. But a similar approach would be wrong and naive, because only the dual observations are actually unbounded.

[11] Moreover, when maximum likelihood is used, the estimate of $\xi$ would correspond to $1/\alpha$, where $\alpha$ is estimated according to [58].

Working with the dual observations, we find out that the best GPD fit threshold is around 200K victims, with 34.7% of the observations lying above. For what concerns the GPD parameters, we estimate $\xi = 1.62$ (standard error 0.52), and $\beta = 1'174.7K$ (standard error $536.5K$). As expected $\xi > 1$ once again supporting the idea of an infinite first moment[12]. Visual inspections and statistical tests [6, 101] support the goodness-of-fit for the exceedance distribution and the tail.

Given $\xi$ and $\beta$, we can use Equations (19.2) and (19.6) to compute the shadow mean of the numbers of victims in pandemics. For actual data we get a shadow mean of 20.1M, which is definitely larger (almost 1.5 times) than the corresponding sample tail mean of 13.9M (this is the mean of all the actual numbers above the 200K threshold.). Combining the shadow mean with the sample mean below the $200K$ threshold, we get an overall mean of 7M instead of the naive 4.9M we have computed initially. It is therefore important to stress that a naive use of the sample mean would induce an underestimation of risk, and would also be statistically incorrect.

## 19.4 DATA RELIABILITY ISSUES

As observed in [51, 249, 268] for war casualties, but the same reasoning applies to pandemics of the past, the estimates of the number of victims are not at all unique and precise. Figures are very often anecdotal, based on citations and vague reports, and usually dependent on the source of the estimate. In Table 1, it is evident that some events vary considerably in estimates.

Natural questions thus arise: are the tail risk estimates of Section 19.5 robust? What happens if some of the casualties estimates change? What is the impact of ignoring some events in our collection? The use of extreme value statistics in studying tail risk already guarantees the robustness of our estimates to changes in the underlying data, when these lie below the threshold $u$. However, to verify robustness more rigorously and thoroughly, we have decided to stress the data, to study how the tails potentially vary.

First of all, we have generated 10K distorted copies of our dual data. Each copy contains exactly the same number of observations as per Table 1, but every data point has been allowed to vary between 80% and 120% of its recorded value before imposing the log-transformation of Equation (19.1). In other words, each of the 10K new samples contains 72 observations, and each observation is a (dual) perturbation ($\pm 20\%$) of the corresponding observation in Table 1.

---

[12] Looking at the standard error of $\xi$, one could argue that, with more data from the upper tail, the first moment could possibly become finite, yet there would be no discussion about the non existence of the second moment, and thus the unreliability of the sample mean [267]. Pandemic fatalities would still be an extremely erratic phenomenon, with substantial tail risk in the number of fatalities. In any case, Figures 19.1 and 19.5 make us prefer to consider the first moment as infinite, and not to trust sample averages.

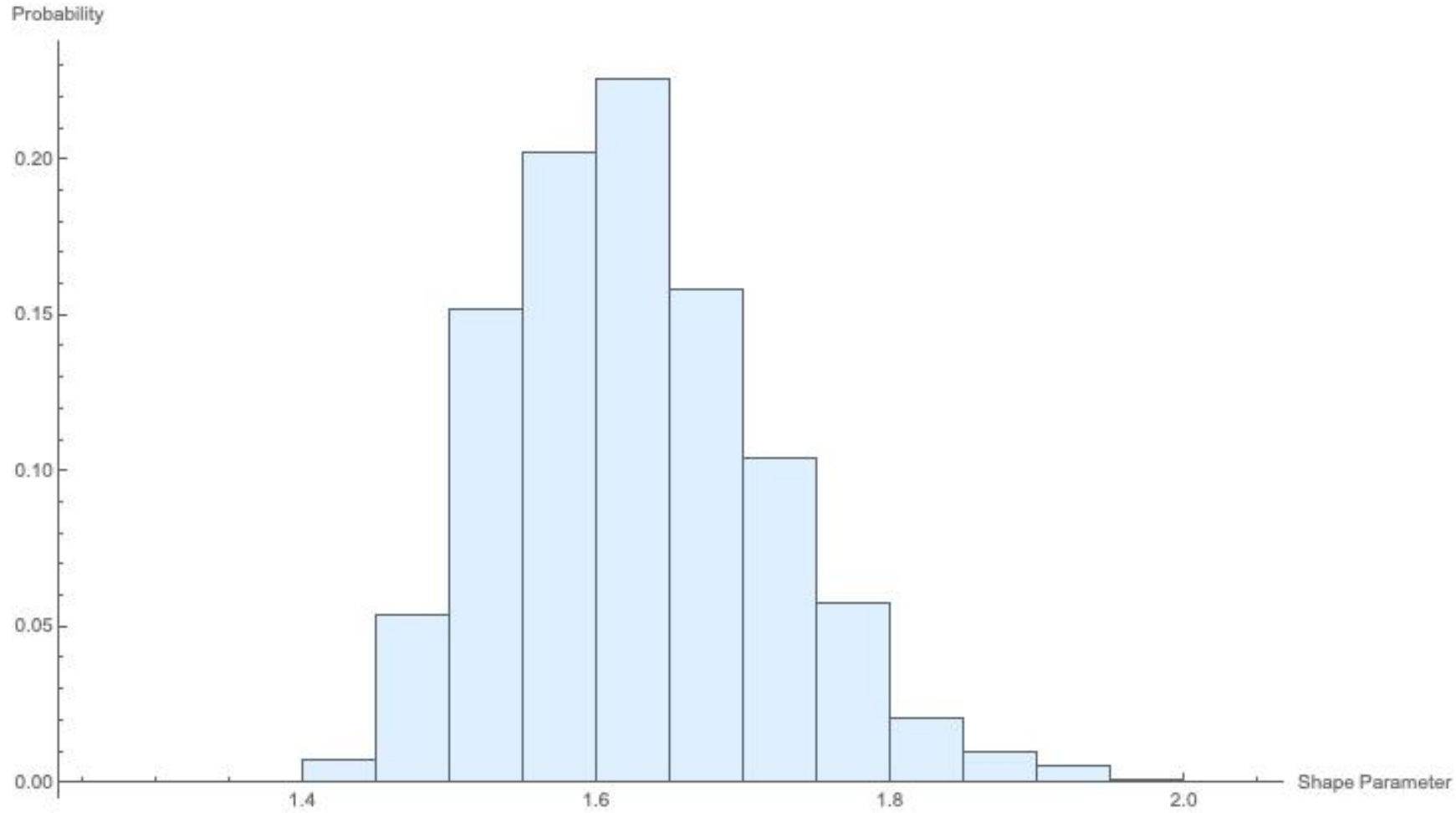

Figure 19.7: *Values of the shape parameter $\xi$ over 10,000 distorted copies of the the dual versions of the average deaths in Table 1, allowing for a random variation of $\pm 20\%$ for each single observation. The $\xi$ parameter consistently indicates an apparently infinite-mean phenomenon.*

Figure 19.7 contains the histogram of the $\xi$ parameter over the 10K distorted copies of the dual numbers. The values are always above 1, indicating an apparently infinite mean, and the average value is 1.62 (standard deviation 0.10), in line with our previous findings. Our tail estimates are thus robust to imprecise observations. Consistent results hold for the $\beta$ parameter.

But it also true that our data set is likely to be incomplete, not containing all epidemics and pandemics with more than 1K victims, or that some of the events we have collected are too biased to be reliable and should be discarded anyway. To account for this, we have once again generated 10K copies of our sample via jackknife. Each new dual sample is obtained by removing from 1 to 7 observations at random, so that one sample could not contain the Spanish flu, while another could ignore the Yellow Fever and AIDS. In Figure 19.8 we show the impact of such a procedure on the $\xi$ parameter. Once again, the main message of this work remains unchanged: we are looking at a very fat-tailed phenomenon, with an extremely large tail risk and potentially destructive consequences, which should not be downplayed in any serious policy discussion.

Table 19.1: *The data set used for the analysis. All estimates in thousands, apart from coeval population, which is expressed in millions. For Covid-19 [215], the upper estimate includes the supposed number of Chinese victims (42K) for some Western media. (All are in 000 except population in millions.)*

| Name | Start | End | Lower | Avg(k) | Upper | Resc Avg | Pop |
|---|---|---|---|---|---|---|---|
| **Plague of Athens** | -429 | -426 | 75 | 88 | 100 | 13376 | 50 |

| Name | Start | End | Lower | Avg(k) | Upper | Resc Avg | Pop |
|---|---|---|---|---|---|---|---|
| **Antonine Plague** | 165 | 180 | 5000 | 7500 | 10000 | 283355 | 202 |
| **Plague of Cyprian** | 250 | 266 | 1000 | 1000 | 1000 | 37227 | 205 |
| **Plague of Justinian** | 541 | 542 | 25M | 62.5M | 100M | 2,246M | 213 |
| **Plague of Amida** | 562 | 562 | 30 | 30 | 30 | 1078 | 213 |
| **Roman Plague of 590** | 590 | 590 | 10 | 20 | 30 | 719 | 213 |
| **Plague of Sheroe** | 627 | 628 | 100 | 100 | 100 | 3594 | 213 |
| **British Isles** | 664 | 689 | 150 | 175 | 200 | 6290 | 213 |
| **Plague of Basra** | 688 | 689 | 200 | 200 | 200 | 7189 | 213 |
| **Japanese smallpox** | 735 | 737 | 2000 | 2000 | 2000 | 67690 | 226 |
| **Black Death** | 1331 | 1353 | 75000 | 137500 | 200000 | 2678283 | 392 |
| **Sweating sickness** | 1485 | 1551 | 10 | 10 | 10 | 166 | 461 |
| **Mexico Smallpox** | 1520 | 1520 | 5000 | 6500 | 8000 | 107684 | 461 |
| **Cocoliztli Epidemic** | 1545 | 1548 | 5000 | 10000 | 15000 | 165668 | 461 |
| **1563 London plague** | 1562 | 1564 | 20 | 20 | 20 | 277 | 554 |
| **Cocoliztli epidemic** | 1576 | 1580 | 2000 | 2250 | 2500 | 31045 | 554 |
| **1592-93 London plague** | 1592 | 1593 | 20 | 20 | 20 | 275 | 554 |
| **Plague of Malta** | 1592 | 1593 | 3 | 3 | 3 | 41 | 554 |
| **Plague in Spain** | 1596 | 1602 | 600 | 650 | 700 | 8969 | 554 |
| **New England** | 1616 | 1620 | 7 | 7 | 7 | 97 | 554 |
| **Italian plague** | 1629 | 1631 | 280 | 280 | 280 | 3863 | 554 |
| **G. Plague of Sevilla** | 1647 | 1652 | 150 | 150 | 150 | 2070 | 554 |
| **Plague of Naples** | 1656 | 1658 | 1250 | 1250 | 1250 | 15840 | 603 |

| Name | Start | End | Lower | Avg(k) | Upper | Resc Avg | Pop |
|---|---|---|---|---|---|---|---|
| Netherlands | 1663 | 1664 | 24 | 24 | 24 | 306 | 603 |
| G. Plague of London | 1665 | 1666 | 100 | 100 | 100 | 1267 | 603 |
| Plague in France | 1668 | 1668 | 40 | 40 | 40 | 507 | 603 |
| Malta plague epidemic | 1675 | 1676 | 11 | 11 | 11 | 143 | 603 |
| G. Plague of Vienna | 1679 | 1679 | 76 | 76 | 76 | 963 | 603 |
| G. N. War plague | 1700 | 1721 | 176 | 192 | 208 | 2427 | 603 |
| Smallpox in Iceland | 1707 | 1709 | 18 | 18 | 18 | 228 | 603 |
| G. Plague of Marseille | 1720 | 1722 | 100 | 100 | 100 | 1267 | 603 |
| G. Plague of 1738 | 1738 | 1738 | 50 | 50 | 50 | 470 | 814 |
| Russian plague | 1770 | 1772 | 50 | 50 | 50 | 470 | 814 |
| Persian Plague | 1772 | 1772 | 2000 | 2000 | 2000 | 15444 | 990 |
| Ottoman Plague | 1812 | 1819 | 300 | 300 | 300 | 2317 | 990 |
| Caragea’s plague | 1813 | 1813 | 60 | 60 | 60 | 463 | 990 |
| Malta plague epidemic | 1813 | 1814 | 5 | 5 | 5 | 35 | 990 |
| 1st cholera pand. | 1816 | 1826 | 100 | 100 | 100 | 772 | 990 |
| 2nd cholera pand. | 1829 | 1851 | 100 | 100 | 100 | 772 | 990 |
| Canada Typhus | 1847 | 1848 | 20 | 20 | 20 | 154 | 990 |
| Third cholera-pandemic | 1852 | 1860 | 1000 | 1000 | 1000 | 6053 | 1263 |
| Copenhagen Cholera | 1853 | 1853 | 5 | 5 | 5 | 29 | 1263 |
| Third plague pandemic | 1855 | 1960 | 15000 | 18500 | 22000 | 111986 | 1263 |

| Name | Start | End | Lower | Avg(k) | Upper | Resc Avg | Pop |
|---|---|---|---|---|---|---|---|
| B. Columbiam Smallpox | 1862 | 1863 | 3 | 3 | 3 | 18 | 1263 |
| Fourth cholera pandemic | 1863 | 1875 | 600 | 600 | 600 | 3632 | 1263 |
| Fiji Measles outbreak | 1875 | 1875 | 40 | 40 | 40 | 242 | 1263 |
| Yellow Fever | 1880 | 1900 | 100 | 125 | 150 | 757 | 1263 |
| Fifth cholera pandemic | 1881 | 1896 | 9 | 9 | 9 | 42 | 1654 |
| Smallpox in Montreal | 1885 | 1885 | 3 | 3 | 3 | 14 | 1654 |
| Russian flu | 1889 | 1890 | 1000 | 1000 | 1000 | 4620 | 1654 |
| Sixth cholera pandemic | 1899 | 1923 | 800 | 800 | 800 | 3696 | 1654 |
| China plague | 1910 | 1912 | 40 | 40 | 40 | 185 | 1654 |
| Encephalitis lethargica | 1915 | 1926 | 1500 | 1500 | 1500 | 6930 | 1654 |
| American polio epidemic | 1916 | 1916 | 6 | 7 | 7 | 30 | 1654 |
| Spanish flu | 1918 | 1920 | 17000 | 58500 | 100000 | 193789 | 2307 |
| HIV/AIDS pandemic | 1920 | 2020 | 25000 | 30000 | 35000 | 61768 | 3712 |
| Poliomyelitis in USA | 1946 | 1946 | 2 | 2 | 2 | 5 | 2948 |
| Asian flu | 1957 | 1958 | 2000 | 2000 | 2000 | 5186 | 2948 |
| Hong Kong flu | 1968 | 1969 | 1000 | 1000 | 1000 | 2102 | 3637 |
| London flu | 1972 | 1973 | 1 | 1 | 1 | 2 | 3866 |
| Smallpox epidemic of India | 1974 | 1974 | 15 | 15 | 15 | 29 | 4016 |
| Zimbabwe cholera | 2008 | 2009 | 4 | 4 | 4 | 5 | 6788 |
| Swine Flu | 2009 | 2009 | 152 | 364 | 575 | 409 | 6788 |

| **Name** | **Start** | **End** | **Lower** | **Avg(k)** | **Upper** | **Resc Avg** | **Pop** |
|---|---|---|---|---|---|---|---|
| **Haiti cholera outbreak** | 2010 | 2020 | 10 | 10 | 10 | 11 | 7253 |
| **Measles in D.R. Congo** | 2011 | 1018 | 5 | 5 | 5 | 5 | 7253 |
| **Ebola in West Africa** | 2013 | 2016 | 11 | 11 | 11 | 12 | 7176 |
| **Indian swine flu outbreak** | 2015 | 2015 | 2 | 2 | 2 | 2 | 7253 |
| **Yemen cholera outbreak** | 2016 | 2020 | 4 | 4 | 4 | 4 | 7643 |
| **Kivu Ebola** | 2018 | 2020 | 2 | 2 | 3 | 2 | 7643 |
| **COVID-19 Pandemic** | 2019 | 2020 | 117 | 133.5 | 150 | 50 | 7643 |
| **Measles in D.R. Congo** | 2019 | 2020 | 5 | 5 | 5 | 5 | 7643 |
| **Dengue fever** | 2019 | 2020 | 2 | 2 | 2 | 2 | 7643 |

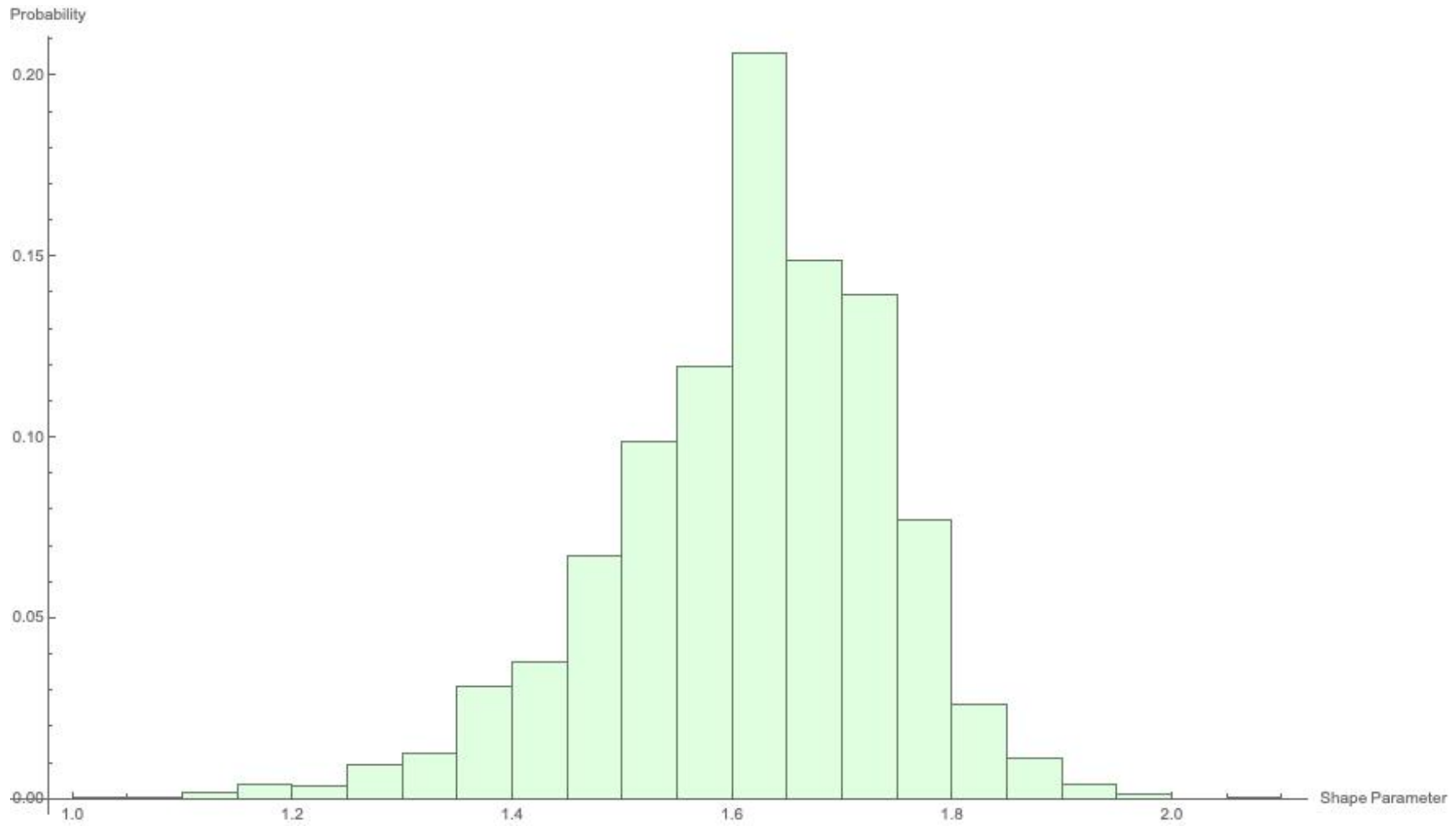

Figure 19.8: *Values of the shape parameter $\xi$ over 10,000 jackknifed versions of the dual versions of the actual average numbers in Table 1, when allowing at least 1% and up to about 10% of the observations to be missing. The $\xi$ parameter consistently indicates an apparently infinite-mean phenomenon.*

Part VI

# METAPROBABILITY PAPERS

# 20 HOW THICK TAILS EMERGE FROM RECURSIVE EPISTEMIC UNCERTAINTY†

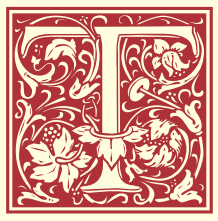HE Opposite of Central Limit: With the Central Limit Theorem, we start with a specific distribution and end with a Gaussian. The opposite is more likely to be true. Recall how we fattened the tail of the Gaussian by stochasticizing the variance? Now let us use the same metaprobability method, putting additional layers of uncertainty.

**The Regress Argument (Error about Error)** The main problem behind *The Black Swan* is the limited understanding of model (or representation) error, and, for those who get it, a lack of understanding of second order errors (about the methods used to compute the errors) and by a regress argument, an inability to continuously reapplying the thinking all the way to its limit ( ***particularly when one provides no reason to stop***). Again, there is no problem with stopping the recursion, provided it is accepted as a declared *a priori* that escapes quantitative and statistical methods.

**Epistemic not statistical re-derivation of power laws** Note that previous derivations of power laws have been statistical (cumulative advantage, preferential attachment, winner-take-all effects, criticality), and the properties derived by Yule, Mandelbrot, Zipf, Simon, Bak, and others result from structural conditions or breaking the independence assumptions in the sums of random variables allowing for the application of the central limit theorem, [106] [251][118] [191] [190] . This work is entirely epistemic, based on standard philosophical doubts and regress arguments.

---

Discussion chapter.
A version of this chapter was presented at Benoit Mandelbrot's Scientific Memorial on April 29, 2011,in New Haven, CT.

## 20.1 METHODS AND DERIVATIONS

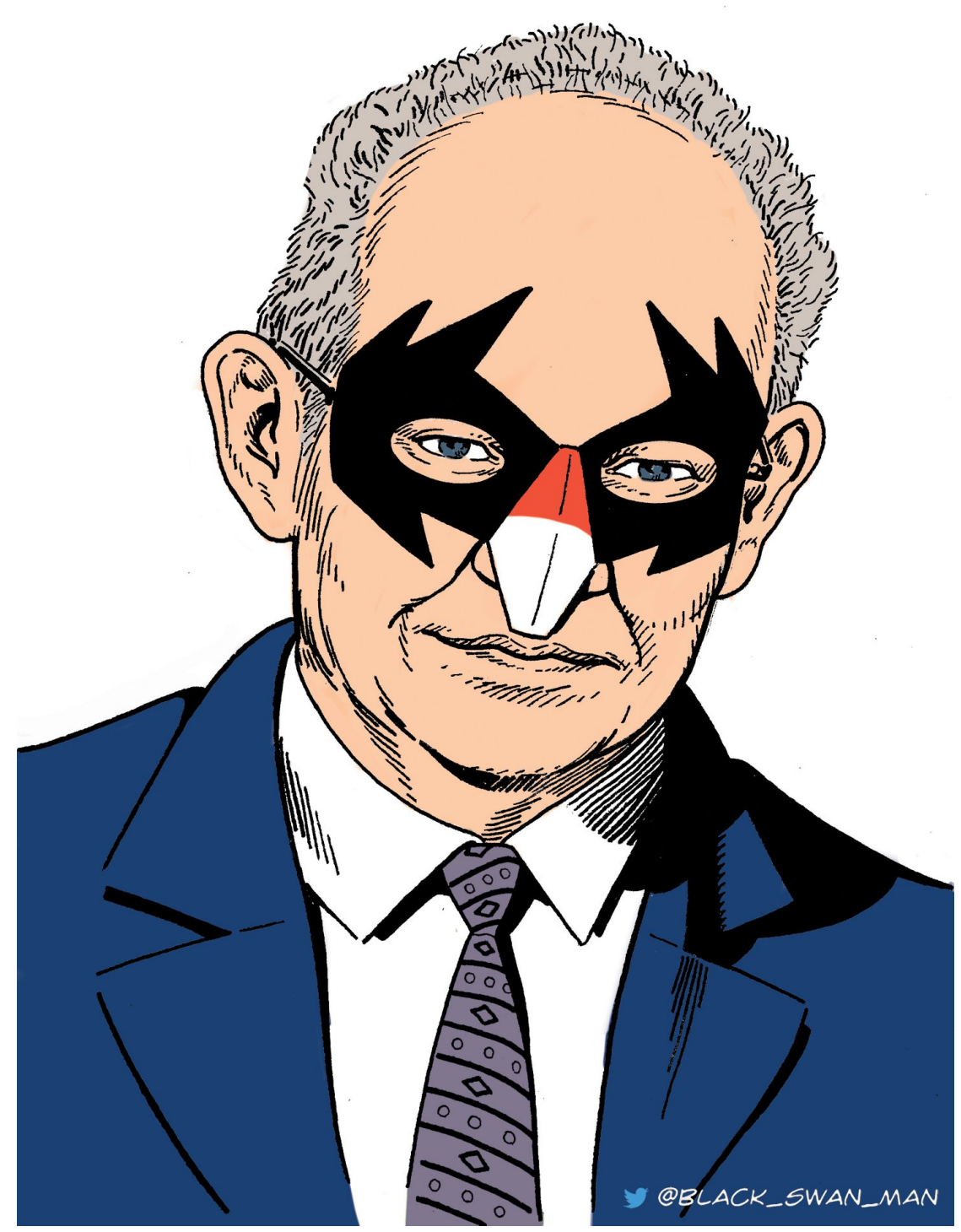

Figure 20.1: *A version of this chapter was presented at Benoit Mandelbrot's memorial.*

### 20.1.1 Layering Uncertainties

Take a standard probability distribution, say the Gaussian. The measure of dispersion, here $\sigma$, is estimated, and we need to attach some measure of dispersion around it. The uncertainty about the rate of uncertainty, so to speak, or higher order parameter, similar to what called the "volatility of volatility" in the lingo of option operators (see Taleb, 1997, Derman, 1994, Dupire, 1994, Hull and White, 1997) –here it would be "uncertainty rate about the uncertainty rate". And there is no reason to stop there: we can keep nesting these uncertainties into higher orders, with the uncertainty rate of the uncertainty rate of the uncertainty rate, and so forth. There is no reason to have certainty anywhere in the process.

### 20.1.2 Higher Order Integrals in the Standard Gaussian Case

We start with the case of a Gaussian and focus the uncertainty on the assumed standard deviation. Define $\phi(\mu,\sigma;x)$ as the Gaussian PDF for value $x$ with mean $\mu$ and standard deviation $\sigma$.

A $2^{nd}$ order stochastic standard deviation is the integral of $\phi$ across values of $\sigma \in \mathbb{R}^+$, under the PDF $f(\bar{\sigma},\sigma_1;\sigma)$, with $\sigma_1$ its scale parameter (our approach to trach the error of the error), not necessarily its standard deviation; the expected value of $\sigma_1$ is $\overline{\sigma_1}$.

$$f(x)_1 = \int_0^\infty \phi(\mu,\sigma,x) f(\bar{\sigma},\sigma_1;\sigma)\, \mathrm{d}\sigma$$

Generalizing to the $N^{\text{th}}$ order, the density function $f(x)$ becomes

$$f(x)_N = \int_0^\infty ... \int_0^\infty \phi(\mu,\sigma,x) f(\bar{\sigma},\sigma_1,\sigma)\, f(\overline{\sigma_1},\sigma_2,\sigma_1) \ldots \\ f(\overline{\sigma_{N-1}},\sigma_N,\sigma_{N-1})\, \mathrm{d}\sigma\, \mathrm{d}\sigma_1\, \mathrm{d}\sigma_2 ... \mathrm{d}\sigma_N \quad (20.1)$$

The problem is that this approach is parameter-heavy and requires the specifications of the subordinated distributions (in finance, the lognormal has been traditionally used for $\sigma^2$ (or Gaussian for the ratio $\text{Log}[\frac{\sigma_t^2}{\sigma^2}]$ since the direct use of a Gaussian allows for negative values). We would need to specify a measure $f$ for each layer of error rate. Instead this can be approximated by using the mean deviation for $\sigma$, as we will see next.

Discretization using nested series of two-states for $\sigma$- a simple multiplicative process

We saw in the last chapter a quite effective simplification to capture the convexity, the ratio of (or difference between) $\phi(\mu,\sigma,x)$ and $\int_0^\infty \phi(\mu,\sigma,x) f(\bar{\sigma},\sigma_1,\sigma)\, \mathrm{d}\sigma$ (the first order standard deviation) by using a weighted average of values of $\sigma$, say, for a simple case of one-order stochastic volatility:

$$\sigma(1 \pm a(1))$$

with $0 \leq a(1) < 1$, where $a(1)$ is the proportional mean absolute deviation for $\sigma$, in other word the measure of the absolute error rate for $\sigma$. We use $\frac{1}{2}$ as the probability of each state. Unlike the earlier situation we are not preserving the variance, rather the STD. Thus the distribution using the first order stochastic standard deviation can be expressed as:

$$f(x)_1 = \frac{1}{2}\Big(\phi(\mu,\sigma\,(1+a(1)),x) + \phi(\mu,\sigma(1-a(1)),x)\Big) \quad (20.2)$$

Now assume uncertainty about the error rate a(1), expressed by a(2), in the same manner as before. Thus in place of a(1) we have $\frac{1}{2}$ a(1)( 1$\pm$ a(2)).

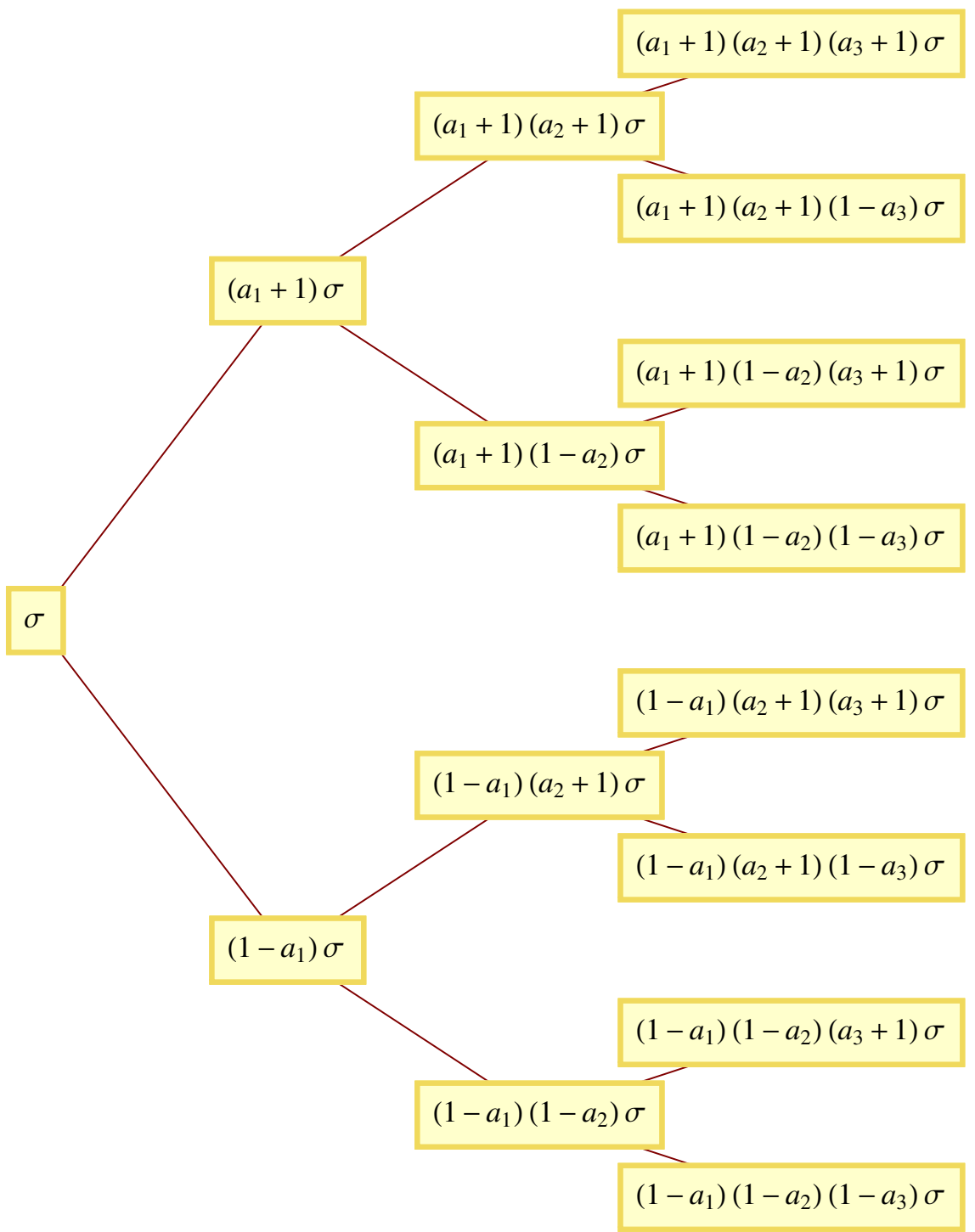

Figure 20.2: *Three levels of error rates for $\sigma$ following a multiplicative process*

The second order stochastic standard deviation:

$$
\begin{aligned}
f(x)_2 = \frac{1}{4}\Bigg( & \phi\Big(\mu, \sigma(1+a(1)(1+a(2))), x\Big) + \\
& \phi\Big(\mu, \sigma(1-a(1)(1+a(2))), x) + \phi(\mu, \sigma(1+a(1)(1-a(2)), x\Big) \\
& \qquad\qquad error + \phi\Big(\mu, \sigma(1-a(1)(1-a(2))), x\Big)\Bigg)
\end{aligned}
\tag{20.3}
$$

and the $N^{\text{th}}$ order:

$$f(x)_N = \frac{1}{2^N}\sum_{i=1}^{2^N}\phi(\mu, \sigma M_i^N, x)$$

where $M_i^N$ is the $i^{\text{th}}$ scalar (line) of the matrix $M^N\ (2^N \times 1)$

$$M^N = \left( \prod_{j=1}^{N} (a(j)\mathbf{T}_{i,j} + 1) \right)_{i=1}^{2^N}$$

and $\mathbf{T}_{i,j}$ the element of $i^{\text{th}}$line and $j^{\text{th}}$column of the matrix of the exhaustive combination of $n$-Tuples of the set $\{-1, 1\}$,that is the sequences of $n$ length $(1, 1, 1, ...)$ representing all combinations of 1 and $-1$.

for N=3,

$$T = \begin{pmatrix} 1 & 1 & 1 \\ 1 & 1 & -1 \\ 1 & -1 & 1 \\ 1 & -1 & -1 \\ -1 & 1 & 1 \\ -1 & 1 & -1 \\ -1 & -1 & 1 \\ -1 & -1 & -1 \end{pmatrix}$$

and

$$M^3 = \begin{pmatrix} (1 - a(1))(1 - a(2))(1 - a(3)) \\ (1 - a(1))(1 - a(2))(a(3) + 1) \\ (1 - a(1))(a(2) + 1)(1 - a(3)) \\ (1 - a(1))(a(2) + 1)(a(3) + 1) \\ (a(1) + 1)(1 - a(2))(1 - a(3)) \\ (a(1) + 1)(1 - a(2))(a(3) + 1) \\ (a(1) + 1)(a(2) + 1)(1 - a(3)) \\ (a(1) + 1)(a(2) + 1)(a(3) + 1) \end{pmatrix}$$

So $M_1^3 = \{(1 - a(1))(1 - a(2))(1 - a(3))\}$, etc.

Note that the various error rates $a(i)$ are not similar to sampling errors, but rather projection of error rates into the future. They are, to repeat, *epistemic*.

**The Final Mixture Distribution** The mixture weighted average distribution (recall that $\phi$ is the ordinary Gaussian PDF with mean $\mu$, std $\sigma$ for the random variable $x$).

$$f(x|\mu, \sigma, M, N) = 2^{-N} \sum_{i=1}^{2^N} \phi \left( \mu, \sigma M_i^N, x \right)$$

It could be approximated by a lognormal distribution for $\sigma$ and the corresponding V as its own variance. But it is precisely the V that interest us, and V depends on how higher order errors behave.

Next let us consider the different regimes for higher order errors.

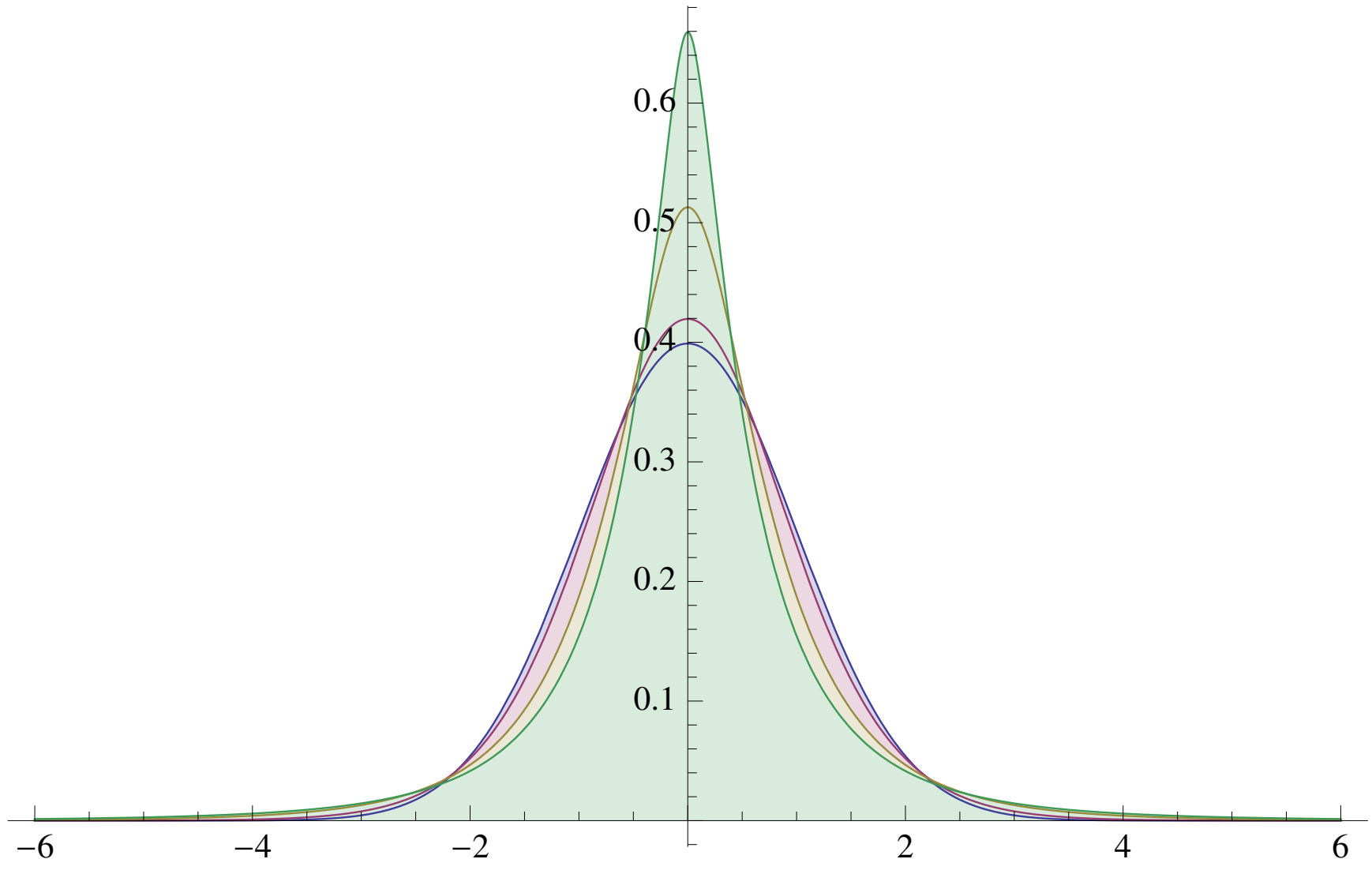

Figure 20.3: *Thicker tails (higher peaks) for higher values of N; here* $N = 0, 5, 10, 25, 50$, *all values of* $a = \frac{1}{10}$

## REGIME 1 (EXPLOSIVE): CASE OF A CONSTANT PARAMETER $a$

**Special case of constant $a$**: Assume that a(1)=a(2)=...a(N)=a, i.e. the case of flat proportional error rate $a$. The Matrix $M$ collapses into a conventional binomial tree for the dispersion at the level $N$.

$$f(x|\mu, \sigma, M, N) = 2^{-N} \sum_{j=0}^{N} \binom{N}{j} \phi\left(\mu, \sigma(a+1)^{j}(1-a)^{N-j}, x\right) \tag{20.4}$$

Because of the linearity of the sums, when a is constant, we can use the binomial distribution as weights for the moments (note again the artificial effect of constraining the first moment $\mu$ in the analysis to a set, certain, and known *a priori*).

$$\begin{pmatrix} & \text{Moment} \\ 1 & \mu \\ 2 & \sigma^2\left(a^2+1\right)^N+\mu^2 \\ 3 & 3\mu\sigma^2\left(a^2+1\right)^N+\mu^3 \\ 4 & 6\mu^2\sigma^2\left(a^2+1\right)^N+\mu^4+3\left(a^4+6a^2+1\right)^N\sigma^4 \end{pmatrix}$$

Note again the oddity that in spite of the explosive nature of higher moments, the expectation of the absolute value of $x$ is both independent of $a$ and $N$, since the perturbations of $\sigma$ do not affect the first absolute moment = $\sqrt{\frac{2}{\pi}}\sigma$ (that is, the initial assumed $\sigma$). The situation would be different under addition of $x$.

Every recursion multiplies the variance of the process by $(1 + a^2)$. The process is similar to a stochastic volatility model, with the standard deviation (not the variance) following a lognormal distribution, the volatility of which grows with M, hence will reach infinite variance at the limit.

Consequences

For a constant $a > 0$, and in the more general case with variable a where a(n) $\geq$ a(n-1), the moments explode.

A- Even the smallest value of $a > 0$, since $\left(1+a^2\right)^N$ is unbounded, which leads to the second moment going to infinity (though not the first) when $N \to \infty$. So something as small as a .001% error rate will still lead to explosion of moments and invalidation of the use of the class of $\mathcal{L}^2$ distributions.

B- In these conditions, we need to use power laws for epistemic reasons, or, at least, distributions outside the $\mathcal{L}^2$ norm, regardless of observations of past data.

Note that we need an *a priori* reason (in the philosophical sense) to cutoff the N somewhere, hence bound the expansion of the second moment.

Convergence to Properties Similar to Power Laws

We can see on the Log-Log plot how, at higher orders of stochastic volatility, with equally proportional stochastic coefficient, (where $a(1) = a(2) = ... = a(N) = \frac{1}{10}$) how the density approaches that of a Power Law (just as the Lognormal distribution does when set at higher variance), as shown in flatter density on the LogLog plot. The probabilities keep rising in the tails as we add layers of uncertainty until they seem to reach the boundary of the power law, while ironically the first moment remains invariant.

The same effect takes place as a increases towards 1, as at the limit the tail exponent P>x approaches 1 but remains >1.

### 20.1.3 Effect on Small Probabilities

Next we measure the effect on the thickness of the tails. The obvious effect is the rise of small probabilities.

Take the exceedant probability, that is, the probability of exceeding K, given N, for parameter a constant:

$$P_{>K|N} = \sum_{j=0}^{N} 2^{-N-1} \binom{N}{j} \operatorname{erfc}\left(\frac{K}{\sqrt{2}\sigma(a+1)^j(1-a)^{N-j}}\right) \quad (20.5)$$

where erfc(.) is the complementary of the error function, 1-erf(.), $\operatorname{erf}(z) = \frac{2}{\sqrt{\pi}} \int_0^z e^{-t^2} dt$.

**Convexity effect** The next Table shows the ratio of exceedant probability under different values of N divided by the probability in the case of a standard Gaussian.

Table 20.1: *Case of* $a = \frac{1}{10}$

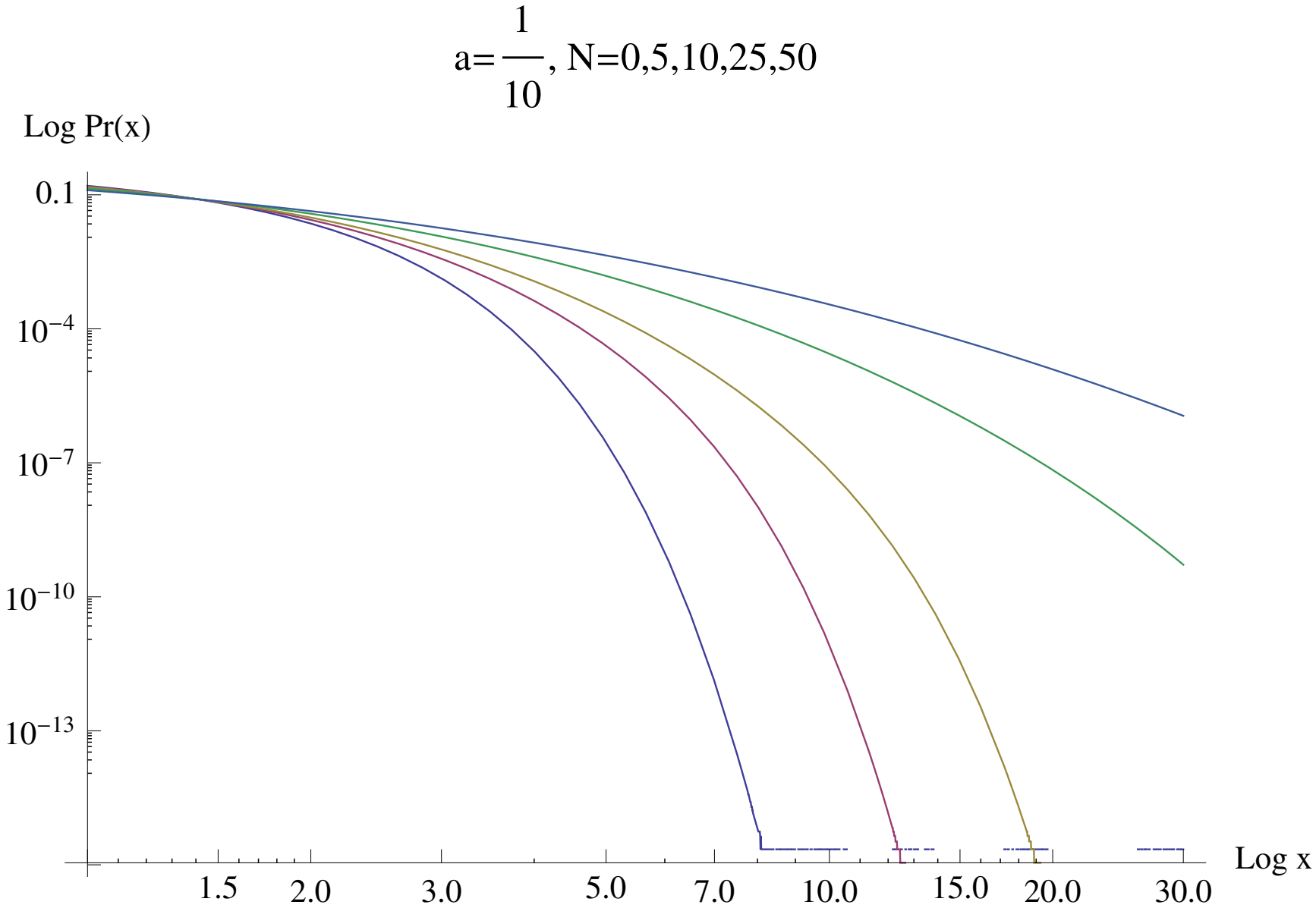

Figure 20.4: *LogLog Plot of the probability of exceeding x showing power law-style flattening as N rises. Here all values of a= 1/10*

| $N$ | $\frac{P>3,N}{P>3,N=0}$ | $\frac{P>5,N}{P>5,N=0}$ | $\frac{P>10,N}{P>10,N=0}$ |
|---|---|---|---|
| 5 | 1.01724 | 1.155 | 7 |
| 10 | 1.0345 | 1.326 | 45 |
| 15 | 1.05178 | 1.514 | 221 |
| 20 | 1.06908 | 1.720 | 922 |
| 25 | 1.0864 | 1.943 | 3347 |

Table 20.2: *Case of* $a = \frac{1}{100}$

| N | $\frac{P>3,N}{P>3,N=0}$ | $\frac{P>5,N}{P>5,N=0}$ | $\frac{P>10,N}{P>10,N=0}$ |
|---|---|---|---|
| 5 | 2.74 | 146 | $1.09 \times 10^{12}$ |
| 10 | 4.43 | 805 | $8.99 \times 10^{15}$ |
| 15 | 5.98 | 1980 | $2.21 \times 10^{17}$ |
| 20 | 7.38 | 3529 | $1.20 \times 10^{18}$ |
| 25 | 8.64 | 5321 | $3.62 \times 10^{18}$ |

## 20.2 REGIME 2: CASES OF DECAYING PARAMETERS $a(n)$

As we said, we may have (actually we need to have) *a priori* reasons to decrease the parameter $a$ or stop $N$ somewhere. When the higher order of $a$(i) decline, then the moments tend to be capped (the inherited tails will come from the lognormality of $\sigma$).

### 20.2.1 Regime 2-a;"Bleed" of Higher Order Error

Take a "bleed" of higher order errors at the rate $\lambda$, $0 \leq \lambda < 1$, such as $a(N) = \lambda$ a(N-1), hence $a$(N) $=\lambda^N$ $a(1)$, with $a(1)$ the conventional intensity of stochastic standard deviation. Assume $\mu$=0.

With $N$=2 , the second moment becomes:

$$M_2(2) = \left(a(1)^2 + 1\right) \sigma^2 \left(a(1)^2 \lambda^2 + 1\right)$$

With $N$=3,

$$M_2(3) = \sigma^2 \left(1 + a(1)^2\right) \left(1 + \lambda^2 a(1)^2\right) \left(1 + \lambda^4 a(1)^2\right)$$

finally, for the general N:

$$M_3(N) = \left(a(1)^2 + 1\right) \sigma^2 \prod_{i=1}^{N-1} \left(a(1)^2 \lambda^{2i} + 1\right) \tag{20.6}$$

We can reexpress 20.6 using the Q-Pochhammer symbol $(a;q)_N = \prod_{i=1}^{N-1} \left(1 - aq^i\right)$

$$M_2(N) = \sigma^2 \left(-a(1)^2; \lambda^2\right)_N$$

Which allows us to get to the limit

$$\lim_{N\to\infty} M_2(N) = \sigma^2 \frac{\left(\lambda^2;\lambda^2\right)_2 \left(a(1)^2;\lambda^2\right)_\infty}{\left(\lambda^2 - 1\right)^2 \left(\lambda^2 + 1\right)}$$

As to the fourth moment:

By recursion:

$$M_4(N) = 3\sigma^4 \prod_{i=0}^{N-1} \left(6a(1)^2 \lambda^{2i} + a(1)^4 \lambda^{4i} + 1\right)$$

$$M_4(N) = 3\sigma^4 \left(\left(2\sqrt{2} - 3\right) a(1)^2; \lambda^2\right)_N \left(-\left(3 + 2\sqrt{2}\right) a(1)^2; \lambda^2\right)_N \tag{20.7}$$

$$\lim_{N\to\infty} M_4(N) = 3\sigma^4 \left(\left(2\sqrt{2} - 3\right) a(1)^2; \lambda^2\right)_\infty \left(-\left(3 + 2\sqrt{2}\right) a(1)^2; \lambda^2\right)_\infty \tag{20.8}$$

So the limiting second moment for $\lambda$=.9 and a(1)=.2 is just 1.28 $\sigma^2$, a significant but relatively benign convexity bias. The limiting fourth moment is just $9.88\sigma^4$, more than 3 times the Gaussian's (3 $\sigma^4$), but still finite fourth moment. For small values of a and values of $\lambda$ close to 1, the fourth moment collapses to that of a Gaussian.

### 20.2.2 Regime 2-b; Second Method, a Non Multiplicative Error Rate

For $N$ recursions,

$$\sigma(1 \pm (a(1)(1 \pm (a(2)(1 \pm a(3)( \ldots )))$$

$$\mathbb{P}(X, \mu, \sigma, N) = \frac{1}{L} \sum_{i=1}^{L} f\left(x, \mu, \sigma\left(1 + \left(\mathbf{T}^N . \mathbf{A}^N\right)_i\right)\right.$$

$(\mathbf{M}^N . \mathbf{T} + 1)_i)$ is the $i^t h$ component of the $(N \times 1)$ dot product of $\mathbf{T}^N$ the matrix of Tuples in (xx) , $L$ the length of the matrix, and $A$ contains the parameters

$$A^N = \left(a^j\right)_{j=1,\ldots N}$$

So for instance, for $N = 3$, $\mathbf{T} = \left(1, a, a^2, a^3\right)$

$$\mathbf{A}^3\ \mathbf{T}^3 = \begin{pmatrix} a^3 + a^2 + a \\ -a^3 + a^2 + a \\ a^3 - a^2 + a \\ -a^3 - a^2 + a \\ a^3 + a^2 - a \\ -a^3 + a^2 - a \\ a^3 - a^2 - a \\ -a^3 - a^2 - a \end{pmatrix}$$

The moments are as follows:

$$M_1(N) = \mu$$

$$M_2(N) = \mu^2 + 2\sigma$$

$$M_4(N) = \mu^4 + 12\mu^2\sigma + 12\sigma^2 \sum_{i=0}^{N} a^{2i}$$

At the limit:

$$\lim_{N \to \infty} M_4(N) = \frac{12\sigma^2}{1 - a^2} + \mu^4 + 12\mu^2\sigma$$

which is very mild.

## 20.3 LIMIT DISTRIBUTION

See Taleb and Cirillo [291] for the treatment of the limit distribution which will be a lognormal under the right conditions. In fact lognormal approximations work well when errors on errors are in constant proportion.

# 21 | STOCHASTIC TAIL EXPONENT FOR ASYMMETRIC POWER LAWS†

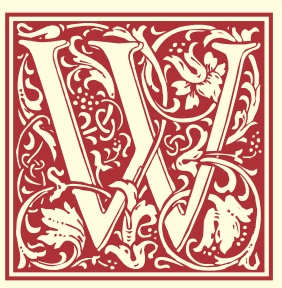

E EXAMINE random variables in the power law/slowly varying class with stochastic tail exponent , the exponent $\alpha$ having its own distribution. We show the effect of stochasticity of $\alpha$ on the expectation and higher moments of the random variable. For instance, the moments of a right-tailed or right-asymmetric variable, when finite, increase with the variance of $\alpha$; those of a left-asymmetric one decreases. The same applies to conditional shortfall (CVar), or mean-excess functions.

We prove the general case and examine the specific situation of lognormally distributed $\alpha \in [b, \infty), b > 1$.

The stochasticity of the exponent induces a significant bias in the estimation of the mean and higher moments in the presence of data uncertainty. This has consequences on sampling error as uncertainty about $\alpha$ translates into a higher expected mean.

The bias is conserved under summation, even upon large enough a number of summands to warrant convergence to the stable distribution. We establish inequalities related to the asymmetry.

We also consider the situation of capped power laws (i.e. with compact support), and apply it to the study of violence by Cirillo and Taleb (2016). We show that uncertainty concerning the historical data increases the true mean.

---

## 21.1 BACKGROUND

Stochastic volatility has been introduced heuristically in mathematical finance by traders looking for biases on option valuation, where a Gaussian distribution is considered to have several possible variances, either locally or at some specific future date. Options far from the money (i.e. concerning tail events) increase in value with uncertainty on the variance of the distribution, as they are convex to the standard deviation.

This led to a family of models of Brownian motion with stochastic variance (see review in Gatheral [120]) and proved useful in tracking the distributions of the underlying and the effect of the nonGaussian character of random processes on functions of the process (such as option prices).

Just as options are convex to the scale of the distribution, we find many situations where expectations are convex to the Power Law tail exponent . This note examines two cases:

- The standard power laws, one-tailed or asymmetric.
- The pseudo-power law, where a random variable appears to be a Power law but has compact support, as in the study of violence [55] where wars have the number of casualties capped at a maximum value.

## 21.2 ONE TAILED DISTRIBUTIONS WITH STOCHASTIC ALPHA

### 21.2.1 General Cases

**Definition 21.1**
*Let X be a random variable belonging to the class of distributions with a "power law" right tail, that is support in* $[x_0, +\infty), \in \mathbb{R}$*:*

*Subclass* $\mathfrak{P}_1$*:*

$$\{X : \mathbb{P}(X > x) = L(x)x^{-\alpha}, \frac{\partial^q L(x)}{\partial x^q} = 0 \text{ for } q \geq 1\} \tag{21.1}$$

*We note that x_o can be negative by shifting, so long as* $x_0 > -\infty$*.*

*Class* $\mathfrak{P}$*:*

$$\{X : \mathbb{P}(X > x) = L(x)\, x^{-\alpha}\} \tag{21.2}$$

*where* $\sim$ *means that the limit of the ratio or rhs to lhs goes to 1 as* $x \to \infty$*.* $L : [x_{\min}, +\infty) \to (0, +\infty)$ *is a slowly varying function, defined as* $\lim_{x\to+\infty} \frac{L(kx)}{L(x)} = 1$ *for any* $k > 0$*.* $L'(x)$ *is monotone. The constant* $\alpha > 0$*.*

*We further assume that:*

$$\lim_{x\to\infty} L'(x)\, x = 0 \tag{21.3}$$

$$\lim_{x\to\infty} L''(x)\, x = 0 \tag{21.4}$$

*We have*

$$\mathfrak{P}_1 \subset \mathfrak{P}$$

We note that the first class corresponds to the Pareto distributions (with proper shifting and scaling), where $L$ is a constant and $\mathfrak{P}$ to the more general one-sided power laws.

### 21.2.2 Stochastic Alpha Inequality

Throughout the rest of the paper we use for notation $X'$ for the stochastic alpha version of $X$, the constant $\alpha$ case.

**Proposition 21.1**
*Let $p = 1, 2, ...,$ $X'$ be the same random variable as X above in $\mathfrak{P}_1$ (the one-tailed regular variation class), with $x_0 \geq 0$, except with stochastic $\alpha$ with all realizations $> p$ that preserve the mean $\bar{\alpha}$,*

$$\mathbb{E}(X'^p) \geq \mathbb{E}(X^p).$$

**Proposition 21.2**
*Let K be a threshold. With X in the $\mathfrak{P}$ class, we have the expected conditional shortfall (CVar):*

$$\lim_{K \to \infty} \mathbb{E}(X'|_{X'>K}) \geq \lim_{K \to \infty} \mathbb{E}(X|_{X>K}).$$

The sketch of the proof is as follows.

We remark that $\mathbb{E}(X^p)$ is convex to $\alpha$, in the following sense. Let $X_{\alpha_i}$ be the random variable distributed with constant tail exponent $\alpha_i$, with $\alpha_i > p, \forall i$, and $\omega_i$ be the normalized positive weights: $\sum_i \omega_i = 1$, $0 \leq |\omega_i| \leq 1$, $\sum_i \omega_i \alpha_i = \bar{\alpha}$. By Jensen's inequality:

$$\omega_i \sum_i \mathbb{E}(X^p_{\alpha_i}) \geq \mathbb{E}(\sum_i (\omega_i X^p_{\alpha_i})).$$

As the classes are defined by their survival functions, we first need to solve for the corresponding density: $\varphi(x) = \alpha x^{-\alpha-1} L(x, \alpha) - x^{-\alpha} L^{(1,0)}(x, \alpha)$ and get the normalizing constant.

$$L(x_0, \alpha) = x_0^{\alpha} - \frac{2x_0 L^{(1,0)}(x_0, \alpha)}{\alpha - 1} - \frac{2x_0^2 L^{(2,0)}(x_0, \alpha)}{(\alpha - 1)(\alpha - 2)}, \tag{21.5}$$

$\alpha \neq 1, 2$ when the first and second derivative exist, respectively. The slot notation $L^{(p,0)}(x_0, \alpha)$ is short for $\frac{\partial^p L(x,\alpha)}{\partial x^p}|_{x=x_0}$.

By the Karamata representation theorem, [26],[300], a function $L$ on $[x_0, +\infty)$ is slowly moving (Definition) if and only if it can be written in the form

$$L(x) = \exp\left( \int_{x_0}^{x} \frac{\epsilon(t)}{t} \, dt + \eta(x) \right)$$

where $\eta(.)$ is a bounded measurable function converging to a finite number as $x \to +\infty$, and $\epsilon(x)$ is a bounded measurable function converging to zero as $x \to +\infty$.

Accordingly, $L'(x)$ goes to 0 as $x \to \infty$. (We further assumed in 21.3 and 21.4 that $L'(x)$ goes to 0 faster than $x$ and $L''(x)$ goes to 0 faster than $x^2$). Integrating by parts,

$$\mathbb{E}(X^p) = x_0^p + p \int_{x_0}^{\infty} x^{p-1}\, \mathrm{d}\bar{F}(x)$$

where $\bar{F}$ is the survival function in Eqs. 27.1 and 21.2. Integrating by parts three additional times and eliminating derivatives of $L(.)$ of higher order than 2:

$$\mathbb{E}(X^p) = \frac{x_0^{p-\alpha} L(x_0,\alpha)}{p-\alpha} - \frac{x_0^{p-\alpha+1} L^{(1,0)}(x_0,\alpha)}{(p-\alpha)(p-\alpha+1)} + \frac{x_0^{p-\alpha+2} L^{(2,0)}(x_0,\alpha)}{(p-\alpha)(p-\alpha+1)(p-\alpha+2)} \tag{21.6}$$

which, for the special case of $X$ in $\mathfrak{P}_1$ reduces to:

$$\mathbb{E}(X^p) = x_0^p \frac{\alpha}{\alpha - p} \tag{21.7}$$

As to Proposition 2, we can approach the proof from the property that $\lim_{x\to\infty} L'(x) = 0$. This allows a proof of var der Mijk's law that Paretian inequality is invariant to the threshold in the tail, that is $\frac{\mathbb{E}(X|_{X>K})}{K}$ converges to a constant as $K \to +\infty$. Equation 21.6 presents the exact conditions on the functional form of $L(x)$ for the convexity to extend to sub-classes between $\mathfrak{P}_1$ and $\mathfrak{P}$.

Our results hold to distributions that are transformed by shifting and scaling, of the sort:

$x \mapsto x - \mu + x_0$ (Pareto II), or with further transformations to Pareto types II and IV.

We note that the representation $\mathfrak{P}_1$ uses the same parameter, $x_0$, for both scale and minimum value, as a simplification.

We can verify that the expectation from Eq. 21.7 is convex to $\alpha$: $\frac{\partial \mathbb{E}(X^p)}{\partial \alpha^2} = x_0^p \frac{2}{(\alpha-1)^3}$.

### 21.2.3 Approximations for the Class $\mathfrak{P}$

For $\mathfrak{P} \setminus \mathfrak{P}_1$, our results hold when we can write an approximation the expectation of $X$ as a constant multiplying the integral of $x^{-\alpha}$, namely

$$\mathbb{E}(X) \approx k \frac{\nu(\alpha)}{\alpha - 1} \tag{21.8}$$

where $k$ is a positive constant that does not depend on $\alpha$ and $\nu(.)$ is approximated by a linear function of $\alpha$ (plus a threshold). The expectation will be convex to $\alpha$.

**Example: Student T Distribution** For the Student T distribution with tail $\alpha$, the "sophisticated" slowly varying function in common use for symmetric power laws in quantitative finance, the half-mean or the mean of the one-sided distribution (i.e. with support on $\mathbb{R}^+$ becomes

$$2\nu(\alpha) = 2\frac{\sqrt{\alpha}\Gamma\left(\frac{\alpha+1}{2}\right)}{\sqrt{\pi}\Gamma\left(\frac{\alpha}{2}\right)} \approx \alpha\frac{(1+\log(4))}{\pi},$$

where $\Gamma(.)$ is the gamma function.

## 21.3 SUMS OF POWER LAWS

As we are dealing from here on with convergence to the stable distribution, we consider situations of $1 < \alpha < 2$, hence $p = 1$ and will be concerned solely with the mean.

We observe that the convexity of the mean is invariant to summations of Power Law distributed variables as $X$ above. The Stable distribution has a mean that in conventional parameterizations does not appear to depend on $\alpha$ –but in fact depends on it.

Let $Y$ be distributed according to a Pareto distribution with density $f(y) \triangleq \alpha\lambda^\alpha y^{-\alpha-1}, y \geq \lambda > 0$ and with its tail exponent $1 < \alpha < 2$. Now, let $Y_1, Y_2, \ldots Y_n$ be identical and independent copies of $Y$. Let $\chi(t)$ be the characteristic function for $f(y)$. We have $\chi(t) = \alpha(-it)^\alpha\Gamma(-\alpha, -it)$, where $\gamma(.,.)$ is the incomplete gamma function. We can get the mean from the characteristic function of the average of $n$ summands $\frac{1}{n}(Y_1 + Y_2 + ...Y_n)$, namely $\chi(\frac{t}{n})^n$. Taking the first derivative:

$$-i\frac{\partial\chi(\frac{t}{n})^n}{\partial t} = (-i)^{\alpha(n-1)}n^{1-\alpha n}\alpha^n\lambda^{\alpha(n-1)}t^{\alpha(n-1)-1}\Gamma\Bigg(-\alpha, -\frac{it\lambda}{n}\Bigg)^{n-1}\left((-i)^\alpha\alpha\lambda^\alpha t^\alpha\Gamma\left(-\alpha, -\frac{it\lambda}{n}\right) - n^\alpha e^{\frac{i\lambda t}{n}}\right) \tag{21.9}$$

and

$$\lim_{n\to\infty} -i\frac{\partial\chi(\frac{t}{n})^n}{\partial t}\Bigg|_{t=0} = \lambda\frac{\alpha}{\alpha-1} \tag{21.10}$$

Thus we can see how the converging asymptotic distribution for the average will have for mean the scale times $\frac{\alpha}{\alpha-1}$, which does not depends on $n$.

Let $\chi^S(t)$ be the characteristic function of the corresponding stable distribution $S_{\alpha,\beta,\mu,\sigma}$, from the distribution of an infinitely summed copies of $Y$. By the Lévy continuity theorem, we have

- $\frac{1}{n}\Sigma_{i\leq n}Y_i \xrightarrow{\mathcal{D}} S$, with distribution $S_{\alpha,\beta,\mu,\sigma}$, where $\xrightarrow{\mathcal{D}}$ denotes convergence in distribution

  and

- $\chi^S(t) = \lim_{n\to\infty}\chi(t/n)^n$

are equivalent.

So we are dealing with the standard result [327],[246], for exact Pareto sums [324], replacing the conventional $\mu$ with the mean from above:

$$\chi^S(t) = \exp\left(i\left(\lambda\frac{\alpha t}{\alpha-1} + |t|^\alpha\left(\beta\tan\left(\frac{\pi\alpha}{2}\right)\operatorname{sgn}(t) + i\right)\right)\right).$$

## 21.4 ASYMMETRIC STABLE DISTRIBUTIONS

We can verify by symmetry that, effectively, flipping the distribution in subclasses $\mathfrak{P}_1$ and $\mathfrak{P}_2$ around $y_0$ to make it negative yields a negative value of the mean d higher moments, hence degradation from stochastic $\alpha$.

The central question becomes:

**Remark 21.1: Preservation of Asymmetry**

*A normalized sum in $\mathfrak{P}_1$ one-tailed distribution with expectation that depends on $\alpha$ of the form in Eq. 21.8 will necessarily converge in distribution to an asymmetric stable distribution $S_{\alpha,\beta,\mu,1}$, with $\beta \neq 0$.*

**Remark 21.2**

*Let $Y'$ be $Y$ under mean-preserving stochastic $\alpha$. The convexity effect becomes*

$$sgn\left(\mathbb{E}(Y') - \mathbb{E}(Y)\right) = sgn(\beta).$$

The sketch of the proof is as follows. Consider two slowly varying functions as in 27.1, each on one side of the tails. We have $L(y) = \mathbb{1}_{y<y_\theta}L^-(y) + \mathbb{1}_{y\geq y_\theta}L^+(y)$:

$$\begin{cases} L^+(y), L:[y_\theta,+\infty], & \lim_{y\to\infty} L^+(y) = c \\ \\ L^-(y), L:[-\infty,y_\theta], & \lim_{y\to-\infty} L^-(y) = d. \end{cases}$$

From [246],

$$\text{if} \begin{cases} \mathbb{P}(X > x) \sim cx^{-\alpha}, x \to +\infty \\ \\ \mathbb{P}(X < x) \sim d|x|^{-\alpha}, x \to +\infty, \end{cases} \quad \text{then } Y \text{ converges in distribution to } S_{\alpha,\beta,\mu,1}$$

with the coefficient $\beta = \frac{c-d}{c+d}$.

We can show that the mean can be written as $(\lambda_+ - \lambda_-)\frac{\alpha}{\alpha-1}$ where:

$$\lambda_+ \geq \lambda_- \text{ if } \int_{y_\theta}^{\infty} L^+(y)\mathrm{d}y, \geq \int_{-\infty}^{y_\theta} L^-(y)\mathrm{d}y$$

## 21.5 PARETO DISTRIBUTION WITH LOGNORMALLY DISTRIBUTED $\alpha$

Now assume $\alpha$ is following a shifted Lognormal distribution with mean $\alpha_0$ and minimum value $b$, that is, $\alpha - b$ follows a Lognormal $\mathcal{L}\left(\log(\alpha_0) - \frac{\sigma^2}{2}, \sigma\right)$. The parameter $b$ allows us to work with a lower bound on the tail exponent in order to satisfy finite expectation. We know that the tail exponent will eventually converge to $b$ but the process may be quite slow.

**Proposition 21.3**
*Assuming finite expectation for X′ and for exponent the lognormally distributed shifted variable $\alpha - b$ with law $\mathcal{L}\left(\log(\alpha_0) - \frac{\sigma^2}{2}, \sigma\right)$, $b \geq 1$ mininum value for $\alpha$, and scale $\lambda$:*

$$\mathbb{E}(Y') = \mathbb{E}(Y) + \lambda \frac{(e^{\sigma^2} - b)}{\alpha_0 - b} \tag{21.11}$$

We need $b \geq 1$ to avoid problems of infinite expectation.

Let $\phi(y, \alpha)$ be the density with stochastic tail exponent. With $\alpha > 0, \alpha_0 > b, b \geq 1, \sigma > 0, Y \geq \lambda > 0$ ,

$$\begin{aligned}
\mathbb{E}(Y) &= \int_b^\infty \int_L^\infty y \phi(y; \alpha)\, \mathrm{d}y\, \mathrm{d}\alpha \\
&= \int_b^\infty \lambda \frac{\alpha}{\alpha - 1} \frac{1}{\sqrt{2\pi}\sigma(\alpha - b)} \\
&\quad \exp\left( - \frac{\left(\log(\alpha - b) - \log(\alpha_0 - b) + \frac{\sigma^2}{2}\right)^2}{2\sigma^2} \right) \mathrm{d}\alpha \\
&= \frac{\lambda\left(\alpha_0 + e^{\sigma^2} - b\right)}{\alpha_0 - b} .
\end{aligned} \tag{21.12}$$

### Approximation of the Density

With $b = 1$ (which is the lower bound for $b$),we get the density with stochastic $\alpha$:

$$\phi(y; \alpha_0, \sigma) = \lim_{k \to \infty} \frac{1}{Y^2} \sum_{i=0}^{k} \frac{1}{i!} L(\alpha_0 - 1)^i e^{\frac{1}{2} i(i-1)\sigma^2} (\log(\lambda) - \log(y))^{i-1} (i + \log(\lambda) - \log(y)) \tag{21.13}$$

This result is obtained by expanding $\alpha$ around its lower bound $b$ (which we simplified to $b = 1$) and integrating each summand.

## 21.6 PARETO DISTRIBUTION WITH GAMMA DISTRIBUTED ALPHA

**Proposition 21.4**
*Assuming finite expectation for $X'$ scale $\lambda$, and for exponent a gamma distributed shifted variable $\alpha - 1$ with law $\varphi(.)$, mean $\alpha_0$ and variance $s^2$, all values for $\alpha$ greater than 1:*

$$\mathbb{E}(X') = \mathbb{E}(X') + \frac{s^2}{(\alpha_0 - 1)(\alpha_0 - s - 1)(\alpha_0 + s - 1)} \tag{21.14}$$

*Proof.*

$$\varphi(\alpha) = \frac{e^{-\frac{(\alpha-1)(\alpha_0-1)}{s^2}} \left( \frac{s^2}{(\alpha-1)(\alpha_0-1)} \right)^{-\frac{(\alpha_0-1)^2}{s^2}}}{(\alpha-1)\Gamma\left( \frac{(\alpha_0-1)^2}{s^2} \right)}, \quad \alpha > 1 \tag{21.15}$$

$$\int_1^\infty \alpha \lambda^\alpha x^{-\alpha-1} \varphi(\alpha)\, d\alpha \tag{21.16}$$

$$= \int_1^\infty \frac{\alpha \left( e^{-\frac{(\alpha-1)(\alpha_0-1)}{s^2}} \left( \frac{s^2}{(\alpha-1)(\alpha_0-1)} \right)^{-\frac{(\alpha_0-1)^2}{s^2}} \right)}{(\alpha - 1)\left( (\alpha - 1)\Gamma\left( \frac{(\alpha_0-1)^2}{s^2} \right) \right)} d\alpha$$

$$= \frac{1}{2}\left( \frac{1}{\alpha_0 + s - 1} + \frac{1}{\alpha_0 - s - 1} + 2 \right)$$

□

## 21.7 THE BOUNDED POWER LAW IN CIRILLO AND TALEB (2016)

In [55] and [54], the studies make use of bounded power laws, applied to violence and operational risk, respectively. Although with $\alpha < 1$ the variable $Z$ has finite expectations owing to the upper bound.

The methods offered were a smooth transformation of the variable as follows: we start with $z \in [L, H), L > 0$ and transform it into $x \in [L, \infty)$, the latter legitimately being Power Law distributed.

So the smooth logarithmic transformation):

$$x = \varphi(z) = L - H \log\left( \frac{H - z}{H - L} \right),$$

and

$$f(x) = \frac{\left( \frac{x-L}{\alpha\sigma} + 1 \right)^{-\alpha-1}}{\sigma}.$$

We thus get the distribution of $Z$ which will have a finite expectation for all positive values of $\alpha$.

$$\frac{\partial^2 \mathbb{E}(Z)}{\partial \alpha^2} = \frac{1}{H^3}(H-L)\left(e^{\frac{\alpha\sigma}{H}}\left(2H^3 G_{3,4}^{4,0}\left(\frac{\alpha\sigma}{H}\Big|\begin{array}{c}\alpha+1,\alpha+1,\alpha+1\\ 1,\alpha,\alpha,\alpha\end{array}\right) - 2H^2(H+\sigma)G_{2,3}^{3,0}\left(\frac{\alpha\sigma}{H}\Big|\begin{array}{c}\alpha+1,\alpha+1\\ 1,\alpha,\alpha\end{array}\right) + \sigma\left(\alpha\sigma^2+(\alpha+1)H^2+2\alpha H\sigma\right)E_\alpha\left(\frac{\alpha\sigma}{H}\right)\right) - H\sigma(H+\sigma)\right) \tag{21.17}$$

which appears to be positive in the range of numerical perturbations in [55].[3] At such a low level of $\alpha$, around $\frac{1}{2}$, the expectation is extremely convex and the bias will be accordingly extremely pronounced.

This convexity has the following practical implication. Historical data on violence over the past two millennia, is fundamentally unreliable [55]. Hence an imprecision about the tail exponent , from errors embedded in the data, need to be present in the computations. The above shows that uncertainty about $\alpha$, is more likely to make the "true" statistical mean (that is the mean of the process as opposed to sample mean) higher than lower, hence supports the statement that more uncertainty increases the estimation of violence.

## 21.8 ADDITIONAL COMMENTS

The bias in the estimation of the mean and shortfalls from uncertainty in the tail exponent can be added to analyses where data is insufficient, unreliable, or simply prone to forgeries.

In additional to statistical inference, these result can extend to processes, whether a compound Poisson process with power laws subordination [257] (i.e. a Poisson arrival time and a jump that is Power Law distributed) or a Lévy process. The latter can be analyzed by considering successive "slice distributions" or discretization of the process [60]. Since the expectation of a sum of jumps is the sum of expectation, the same convexity will appear as the one we got from Eq. 21.8.

## 21.9 ACKNOWLEDGMENTS

Marco Avellaneda, Robert Frey, Raphael Douady, Pasquale Cirillo.

---

[3] $G_{3,4}^{4,0}\left(\frac{\alpha\sigma}{H}\Big|\begin{array}{c}\alpha+1,\alpha+1,\alpha+1\\ 1,\alpha,\alpha,\alpha\end{array}\right)$ is the Meijer G function.

# 22 META-DISTRIBUTION OF P-VALUES AND P-HACKING‡

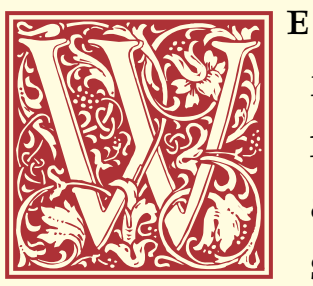e present an exact probability distribution (meta-distribution) for p-values across ensembles of statistically identical phenomena, as well as the distribution of the minimum p-value among $m$ independents tests. We derive the distribution for small samples $2 < n \leq n^* \approx 30$ as well as the limiting one as the sample size $n$ becomes large. We also look at the properties of the "power" of a test through the distribution of its inverse for a given p-value and parametrization.

P-values are shown to be extremely skewed and volatile, regardless of the sample size $n$, and vary greatly across repetitions of exactly same protocols under identical stochastic copies of the phenomenon; such volatility makes the minimum $p$ value diverge significantly from the "true" one. Setting the power is shown to offer little remedy unless sample size is increased markedly or the p-value is lowered by at least one order of magnitude.

The formulas allow the investigation of the stability of the reproduction of results and "p-hacking" and other aspects of meta-analysis –including a metadistribution of p-hacked results.

From a probabilistic standpoint, neither a p-value of .05 nor a "power" at .9 appear to make the slightest sense.

Assume that we know the "true" p-value, $p_s$, what would its realizations look like across various attempts on statistically identical copies of the phenomena? By true value $p_s$, we mean its expected value by the law of large numbers across an $m$ ensemble of possible samples for the phenomenon under scrutiny, that is $\frac{1}{m}\sum_{\leq m} p_i \xrightarrow{P} p_s$ (where $\xrightarrow{P}$ denotes convergence in probability). A similar convergence argument can be also made for the corresponding "true median" $p_M$. The main result of the paper is that the the distribution of $n$ small samples can be made explicit (albeit with special inverse functions), as well as its parsimonious limiting one for $n$ large, with no other parameter than the median value $p_M$. We were unable to get an explicit form for $p_s$ but we go around it with the use of the median. Finally, the distribution of the minimum p-value under can be made explicit, in a parsimonious formula allowing for the understanding of biases in scientific studies.

‡ Research chapter.

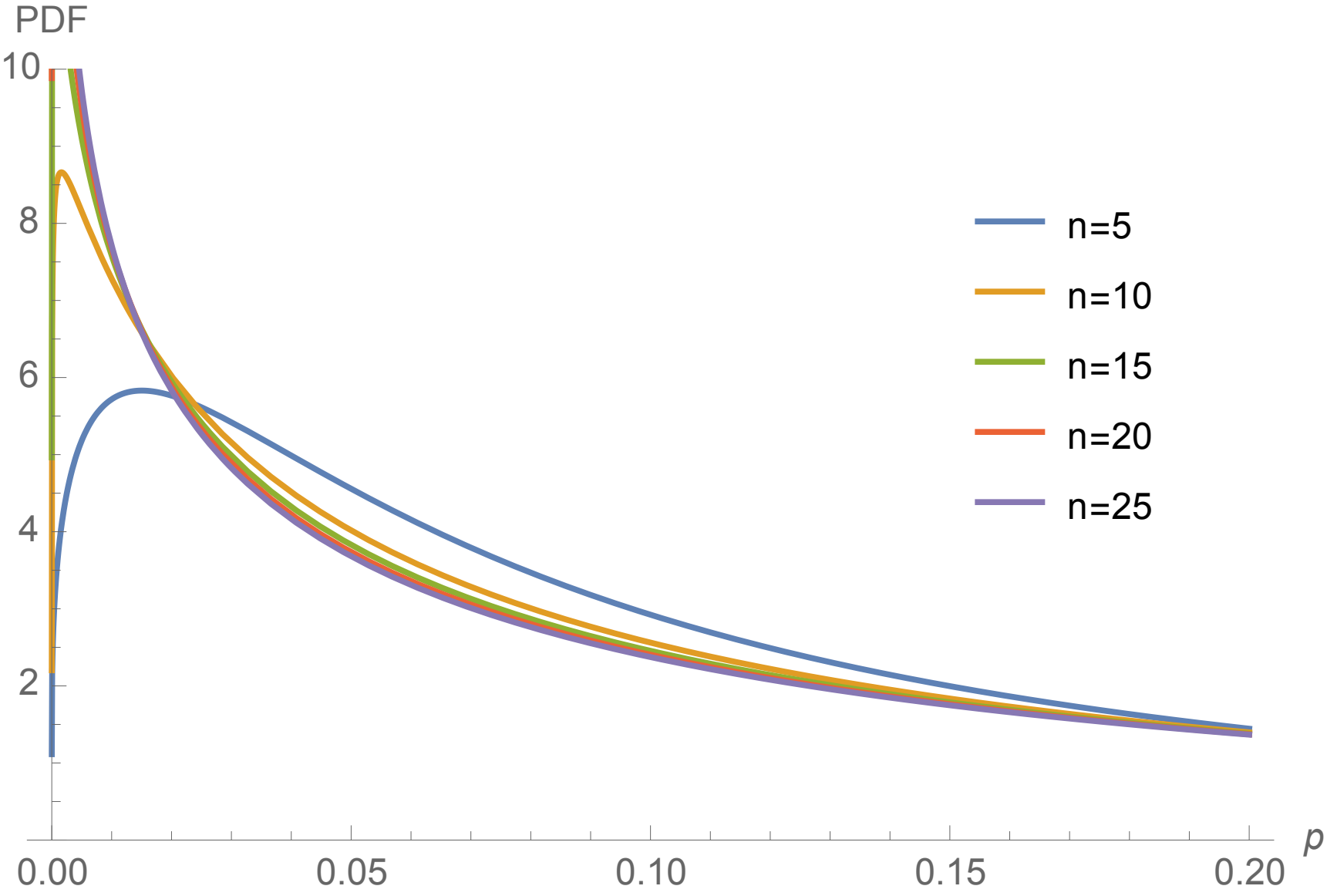

Figure 22.1: *The different values for Equ. 22.1 showing convergence to the limiting distribution.*

It turned out, as we can see in Figure 22.2 the distribution is extremely asymmetric (right-skewed), to the point where 75% of the realizations of a "true" p-value of .05 will be <.05 (a borderline situation is 3× as likely to pass than fail a given protocol), and, what is worse, 60% of the true p-value of .12 will be below .05.

Although with compact support, the distribution exhibits the attributes of extreme fat-tailedness. For an observed p-value of, say, .02, the "true" p-value is likely to be >.1 (and very possibly close to .2), with a standard deviation >.2 (sic) and a mean deviation of around .35 (sic, sic). Because of the excessive skewness, measures of dispersion in $\mathcal{L}^1$ and $\mathcal{L}^2$ (and higher norms) vary hardly with $p_s$, so the standard deviation is not proportional, meaning an in-sample .01 p-value has a significant probability of having a true value $> .3$.

**So clearly we don't know what we are talking about when we talk about p-values.**

Earlier attempts for an explicit meta-distribution in the literature were found in [154] and [245], though for situations of Gaussian subordination and less parsimonious parametrization. The severity of the problem of *significance of the so-called "statistically significant"* has been discussed in [123] and offered a remedy via Bayesian methods in [164], which in fact recommends the same tightening of standards to p-values $\approx$ .01. But the gravity of the extreme skewness of the distribution of p-values is only apparent when one looks at the meta-distribution.

For notation, we use $n$ for the sample size of a given study and $m$ the number of trials leading to a p-value.

## 22.1 PROOFS AND DERIVATIONS

**Proposition 22.1**
*Let $P$ be a random variable $\in [0,1]$) corresponding to the sample-derived one-tailed p-value from the paired T-test statistic (unknown variance) with median value $\mathbb{M}(P) = p_M \in [0,1]$ derived from a sample of $n$ size. The distribution across the ensemble of statistically identical copies of the sample has for PDF*

$$\varphi(p; p_M) = \begin{cases} \varphi(p; p_M)_L & \text{for } p < \frac{1}{2} \\ \varphi(p; p_M)_H & \text{for } p > \frac{1}{2} \end{cases}$$

$$\varphi(p; p_M)_L = \lambda_p^{\frac{1}{2}(-n-1)} \sqrt{-\frac{\lambda_p \left(\lambda_{p_M} - 1\right)}{\left(\lambda_p - 1\right)\lambda_{p_M} - 2\sqrt{\left(1-\lambda_p\right)\lambda_p}\sqrt{\left(1-\lambda_{p_M}\right)\lambda_{p_M}} + 1}} \left(\frac{1}{\frac{1}{\lambda_p} - \frac{2\sqrt{1-\lambda_p}\sqrt{\lambda_{p_M}}}{\sqrt{\lambda_p}\sqrt{1-\lambda_{p_M}}} + \frac{1}{1-\lambda_{p_M}} - 1}\right)^{n/2}$$

$$\varphi(p; p_M)_H = \left(1-\lambda'_p\right)^{\frac{1}{2}(-n-1)} \left(\frac{\left(\lambda'_p - 1\right)\left(\lambda_{p_M} - 1\right)}{\lambda'_p\left(-\lambda_{p_M}\right) + 2\sqrt{\left(1-\lambda'_p\right)\lambda'_p}\sqrt{\left(1-\lambda_{p_M}\right)\lambda_{p_M}} + 1}\right)^{\frac{n+1}{2}} \tag{22.1}$$

where $\lambda_p = I^{-1}_{2p}\left(\frac{n}{2}, \frac{1}{2}\right)$, $\lambda_{p_M} = I^{-1}_{1-2p_M}\left(\frac{1}{2}, \frac{n}{2}\right)$, $\lambda'_p = I^{-1}_{2p-1}\left(\frac{1}{2}, \frac{n}{2}\right)$, and $I^{-1}_{(.)}(.,.)$ is the inverse beta regularized function.

**Remark 22.1**
*For $p=\frac{1}{2}$ the distribution doesn't exist in theory, but does in practice and we can work around it with the sequence $p_{m_k} = \frac{1}{2} \pm \frac{1}{k}$, as in the graph showing a convergence to the Uniform distribution on $[0,1]$ in Figure 22.3. Also note that what is called the "null" hypothesis is effectively a set of measure 0.*

*Proof.* Let Z be a random normalized variable with realizations $\zeta$, from a vector $\vec{v}$ of $n$ realizations, with sample mean $m_v$, and sample standard deviation $s_v$, $\zeta = \frac{m_v - m_h}{\frac{s_v}{\sqrt{n}}}$ (where $m_h$ is the level it is tested against), hence assumed to $\sim$ Student T with $n$ degrees of freedom, and, crucially, supposed to deliver a mean of $\bar{\zeta}$,

$$f(\zeta;\bar{\zeta}) = \frac{\left(\frac{n}{(\bar{\zeta}-\zeta)^2+n}\right)^{\frac{n+1}{2}}}{\sqrt{n}B\left(\frac{n}{2},\frac{1}{2}\right)}$$

where B(.,.) is the standard beta function. Let $g(.)$ be the one-tailed survival function of the Student T distribution with zero mean and $n$ degrees of freedom:

$$g(\zeta) = \mathbb{P}(Z > \zeta) = \begin{cases} \frac{1}{2} I_{\frac{n}{\zeta^2+n}}\left(\frac{n}{2},\frac{1}{2}\right) & \zeta \geq 0 \\ \frac{1}{2}\left(I_{\frac{\zeta^2}{\zeta^2+n}}\left(\frac{1}{2},\frac{n}{2}\right)+1\right) & \zeta < 0 \end{cases}$$

where $I_{(.,.)}$ is the incomplete Beta function.

We now look for the distribution of $g \circ f(\zeta)$. Given that g(.) is a legit Borel function, and naming $p$ the probability as a random variable, we have by a standard result for the transformation:

$$\varphi(p,\bar{\zeta}) = \frac{f\left(g^{(-1)}(p)\right)}{|g'\left(g^{(-1)}(p)\right)|}$$

We can convert $\bar{\zeta}$ into the corresponding median survival probability because of symmetry of $Z$. Since one half the observations fall on either side of $\bar{\zeta}$, we can ascertain that the transformation is median preserving: $g(\bar{\zeta}) = \frac{1}{2}$, hence $\varphi(p_M,.) = \frac{1}{2}$. Hence we end up having $\{\bar{\zeta} : \frac{1}{2} I_{\frac{n}{\bar{\zeta}^2+n}}\left(\frac{n}{2},\frac{1}{2}\right) = p_M\}$ (positive case) and $\{\bar{\zeta} : \frac{1}{2}\left(I_{\frac{\bar{\zeta}^2}{\bar{\zeta}^2+n}}\left(\frac{1}{2},\frac{n}{2}\right)+1\right) = p_M\}$ (negative case). Replacing we get Eq.22.1 and Proposition 22.1 is done.

□

We note that $n$ does not increase significance, since p-values are computed from normalized variables (hence the universality of the meta-distribution); a high $n$ corresponds to an increased convergence to the Gaussian. For large $n$, we can prove the following proposition:

**Proposition 22.2**
*Under the same assumptions as above, the limiting distribution for $\varphi(.)$:*

$$\lim_{n\to\infty} \varphi(p;p_M) = e^{-erfc^{-1}(2p_M)\left(erfc^{-1}(2p_M)-2erfc^{-1}(2p)\right)} \tag{22.2}$$

*where erfc(.) is the complementary error function and $erfc(.)^{-1}$ its inverse.*

*The limiting CDF* $\Phi(.)$

$$\Phi(k; p_M) = \frac{1}{2} erfc\left( erf^{-1}(1-2k) - erf^{-1}(1-2p_M) \right) \tag{22.3}$$

*Proof.* For large $n$, the distribution of $Z = \frac{m_v}{\frac{s_v}{\sqrt{n}}}$ becomes that of a Gaussian, and the one-tailed survival function $g(.) = \frac{1}{2}\text{erfc}\left(\frac{\zeta}{\sqrt{2}}\right)$, $\zeta(p) \to \sqrt{2}\text{erfc}^{-1}(p)$. □

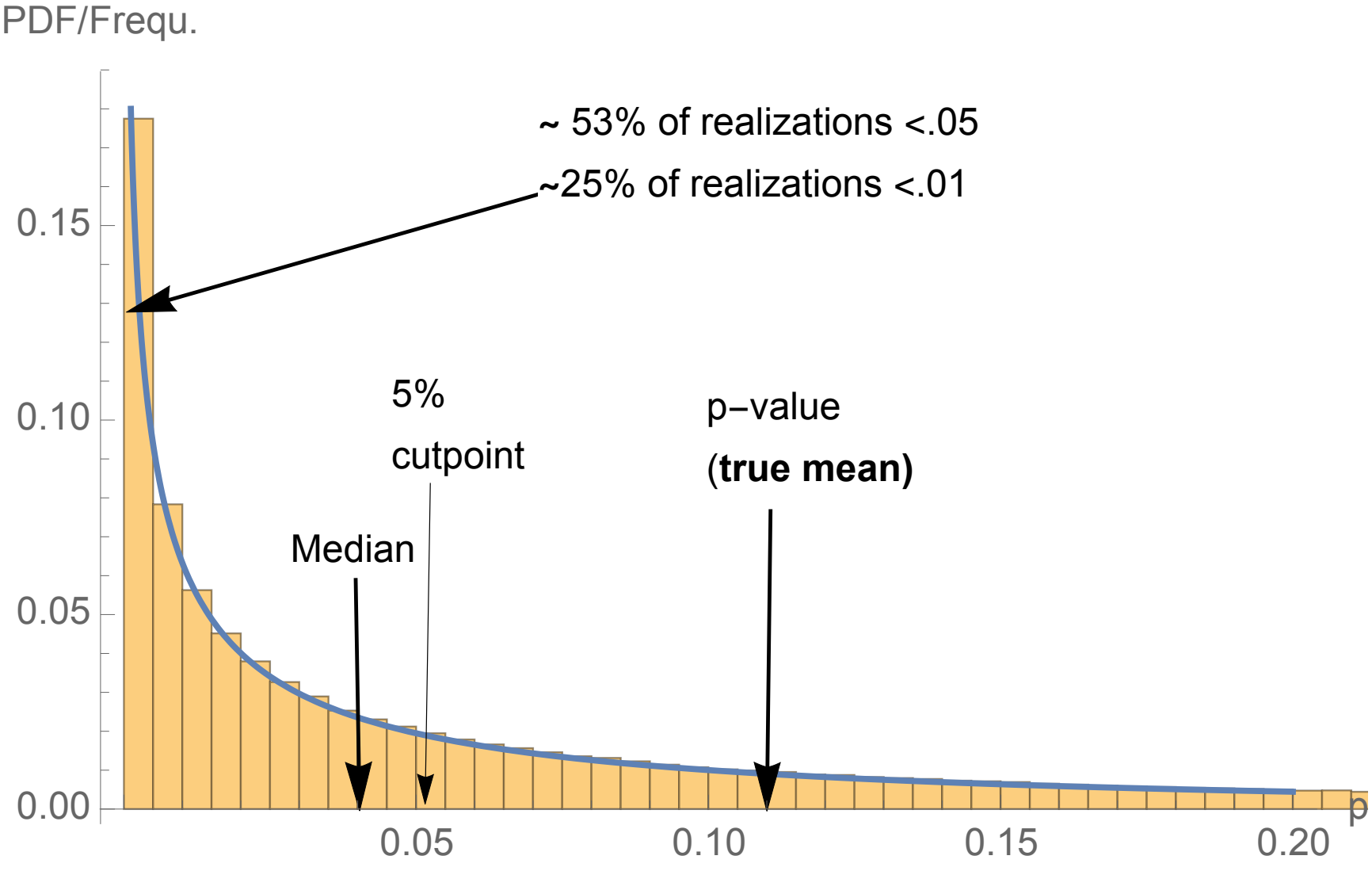

Figure 22.2: *The probability distribution of a one-tailed p-value with expected value .11 generated by Monte Carlo (histogram) as well as analytically with* $\varphi(.)$ *(the solid line). We draw all possible sub-samples from an ensemble with given properties. The excessive skewness of the distribution makes the average value considerably higher than most observations, hence causing illusions of "statistical significance".*

This limiting distribution applies for paired tests with known or assumed sample variance since the test becomes a Gaussian variable, equivalent to the convergence of the T-test (Student T) to the Gaussian when $n$ is large.

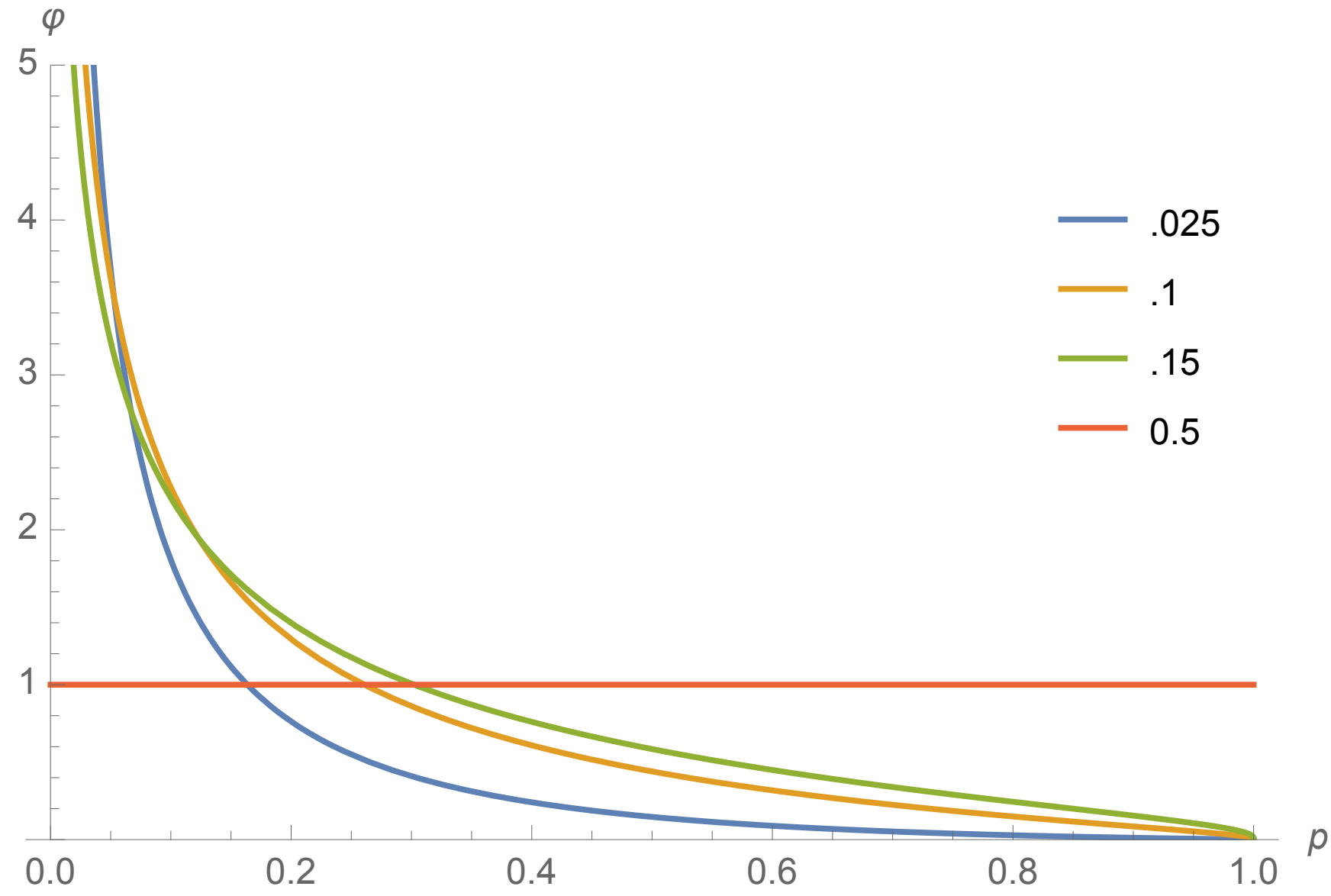

Figure 22.3: *The probability distribution of p at different values of $p_M$. We observe how $p_M = \frac{1}{2}$ leads to a uniform distribution.*

**Remark 22.2**

*For values of p close to 0, φ in Equ. 22.2 can be usefully calculated as:*

$$\varphi(p; p_M) = \sqrt{2\pi}\, p_M \sqrt{\log\left(\frac{1}{2\pi p_M^2}\right)}\; e^{\sqrt{-\log\left(2\pi \log\left(\frac{1}{2\pi p^2}\right)\right) - 2\log(p)}\sqrt{-\log\left(2\pi \log\left(\frac{1}{2\pi p_M^2}\right)\right) - 2\log(p_M)}} + O(p^2). \quad (22.4)$$

*The approximation works more precisely for the band of relevant values* $0 < p < \frac{1}{2\pi}$.

From this we can get numerical results for convolutions of $\varphi$ using the Fourier Transform or similar methods.

We can and get the distribution of the minimum p-value per $m$ trials across statistically identical situations thus get an idea of "p-hacking", defined as attempts by researchers to get the lowest p-values of many experiments, or try until one of the tests produces statistical significance.

**Proposition 22.3**
*The distribution of the minimum of m observations of statistically identical p-values becomes (under the limiting distribution of proposition 22.2):*

$$\varphi_m(p; p_M) = m\, e^{erfc^{-1}(2p_M)\left(2erfc^{-1}(2p)-erfc^{-1}(2p_M)\right)}$$
$$\left(1-\frac{1}{2}erfc\left(erfc^{-1}(2p)-erfc^{-1}(2p_M)\right)\right)^{m-1} \tag{22.5}$$

*Proof.* $P\left(p_1 > p, p_2 > p, \ldots, p_m > p\right) = \bigcap_{i=1}^{n} \Phi(p_i) = \bar{\Phi}(p)^m$. Taking the first derivative we get the result. □

Outside the limiting distribution: we integrate numerically for different values of $m$ as shown in Figure 22.4. So, more precisely, for $m$ trials, the expectation is calculated as:

$$\mathbb{E}(p_{min}) = \int_0^1 -m\ \varphi(p; p_M) \left(\int_0^p \varphi(u,.)\ \mathrm{d}u\right)^{m-1} \mathrm{d}p$$

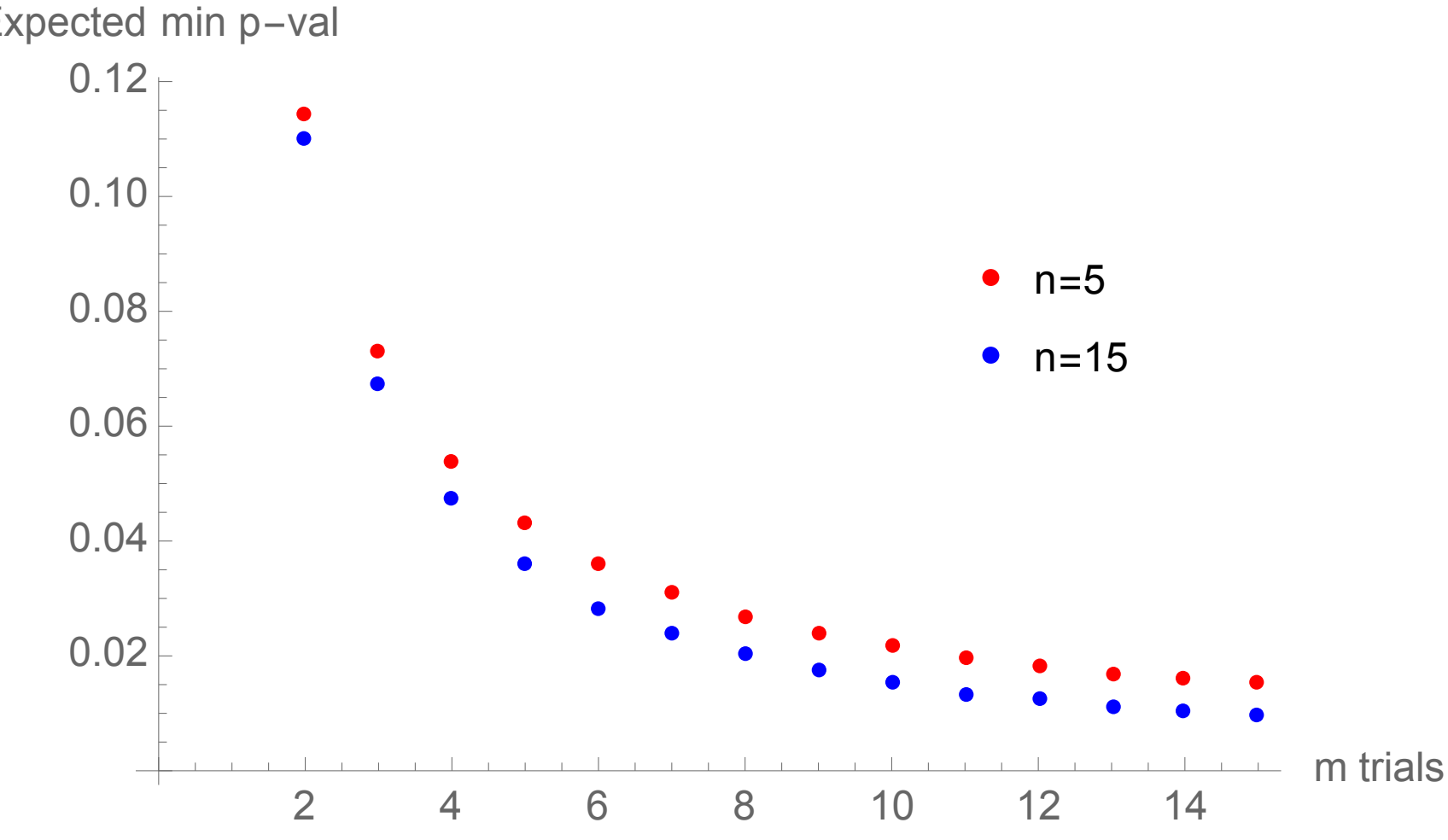

Figure 22.4: *The "p-hacking" value across m trials for* $p_M = .15$ *and* $p_s = .22$.

## 22.2 INVERSE POWER OF TEST

Let $\beta$ be the power of a test for a given p-value $p$, for random draws $X$ from unobserved parameter $\theta$ and a sample size of $n$. To gauge the reliability of $\beta$ as a true measure of power, we perform an inverse problem:

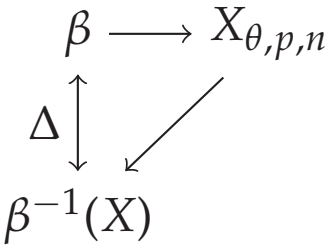

**Proposition 22.4**
*Let $\beta_c$ be the projection of the power of the test from the realizations assumed to be student T distributed and evaluated under the parameter $\theta$. We have*

$$\Phi(\beta_c) = \begin{cases} \Phi(\beta_c)_L & \text{for } \beta_c < \frac{1}{2} \\ \Phi(\beta_c)_H & \text{for } \beta_c > \frac{1}{2} \end{cases}$$

*where*

$$\Phi(\beta_c)_L = \sqrt{1-\gamma_1}\gamma_1^{-\frac{n}{2}} \frac{\left( -\frac{\gamma_1}{2\sqrt{\frac{1}{\gamma_3}-1}\sqrt{-(\gamma_1-1)\gamma_1}-2\sqrt{-(\gamma_1-1)\gamma_1}+\gamma_1\left(2\sqrt{\frac{1}{\gamma_3}-1}-\frac{1}{\gamma_3}\right)-1} \right)^{\frac{n+1}{2}}}{\sqrt{-(\gamma_1-1)\gamma_1}} \tag{22.6}$$

$$\Phi(\beta_c)_H = \sqrt{\gamma_2}\,(1-\gamma_2)^{-\frac{n}{2}} B\left(\frac{1}{2},\frac{n}{2}\right) \frac{\left( \frac{1}{\frac{-2\left(\sqrt{-(\gamma_2-1)\gamma_2}+\gamma_2\right)\sqrt{\frac{1}{\gamma_3}-1}+2\sqrt{\frac{1}{\gamma_3}-1}+2\sqrt{-(\gamma_2-1)\gamma_2}-1}{\gamma_2-1}+\frac{1}{\gamma_3}} \right)^{\frac{n+1}{2}}}{\sqrt{-(\gamma_2-1)\gamma_2}B\left(\frac{n}{2},\frac{1}{2}\right)} \tag{22.7}$$

*where* $\gamma_1 = I^{-1}_{2\beta_c}\left(\frac{n}{2},\frac{1}{2}\right)$, $\gamma_2 = I^{-1}_{2\beta_c-1}\left(\frac{1}{2},\frac{n}{2}\right)$, *and* $\gamma_3 = I^{-1}_{(1,2p_s-1)}\left(\frac{n}{2},\frac{1}{2}\right)$.

## 22.3 APPLICATION AND CONCLUSION

- One can safely see that under such stochasticity for the realizations of p-values and the distribution of its minimum, to get what people mean by 5% confidence (and the inferences they get from it), they need a p-value of at least one order of magnitude smaller.
- Attempts at replicating papers, such as the open science project [59], should consider a margin of error in *its own* procedure and a pronounced bias towards favorable results (Type-I error). There should be no surprise that a previously deemed significant test fails during replication –in fact it is the replication of results deemed significant at a close margin that should be surprising.
- The "power" of a test has the same problem unless one either lowers p-values or sets the test at higher levels, such at .99.

### ACKNOWLEDGMENT

Marco Avellaneda, Pasquale Cirillo, Yaneer Bar-Yam, friendly people on twitter
...

# I SOME CONFUSIONS IN BEHAVIORAL ECONOMICS

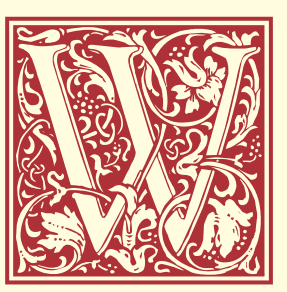

We saw earlier (Chapters 3 and 11) that the problem of "overestimation of the tails" by agents is more attributable to the use of a wrong "normative" model by psychologists and decision scientists who are innocent of fat tails. Here we use two cases illustrative of such improper use of probability, uncovered with our simple heuristic of inducing a second order effect and seeing the effect of Jensens's inequality on the expectation operator.

One such unrigorous use of probability (the equity premium puzzle) involves the promoter of "nudging", an invasive and sinister method devised by psychologists that aim at manipulating decisions by citizens.

## I.1 CASE STUDY: HOW THE MYOPIC LOSS AVERSION IS MISSPECIFIED

The so-called "equity premium puzzle", originally detected by Mehra and Prescott [198], is called so because equities have historically yielded too high a return over fixed income investments; the puzzle is why it isn't arbitraged away.

We can easily figure out that the analysis misses the absence of ergodicity in such domain, as we saw in Chapter 3: agents do not really capture market returns unconditionally; it is foolish to use ensemble probabilities and the law of large numbers for individual investors who only have one life. Also "positive expected returns" for a market is not a sufficient condition for an investor to obtain a positive expectation; a certain Kelly-style path scaling strategy, or path dependent dynamic hedging is required.

Benartzi and Thaler [20] claims that the Kahneman-Tversky prospect theory [165] explains such behavior owing to myopia. This might be true but such an analysis falls apart under thick tails.

So here we fatten tails of the distribution with stochasticity of, say, the scale parameter, and can see what happens to some results in the literature that seem absurd at face value, and in fact are absurd under more rigorous use of probabilistic analyses.

## Myopic loss aversion

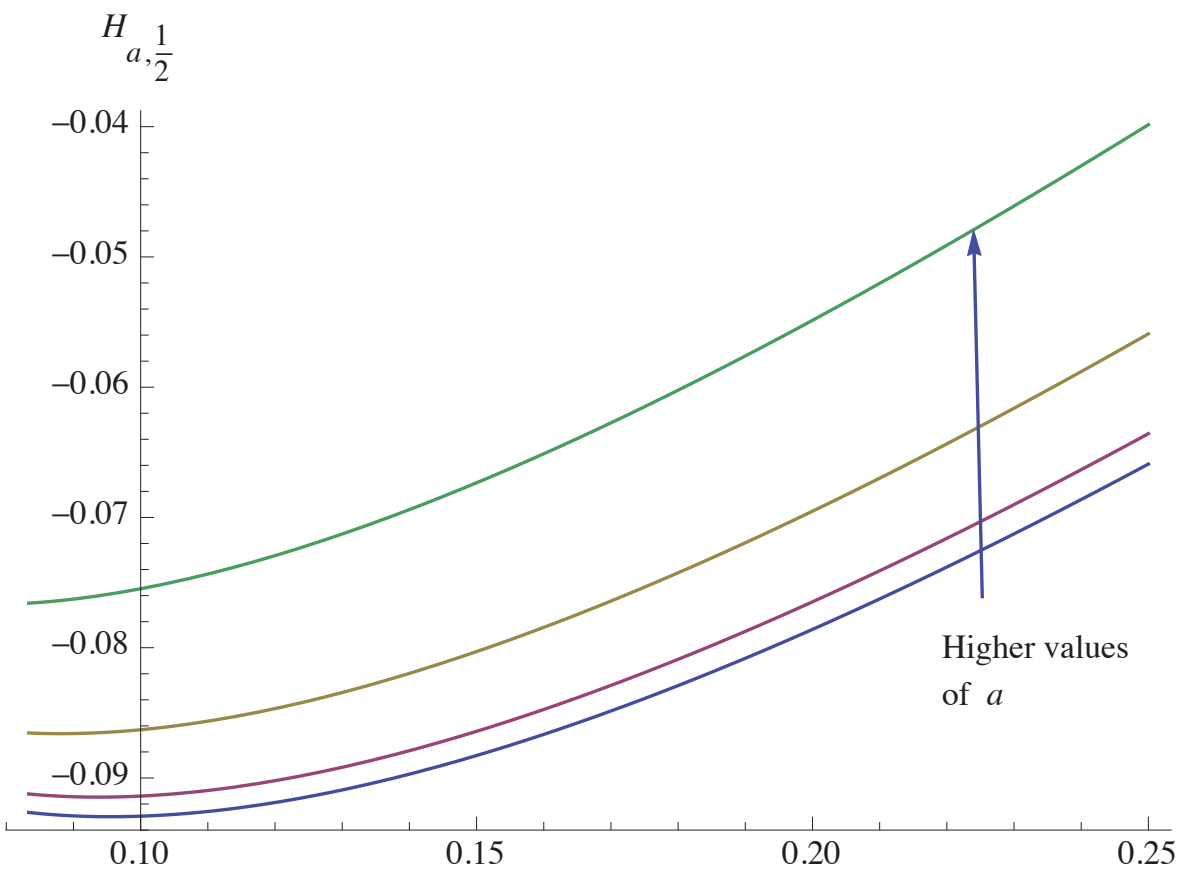

Figure I.1: *The effect of $H_{a,p}(t)$ "utility" or prospect theory of under second order effect on variance. Here $\sigma = 1, \mu = 1$ and $t$ variable.*

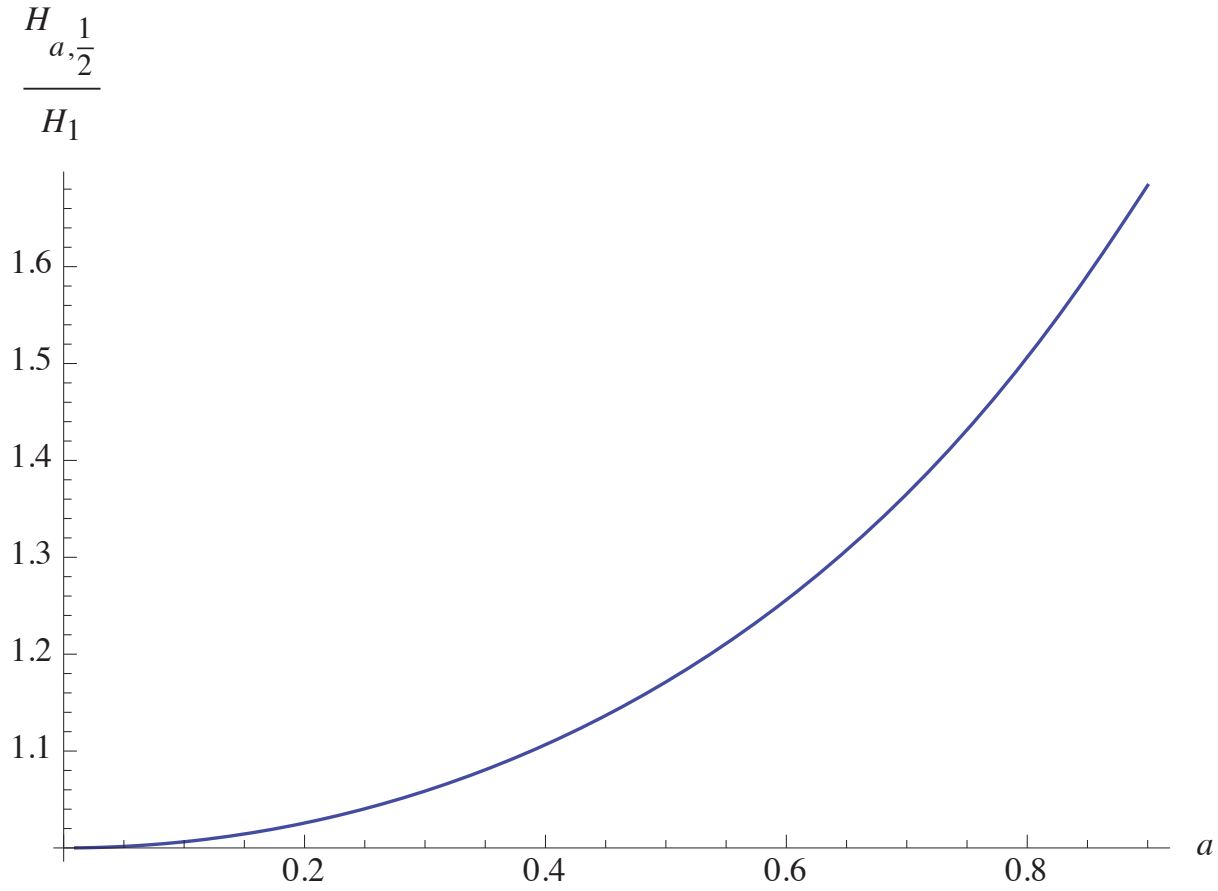

Figure I.2: *The ratio $\frac{H_{a,\frac{1}{2}}(t)}{H_0}$ or the degradation of "utility" under second order effects.*

Take the prospect theory valuation $w$ function for $x$ changes in wealth $x$, parameterized with $\lambda$ and $\alpha$.

$$w_{\lambda,\alpha}(x) = x^{\alpha}\, \mathbb{1}_{x\geq 0} - \lambda(-x^{\alpha})\, \mathbb{1}_{x<0}$$

Let $\phi_{\mu t,\sigma\sqrt{t}}(x)$ be the Normal Distribution density with corresponding mean and standard deviation (scaled by $t$)

The expected "utility" (in the prospect sense):

$$H_0(t) = \int_{-\infty}^{\infty} w_{\lambda,\alpha}(x)\phi_{\mu t,\sigma\sqrt{t}}(x)\, \mathrm{d}x \tag{I.1}$$

$$= \frac{1}{\sqrt{\pi}} 2^{\frac{\alpha}{2}-2} \left(\frac{1}{\sigma^2 t}\right)^{-\frac{\alpha}{2}} \left(\Gamma\left(\frac{\alpha+1}{2}\right)\left(\sigma^{\alpha} t^{\alpha/2}\left(\frac{1}{\sigma^2 t}\right)^{\alpha/2} - \lambda\sigma\sqrt{t}\sqrt{\frac{1}{\sigma^2 t}}\right) {}_1F_1\left(-\frac{\alpha}{2};\frac{1}{2};-\frac{t\mu^2}{2\sigma^2}\right) + \frac{1}{\sqrt{2}\sigma}\mu\Gamma\left(\frac{\alpha}{2}+1\right)\left(\sigma^{\alpha+1}t^{\frac{\alpha}{2}+1}\left(\frac{1}{\sigma^2 t}\right)^{\frac{\alpha+1}{2}} + \sigma^{\alpha} t^{\frac{\alpha+1}{2}}\left(\frac{1}{\sigma^2 t}\right)^{\alpha/2} + 2\lambda\sigma t\sqrt{\frac{1}{\sigma^2 t}}\right) {}_1F_1\left(\frac{1-\alpha}{2};\frac{3}{2};-\frac{t\mu^2}{2\sigma^2}\right)\right) \tag{I.2}$$

We can see from I.2 that the more frequent sampling of the performance translates into worse utility. So what Benartzi and Thaler did was try to find the sampling period "myopia" that translates into the sampling frequency that causes the "premium" —the error being that they missed second order effects.

Now under variations of $\sigma$ with stochatic effects, heuristically captured, the story changes: what if there is a very small probability that the variance gets multiplied by a large number, with the total variance remaining the same? The key here is that we are not even changing the variance at all: we are only shifting the distribution to the tails. We are here generously assuming that by the law of large numbers it was established that the "equity premium puzzle" was true and that stocks *really* outperformed bonds.

So we switch between two states, $(1+a)\,\sigma^2$ w.p. $p$ and $(1-a)$ w.p. $(1-p)$.

Rewriting I.1

$$H_{a,p}(t) = \int_{-\infty}^{\infty} w_{\lambda,\alpha}(x)\left(p\,\phi_{\mu\, t,\sqrt{1+a}\,\sigma\sqrt{t}}(x) + (1-p)\,\phi_{\mu\, t,\sqrt{1-a}\,\sigma\sqrt{t}}(x)\right)\,\mathrm{d}x \tag{I.3}$$

**Result** Conclusively, as can be seen in figures I.1 and I.2, second order effects cancel the statements made from "myopic" loss aversion. This doesn't mean that myopia doesn't have effects, rather that it cannot explain the "equity premium", not from the outside (i.e. the distribution might have different returns, but from the inside, owing to the structure of the Kahneman-Tversky value function $v(x)$.

**Comment** We used the $(1+a)$ heuristic largely for illustrative reasons; we could use a full distribution for $\sigma^2$ with similar results. For instance the gamma distribution with density $f(v) = \frac{v^{\gamma-1}e^{-\frac{\alpha v}{V}}\left(\frac{V}{\alpha}\right)^{-\gamma}}{\Gamma(\gamma)}$ with expectation $V$ matching the variance used in the "equity premium" theory.

Rewriting I.3 under that form,

$$\int_{-\infty}^{\infty}\int_{0}^{\infty} w_{\lambda,\alpha}(x)\phi_{\mu\, t,\,\sqrt{v\, t}}(x)\, f(v)\,\mathrm{d}v\,\mathrm{d}x$$

Which has a closed form solution (though a bit lengthy for here).

**True problem with Benartzi and Thaler** Of course the problem has to do with thick tails and the convergence under LLN, which we treat separately.

### Time Preference Under Model Error

Another example of the effect of the randomization of a parameter –the creation of an additional layer of uncertainty so to speak.

This author once watched with a great deal of horror one Laibson [179] at a conference in Columbia University present the idea that having one massage today to two tomorrow, but reversing in a year from now is irrational (or something of the sort) and we need to remedy it with some policy. (For a review of time discounting and intertemporal preferences, see [114], as economists tend to impart to agents what seems to be a varying "discount rate", derived in a simplified model).[1]

Intuitively, what if I introduce the probability that the person offering the massage is full of balloney? It would clearly make me both prefer immediacy at almost any cost and conditionally on his being around at a future date, reverse the preference. This is what we will model next.

First, time discounting has to have a geometric form, so preference doesn't become negative: linear discounting of the form $Ct$, where $C$ is a constant ant $t$ is time into the future is ruled out: we need something like $C^t$ or, to extract the rate, $(1+k)^t$ which can be mathematically further simplified into an exponential, by taking it to the continuous time limit. Exponential discounting has the form $e^{-k\,t}$. Effectively, such a discounting method using a shallow model prevents "time inconsistency", so with $\delta < t$:

$$\lim_{t\to\infty} \frac{e^{-k\,t}}{e^{-k\,(t-\delta)}} = e^{-k\,\delta}$$

Now add another layer of stochasticity: the discount parameter, for which we use the symbol $\lambda$, is now stochastic.

So we now can only treat $H(t)$ as

$$H(t) = \int e^{-\lambda\,t} \phi(\lambda)\, \mathrm{d}\lambda.$$

It is easy to prove the general case that under symmetric stochasticization of intensity $\Delta\lambda$ (that is, with probabilities $\frac{1}{2}$ around the center of the distribution) using the same technique we did in 4.1:

$$H'(t, \Delta\lambda) = \frac{1}{2}\left(e^{-(\lambda-\Delta\lambda)t} + e^{-(\lambda+\Delta\lambda)t}\right)$$

$$\frac{H'(t,\Delta\lambda)}{H'(t,0)} = \frac{1}{2} e^{\lambda t}\left(e^{(-\Delta\lambda-\lambda)t} + e^{(\Delta\lambda-\lambda)t}\right) = \cosh(\Delta\,\lambda t)$$

---

[1] Farmer and Geanakoplos [105] have applied a similar approach to Hyperbolic discounting.

Where cosh is the cosine hyperbolic function — which will converge to a certain value where intertemporal preferences are flat in the future.

**Example: Gamma Distribution** Under the gamma distribution with support in $\mathbb{R}^+$, with parameters $\alpha$ and $\beta$, $\phi(\lambda) = \frac{\beta^{-\alpha}\lambda^{\alpha-1}e^{-\frac{\lambda}{\beta}}}{\Gamma(\alpha)}$
we get:

$$H(t,\alpha,\beta) = \int_0^\infty e^{-\lambda t}\frac{\left(\beta^{-\alpha}\lambda^{\alpha-1}e^{-\frac{\lambda}{\beta}}\right)}{\Gamma(\alpha)}\,d\lambda = \beta^{-\alpha}\left(\frac{1}{\beta}+t\right)^{-\alpha},$$

so

$$\lim_{t\to\infty}\frac{H(t,\alpha,\beta)}{H(t-\delta,\alpha,\beta)} = 1$$

Meaning that preferences become flat in the future no matter how steep they are in the present, which explains the drop in discount rate in the economics literature.

Further, fudging the distribution and normalizing it, when

$$\phi(\lambda)=\frac{e^{-\frac{\lambda}{k}}}{k},$$

we get the *normatively obtained* so-called hyperbolic discounting:

$$H(t) = \frac{1}{1+k\,t},$$

which turns out to not be the empirical "pathology" that naive researchers have claimed it to be. It is just that their model missed a layer of uncertainty.

# 23 LINDY AS DISTANCE FROM AN ABSORBING BARRIER

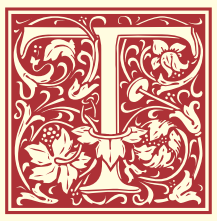he Lindy effect (or law) has been investigated through various traditions. In short it corresponds to situations where the conditional expectation of additional life expectancy increases with time, which requires the survival function *of survival time* to be that of a power law[a]. This maps to a declining force of mortality under the standard hazard rate representation.

Here we model it as the exit time of a stochastic process (arithmetic) with drift $\mu$ and show how the force of mortality behaves with respect to the distance from absorption. We show that, while a process with a drift $\mu = 0$ produces a Lindy survival function, any amount of negative drift makes it exit that class.

[a] That is, where $X$ is an r.v. in $\mathbb{R}$, $\mathbb{P}(X > x) = L(x)x^{-\alpha}$, where $\alpha > 0$ and $L(.)$ is slowly varying function.

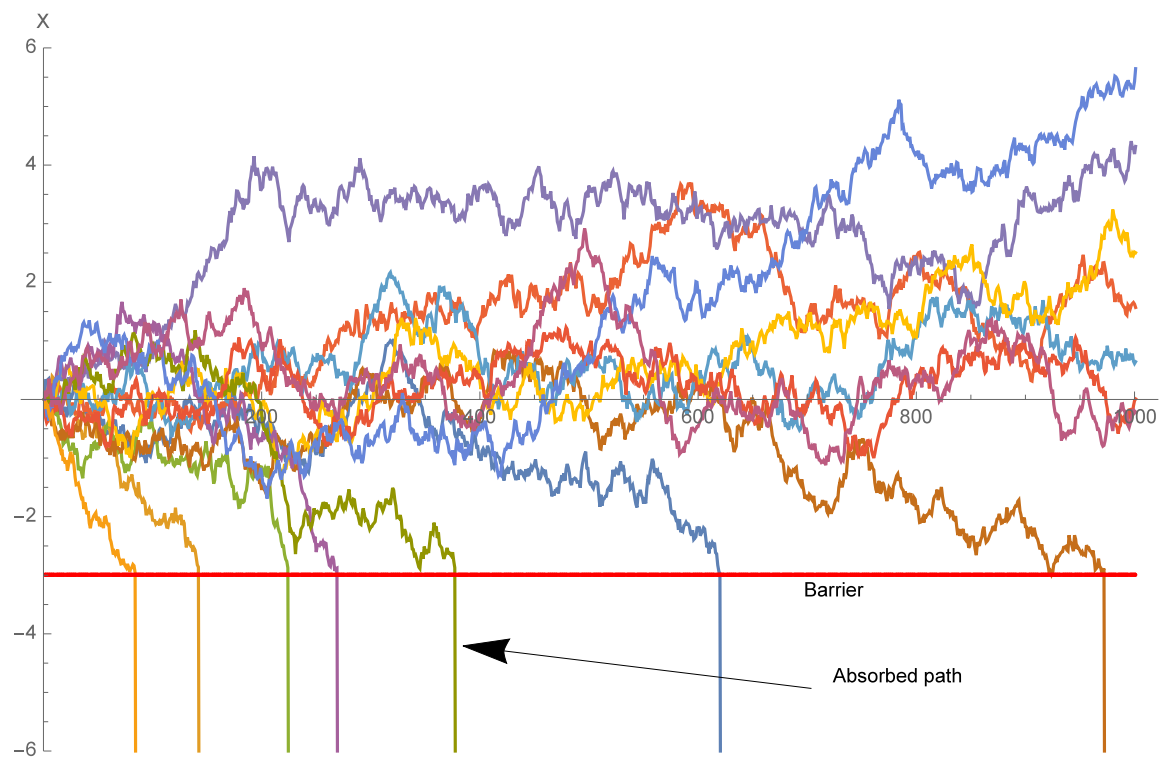

Figure 23.1: *Sample paths for an arithmetic Brownian motion (ABM) subjected to an absorbing barrier (from above). An absorbed path is effectively permanently terminated and is allowed no recovery.*

Discussion chapter with research potential.

## 23.1 FORCE OF MORTALITY

Let $\tau \in \mathbb{R}^+$ be the stopping time random variable and $X = (X_t)_{t \in T}$ be a Brownian motion. Let $L \in \mathbb{R}$ be a barrier below current or initial level; we define the stopping time "from above" as $\tau \triangleq \inf\{t > 0; X_t < L\}$.

Next we look at metrics used in the sciences, insurance industry, and engineering applications. Let us compute the distribution of the force of mortality (a.k.a. hazard rate), which maps, in our representation, to the instantaneous probability of absorption. Consider the probability of conditional stopping time (conditional on survival) in the interval $(t, t + \Delta t)$:

$$\mathbb{P}\Big(\tau \in (t, t+\Delta\, t) \mid \; \tau > t\Big) \overset{Bayes}{=} \frac{1}{\mathbb{P}(\tau > t)} \underbrace{\mathbb{P}\Big(\tau > t \mid \; \tau \in (t, t+\Delta t)\Big)}_{=1} \times \mathbb{P}\Big(\tau \in (t, t+\Delta t)\Big) \tag{23.1}$$

Now let $\varphi(.)$ be the density function for $\tau$:

$$\mathbb{P}\Big(\tau \in (t, t+\Delta\, t) \mid \; \tau > t\Big) = \frac{1}{\int_t^{+\infty} \varphi(\tau) d\tau} \int_t^{t+\Delta t} \varphi(\tau) d\tau \tag{23.2}$$

Allora let's define the force of mortality as the instantaneous rate at $\tau$,

$$\mu(\tau) \triangleq \lim_{\Delta t \to 0} \frac{1}{\Delta t} \mathbb{P}\Big(\tau \in (t, t+\Delta\, t) \mid \; \tau > t\Big),$$

which maps to

$$\mu(\tau) = \frac{1}{\int_\tau^{+\infty} \varphi(t) dt} \varphi(\tau). \tag{23.3}$$

The insurance letteratura works with survival functions $S(\tau)$ or exceedance probabilities, allora:

$$\mu(\tau) = -\frac{S'(\tau)}{S(\tau)} = -\frac{d}{d\tau} \log(S(\tau)). \tag{23.4}$$

With $\mu(.)$ we have a way to integrate over periods. Hence by 23.4 :

$$-\int_t^{t+k} \mu(\tau) d\tau = \log\Big(\frac{S(t+k)}{S(t)}\Big), \tag{23.5}$$

and exponentiating we get the behavior of the survival function over the time increment $[t, t+k]$:

$$S(t+k) = e^{-\int_t^{t+k} \mu(\tau) d\tau} S(t). \tag{23.6}$$

We will return to the mortality business. For now let us examine Lindy and distributions. Let us look at the demarcation properties.

Let $X$ be a random variable that lives in either $(0, \infty)$ or $(-\infty, \infty)$ and $\mathbb{E}$ the expectation operator under "real world" (physical) distribution. By classical results [97]:

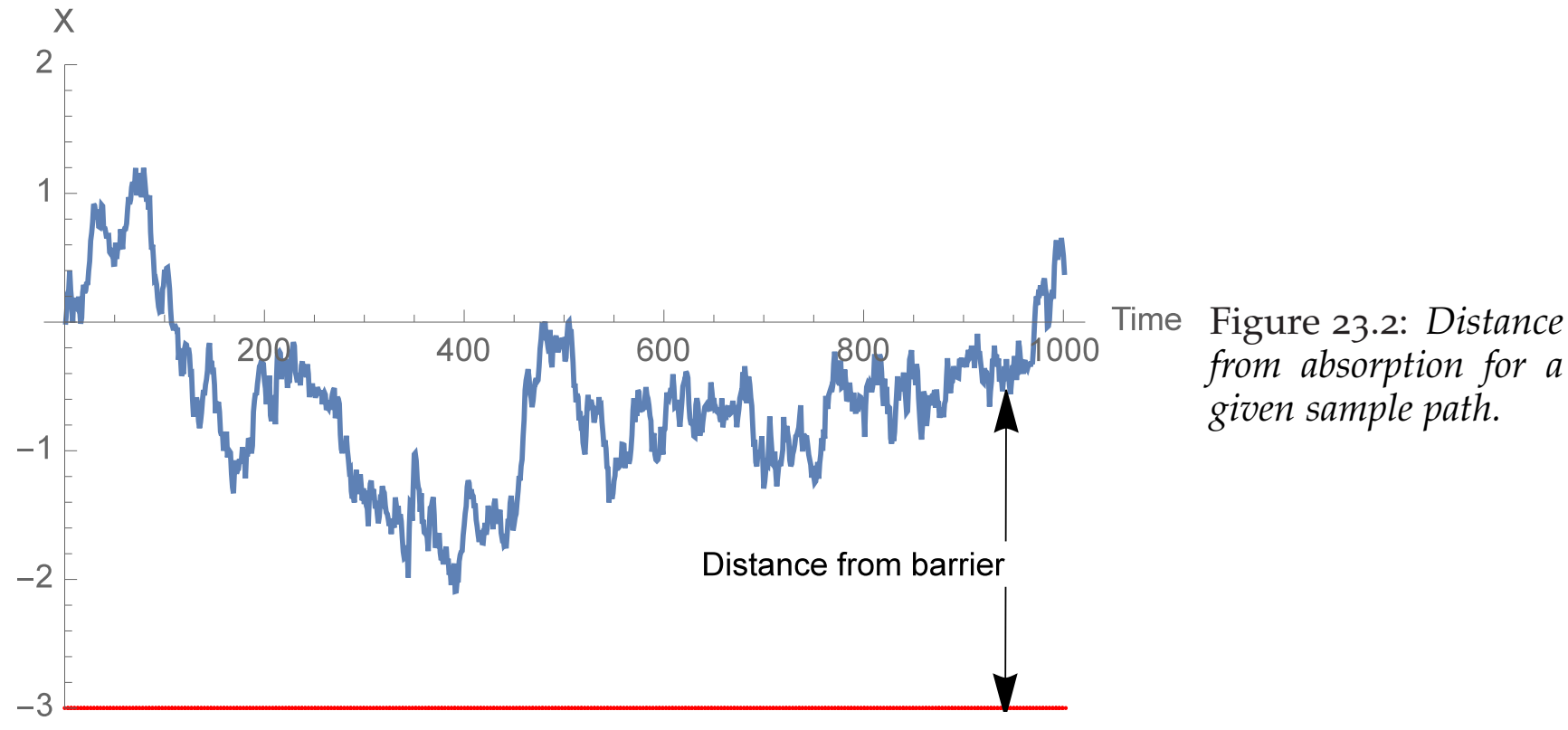

Figure 23.2: *Distance from absorption for a given sample path.*

$$\lim_{K\to\infty} \frac{1}{K} \mathbb{E}(X|_{X>K}) = \lambda \tag{23.7}$$

- If $\lambda = 1$ , $X$ is said to be in the thin tailed class $\mathcal{D}_1$ and has a characteristic scale
- If $\lambda > 1$ , $X$ is said to be in the fat tailed regular variation class $\mathcal{D}_2$ and has no characteristic scale.
- If
$$\lim_{K\to\infty} \mathbb{E}(X|_{X>K}) - K = \lambda$$
where $\lambda > 0$, then $X$ is in the borderline exponential class

We can reexpress the above in terms of the force of mortality at a future period $t$, $\mu(t)$.

- If the time random variable belongs to the power law class with pdf $\xrightarrow{t} t^{-\alpha-1}$ then $\mu(t) = \frac{\alpha}{t}$, which *declines* with time.
- If the time random variable is in the borderline exponential class with pdf $\varphi(t) = e^{-\lambda t}$, $t > 0$ then $\mu(t) = \lambda$, which is independent of $t$.
- If the r.v. is in the thin-tailed classes, then $\mu(.)$ should increase with time.

## 23.2 THE PROCESS

Feller's second volume [107] (page 174) showed that the distribution of the time to first passage for a Brownian motion is "stable with exponent $\frac{1}{2}$", using scalability arguments under a Markov property –in fact this would be "Lindy", applied to time to hit any point for a one-dimensional Brownian motion (the distribution is Cauchy for two dimensions). We will first re-derive then add negative drifts, thanks to Girsanov's theorem , in order to gauge the properties of the distribution.

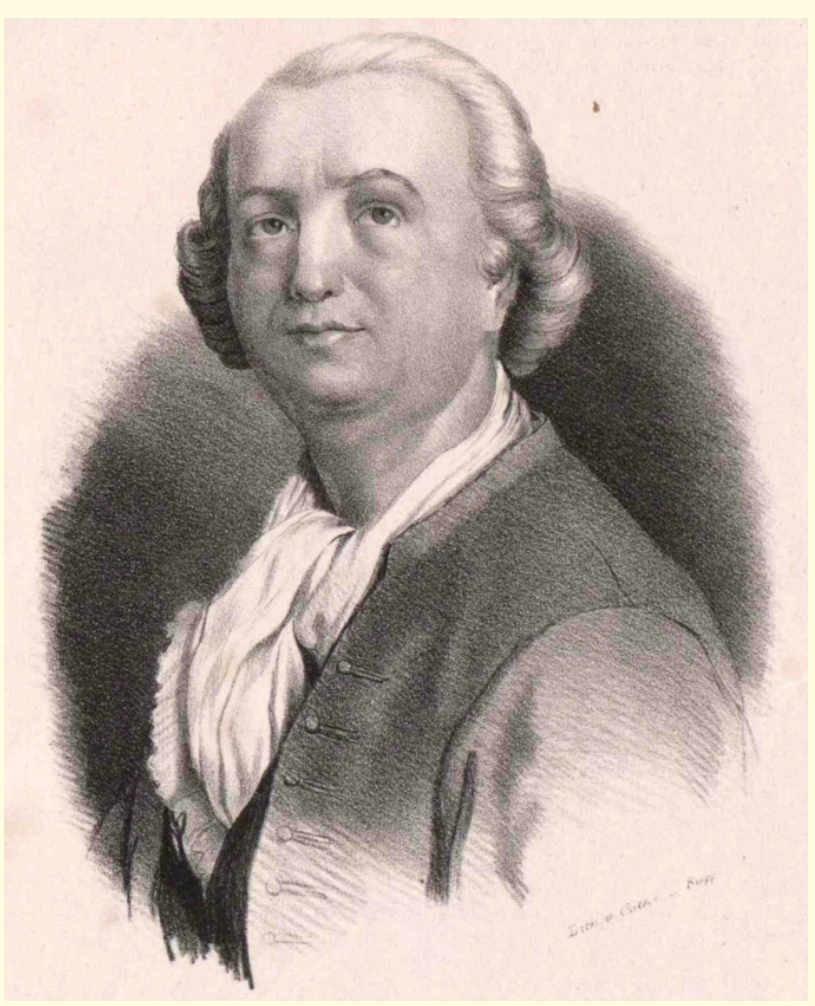

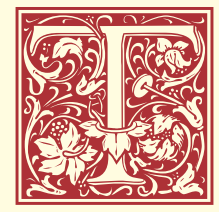The Sicilian Adventurer and occultist Giuseppe Balsamo[a] who went by the pseudonym of Count Cagliostro claimed to be immortal. He was frozen in middle age at the height of his physical and intellectual powers (although various representations such as the one above show him plump, soft and sarcopenic, that is, unacquainted with the deadlifting barbell.)

In a (fully or perhaps only partially) imagined dinner by Alexandre Dumas in his historical novel *Le collier de la reine (the Queen's Necklace)*, Cagliostro claimed to be 3,000 years old and having partaken of all manner of historical battles.

The Marquis de Condorcet (of the famous paradox) busted his reasoning, showing the difference between immortality and absence of physical aging. The slightest probability of death would cumulate to certainty, and life expectancy must be finite –and (relatively) short.

However, Cagliostro appeared to have some counterargument up his sleeve. He explained the notion of experience: $3,000$ years allow someone to learn to spot the signs of danger, even predict the trajectory of bullets, which does reduce the probability of absorption and causes the decline of the hazard rate over time. Effectively, people who do not age can be Lindy, hence their survival is power-law distributed.

It is quite remarkable that Dumas, almost two centuries ago, captured these probabilistic subtleties which have not yet reached decision-theory.

[a] Creative Commons.

### 23.2.1 Process without a drift

Let $X_t$ be a stochastic process with no drift satisfying $dX = \sigma dB$, where $B$ is a Brownian motion $X_0 = 0$, so by standard result $X_t = X_0 + \sigma\sqrt{t}W_{0,1}$ and $W_{0,1}$ is a standardized Gaussian r.v.

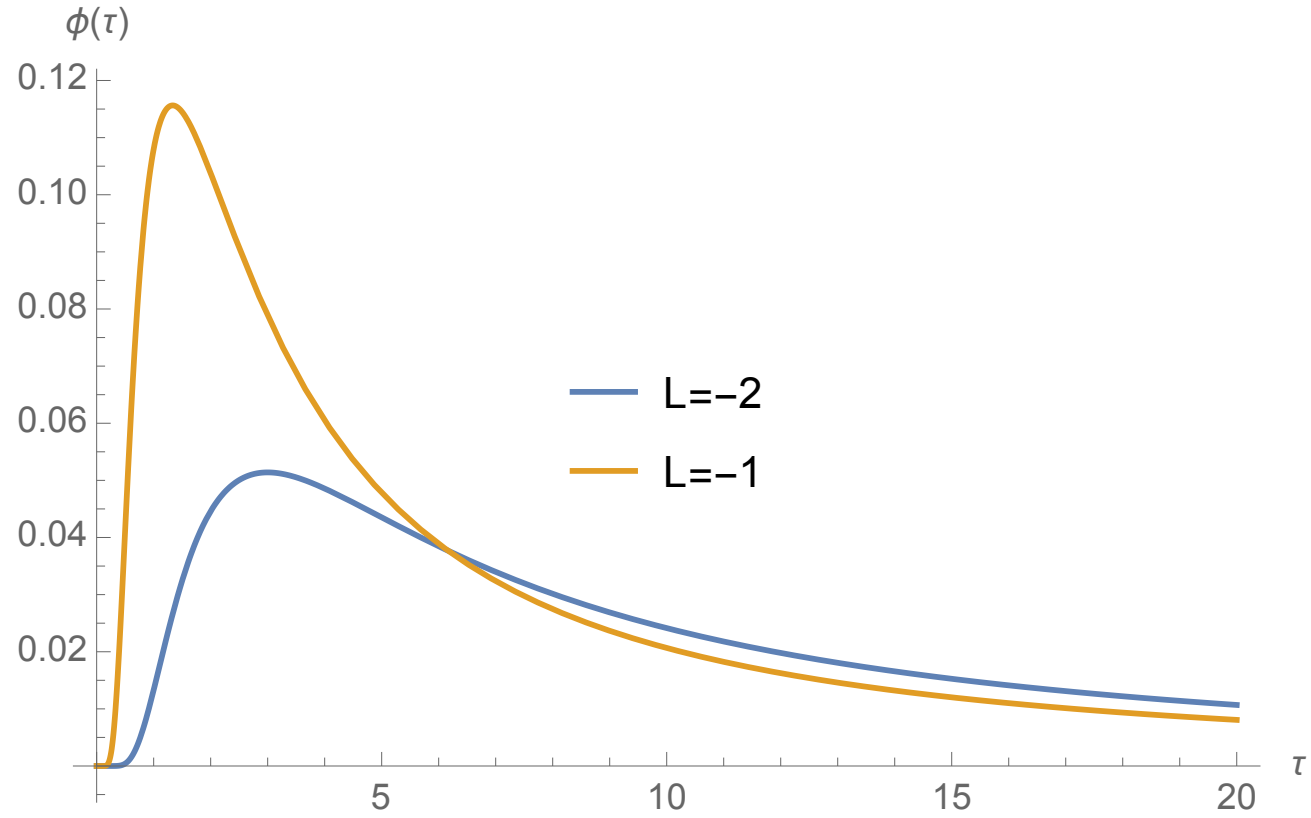

Figure 23.3: *Stopping time distributions at various absorption levels.*

Let $L < 0$ be the level of the absorbing barrier (constant) hitting from above and $\tau$ the first passage time $\tau = \inf\{t : X_t = L\}\, z$ We have, using the reflection principle (Fig. 23.2.1), the following distribution of $\tau$:

Consider the probability of $X$ below $L$:

$$\mathbb{P}\left(X_t < L\right) = \mathbb{P}\left(X < L|_{\tau<t}\right)\mathbb{P}(\tau < t) + \underbrace{\mathbb{P}\left(X < L|_{\tau>t}\right)}_{=0}\mathbb{P}(\tau > t), \tag{23.8}$$

since by definition one cannot have a conditional probability of hitting L with $\tau > t$, $\mathbb{P}\left(X < L|_{\tau>t}\right) = 0$ (in other words, the path is continous, one cannot be below L without having first hit $L$, even in Saudi Arabia).

$$\mathbb{P}\left(X_t < L\right) = \mathbb{P}\left(X < L|_{\tau<t}\right)\mathbb{P}(\tau < t) \tag{23.9}$$

Now by symmetry, at (or, rather "in the neighborhood of") $\tau$, $X$ is as likely to be above $L$ as to be below $L$, so

$$\mathbb{P}\left(X < L|_{\tau<t}\right) = \mathbb{P}\left(X > L|_{\tau<t}\right) = \frac{1}{2}.$$

Allora the cumulative $\mathbb{P}(\tau < t)) = 2\left(P\left(X_t < L\right)\right) = 2\text{CDF}(L) = \text{erf}\left(\frac{L}{\sqrt{2}\sigma\sqrt{\tau}}\right) + 1$ and the distribution of $\tau$ becomes (assuming $X = 0$, so it is expressed as arithmetic distance from $X$)

$$\varphi(\tau) = \frac{\partial\mathbb{P}(\tau < t)}{\partial\tau} = -\frac{L\, e^{-\frac{L^2}{2\sigma^2\tau}}}{\sqrt{2\pi}\,\sigma\tau^{3/2}} \tag{23.10}$$

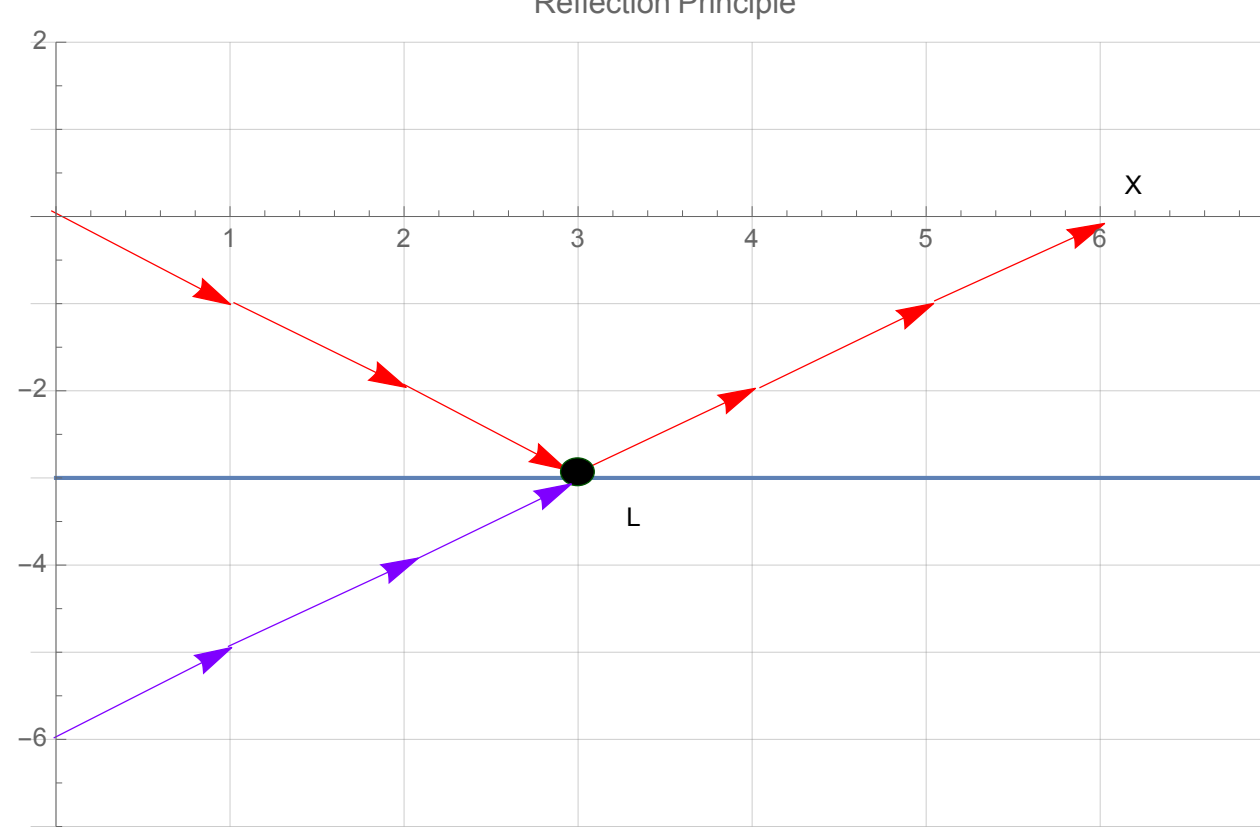

Figure 23.4: *The reflection principle, illustrated with a random walk of step size $(+1,-1)$. The number of paths from 0 (the origin) to X via L are equal to those from $-2L$ to X.*

Noting the CDF for the stopping time, $\Phi(\tau) = \text{erf}\left(\frac{L}{\sqrt{2}\sigma\sqrt{\tau}}\right) + 1$, we get:

$$\lim_{\tau\to\infty} \frac{\log\left(1-\Phi(\tau)\right)}{\log(\tau)} = -\frac{1}{2},$$

meaning that the survival function $\to Kt^{-\frac{1}{2}}$, where $K$ is a constant.[2]

In fact we can see that $\varphi(\tau)$ maps to a Lévy distribution with parameters $(0, \frac{1}{\sigma^2})$. As it has no moments, we can get an idea of its median $\frac{1}{2\,\text{erfc}^{-1}\left(\frac{1}{2}\right)^2\sigma^2}$

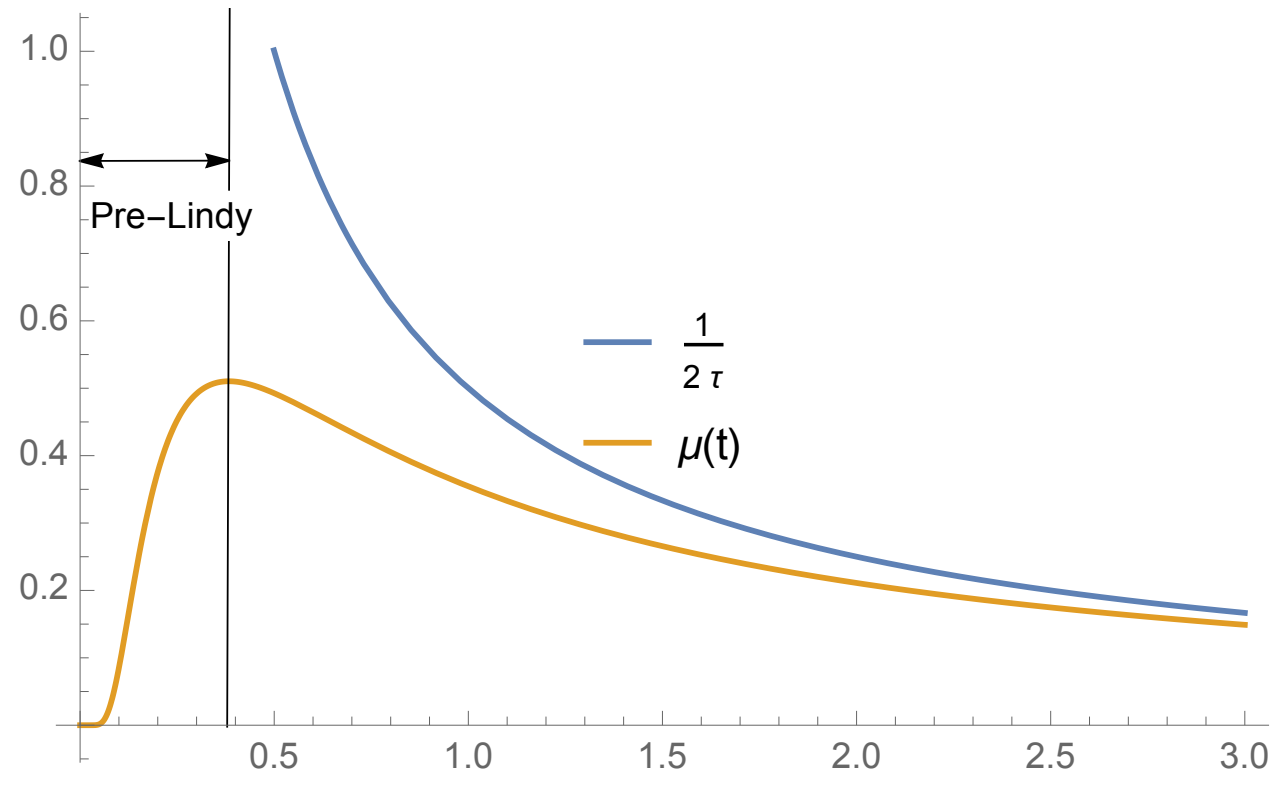

Figure 23.5: *The force of mortality for a driftless Brownian Motion is not declining monotonically as the "strong Pareto property" is not reached until later in the game.*

[2] Feller's reasoning using dimensionless analysis is as follows. The residual time to reach $L + L'$ after having hit $L$, which we write as $\tau_{L+L'}$- $\tau_L$ is independent of $\tau_L$ and has the same distribution as $\tau_{L'}$. As we can see in the graph in Fig. 23.2.1, the transition probability, given that is is a standard random walk, is proportional to $\frac{L^2}{t}$ (assuming a time increment of $t$).

We note the hazard function:

$$\mu(\tau) = -\frac{\partial \log(1-\Phi(\tau))}{\partial \tau} = \frac{L e^{-\frac{L^2}{2\sigma^2\tau}}}{\sqrt{2\pi}\sigma\tau^{3/2}\operatorname{erf}\left(\frac{L}{\sqrt{2}\sigma\sqrt{\tau}}\right)} \xrightarrow{\tau} \frac{1}{2\tau}.$$

**Remark 23.1**

*The force of mortality is not "Lindy" over the pre-asymptotic zone in the power law where the distribution* $\frac{\log(S(x))}{\log(x)}$ *(with* $S(.)$ *a survival function) is not a constant.*

From the standard definition of an r.v. in the power law class, one must have cleared the "Karamata point", that is the area where the slowly varying function wanes and reach the "strong Pareto property". The standard representation is $\mathbb{P}(X > x) = L(x)x^{-\alpha}$, where $\alpha > 0$ and $L : [x_{\min}, +\infty) \to (0, +\infty)$ is a slowly varying function, defined as

$$\lim_{x \to +\infty} \frac{L(kx)}{L(x)} = 1$$

for any $k > 0$. In other words, we must have $L(x)$ a constant. If $L(x)$ is not a constant, then the force of mortality

The arguments used in this section do not adapt to an arithmetic Brownian motion with drift (it would cancel the symmetry of the reflection principle), nor, besides, do they adapt to a geometric Brownian motion (see *Dynamic Hedging*, [271]). For that, one needs a change of measure as we will do next.

### 23.2.2 Adding a drift via Girsanov's theorem

Let $\mu < 0$ be a negative drift. We have the Radon-Nikodym derivative for $\hat{X} = \mu t + X$, where $X$ and$\hat{X}$ are martingales under the $\mathbb{P}$ and $\mathbb{Q}$ measures, respectively:

$$\frac{\mathrm{d}\mathbb{Q}}{\mathrm{d}\mathbb{P}} = e^{\frac{\mu^2 t}{2} - \mu X_t} \tag{23.11}$$

$$\mathbb{P}(t \in dt) = \mathbb{E}_{\mathbb{P}}(\mathbb{1}_{t\in dt}) = \mathbb{E}_{\mathbb{Q}}\left(\left(\frac{d\mathbb{Q}}{d\mathbb{P}}\right)^{-1}\mathbb{1}_{t\in dt}\right) = \mathbb{E}_{\mathbb{Q}}\left(e^{\mu X_t - \frac{\mu^2 t}{2}}\mathbb{1}_{t\in dt}\right) \tag{23.12}$$

Now $\mathbb{1}_{t\in dt} = \mathbb{1}_{\hat{X}=L}$, allora (adjusting for $\sigma^2$ since we did not normalize 23.10:

$$\mathbb{P}(t \in dt) = e^{\mu L - \frac{\mu^2\sigma^2 t}{2}}\mathbb{Q}(t \in dt) \tag{23.13}$$

Since, from the earlier result 23.10 we have $\mathbb{P}(t \in dt) = -\frac{\left(Le^{-\frac{L^2}{2t}}\right)e^{\mu L-\frac{\mu^2 t}{2}}}{\sqrt{2\pi}t^{3/2}}$, and since $L$ is negative, we focus on $\mu$ negative as they must be of the same sign for $\mathbb{P}(t \in dt)$ to represent a density "from above":

$$\varphi_\mu(\tau) = -\frac{L\, e^{-\frac{\left(L-\mu\sigma^2\tau\right)^2}{2\sigma^2\tau}}}{\sqrt{2\pi}\,\sigma\tau^{3/2}},\ \mu < 0. \tag{23.14}$$

**Remark 23.2**

*Any amount of negative drift causes the stopping time distribution to exit the power-law class, hence lose the "Lindy" attribute.*

*Proof.* The moments of order $p$ are finite for all values of $\mu < 0$:

$$\mathbb{M}(p) = \sqrt{\frac{2}{\pi}}\left(-e^{\mu L}\right)L^{p+\frac{1}{2}}\mu^{\frac{1}{2}-p}\sigma^{-2p}K_{p-\frac{1}{2}}(L\mu) \tag{23.15}$$

□

and the Bessel function $K_{p-\frac{1}{2}}(0) = -\infty$. We note the finite mean

$$\mathbb{M}(1) = \frac{L}{\mu\sigma^2},$$

which is positive for all $\mu < 0$.

**Remark 23.3**

*Why do we need $\mu$ to be $\leq 0$ to get $\varphi(.)$ to be a density?*

*Simply:*

$$\mathbb{P}(t < +\infty) = \int_0^\infty -\frac{Le^{-\frac{\left(L-\mu\sigma^2\tau\right)^2}{2\sigma^2\tau}}}{\sqrt{2\pi}\sigma\tau^{3/2}}\,d\tau = e^{\left(\sqrt{\mu^2}+\mu\right)L}.$$

*Should $\mu$ be positive, it is worse than when $\mu = 0$ where no moment $> 0$ exists. Here the $0^{th}$ moment would not exist as there is $e^{2\mu L} - 1$ probability of the hitting time never taking place. This is, literally, immortality.*

## 23.3 SOME CONSIDERATIONS ON LIFE EXPECTANCY AND AGING

Is human life bounded? Soft-bounded (i.e. have an accelerating force of mortality, with exponential decline for large deviations)? Or is it exponential, reaching a high but constant force of mortality at some stage? In "Human life is unlimited but short" [236], Rootzen and Zholud, using the International Database of Longevity containing validated life lengths of 668 supercentenarians (i.e., >110) from 15 countries, found, in spite of the small sample size in the tails, data in agreement with an exponential distribution. That is, supercentenarians seem to have a high probability of death (around .47 in any year), but such probability remains flat.

Their method was made more convincing in their use of flatness men/women beyond a point. Women outnumber men in the tails, but they may converge to the same mortality.

> The fact that data doesn't shows any difference in survival at extreme age between women and men; between persons with different lifestyles or genetic backgrounds; between different time periods; or between different ages, say between 110-year-olds and 115-year-olds is surprising and remarkable and can inform the search for demographic or biological theories of aging. In particular, the fact that lifestyle factors which are important for survival at younger ages ceases to play a role after age 110 is of both biological and popular interest.[236]

Now there has been a debate, with [169], [110], and a rejoinder [237]).

> This debate has remarkable relevance to medicine and fragility as it appears that, if the exponential distribution is remotely true, some organs suffer damage, become fragile, then cease to age! Identifying *what is* aging becomes automatic.

# Part VII

# QUANTITATIVE FINANCE AND FAT TAILS

# 24 FINANCIAL THEORY'S FAILURES WITH OPTION PRICING†

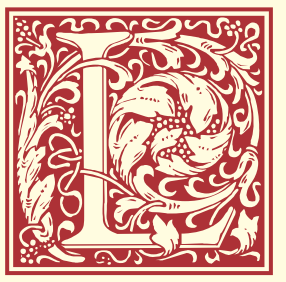ET US DISCUSS why option theory, as seen according to the so-called "neoclassical economics", fails in the real world. How does financial theory price financial products? The principal difference in paradigm between the one presented by Bachelier in 1900, [9] and the modern finance one known as Black-Scholes-Merton [28] and [200] lies in a few central assumptions by which Bachelier was closer to reality and the way traders do business and have done business for centuries.

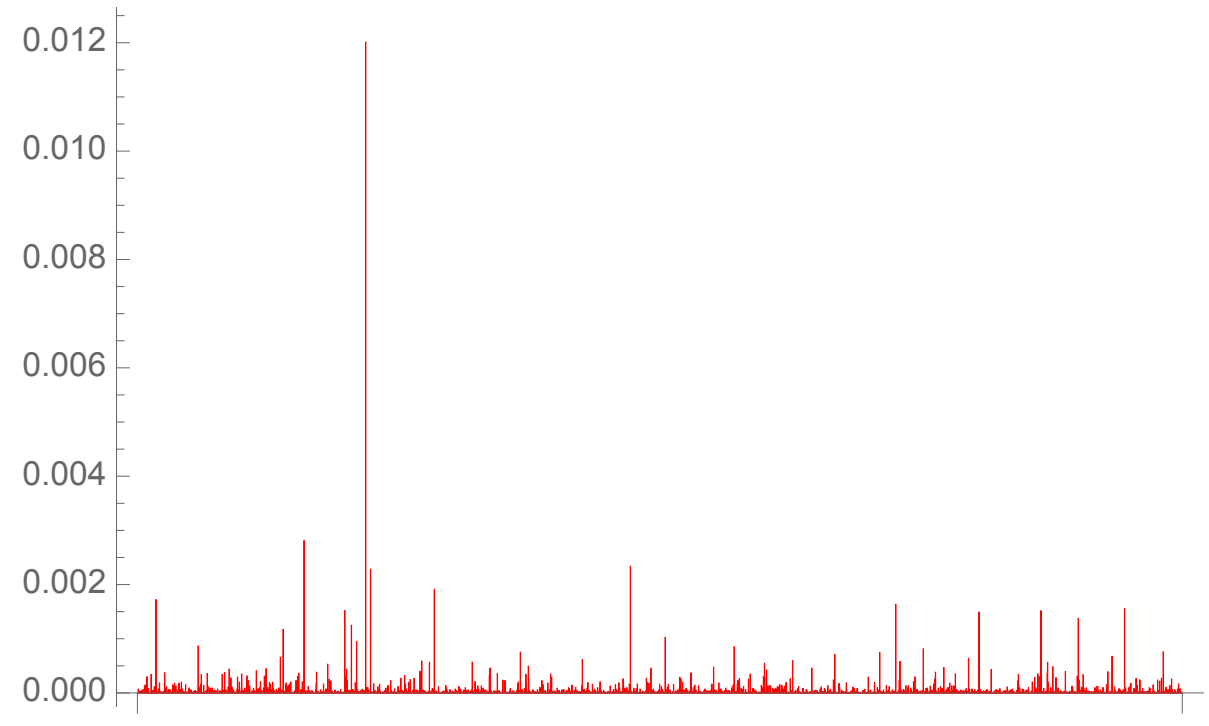

Figure 24.1: *The hedging errors for an option portfolio (under a daily revision regime) over 3000 days, under a constant volatility Student T with tail exponent* $\alpha = 3$*. Technically the errors should not converge in finite time as their distribution has infinite variance.*

## 24.1 BACHELIER NOT BLACK-SCHOLES

Bachelier's model is based on an actuarial expectation of final payoffs –not dynamic hedging. It means you can use any distribution! A more formal proof using measure theory is provided in Chapter 25 so for now let us just get the intuition without too much mathematics.

† Discussion chapter.

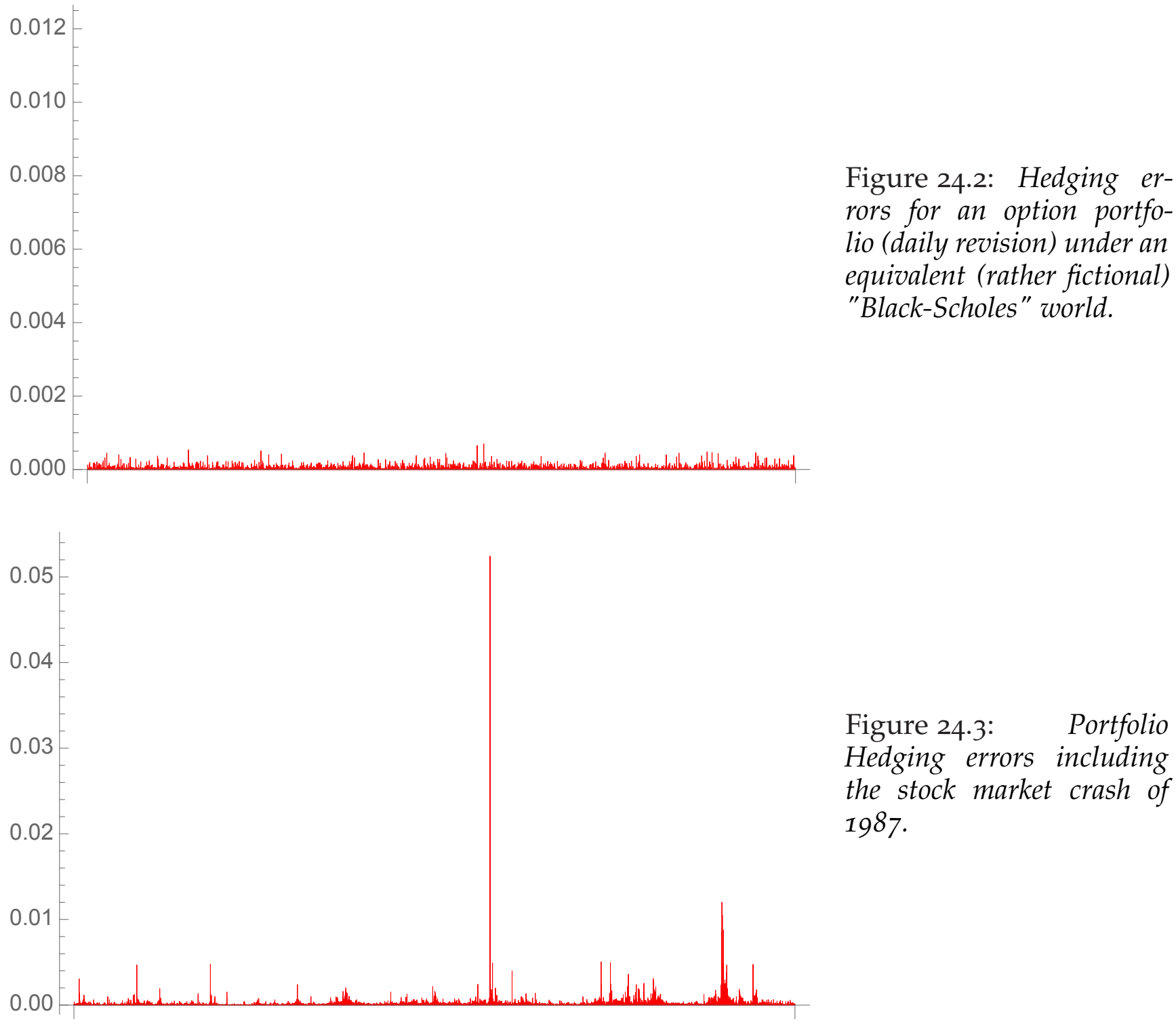

Figure 24.2: *Hedging errors for an option portfolio (daily revision) under an equivalent (rather fictional) "Black-Scholes" world.*

Figure 24.3: *Portfolio Hedging errors including the stock market crash of 1987.*

The same method was later used by a series of researchers, such as Sprenkle [256] in 1964, Boness, [30] in 1964, Kassouf and Thorp, [**?** ] in 1967, Thorp, [301] (only published in 1973).

They all encountered the following problem: how to produce a risk parameter –a risky asset discount rate – to make it compatible with portfolio theory? The Capital Asset Pricing Model requires that securities command an expected rate of return in proportion to their riskiness. In the Black-Scholes-Merton approach, an option price is derived from continuous-time dynamic hedging, and only in properties obtained from continuous time dynamic hedging –we will describe dynamic hedging in some details further down. Thanks to such a method, an option collapses into a deterministic payoff and provides returns independent of the market; hence it does not require any risk premium.

### 24.1.1 Distortion from Idealization

The problem we have with the Black-Scholes-Merton approach is that the requirements for dynamic hedging are extremely idealized, requiring the following strict conditions. The operator is assumed to be able to buy and sell in a frictionless market, incurring no transaction costs. The procedure does not allow for the price impact of the order flow –if an operator sells a quantity of shares, it should not have consequences on the subsequent price. The operator knows the probability distribution, which is the Gaussian, with fixed and constant parameters through time (all parameters do not change). Finally, the most significant restriction: no scalable jumps. In a subsequent revision [Merton, 1976] allows for jumps but these are deemed to be Poisson arrival time, and fixed or, at the worst, Gaussian. The framework does not allow the use of power laws both in practice and mathematically. Let us examine the mathematics behind the stream of dynamic hedges in the Black-Scholes-Merton equation.

Assume the risk-free interest rate $r = 0$ with no loss of generality. The canonical Black-Scholes-Merton model consists in selling a call and purchasing shares of stock that provide a hedge against instantaneous moves in the security. Thus the portfolio $\pi$ locally "hedged" against exposure to the first moment of the distribution is the following:

$$\pi = -C + \frac{\partial C}{\partial S} S \tag{24.1}$$

where $C$ is the call price, and $S$ the underlying security.

Take the change in the values of the portfolio

$$\Delta \pi = -\Delta C + \frac{\partial C}{\partial S} \Delta S \tag{24.2}$$

By expanding around the initial values of S, we have the changes in the portfolio in discrete time. Conventional option theory applies to the Gaussian in which all orders higher than $(\Delta \mathrm{S})^2$ and $\Delta \mathrm{t}$ disappears rapidly.

$$\Delta \pi = -\frac{\partial C}{\partial t} \Delta \mathrm{t} - \frac{1}{2} \frac{\partial^2 C}{\partial S^2} \Delta \mathrm{S}^2 + O\left(\Delta \mathrm{S}^3\right) \tag{24.3}$$

Taking expectations on both sides, we can see from (3) very strict requirements on moment finiteness: *all* moments need to converge. If we include another term, $-\frac{1}{6} \frac{\partial^3 C}{\partial S^3} \Delta \mathrm{S}^3$, it may be of significance in a probability distribution with significant cubic or quartic terms. Indeed, although the $n^{\text{th}}$ derivative with respect to $S$ can decline very sharply, for options that have a strike $K$ away from the center ot the distribution, it remains that the moments are rising disproportionately fast for that to carry a mitigating effect.

So here we mean all moments need to be finite and losing in impact –no approximation. Note here that the jump diffusion model (Merton,1976) does not cause much trouble since it has all the moments. And the annoyance is that a power law will have every moment higher than $\alpha$ infinite, causing the equation of the Black-Scholes-Merton portfolio to fail.

As we said, the logic of the Black-Scholes-Merton so-called solution thanks to Itô's lemma was that the portfolio collapses into a deterministic payoff. But let us see how quickly or effectively this works in practice.

### 24.1.2 The Actual Replication Process:

The payoff of a call should be replicated with the following stream of dynamic hedges, the limit of which can be seen here, between t and T

$$\lim_{\Delta t \to 0} \left( \sum_{i=1}^{n=T/\Delta t} \frac{\partial C}{\partial S} |_{S=S_{t+(i-1)\Delta \mathrm{t}}, t=t+(i-1)\Delta \mathrm{t},} \left( S_{t+i\Delta \mathrm{t}} - S_{t+(i-1)\Delta \mathrm{t}} \right) \right) \tag{24.4}$$

We break up the period into n increments Δt. Here the hedge ratio $\frac{\partial C}{\partial S}$ is computed as of time *t +(i-1)* $\Delta t$, but we get the nonanticipating difference between the price at the time the hedge was initiatied and the resulting price at $t+ i \; \Delta t$.

This is supposed to make the payoff deterministic *at the limit of* $\Delta t \to 0$. In the Gaussian world, this would be an Itô-McKean integral.

### 24.1.3 Failure: How Hedging Errors Can Be Prohibitive.

As a consequence of the mathematical property seen above, hedging errors in an cubic $\alpha$ appear to be indistinguishable from those from an infinite variance process. Furthermore such error has a disproportionately large effect on strikes away from the money.

In short: dynamic hedging in a power law world removes no risk.

## NEXT

The next chapter will use measure theory to show why options can still be risk-neutral.

# 25 UNIQUE OPTION PRICING MEASURE (NO DYNAMIC HEDGING/COMPLETE MARKETS)‡

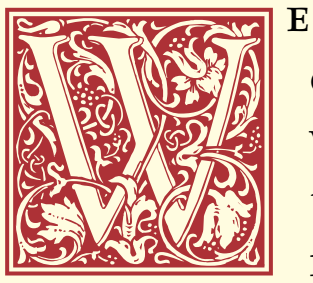E PRESENT THE proof that under simple assumptions, such as constraints of Put-Call Parity, the probability measure for the valuation of a European option has the mean derived from the forward price which can, but does not have to be the risk-neutral one, under any general probability distribution, bypassing the Black-Scholes-Merton dynamic hedging argument, and without the requirement of complete markets and other strong assumptions. We confirm that the heuristics used by traders for centuries are both more robust, more consistent, and more rigorous than held in the economics literature. We also show that options can be priced using infinite variance (finite mean) distributions.

## 25.1 BACKGROUND

Option valuations methodologies have been used by traders for centuries, in an effective way (Haug and Taleb, [147]). In addition, valuations by expectation of terminal payoff forces the mean of the probability distribution used for option prices to be that of the forward, thanks to Put-Call Parity and, should the forward be risk-neutrally priced, so will the option be. The Black-Scholes argument (Black and Scholes, 1973, Merton, 1973) is held to allow risk-neutral option pricing thanks to dynamic hedging, as the option becomes redundant (since its payoff can be built as a linear combination of cash and the underlying asset dynamically revised through time). This is a puzzle, since: 1) Dynamic Hedging is not operationally feasible in financial markets owing to the dominance of portfolio changes resulting from jumps, 2) The dynamic hedging argument doesn't stand mathematically under fat tails; it requires a very specific "Black-Scholes world" with many impossible assumptions, one of which requires finite quadratic variations, 3) Traders use the same Black-Scholes "risk neutral argument" for the valuation of options on assets that do not allow dynamic replication, 4) Traders trade options consistently in domain where the risk-neutral arguments do not apply 5) There are fundamental informational limits preventing the convergence of the stochastic integral.[2]

‡ Research chapter.

There have been a couple of predecessors to the present thesis that Put-Call parity is sufficient constraint to enforce some structure at the level of the mean of the underlying distribution, such as Derman and Taleb (2005), Haug and Taleb (2010). These approaches were heuristic, robust though deemed hand-waving (Ruffino and Treussard, [244]). In addition they showed that operators need to use the risk-neutral mean. What this chapter does is

- It goes beyond the "handwaving" with formal proofs.
- It uses a completely distribution-free, expectation-based approach and proves the risk-neutral argument without dynamic hedging, and without any distributional assumption.
- Beyond risk-neutrality, it establishes the case of a unique pricing distribution for option prices in the absence of such argument. The forward (or future) price can embed expectations and deviate from the arbitrage price (owing to, say, regulatory or other limitations) yet the options can still be priced at a distibution corresponding to the mean of such a forward.
- It shows how one can *practically* have an option market without "completeness" and without having the theorems of financial economics hold.

These are done with solely two constraints: "horizontal", i.e. put-call parity, and "vertical", i.e. the different valuations across strike prices deliver a probability measure which is shown to be unique. The only economic assumption made here is that the forward exits, is tradable — in the absence of such unique forward price it is futile to discuss standard option pricing. We also require the probability measures to correspond to distributions with finite first moment.

Preceding works in that direction are as follows. Breeden and Litzenberger [35] and Dupire [85], show how option spreads deliver a unique probability measure; there are papers establishing broader set of arbitrage relations between options such as Carr and Madan [41][3].

However 1) none of these papers made the bridge between calls and puts via the forward, thus translating the relationships from arbitrage relations between options delivering a probability distribution into the necessity of lining up to the mean of the distribution of the forward, hence the risk-neutral one (in case the forward is arbitraged.) 2) Nor did any paper show that in the absence of second moment (say, infinite variance), we can price options very easily. Our methodology and proofs make no use of the variance. 3) Our method is vastly simpler, more direct, and robust to changes in assumptions.

---

[2] Further, in a case of scientific puzzle, the exact formula called "Black-Scholes-Merton" was written down (and used) by Edward Thorp in a heuristic derivation by expectation that did not require dynamic hedging, see Thorpe [304].

[3] See also Green and Jarrow [134] and Nachman [204]. We have known about the possibility of risk neutral pricing without dynamic hedging since Harrison and Kreps [144] but the theory necessitates extremely strong –and severely unrealistic –assumptions, such as strictly complete markets and a multiperiod pricing kernel

We make no assumption of general market completeness. Options are not redundant securities and remain so. Table 1 summarizes the gist of the paper.[4] [5]

## 25.2 PROOF

Define $C(S_{t_0}, K, t)$ and $P(S_{t_0}, K, t)$ as European-style call and put with strike price K, respectively, with expiration $t$, and $S_0$ as an underlying security at times $t_0$, $t \geq t_0$, and $S_t$ the possible value of the underlying security at time t.

### 25.2.1 Case 1: Forward as risk-neutral measure

Define $r = \frac{1}{t-t_0} \int_{t_0}^{t} r_s \mathrm{d}s$, the return of a risk-free money market fund and $\delta = \frac{1}{t-t_0} \int_{t_0}^{t} \delta_s \mathrm{d}s$ the payout of the asset (continuous dividend for a stock, foreign interest for a currency).

We have the arbitrage forward price $F_t^Q$:

$$F_t^Q = S_0 \frac{(1+r)^{(t-t_0)}}{(1+\delta)^{(t-t_0)}} \approx S_0 \, e^{(r-\delta)(t-t_0)} \tag{25.1}$$

by arbitrage, see Keynes 1924. We thus call $F_t^Q$ the future (or forward) price obtained by arbitrage, at the risk-neutral rate. Let $F_t^P$ be the future requiring a risk-associated "expected return" $m$, with expected forward price:

$$F_t^P = S_0 (1+m)^{(t-t_0)} \approx S_0 \, e^{m\,(t-t_0)}. \tag{25.2}$$

**Remark:** *By arbitrage, all tradable values of the forward price given $S_{t_0}$ need to be equal to $F_t^Q$.*

"Tradable" here does not mean "traded", only subject to arbitrage replication by "cash and carry", that is, borrowing cash and owning the secutity yielding $d$ if the embedded forward return diverges from $r$.

### 25.2.2 Derivations

In the following we take $F$ as having dynamics on its own –irrelevant to whether we are in case 1 or 2 –hence a unique probability measure $Q$.

Define $\Omega = [0, \infty) = A_K \cup A_K^c$ where $A_K = [0, K]$ and $A_K^c = (K, \infty)$.

[4] The famed Hakkanson paradox is as follows: if markets are complete and options are redudant, why would someone need them? If markets are incomplete, we may need options but how can we price them? This discussion may have provided a solution to the paradox: markets are incomplete *and* we can price options.

[5] Option prices are not unique in the absolute sense: the premium over intrinsic can take an entire spectrum of values; it is just that the put-call parity constraints forces the measures used for puts and the calls to be the same and to have the same expectation as the forward. As far as securities go, options are securities on their own; they just have a strong link to the forward.

Table 25.1: *Main practical differences between the dynamic hedging argument and the static Put-Call parity with spreading across strikes.*

| | **Black-Scholes Merton** | **Put-Call Parity with Spreading** |
|---|---|---|
| **Type** | Continuous rebalancing. | Interpolative static hedge. |
| **Limit** | Law of large numbers in time (horizontal). | Law of large numbers across strikes (vertical). |
| **Market Assumptions** | 1) Continuous Markets, no gaps, no jumps.<br><br>2) Ability to borrow and lend underlying asset for all dates.<br><br>3) No transaction costs in trading asset. | 1) Gaps and jumps acceptable. Possibility of continuous Strikes, or acceptable number of strikes.<br><br>2) Ability to borrow and lend underlying asset for single forward date.<br><br>3) Low transaction costs in trading options. |
| **Probability Distribution** | Requires all moments to be finite. Excludes the class of slowly varying distributions | Requires finite $1^{st}$ moment (infinite variance is acceptable). |
| **Market Completeness** | Achieved through dynamic completeness | Not required (in the traditional sense) |
| **Realism of Assumptions** | Low | High |
| **Convergence** | Uncertain; one large jump changes expectation | Robust |
| **Fitness to Reality** | Only used after "fudging" standard deviations per strike. | Portmanteau, using specific distribution adapted to reality |

Consider a class of standard (simplified) probability spaces $(\Omega, \mu_i)$ indexed by $i$, where $\mu_i$ is a probability measure, i.e., satisfying $\int_\Omega d\mu_i = 1$.

**Theorem 6**

*For a given maturity T, there is a unique measure $\mu_Q$ that prices European puts and calls by expectation of terminal payoff.*

This measure can be risk-neutral in the sense that it prices the forward $F_t^Q$, but does not have to be and imparts rate of return to the stock embedded in the forward.

**Lemma 25.1**
*For a given maturity T, there exist two measures $\mu_1$ and $\mu_2$ for European calls and puts of the same maturity and same underlying security associated with the valuation by expectation of terminal payoff, which are unique such that, for any call and put of strike K, we have:*

$$C = \int_\Omega f_C \, \mathrm{d}\mu_1 , \tag{25.3}$$

*and*

$$P = \int_\Omega f_P \, \mathrm{d}\mu_2 , \tag{25.4}$$

*respectively, and where $f_C$ and $f_P$ are $(S_t - K)^+$ and $(K - S_t)^+$ respectively.*

*Proof.* For clarity, set $r$ and $\delta$ to 0 without a loss of generality. By Put-Call Parity Arbitrage, a positive holding of a call ("long") and negative one of a put ("short") replicates a tradable forward; because of P/L variations, using positive sign for long and negative sign for short:

$$C(S_{t_0}, K, t) - P(S_{t_0}, K, t) + K = F_t^P \tag{25.5}$$

necessarily since $F_t^P$ is tradable.

Put-Call Parity holds for all strikes, so:

$$C(S_{t_0}, K + \Delta K, t) - P(S_{t_0}, K + \Delta K, t) + K + \Delta K = F_t^P \tag{25.6}$$

for all $K \in \Omega$

Now a Call spread in quantities $\frac{1}{\Delta K}$, expressed as

$$C(S_{t_0}, K, t) - C(S_{t_0}, K + \Delta K, t),$$

delivers \$1 if $S_t > K + \Delta K$ (that is, corresponds to the indicator function $\mathbf{1}_{S > K + \Delta K}$), 0 if $S_t \leq K$ (or $\mathbf{1}_{S > K}$), and the quantity times $S_t - K$ if $K < S_t \leq K + \Delta K$, that is, between 0 and \$1 (see Breeden and Litzenberger, 1978[35]). Likewise, consider the converse argument for a put, with $\Delta K < S_t$.

At the limit, for $\Delta K \to 0$

$$\frac{\partial C(S_{t_0}, K, t)}{\partial K} = -P(S_t > K) = -\int_{A_K^c} \mathrm{d}\mu_1 . \tag{25.7}$$

By the same argument:

$$\frac{\partial P(S_{t_0}, K, t)}{\partial K} = \int_{A_K} \mathrm{d}\mu_2 = 1 - \int_{A_K^c} \mathrm{d}\mu_2. \tag{25.8}$$

As semi-closed intervals generate the whole Borel $\sigma$-algebra on $\Omega$, this shows that $\mu_1$and $\mu_2$ are unique.

□

**Lemma 25.2**
*The probability measures of puts and calls are the same, namely for each Borel set $A$ in $\Omega$, $\mu_1(A) = \mu_2(A)$.*

*Proof.* Combining Equations 25.5 and 25.6, dividing by $\frac{1}{\Delta K}$ and taking $\Delta K \to 0$:

$$-\frac{\partial C(S_{t_0}, K, t)}{\partial K} + \frac{\partial P(S_{t_0}, K, t)}{\partial K} = 1 \tag{25.9}$$

for all values of $K$, so

$$\int_{A_K^c} \mathrm{d}\mu_1 = \int_{A_K^c} \mathrm{d}\mu_2, \tag{25.10}$$

hence $\mu_1(A_K) = \mu_2(A_K)$ for all $K \in [0, \infty)$. This equality being true for any semi-closed interval, it extends to any Borel set.

□

**Lemma 25.3**
*Puts and calls are required, by static arbitrage, to be evaluated at same as risk-neutral measure $\mu_Q$ as the tradable forward.*

*Proof.*

$$F_t^P = \int_{\Omega} F_t \, \mathrm{d}\mu_Q; \tag{25.11}$$

from Equation 25.5

$$\int_{\Omega} f_C(K) \, \mathrm{d}\mu_1 - \int_{\Omega} f_P(K) \, \mathrm{d}\mu_1 = \int_{\Omega} F_t \, \mathrm{d}\mu_Q - K \tag{25.12}$$

Taking derivatives on both sides, and since $f_C - f_P = S_0 + K$, we get the Radon-Nikodym derivative:

$$\frac{\mathrm{d}\mu_Q}{\mathrm{d}\mu_1} = 1 \tag{25.13}$$

for all values of K.

□

## 25.3 CASE WHERE THE FORWARD IS NOT RISK NEUTRAL

Consider the case where $F_t$ is observable, tradable, and use it solely as an underlying security with dynamics on its own. In such a case we can completely ignore the dynamics of the nominal underlying $S$, or use a non-risk neutral "implied" rate linking cash to forward, $m^* = \frac{\log\left(\frac{F}{S_0}\right)}{t-t_0}$. the rate $m$ can embed risk premium, difficulties in financing, structural or regulatory impediments to borrowing, with no effect on the final result.

In that situation, it can be shown that the exact same results as before apply, by remplacing the measure $\mu_Q$ by another measure $\mu_{Q^*}$. Option prices remain unique [6].

## 25.4 COMMENT

We have replaced the complexity and intractability of dynamic hedging with a simple, more benign interpolation problem, and explained the performance of pre-Black-Scholes option operators using simple heuristics and rules, bypassing the structure of the theorems of financial economics.

Options can remain non-redundant and markets incomplete: we are just arguing here for a form of arbitrage pricing (which includes risk-neutral pricing at the level of the expectation of the probability measure), nothing more. But this is sufficient for us to use any probability distribution with finite first moment, which includes the Lognormal, which recovers Black-Scholes.

A final comparison. In dynamic hedging, missing a single hedge, or encountering a single gap (a tail event) can be disastrous —as we mentioned, it requires a series of assumptions beyond the mathematical, in addition to severe and highly unrealistic constraints on the mathematical. Under the class of fat tailed distributions, increasing the frequency of the hedges does not guarantee reduction of risk. Further, the standard dynamic hedging argument requires the exact specification of the *risk-neutral* stochastic process between $t_0$ and $t$, something econometrically unwieldy, and which is generally reverse engineered from the price of options, as an arbitrage-oriented interpolation tool rather than as a representation of the process.

Here, in our Put-Call Parity based methodology, our ability to track the risk neutral distribution is guaranteed by adding strike prices, and since probabilities add up to 1, the degrees of freedom that the recovered measure $\mu_Q$ has in the gap area between a strike price $K$ and the next strike up, $K + \Delta K$, are severely reduced, since the measure in the interval is constrained by the difference $\int_{A_K}^{c} d\mu - \int_{A_{K+\Delta K}}^{c} d\mu$. In other words, no single gap between strikes can significantly affect the probability measure, even less the first moment, unlike with dynamic hedging. In fact it is no different from standard kernel smoothing methods for statistical samples, but applied to the distribution across strikes.[7]

[6] We assumed 0 discount rate for the proofs; in case of nonzero rate, premia are discounted at the rate of the arbitrage operator

The assumption about the presence of strike prices constitutes a natural condition: conditional on having a *practical* discussion about options, options strikes need to exist. Further, as it is the experience of the author, market-makers can add over-the-counter strikes at will, should they need to do so.

## ACKNOWLEDGMENTS

Peter Carr, Marco Avellaneda, Hélyette Geman, Raphael Douady, Gur Huberman, Espen Haug, and Hossein Kazemi.

7 For methods of interpolation of implied probability distribution between strikes, see Avellaneda et al.[7].

# 26 OPTION TRADERS NEVER USE THE BLACK-SCHOLES FORMULA[*,‡]

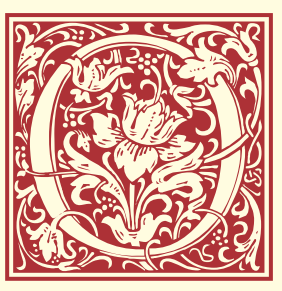

PTION TRADERS use a heuristically derived pricing formula which they adapt by fudging and changing the tails and skewness by varying one parameter, the standard deviation of a Gaussian. Such formula is popularly called "Black-Scholes-Merton" owing to an attributed eponymous discovery (though changing the standard deviation parameter is in contradiction with it). However, we have historical evidence that: (1) the said Black, Scholes and Merton did not invent any formula, just found an argument to make a well known (and used) formula compatible with the economics establishment, by removing the risk parameter through dynamic hedging, (2) option traders use (and evidently have used since 1902) sophisticated heuristics and tricks more compatible with the previous versions of the formula of Louis Bachelier and Edward O. Thorp (that allow a broad choice of probability distributions) and removed the risk parameter using put-call parity, (3) option traders did not use the Black-Scholes-Merton formula or similar formulas after 1973 but continued their bottom-up heuristics more robust to the high impact rare event. The chapter draws on historical trading methods and 19th and early 20th century references ignored by the finance literature. It is time to stop using the wrong designation for option pricing.

## 26.1 BREAKING THE CHAIN OF TRANSMISSION

For us, practitioners, theories should arise from practice [2]. This explains our concern with the "scientific" notion that practice should fit theory. Option hedging, pricing, and trading is neither philosophy nor mathematics. It is a rich craft with traders learning from traders (or traders copying other traders) and tricks

Research chapter.

2 For us, in this discussion, a "practitioner" is deemed to be someone involved in repeated decisions about option hedging, that is with a risk-P/L and skin in the game, not a support quant who writes pricing software or an academic who provides consulting advice.

developing under evolution pressures, in a bottom-up manner. It is techne, not episteme. Had it been a science it would not have survived for the empirical and scientific fitness of the pricing and hedging theories offered are, we will see, at best, defective and unscientific (and, at the worst, the hedging methods create more risks than they reduce). Our approach in this chapter is to ferret out historical evidence of techne showing how option traders went about their business in the past.

Options, we will show, have been extremely active in the pre-modern finance world. Tricks and heuristically derived methodologies in option trading and risk management of derivatives books have been developed over the past century, and used quite effectively by operators. In parallel, many derivations were produced by mathematical researchers. The economics literature, however, did not recognize these contributions, substituting the rediscoveries or subsequent re formulations done by (some) economists. There is evidence of an attribution problem with Black-Scholes-Merton option formula , which was developed, used, and adapted in a robust way by a long tradition of researchers and used heuristically by option book runners. Furthermore, in a case of scientific puzzle, the exact formula called Black-Sholes-Merton was written down (and used) by Edward Thorp which, paradoxically, while being robust and realistic, has been considered unrigorous. This raises the following: 1) The Black-Scholes-Merton innovation was just a neoclassical finance argument, no more than a thought experiment [3], 2) We are not aware of traders using their argument or their version of the formula.

It is high time to give credit where it belongs.

## 26.2 INTRODUCTION/SUMMARY

### 26.2.1 Black-Scholes was an argument

Option traders call the formula they use the Black-Scholes-Merton formula without being aware that by some irony, of all the possible options formulas that have been produced in the past century, what is called the Black-Scholes-Merton formula (after Black and Scholes, 1973, and Merton, 1973) is the one the furthest away from what they are using. In fact of the formulas written down in a long history it is the only formula that is fragile to jumps and tail events.

First, something seems to have been lost in translation: Black and Scholes [29] and Merton [201] actually never came up with a new option formula, but only an theoretical economic argument built on a new way of deriving, rather re-deriving, an already existing and well known formula. The argument, we will see, is extremely fragile to assumptions. The foundations of option hedging and pricing were already far more firmly laid down before them. The Black-Scholes-Merton argument, simply, is that an option can be hedged using a certain methodology called dynamic hedging and then turned into a risk-free instrument, as the portfolio would no longer be stochastic. Indeed what Black, Scholes and Merton did was marketing, finding a way to make a well-known formula palatable to the economics establishment of the time, little else, and in fact distorting its essence.

[3] Here we question the notion of confusing thought experiments in a hypothetical world, of no predictive power, with either science or practice. The fact that the Black-Scholes-Merton argument works in a Platonic world and appears to be elegant does not mean anything since one can always produce a Platonic world in which a certain equation works, or in which a rigorous proof can be provided, a process called reverse-engineering.

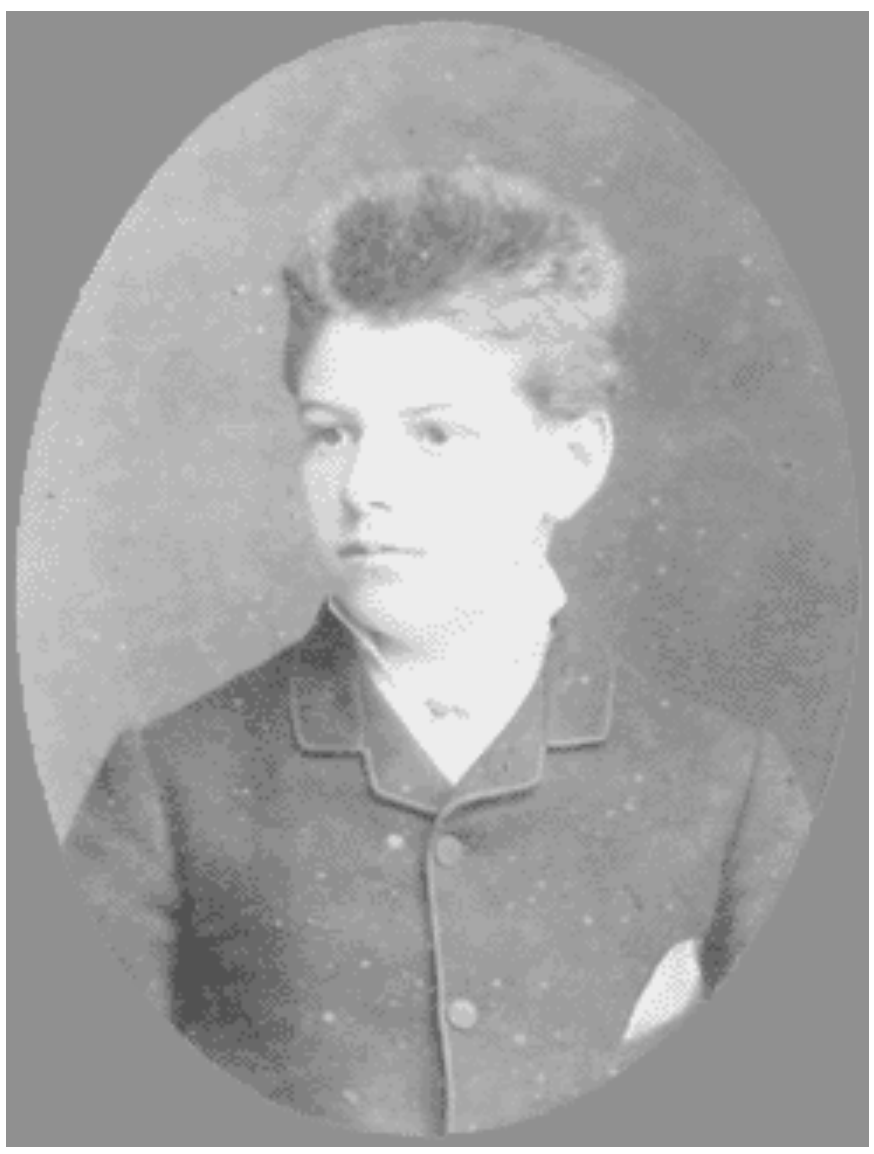

Figure 26.1: *Louis Bachelier, who came up with an option formula based on expectation. It is based on more rigorous foundations than the Black-Scholes dynamic hedging argument as it does not require a thin-tailed distribution. Few people are aware of the fact that the Black-Scholes so-called discovery was an argument to remove the expectation of the underlying security, not the derivation of a new equation.*

Such argument requires strange far-fetched assumptions: some liquidity at the level of transactions, knowledge of the probabilities of future events (in a neoclassical Arrow-Debreu style) , and, more critically, a certain mathematical structure that requires thin-tails, or mild randomness, on which, later[4]. The entire argument is indeed, quite strange and rather inapplicable for someone clinically and observation-driven standing outside conventional neoclassical economics. Simply, the dynamic hedging argument is dangerous in practice as it subjects you to blowups; it makes no sense unless you are concerned with neoclassical economic theory. The Black-Scholes-Merton argument and equation flow a top-down general equilibrium theory, built upon the assumptions of operators working in full knowledge of the probability distribution of future outcomes in addition to a collection of assumptions that, we will see, are highly invalid mathematically, the main one being the ability to cut the risks using continuous trading which only works in the very narrowly special case of thin-tailed distributions. But it is not just these flaws that make it inapplicable: option traders do not buy theories , particularly speculative general equilibrium ones, which they find too risky for them and extremely lacking in standards of reliability. A normative theory is, simply, not good for decision-making under uncertainty (particularly if it is in chronic disagreement with empirical evidence). People may take decisions based on speculative theories, but avoid the fragility of theories in running their risks.

[4] Of all the misplaced assumptions of Black Scholes that cause it to be a mere thought experiment, though an extremely elegant one, a flaw shared with modern portfolio theory, is the certain knowledge of future delivered variance for the random variable (or, equivalently, all the future probabilities). This is what makes it clash with practice the rectification by the market fattening the tails is a negation of the Black-Scholes thought experiment.

Yet professional traders, including the authors (and, alas, the Swedish Academy of Science) have operated under the illusion that it was the Black-Scholes-Merton formula they actually used we were told so. This myth has been progressively reinforced in the literature and in business schools, as the original sources have been lost or frowned upon as anecdotal (Merton [203]).

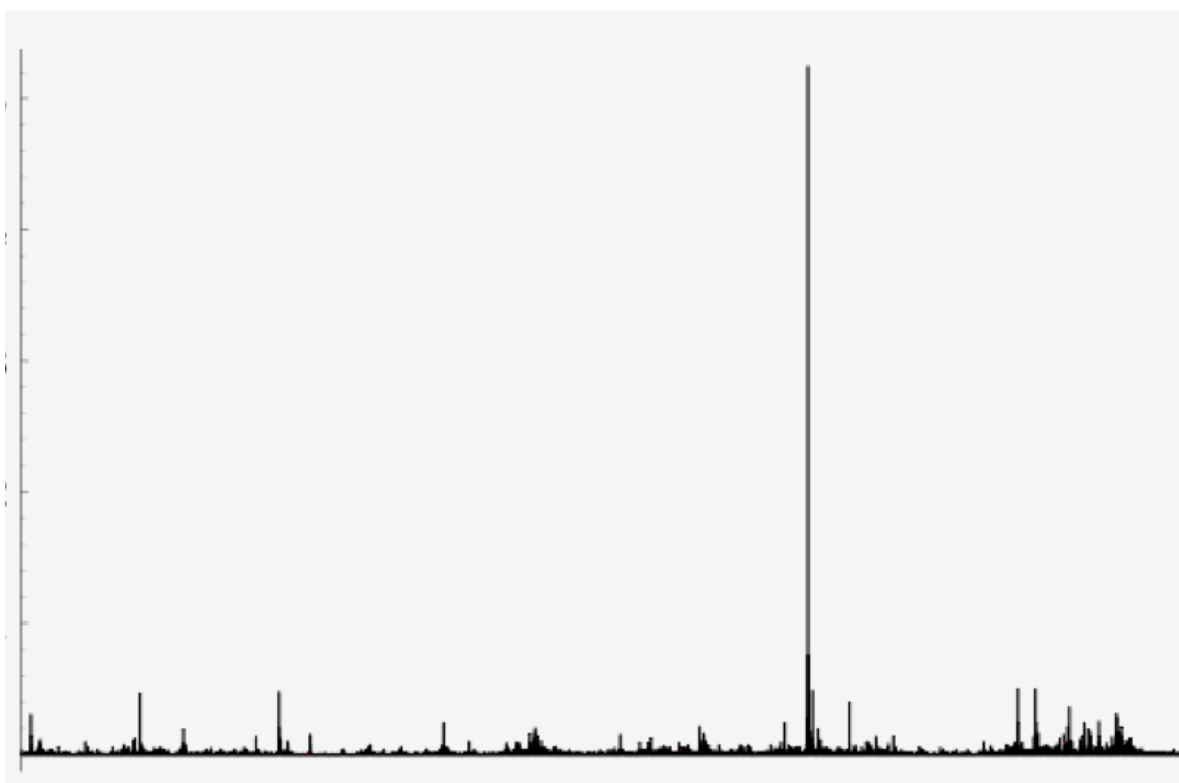

Figure 26.2: *The typical "risk reduction" performed by the Black-Scholes-Merton argument. These are the variations of a dynamically hedged portfolio (and a quite standard one). BSM indeed "smoothes" out variations but exposes the operator to massive tail events reminiscent of such blowups as LTCM. Other option formulas are robust to the rare event and make no such claims.*

This discussion will present our real-world, ecological understanding of option pricing and hedging based on what option traders actually do and did for more than a hundred years.

This is a very general problem. As we said, option traders develop a chain of transmission of techne, like many professions. But the problem is that the chain is often broken as universities do not store the acquired skills by operators. Effectively plenty of robust heuristically derived implementations have been developed over the years, but the economics establishment has refused to quote them or acknowledge them. This makes traders need to relearn matters periodically. Failure of dynamic hedging in 1987, by such firm as Leland O'ŹBrien Rubinstein, for instance, does not seem to appear in the academic literature published after the event (Merton, [203], Rubinstein,[242], Ross [240]); to the contrary dynamic hedging is held to be a standard operation [5].

There are central elements of the real world that can escape them academic research without feedback from practice (in a practical and applied field) can cause the diversions we witness between laboratory and ecological frameworks. This explains why so many finance academics have had the tendency to produce smooth returns, then blow up using their own theories[6]. We started the other way around, first by years of option trading doing million of hedges and thousands of option trades. This in combination with investigating the forgotten and ignored ancient knowledge in option pricing and trading we will explain some common myths about option pricing and hedging. There are indeed two myths:

- That we had to wait for the Black-Scholes-Merton options formula to trade the product, price options, and manage option books. In fact the introduction of the Black, Scholes and Merton argument increased our risks and set us back in risk management. More generally, it is a myth that traders rely on theories, even less a general equilibrium theory, to price options.
- That we use the Black-Scholes-Merton options pricing formula. We, simply don't.

[5] For instance how mistakes never resurface into the consciousness, Mark Rubinstein was awarded in 1995 the Financial Engineer of the Year award by the International Association of Financial Engineers. There was no mention of portfolio insurance and the failure of dynamic hedging.

[6] For a standard reaction to a rare event, see the following: "Wednesday is the type of day people will remember in quant-land for a very long time," said Mr. Rothman, a University of Chicago Ph.D. who ran a quantitative fund before joining Lehman Brothers. "Events that models only predicted would happen once in 10,000 years happened every day for three days." One "Quant Sees Shakeout For the Ages – '10,000 Years" By Kaja Whitehouse,*Wall Street Journal* August 11, 2007; Page B3.

In our discussion of these myth we will focus on the bottom-up literature on option theory that has been hidden in the dark recesses of libraries. And that addresses only recorded matters not the actual practice of option trading that has been lost.

## 26.3 myth 1: traders did not price options before bsm

It is assumed that the Black-Scholes-Merton theory is what made it possible for option traders to calculate their delta hedge (against the underlying) and to price options. This argument is highly debatable, both historically and analytically.

Options were actively trading at least already in the 1600 as described by Joseph De La Vega implying some form of techneń, a heuristic method to price them and deal with their exposure. De La Vega describes option trading in the Netherlands, indicating that operators had some expertise in option pricing and hedging. He diffusely points to the put-call parity, and his book was not even meant to teach people about the technicalities in option trading. Our insistence on the use of Put-Call parity is critical for the following reason: The Black-Scholes-Merton Źs claim to fame is removing the necessity of a risk-based drift from the underlying security to make the trade risk-neutral. But one does not need dynamic hedging for that: simple put call parity can suffice (Derman and Taleb, 2005), as we will discuss later. And it is this central removal of the risk-premium that apparently was behind the decision by the Nobel committee to grant Merton and Scholes the (then called) Bank of Sweden Prize in Honor of Alfred Nobel: Black, Merton and Scholes made a vital contribution by showing that it is in fact not necessary to use any risk premium when valuing an option. This does not mean that the risk premium disappears; instead it is already included in the stock price. It is for having removed the effect of the drift on the value of the option, using a thought experiment, that their work was originally cited, something that was mechanically present by any form of trading and converting using far simpler techniques.

Options have a much richer history than shown in the conventional literature. Forward contracts seems to date all the way back to Mesopotamian clay tablets dating all the way back to 1750 B.C. Gelderblom and Jonker [122] show that Amsterdam grain dealers had used options and forwards already in 1550.

In the late 1800 and the early 1900 there were active option markets in London and New York as well as in Paris and several other European exchanges. Markets it seems, were active and extremely sophisticated option markets in 1870. Kairys and Valerio (1997) discuss the market for equity options in USA in the 1870s, indirectly showing that traders were sophisticated enough to price for tail events[7].

There was even active option arbitrage trading taking place between some of these markets. There is a long list of missing treatises on option trading: we traced at least ten German treatises on options written between the late 1800s and the hyperinflation episode[8].

## 26.4 METHODS AND DERIVATIONS

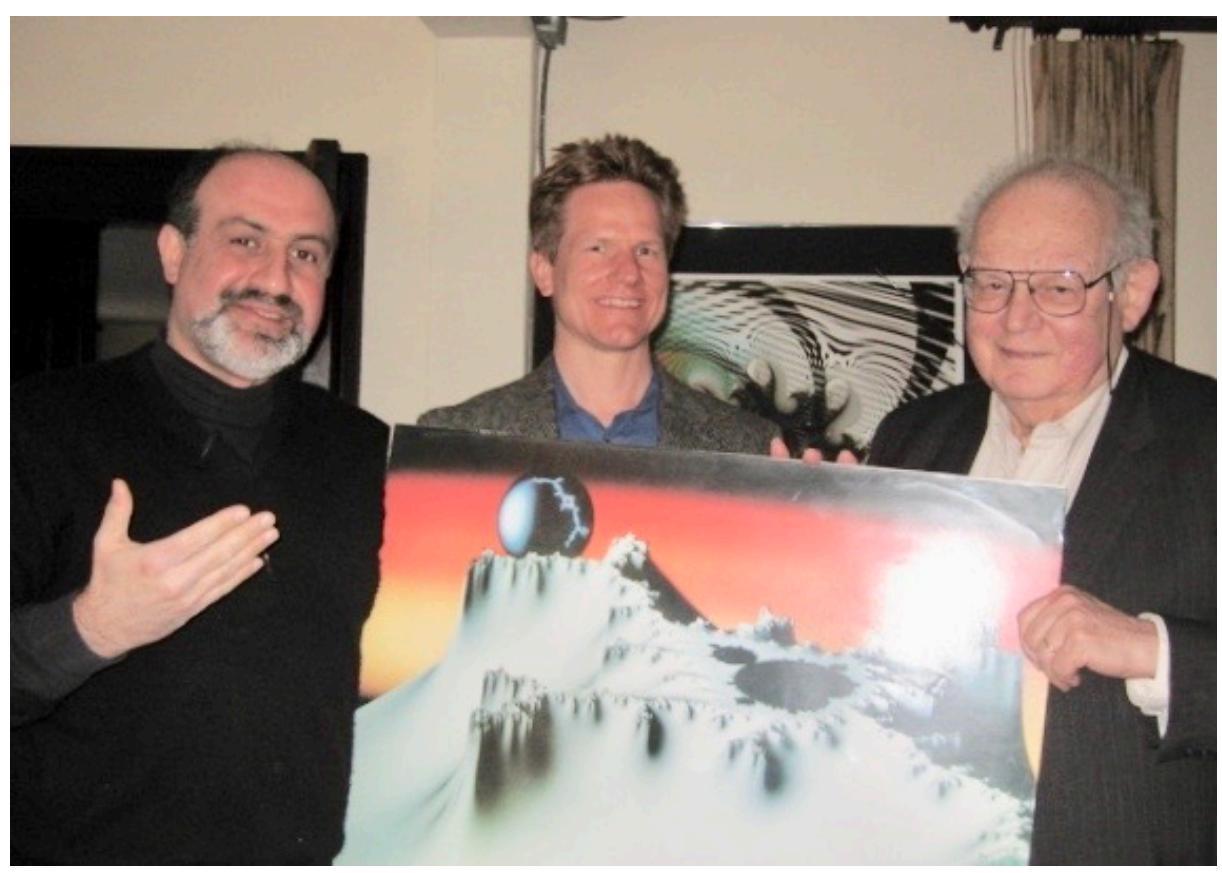

Figure 26.3: *Espen Haug (coauthor of chapter) with Mandelbrot and this author in 2007.*

One informative extant source, Nelson [206], speaks volumes: An option trader and arbitrageur, S.A. Nelson published a book The A B C of Options and Arbi-

[7] The historical description of the market is informative until Kairys and Valerio [166] try to gauge whether options in the 1870s were underpriced or overpriced (using Black-Scholes-Merton style methods). There was one tail-event in this period, the great panic of September 1873. Kairys and Valerio find that holding puts was profitable, but deem that the market panic was just a one-time event :

"However, the put contracts benefit from the financial panic that hit the market in September, 1873. Viewing this as a one-time event, we repeat the analysis for puts excluding any unexpired contracts written before the stock market panic."

Using references to the economic literature that also conclude that options in general were overpriced in the 1950s 1960s and 1970s they conclude: "Our analysis shows that option contracts were generally overpriced and were unattractive for retail investors to purchase. They add: İEmpirically we find that both put and call options were regularly overpriced relative to a theoretical valuation model." These results are contradicted by the practitioner Nelson (1904): "The majority of the great option dealers who have found by experience that it is the givers, and not the takers, of option money who have gained the advantage in the long run."

[8] Here is a partial list: Bielschowsky, R (1892): *Ueber die rechtliche Natur der Prämiengeschäfte*, Bresl. Genoss.-Buchdr; Granichstaedten-Czerva, R (1917): *Die Prämiengeschäfte an der Wiener Börse*, Frankfurt am Main; Holz, L. (1905) *Die Prämiengeschäfte*, Thesis (doctoral)–Universität Rostock; Kitzing, C. (1925):*Prämiengeschäfte : Vorprämien-, Rückprämien-, Stellagen- u. Nochgeschäfte ; Die solidesten Spekulationsgeschäfte mit Versicherg auf Kursverlust*, Berlin; Leser, E, (1875): *Zur Geschichte der Prämiengeschäfte*; Szkolny, I. (1883): *Theorie und praxis der prämiengeschäfte nach einer originalen methode dargestellt.*, Frankfurt am Main; Author Unknown (1925): *Das Wesen der Prämiengeschäfte*, Berlin : Eugen Bab & Co., Bankgeschäft.

trage based on his observations around the turn of the twentieth century. According to Nelson (1904) up to 500 messages per hour and typically 2000 to 3000 messages per day were sent between the London and the New York market through the cable companies. Each message was transmitted over the wire system in less than a minute. In a heuristic method that was repeated in *Dynamic Hedging* [271] , Nelson, describe in a theory-free way many rigorously clinical aspects of his arbitrage business: the cost of shipping shares, the cost of insuring shares, interest expenses, the possibilities to switch shares directly between someone being long securities in New York and short in London and in this way saving shipping and insurance costs, as well as many more tricks etc.

The formal financial economics canon does not include historical sources from outside economics, a mechanism discussed in Taleb (2007)[273]. The put-call parity was according to the formal option literature first fully described by Stoll [259], but neither he nor others in the field even mention Nelson. Not only was the put-call parity argument fully understood and described in detail by Nelson, but he, in turn, makes frequent reference to Higgins (1902) [151]. Just as an example Nelson (1904) referring to Higgins (1902) writes:

> It may be worthy of remark that calls are more often dealt than puts the reason probably being that the majority of punters in stocks and shares are more inclined to look at the bright side of things, and therefore more often see a rise than a fall in prices.
>
> This special inclination to buy calls and to leave the puts severely alone does not, however, tend to make calls dear and puts cheap, for it can be shown that the adroit dealer in options can convert a put into a call, a call into a put, a call or more into a put-and-call, in fact any option into another, by dealing against it in the stock. We may therefore assume, with tolerable accuracy, that the call of a stock at any moment costs the same as the put of that stock, and half as much as the Put-and-Call.

The Put-and-Call was simply a put plus a call with the same strike and maturity, what we today would call a straddle. Nelson describes the put-call parity over many pages in full detail. Static market neutral delta hedging was also known at that time, in his book Nelson for example writes:

> Sellers of options in London as a result of long experience, if they sell a Call, straightway buy half the stock against which the Call is sold; or if a Put is sold; they sell half the stock immediately.

We must interpret the value of this statement in the light that standard options in London at that time were issued at-the-money (as explicitly pointed out by Nelson); furthermore, all standard options in London were European style. In London in- or out-of-the-money options were only traded occasionally and were known as fancy options. It is quite clear from this and the rest of Nelson's book that the option dealers were well aware that the delta for at-the-money options was approximately 50%. As a matter of fact, at-the-money options trading in London at that time were adjusted to be struck to be at-the-money forward, in order to make puts and calls of the same price. We know today that options that are at-the-money forward and do not have very long time to maturity have a delta very close to 50% (naturally minus 50% for puts). The options in London at that time typically had one month to maturity when issued.

Nelson also diffusely points to dynamic delta hedging, and that it worked better in theory than practice (see Haug [146]. It is clear from all the details described by Nelson that options in the early 1900 traded actively and that option traders at that time in no way felt helpless in either pricing or in hedging them.

Herbert Filer was another option trader that was involved in option trading from 1919 to the 1960s. Filer(1959) describes what must be considered a reasonable active option market in New York and Europe in the early 1920s and 1930s. Filer mentions however that due to World War II there was no trading on the European Exchanges, for they were closed. Further, he mentions that London option trading did not resume before 1958. In the early 1900, option traders in London were considered to be the most sophisticated, according to [207]. It could well be that World War II and the subsequent shutdown of option trading for many years was the reason known robust arbitrage principles about options were forgotten and almost lost, to be partly re-discovered by finance professors such as Stoll.

Earlier, in 1908, Vinzenz Bronzin published a book deriving several option pricing formulas, and a formula very similar to what today is known as the Black-Scholes-Merton formula, see also Hafner and Zimmermann (2007, 2009) [137]. Bronzin based his risk-neutral option valuation on robust arbitrage principles such as the put-call parity and the link between the forward price and call and put options in a way that was rediscovered by Derman and Taleb (2005)[9]. Indeed, the put-call parity restriction is sufficient to remove the need to incorporate a future return in the underlying security it forces the lining up of options to the forward price[10].

Again, in 1910 Henry Deutsch describes put-call parity but in less detail than Higgins and Nelson. In 1961 Reinach again described the put-call parity in quite some detail (another text typically ignored by academics). Traders at New York stock exchange specializing in using the put-call parity to convert puts into calls or calls into puts was at that time known as Converters. Reinach (1961) [231]:

> Although I have no figures to substantiate my claim, I estimate that over 60 percent of all Calls are made possible by the existence of Converters.

In other words the converters (dealers) who basically operated as market makers were able to operate and hedge most of their risk by statically hedging options with options. Reinach wrote that he was an option trader (Converter) and gave

[9] The argument Derman Taleb(2005) [75] was present in [271] but remained unnoticed.

[10] Ruffino and Treussard (2006) [241] accept that one could have solved the risk-premium by happenstance, not realizing that put-call parity was so extensively used in history. But they find it insufficient. Indeed the argument may not be sufficient for someone who subsequently complicated the representation of the world with some implements of modern finance such as "stochastic discount rates" while simplifying it at the same time to make it limited to the Gaussian and allowing dynamic hedging. They write that the use of a non-stochastic discount rate common to both the call and the put options is inconsistent with modern equilibrium capital asset pricing theory. Given that we have never seen a practitioner use stochastic discount rate, we, like our option trading predecessors, feel that put-call parity is sufficient & does the job.

The situation is akin to that of scientists lecturing birds on how to fly, and taking credit for their subsequent performance except that here it would be lecturing them the wrong way.

examples on how he and his colleagues tended to hedge and arbitrage options against options by taking advantage of options embedded in convertible bonds:

Writers and traders have figured out other procedures for making profits writing Puts & Calls. Most are too specialized for all but the seasoned professional. One such procedure is the ownership of a convertible bond and then writing of Calls against the stock into which the bonds are convertible. If the stock is called converted and the stock is delivered.

Higgins, Nelson and Reinach all describe the great importance of the put-call parity and to hedge options with options. Option traders were in no way helpless in hedging or pricing before the Black-Scholes-Merton formula. Based on simple arbitrage principles they were able to hedge options more robustly than with Black- Scholes-Merton. As already mentioned static market-neutral delta hedging was described by Higgins and Nelson in 1902 and 1904. Also, W. D. Gann (1937) discusses market neutral delta hedging for at-the-money options, but in much less details than Nelson (1904). Gann also indicates some forms of auxiliary dynamic hedging.

Mills (1927) illustrates how jumps and fat tails were present in the literature in the pre-Modern Portfolio Theory days. He writes: "(...) distribution may depart widely from the Gaussian type because the influence of one or two extreme price changes".

### 26.4.1 Option formulas and Delta Hedging

Which brings us to option pricing formulas. The first identifiable one was Bachelier (1900) [8]. Sprenkle in 1961 [255] extended Bacheliers work to assume lognormal rather than normal distributed asset price. It also avoids discounting (to no significant effect since many markets, particularly the U.S., option premia were paid at expiration).

James Boness (1964) [30] also assumed a lognormal asset price. He derives a formula for the price of a call option that is actually identical to the Black-Scholes-Merton 1973 formula, but the way Black, Scholes and Merton derived their formula based on continuous dynamic delta hedging or alternatively based on CAPM they were able to get independent of the expected rate of return. It is in other words not the formula itself that is considered the great discovery done by Black, Scholes and Merton, but how they derived it. This is among several others also pointed out by Rubinstein (2006) [243]:

The real significance of the formula to the financial theory of investment lies not in itself, but rather in how it was derived. Ten years earlier the same formula had been derived by Case M. Sprenkle [255] and A. James Boness [30].

Samuelson (1969) and Thorp (1969) published somewhat similar option pricing formulas to Boness and Sprenkle. Thorp (2007) claims that he actually had an identical formula to the Black-Scholes-Merton formula programmed into his computer years before Black, Scholes and Merton published their theory.

Now, delta hedging. As already mentioned static market-neutral delta hedging was clearly described by Higgins and Nelson 1902 and 1904. Thorp and Kassouf (1967) presented market neutral static delta hedging in more details, not only for at-the-money options, but for options with any delta. In his 1969 paper Thorp is shortly describing market neutral static delta hedging, also briefly pointed in the direction of some dynamic delta hedging, not as a central pricing device, but a risk-management tool. Filer also points to dynamic hedging of options, but without showing much knowledge about how to calculate the delta. Another ignored and forgotten text is a book/booklet published in 1970 by Arnold Bernhard & Co. The authors are clearly aware of market neutral static delta hedging or what they name balanced hedge for any level in the strike or asset price. This book has multiple examples of how to buy warrants or convertible bonds and construct a market neutral delta hedge by shorting the right amount of common shares. Arnold Bernhard & Co also published deltas for a large number of warrants and convertible bonds that they distributed to investors on Wall Street.

Referring to Thorp and Kassouf (1967)[302], Black, Scholes and Merton took the idea of delta hedging one step further, with Black and Scholes (1973) [29]:

If the hedge is maintained continuously, then the approximations mentioned above become exact, and the return on the hedged position is completely independent of the change in the value of the stock. In fact, the return on the hedged position becomes certain. This was pointed out to us by Robert Merton.

This may be a brilliant mathematical idea, but option trading is not mathematical theory. It is not enough to have a theoretical idea so far removed from reality that is far from robust in practice. What is surprising is that the only principle option traders do not use and cannot use is the approach named after the formula, which is a point we discuss next.

## 26.5 myth 2: traders today use black-scholes

Traders don't do Valuation.

First, operationally, a price is not quite valuation. Valuation requires a strong theoretical framework with its corresponding fragility to both assumptions and the structure of a model. For traders, a price produced to buy an option when one has no knowledge of the probability distribution of the future is not valuation, but an expedient. Such price could change. Their beliefs do not enter such price. It can also be determined by his inventory.

This distinction is critical: traders are engineers, whether boundedly rational (or even non interested in any form of probabilistic rationality), they are not privy to informational transparency about the future states of the world and their probabilities. So they do not need a general theory to produce a price merely the avoidance of Dutch-book style arbitrages against them, and the compatibility with some standard restriction: In addition to put-call parity, a call of a certain strike $K$ cannot trade at a lower price than a call $K + \Delta K$ (avoidance of negative call and put spreads), a call struck at $K$ and a call struck at $K + 2\Delta K$ cannot be more expensive than twice the price of a call struck at $K + \Delta$ (negative butterflies), horizontal calendar spreads cannot be negative (when interest rates are low), and so forth. The degrees of freedom for traders are thus reduced: they need to abide by put-call parity and compatibility with other options in the market.

In that sense, traders do not perform valuation with some pricing kernel until the expiration of the security, but, rather, produce a price of an option compatible with other instruments in the markets, with a holding time that is stochastic. They do not need top-down science.

### 26.5.1 When do we value?

If you find traders operated solo, in a desert island, having for some to produce an option price and hold it to expiration, in a market in which the forward is absent, then some valuation would be necessary but then their book would be minuscule. And this thought experiment is a distortion: people would not trade options unless they are in the business of trading options, in which case they would need to have a book with offsetting trades. For without offsetting trades, we doubt traders would be able to produce a position beyond a minimum (and negligible) size as dynamic hedging not possible. (Again we are not aware of many non-blownup option traders and institutions who have managed to operate in the vacuum of the Black Scholes-Merton argument). It is to the impossibility of such hedging that we turn next.

## 26.6 ON THE MATHEMATICAL IMPOSSIBILITY OF DYNAMIC HEDGING

Finally, we discuss the severe flaw in the dynamic hedging concept. It assumes, nay, requires all moments of the probability distribution to exist[11].

Assume that the distribution of returns has a scale-free or fractal property that we can simplify as follows: for $x$ large enough, (i.e. in the tails), $\frac{\mathbb{P}(X>nx)}{\mathbb{P}(X>x)}$ depends on $n$, not on $x$. In financial securities, say, where $X$ is a daily return, there is no reason for P(X>20%)/P(X>10%) to be different from P(X>15%)/P(X>7.5%). This self-similarity at all scales generates power law, or Paretian, tails, i.e., above a crossover point, $\mathbb{P}(X > x) = Kx^{\alpha}$. It happens, looking at millions of pieces of data, that such property holds in markets all markets, baring sample error. For overwhelming empirical evidence, see Mandelbrot (1963), which predates Black-Scholes-Merton (1973) and the jump-diffusion of Merton (1976); see also Stanley et al. (2000), and Gabaix et al. (2003). The argument to assume the scale-free is as follows: the distribution might have thin tails at some point (say above some value of X). But we do not know where such point is we are epistemologically in the dark as to where to put the boundary, which forces us to use infinity.

Some criticism of these "true fat-tails" accept that such property might apply for daily returns, but, owing to the Central Limit Theorem, the distribution is held to become Gaussian under aggregation for cases in which $\alpha$ is deemed higher than 2. Such argument does not hold owing to the preasymptotics of scalable distributions: Bouchaud and Potters (2003) and Mandelbrot and Taleb (2007) argue that the presasymptotics of fractal distributions are such that the effect of the Central Limit Theorem are exceedingly slow in the tails in fact irrelevant. Furthermore, there is sampling error as we have less data for longer periods, hence fewer tail episodes, which give an in-sample illusion of thinner tails. In addition, the point that aggregation thins out the tails does not hold for dynamic hedging in which the operator depends necessarily on high frequency data and their statistical properties. So long as it is scale-free at the time period of dynamic hedge, higher moments become explosive, infinite to disallow the formation of a dynamically hedge portfolio. Simply a Taylor expansion is impossible as moments of higher order that 2 matter critically one of the moments is going to be infinite.

[11] Merton (1992) seemed to accept the inapplicability of dynamic hedging but he perhaps thought that these ills would be cured thanks to his prediction of the financial world "spiraling towards dynamic completeness". Fifteen years later, we have, if anything, spiraled away from it.

The mechanics of dynamic hedging are as follows. Assume the risk-free interest rate of 0 with no loss of generality. The canonical Black-Scholes-Merton package consists in selling a call and purchasing shares of stock that provide a hedge against instantaneous moves in the security. Thus the portfolio $\pi$ locally "hedged" against exposure to the first moment of the distribution is the following:

$$\pi = -C + \frac{\partial C}{\partial S} S$$

where $C$ is the call price, and $S$ the underlying security. Take the discrete time change in the values of the portfolio

$$\Delta\pi = -\Delta C + \frac{\partial C}{\partial S} \Delta S$$

By expanding around the initial values of $S$, we have the changes in the portfolio in discrete time. Conventional option theory applies to the Gaussian in which all orders higher than $\Delta S^2$ disappear rapidly.

Taking expectations on both sides, we can see here very strict requirements on moment finiteness: all moments need to converge. If we include another term, of order $\Delta S^3$, such term may be of significance in a probability distribution with significant cubic or quartic terms. Indeed, although the nth derivative with respect to S can decline very sharply, for options that have a strike $K$ away from the center of the distribution, it remains that the delivered higher orders of $S$ are rising disproportionately fast for that to carry a mitigating effect on the hedges. So here we mean all moments–no approximation. The logic of the Black-Scholes-Merton so-called solution thanks to Ito's lemma was that the portfolio collapses into a deterministic payoff. But let us see how quickly or effectively this works in practice. The Actual Replication process is as follows: The payoff of a call should be replicated with the following stream of dynamic hedges, the limit of which can be seen here, between $t$ and $T$:

$$\lim_{\Delta t \to 0} \left( \sum_{i=1}^{n=T/\Delta t} \frac{\partial C}{\partial S} \Big|_{S=S_{t+(i-1)\Delta \mathrm{t}}, t=t+(i-1)\Delta \mathrm{t},} \left( S_{t+i\Delta \mathrm{t}} - S_{t+(i-1)\Delta \mathrm{t}} \right) \right) \tag{26.1}$$

Such policy does not match the call value: the difference remains stochastic (while according to Black Scholes it should shrink), unless one lives in a fantasy world in which such risk reduction is possible.

Further, there is an inconsistency in the works of Merton making us confused as to what theory finds acceptable: in Merton (1976) he agrees that we can use Bachelier-style option derivation in the presence of jumps and discontinuities, no dynamic hedging but only when the underlying stock price is uncorrelated to the market. This seems to be an admission that dynamic hedging argument applies only to some securities: those that do not jump and are correlated to the market.

### 26.6.1 The (confusing) Robustness of the Gaussian

The success of the formula last developed by Thorp, and called Black-Scholes-Merton was due to a simple attribute of the Gaussian: you can express any probability distribution in terms of Gaussian, even if it has fat tails, by varying the standard deviation $\sigma$ at the level of the density of the random variable. It does not mean that you are using a Gaussian, nor does it mean that the Gaussian is particularly parsimonious (since you have to attach a $\sigma$ for every level of the price). It simply mean that the Gaussian can express anything you want if you add a function for the parameter $\sigma$, making it a function of strike price and time to expiration.

This volatility smile, i.e., varying one parameter to produce $\sigma(K)$, or volatility surface, varying two parameter, $\sigma(S,t)$ is effectively what was done in different ways by Dupire (1994, 2005) [85, 86] and Derman [73, 76] see Gatheral (2006 [121]). They assume a volatility process not because there is necessarily such a thing only as a method of fitting option prices to a Gaussian. Furthermore, although the Gaussian has finite second moment (and finite all higher moments as well), you can express a scalable with infinite variance using Gaussian volatility surface. One strong constrain on the $\sigma$ parameter is that it must be the same for a put and call with same strike (if both are European-style), and the drift should be that of the forward.

Indeed, ironically, the volatility smile is inconsistent with the Black-Scholes-Merton theory. This has lead to hundreds if not thousands of papers trying extend (what was perceived to be) the Black-Scholes-Merton model to incorporate stochastic volatility and jump-diffusion. Several of these researchers have been surprised that so few traders actually use stochastic volatility models. It is not a model that says how the volatility smile should look like, or evolves over time; it is a hedging method that is robust and consistent with an arbitrage free volatility surface that evolves over time.

In other words, you can use a volatility surface as a map, not a territory. However it is foolish to justify Black-Scholes-Merton on grounds of its use: we repeat that the Gaussian bans the use of probability distributions that are not Gaussian whereas non-dynamic hedging derivations (Bachelier, Thorp) are not grounded in the Gaussian.

### 26.6.2 Order Flow and Options

It is clear that option traders are not necessarily interested in the probability distribution at expiration time given that this is abstract, even metaphysical for them. In addition to the put-call parity constrains that according to evidence was fully developed already in 1904, we can hedge away inventory risk in options with other options. One very important implication of this method is that if you hedge options with options then option pricing will be largely demand and supply based. This in strong contrast to the Black-Scholes-Merton (1973) theory that based on the idealized world of geometric Brownian motion with continuous-time delta hedging then demand and supply for options simply should not affect the price of options. If someone wants to buy more options the market makers can simply manufacture them by dynamic delta hedging that will be a perfect substitute for the option itself.

This raises a critical point: option traders do not estimate the odds of rare events by pricing out-of-the-money options. They just respond to supply and demand. The notion of implied probability distribution is merely a Dutch-book compatibility type of proposition.

### 26.6.3 Bachelier-Thorp

The argument often casually propounded attributing the success of option volume to the quality of the Black-Scholes formula is rather weak. It is particularly weakened by the fact that options had been so successful at different time periods and places.

Furthermore, there is evidence that while both the Chicago Board Options Exchange and the Black-Scholes-Merton formula came about in 1973, the model was "rarely used by traders" before the 1980s (O'Connell, 2001). When one of the authors (Taleb) became a pit trader in 1992, almost two decades after Black-Scholes-Merton, he was surprised to find that many traders still priced options sheets free, pricing off the butterfly, and off the conversion, without recourse to any formula.

Even a book written in 1975 by a finance academic appears to credit Thorpe and Kassouf (1967) – rather than Black-Scholes (1973), although the latter was present in its bibliography. Auster (1975):

Sidney Fried wrote on warrant hedges before 1950, but it was not until 1967 that the book Beat the Market by Edward O. Thorp and Sheen T. Kassouf rigorously, but simply, explained the short warrant/long common hedge to a wide audience.

We conclude with the following remark. Sadly, all the equations, from the first (Bachelier), to the last pre-Black-Scholes-Merton (Thorp) accommodate a scale-free distribution. The notion of explicitly removing the expectation from the forward was present in Keynes (1924) and later by Blau (1944) â and long a Call short a put of the same strike equals a forward. These arbitrage relationships appeared to be well known in 1904.

One could easily attribute the explosion in option volume to the computer age and the ease of processing transactions, added to the long stretch of peaceful economic growth and absence of hyperinflation. From the evidence (once one removes the propaganda), the development of scholastic finance appears to be an epiphenomenon rather than a cause of option trading. Once again, lecturing birds how to fly does not allow one to take subsequent credit.

This is why we call the equation Bachelier-Thorp. We were using it all along and gave it the wrong name, after the wrong method and with attribution to the wrong persons. It does not mean that dynamic hedging is out of the question; it is just not a central part of the pricing paradigm. It led to the writing down of a certain stochastic process that may have its uses, some day, should markets spiral towards dynamic completeness. But not in the present.

# 27 OPTION PRICING UNDER POWER LAWS: A ROBUST HEURISTIC*,‡

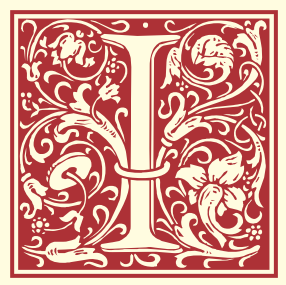

In this (research) chapter, we build a heuristic that takes a given option price in the tails with strike $K$ and extends (for calls, all strikes $> K$, for puts all strikes $< K$) assuming the continuation falls into what we define as "Karamata constant" or "Karamata point" beyond which the strong Pareto law holds. The heuristic produces relative prices for options, with for sole parameter the tail index $\alpha$ under some mild arbitrage constraints.

Usual restrictions such as finiteness of variance are not required.

The heuristic allows us to scrutinize the volatility surface and test theories of relative tail option mispricing and overpricing usually built on thin tailed models and modification of the Black-Scholes formula.

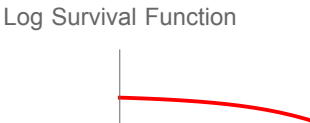

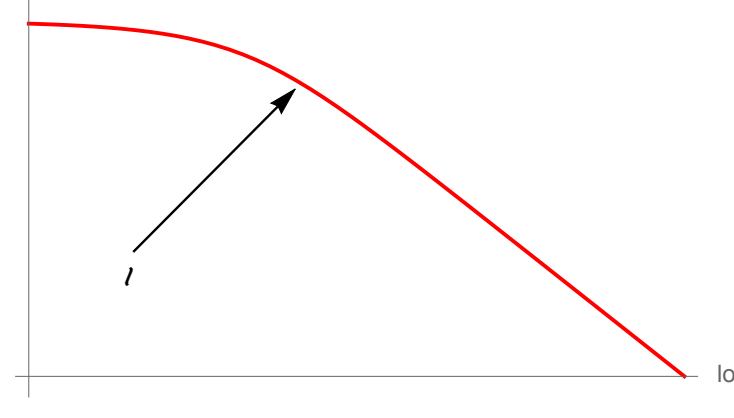

Figure 27.1: *The Karamata point where the slowly moving function is safely replaced by a constant $L(S) = l$. The constant varies whether we use the price $S$ or its geometric return –but not the asymptotic slope which corresponds to the tail index $\alpha$.*

Research chapter, with the Universa team: Brandon Yarckin, Chitpuneet Mann, Damir Delic, and Mark Spitznagel.

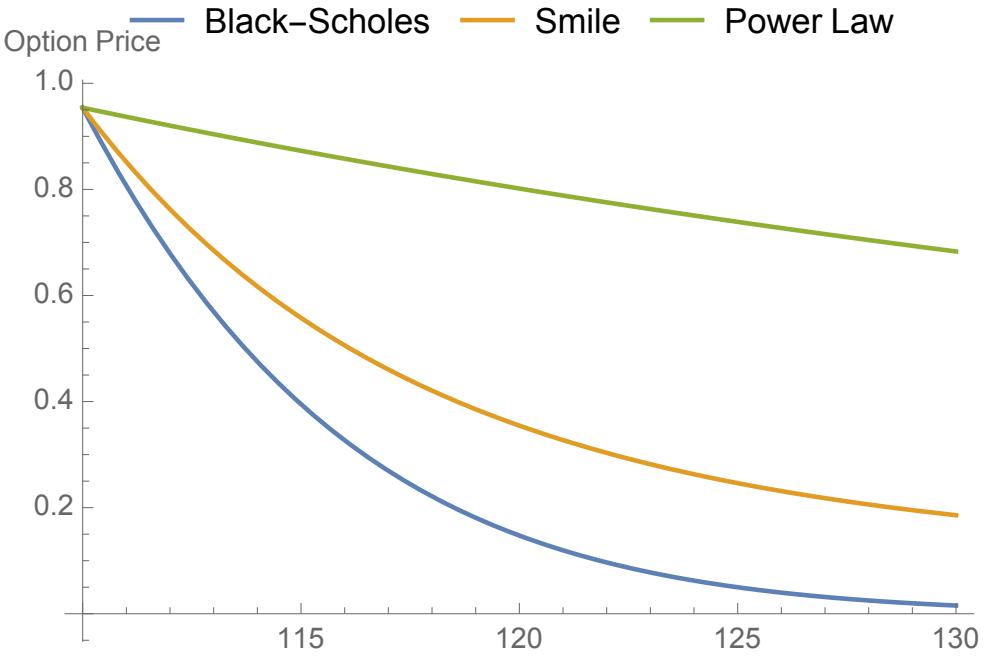

Figure 27.2: *We show a straight Black-Scholes option price (constant volatility), one with a volatility "smile", i.e. the scale increases in the tails, and power law option prices. Under the simplified case of a power law distribution for the underlying, option prices are linear to strike.*

## 27.1 INTRODUCTION

The power law class is conventionally defined by the property of the survival function, as follows. Let $X$ be a random variable belonging to the class of distributions with a "power law" right tail, that is:

$$\mathbb{P}(X > x) = L(x)\, x^{-\alpha} \tag{27.1}$$

where $L : [x_{\min}, +\infty) \to (0, +\infty)$ is a slowly varying function, defined as $\lim_{x\to+\infty} \frac{L(kx)}{L(x)} = 1$ for any $k > 0$ [26].

The survival function of $X$ is called to belong to the "regular variation" class $RV_\alpha$. More specifically, a function $f : \mathbb{R}^+ \to \mathbb{R}^+$ is index varying at infinity with index $\rho$ ($f \in RV_\rho$) when

$$\lim_{t\to\infty} \frac{f(tx)}{f(t)} = x^\rho.$$

More practically, there is a point where $L(x)$ approaches its limit, $l$, becoming a constant as in Figure 27.1–we call it the "Karamata constant". Beyond such value the tails for power laws are calibrated using such standard techniques as the Hill estimator. The distribution in that zone is dubbed the strong Pareto law by B. Mandelbrot [191],[88].

## 27.2 CALL PRICING BEYOND THE KARAMATA CONSTANT

Now define a European call price $C(K)$ with a strike $K$ and an underlying price $S$, $K, S \in (0, +\infty)$, as $(S - K)^+$, with its valuation performed under some probability measure $\mathbb{P}$, thus allowing us to price the option as $\mathbb{E}_P(S-K)^+ = \int_K^\infty (S-K)dP$. This allows us to immediately prove the following.

### 27.2.1 First approach, $S$ is in the regular variation class

We start with a simplified case, to build the intuition. Let $S$ have a survival function in the regular variation class $RV_\alpha$ as per 27.1. For all $K > l$ and $\alpha > 1$,

$$C(K) = \frac{K^{1-\alpha} l^\alpha}{\alpha - 1} \tag{27.2}$$

> **Remark 27.1**
> *We note that the parameter $l$, when derived from an existing option price, contains all necessary information about the probability distribution below $S = l$, which under a given $\alpha$ parameter makes it unnecessary to estimate the mean, the "volatility" (that is, scale) and other attributes.*

Let us assume that $\alpha$ is exogenously set (derived from fitting distributions, or, simply from experience, in both cases $\alpha$ is supposed to fluctuate minimally [285] ). We note that $C(K)$ is invariant to distribution calibrations and the only parameters needed $l$ which, being constant, disappears in ratios. Now consider as set the market price of an "anchor" tail option in the market is $C_m$ with strike $K_1$, defined as an option for the strike of which other options are priced in relative value. We can simply generate all further strikes from $l = \left((\alpha - 1) C_m K_1^{\alpha-1}\right)^{1/\alpha}$ and applying Eq. 27.2.

> **Result 1: Relative Pricing under Distribution for $S$**
> *For $K_1, K_2 \geq l$,*
> $$C(K_2) = \left(\frac{K_2}{K_1}\right)^{1-\alpha} C(K_1). \tag{27.3}$$

The advantage is that all parameters in the distributions are eliminated: all we need is the price of the tail option and the $\alpha$ to build a unique pricing mechanism.

> **Remark 27.2: Avoiding confusion about $L$ and $\alpha$**
> *The tail index $\alpha$ and Karamata constant $l$ should correspond to the assigned distribution for the specific underlying. A tail index $\alpha$ for $S$ in the regular variation class as as per 27.1 leading to Eq. 27.2 is different from that for $r = \frac{S - S_0}{S_0} \in RV_\alpha$ . For consistency, each should have its own Zipf plot and other representations.*
>
> 1. *If $\mathbb{P}(X > x) = L_a(x) x^{-\alpha}$, and $\mathbb{P}(\frac{X - X_0}{X_0} > \frac{x - X_0}{X_0}) = L_b(x) x^{-\alpha}$, the $\alpha$ constant will be the same, but the the various $L_{(.)}$ will be reaching their constant level at a different rate.*
> 2. *If $r_c = \log \frac{S}{S_0}$, it is not in the regular variation class, see theorem.*

The reason $\alpha$ stays the same is owing to the scale-free attribute of the tail index.

**Theorem 7: Log returns**

*Let $S$ be a random variable with survival function $\varphi(s) = L(s)s^{-\alpha} \in RV_\alpha$, where $L(.)$ is a slowly varying function. Let $r_l$ be the log return $r_l = \log \frac{s}{s_0}$. $\varphi_{r_l}(r_l)$ is not in the $RV_\alpha$ class.*

*Proof.* Immediate. The transformation $\varphi_{r_l}(r_l) = L(s)s^{-\frac{\log(\log^\alpha(s))}{\log(s)}}$. □

We note, however, that in practice, although we may need continuous compounding to build dynamics [275], our approach assumes such dynamics are contained in the anchor option price selected for the analysis (or $l$). Furthermore there is no tangible difference, outside the far tail, between $\log \frac{S}{S_0}$ and $\frac{S-S_0}{S_0}$.

### 27.2.2 Second approach, $S$ has geometric returns in the regular variation class

Let us now apply to real world cases where the returns $\frac{S-S_0}{S_0}$ are Paretian. Consider, for $r > l$, $S = (1+r)S_0$, where $S_0$ is the initial value of the underlying and $r \sim \mathcal{P}(l, \alpha)$ (Pareto I distribution) with survival function

$$\left(\frac{K-S_0}{lS_0}\right)^{-\alpha}, \; K > S_0(1+l) \tag{27.4}$$

and fit to $C_m$ using $l = \frac{(\alpha-1)^{1/\alpha} C_m^{1/\alpha} (K-S_0)^{1-\frac{1}{\alpha}}}{S_0}$, which, as before shows that practically all information about the distribution is embedded in $l$.

Let $\frac{S-S_0}{S_0}$ be in the regular variation class. For $S \geq S_0(1+l)$,

$$C(K, S_0) = \frac{(l\; S_0)^\alpha (K-S_0)^{1-\alpha}}{\alpha-1} \tag{27.5}$$

We can thus rewrite Eq. 27.3 to eliminate $l$:

**Result 2: Relative Pricing under Distribution for $\frac{S-S_0}{S_0}$**

*For $K_1, K_2 \geq (1+l)S_0$,*

$$C(K_2) = \left(\frac{K_2-S_0}{K_1-S_0}\right)^{1-\alpha} C(K_1). \tag{27.6}$$

**Remark 27.3**

*Unlike the pricing methods in the Black-Scholes modification class (stochastic and local volatility models, (see the expositions of Dupire, Derman and Gatheral, [87] [120], [72], finiteness of variance is not required for our model or option pricing in general, as shown in [275]. The only requirement is $\alpha > 1$, that is, finite first moment.*

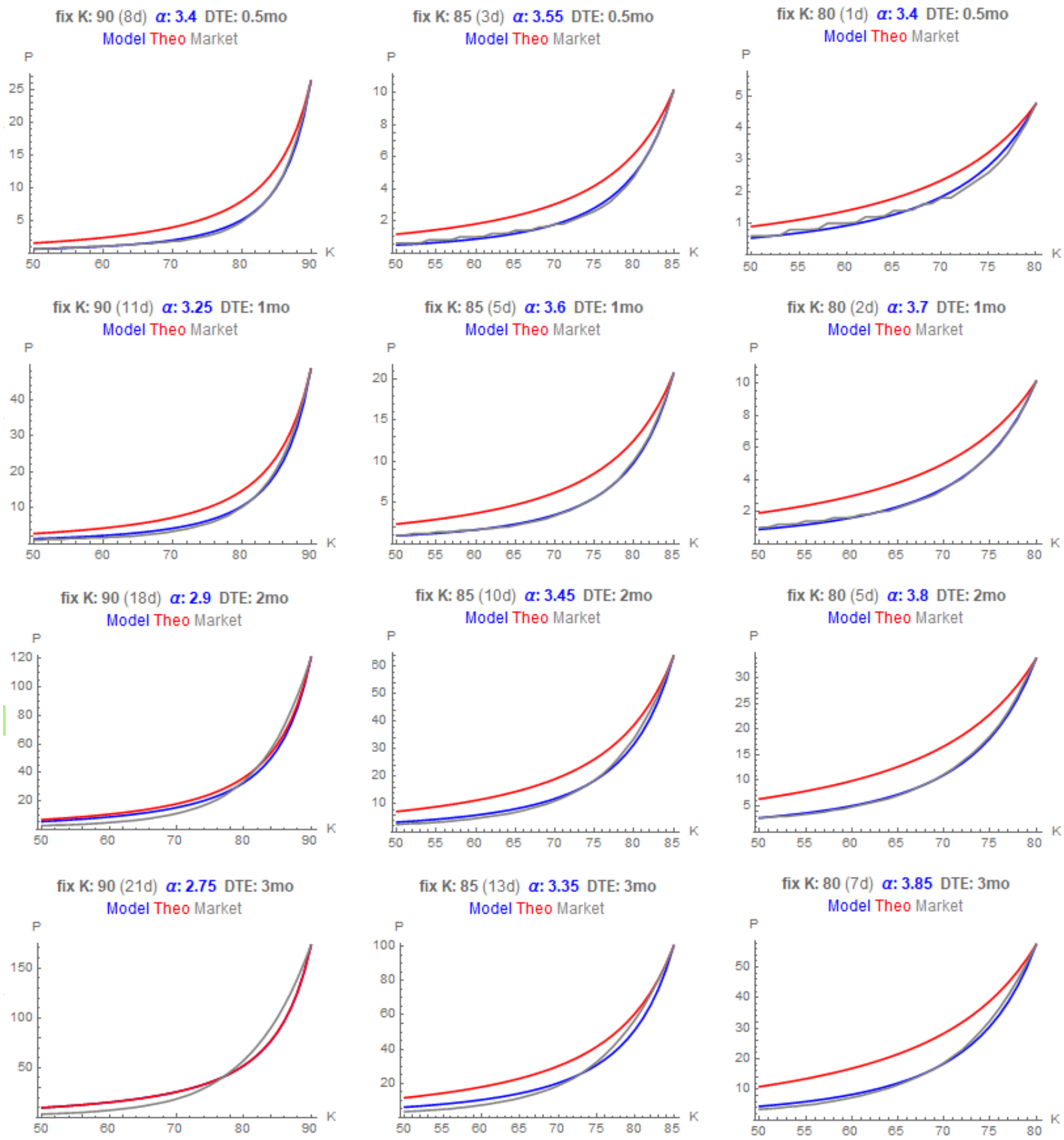

Figure 27.3: *Put Prices in the SP500 using "fix K" as anchor (from Dec 31, 2018 settlement), and generating an option prices using a tail index α that matches the market (blue) ("model), and in red prices for* $\alpha = 2.75$. *We can see that market prices tend to 1) fit a power law (matches stochastic volatility with fudged parameters), 2) but with an α that thins the tails. This shows how models claiming overpricing of tails are grossly misspecified.*

## 27.3 PUT PRICING

We now consider the put strikes (or the corresponding calls in the negative tail, which should be priced via put-call parity arbitrage). Unlike with calls, we can only consider the variations of $\frac{S-S_0}{S_0}$, not the logarithmic returns (nor those of $S$ taken separately).

We construct the negative side with a negative return for the underlying. Let $r$ be the rate of return $S = (1-r)S_0$, and Let $r > l > 0$ be Pareto distributed in the positive domain, with density $f_r(r) = \alpha\, l^\alpha r^{-\alpha-1}$. We have by probabilistic transformation and rescaling the PDF of the underlying:

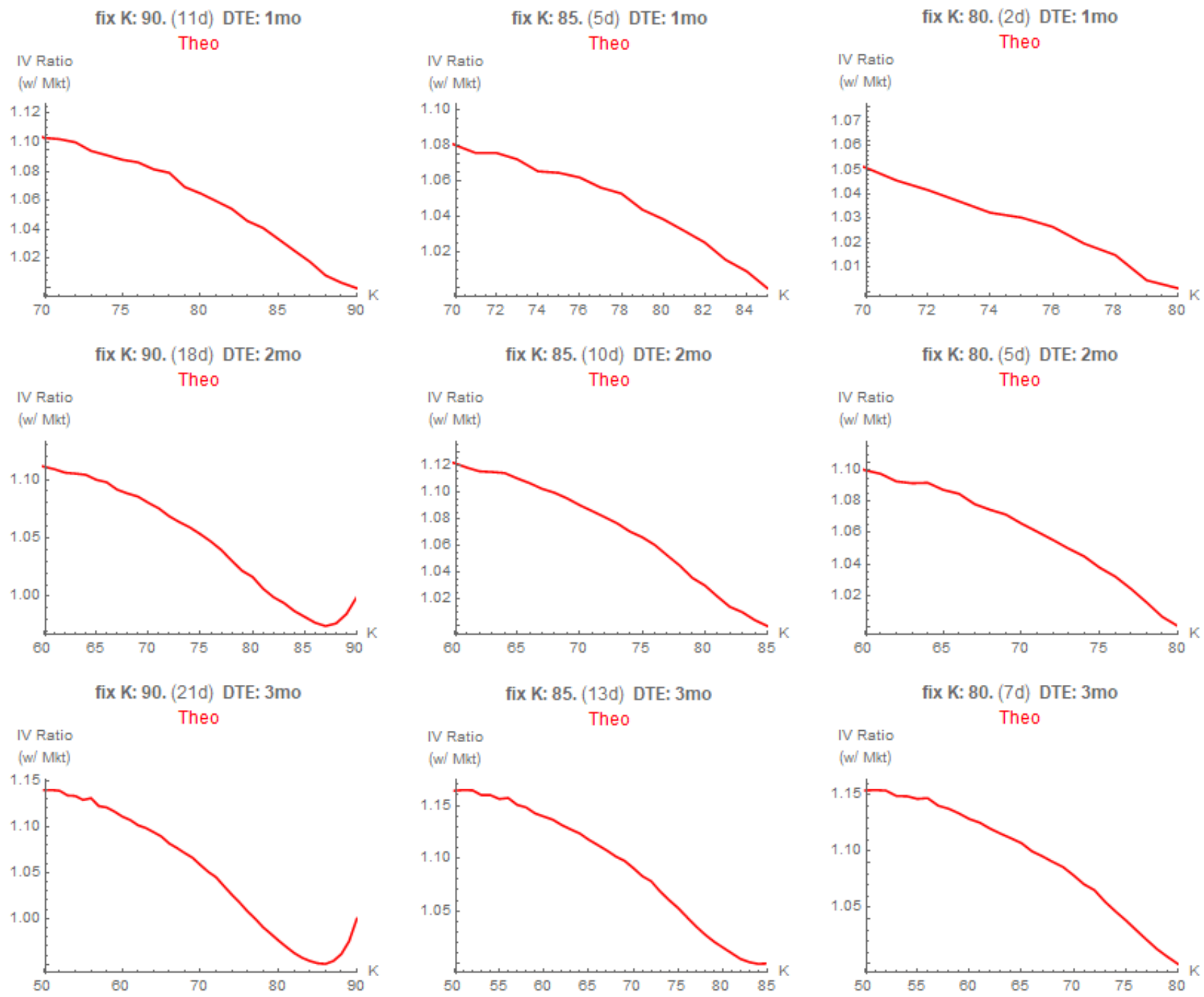

Figure 27.4: *Same results as in Fig 27.3 but expressed using implied volatility. We match the price to implied volatility for downside strikes (anchor* 90, 85, *and* 80*) using our model vs market, in ratios. We assume* $\alpha = 2.75$.

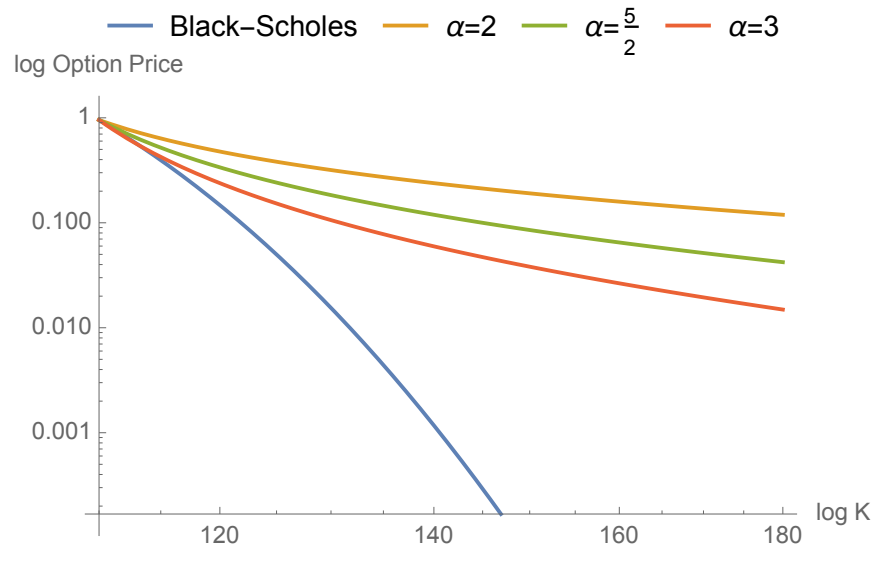

Figure 27.5: *The intuition of the Log log plot for the second calibration*

$$f_S(S) = -\frac{\alpha\left(-\frac{S-S_0}{l S_0}\right)^{-\alpha-1}}{l S_0}\lambda \qquad S \in [0, (1-l)S_0)$$

where the scaling constant $\lambda = \left(\frac{1}{(-1)^{\alpha+1}(l^{\alpha}-1)}\right)$ is set in a way to get $f_s(S)$ to integrate to 1. The parameter $\lambda$, however, is close to 1, making the correction negligible, in applications where $\sigma\sqrt{t} \leq \frac{1}{2}$ ($\sigma$ being the Black-Scholes equivalent implied volatility and $t$ the time to expiration of the option).

Remarkably, both the parameters $l$ and the scaling $\lambda$ are eliminated.

**Result 3: Put Pricing**

*For* $K_1, K_2 \leq (1-l)S_0$,

$$P(K_2) = P(K_1) \frac{(K_2 - S_0)^{1-\alpha} - S_0^{1-\alpha}\left((\alpha-1)K_2 + S_0\right)}{(K_1 - S_0)^{1-\alpha} - S_0^{1-\alpha}\left((\alpha-1)K_1 + S_0\right)} \tag{27.7}$$

## 27.4 ARBITRAGE BOUNDARIES

Obviously, there is no arbitrage for strikes higher than the baseline one $K_1$ in previous equations. For we can verify the Breeden-Litzenberger result [36], where the density is recovered from the second derivative of the option with respect to the strike $\frac{\partial^2 C(K)}{\partial K^2}|_{K \geq K_1} = \alpha K^{-\alpha-1} L^{\alpha} \geq 0$.

However there remains the possibility of arbitrage between strikes $K_1 + \Delta K$, $K_1$, and $K_1 - \Delta K$ by violating the following boundary: let $BSC(K, \sigma(K))$ be the Black-Scholes value of the call for strike $K$ with volatility $\sigma(K)$ a function of the strike and $t$ time to expiration. We have

$$C(K_1 + \Delta K) + BSC(K_1 - \Delta K) \geq 2\, C(K_1), \tag{27.8}$$

where $BSC(K_1, \sigma(K_1)) = C(K_1)$. For inequality 27.8 to be satisfied, we further need an inequality of call spreads, taken to the limit:

$$\frac{\partial BSC(K, \sigma(K))}{\partial K}|_{K=K_1} \geq \frac{\partial C(K)}{\partial K}|_{K=K_1} \tag{27.9}$$

Such an arbitrage puts a lower bound on the tail index $\alpha$. Assuming 0 rates to simplify:

$$\alpha \geq \frac{1}{-\log(K - S_0) + \log(l) + \log(S_0)}$$
$$\log\left(\frac{1}{2}\operatorname{erfc}\left(\frac{t\sigma(K)^2 + 2\log(K) - 2\log(S_0)}{2\sqrt{2}\sqrt{t}\sigma(K)}\right)\right.$$
$$\left. - \frac{\sqrt{S_0}\sqrt{t}\sigma'(K) K^{\frac{\log(S_0)}{t\sigma(K)^2} + \frac{1}{2}} \exp\left(-\frac{\log^2(K) + \log^2(S_0)}{2t\sigma(K)^2} - \frac{1}{8}t\sigma(K)^2\right)}{\sqrt{2\pi}}\right) \tag{27.10}$$

## 27.5 COMMENTS

As we can see in Figure 27.5, stochastic volatility models and similar adaptations (say, jump-diffusion or standard Poisson variations) eventually fail "out in the tails" outside the zone for which they were calibrated. There has been poor attempts to extrapolate the option prices using a fudged thin-tailed probability distribution rather than a Paretian one –hence the numerous claims in the finance literature on "overpricing" of tail options combined with some psycholophastering on "dread risk" are unrigorous on that basis. The proposed methods allow us to approach such claims with more realism.

Finaly, note that our approach isn't about absolute mispricing of tail options, but relative to a given strike closer to the money.

## ACKNOWLEDGMENTS

Bruno Dupire, Peter Carr, students at NYU Tandon School of Engineering.

# 28 FOUR MISTAKES IN QUANTITATIVE FINANCE*,‡

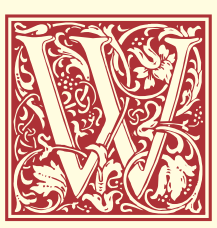
E DISCUSS Jeff Holman's comments in *Quantitative Finance* to illustrate four critical errors students should learn to avoid. (Holman was, surprisingly, at the time, a senior risk officer for a large hedge fund.) :

1. Mistaking tails (4th moment and higher) for volatility (2nd moment)
2. Missing Jensen's Inequality when calculating return potential
3. Analyzing the hedging results without the performance of the underlying
4. The necessity of a numéraire in finance.

The review of *Antifragile* by Mr Holman (Dec 4, 2013) is replete with factual, logical, and analytical errors. We will only list here the critical ones, and ones with generality to the risk management and quantitative finance communities; these should be taught to students in quantitative finance as central mistakes to avoid, so beginner quants and risk managers can learn from these fallacies.

## 28.1 CONFLATION OF SECOND AND FOURTH MOMENTS

It is critical for beginners not to fall for the following elementary mistake. Mr Holman gets the relation of the VIX (volatility contract) to betting on "tail events" backwards. Let us restate the notion of "tail events" (we saw earlier in the book): it means a disproportionate role of the tails in determining the properties of distribution, which, mathematically, means a smaller one for the "body".[2]

Mr Holman seems to get the latter part of the attributes of fattailedness in reverse. It is an error to mistake the VIX for tail events. The VIX is mostly affected by

Discussion chapter.

[2] The point is staring at every user of spreadsheets: kurtosis, or scaled fourth moment, the standard measure of fattailedness, entails normalizing the fourth moment by the square of the variance.

at-the-money options which corresponds to the center of the distribution, closer to the second moment not the fourth (at-the-money options are actually linear in their payoff and correspond to the conditional first moment). As explained about seventeen years ago in *Dynamic Hedging* (Taleb, 1997) (see appendix), in the discussion on such tail bets, or "fourth moment bets", betting on the disproportionate role of tail events of fattailedness is done by *selling* the around-the-money-options (the VIX) and purchasing options in the tails, in order to extract the second moment and achieve neutrality to it (sort of becoming "market neutral"). Such a neutrality requires some type of "short volatility" in the body because higher kurtosis means lower action in the center of the distribution.

A more mathematical formulation is in the technical version of the *Incerto* : fat tails means "higher peaks" for the distribution as, the fatter the tails, the more markets spend time between $\mu - \sqrt{\frac{1}{2}\left(5-\sqrt{17}\right)}\sigma$ and $\mu + \sqrt{\frac{1}{2}\left(5-\sqrt{17}\right)}\sigma$ where $\sigma$ is the standard deviation and $\mu$ the mean of the distribution (we used the Gaussian here as a base for ease of presentation but the argument applies to all unimodal distributions with "bell-shape" curves, known as semiconcave). And "higher peaks" means less variations that are not tail events, more quiet times, not less. For the consequence on option pricing, the reader might be interested in a quiz I routinely give students after the first lecture on derivatives: "What happens to at-the-money options when one fattens the tails?", the answer being that they should drop in value. [3]

Effectively, but in a deeper argument, in the QF paper (Taleb and Douady 2013), our measure of fragility has an opposite sensitivity to events around the center of the distribution, since, by an argument of survival probability, what is fragile is sensitive to tail shocks and, critically, should not vary in the body (otherwise it would be broken).

## 28.2 MISSING JENSEN'S INEQUALITY IN ANALYZING OPTION RETURNS

Here is an error to avoid at all costs in discussions of volatility strategies or, for that matter, anything in finance. Mr Holman seems to miss the existence of Jensen's inequality, which is the entire point of owning an option, a point that has been belabored in *Antifragile*. One manifestation of missing the convexity effect is a critical miscalculation in the way one can naively assume options respond to the VIX.

> "A $1 investment on January 1, 2007 in a strategy of buying and rolling short-term VIX futures would have peaked at $4.84 on November 20, 2008 -and then

[3] **Technical Point: Where Does the Tail Start?** As we saw in 4.3, for a general class of symmetric distributions with power laws, the tail starts at: $\pm \frac{\sqrt{\frac{5\alpha+\sqrt{(\alpha+1)(17\alpha+1)}+1}{\alpha-1}}\, s}{\sqrt{2}}$, with $\alpha$ infinite in the stochastic volatility Gaussian case and $s$ the standard deviation. The "tail" is located between around 2 and 3 standard deviations. This flows from the heuristic definition of fragility as second order effect: the part of the distribution is convex to errors in the estimation of the scale. But in practice, because historical measurements of STD will be biased lower because of small sample effects (as we repeat fat tails accentuate small sample effects), the deviations will be > 2-3 STDs.

subsequently lost 99% of its value over the next four and a half years, finishing under $0.05 as of May 31, 2013." [4]

This mistake in the example given underestimates option returns by up to... several orders of magnitude. Mr Holman analyzes the performance a tail strategy using investments in financial options by using the VIX (or VIX futures) as proxy, which is mathematically erroneous owing to second- order effects, as the link is tenuous (it would be like evaluating investments in ski resorts by analyzing temperature futures). Assume a periodic rolling of an option strategy: an option 5 STD away from the money [5] gains 16 times in value if its implied volatility goes up by 4, but only loses its value if volatility goes to 0. For a 10 STD it is 144 times. And, to show the acceleration, assuming these are traded, a 20 STD options by around 210,000 times[6]. There is a second critical mistake in the discussion: Mr Holman's calculations here exclude the payoff from actual in-the-moneyness.

One should remember that the VIX is not a price, but an inverse function, an index derived from a price: one does not buy "volatility" like one can buy a tomato; operators buy options correponding to such inverse function and there are severe, very severe nonlinearities in the effect. Although more linear than tail options, the VIX is still convex to actual market volatility, somewhere between variance and standard deviation, since a strip of options spanning all strikes should deliver the variance (Gatheral,2006). The reader can go through a simple exercise. Let's say that the VIX is "bought" at 10% -that is, the component options are purchased at a combination of volatilities that corresponds to a VIX at that level. Assume returns are in squares. Because of nonlinearity, the package could benefit from an episode of 4% volatility followed by an episode of 15%, for an average of 9.5%; Mr Holman believes or wants the reader to believe that this 0.5 percentage point should be treated as a loss when in fact second order un-evenness in volatility changes are more relevant than the first order effect.

## 28.3 THE INSEPARABILITY OF INSURANCE AND INSURED

One should never calculate the cost of insurance without offsetting it with returns generated from packages than one would not have purchased otherwise.

Even had he gotten the sign right on the volatility, Mr Holman in the example above analyzes the performance of a strategy buying options to protect a tail event without adding the performance of the portfolio itself, like counting the cost side of the insurance without the performance of what one is insuring that would not have been bought otherwise. Over the same period he discusses the market rose more than 100%: a healthy approach would be to compare dollar-for-dollar what an investor would have done (and, of course, getting rid of this "VIX" business and focusing on very small dollars invested in tail options that would allow such an aggressive stance). Many investors (such as this author) would have stayed

[4] In the above discussion Mr Holman also shows evidence of dismal returns on index puts which, as we said before, respond to volatility not tail events. These are called, in the lingo, "sucker puts".

[5] We are using implied volatility as a benchmark for its STD.

[6] An event this author witnessed, in the liquidation of Victor Niederhoffer, options sold for $.05 were purchased back at up to $38, which bankrupted Refco, and, which is remarkable, without the options getting close to the money: it was just a panic rise in implied volatility.

out of the market, or would not have added funds to the market, without such an insurance.

## 28.4 THE NECESSITY OF A NUMÉRAIRE IN FINANCE

There is a deeper analytical error.

A barbell is defined as a bimodal investment strategy, presented as investing a portion of your portfolio in what is explicitly defined as a "numéraire repository of value" (*Antifragile*), and the rest in risky securities (*Antifragile* indicates that such numéraire would be, among other things, inflation protected). Mr Holman goes on and on in a nihilistic discourse on the absence of such riskless numéraire (of the sort that can lead to such sophistry as "he is saying one is safer on *terra firma* than at sea, but what if there is an earthquake?").

The familiar Black and Scholes derivation uses a riskless asset as a baseline; but the literature since around 1977 has substituted the notion of "cash" with that of a numéraire , along with the notion that one can have different currencies, which technically allows for changes of probability measure. A numéraire is defined as *the unit to which all other units relate*. ( Practically, the numéraire is a basket the variations of which do not affect the welfare of the investor.) Alas, without numéraire, there is no probability measure, and no quantitative in quantitative finance, as one needs a unit to which everything else is brought back to. In this (emotional) discourse, Mr Holton is not just rejecting the barbell per se, but any use of the expectation operator with any economic variable, meaning he should go attack the tens of thousand research papers and the existence of the journal *Quantitative Finance* itself.

Clearly, there is a high density of other mistakes or incoherent statements in the outpour of rage in Mr Holman's review; but I have no doubt these have been detected by the *Quantitative Finance* reader and, as we said, the object of this discussion is the prevention of analytical mistakes in quantitative finance.

To conclude, this author welcomes criticism from the finance community that are not straw man arguments, or, as in the case of Mr Holmam, violate the foundations of the field itself.

## 28.5 APPENDIX (BETTING ON TAILS OF DISTRIBUTION)

From *Dynamic Hedging*, pages 264-265:

> *A fourth moment bet is long or short the volatility of volatility. It could be achieved either with out-of-the-money options or with calendars. Example: A ratio "backspread" or reverse spread is a method that includes the buying of out-of-the-money options in large amounts and the selling of smaller amounts of at-the-money but making sure the trade satisfies the "credit" rule (i.e., the trade initially generates a positive cash flow). The credit rule is more difficult to interpret when one uses in-the-money options. In that case, one should deduct the present value of the intrinsic part of every option using the put-call parity rule to equate them with out-of-the-money.*

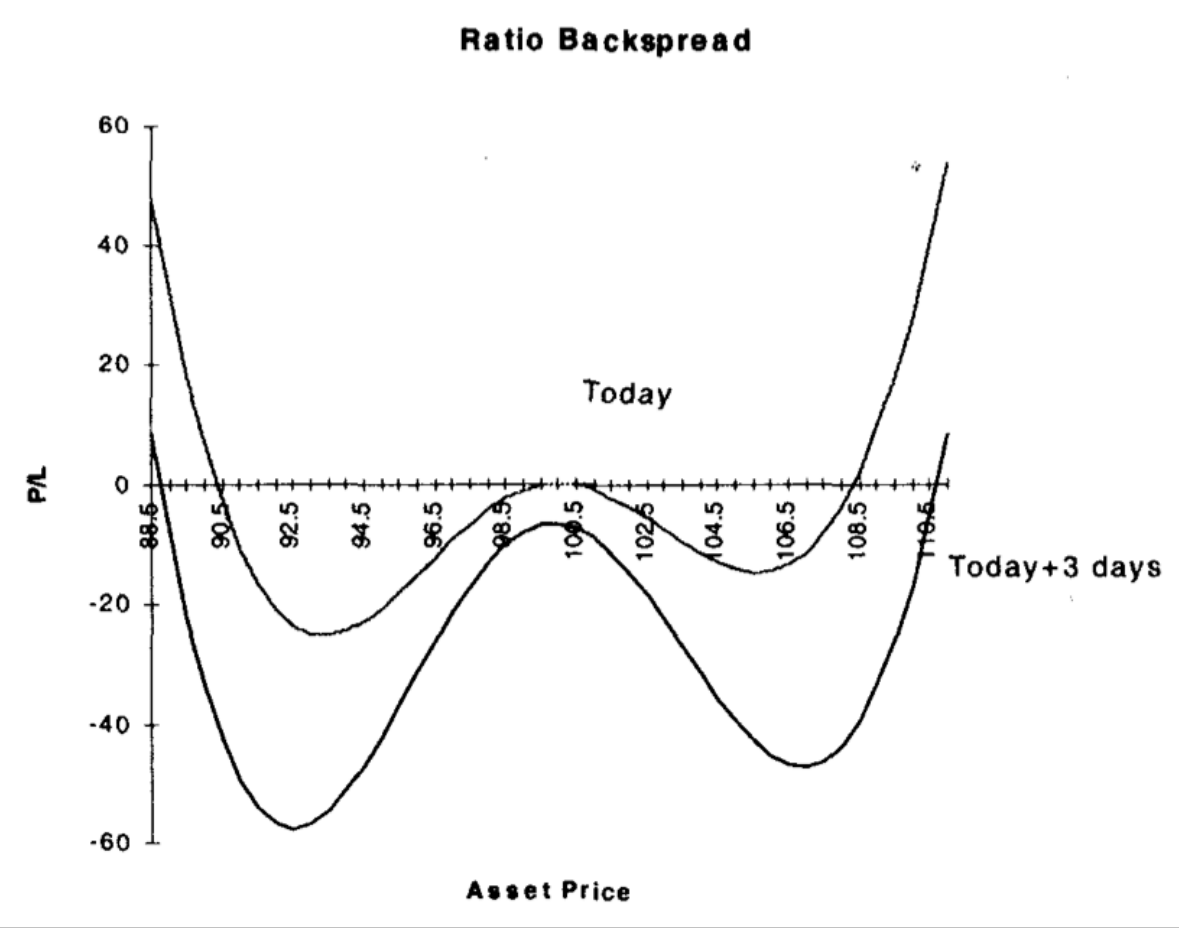

Figure 28.1: *First Method to Extract the Fourth Moment, from Dynamic Hedging, 1997.*

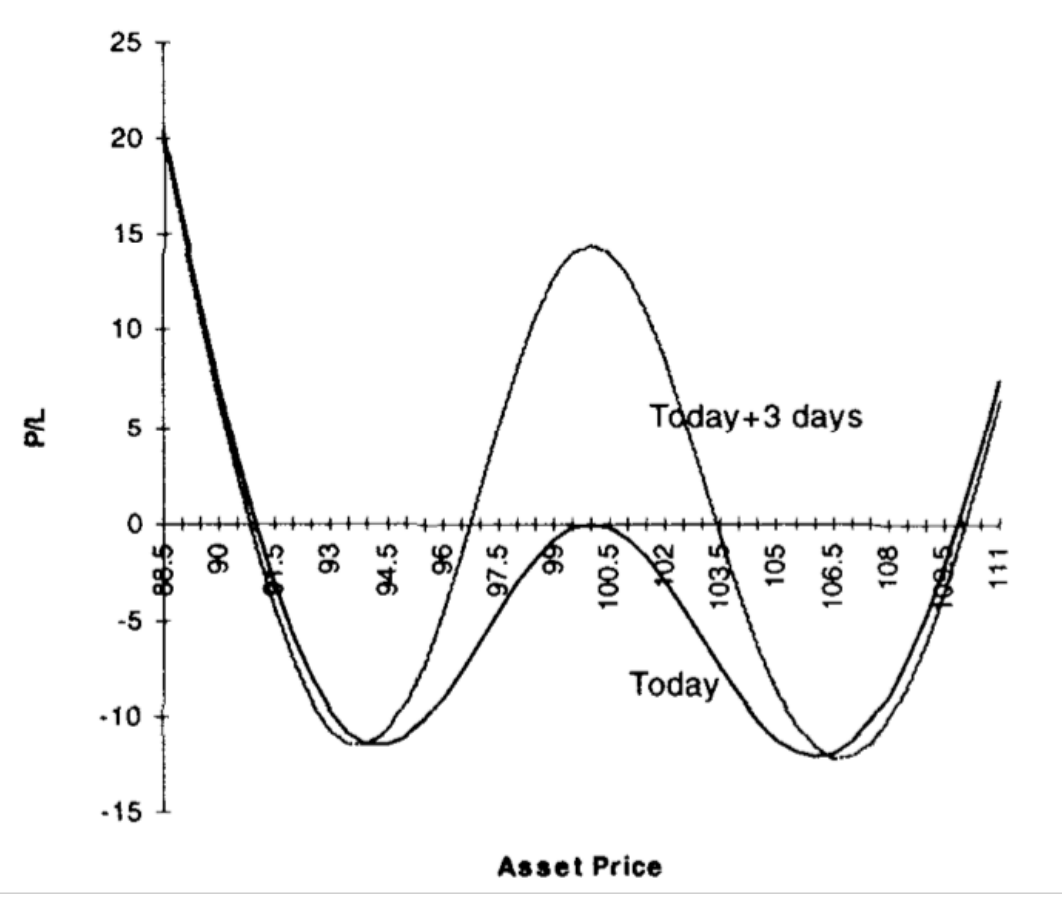

Figure 28.2: *Second Method to Extract the Fourth Moment , from Dynamic Hedging, 1997.*

*The trade shown in Figure 28.1 was accomplished with the purchase of both out-of-the-money puts and out-of-the-money calls and the selling of smaller amounts of at-the-money straddles of the same maturity.*

*Figure 28.2 shows the second method, which entails the buying of 60- day options in some amount and selling 20-day options on 80% of the amount. Both trades show the position benefiting from the fat tails and the high peaks. Both trades, however, will have different vega sensitivities, but close to flat modified vega.*

See The Body, The Shoulders, and The Tails from section 4.3 where we assume tails start at the level of convexity of the segment of the probability distribution to the scale of the distribution.

# 29 PORTFOLIOS SHOULD NEVER RELY ON CORRELATION

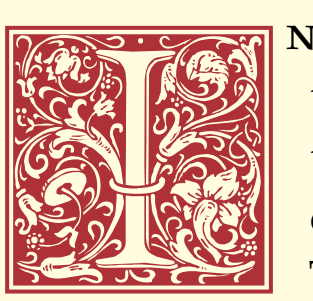

n this (research) chapter we address the practical implication of a central flaw in the application of theoretical finance to the construction of portfolios, linked to non-ellipticality discussed in Chapter 6, with application to pension funds.

The central decision for such a fund with pension responsibilities is the allocation between stocks and bonds, often relying, for intellectual backup, on metrics and methods from Modern Portfolio Theory (MPT). We show how, historically, such an "optimal" portfolio is in effect the *least* optimal one, as it fails to protect against tail risk and under-allocates to the high-returning asset class. MPT fails in both risk control and real-world investment optimization.

## MODERN PORTFOLIO THEORY

Modern Portfolio Theory (MPT) originated about 70 years ago with the publication of "Portfolio Selection" , by Harry Markowitz [195]. The mean-variance tools of analysis accompanying the Markowitz framework was a cornerstone supporting the Capital Asset Pricing Model (CAPM), which intended to explain the relationship between systemic risk and asset returns. In spite of its adoption in business schools, CAPM has never been supported by empirical evidence –in fact changing assumptions about joint asset returns makes the mathematical framework crumble. Accordingly, it is not our intent to make the case here. Instead, we consider the pitfalls of MPT when applied by pension funds in asset allocation and portfolio management, focusing specifically on the role of correlation.

Why is this of concern? Blind deference to MPT, which we repeat made neither empirical nor theoretical sense, caused most institutional investors to misallocate assets. The average funded status of US state-managed pension plans dropped from almost 100% in 2001 to about 74% in 2019. With combined liabilities of

Research chapter, with Ron Lagnado at Universa Investments.

over $4 trillion, the shortfall stands grimly at over $1 trillion despite the longest economic expansion in U.S. history spanning over a decade from 2009 to 2020.

MPT with its central tool of mean-variance analysis supposedly provides a formal argument for the benefits of diversification in investing. It leads investors seeking to determine the so-called "optimal" allocation to broad asset classes in top-down portfolio construction. But there are two major fallacies.

The first fallacy is that it is possible to forecast asset returns, volatilities, and correlations with sufficient accuracy ex ante to facilitate the optimal allocation and the maximum risk-adjusted return ex post.

**Remark 29.1: First Fallacy of MPT**

*Future returns and correlation —the two central parameters in MPT —are not observable in the real world.*

The second misconception is that attaining the objective of maximizing risk-adjusted returns, i.e., maximizing returns subject to a constraint on volatility, is beneficial. Does suppression of volatility across all market environments provide cost effective protection against tail risk and ultimately better long-run compounded returns? Clearly, it does not.

**Remark 29.2: Second Fallacy of MPT**

*Volatility of returns (as expressed in portfolio standard deviation) does not map to risk.*

## CORRELATION AND RISK PARITY

The notion of stable and estimable expected returns is the easiest aspect of MPT to debunk. Nevertheless, extensions of the theory to risk parity strategies obviate the estimation of expected return but still require accuracy in estimating risk parameters.

The key risk parameter – correlation –is neither robust nor reliable a measure of dependence when asset returns have fat-tailed distributions or when the relationship between the variables is not linear —both cases seem to apply to stocks and bonds. Let's never ignore that stocks and bonds have patently non-Gaussian distributions, and the relationship between the two variables, when regressed, exhibits a severe nonlinearity.

**Remark 29.3: Correlation and association**

*Correlation, even if predictable (it's not), is a poor measure of association between nonGaussian variables or in the presence of nonlinear dependence.*

To illustrate the problems associated with application of correlation in MPT, consider a risk averse investor who constructs a portfolio with allocation to the S&P

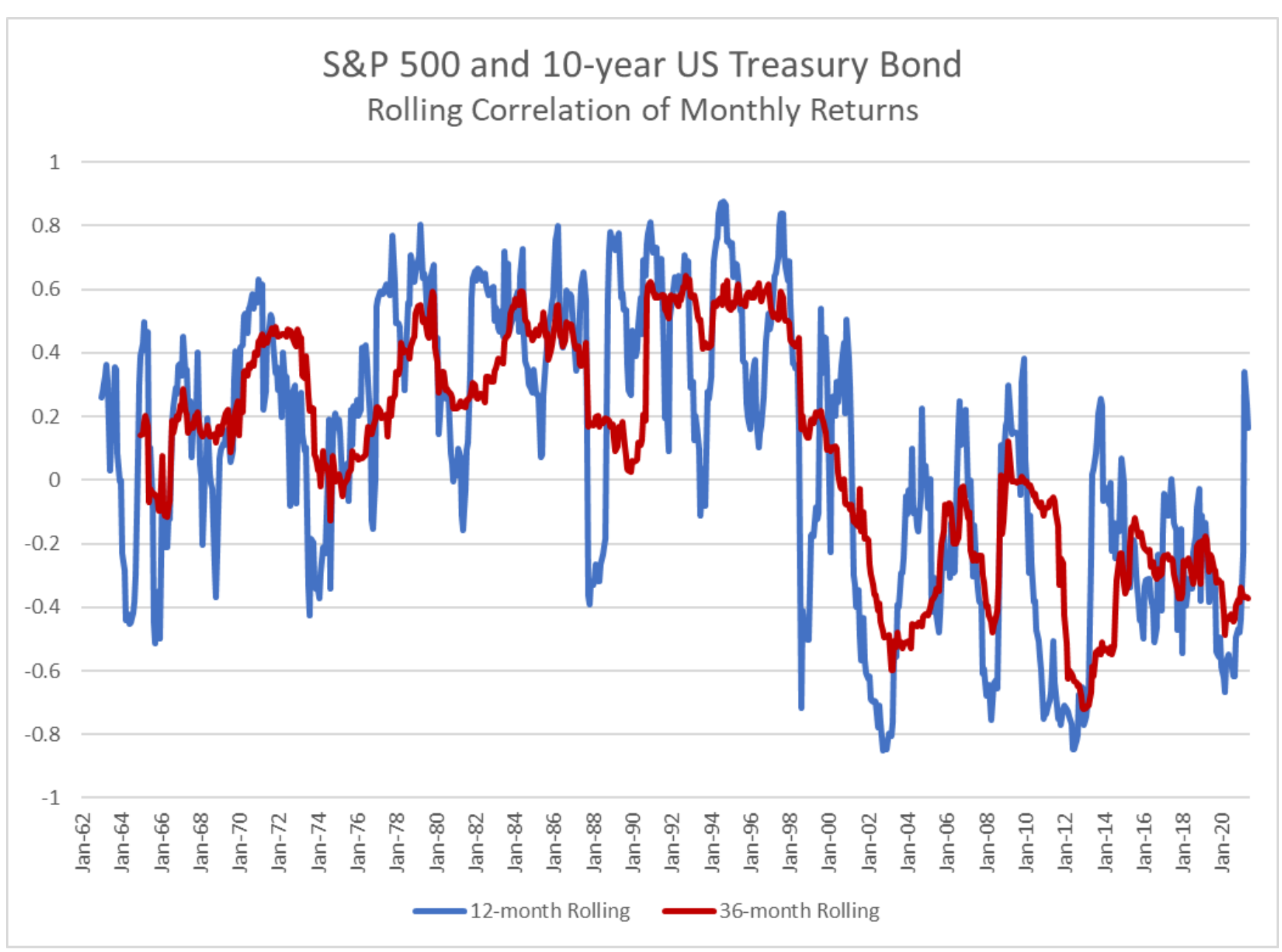

Figure 29.1: *Historical correlation swings between* $-1$ *and* $1$*; there is less than* $10^{-12}$ *chance that these variations would come from simple sampling error under standard assumptions. The technical appendix explains why this is sufficient to invalidate the risk-reduction claims of MPT.*

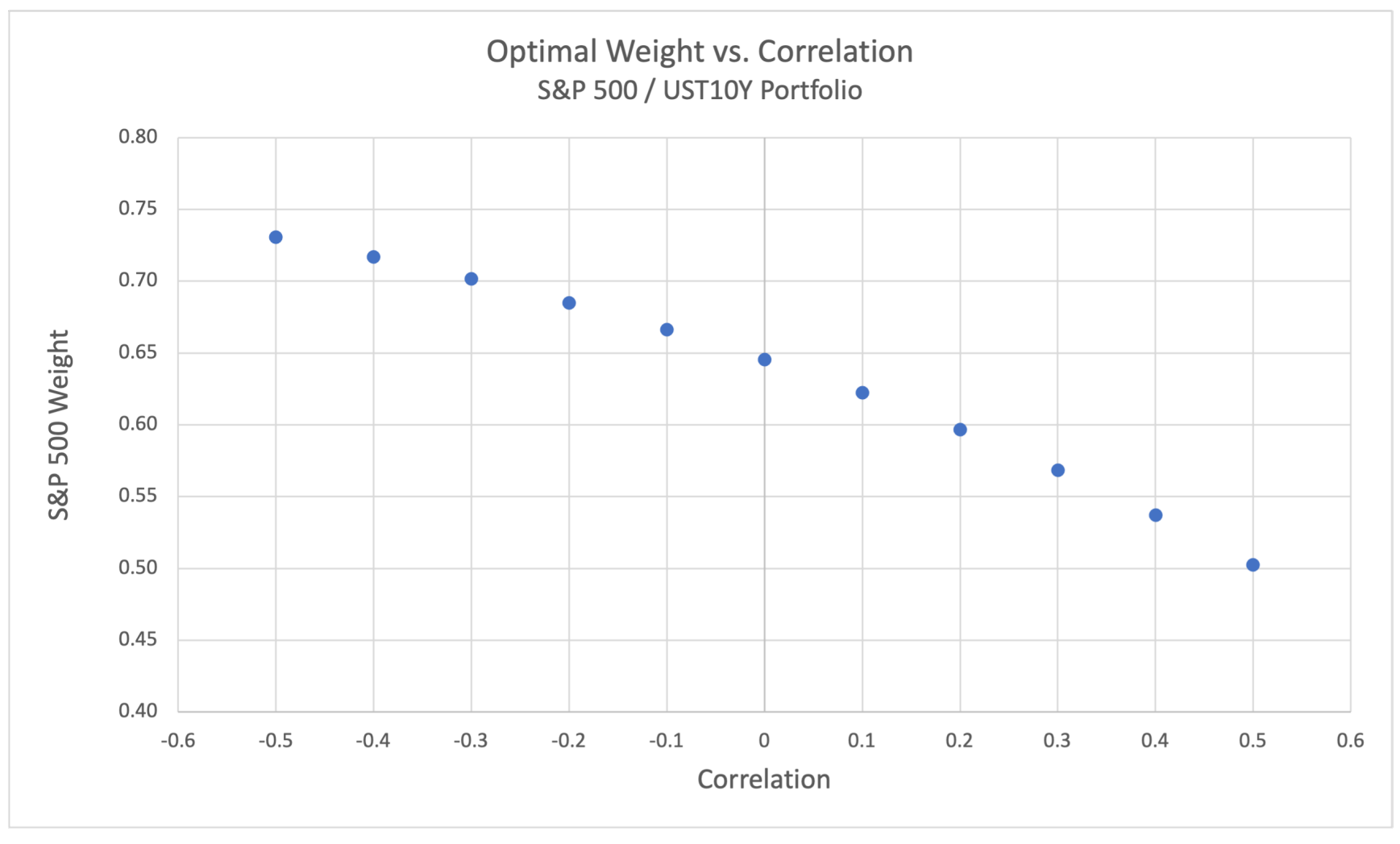

Figure 29.2: *Optimal weights.*

500 index and a U.S 10-year Treasury Bond index. Over the approximately 60-year period from January 1962 to June 2021, the annualized total return and volatility for the S&P 500 was 11% and 14.9%, respectively. The return and volatility of the constant maturity U.S. 10-year Treasury Bond were 7.1% and 7.8%, respectively. (Note that since we believe that, owing to the nonGaussian attributes of the returns, standard deviation is not a fit measure of expected future variations, we will only use it in this exercise insofar as it is illustrative of that specific shortcoming.)

Figure 1 shows the correlation of monthly returns over this period with 12 and 36-month rolling windows. First, we can verify that correlation is far from being constant, violating the requirements for hedging and risk reduction. Second, more technically, it does appear to have a random structure, making the joint distribution between the two variables non-elliptical, thus violating a central assumption at the foundation of portfolio theory. The mathematical consequence is that correlation can be deemed, at best, as a non-informative metric and an unreliable indicator of the association. See Taleb (2020) [286] for the mathematical derivations.

Suppose the investor has perfect foresight with respect to future returns and volatilities, and both assets follow the 1962-2021 trajectories over the subsequent 60-year period. The initial allocation (with full investment and no leverage) is determined by maximizing expected return subject to a typical 10% volatility constraint; the portfolio is rebalanced monthly with zero cost. The optimal weight for the S&P 500 as a function of a correlation estimate ranging from $-0.5$ to +0.5 is shown in Figure 2. Depending on the ex-ante assumption for correlation the investor would have selected a portfolio ranging from 73/27 to 50/50 in composition with the conviction that the "optimal" choice had been made.

Table 1 presents the range in realized return over the subsequent 60-year period shown in Figure 3; we can see extremely widespread outcomes. The accumulated wealth would have been about 1.5 times greater if the correlation had been assumed to be $-0.5$ rather than 0.5. Assuming that bonds are consistently negatively correlated to stocks would have resulted in a higher stock allocation of 73% versus 50% (and a better outcome). However, the average realized correlation between stocks and bonds during the decades prior to the DotCom Bubble collapse in 2001-2002 was about 0.3. There would have been no empirical grounds to support an assumption of negative correlation, and optimization would have resulted in an under-allocation to stocks.

To summarize, all portfolio allocations represented in Table 1 are mean-variance "optimal" for different assumed correlations. Choosing a low correlation of $-0.5$ enabled the investor to reap the greatest reward on this path through time despite the worst risk-adjusted return and a volatility exceeding the target of 10%. Choosing a high correlation of 0.5 (supported by so-called empirical evidence) resulted in the best "risk adjusted" return but the worst cumulative gain. This sensitivity to an unpredictable parameter — correlation — highlights the fallacy of the label portfolio "optimization" and utter uselessness of MPT.

Table 29.1: *Varying Correlation*

| Assumed Correlation | Equity Weight | Compound Return | Annualized Return | Vol. | Ret/Vol |
|---|---|---|---|---|---|
| −0.5 | 0.73 | 247.63 | 0.1 | 0.11 | 0.86 |
| −0.4 | 0.72 | 242.57 | 0.1 | 0.11 | 0.87 |
| −0.3 | 0.7 | 237.03 | 0.1 | 0.11 | 0.88 |
| −0.2 | 0.68 | 230.95 | 0.1 | 0.11 | 0.9 |
| −0.1 | 0.67 | 224.29 | 0.1 | 0.1 | 0.91 |
| 0. | 0.65 | 216.98 | 0.09 | 0.1 | 0.93 |
| 0.1 | 0.62 | 208.96 | 0.09 | 0.1 | 0.94 |
| 0.2 | 0.6 | 200.19 | 0.09 | 0.1 | 0.96 |
| 0.3 | 0.57 | 190.65 | 0.09 | 0.09 | 0.98 |
| 0.4 | 0.54 | 180.37 | 0.09 | 0.09 | 1.01 |
| 0.5 | 0.5 | 169.44 | 0.09 | 0.09 | 1.03 |

## CHASING CORRELATION

Would a real pension fund manager simply allocate based on a point estimate of correlation and maintain the allocation indefinitely? As preposterous as that may sound, it has been the actual practice at some funds that subscribe to the 60/40 approach to portfolio construction. Alternatively, it would seem sensible to update correlation and other parameters periodically and recompute the "optimal" allocation.

Pension funds typically perform a review of asset allocation and revise the apportionment every 2 to 5 years. Suppose instead of maintaining a constant assumed correlation, the investor estimates correlation using historical data in a 3-year lookback period and recomputes and maintains the mean-variance optimal allocation for the subsequent 3-year period. The history of the estimated correlation over the period from 1965 to 2021 is shown in Figure 4. Figure 5 then compares the performance of the dynamic "optimal" portfolio using 3-year lookback correlation with static portfolios based on assumptions of +0.3 and −0.3 correlation —average values in sub-periods before and after the year 2000. The performance of the dynamic portfolio is no better and, in fact, much worse than would be attained under assumptions leading to higher equity allocations.

## CONCLUSION

The single most consequential investment decision for a pension is the allocation between stocks and bonds (or more generally between growth and income investments). The influence of MPT on pension fund managers is apparent in that a 60/40 stock-bond portfolio is representative of pension funds on average. Correlation is the driving force behind the allocation decision — while both the mathematical and empirical justifications of such an application have been lacking.

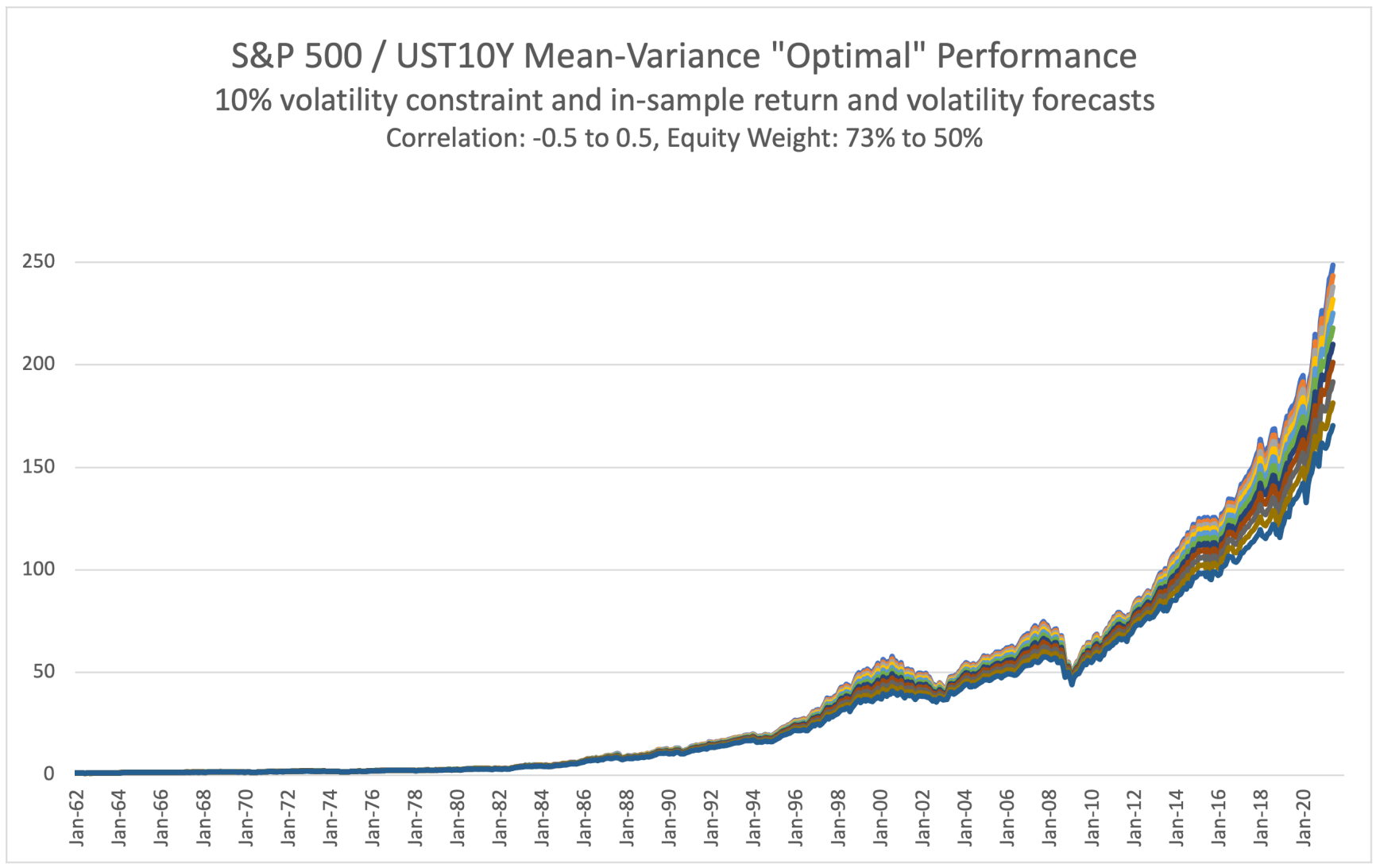

Figure 29.3: *Resulting Performance.*

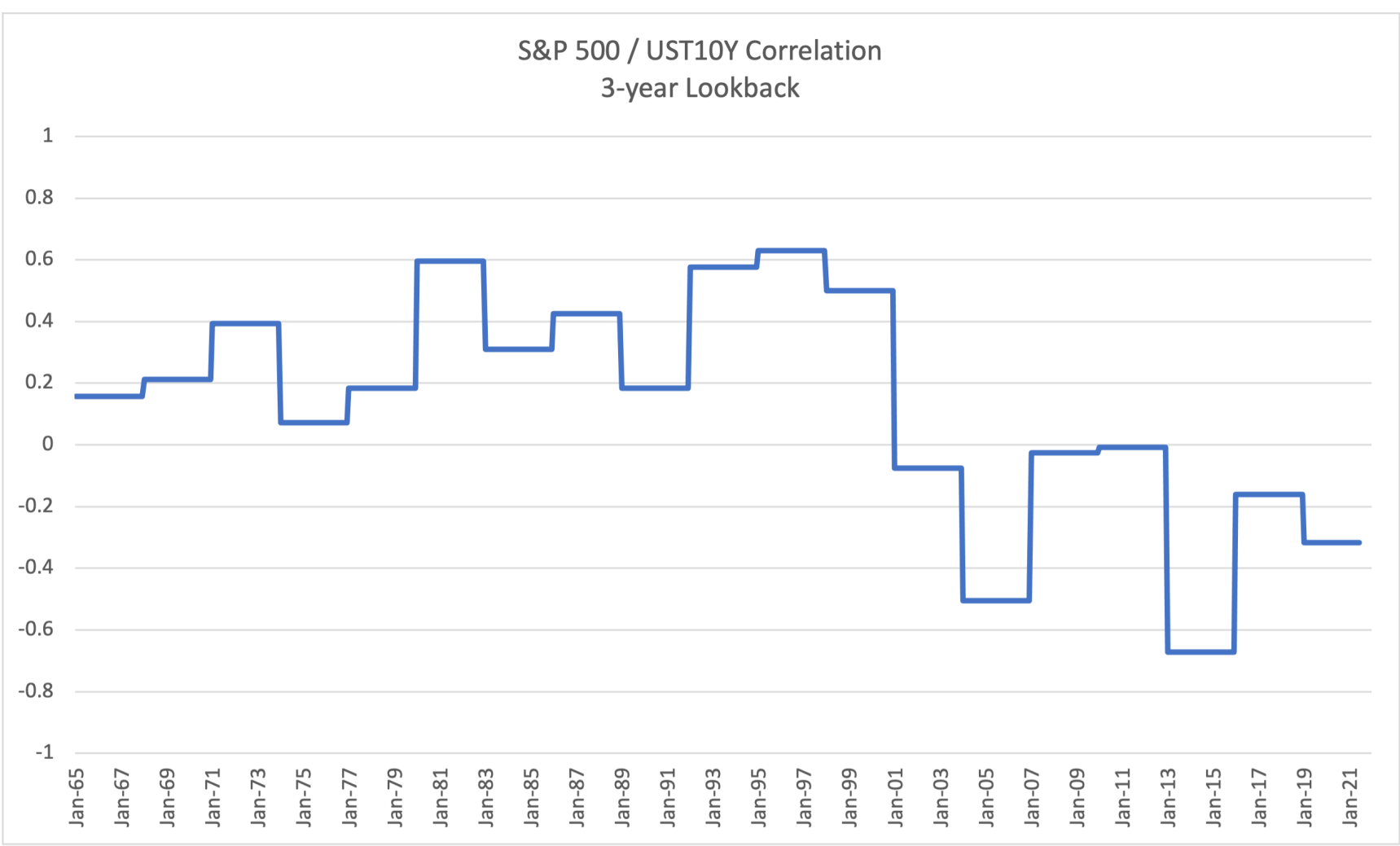

Figure 29.4: *Lookback correlation levels.*

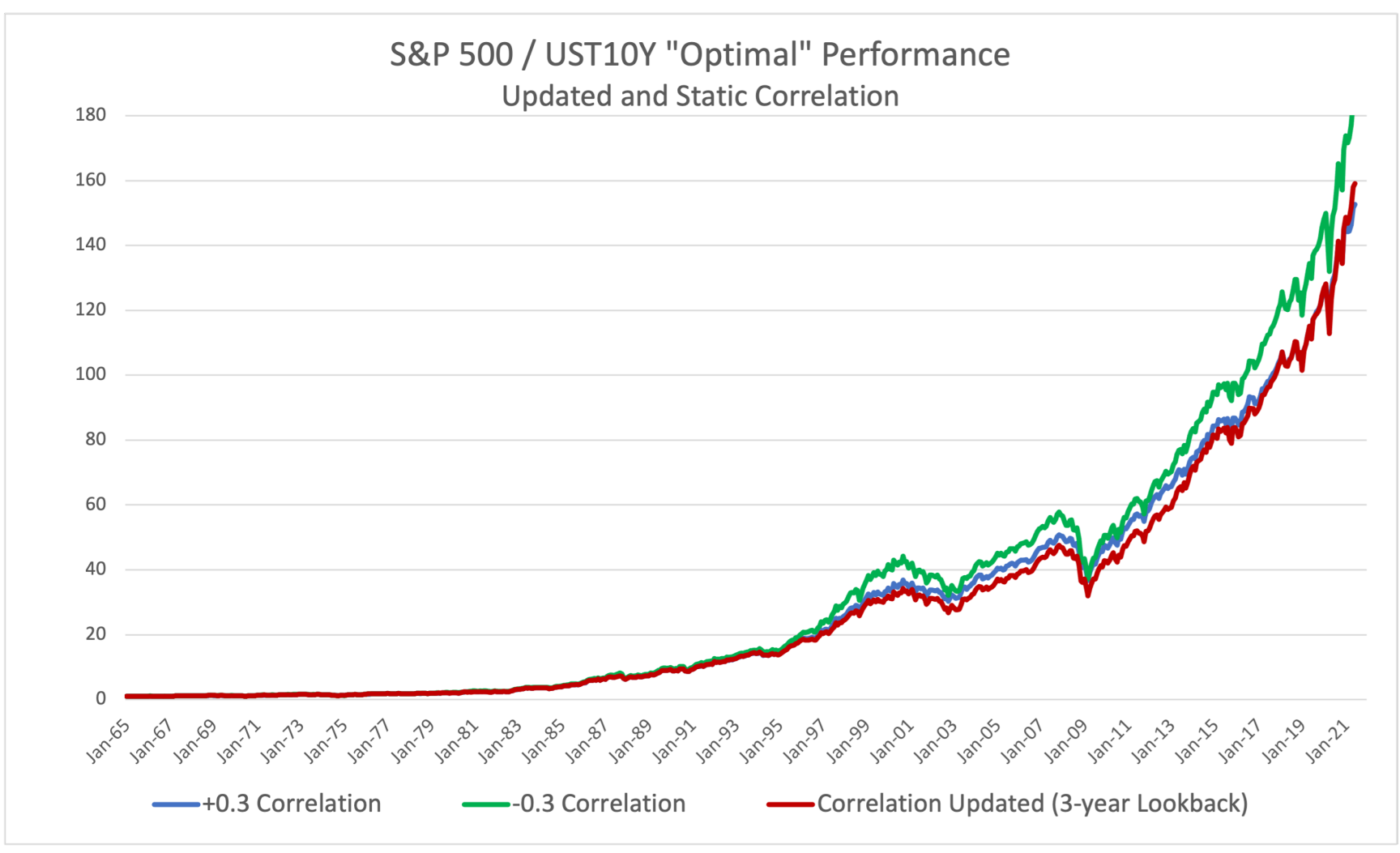

Figure 29.5: *"Optimal Performance"*

We believe that the deficiencies of MPT and the reliance on correlation estimates have been responsible for tepid performance of pension funds since the Global Financial Crisis. It is never possible to reliably construct the "optimal" portfolio that will deliver the best possible risk-adjusted return. Even if it were possible, it would not be beneficial. Following this course limits portfolio volatility in benign market environments over the short term while making huge sacrifices in long-run performance. The so-called "optimal" portfolio is in effect the worst of all worlds. It offers scant protection against tail risk and, at the same time, achieves an under-allocation to the riskier assets with higher returns in the long periods of economic expansion such as the past decade.

## TECHNICAL APPENDIX

### What is the failure of Ellipticality?

From the standard definition, [104], $\mathbf{X}$, a $p \times 1$ random vector is said to have an elliptical (or elliptical contoured) distribution with location parameters $\mu$, a non-negative matrix $\Sigma$, and some scalar function $\Psi$ if its characteristic function $\varphi$ is of the form

$$\varphi(t) = \exp(it'\mu)\Psi(t\Sigma t'). \tag{29.1}$$

There are equivalent definitions focusing on the density; consider that the central attribute is that $\Psi$ is a function of *a single* covariance matrix $\Sigma$.

The main property of the class of elliptical distribution is that it is closed under linear transformation. Intuitively, it means (in a bivariate situation) that tail effects are less likely to come from one than two marginal deviations.

This closure under linear transformation leads to attractive properties in the building of portfolios, and in the results of portfolio theory (in fact one cannot have portfolio theory without ellipticality of distributions). For under ellipticality, all portfolios can be characterized completely by their location and scale and any two portfolios with identical location and scale (in return space) have identical distributions returns.

Accordingly, ellipticality (under the condition of finite variance) allows the extension of the results of modern portfolio theory (MPT) under the so-called "non-normality", an argument initally discovered by [216], also see [141]. However it appears (from those of us who work with stochastic correlations) that returns are not elliptical by any conceivable measure, see Chicheportiche and Bouchaud [46] and simple visual graphs of stability of correlation as in Fig.1.

In other words, the characteristic function under stochastic correlation becomes:

$$\varphi(t) = \sum_{j}^{n} \omega_j \exp(it'\mu)\Psi(t\Sigma_j t'). \tag{29.2}$$

where $\Sigma_j$ represents different covariance matrices indexed by the state $j$, and $\omega_j$ are the weights of the distributions under consideration mapping to the frequencies of states. So if, as we saw in Fig. 1, one needs to average across correlations, the condition in equation 29.2 fails and, given that market returns have fat tails, and there is no mathematical justification for MPT.

### Distinguishing between sample error and random correlation

Next we derive the distribution of the second moment of a sample of correlation coefficients under the null assumption that variations are due to sampling error off a constant correlation. We focus on non-overlapping observations from a sample size of 713 months of underlying pairs (of stocks and bonds).

Let $X$ and $Y$ be $n$ independent Gaussian variables centered to a mean 0 and scaled to a unitary variance. Let $\rho_n(.)$ be the operator.

$$\rho_n(\tau) = \frac{X_\tau Y_\tau + X_{\tau+1}Y_{\tau+1} \ldots + X_{\tau+n-1}Y_{\tau+n-1}}{\sqrt{(X_\tau^2 + X_{\tau+1}^2 \ldots + X_{\tau+n-1}^2)(Y_\tau^2 + Y_{\tau+1}^2 \ldots + Y_{\tau+n-1}^2)}}. \tag{29.3}$$

First, we consider the distribution of the Pearson correlation for $n$ observations of pairs assuming $\mathbb{E}(\rho) \approx 0$ (the mean is of small relevance as we are focusing on the second moment, with tracking of $O\left(\frac{1}{n^2}\right)$):

$$f_n(\rho) = \frac{\left(1-\rho^2\right)^{\frac{n-4}{2}}}{B\left(\frac{1}{2}, \frac{n-2}{2}\right)}, \tag{29.4}$$

with characteristic function:

$$\chi_n(\omega) = 2^{\frac{n-1}{2}-1} \omega^{\frac{3-n}{2}} \Gamma\left(\frac{n}{2} - \frac{1}{2}\right) J_{\frac{n-3}{2}}(\omega),$$

where $J_{(.)}(.)$ is the Bessel J function.

We can assert that, for $n$ sufficiently large:

$$2^{\frac{n-1}{2}-1} \omega^{\frac{3-n}{2}} \Gamma\left(\frac{n}{2} - \frac{1}{2}\right) J_{\frac{n-3}{2}}(\omega) \approx e^{-\frac{\omega^2}{2(n-1)}},$$

the corresponding characteristic function of the Gaussian.

Moments of order $p$ become:

$$M(p) = \frac{\left((-1)^p + 1\right) \Gamma\left(\frac{n}{2} - 1\right) \Gamma\left(\frac{p+1}{2}\right)}{2B\left(\frac{1}{2}, \frac{n-2}{2}\right) \Gamma\left(\frac{1}{2}(n+p-1)\right)} \tag{29.5}$$

where $B(.,.)$ is the Beta function. The standard deviation is $\sigma_n = \sqrt{\frac{1}{n-1}}$ and the kurtosis $\kappa_n = 3 - \frac{6}{n+1}$.

This allows us to treat the distribution of $\rho$ as Gaussian, and given infinite divisibility, derive the variation of the components, again of $O(\frac{1}{n^2})$ (hence simplify by using the second moment in place of the variance):

$$\rho_n \sim \mathcal{N}\left(0, \sqrt{\frac{1}{n-1}}\right),$$

To test how the second moment of the sample coefficient compares to that of a random series, and thanks to the assumption of a mean of 0, define the squares for nonoverlapping correlations:

$$\Delta_{n,m} = \frac{1}{m} \sum_{i=1}^{\lfloor m/n \rfloor} \rho_n^2(i\, n),$$

where $m$ is the sample size and $n$ is the correlation window.

Now we can show that:

$$\Delta_{n,m} \sim \mathcal{G}\left(\frac{p}{2}, \frac{2}{(n-1)p}\right),$$

where $p = \lfloor m/n \rfloor$ and $\mathcal{G}$ is the gamma distribution with PDF:

$$f(\Delta) = \frac{2^{-\frac{p}{2}} \left(\frac{1}{(n-1)p}\right)^{-\frac{p}{2}} \Delta^{\frac{p}{2}-1} e^{-\frac{1}{2}\Delta(n-1)p}}{\Gamma\left(\frac{p}{2}\right)},$$

and survival function:

$$S(\Delta) = Q\left(\frac{p}{2}, \frac{1}{2}\Delta(n-1)p\right),$$

which allows us to obtain p-values below, using $m = 714$ observations (and using the leading order $O(.)$:

| $n$ | Sample $\Delta_n$ | p-values $O(.)$ |
|---|---|---|
| 12 | 0.2049 | $10^{-8}$ |
| 15 | 0.1735 | $10^{-8}$ |
| 20 | 0.1535 | $10^{-9}$ |
| 25 | 0.1324 | $10^{-9}$ |
| 36 | 0.1370 | $10^{-13}$ |

Such low p-values exclude any controversy as to their effectiveness [278].

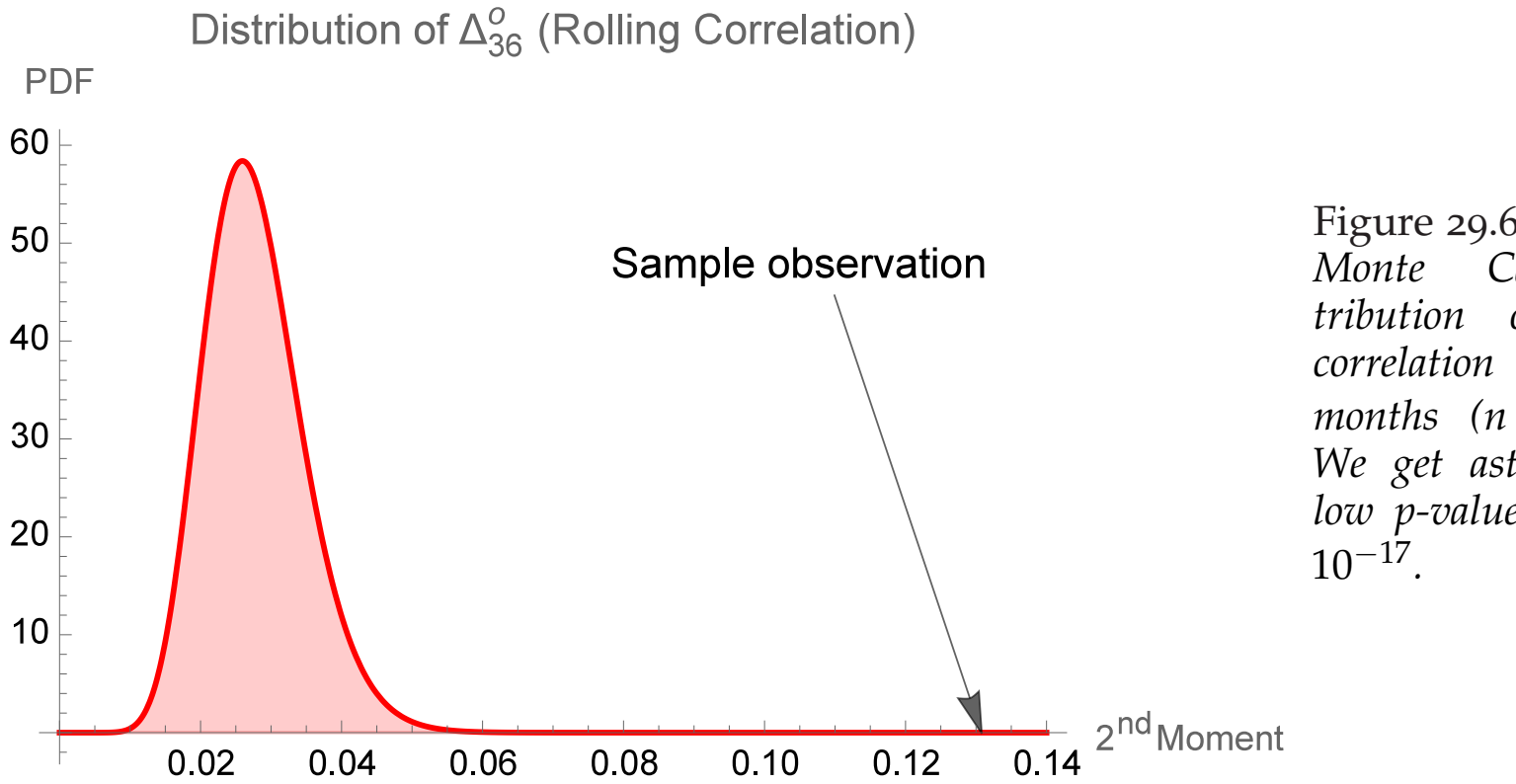

Figure 29.6: *The Monte Carlo distribution of rolling correlation for 36 months ($n = 10^6$). We get astonishingly low p-values of order $10^{-17}$.*

We can also compare rolling correlations using a Monte Carlo for the null with practically the same results (given the exceedingly low p-values). We simulate $\Delta^o_{n,m}$ with overlapping observations:

$$\Delta^o_{n,m} = \frac{1}{m} \sum_{i=1}^{m-n-1} \rho_n^2(i),$$

Rolling windows have the same second moment, but a mildly more compressed distribution since the observations of $\rho$ over overlapping windows of length $n$ are autocorrelated (with, we note, an autocorrelation between two observations $i$ orders apart of $\approx 1 - \frac{1}{n-i}$). As shown in Figure 29.6, for $n = 36$ we get exceedingly low p-values of order $10^{-17}$.

# 30 | TAIL RISK CONSTRAINTS AND MAXIMUM ENTROPY (W. D.& H. GEMAN)‡

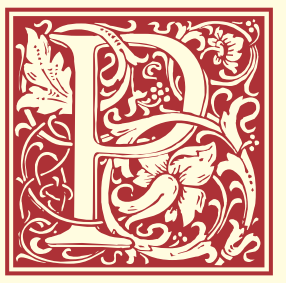

ORTFOLIO SELECTION in the financial literature has essentially been analyzed under two central assumptions: full knowledge of the joint probability distribution of the returns of the securities that will comprise the target portfolio; and investors' preferences are expressed through a utility function. In the real world, operators build portfolios under risk constraints which are expressed both by their clients and regulators and which bear on the maximal loss that may be generated over a given time period at a given confidence level (the so-called *V*alue at Risk of the position). Interestingly, in the finance literature, a serious discussion of how much or little is known from a probabilistic standpoint about the multi-dimensional density of the assets' returns seems to be of limited relevance.

Our approach in contrast is to highlight these issues and then adopt throughout a framework of entropy maximization to represent the real world ignorance of the "true" probability distributions, both univariate and multivariate, of traded securities' returns. In this setting, we identify the optimal portfolio under a number of downside risk constraints. Two interesting results are exhibited: (i) the left-tail constraints are sufficiently powerful to override all other considerations in the conventional theory; (ii) the "barbell portfolio" (maximal certainty/ low risk in one set of holdings, maximal uncertainty in another), which is quite familiar to traders, naturally emerges in our construction.

## 30.1 LEFT TAIL RISK AS THE CENTRAL PORTFOLIO CONSTRAINT

Customarily, when working in an institutional framework, operators and risk takers principally use regulatorily mandated tail-loss limits to set risk levels in their portfolios (obligatorily for banks since Basel II). They rely on stress tests, stop-losses, value at risk (VaR) , expected shortfall (–i.e., the expected loss conditional on the loss exceeding VaR, also known as CVaR), and similar loss curtailment methods, rather than utility. In particular, the margining of financial transactions is calibrated by clearing firms and exchanges on tail losses, seen both probabilistically and through stress testing. (In the risk-taking terminology, a stop loss is a mandatory order that attempts to terminate all or a portion of the exposure upon a trigger, a certain pre-defined nominal loss. Basel II is a generally used name for recommendations on banking laws and regulations issued by the Basel Committee on Banking Supervision. The Value-at-risk, VaR, is defined as a threshold loss value $K$ such that the probability that the loss on the portfolio over the given time horizon exceeds this value is $\epsilon$. A stress test is an examination of the performance upon an arbitrarily set deviation in the underlying variables.) The information embedded in the choice of the constraint is, to say the least, a meaningful statistic about the appetite for risk and the shape of the desired distribution.

‡ Research chapter.

Operators are less concerned with portfolio variations than with the drawdown they may face over a time window. Further, they are in ignorance of the joint probability distribution of the components in their portfolio (except for a vague notion of association and hedges), but can control losses organically with allocation methods based on maximum risk. (The idea of substituting variance for risk can appear very strange to practitioners of risk-taking. The aim by Modern Portfolio Theory at lowering variance is inconsistent with the preferences of a rational investor, regardless of his risk aversion, since it also minimizes the variability in the profit domain –except in the very narrow situation of certainty about the future mean return, and in the far-fetched case where the investor can only invest in variables having a symmetric probability distribution, and/or only have a symmetric payoff. Stop losses and tail risk controls violate such symmetry.) The conventional notions of utility and variance may be used, but not directly as information about them is embedded in the tail loss constaint.

Since the stop loss, the VaR (and expected shortfall) approaches and other risk-control methods concern only one segment of the distribution, the negative side of the loss domain, we can get a dual approach akin to a portfolio separation, or "barbell-style" construction, as the investor can have opposite stances on different parts of the return distribution. Our definition of barbell here is the mixing of two extreme properties in a portfolio such as a linear combination of maximal conservatism for a fraction $w$ of the portfolio, with $w \in (0, 1)$, on one hand and maximal (or high) risk on the $(1 - w)$ remaining fraction.

Historically, finance theory has had a preference for parametric, less robust, methods. The idea that a decision-maker has clear and error-free knowledge about the distribution of future payoffs has survived in spite of its lack of practical and theoretical validity –for instance, correlations are too unstable to yield precise measurements. It is an approach that is based on distributional and parametric certainties, one that may be useful for research but does not accommodate responsible risk taking. (Correlations are unstable in an unstable way, as joint returns for assets are not elliptical, see Bouchaud and Chicheportiche (2012) [46].)

There are roughly two traditions: one based on highly parametric decision-making by the economics establishment (largely represented by Markowitz [195]) and the other based on somewhat sparse assumptions and known as the Kelly criterion (Kelly, 1956 [170], see Bell and Cover, 1980 [18].) (In contrast to the minimum-variance approach, Kelly's method, developed around the same period as Markowitz, requires no joint distribution or utility function. In practice one needs the ratio of expected profit to worst-case return dynamically adjusted to avoid ruin. Obviously, model error is of smaller consequence under the Kelly criterion: Thorp (1969) [303], Haigh (2000) [139], Mac Lean, Ziemba and Blazenko [186]. For a discussion of the differences between the two approaches, see Samuelson's objection to the Kelly criterion and logarithmic sizing in Thorp 2010 [305].) Kelly's method is also related to left-tail control due to proportional investment, which automatically reduces the portfolio in the event of losses; but the original method requires a hard, nonparametric worst-case scenario, that is, securities that have a lower bound in their variations, akin to a gamble in a casino, which is something that, in finance, can only be accomplished through binary options. The Kelly criterion, in addition, requires some precise knowledge of future returns such as the mean. Our approach goes beyond the latter method in accommodating more uncertainty about the returns, whereby an operator can only control his left-tail via derivatives and other forms of insurance or dynamic portfolio construction based on stop-losses. (Xu, Wu, Jiang, and Song (2014) [321] contrast mean variance to maximum entropy and uses entropy to construct robust portfolios.) In a nutshell, we hardwire the curtailments on loss but otherwise assume maximal uncertainty about the returns. More precisely, we equate the return distribution with the *m*aximum entropy extension of constraints expressed as statistical expectations on the left-tail behavior as well as on the expectation of the return or log-return in the non-danger zone. (Note that we use *S*hannon entropy throughout. There are other information measures, such as Tsallis entropy [309] , a generalization of Shannon entropy, and Renyi entropy [163], some of which may be more convenient computationally in special cases. However, Shannon entropy is the best known and has a well-developed maximization framework.)

Here, the "left-tail behavior" refers to the hard, explicit, institutional constraints discussed above. We describe the shape and investigate other properties of the resulting so-called *m*axent distribution. In addition to a mathematical result revealing the link between acceptable tail loss (VaR) and the expected return in the Gaussian mean-variance framework, our contribution is then twofold: 1) an investigation of the shape of the distribution of returns from portfolio construction under more natural constraints than those imposed in the mean-variance method, and 2) the use of stochastic entropy to represent residual uncertainty.

VaR and CVaR methods are not error free –parametric VaR is known to be ineffective as a risk control method on its own. However, these methods can be made robust using constructions that, upon paying an insurance price, no longer depend on parametric assumptions. This can be done using derivative contracts or by organic construction (clearly if someone has 80% of his portfolio in numéraire securities, the risk of losing more than 20% is zero independent from all possible models of returns, as the fluctuations in the numéraire are not considered risky). We use "pure robustness" or both VaR and zero shortfall via the "hard stop" or insurance, which is the special case in our paper of what we called earlier a "barbell" construction.

It is worth mentioning that it is an old idea in economics that an investor can build a portfolio based on two distinct risk categories, see Hicks (1939) [150].

Modern Portfolio Theory proposes the mutual fund theorem or "separation" theorem, namely that all investors can obtain their desired portfolio by mixing two mutual funds, one being the riskfree asset and one representing the optimal mean-variance portfolio that is tangent to their constraints; see Tobin [306], Markowitz[196], and the variations in Merton[199], Ross d d com[239]. In our case a riskless asset is the part of the tail where risk is set to exactly zero. Note that the risky part of the portfolio needs to be minimum variance in traditional financial economics; for our method the exact opposite representation is taken for the risky one.

### 30.1.1 The Barbell as seen by E.T. Jaynes

Our approach to constrain only what can be constrained (in a robust manner) and to maximize entropy elsewhere echoes a remarkable insight by E.T. Jaynes in "How should we use entropy in economics?" [160]:

> "It may be that a macroeconomic system does not move in response to (or at least not solely in response to) the forces that are supposed to exist in current theories; it may simply move in the direction of increasing entropy as constrained by the conservation laws imposed by Nature and Government."

## 30.2 REVISITING THE MEAN VARIANCE SETTING

Let $\vec{X} = (X_1, ..., X_m)$ denote $m$ asset returns over a given single period with joint density $g(\vec{x})$, mean returns $\vec{\mu} = (\mu_1, ..., \mu_m)$ and $m \times m$ covariance matrix $\Sigma$: $\Sigma_{ij} = \mathbb{E}(X_i X_j) - \mu_i \mu_j, 1 \leq i, j \leq m$. Assume that $\vec{\mu}$ and $\Sigma$ can be reliably estimated from data.

The return on the portolio with weights $\vec{w} = (w_1, ..., w_m)$ is then

$$X = \sum_{i=1}^{m} w_i X_i,$$

which has mean and variance

$$\mathbb{E}(X) = \vec{w}\vec{\mu}^T, \quad V(X) = \vec{w}\Sigma\vec{w}^T.$$

In standard portfolio theory one minimizes $V(X)$ over all $\vec{w}$ subject to $\mathbb{E}(X) = \mu$ for a fixed desired average return $\mu$. Equivalently, one maximizes the expected return $\mathbb{E}(X)$ subject to a fixed variance $V(X)$. In this framework variance is taken as a substitute for risk.

To draw connections with our entropy-centered approach, we consider two standard cases:

(1) **N**ormal World: The joint distribution $g(\vec{x})$ of asset returns is multivariate Gaussian $N(\vec{\mu}, \Sigma)$. Assuming normality is equivalent to assuming $g(\vec{x})$ has maximum (Shannon) entropy among all multivariate distributions with the given first- and second-order statistics $\vec{\mu}$ and $\Sigma$. Moreover, for a fixed mean $\mathbb{E}(X)$, minimizing the variance $V(X)$ is equivalent to minimizing the entropy (uncertainty) of $X$. (This is true since joint normality implies that $X$ is univariate normal for any choice of weights and the entropy of a $\mathcal{N}(\mu, \sigma^2)$ variable is $H = \frac{1}{2}(1 + \log(2\pi\sigma^2))$.) This is natural in a world with complete information. ( The idea of entropy as mean uncertainty is in Philippatos and Wilson (1972) [221]; see Zhou –et al. (2013) [325] for a review of entropy in financial economics and Georgescu-Roegen (1971) [125] for economics in general.)

(2) **U**nknown Multivariate Distribution: Since we assume we can estimate the second-order structure, we can still carry out the Markowitz program, –i.e., choose the portfolio weights to find an optimal mean-variance performance, which determines $\mathbb{E}(X) = \mu$ and $V(X) = \sigma^2$. However, we do not know the distribution of the return $X$. Observe that *a*ssuming $X$ is normally distributed $\mathcal{N}(\mu, \sigma^2)$ is equivalent to assuming the entropy of $X$ is *m*aximized since, again, the normal maximizes entropy at a given mean and variance, see [221].

Our strategy is to generalize the second scenario by replacing the variance $\sigma^2$ by two left-tail value-at-risk constraints and to model the portfolio return as the maximum entropy extension of these constraints together with a constraint on the overall performance or on the growth of the portfolio in the non-danger zone.

### 30.2.1 Analyzing the Constraints

Let $X$ have probability density $f(x)$. In everything that follows, let $K < 0$ be a normalizing constant chosen to be consistent with the risk-taker's wealth. For any $\epsilon > 0$ and $\nu_- < K$, the *v*alue-at-risk constraints are:

(1) **T**ail probability:

$$\mathbb{P}(X \leq K) = \int_{-\infty}^{K} f(x)\, \mathrm{d}x = \epsilon.$$

(2) **E**xpected shortfall (CVaR):

$$\mathbb{E}(X | X \leq K) = \nu_-.$$

Assuming (1) holds, constraint (2) is equivalent to

$$\mathbb{E}(X I_{(X \leq K)}) = \int_{-\infty}^{K} x f(x)\, \mathrm{d}x = \epsilon \nu_-.$$

Given the value-at-risk parameters $\theta = (K, \epsilon, \nu_-)$, let $\Omega_{var}(\theta)$ denote the set of probability densities $f$ satisfying the two constraints. Notice that $\Omega_{var}(\theta)$ is convex: $f_1, f_2 \in \Omega_{var}(\theta)$ implies $\alpha f_1 + (1 - \alpha) f_2 \in \Omega_{var}(\theta)$. Later we will add another constraint involving the overall mean.

## 30.3 REVISITING THE GAUSSIAN CASE

Suppose we assume $X$ is Gaussian with mean $\mu$ and variance $\sigma^2$. In principle it should be possible to satisfy the VaR constraints since we have two free parameters. Indeed, as shown below, the left-tail constraints determine the mean and variance; see Figure 30.1. However, satisfying the VaR constraints imposes interesting restrictions on $\mu$ and $\sigma$ and leads to a natural inequality of a "no free lunch" style.

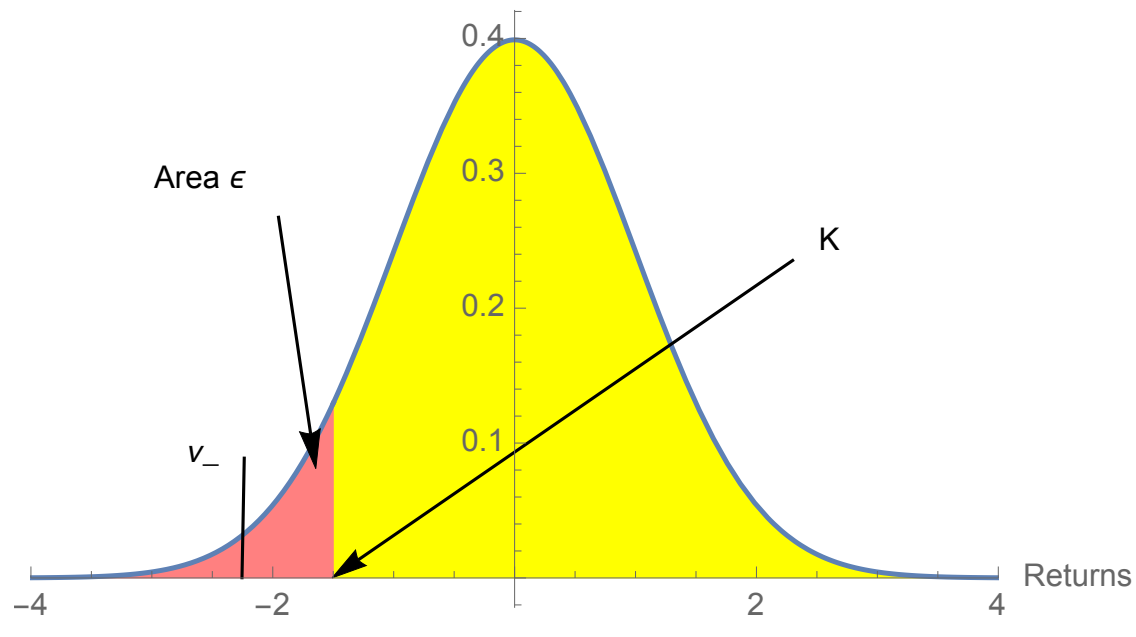

Figure 30.1: *By setting K (the value at risk), the probability ϵ of exceeding it, and the shortfall when doing so, there is no wiggle room left under a Gaussian distribution: σ and μ are determined, which makes construction according to portfolio theory less relevant.*

Let $\eta(\epsilon)$ be the $\epsilon$-quantile of the standard normal distribution, –i.e., $\eta(\epsilon) = \Phi^{-1}(\epsilon)$, where $\Phi$ is the c.d.f. of the standard normal density $\phi(x)$. In addition, set

$$B(\epsilon) = \frac{1}{\epsilon\eta(\epsilon)}\phi(\eta(\epsilon)) = \frac{1}{\sqrt{2\pi}\epsilon\eta(\epsilon)} \exp\{-\frac{\eta(\epsilon)^2}{2}\}.$$

**Proposition 30.1**
*If $X \sim N(\mu, \sigma^2)$ and satisfies the two VaR constraints, then the mean and variance are given by:*

$$\mu = \frac{\nu_- + KB(\epsilon)}{1 + B(\epsilon)}, \quad \sigma = \frac{K - \nu_-}{\eta(\epsilon)(1 + B(\epsilon))}.$$

*Moreover, $B(\epsilon) < -1$ and $\lim_{\epsilon \downarrow 0} B(\epsilon) = -1$.*

The proof is in the Appendix. The VaR constraints lead directly to two linear equations in $\mu$ and $\sigma$:

$$\mu + \eta(\epsilon)\sigma = K, \quad \mu - \eta(\epsilon)B(\epsilon)\sigma = \nu_-.$$

Consider the conditions under which the VaR constraints allow a *positive* mean return $\mu = \mathbb{E}(X) > 0$. First, from the above linear equation in $\mu$ and $\sigma$ in terms of $\eta(\epsilon)$ and $K$, we see that $\sigma$ increases as $\epsilon$ increases for any fixed mean $\mu$, and that $\mu > 0$ if and only if $\sigma > \frac{K}{\eta(\epsilon)}$, –i.e., we must accept a lower bound on the variance which increases with $\epsilon$, which is a reasonable property. Second, from the expression for $\mu$ in Proposition 1, we have

$$\mu > 0 \iff |\nu_-| > KB(\epsilon).$$

Consequently, the only way to have a positive expected return is to accommodate a sufficiently large risk expressed by the various tradeoffs among the risk parameters $\theta$ satisfying the inequality above. (This type of restriction also applies more generally to symmetric distributions since the left tail constraints impose a structure on the location and scale. For instance, in the case of a Student T distribution with scale $s$, location $m$, and tail exponent $\alpha$, the same linear relation between $s$ and $m$ applies: $s = (K - m)\kappa(\alpha)$, where $\kappa(\alpha) = -\frac{i\sqrt{I_{2\epsilon}^{-1}\left(\frac{\alpha}{2},\frac{1}{2}\right)}}{\sqrt{\alpha}\sqrt{I_{2\epsilon}^{-1}\left(\frac{\alpha}{2},\frac{1}{2}\right)-1}}$, where $I^{-1}$ is the inverse of the regularized incomplete beta function $I$, and $s$ the solution of $\epsilon = \frac{1}{2} I_{\frac{\alpha s^2}{(k-m)^2+\alpha s^2}}\left(\frac{\alpha}{2}, \frac{1}{2}\right)$.

### 30.3.1 A Mixture of Two Normals

In many applied sciences, a mixture of two normals provides a useful and natural extension of the Gaussian itself; in finance, the Mixture Distribution Hypothesis (denoted as MDH in the literature) refers to a mixture of two normals and has been very widely investigated (see for instance Richardson and Smith (1995) [233]). H. Geman and T. Ané (1996) [4] exhibit how an infinite mixture of normal distributions for stock returns arises from the introduction of a "stochastic clock" accounting for the uneven arrival rate of information flow in the financial markets. In addition, option traders have long used mixtures to account for fat tails, and to examine the sensitivity of a portfolio to an increase in kurtosis ("DvegaDvol"); see Taleb (1997) [271]. Finally, Brigo and Mercurio (2002) [38] use a mixture of two normals to calibrate the skew in equity options.

Consider the mixture

$$f(x) = \lambda N(\mu_1, \sigma_1^2) + (1 - \lambda) N(\mu_2, \sigma_2^2).$$

An intuitively simple and appealing case is to fix the overall mean $\mu$, and take $\lambda = \epsilon$ and $\mu_1 = \nu_-$, in which case $\mu_2$ is constrained to be $\frac{\mu - \epsilon\nu_-}{1-\epsilon}$. It then follows that the left-tail constraints are approximately satisfied for $\sigma_1, \sigma_2$ sufficiently small. Indeed, when $\sigma_1 = \sigma_2 \approx 0$, the density is effectively composed of two spikes (small variance normals) with the left one centered at $\nu_-$ and the right one centered at at $\frac{\mu - \epsilon\nu_-}{1-\epsilon}$. The extreme case is a Dirac function on the left, as we see next.

**Dynamic Stop Loss, A Brief Comment** One can set a level $K$ below which there is no mass, with results that depend on accuracy of the execution of such a stop. The distribution to the right of the stop-loss no longer looks like the standard Gaussian, as it builds positive skewness in accordance to the distance of the stop from the mean. We limit any further discussion to the illustrations in Figure 30.2.

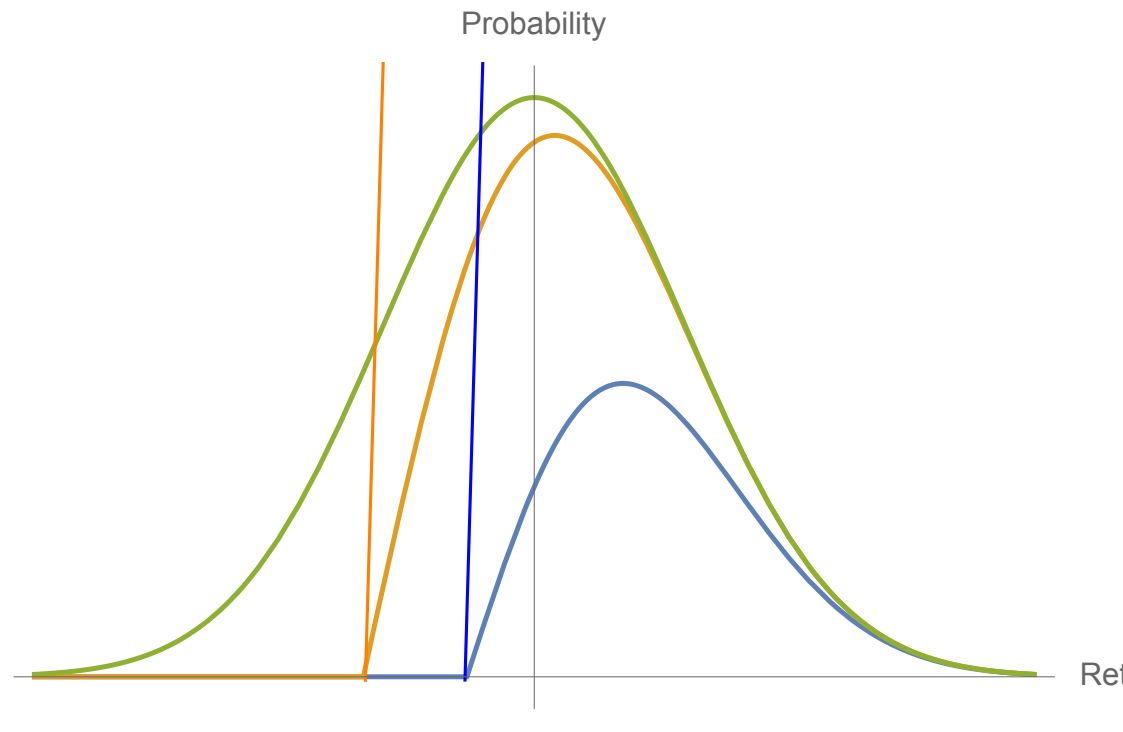

Figure 30.2: *A dynamic stop loss acts as an absorbing barrier, with a Dirac function at the executed stop.*

## 30.4 MAXIMUM ENTROPY

From the comments and analysis above, it is clear that, in practice, the density $f$ of the return $X$ is unknown; in particular, no theory provides it. Assume we can adjust the portfolio parameters to satisfy the VaR constraints, and perhaps another constraint on the expected value of some function of $X$ (e.g., the overall mean). We then wish to compute probabilities and expectations of interest, for example $\mathbb{P}(X > 0)$ or the probability of losing more than $2K$, or the expected return given $X > 0$. One strategy is to make such estimates and predictions under the most unpredictable circumstances consistent with the constraints. That is, use the *m*aximum entropy extension (MEE) of the constraints as a model for $f(x)$.

The "differential entropy" of $f$ is $h(f) = -\int f(x) \ln f(x)\, dx$. (In general, the integral may not exist.) Entropy is concave on the space of densities for which it is defined. In general, the MEE is defined as

$$f_{MEE} = \arg\max_{f \in \Omega} h(f)$$

where $\Omega$ is the space of densities which satisfy a set of constraints of the form $E\phi_j(X) = c_j, j = 1, ..., J$. Assuming $\Omega$ is non-empty, it is well-known that $f_{MEE}$ is unique and (away from the boundary of feasibility) is an exponential distribution in the constraint functions, –i.e., is of the form

$$f_{MEE}(x) = C^{-1} \exp\left(\sum_j \lambda_j \phi_j(x)\right)$$

where $C = C(\lambda_1, ..., \lambda_M)$ is the normalizing constant. (This form comes from differentiating an appropriate functional $J(f)$ based on entropy, and forcing the integral to be unity and imposing the constraints with Lagrange mult1ipliers.) In the special cases below we use this characterization to find the *MEE* for our constraints.

In our case we want to maximize entropy subject to the VaR constraints together with any others we might impose. Indeed, the VaR constraints alone do not admit an MEE since they do not restrict the density $f(x)$ for $x > K$. The entropy can be made arbitrarily large by allowing $f$ to be identically $C = \frac{1-\epsilon}{N-K}$ over $K < x < N$ and letting $N \to \infty$. Suppose, however, that we have adjoined one or more constraints on the behavior of $f$ which are compatible with the VaR constraints in the sense that the set of densities $\Omega$ satisfying all the constraints is non-empty. Here $\Omega$ would depend on the VaR parameters $\theta = (K, \epsilon, \nu_-)$ together with those parameters associated with the additional constraints.

### 30.4.1 Case A: Constraining the Global Mean

The simplest case is to add a constraint on the mean return, –i.e., fix $\mathbb{E}(X) = \mu$. Since $\mathbb{E}(X) = \mathbb{P}(X \leq K)\mathbb{E}(X|X \leq K) + \mathbb{P}(X > K)\mathbb{E}(X|X > K)$, adding the mean constraint is equivalent to adding the constraint

$$\mathbb{E}(X|X > K) = \nu_+$$

where $\nu_+$ satisfies $\epsilon\nu_- + (1-\epsilon)\nu_+ = \mu$.

Define

$$f_-(x) = \begin{cases} \frac{1}{(K-\nu_-)} \exp\left[-\frac{K-x}{K-\nu_-}\right] & \text{if } x < K, \\ 0 & \text{if } x \geq K. \end{cases}$$

and

$$f_+(x) = \begin{cases} \frac{1}{(\nu_+-K)} \exp\left[-\frac{x-K}{\nu_+-K}\right] & \text{if } x > K, \\ 0 & \text{if } x \leq K. \end{cases}$$

It is easy to check that both $f_-$ and $f_+$ integrate to one. Then

$$f_{MEE}(x) = \epsilon f_-(x) + (1-\epsilon) f_+(x)$$

is the MEE of the three constraints. First, evidently

1. $\int_{-\infty}^{K} f_{MEE}(x)\,\mathrm{d}x = \epsilon$;
2. $\int_{-\infty}^{K} x f_{MEE}(x)\,\mathrm{d}x = \epsilon\nu_-$;
3. $\int_{K}^{\infty} x f_{MEE}(x)\,\mathrm{d}x = (1-\epsilon)\nu_+$.

Hence the constraints are satisfied. Second, $f_{MEE}$ has an exponential form in our constraint functions:

$$f_{MEE}(x) = C^{-1} \exp\left[-(\lambda_1 x + \lambda_2 I_{(x \leq K)} + \lambda_3 x I_{(x \leq K)})\right].$$

The shape of $f_-$ depends on the relationship between $K$ and the expected shortfall $\nu_-$. The closer $\nu_-$ is to $K$, the more rapidly the tail falls off. As $\nu_- \to K$, $f_-$ converges to a unit spike at $x = K$ (Figures 30.3 and 30.4).

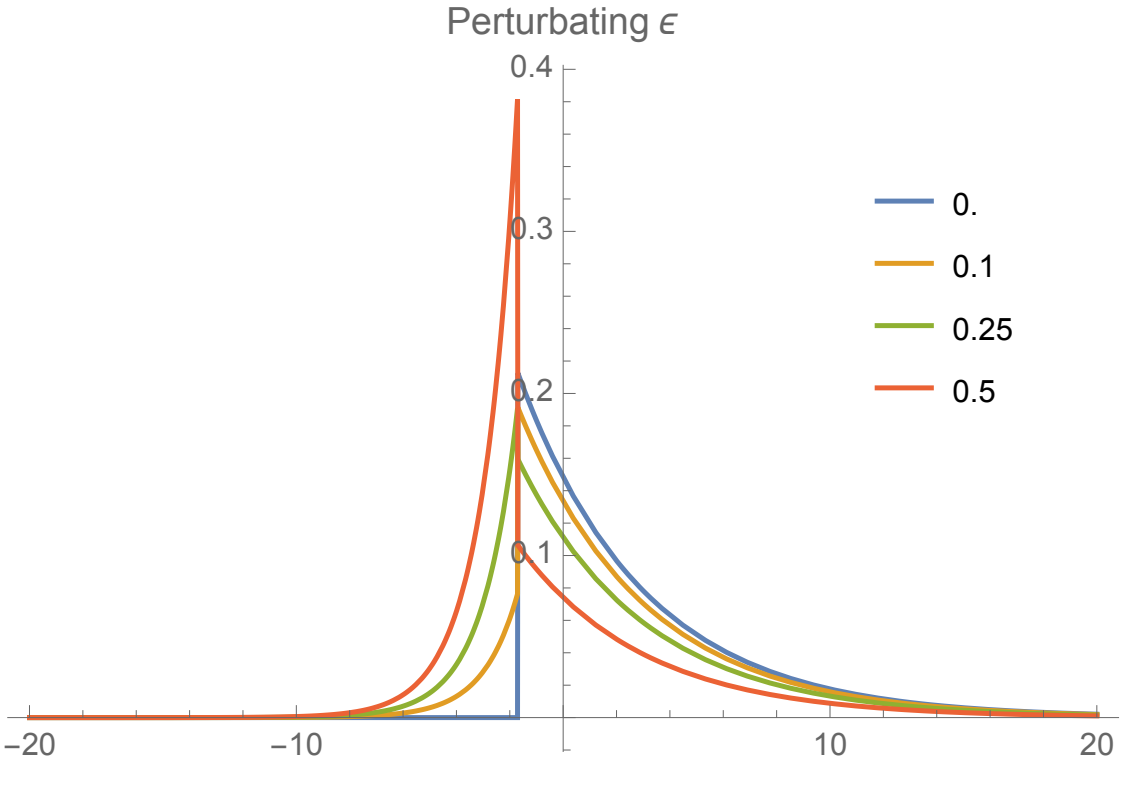

Figure 30.3: *Case A: Effect of different values of $\epsilon$ on the shape of the distribution.*

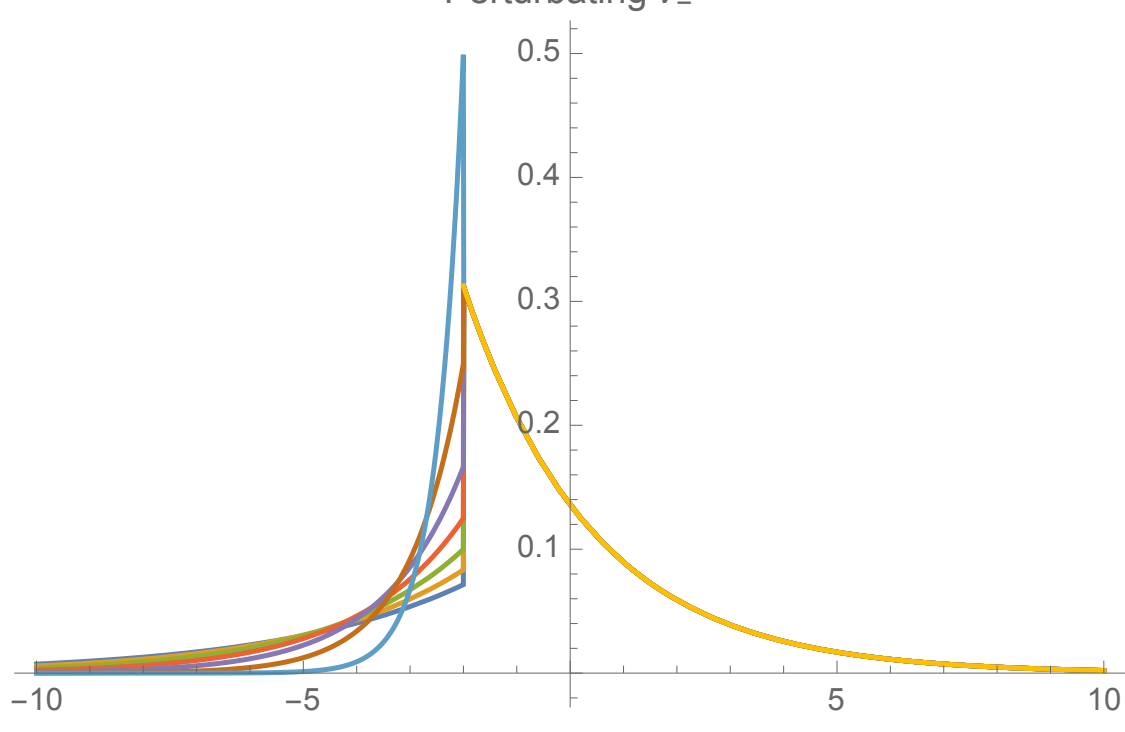

Figure 30.4: *Case A: Effect of different values of $\nu_-$ on the shape of the distribution.*

### 30.4.2 Case B: Constraining the Absolute Mean

If instead we constrain the absolute mean, namely

$$E|X|= \int |x| f(x)\, \mathrm{d}x = \mu,$$

then the MEE is somewhat less apparent but can still be found. Define $f_-(x)$ as above, and let

$$f_+(x) = \begin{cases} \frac{\lambda_1}{2-\exp(\lambda_1 K)} \exp(-\lambda_1 |x|) & \text{if } x \geq K, \\ 0 & \text{if } x < K. \end{cases}$$

Then $\lambda_1$ can be chosen such that

$$\epsilon \nu_- + (1-\epsilon) \int_K^\infty |x| f_+(x)\, \mathrm{d}x = \mu.$$

### 30.4.3 Case C: Power Laws for the Right Tail

If we believe that actual returns have "fat tails," in particular that the right tail decays as a Power Law rather than exponentially (as with a normal or exponential density), than we can add this constraint to the VaR constraints instead of working with the mean or absolute mean. In view of the exponential form of the MEE, the density $f_+(x)$ will have a power law, namely

$$f_+(x) = \frac{1}{C(\alpha)}(1+|x|)^{-(1+\alpha)}, x \geq K,$$

for $\alpha > 0$ if the constraint is of the form

$$E\left(\log(1+|X|)|X > K\right) = A.$$

Moreover, again from the MEE theory, we know that the parameter is obtained by minimizing the logarithm of the normalizing function. In this case, it is easy to show that

$$C(\alpha) = \int_K^{\infty}(1+|x|)^{-(1+\alpha)}\,\mathrm{d}x = \frac{1}{\alpha}(2-(1-K)^{-\alpha}).$$

It follows that $A$ and $\alpha$ satisfy the equation

$$A = \frac{1}{\alpha} - \frac{\log(1-K)}{2(1-K)^{\alpha}-1}.$$

We can think of this equation as determining the decay rate $\alpha$ for a given $A$ or, alternatively, as determining the constraint value $A$ necessary to obtain a particular Power Law $\alpha$.

The final MEE extension of the VaR constraints together with the constraint on the log of the return is then:

$$f_{MEE}(x) = \epsilon I_{(x \leq K)} \frac{1}{(K-\nu_-)}\exp\left[-\frac{K-x}{K-\nu_-}\right] + (1-\epsilon)I_{(x>K)}\frac{(1+|x|)^{-(1+\alpha)}}{C(\alpha)},$$

(see Figures 30.5 and 30.6).

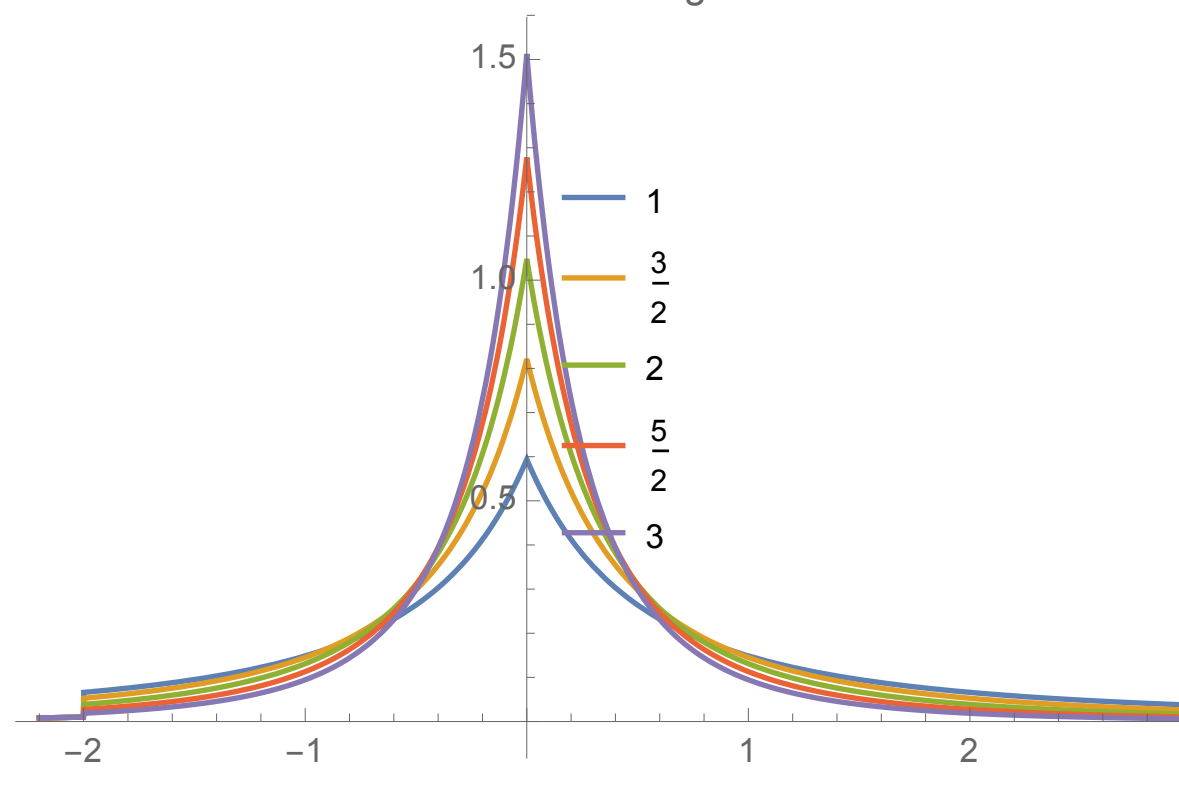

Figure 30.5: *Case C: Effect of different values of on the shape of the fat-tailed maximum entropy distribution.*

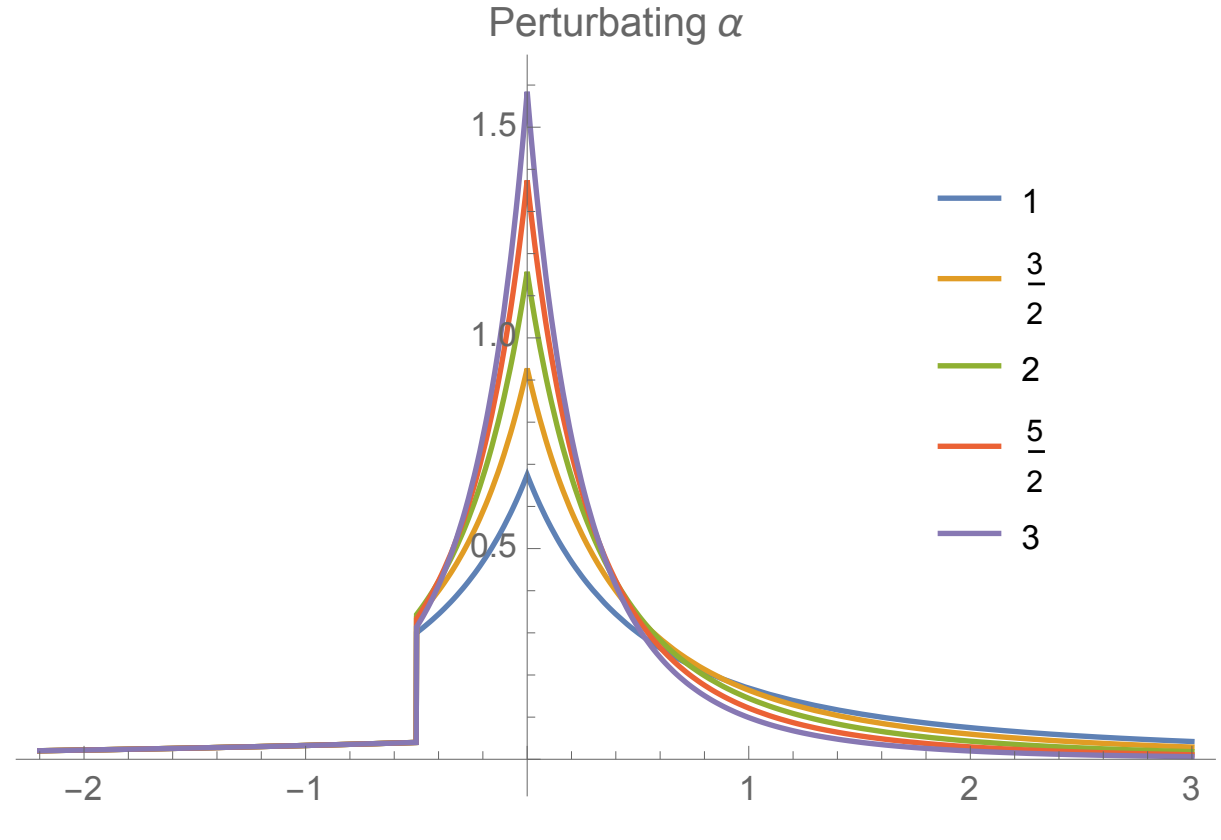

Figure 30.6: *Case C: Effect of different values of on the shape of the fat-tailed maximum entropy distribution (closer K).*

### 30.4.4 Extension to a Multi-Period Setting: A Comment

Consider the behavior in multi-periods. Using a naive approach, we sum up the performance as if there was no response to previous returns. We can see how Case A approaches the regular Gaussian, but not Case C (Figure 30.7).

For case A the characteristic functioncan be written:

$$\Psi^A(t) = \frac{e^{iKt}(t(K - \nu_- \epsilon + \nu_+(\epsilon - 1)) - i)}{(Kt - \nu_- t - i)(-1 - it(K - \nu_+))}$$

So we can derive from convolutions that the function $\Psi^A(t)^n$ converges to that of an $n$-summed Gaussian. Further, the characteristic function of the limit of the average of strategies, namely

$$\lim_{n \to \infty} \Psi^A(t/n)^n = e^{it(\nu_+ + \epsilon(\nu_- - \nu_+))}, \tag{30.1}$$

is the characteristic function of the Dirac delta, visibly the effect of the law of large numbers delivering the same result as the Gaussian with mean $\nu_+ + \epsilon(\nu_- - \nu_+)$ .

As to the Power Law in Case C, convergence to Gaussian only takes place for $\alpha \geq 2$, and rather slowly.

## 30.5 COMMENTS AND CONCLUSION

We note that the stop loss plays a larger role in determining the stochastic properties than the portfolio composition. Simply, the stop is not triggered by individual components, but by variations in the total portfolio. This frees the analysis from focusing on individual portfolio components when the tail –via derivatives or organic construction– is all we know and can control.

To conclude, most papers dealing with entropy in the mathematical finance literature have used minimization of entropy as an optimization criterion. For instance, Fritelli (2000) [116] exhibits the unicity of a "minimal entropy martin-gale measure" under some conditions and shows that minimization of entropy is equivalent to maximizing the expected exponential utility of terminal wealth. We have, instead, and outside any utility criterion, proposed entropy maximization as the recognition of the uncertainty of asset distributions. Under VaR and Expected Shortfall constraints, we obtain in full generality a "barbell portfolio" as the optimal solution, extending to a very general setting the approach of the two-fund separation theorem.

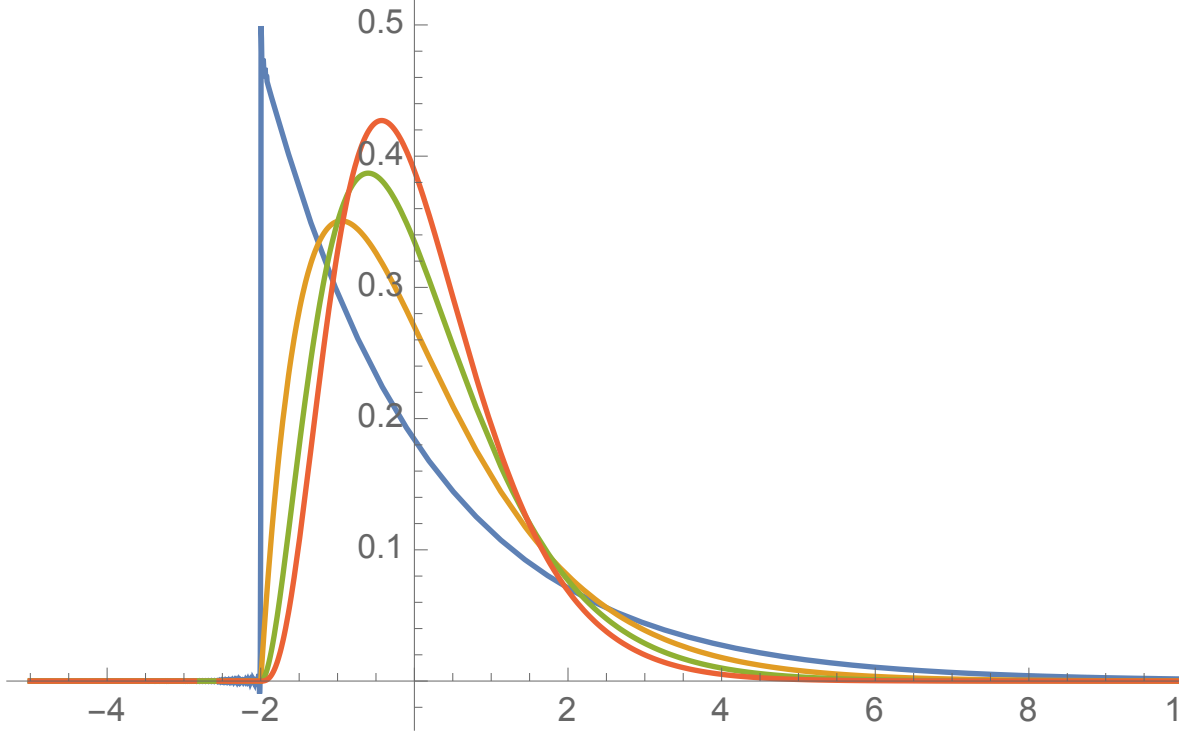

Figure 30.7: *Average return for multiperiod naive strategy for Case A, that is, assuming independence of "sizing", as position size does not depend on past performance. They aggregate nicely to a standard Gaussian, and (as shown in Equation (30.1)), shrink to a Dirac at the mean value.*

## 30.6 APPENDIX/PROOFS

Proof of Proposition 1: Since $X \sim N(\mu, \sigma^2)$, the tail probability constraint is

$$\epsilon = \mathbb{P}(X < K) = \mathbb{P}(Z < \frac{K-\mu}{\sigma}) = \Phi(\frac{K-\mu}{\sigma}).$$

By definition, $\Phi(\eta(\epsilon)) = \epsilon$. Hence,

$$K = \mu + \eta(\epsilon)\sigma \tag{30.2}$$

For the shortfall constraint,

$$\begin{aligned} \mathbb{E}(X; X < k) &= \int_{-\infty}^{K} \frac{x}{\sqrt{2\pi}\sigma} \exp -\frac{(x-\mu)^2}{2\sigma^2} \, dx \\ &= \mu\epsilon + \sigma \int_{-\infty}^{(K-\mu)/\sigma} x\phi(x) \, dx \\ &= \mu\epsilon - \frac{\sigma}{\sqrt{2\pi}} \exp -\frac{(K-\mu)^2}{2\sigma^2} \end{aligned}$$

Since, $\mathbb{E}(X; X < K) = \epsilon\nu_-$, and from the definition of $B(\epsilon)$, we obtain

$$\nu_- = \mu - \eta(\epsilon)B(\epsilon)\sigma \tag{30.3}$$

Solving (30.2) and (30.3) for $\mu$ and $\sigma^2$ gives the expressions in Proposition 1.

Finally, by symmetry to the "upper tail inequality" of the standard normal, we have, for $x < 0$, $\Phi(x) \leq \frac{\phi(x)}{-x}$. Choosing $x = \eta(\epsilon) = \Phi^{-1}(\epsilon)$ yields $\epsilon = \mathbb{P}(X < \eta(\epsilon)) \leq -\epsilon B(\epsilon)$ or $1 + B(\epsilon) \leq 0$. Since the upper tail inequality is asymptotically exact as $x \to \infty$ we have $B(0) = -1$, which concludes the proof.

# BIBLIOGRAPHY AND INDEX